\documentclass[seceqn]{elsart}
\usepackage{epsf,graphicx}
\usepackage{times,latexsym,euscript,amstext,amssymb,amsbsy,amsmath,amsfonts}
\usepackage{epsfig}
\usepackage{slashbox}

\usepackage[normalem]{ulem} 
\usepackage[dvips]{color} 
\renewcommand\sout{\bgroup \color{red} \ULdepth=-.5ex \ULset}

\hoffset   =- .5 cm \textwidth = 16. cm
\def\beq{\begin{eqnarray}}
\def\eeq{\end{eqnarray}}
\def\bea{\begin{eqnarray}}
\def\eea{\end{eqnarray}}
\def\beqa{\begin{eqnarray}\begin{array}{l}}
\def\eeqa{\end{array}\end{eqnarray}}

\def\barr{\left(\begin{array}{c}}
\def\earr{\end{array}\right)}
\def\bmat{\left(\begin{array}{cc}}
\def\emat{\end{array}\right)}

\def\mathscr{\mathcal}

\def\3d{3-D}

\newcommand{\norm}{\frac{2}{(2\pi)^3}}
\newcommand{\qd}[1]{\left[ #1 \right]}
\newcommand{\td}[1]{\left( #1 \right)}

\newcommand{\itg}[1]{\norm \int d^3k f_{#1}(k)}

\newcommand{\itau}{\itg{\tau}g(k)}

\newcommand{\rz}{\rho_{_0}}

\newcommand{\umd}[1]{ \td{ \frac{1}{2}+x_{#1} } }

\newcommand{\inew}[1]{\mathcal{I}_#1 }

\newcommand{\Msun}{$\mathrm{M}_{\odot}$}

\def\ld{\lambda}
\def\Ld{\Lambda}

\def\sg{\sigma}

\def\pp{\partial}
\def\r{\rho}

\begin{document}
\begin{frontmatter}

\title{Recent Progress and New Challenges in Isospin Physics with Heavy-Ion
Reactions}

\author[TAMUc]{Bao-An~Li}\footnote{Email:Bao-An$\_$Li@Tamu-Commerce.edu},
\author[SJTU]{Lie-Wen~Chen}\footnote{Email:Lwchen@Sjtu.edu.cn},
\author[TAMU]{Che~Ming~Ko}\footnote{Email:Ko@Comp.tamu.edu}

\address[TAMUc]{Department of Physics, Texas A\&M University-Commerce, Commerce,
Texas 75429-3011, USA}

\address[SJTU]{Institute of Theoretical Physics, Shanghai Jiao Tong University,
Shanghai 200240, China}

\address[TAMU]{Cyclotron Institute and Physics Department, Texas A\&M University,
College Station, Texas 77843-3366, USA}

\begin{abstract}
The ultimate goal of studying isospin physics via heavy-ion
reactions with neutron-rich, stable and/or radioactive nuclei is to
explore the isospin dependence of in-medium nuclear effective
interactions and the equation of state of neutron-rich nuclear
matter, particularly the isospin-dependent term in the equation of
state, i.e., the density dependence of the symmetry energy. Because
of its great importance for understanding many phenomena in both
nuclear physics and astrophysics, the study of the density
dependence of the nuclear symmetry energy has been the main focus of
the intermediate-energy heavy-ion physics community during the last
decade, and significant progress has been achieved both
experimentally and theoretically. In particular, a number of
phenomena or observables have been identified as sensitive probes to
the density dependence of the nuclear symmetry energy. Experimental
studies have confirmed some of these interesting isospin-dependent
effects and allowed us to constrain relatively stringently the
symmetry energy at sub-saturation densities. The impacts of this
constrained density dependence of the symmetry energy on the
properties of neutron stars have also been studied, and they were
found to be very useful for the astrophysical community. With new
opportunities provided by the various radioactive beam facilities
being constructed around the world, the study of isospin physics is
expected to remain one of the forefront research areas in nuclear
physics. In this report, we review the major progress achieved
during the last decade in isospin physics with heavy ion reactions
and discuss future challenges to the most important issues in this
field.
\end{abstract}
\end{frontmatter}
\tableofcontents


\section{Introduction}
\label{introduction}

Besides the many radioactive beam facilities that already exist in
the world, a number of next-generation radioactive beam facilities
are being constructed or planned. At these facilities, nuclear
reactions involving nuclei with large neutron or proton excess can
be studied, thus providing a great opportunity to study both the
structure of rare isotopes and the properties of isospin asymmetric
nuclear matter that has a large neutron to proton ratio. This has
stimulated much interest and a lot of activities in a new research
direction in nuclear physics, namely the isospin physics. Many
extensive reviews on the nuclear structure aspect of this exciting
new field can be found in the literature. Complementary to the
nuclear structure studies but being equally important and exciting
are reaction studies with radioactive beams. In this review, we
focus on the reaction aspect of isospin physics, especially
heavy-ion reactions induced by neutron-rich beams at intermediate
energies. The ultimate goal of this branch of isospin physics is to
determine the isospin dependence of the in-medium nuclear effective
interactions and the equation of state (EOS) of isospin asymmetric
nuclear matter, particularly its isospin-dependent term, i.e., the
density dependence of the nuclear symmetry energy. The latter has
been identified explicitly as one of the most outstanding questions
in the 2007 US Nuclear Physics Long Range Plan by the NSF/DOE's
Nuclear Science Advisory Committee \cite{DOE07}. A number of earlier
reviews on isospin physics with heavy-ion reactions can be found in,
e.g., Refs.~\cite{LiBA98,LiBA01b,Dan02a,Lat04,Bar05,Ste05a}.
Impressive progress has since been made both experimentally and
theoretically. With the new opportunity provided by next-generation
radioactive beam facilities, it is timely to review the recent
progress and discuss new challenges in this rapidly growing field.

Knowledge on the nuclear symmetry energy is essential for
understanding not only many problems in nuclear physics, such as the
dynamics of heavy-ion collisions induced by radioactive beams and
the structure of exotic nuclei, but also a number of important
issues in astrophysics, such as the nucleosynthesis during
pre-supernova evolution of massive stars and the cooling of
protoneutron stars. Although the nuclear symmetry energy at normal
nuclear matter density is known to be around $30$ MeV from the
empirical liquid-drop mass formula \cite{Mey66,Pom03}, its values at
other densities, especially at supra-normal densities, are poorly
known \cite{LiBA98,LiBA01b}. This is in contrast to our knowledge on
the symmetric part of the nuclear EOS. Through the efforts in both
the nuclear structure and the heavy-ion reaction community for over
three decades \cite{Dan02a}, the incompressibility of symmetric
nuclear matter at its saturation density $\rho _{0}\approx 0.16$
fm$^{-3}$ has been determined to be $K_0=231\pm 5 $ MeV from nuclear
giant monopole resonances (GMR) \cite{You99}, and the EOS of nuclear
matter at densities $2\rho _{0}<\rho <5\rho _{0}$ has also been
constrained by measurements of collective flows \cite{Dan02a} and of
subthreshold kaon production \cite{Fuc06a} in relativistic
nucleus-nucleus collisions. As pointed out in
Refs.~\cite{Dan02a,Pie05,Col05}, remaining uncertainties in the
determination of both the $K_0$ and the EOS of symmetric nuclear
matter are mainly related to those in the density dependence of the
nuclear symmetry energy. It is important to mention that there are
many interesting studies in the literatures on the surface symmetry
energy of finite nuclei as well as its relation to the bulk symmetry
energy and the structure of rare isotopes, see, e.g., Refs.~
\cite{Ste05a,Sat01,Dan03,Die05,Rei06,Ban06,Die07}. In the present
review, however, we will focus on the recent progress and new
challenges in determining the density dependence of the bulk nuclear
symmetry energy in neutron-rich nuclear matter with heavy-ion
reactions.

Theoretical studies of the EOS of isospin asymmetric nuclear matter
were pioneered by Brueckner \textit{et al.} \cite{Bru67} and many
others in the late 1960's. Since then, there have been many studies
on this subject based on different many-body theories using various
two-body and three-body forces or interaction Lagrangians. However,
because of our poor knowledge about the isospin dependence of the
in-medium nucleon-nucleon interactions and the difficulties in
solving the nuclear many-body problems, predictions on the EOS of
isospin asymmetric nuclear matter based on various many-body
theories differ widely at both low and high densities
\cite{Bom01,Die03}. On the other hand, heavy-ion reactions provide a
unique opportunity to investigate in terrestrial laboratories the
EOS of isospin asymmetric nuclear matter.  During the last decade,
there have been significant activities in exploring the isospin
asymmetric part of the EOS, namely, the density dependence of the
nuclear symmetry energy
\cite{LiBA98,LiBA01b,Bar05,LiBA95,Mul95,LiBA96,LiBA97a,LiBA97b,Che97,%
Che98,Bar98,Che99a,Zha99,LiBA00,Xu00,Che00,Zha00,Tsa01,Tan01a,Tan01b,%
LiBA01a,LiBA02,Bar02,Che03a,Che03b,Ono03,Liu03,Shi03,LiBA04a,LiBA04b,%
Riz04,Che04,Ono04,Gai04,LiBA05a,LiBA05b,LiBA05c,Zha05,LiQF05a,LiQF05b,%
LiBA05e,Tia05,Fer05,Yon07,Ylc07}.
Very impressively, some important discoveries of novel phenomena and
a significantly constrained nuclear symmetry energy at
sub-saturation densities have been obtained during this short time.

To extract information on the EOS of neutron-rich nuclear matter,
especially the density dependence of the nuclear symmetry energy,
from heavy-ion reactions induced by neutron-rich (stable and/or
radioactive) beams, one needs reliable theoretical tools. For this
purpose, it has been especially useful to have transport models that
include explicitly the isospin degrees of freedom and thus the
isospin-dependent physical quantities, such as the isovector
(symmetry) potential, and isospin-dependent in-medium
nucleon-nucleon (NN) cross sections and Pauli blocking.  Significant
progresses have been made during the past two decades in developing
semi-classical transport models for nuclear reactions. These
semi-classical transport models include mainly the following two
types: the Boltzmann-Uehling-Ulenbeck (BUU) model \cite{Ber88b} and
the quantum molecular dynamical (QMD) model \cite{Aic91}. While it
is important to develop practically implementable quantum transport
theories, applications of the semi-classical transport models have
enabled us to learn a great deal of interesting physics from
heavy-ion reactions, especially the EOS of symmetric nuclear matter.
With the development of radioactive ion beam physics, several rather
comprehensive isospin-dependent, but mostly semi-classical transport
models \cite{LiBA95,LiBA96,LiBA97a,Che98,Bar98,Riz04,LiBA05c,Zha05},
have been successfully developed in recent years to describe nuclear
reactions induced by neutron-rich nuclei at intermediate and high
energies.

Also, the identification of experimental observables that are
sensitive to the density dependence of the nuclear symmetry energy
is required to extract the properties of isospin asymmetric nuclear
matter from heavy-ion reactions induced by neutron-rich nuclei.
Since the symmetry potentials for neutrons and protons have opposite
signs and they are generally weaker than the nuclear isoscalar
potential at same density, most isospin sensitive observables are
usually based on differences or ratios of isospin multiplets of
baryons, mirror nuclei and/or mesons, such as the neutron/proton
ratio of emitted nucleons \cite{LiBA97a}, the neutron-proton
differential flow \cite{LiBA00}, the neutron-proton correlation
function \cite{Che03a}, the $t$/$^{3}$He \cite{Che03b,Zha05}, $\pi
^{-}/\pi ^{+}$ \cite{LiBA02,Gai04,LiBA05a,LiQF05b}, $\Sigma
^{-}/\Sigma ^{+}$ \cite{LiQF05a} and $K^{0}/K^{+}$ ratios
\cite{Fer05}, etc.. In addition, to reduce the systematical errors
and the effects of the Coulomb force which acts against the symmetry
potentials, double ratios and/or differences taken from several
reaction systems using different isotopes of the same element have
also been proposed \cite{LiBA06b,Yon06a,Yon06b}.

Among the many exciting results, it is of particular interest to
mention the recent isospin diffusion experiments at the National
Superconducting Cyclotron Laboratory (NSCL) at Michigan State
University and the associated theoretical analysis that have led
to a relatively stringent constraint on the nuclear symmetry
energy at subnormal densities \cite{LiBA05c,Tsa04,Che05a}. This
result has already had some important impacts on both nuclear
physics and astrophysics. For instance, within the Hatree-Fock
approach this has allowed one to exclude many popular Skyrme
density functionals used extensively in nuclear structure studies
\cite{Che05b}. Using the Skyrme interactions allowed by the
isospin diffusion data, the neutron-skin thickness in $^{208}$Pb
calculated within the Hartree-Fock approach was found to be
consistent with available experimental data
\cite{Che05b,Ste05b,LiBA06a}. Also, a rather consistent conclusion
regarding the symmetry energy at sub-saturation densities has been
reached using several complementary observables. In particular,
the symmetry energy most favored by the isospin diffusion data
coincides with that from a relativistic mean-field model using an
accurately calibrated parameter set that reproduces the giant
monopole resonances in both $^{90}$Zr and $^{208}$Pb, and the
isovector giant dipole resonance of $^{208}$Pb \cite{Yon07,Tod05}.
It further agrees with the symmetry energy recently obtained from
the isoscaling analyses of the isotope ratios in
intermediate-energy heavy ion collisions \cite{She07}. These
different but complementary studies have provided so far the best
phenomenological constraints on the symmetry energy at sub-normal
densities. However, the situation at supra-normal densities is
very different. Widely different high density behaviors of the
symmetry energy have been predicted using different many-body
theories with various interactions. On the other hand, a number of
experimental observables, that are sensitive to the high density
behavior of the symmetry energy in heavy-ion reactions induced by
high energy radioactive beams, have been identified using
transport model simulations. Unfortunately, essentially no
experimental information about the symmetry energy at higher
densities is available at present. Nevertheless, high energy
radioactive beam facilities under construction at the CSR/China
\cite{CSR}, FAIR/Germany \cite{FAIR}, RIKEN/Japan \cite{Yan07},
SPIRAL2/GANIL in France \cite{SPIRAL2}, and the planned Facility
for Rare Isotope Beams (FRIB) in the USA \cite{RIA} give us the
great hope that the high density behavior of the symmetry energy
can be studied experimentally in the near future.

Assuming that the nuclear effective interactions used in the
transport model to study the isospin diffusion data are valid within
a broad density range, the density dependence of the symmetry energy
constrained at sub-saturation densities has been used in studying
several global properties of neutron stars.  Especially, it has
allowed one to constrain significantly the radii and cooling
mechanisms of neutron stars as well as the possible changing rate of
the gravitational constant G \cite{LiBA06a,Kra07a}. Of course,
explicit constraints on the high density behavior of the symmetry
energy from nuclear reactions with high energy radioactive beams
will more tightly restrict these predictions and advance further our
understanding of the neutron stars.

Besides the density dependence of the symmetry energy and the
underlying isovector nucleon-nucleon interaction, there are also
other interesting novel phenomena in isospin asymmetric nuclear
matter, which may offer further opportunity to better understand
the properties of neutron-rich nuclear matter. Of particularly
interesting to the heavy-ion community are the special features of
the liquid-gas (LG) phase transition in dilute isospin asymmetric
nuclear matter. For instance, the order of the LG phase transition
in asymmetric matter might be different than that in symmetric
matter \cite{Mul95}. A new feature associated with the LG phase
transition in isospin asymmetric nuclear matter is the isospin
fractionation, namely, an unequal partition of the system's
isospin asymmetry between the liquid and gas phases, with the low
density gas phase normally more neutron-rich than the liquid
phase. This has been found as a general phenomenon within
essentially all thermodynamical models and dynamic transport model
simulations \cite{Cho04}. The experimental manifestation of the
isospin fractionation in aymmetric nuclear matter has also been
unambiguously observed in several experiments \cite{Das05}. More
recently, the new concept of differential isospin fractionation as
a function of nucleon momentum was introduced \cite{LiBA07a}. The
fine structure in the differential isospin fractionation can
reveal some novel features of the LG phase transition in isospin
asymmetric nuclear matter. Depending on the momentum dependence of
the symmetry potential, it is possible to have a transition from
the normal isospin fractionation for low energy nucleons to an
opposite isospin fractionation, i.e., the gas phase is less
neutron-rich than the liquid phase, for more energetic nucleons.
It will be of great interest to study this phenomenon
experimentally to see if this prediction can be verified.

Isospin physics with heavy-ion reactions is a fast growing field,
there are many interesting studies in the literature. In this
article, major progresses in several selected areas of isospin
physics as outlined above will be reviewed. Specifically, various
theoretical predictions on the EOS of asymmetric nuclear matter
are reviewed in Chapter~\ref{chapter_eos}. They are followed by
discussions on the momentum dependence of the nucleon isovector
potential in Chapter~\ref{chapter_ria}. In
Chapter~\ref{chapter_rmf}, we review the predictions from various
relativistic mean-field models on the isospin-dependent bulk and
single-particle properties in asymmetric nuclear matter. The
properties of asymmetric nuclear matter at finite temperature are
then reviewed in Chapter~\ref{chapter_temperature}, while the
in-medium NN cross sections are discussed in
Chapter~\ref{chapter_crosssection}. In
Chapter~\ref{chapter_observables}, we give an extensive review on
experimental observables that have been proposed for studying the
properties of asymmetric nuclear matter. The role of nuclear
symmetry energy on our understanding of the neutron skin thickness
of neutron-rich nuclei and the properties of neutron stars are
reviewed in Chapters~\ref{chapter_neutronskin} and
\ref{chapter_neutronstars}, respectively. Finally, we present in
Chapter~\ref{chapter_summary} a brief summary of recent
accomplishments in understanding the nuclear symmetry energy and
future challenges in isospin physics with radioactive nuclear
beams.

Because of the limitations of our knowledge and the scope of this
article, it is impossible for us to cover all topics in isospin
physics with heavy-ion reactions. We apologize to those whose work
may have not been cited here.


\section{The equation of state of isospin-asymmetric
nuclear matter}
\label{chapter_eos}

The EOS of isospin asymmetric nuclear matter is a longstanding
problem in both nuclear physics and astrophysics, and has received
much attention in the past. Because of the development of
radioactive beam facilities around the world during the last decade,
which make it possible to study experimentally the properties of
nuclear matter or nuclei under the extreme condition of large
isospin asymmetry in terrestrial laboratories, there has been a
surge of research activities on this problem. Theoretically, since
the early work pioneered especially by Brueckner \textit{et al.}
\cite{Bru67} and many others in the late 1960's, various approaches
involving different physical approximations and numerical techniques
have been developed to deal with the many-body problem of
isospin-asymmetric nuclear matter. These approaches can be roughly
classified into three categories: the microscopic many-body
approach, the effective-field theory approach, and the
phenomenological approach. Instead of discussing the details of
these theoretical approaches, we provide here a brief introduction
to each of them. In particular, we concentrate on the predictions of
these approaches on the most important, common features of the EOS
of isospin-asymmetric nuclear matter.

\subsection{Microscopic and phenomenological many-body approaches}

\subsubsection{The microscopic many-body approach}

In the microscopic many-body approach, the nuclear many-body problem
is treated microscopically using the bare NN interactions obtained
from fitting the experimental NN scattering phase shifts and
deuteron properties, and the empirical three-nucleon (3N) forces.
During past few decades, significant progress has been achieved in
the development of the microscopic many-body approach and its
applications in nuclear physics. These microscopic many-body
approaches mainly include the non-relativistic
Brueckner-Hartree-Fock (BHF) approach
\cite{Sjo74,Cug87,Bom91,Zuo02}, the relativistic
Dirac-Brueckner-Hartree-Fock (DBHF) approach
\cite{Bro84,Hor87,Mut87,Har87,Bro90,Sum92,Hub93,Seh97,Fuc98,Jon98,Gro99,Fuc04,Ma04,Sam05a},
the self-consistent Green's function (SCGF) approach
\cite{Die03,Mut00,Dew02,Car03,Dic04}, and the variational many-body
(VMB) approach \cite{Fri81,Lag81,Wir88a,Akm98,Muk07}.

Among the many non-relativistic microscopic many-body approaches,
the non-relativistic BHF approach is particularly suited for
nuclear systems. As described in many review articles in the
literature, see, e.g., Refs. \cite{Bal99,Lom01,Bal07}, this
approach has been extensively used to study homogeneous cold/hot
nuclear matter. The non-relativistic BHF approach can be
interpreted as a mean-field theory in the lowest order
non-relativistic Brueckner-Bethe-Goldstone (BBG) theory. The
latter is based on the linked cluster expansion of the
ground-state energy by means of the G-matrix, which plays the role
of the in-medium effective NN interaction (in-medium two-body
scattering matrix) and renormalizes the short-range nuclear
repulsion. The perturbation expansion of the energy per particle
in this approach can be ordered according to the number of hole
lines in the corresponding diagrams, and it shows a rapid
convergence at low densities. The diagrams with a given number $n$
of hole lines describe the $n$-particle correlations in the
system. At the two hole-line approximation, the corresponding
summation of diagrams leads to the BHF approximation that
incorporates in an exact way the two-particle correlations.
Another essential ingredient of the BHF approximation is that it
includes a self-consistent procedure for determining the
single-particle auxiliary potential. The convergence of the
hole-line expansion was proved by Day and Wiringa \cite{Day85}
within the framework of the BBG theory with the `standard choice'
or `gap choice' for the single-particle auxiliary potential, which
assumes that the auxiliary potential is zero above the Fermi
momentum. In the BHF approach, the definition of the
single-particle auxiliary potential is, however, not unique.
Although the final result of a hypothetically exact BBG
calculation is independent of the auxiliary single-particle
potential, the convergence rate depends on the particular choice
adopted in the calculations. Besides the gap choice of the
auxiliary potential, there is another popular choice, i.e., the
continuous choice \cite{Jeu76}, in which the definition of the
potential is extended to momenta larger than the Fermi momentum,
thus making the potential a continuous function through the Fermi
surface. During the last decade, an important progress in BBG
theory has been made by calculating the three hole-line
contributions which requires to solve the Faddeev equation for the
in-medium three-body problem, i.e., the Bethe-Faddeev equation
\cite{Son98}. The resulting equation of state of the nuclear
matter is found, to a high degree of accuracy, to be independent
of the choice of the auxiliary potential. Furthermore, it has been
shown that the continuous choice BHF calculations give results
much closer to those from the BBG calculations with the three
hole-line contributions than the gap choice BHF calculations
\cite{Bal01}, indicating that the continuous choice is more
optimal than the gap choice in BHF calculations. Nowadays, the
continuous choice is thus usually used \cite{LiZH06}.

In the self-consistent Green's function approach, the binding
energy as well as all single-particle observables in the nuclear
matter are calculated from the exact in-medium single-particle
propagator. The latter is obtained from the Dyson equation, where
medium effects are taken into account by the irreducible
self-energy that is obtained from an expansion in terms of the
effective interaction obtained from the sum of all ladder
diagrams. One important feature of the SCGF approach is that
particles and holes are treated on an equal footing, whereas in
BHF only intermediate particle ($k>k_{\mathrm{F}}$) states are
included in the ladder diagrams. This feature assures that the
thermodynamic consistency is satisfied in the SCGF approach, e.g.,
the Fermi energy or chemical potential of the nucleons equals the
binding energy at saturation (i.e., the Hugenholz-van Hove
theorem). For example, in a recent study with the SCGF approach
using the Reid 93 interaction \cite{Dew02}, it has been shown that
the Hugenholz--van Hove theorem is satisfied within less than $1$
MeV, which is in contrast with the BHF scheme where the Fermi
energy is more than $15$ MeV below the binding energy at
saturation. In the low-density limit, the BHF approach and the
SCGF approach coincide. As the density increases, the phase space
for hole-hole propagation is no longer negligible, and this leads
to an important repulsive effect on the total energy. Since
particle-particle (pp) and hole-hole (hh) ladders are treated in a
completely symmetrical way in the SCGF approach, the Green's
function scheme is also suited for calculations at higher
densities. Furthermore, the SCGF generates realistic spectral
functions, which can be used to evaluate the effective interaction
and corresponding nucleon self-energy. The spectral functions
include a depletion of the quasiparticle peak and the appearance
of the single-particle strength at other values of energy and
momentum, which is in contrast with the BHF approach where all the
single-particle strength is concentrated at the BHF energy. In
practice, however, it is not at all easy to calculate the
effective interaction with completely dressed Green's functions
since the corresponding dressed spectral functions show a
complicated energy dependence, containing both sharp peaks
(reflecting the quasiparticle behavior) and a broad background
distribution. At finite temperatures, calculations are somewhat
easier since the quasiparticle peaks acquire a considerable width,
due to thermal broadening, even close to the Fermi momentum. In
recent years, significant progress has been made in the SCGF
approach and its applications in nuclear matter calculations. A
recent review can be found in Ref. \cite{Dic04}.

Another popular non-relativistic microscopic many-body approach is
the variational approach \cite{Fri81,Lag81,Wir88a,Akm98,Muk07}. In
this method, the expectation value of the Hamiltonian is minimized
in the subspace of the Hilbert space that is spanned by a trial
many-body wave function of the form $\Psi =F\Phi $ with $\Phi $
denoting the wave function of non-interacting particles. The key
quantity in this method is thus the correlation function $F$. A
cluster expansion of the variational energy such as the hypernetted
chain expansion provides an upper bound for the ground-state energy
of a many-body system. Comparing to its applications in atomic and
molecular systems, where the Jastrow-like trial wave functions are
usually used, more complex correlation functions are needed in the
variational approach to the nuclear many-body problem due to the
complexity of the NN interaction. Many review papers exist in the
literature on the variational method and its extensive applications
to nuclear matter calculations, see, e.g. Refs. \cite{Pan79,Nav02}.

The microscopic many-body approaches mentioned above are all based
on the non-relativistic framework. The nuclear EOS and the
properties of nuclei have also been studied in relativistic
framework. The most successful and popular relativistic
microscopic many-body approach developed so far is the
Dirac-Brueckner approach, which is a relativistic extension of the
non-relativistic Brueckner theory. The formalism of this approach
is based on an effective quantum field theory for mesons and
nucleons, and the bare NN (ladder) interaction is described by a
meson-exchange model of nuclear potential (one boson exchange
potentials) while the single-particle motion is determined by the
in-medium Dirac equation. In the Dirac-Brueckner approach, the
in-medium T matrix, which serves as an effective in-medium
two-body interaction, is determined by a self-consistent summation
of the ladder diagrams in the quasipotential approximation
(Thompson equation) to the relativistic Bethe-Salpeter (BS)
equation, i.e., the relativistic counterpart of the
non-relativistic Bethe-Goldstone equation, see, e.g., Ref.
\cite{Bro76}. The in-medium T matrix thus contains all short-range
and many-body correlations in the ladder approximation. In solving
the BS equation, the Pauli principle is respected by projecting
the intermediate scattering states out of the Fermi sea. The
summation of the T matrix over the occupied states inside the
Fermi sea yields finally the self-energy in the Hartree-Fock
(DBHF) approximation. The Dirac--Brueckner approach was proposed
by the Brooklyn group \cite{Ana81,Ana83} in 1980's based on the
first-order perturbation theory. This was followed by a covariant
formulation of a self-consistent treatment of the Thompson
equation developed by Horowitz and Serot \cite{Hor84}, which is
discussed in detail in Ref. \cite{Hor87}. While the NN
interactions in these works were described within the framework of
the $\sigma $-$\omega $ model, calculations with realistic NN
interactions have been performed by Brockmann and Machleidt
\cite{Bro84,Bro90} and later by ter Haar and Malfliet
\cite{Har87,Har87b}. A more rigorous derivation of the Brueckner
approach in the framework of the relativistic Green's function can
be found in Refs. \cite{Mal88,Bot90,Jon91}. There exist several
excellent review articles in the literature, see, e.g., Refs.
\cite{Mac89,Bro99}.

The relativistic DBHF approach describes reasonably well both the
binding energy per nucleon and the saturation density of symmetric
nuclear matter. This is in distinct contrast with non-relativistic
microscopic many-body approaches, which usually yield either too
large a saturation density or too small a binding energy compared
to the empirical values. In particular, the saturation points
obtained from non-relativistic approaches using different bare NN
interactions are all located on the so-called Coester line in the
binding energy versus density plot \cite{Coe70}. Furthermore,
among non-relativistic approaches with same NN potential, the BHF
results differ from those of variational calculations. The reason
for this discrepancy can be understood from the fact that the BHF
approximation corresponds to the lowest-order term in the
hole-line expansion with the effective interaction calculated by
summing all particle-particle ladder diagrams. Neglecting
hole-hole propagation in the BHF approach might be valid at low
densities where the phase space for hole-hole propagation is much
smaller than for particle-particle propagation, higher orders in
the hole lines need to be included at higher densities. Indeed,
calculations including three-hole line contributions in the BHF
approach give results that \cite{Son98} agree reasonably well with
those from more advanced variational calculations
\cite{Day85,Akm98} and shift the saturation point off the Coester
line. Unfortunately, the shift is still not large enough for a
reproduction of the empirical saturation point. The SCGF approach
also gives a similar result. As pointed out in Ref. \cite{Bal07},
these deficiencies from the non-relativistic approaches are
evidently not due to the many-body treatment but to the adopted
non-relativistic Hamiltonian. To remedy these deficiencies from
the the non-relativistic approaches, one may need to further
consider many-body forces (to be distinguished from many-body
correlations), in particular three-body forces (TBF), and
relativistic effects. Actually, relativistic effects introduced
via the Dirac-Brueckner approach can be interpreted as due to a
particular three-body force \cite{Bro87}.

\begin{figure}[th]
\centering
\includegraphics[scale=1.2]{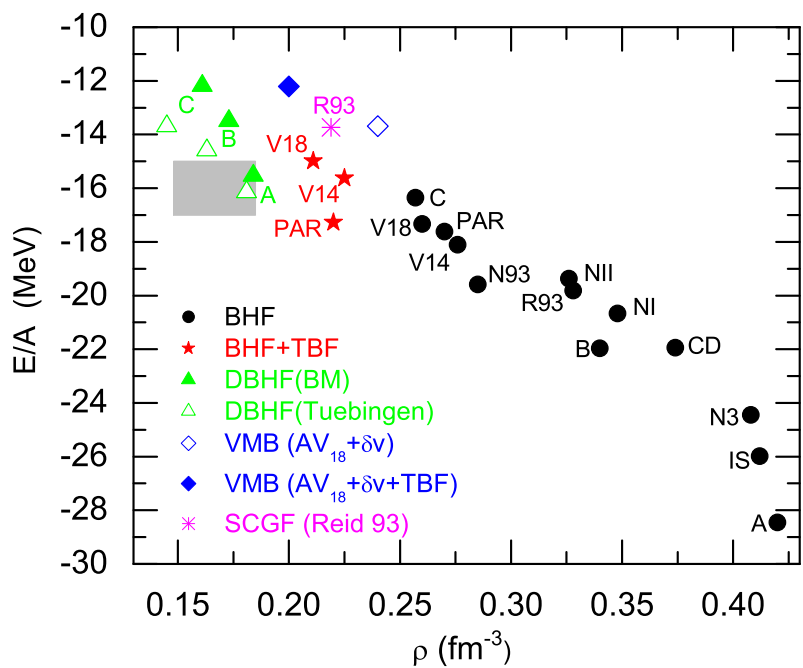}
\caption{(Color online) Saturation points of symmetric nuclear
matter from different microscopic many-body approaches with
various bare NN interaction potentials. Taken from Refs.
\cite{Bro90,Gro99,Dew02,Akm98,LiZH06}} \label{Coester}
\end{figure}

Shown in Fig. \ref{Coester} are the saturation points of symmetric
nuclear matter from different microscopic many-body approaches
with various bare NN interaction potentials. The results from the
BHF approach are taken from a recent work by Li \textit{et al.}
\cite{LiZH06} where a large set of modern and old NN potentials
are used in the continuous choice BHF calculations. Also included
are the results from the DBHF calculations by Brockmann and
Machleidt \cite{Bro90} and more recent calculations based on
improved techniques from Ref. \cite{Gro99} with Bonn potentials,
the VMB calculations based on the latest AV$_{\mathrm{18}}$
version of the Argonne V$_{\mathrm{18}}$ potential with boost
corrections ($\delta v$) from Ref. \cite{Akm98} as well as the
SCGF calculations with Reid93 potential from Ref. \cite{Dew02}.
One sees clearly that the saturation points from the
non-relativistic BHF, VMB, and SCGF approaches without including
TBF are located along a linear band a la the Coester line. The
results from the BHF approach including TBF with Paris, V14, and
V18 as well as the VMB approach including TBF with
AV$_{\mathrm{18}}$ (V18) become closer to the emiprical saturation
point ($E/A=-16\pm 1$ MeV and $\rho _{0}=0.148\sim 0.185$
fm$^{-3}$ corresponding to a Fermi momentum of
$k_{\mathrm{F}}=1.35\pm 0.05$ fm$^{-1}$ as indicated by the shaded
area in Fig. \ref{Coester}) but still significantly deviate from
the empirical point. On the other hand, the relativistic DBHF
calculations with Bonn A potential are seen to fit the empirical
value.

\subsubsection{The effective-field theory approach}

In the effective-field theory (EFT) approach, a systematic
expansion of the EOS of a many-body system in powers of its
density (the Fermi momentum) is obtained from an effective
interaction constructed using the effective-field theory. The
application of EFT as a many-body approach in nuclear structure
studies and nuclear matter calculations has become popular during
the last decade, and for a recent review, see, e.g., Refs.
\cite{Ser97,Fur04}. In nuclear physics, the EFT is based on a
perturbative expansion of the NN interaction or the nuclear mean
field within power-counting schemes, in which a separation of
scales is introduced such that an efficient systematic expansion
is carried out using ratios of these scales as expansion
parameters. In particular, the short-range correlations are
separated from the long- and intermediate-range parts of the NN
interaction with the division between `long' and `short'
characterized by the breakdown scale $\Lambda $ of the EFT. At
present, the effective-field theory approach in nuclear physics is
based on either the density functional theory (DFT)
\cite{Ser97,Fur04} or the chiral perturbation theory (ChPT)
\cite{Pra87,Lut00,Fin04,Vre04,Fri05,Fin06}.

In the DFT formulation of the relativistic nuclear many-body
problem, the Kohn-Sham density functional theory is implemented by
using the framework based on the Lorentz-covariant effective
quantum field theory to approximate the exact nuclear energy
density functional. As a result, the EFT provides the most general
way to parameterize observables that is consistent with the basic
principles of quantum mechanics, special relativity, unitarity,
gauge invariance, cluster decomposition, microscopic causality,
and the required internal symmetries. In Refs. \cite{Ser97,Fur04},
an effective chiral Lagrangian is constructed by including
explicitly the long-range dynamics while parameterizing the
short-range dynamics  generically through fitting experimental
data. The coefficients of the short-range terms in the effective
chiral Lagrangian may eventually be derived from QCD, but at
present they must be fitted by matching calculations with
experimental observables. Fixing these coefficients by fitting the
data implicitly includes short-range dynamics from many-nucleon
forces, fluctuations of the quantum vacuum, and hadron
substructure. Although the effective chiral Lagrangian contains,
in principle, an infinite number of terms, naive dimensional
analysis and naturalness allow one to identify suitable expansion
parameters and to estimate the relative sizes of various terms in
the Lagrangian. Thus, for any desired accuracy, the Lagrangian can
be truncated to a finite number of terms. In particular, it has
been shown that for normal nuclear systems it is possible to
expand the effective chiral Lagrangian systematically in powers of
the meson fields (and their derivatives) and to truncate the
expansion reliably after the first few orders \cite{Ser97,Fur04}.

The ChPT provides another way to treat the nuclear many-body
problem in EFT. In this approach, the long- and intermediate-range
interactions can be treated explicitly within chiral pion-nucleon
dynamics, which allows an expansion of the energy density
functional in powers of $m_{\pi }/M$ or in $k_{F}/M$, where
$m_\pi$ and $M$ are the pion and nucleon masses, respectively, and
$k_F$ is the Fermi momentum of the nuclear matter. On the other
hand, the short-range dynamics, as in the DFT formulation, is not
resolved explicitly but treated by counter-terms (dimensional
regularization) or through a cut-off regularization
\cite{Lut00,Vre04} with the parameters determined by fitting the
empirical saturation point of symmetric nuclear matter or
experimental data of finite nuclei. Significant progress has been
made recently in calculating the energy per particle of
isospin-symmetric nuclear matter from the chiral pion-nucleon
dynamics up to three-loop order by adjusting only one single
parameter (either a coupling or a cut-off $\Lambda $) related to
unresolved short-distance dynamics. As shown in Refs.
\cite{Lut00,Kai02}, the empirical saturation point of nuclear
matter can be reproduced correctly in the CHPT approach. Most
recently, this approach has been extended to study the
isospin-asymmetric nuclear matter by including the effects from
two-pion exchange with single and double virtual $\Delta
(1232)$-isobar excitation \cite{Fri05}. Regularization-dependent
short-range contributions from pion loops are encoded in a few
NN-contact coupling constants. The results indicate that the
isospin-dependent bulk and single-particle properties of
asymmetric nuclear matter are significantly improved by including
the chiral $\pi N\Delta $-dynamics, and they agree well with
sophisticated many-body calculations and (semi)-empirical values.

Since the EFT approach can be linked to low energy QCD and its
symmetry breaking, it has the advantage of having only a small
number of free parameters and a correspondingly higher predictive
power. However, in its present form the validity of this approach is
clearly confined to relatively small values of the Fermi momentum,
i.e., rather low densities.

\subsubsection{The phenomenological approach}

The phenomenological approach is based on effective
density-dependent nuclear forces or effective interaction
Lagrangians. In these approaches, a number of parameters are
adjusted to fit the properties of many finite nuclei and/or nuclear
matter. This type of models mainly includes the RMF theory
\cite{Hor87,Sum92,Ser86,Chi77,Gle82,Hir91,Sug94,Rei89,Rin96},
relativistic and non-relativistic Hartree-Fock approaches
\cite{Mil74,Bro78,Jam81,Hor83,Bou87,Lop88,Ber93,Wer94,Kho96,Vau72,Bra85,Sto07},
Thomas-Fermi approximations \cite{Bra85,Kol85,Ban90}, and
phenomenological potential models based on some particular energy
density functionals. These phenomenological approaches allow the
most precise description of the properties of finite nuclei and/or
nuclear matter. In particular, the non-relativistic Hartree-Fock
with Skyrme forces, i.e., the Skyrme-Hartree-Fock (SHF) method and
the RMF model constitute two main methods in the self-consistent
mean-field approach to nuclear structure studies, and for a review,
see, e.g., Ref. \cite{Ben03}

As a phenomenological approach, the RMF model has been very
successful in describing many nuclear phenomena
\cite{Ser97,Fur04,Ser86,Rei89,Rin96,Ben03,Wal74,Men06}. For example,
it provides a novel saturation mechanism for nuclear matter, an
explanation of the strong spin-orbit interaction in finite nuclei, a
natural energy dependence of the nucleon optical potential, and so
on. The RMF approach is based on effective interaction Lagrangians
with the nucleons interacting via exchanges of mesons. In this
approach, a number of parameters are adjusted to fit the properties
of many nuclei and thus allow the most precise description of the
properties of finite nuclei. Because this approach contains
parameters that are fixed by nuclear properties around the
saturation density, it thus usually gives an excellent description
of the nuclear properties around or below the saturation density.
Since proposed by Walecka more than $30$ years ago \cite{Wal74}, the
original Lagrangian in the RMF model has undergone many adjustments
and extensions, and has also resulted in extensive applications.
Currently, there are three most widely used versions of the RMF
model, namely, the nonlinear models \cite{Ser97,Ser86,Rei89,Rin96},
models with density-dependent meson-nucleon couplings
\cite{Fuc95,She97,Typ99,Hof01}, and point-coupling models
\cite{Fin04,Fin06,Nik92,Bur02,Mad04,Bur04}. For each version of the
RMF model, there are also many different parameter sets with
parameters chosen to fit the binding energies and charge radii of a
large number of nuclei in the periodic table. In particular, by
including isovector mesons in the effective interaction Lagrangians,
the RMF model has also been able to describe successfully the
properties of nuclei far away from the $\beta $-stability line.

In several very recent studies
\cite{Liu04,Ava06,Khv07,Jia07a,Jia07b}, the standard RMF model was
extended by considering the in-medium hadron properties governed
by the Brown-Rho (BR) scaling of hadron in-medium properties
\cite{Bro91} to mimic the chiral symmetry restoration at high
densities. In these studies, both hadron masses and the meson
coupling constants are density-dependent. In particular, the
parameter sets SLC and SLCd constructed in Refs.
\cite{Jia07a,Jia07b} lead to results that are consistent with
current experimental information, including the ground state
properties of finite nuclei. Also, the standard RMF model, which
has an incorrect high energy behavior of the nucleon optical
potential as a result of momentum/energy-independent nucleon self
energies, has been extended to include in the Lagrangian density
the couplings of the meson fields to the derivatives of nucleon
densities \cite{Typ03,Typ05} in order to rectify this deficiency.
The mementum/energy dependent nucleon self-energies can also be
introduced to the RMF model, which is based on the Hartree
approximation, by including the Fock exchange terms by means of
the relativistic Hartree-Fock (RHF) approximation
\cite{Mil74,Bro78,Jam81,Hor83,Bou87,Lop88,Ber93,Blu87,Nie01,Mar04,Lop05}.
The exchange terms further lead to contributions from the pion,
which are absent in the mean-field (Hartree) treatment of an
infinite, spin-saturated medium due to parity conservation. A
density-dependent RHF approach has recently also been developed
\cite{Lon06a,Lon06b,Lon07}, and it can describe the properties of
finite nuclei and nuclear matter comparable to those in standard
RMF models. As an extension of the RMF model to include the quark
degree of freedom, the quark-meson coupling (QMC) model, in which
quarks in non-overlapping nucleon bags interact self-consistently
with (structureless) isoscalar-scalar ($\sigma$) and
isoscalar-vector ($\omega$) mesons in the mean-field
approximation, has been developed \cite{Gui88}. For the most
recent review of the QMC model, see, e.g., Ref.~\cite{Sai07}.

For the non-relativistic Hartree-Fock approach, it has a very long
history. In particular, those with Skyrme \cite{Vau72,Sky56} (SHF)
or Gogny \cite{Dec80} forces have been very successful in describing
the ground-state and low-energy excitation properties of finite
nuclei and/or nuclear matter. As a self-consistent mean-field model,
the SHF method is based on effective energy-density functionals,
often formulated in terms of effective density-dependent
nucleon-nucleon interactions with parameters of the functional
adjusted to fit the experimental data. For the most recent review,
see, e.g., Ref. \cite{Sto07}.

The Thomas-Fermi (TF) approximation, which is based on the
semi-classical method, is useful for evaluating the smooth part of
the energy and has been used widely in atomic, nuclear and
metallic clusters physics. The semi-classical methods of the
TF-type are usually based on the Wigner-Kirkwood (WK) $\hbar $
expansion of the density matrix. Since the single-particle density
and the kinetic energy density can be expressed by means of
functionals of the one-body single-particle mean-field potential,
the $\hbar ^{2}$ or $\hbar ^{4}$ corrections to the lowest-order
TF term thus contain gradients of the one-body single-particle
mean-field potential of second or fourth order that arise from the
non-commutativity between the momentum and position operators.
These TF methods, like the liquid droplet or Strutinsky
calculations, smooth the quantal shell effects and give an
estimate of the average part of the HF energy
\cite{Bha71,Jen75,Mye96}. One of the most popular and successful
semi-classical TF-type approaches is the extended Thomas-Fermi
(ETF) method, which is based on the DFT and is developed together
with the use of the Skyrme forces. In the ETF approach, the WK
$\hbar$ expansion of the density is inverted to recast the kinetic
energy density as a functional of the local density and its
derivatives \cite{Bra85}. If the potential part of the energy
density is also a known functional of the local density as it
happens for the Skyrme forces, the approximate energy density
functional can be minimized to obtain an Euler-Lagrange equation.
The solution of this equation then provides the ground-state
particle density and energy. Although the quantum shell
oscillations are absent in the ETF model, the average densities
and energies are obtained with good accuracy. For the most recent
review, see, e.g., Ref.~\cite{Cen07}.

The potential model provides a simple and useful phenomenological
approach to study the properties of nuclear matter. It is usually
based on some particular energy density functionals with parameters
adjusted to reproduce results obtained from more microscopic
approaches or to fit the empirical properties of nuclear matter. One
typical example of the potential model is given in Ref.
\cite{Pra88b} where a momentum-dependent single-particle potential
is used, and the resulting EOS reproduces the results of microscopic
VMB approach \cite{Wir88a} and is further used to study the
properties of neutron stars. An important feature of the potential
model is that it can be easily and directly used in transport models
for heavy-ion collisions
\cite{Gal87,Pra88,Wel88,Gal90,Far91,Pan93,Zha94,Gre99,Dan00,Per02}.
In particular, an isospin- and momentum-dependent potential model
has been recently proposed based on an isospin- and
momentum-dependent MDI interaction, which is derived from the
Hartree-Fock approximation with a modified Gogny effective
interaction \cite{Das03}. In the MDI interaction, the potential
energy density $V(\rho ,\alpha )$ of an asymmetric nuclear matter at
total density $\rho $ and isospin asymmetry $\alpha $ is given by
\cite{Che05a,Das03}
\begin{eqnarray}
V(\rho ,\alpha ) &=&\frac{A_{u}\rho _{n}\rho _{p}}{\rho _{0}}+%
\frac{A_{l}}{2\rho _{0}}(\rho _{n}^{2}+\rho
_{p}^{2})+\frac{B}{\sigma +1}
\frac{\rho ^{\sigma +1}}{\rho _{0}^{\sigma }}(1-x\alpha^{2})\notag\\
&+& \frac{1}{\rho _{0}}\sum_{\tau ,\tau ^{\prime }}C_{\tau ,\tau
^{\prime }}\int \int d^{3}pd^{3}p^{\prime }\frac{f_{\tau
}(\vec{r},\vec{p})f_{\tau ^{\prime }}(\vec{r},\vec{p}^{\prime
})}{1+(\vec{p}-\vec{p}^{\prime })^{2}/\Lambda ^{2}}.  \label{MDIV}
\end{eqnarray}
In the above, $\tau =1/2$ ($-1/2$) is for neutrons (protons);
$\sigma =4/3$; $ f_{\tau }(\vec{r},\vec{p})$ is the nucleon
phase-space distribution function at coordinate $\vec{r}$ and
momentum $\vec{p}$; $ A_{u}$, $A_{l}$, $B$, $x$, $C_{\tau
,\tau^\prime }$, and $\Lambda $ are parameters.

In the mean-field approximation, Eq. (\ref{MDIV}) leads to the
following single-particle potential for a nucleon with momentum
$\vec{p}$ and isospin $\tau $ in asymmetric nuclear
matter~\cite{Che05a,Das03}:
\begin{eqnarray}
U(\rho ,\alpha ,\vec{p},\tau ) &=&A_{u}\frac{\rho _{-\tau }}{\rho
_{0}} +A_{l}\frac{\rho _{\tau }}{\rho _{0}}+B\left(\frac{\rho }{\rho
_{0}}\right)^{\sigma }(1-x\alpha ^{2})-8\tau x\frac{B}{\sigma
+1}\frac{\rho ^{\sigma -1}}{\rho _{0}^{\sigma }}\alpha \rho _{-\tau
}\notag \\
&+&\frac{2C_{\tau ,\tau }}{\rho _{0}}\int d^{3}p^{\prime
}\frac{f_{\tau }(\vec{r},\vec{p}^{\prime
})}{1+(\vec{p}-\vec{p}^{\prime })^{2}/\Lambda ^{2}}+\frac{2C_{\tau
,-\tau }}{\rho _{0}}\int d^{3}p^{\prime } \frac{f_{-\tau
}(\vec{r},\vec{p}^{\prime })}{1+(\vec{p}-\vec{p}^{\prime
})^{2}/\Lambda ^{2}}. \label{MDIU}
\end{eqnarray}
The last two terms in Eq. (\ref{MDIU}) contain the momentum
dependence of the single-particle potential, including that of the
symmetry potential if one allows for different interaction strength
parameters $C_{\tau ,-\tau }$ and $C_{\tau ,\tau }$ for a nucleon of
isospin $\tau $ interacting, respectively, with unlike and like
nucleons in the background fields. The difference between the
neutron and proton potentials then gives an accurate estimate for
the strength of the isovector or symmetry potential in asymmetric
nuclear matter, i.e.,
\begin{equation}
U_{\rm sym}=(U_{\rm neutron}-U_{\rm proton})/2\delta,
\end{equation}
which is of particularly interest and importance for nuclear
reactions induced by neutron-rich nuclei.

With $f_{\tau }(\vec{r},\vec{p})$ $=\frac{2}{h^{3}}\Theta
(p_{f}(\tau )-p)$ for nuclear matter at zero temperature, the
integrals in Eqs.~(\ref{MDIV})~and (\ref{MDIU}) can be calculated
analytically, and one finds \cite{Che07}
\begin{eqnarray}
&&\int \int d^{3}pd^{3}p^{\prime }\frac{f_{\tau
}(\vec{r},\vec{p})f_{\tau ^{\prime }}(\vec{r},\vec{p}^{\prime
})}{1+(\vec{p}-\vec{p}^{\prime })^{2}/\Lambda ^{2}}
=\frac{1}{6}\left( \frac{4\pi }{h^{3}}\right) ^{2}\Lambda
^{2}\left\{ p_{f}(\tau )p_{f}(\tau ^{\prime })\left[
3(p_{f}^{2}(\tau )+p_{f}^{2}(\tau
^{\prime }))-\Lambda ^{2}\right] \right.   \nonumber \\
&&+4\Lambda \left[ (p_{f}^{3}(\tau )-p_{f}^{3}(\tau ^{\prime }))\tan
^{-1}\frac{p_{f}(\tau )-p_{f}(\tau ^{\prime })}{\Lambda
}-(p_{f}^{3}(\tau )+p_{f}^{3}(\tau ^{\prime }))\tan
^{-1}\frac{p_{f}(\tau )+p_{f}(\tau
^{\prime })}{\Lambda }\right]   \nonumber \\
&&\left. +\frac{1}{4}\left[ \Lambda ^{4}+6\Lambda
^{2}(p_{f}^{2}(\tau )+p_{f}^{2}(\tau ^{\prime }))-3(p_{f}^{2}(\tau
)-p_{f}^{2}(\tau ^{\prime }))^{2}\right] \ln \frac{(p_{f}(\tau
)+p_{f}(\tau ^{\prime }))^{2}+\Lambda ^{2}}{(p_{f}(\tau
)-p_{f}(\tau ^{\prime }))^{2}+\Lambda ^{2}}\right\}
\end{eqnarray}
and
\begin{eqnarray}
&&\int d^{3}p^{\prime }\frac{f_{\tau }(\vec{r},\vec{p}^{\prime
})}{1+(\vec{p}-\vec{p}^{\prime })^{2}/\Lambda
^{2}}=\frac{2}{h^{3}}\pi \Lambda ^{3}\left[ \frac{p_{f}^{2}(\tau
)+\Lambda^{2}-p^{2}}{2p\Lambda }\ln \frac{(p+p_{f}(\tau ))^{2}
+\Lambda ^{2}}{(p-p_{f}(\tau ))^{2}+\Lambda ^{2}}\right.    \nonumber \\
&&\left. +\frac{2p_{f}(\tau )}{\Lambda }-2\tan
^{-1}\frac{p+p_{f}(\tau )} {\Lambda }-2\tan ^{-1}\frac{p-p_{f}(\tau
)}{\Lambda }\right] .  \nonumber \\ &&
\end{eqnarray}

For a given value of $x$, which is introduced to vary the density
dependence of the nuclear symmetry energy while keeping other
properties of the nuclear equation of state fixed \cite{Che05a},
values of the parameters $ A_{u}$, $A_{l}$, $B$, $C_{\tau ,\tau }$,
$C_{\tau ,-\tau }$ and $\Lambda $ can be obtained by fitting the
momentum dependence of $U(\rho ,\alpha ,\vec{p},\tau )$ to that
predicted by the Gogny Hartree-Fock and/or the
Brueckner-Hartree-Fock calculations, the saturation properties of
symmetric nuclear matter, and the symmetry energy of $31.6$ MeV at
normal nuclear matter density $\rho _{0}=0.16$ fm$^{-3}$
\cite{Das03}. Specifically, $C_{\tau ,-\tau }=-103.4$ MeV and
$C_{\tau ,\tau }=-11.7$ MeV have been obtained. Furthermore,
choosing the incompressibility $K_{0}$ of cold symmetric nuclear
matter at saturation density $\rho _{0}$ to be $211$ MeV leads to
the dependence of the parameters $A_{u}$ and $A_{l}$ on the $x$
parameter according to
\begin{eqnarray}
A_{u}(x)=-95.98-x\frac{2B}{\sigma
+1},~A_{l}(x)=-120.57+x\frac{2B}{\sigma +1},
\end{eqnarray}
with $B=106.35~{\rm MeV}$.

\begin{figure}[tbh]
\centering
\includegraphics[scale=0.9]{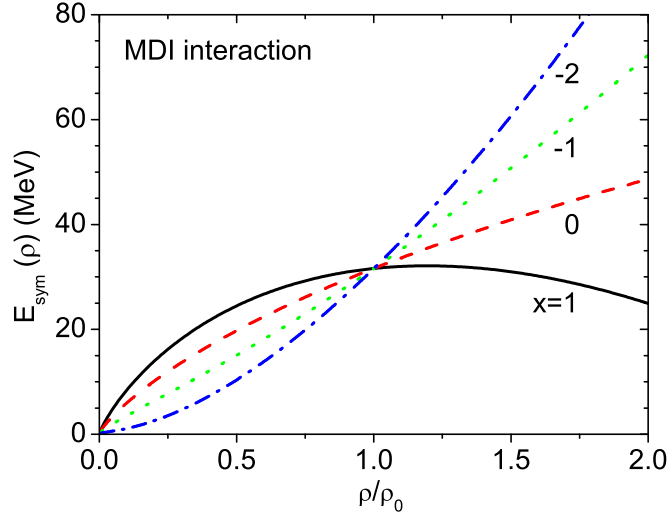}
\caption{(Color online) The density dependence of the nuclear
symmetry energy for different values of the parameter $x$ in the
MDI interaction. Taken from Ref. \cite{Che05a}} \label{MDIsymE}
\end{figure}

\begin{figure}[tbh]
\vspace*{0.5cm}
\includegraphics[height=0.55\textheight] {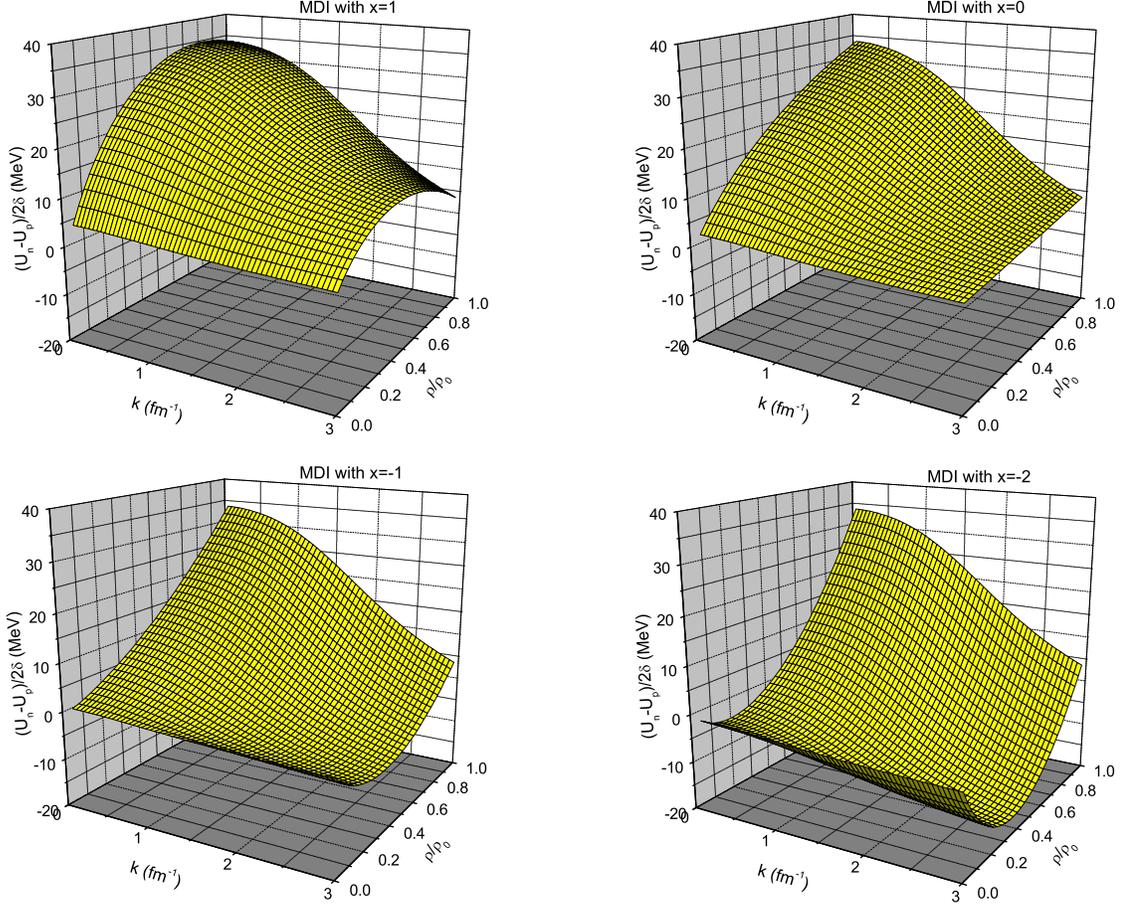} \vspace*{1cm}
\caption{Symmetry potential as a function of momentum and density
for MDI interactions with $x=1,0,-1$ and $-2$. Taken from Ref.
\cite{LiBA05c}.} \label{MDIsymp}
\end{figure}

With above results as well as the well-known contribution from
nucleon kinetic energies in the Fermi gas model, one can easily
obtain the EOS of asymmetric nuclear matter at zero temperature.
As shown in Fig.~\ref{MDIsymE}, adjusting the parameter $x$ in the
MDI interaction leads to a broad range of the density dependence
of the nuclear symmetry energy, similar to those predicted by
various microscopic and/or phenomenological many-body theories.
In Fig.\ \ref{MDIsymp}, the strength of the symmetry potential for
the four $x$ parameters is displayed as a function of momentum and
density. It is noticed that the momentum dependence of the
symmetry potential is independent of the parameter $x$. This is
because by construction the $x$ parameter appears only in the
density-dependent part of the single-nucleon potential as shown in
Eq. (\ref{MDIU}).

Since the MDI interaction can be easily implemented in transport
models for nuclear reactions, one can use it to explore the
effects of the symmetry energy in these reactions. Indeed, the
resulting isospin- and momentum-dependent potential has been
successfully used in the IBUU04 transport model for studying the
isospin effects in intermediate-energy heavy-ion collisions
induced by neutron-rich nuclei
\cite{LiBA04a,Che04,LiBA05a,LiBA05b,LiBA05c,Yon07,LiBA06b,Yon06a,Yon06b,Che05a}.
It has also been used recently to study the thermodynamic
properties of hot isospin-asymmetric nuclear matter
\cite{Xu07,Xu07b}, and this will be discussed in details in
Chapter~\ref{chapter_temperature}.

The above discussions thus indicate that both the phenomenological
and EFT approaches, which contain parameters that are fixed by
nuclear properties around the saturation density, usually give
excellent descriptions of the nuclear properties around or below the
saturation density, although their predictions in the supranormal
density region are probably less reliable. On the other hand, due to
different approximations or techniques used in microscopic many-body
approaches, their predictions on the properties of nuclear matter as
well as those of isospin asymmetric nuclear matter, specially the
density dependence of the nuclear symmetry energy, could be very
different even for the same bare NN interaction \cite{Die03,LiZH06}.

\subsection{The nuclear equation of state and its isospin dependence}

In the following, we review some typical results for the nuclear
matter EOS and its isospin dependence from microscopic many-body
theories and phenomenological approaches. We shall point out the
most obvious, qualitative differences among the model predictions.

\begin{figure}[tbh]
\centering \hspace{-3cm}
\includegraphics[scale=0.5]{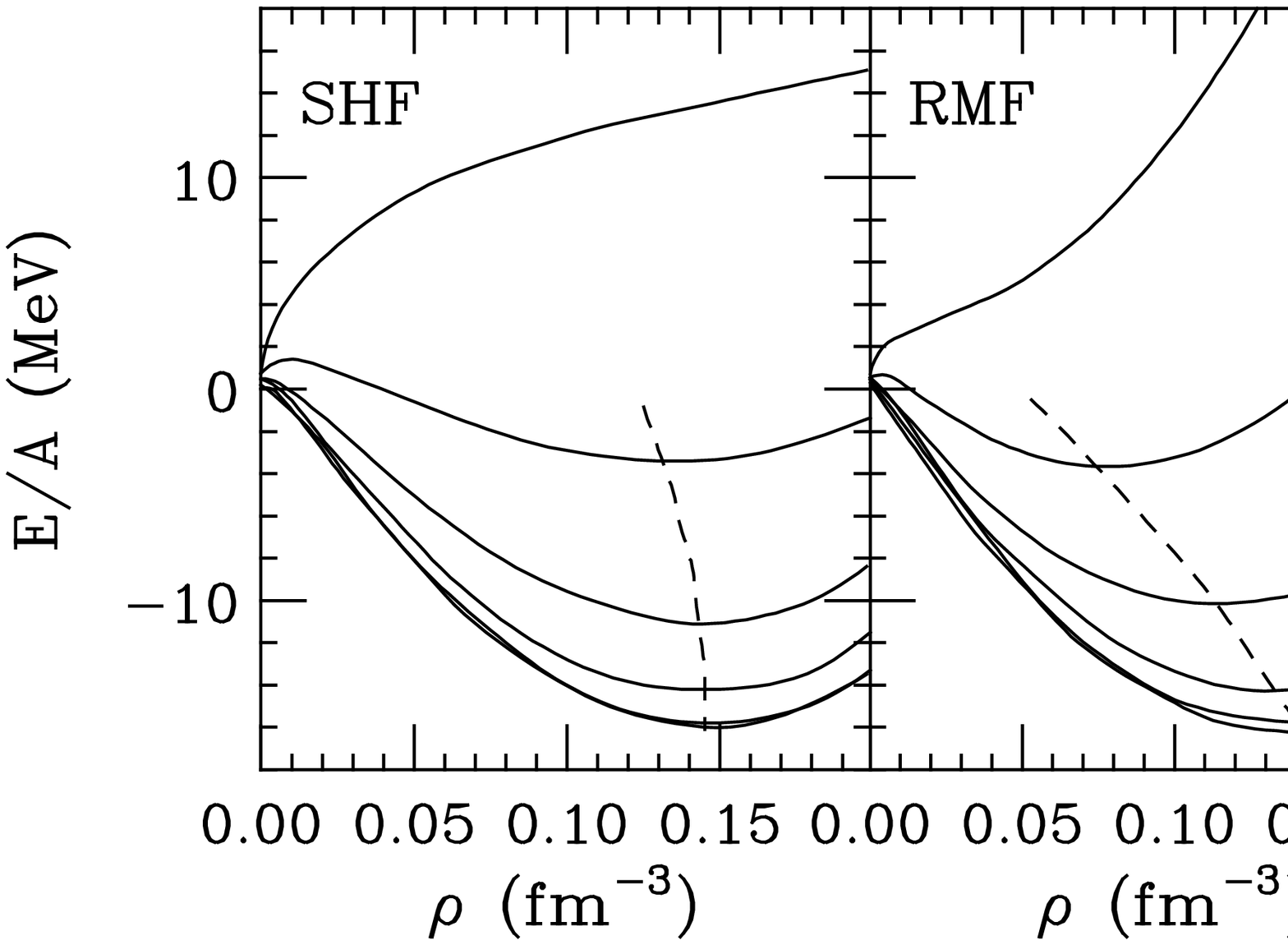}
\caption{The equation of state of asymmetric nuclear matter from the
Skyrme-Hartree-Fock (left panel) and relativistic mean-field (right
panel) model calculations. The solid curves correspond to
proton-to-neutron ratios of 0, 0.2, 0.4, 0.6, 0.8, and 1 (from top
to bottom). Results taken from Ref.~\cite{Tan96}.}
\label{IEOSSHFRMF}
\end{figure}

\begin{figure}[htb]
\centering
\includegraphics[scale=0.8]{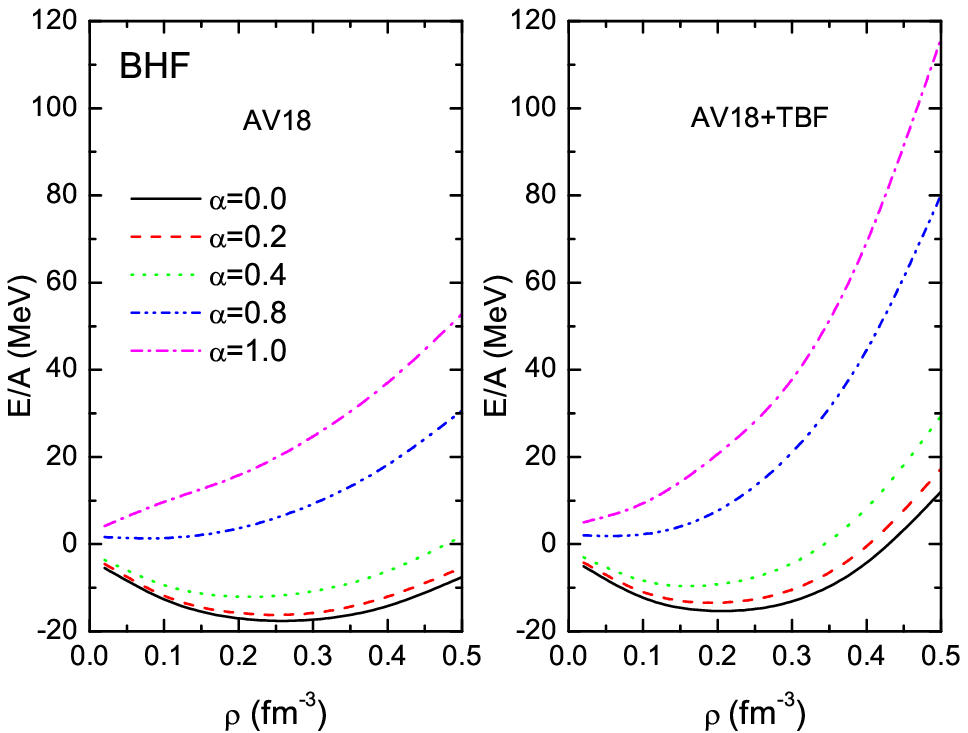}
\hspace{-0.3cm}
\includegraphics[scale=0.81]{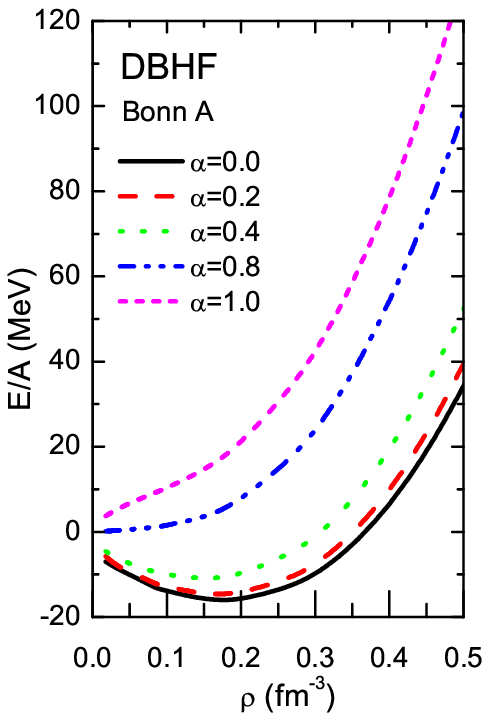}
\caption{(Color online) Same as Fig.~\ref{IEOSSHFRMF} from the
non-relativistic Brueckner-Hartree-Fock calculations \cite{LiZH07}
(left and middle windows) and from the relativistic Dirac
Brueckner-Hartree-Fock calculations \cite{Dal07} (right window).}
\label{IEOSBHFDBHF}
\end{figure}

Figs.~\ref{IEOSSHFRMF} and~\ref{IEOSBHFDBHF} show four typical
predictions for the EOS of asymmetric nuclear matter from the
non-relativistic SHF model using the parameter set SIII
\cite{Tan96}, the RMF model using the parameter set TM1
\cite{Sug94}, the non-relativistic BHF calculation using AV18
interaction with and without three-body force \cite{LiZH07}, and the
most recent calculations using the relativistic DBHF approach
\cite{Dal07}. The isospin asymmetry is indicated for each curve by
the ratio $\rho_p/\rho_n$ of the proton density ($\rho_p$) to that
of neutrons ($\rho_n$) in Fig.\ \ref{IEOSSHFRMF} and the isospin
asymmetry $\alpha=(\rho_n-\rho_p)/(\rho_n-\rho_p)$ in Fig.\
\ref{IEOSBHFDBHF}. A common prediction from these studies is that
the asymmetric nuclear matter is less stiff and bound at saturation.
The minimum in the equation of state, i.e., the energy per nucleon
versus density, disappears before the pure neutron matter limit is
reached, and the compressibility at saturation thus decreases as
nuclear matter becomes more neutron rich. Also, the saturation
density is generally reduced with increasing neutron/proton ratio or
isospin asymmetry.

For the phenomenological SHF and RMF approaches, although they give
the correct saturation properties for symmetric nuclear matter,
their predictions for the EOS of asymmetric nuclear matter, such as
the saturation density, are quantitatively different. In the SHF
model the saturation density depends weakly on the isospin
asymmetry, while in the RMF model the dependence is much stronger.
These different behaviors, which are related directly to the slope
parameter of the symmetry energy and the incompressibility of
symmetric nuclear matter \cite{Lop88}, result in significant
differences in the predicted nucleon density profiles and neutron
skin thickness in radioactive nuclei \cite{LiBA98}.

Including the three-body force in the non-relativistic BHF approach
significantly enhances the binding energy per nucleon of the
asymmetric nuclear matter at higher densities. It also makes the
predictions from the non-relativistic BHF approach more consistent
with the results from the relativistic DBHF approach. A detailed
comparison among results from the Skyrme-Hartree-Fock approach, the
relativistic mean-field theory, the non-relativistic BHF
calculation, and the relativistic DBHF approach can be found in Ref.
\cite{Fuc06b}.

\subsection{The nuclear symmetry energy and the empirical parabolic law}

For asymmetric nuclear matter, various theoretical studies have
shown that the energy per nucleon can be well approximated by
\begin{eqnarray}
E(\rho ,\alpha )=E(\rho ,\alpha =0)+E_{\mathrm{sym}}(\rho )\alpha
^{2}+O(\alpha ^{4}),  \label{EsymPara}
\end{eqnarray}
in terms of the baryon density $\rho =\rho _{n}+\rho _{p}$, the
isospin asymmetry $\alpha$, the energy per nucleon in symmetric
nuclear matter $ E(\rho ,\alpha =0)$, and the bulk nuclear
symmetry energy
\begin{eqnarray}
E_{\mathrm{sym}}(\rho )=\frac{1}{2}\frac{\partial ^{2}E(\rho
,\alpha )}{\partial \alpha ^{2}}|_{\alpha =0}.
\end{eqnarray}
In Eq. (\ref{EsymPara}), there are no odd-order $\alpha $ terms due
to the exchange symmetry between protons and neutrons in nuclear
matter (the charge symmetry of nuclear forces). Higher-order terms
in $\alpha $ are generally negligible for most purposes. For
example, the magnitude of the $\alpha ^{4}$ term at $\rho _{0}$ has
been estimated to be less than $1$ MeV, compared to the value of the
quadratic term $E_{\rm sys}(\rho_0)\sim 30~{\rm MeV}$ at same
density \cite{Sjo74,Bom91,Lag81,Sie70,Lee98}. Nevertheless, it
should be mentioned that the presence of higher-order terms in $
\alpha$ at supra-normal densities can significantly modify the
proton fraction in $\beta$-equilibrium neutron-star matter and the
critical density for the direct Urca process which can lead to
faster cooling of neutron stars \cite{Zha01,Ste06}. Eq.\
(\ref{EsymPara}) is known as the empirical parabolic law for the EOS
of asymmetric nuclear matter and is considered to be valid only at
small isospin asymmetries. However, many non-relativistic and
relativistic calculations have shown that it is actually valid up to
$\alpha =1$, at least for densities up to moderate values.

\begin{figure}[tbh]
\centering
\includegraphics[scale=0.8]{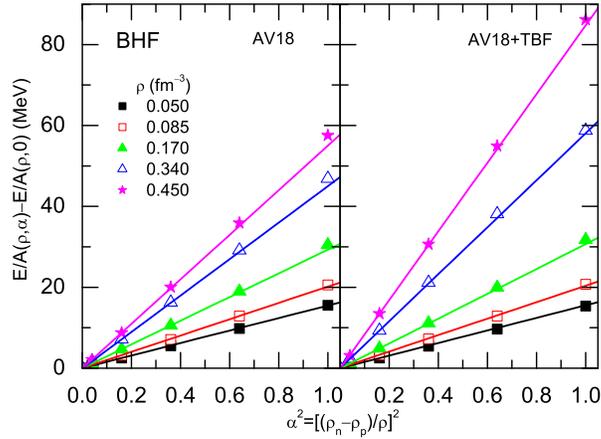}
\caption{(Color online) Energy per nucleon of asymmetric nuclear
matter with isospin asymmetry in the range $0\leq \protect\alpha
^{2}\leq 1$ at five densities as compared to the parabolic fits
(straight lines) obtained from the first three values of
$\protect\alpha $ (0.0, 0.2, 0.4). \ Left panel: BHF results with
only pure $AV_{18}$ 2BF. Right panel: BHF predictions using
$AV_{18}$ plus the 3BF. Data are taken from Ref. \cite{Zuo02}.}
\label{EsymParaBHF}
\end{figure}

\begin{figure}[tbh]
\centering
\includegraphics[scale=0.8]{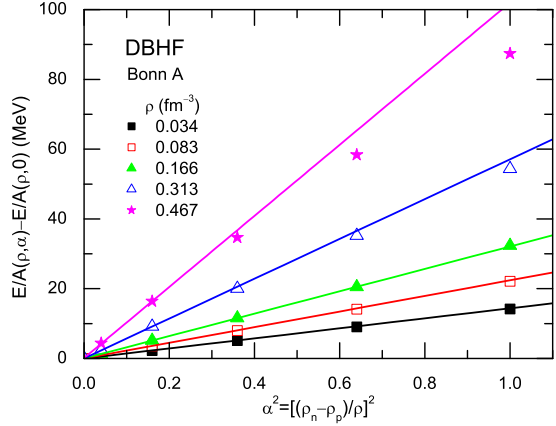}
\caption{(Color online) Same as Fig. \ref{EsymParaBHF} obtained from
the relativistic DBHF approach. Data are taken from Ref.
\cite{Dal07}.} \label{EsymParaDBHF}
\end{figure}

In Fig.\ \ref{EsymParaBHF} and Fig.\ \ref{EsymParaDBHF}, two
examples from recent calculations based on the non-relativistic BHF
approach \cite{Zuo02} and the relativistic DBHF approach
\cite{Dal07} are shown for the total binding energy as a function of
isospin asymmetry at several densities $\rho$. In both cases, the
fit using the parabolic law shown by solid lines is indeed valid in
the whole range of $\alpha $ at least for densities up to moderate
values. At high densities (about three times normal density), the
results from the relativistic DBHF approach deviate somewhat from
the parabolic law, indicating that the higher-order terms in $\alpha
$ become non-negligible.

Using the empirical parabolic law, one can easily extract the
symmetry energy $E_{\mathrm{sym}}(\rho )$ from microscopic
calculations. According to Eq.\ (\ref{EsymPara}) the bulk symmetry
energy $E_{\mathrm{sym}}(\rho )$ can be evaluated approximately
from the two extreme cases of pure neutron matter and symmetric
nuclear matter via
\begin{eqnarray} E_{\mathrm{sym}}(\rho )\approx
E(\rho ,1)-E(\rho ,0),
\end{eqnarray}
which implies that the symmetry energy $E_{\mathrm{sym}}(\rho )$
is an estimate of the energy cost to convert all protons in a
symmetric nuclear matter to neutrons at the fixed density $\rho $.
Furthermore, around the nuclear matter saturation density $\rho
_{0}$, the nuclear symmetry energy $E_{\mathrm{sym}}(\rho )$\ can
be further expanded to second-order as
\begin{eqnarray}
E_{\mathrm{sym}}(\rho )=E_{\mathrm{sym}}(\rho
_{0})+\frac{L}{3}\left( \frac{\rho -\rho _{0}}{\rho _{0}}\right)
+\frac{K_{\mathrm{sym}}}{18}\left( \frac{\rho -\rho _{0}}{\rho
_{0}}\right) ^{2},
\end{eqnarray}
where $L$ and $K_{\mathrm{sym}}$ are the slope and curvature
parameters of the nuclear symmetry energy at $\rho _{0}$, i.e.,
\begin{eqnarray}\label{lsyme}
L=3\rho _{0}\frac{\partial E_{\mathrm{sym}}(\rho )}{\partial \rho
}|_{\rho =\rho _{0}}, \qquad{\rm and}\qquad K_{\mathrm{sym}}=9\rho
_{0}^{2}\frac{\partial ^{2}E_{\mathrm{sym}}(\rho )}{\partial
^{2}\rho }|_{\rho =\rho _{0}}.
\end{eqnarray}
The $L$ and $K_{\mathrm{sym}}$ characterize the density dependence
of the nuclear symmetry energy around normal nuclear matter
density, and thus provide important information on the behaviors
of the nuclear symmetry energy at both high and low densities. In
particular, the slope parameter $L$ has been found to be
correlated linearly with the neutron-skin thickness of heavy
nuclei, and information on the slope parameter $L$ can thus in
principle be obtained from the thickness of the neutron skin in
heavy nuclei
\cite{Die03,Che05b,Ste05b,Bro00,Hor01a,Typ01,Fur02,Kar02}.
Unfortunately, because of the large uncertainties in measured
neutron skin thickness of heavy nuclei, this has so far not been
possible. As to be discussed later, the value of $L$ can be
extracted from studying isospin-sensitive observables in heavy-ion
reactions.

At the nuclear matter saturation density $\rho _{0}$ and around
$\alpha =0$, the isobaric incompressibility of asymmetric nuclear
matter can also be expressed to the second-order of $\alpha $ as
\cite{Lop88,Pra85}
\begin{eqnarray}
K(\alpha )=K_{0}+K_{\mathrm{asy}}\alpha ^{2}
\end{eqnarray}
where $K_{0}$ is the incompressibility of symmetric nuclear matter
at the nuclear matter saturation density $\rho _{0}$. The $K_{\rm
asy}$ in the isospin-dependent part \cite{Bar02}
\begin{eqnarray}
K_{\mathrm{asy}}\approx K_{\mathrm{sym}}-6L
\end{eqnarray}
characterizes the density dependence of the nuclear symmetry energy.
In principle, the information on $K_{\mathrm{asy}}$ can be extracted
experimentally by measuring the giant monopole resonance (GMR) of
neutron-rich nuclei. Earlier attempts to extract the value of
$K_{\mathrm{asy}}$ from experimental GMR data resulted in widely
different values. For example, a value of $K_{\mathrm{asy}}=-320\pm
180$ MeV was obtained in Ref. \cite{Sha88} from a study of the
systematics of GMR in the isotopic chains of Sn and Sm while the
$K_{0}$ was found to be $300\pm 25$ MeV, in contrast with the
commonly accepted value of $230\pm 10$ MeV. A subsequent systematic
study of the GMR of finite nuclei leads to a constraint of $-566\pm
1350<K_{\mathrm{asy}}<139\pm 1617$ MeV, depending on the mass region
of nuclei and the number of parameters used in parameterizing the
incompressibility of finite nuclei \cite{Shl93}. The large
uncertainties in the extracted $K_{\mathrm{asy}}$ thus does not
allow one to distinguish the different nuclear symmetry energies
from theoretical models. Very recently, from measurements of the
isotopic dependence of GMR in the even-A Sn isotopes a more
stringent value of $K_{\mathrm{asy}}=-550\pm 100$ MeV was obtained
in Ref. \cite{Gar07}. This result is consistent with that extracted
from the analysis of the isospin diffusion data
\cite{LiBA05c,Che05a}.

\begin{figure}[htb]
\centering
\includegraphics[scale=0.8]{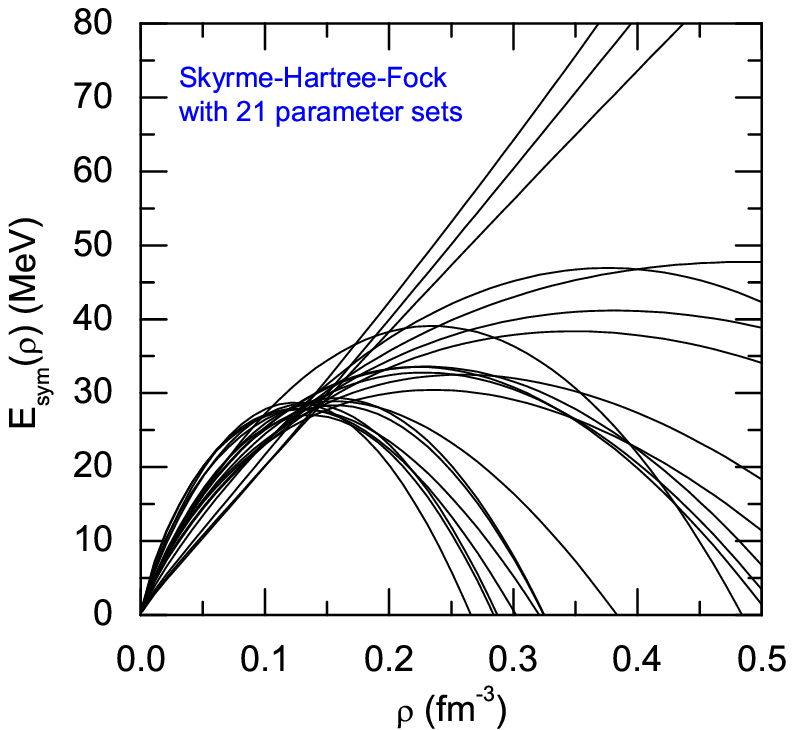}
\includegraphics[scale=0.8]{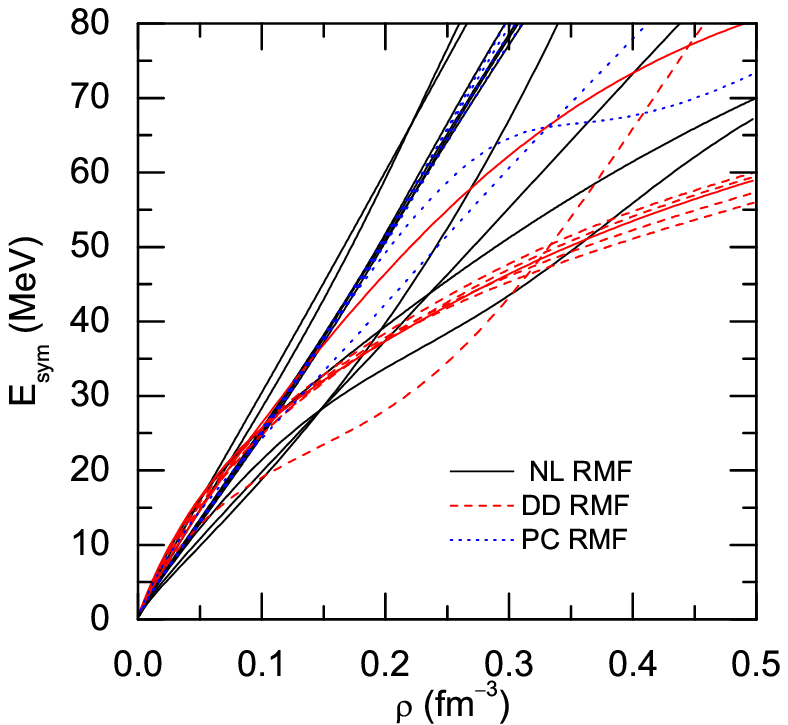}
\caption{(Color online) Left window: Density dependence of the
nuclear symmetry energy $E_{\mathrm{sym}}(\protect\rho )$ from SHF
with 21 sets of Skyrme interaction parameters \cite{Che05a}. Right
window: Same as left panel from the RMF model for the parameter
sets NL1, NL2, NL3, NL-SH, TM1, PK1, FSU-Gold, HA, NL$\protect\rho
$, and NL$\protect\rho \protect\delta $ in the nonlinear RMF model
(solid curves); TW99, DD-ME1, DD-ME2, PKDD, DD, DD-F, and
DDRH-corr in the density-dependent RMF model (dashed curves); and
PC-F1, PC-F2, PC-F3, PC-F4, PC-LA, and FKVW in the point-coupling
RMF model (dotted curves) \cite{Che07}.} \label{EsymSHFRMF}
\end{figure}

\begin{figure}[htb]
\centering
\includegraphics[width=2.5in,height=3in]{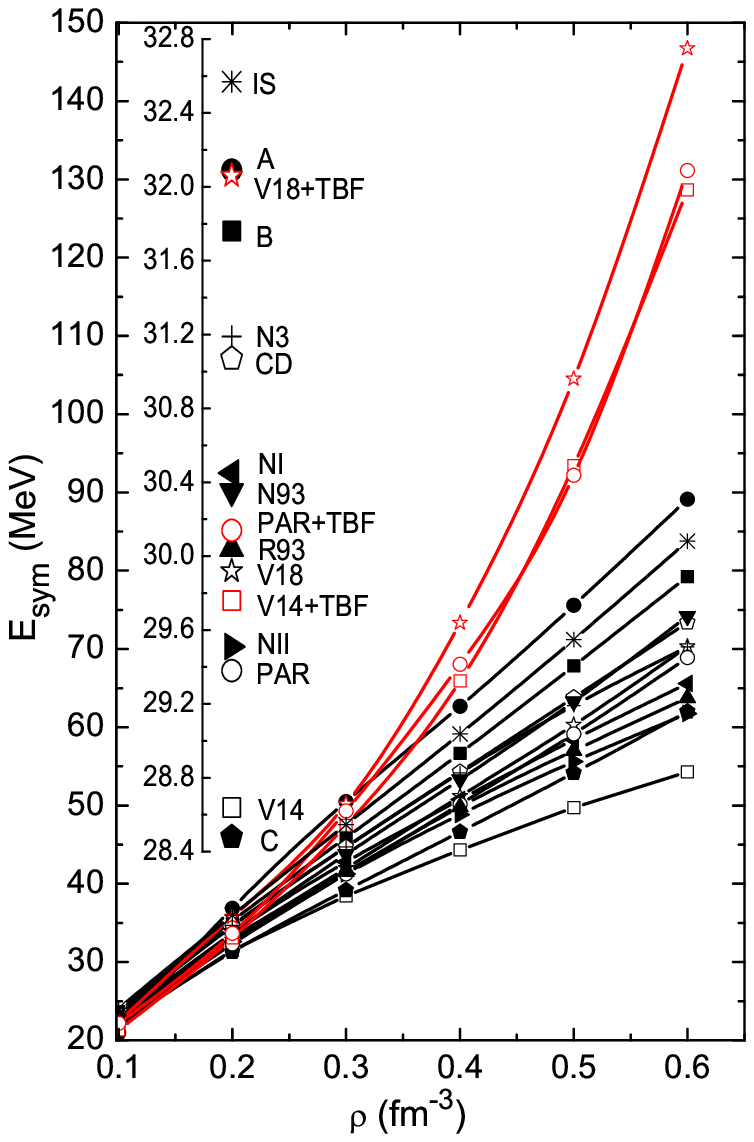}
\includegraphics[width=2.5in,height=3in]{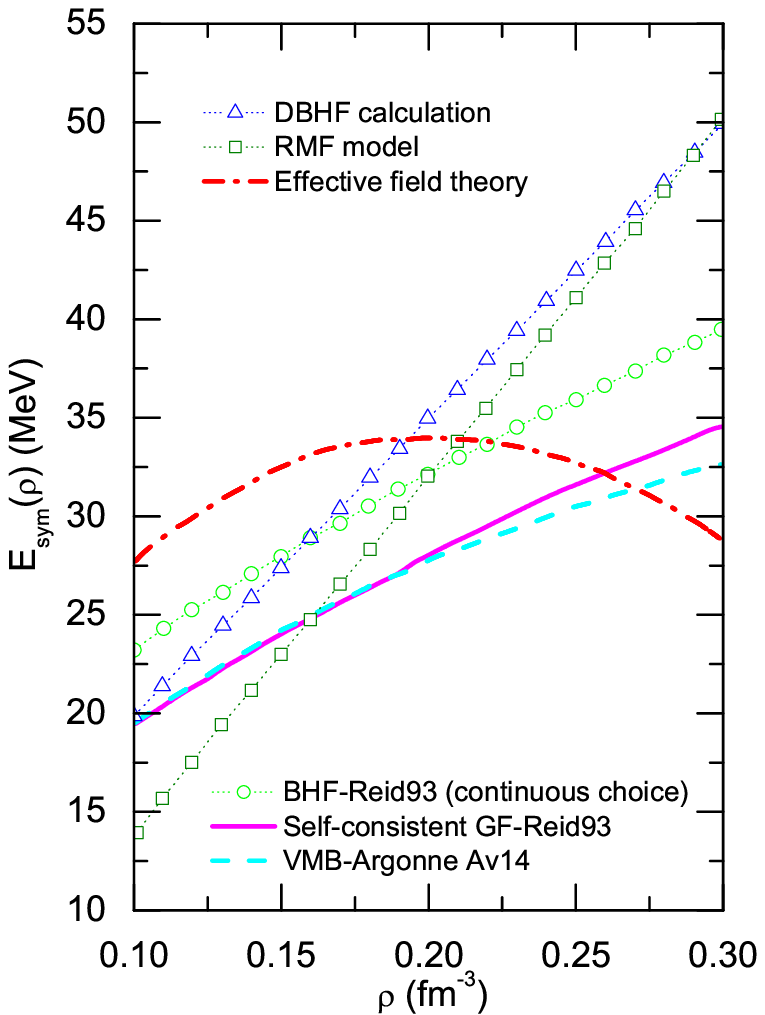}
\caption{(Color online) Left window: Symmetry energy obtained with
different potentials within the BHF approach with (upper red curves)
and without (lower black curves) TBF. The inset shows the values at
normal density $\rho _{0} = 0.17~{\rm fm}^{-3}$ on a magnified
scale. Taken from Ref. \protect\cite{LiZH06}. Right window: Density
dependence of the symmetry energy from the continuous choice
Brueckner-Hartree-Fock with Reid93 potential (circles), the
self-consistent Green's function theory with Reid93 potential (full
line), the variational calculation with Argonne Av14 potential
(dashed line), the Dirac-Brueckner-Hartree-Fock calculation
(triangles), the relativistic mean-field model (squares), and the
effective field theory (dash-dotted line). Results are taken from
Ref. \protect\cite{Die03}.} \label{EsymBHFpiek}
\end{figure}

The symmetry energies at normal nuclear matter density from
various theoretical models are usually tuned to that determined
from the empirical liquid-drop mass formula, which has a value of
$E_{\mathrm{sym}}(\rho _{0})$ around $30$ MeV \cite{Mey66,Pom03}.
For example, in the non-relativistic SHF approach \cite{Che05b},
the predicted values for $E_{\mathrm{sym}}(\rho _{0})$ are between
$26 $ and $35$ MeV depending on the nuclear interactions used in
the calculation while the RMF theory usually gives higher values
of $E_{\mathrm{sym}}(\rho _{0})$ in the range of $30\sim 44$ MeV
\cite{Che07}. In addition, recent calculations from the continuous
choice BHF approach with a large number of modern and old NN
potentials also indicate that the value of $E_{\mathrm{sym}}(\rho
_{0})$ ranges from $28.5$ MeV (Bonn C) to $32.6$ MeV (IS)
\cite{LiZH06}. What distinguishes these models around normal
nuclear matter density are thus the slope $L$ and curvature
$K_{\mathrm{sym}}$. This is clearly illustrated in Figs.\
\ref{EsymSHFRMF} and \ref{EsymBHFpiek} where the density
dependence of the nuclear symmetry energy from several
phenomenological and microscopic approaches are shown. Although
all predict an $E_{\mathrm{sym}}(\rho _{0})$ in the range of
$26\sim 44$ MeV (in agreement with that from the empirical mass
formula), the predicted slope and curvature at $\rho _{0}$ are
very different.

Therefore, despite of the many theoretical efforts, our current
knowledge about the EOS of asymmetric nuclear matter is still
rather limited. In particular, the behavior of the symmetry energy
at supranormal densities, which is essential for understanding the
properties of neutron stars, is most uncertain among all
properties of dense nuclear matter. On the other hand, recent
experimental study of isospin-sensitive observables in
intermediate-energy nuclear reactions involving radioactive beams
has been quite useful in providing some constraints on the density
dependence of the nuclear symmetry energy at subsaturation
densities. Experiments at higher energies in future radioactive
beam facilities is expected to provide the opportunity to study
the nuclear symmetry energy at higher densities.


\section{The momentum dependence of the isovector potential
and the neutron-proton effective mass splitting in neutron-rich
matter} \label{chapter_ria}

Recently, there is a renewed interest in the isovector part of the
nucleon mean-field potential, i.e., the nuclear symmetry potential,
in isospin asymmetric nuclear matter
\cite{Bar05,LiBA04a,Riz04,Che04,Che05a,Bom91,Fuc04,Ma04,Sam05a,Das03,%
Ulr97,LiBA04c,Beh05,Zuo05,Fuc05,Fuc05b,Riz05,Che05c,Ron06,LiZH06b}.
As discussed in Chapter \ref{introduction} and other Chapters of
this review, knowledge on the symmetry potential is important for
understanding not only the structure and reactions of radioactive
nuclei but also many critical issues in astrophysics. Besides
depending on the nuclear density, the symmetry potential also
depends on the momentum or energy of a nucleon. The various
microscopic and phenomenological approaches, such as the
relativistic DBHF \cite{Fuc04,Ma04,Sam05a,Ulr97,Fuc05,Ron06} and
the non-relativistic BHF \cite{Bom91,Zuo05} approach, the chiral
perturbation theory \cite{Fri05}, the RMF approach
\cite{Bar05,Che07}, and the non-relativistic mean-field theory
based on Skyrme-like interactions
\cite{Das03,LiBA04c,Beh05,Xu07c}, that are described in
Chapter~\ref{chapter_eos} for studying the symmetry energy and
potential, as well as the relativistic impulse approximation
\cite{Che05c,LiZH06b} all give widely different predictions for
the momentum dependence of the nuclear symmetry potential, as in
their predictions for the density dependence of the nuclear
symmetry energy. For example, while most models predict a
decreasing symmetry potential with increasing nucleon momentum
albeit at different rates, a few nuclear effective interactions
used in some of the models lead to an opposite conclusion. The
uncertainty on the momentum dependence of the nuclear symmetry
potential further leads to a controversy on the neutron-proton
effective mass splitting in asymmetric nuclear matter.

In this Chapter, we review the present status of the momentum
dependence of the symmetry potential and the neutron-proton
effective mass splitting in neutron-rich matter. In particular, we
will often use the MDI interaction as a reference in comparing the
different interactions, since the MDI single-particle potential in
Eq.~(\ref{MDIU}) has been used extensively in the IBUU04 transport
model \cite{LiBA04a} to study the isospin effects in nuclear
reactions.

\subsection{The nuclear optical potential in the relativistic
impulse approximation}

In the optical model based on the Dirac phenomenology, elastic
nucleon-nucleus scattering is described by the Dirac equation for
the motion of a nucleon in a relativistic potential. For spherical
nuclei, good agreements with experimental data were obtained in the
relativistic approach with a scalar potential (nucleon scalar
self-energy) and the zeroth component of a vector potential (nucleon
vector self-energy), while the standard non-relativistic optical
model using the Schr\"{o}dinger equation failed to describe
simultaneously all experimental observables \cite{Arn79}. Motivated
by the success of the Dirac phenomenology, a microscopic
relativistic model based on the impulse approximation , i.e., the
RIA \cite{Mcn83a,She83,Cla83,Mil83,Ray92}, was developed, and it was
able to fit very well the data from p+$^{40}$Ca and p+$^{208}$Pb
elastic scattering at nucleon energies of both $500$ and $800$ MeV.
A nice feature of the RIA is that it permits very little
phenomenological freedom in deriving the Dirac optical potential in
nuclear matter. The basic ingredients in this approach are the free
invariant NN scattering amplitude and the nuclear scalar and vector
densities in nuclear matter. This is in contrast to the relativistic
DBHF approach, where different approximation schemes and methods
have been introduced for determining the Lorentz and isovector
structure of the nucleon self-energy
\cite{Fuc04,Ma04,Sam05a,Ulr97,Fuc05}.

\subsubsection{The relativistic impulse approximation to the Dirac
optical potential}

Many theoretical studies have suggested that the nucleon-nucleus
scattering at sufficient high energy can be viewed as the
projectile nucleon being scattered from each of the nucleons in
the target nucleus. One thus can describe the process by using the
NN scattering amplitude and the ground state nuclear density
distribution of the target nucleus. For the Lorentz-invariant NN
scattering amplitude, it can be written as
\begin{eqnarray}
\widehat{F}=F_{S}+F_{V}\gamma _{1}^{\mu }\gamma _{2\mu
}+F_{T}\sigma _{1}^{\mu \nu }\sigma _{2\mu \nu }+F_{P}\gamma
_{1}^{5}\gamma _{2}^{5}+F_{A}\gamma _{1}^{5}\gamma _{1}^{\mu
}\gamma _{2}^{5}\gamma _{2\mu }
\end{eqnarray}%
in terms of the scalar $F_{S}$, vector $F_{V}$, tensor $F_{T}$,
pseudoscalar $F_{P}$, and axial vector $F_{A}$ amplitudes. In the
above, subscripts $1$ and $2$ distinguish Dirac operators in the
spinor space of the two scattering nucleons and $\gamma ^{\prime }$s
are the gamma matrices. The five complex amplitudes $F_{S}$,
$F_{V}$, $F_{T}$, $F_{P}$, and $F_{A}$ depend on the squared
momentum transfer $\mathbf{q}^{2}$ and the invariant energy of the
scattering nucleon pair, and they can usually be determined directly
from the NN phase shifts extracted from the NN scattering data
\cite{Mcn83b}. For a spin-saturated nucleus, only the scalar
($F_{S}$) and the zeroth component of the vector ($F_{V}\gamma
_{1}^{0}\gamma _{2}^{0}$) amplitudes dominate the contribution to
the optical potential. In the relativistic impulse approximation,
the optical potential in momentum space is thus obtained by
multiplying each of these two amplitudes with corresponding
momentum-space nuclear scalar ${\tilde{\rho}}_{S}(\mathbf{q})$ and
vector $\tilde{\rho}_{V}(\mathbf{q})$ densities, i.e.,
\begin{eqnarray}
{\tilde{U}}(\mathbf{q})=\frac{-4\pi ip_{\text{lab}}}{M}[F_{S}(q)
{\tilde{\rho}}_{S}(\mathbf{q})+\gamma
_{0}F_{V}(q)\tilde{\rho}_{V}(\mathbf{q})],  \label{trhom}
\end{eqnarray}
where $p_{\text{lab}}$ and $M$ are, respectively, the laboratory
momentum and mass of the incident nucleon. The optical potential
in coordinator space is then given by the Fourier transformation
of ${\tilde{U}}(\mathbf{q})$, similar to the \textquotedblleft
$t\rho $\textquotedblright\ approximation used in the
non-relativistic impulse approximation \cite{Mcn83b}.

Although the $\mathbf{q}$-dependence in the relativistic NN
amplitude is important for calculating observables for nucleons
scattering off finite nuclei within the Dirac phenomenology, only
the forward NN scattering amplitudes, i.e., $F_{S0}\equiv
F_{S}(q=0)$ and $F_{V0}\equiv F_{V}(q=0)$, contribute to the Dirac
optical potential of nucleons in infinite nuclear matter, as the
scalar and vector densities are constant in coordinate space and
thus delta functions in momentum space, i.e., $\sim \delta
^{(3)}(\mathbf{q})$. In this case, the nuclear coordinate-space
optical potential, obtained from the Fourier transform of Eq.
(\ref{trhom}), takes the simple form \cite{Mcn83a}
\begin{eqnarray}
U=\frac{-4\pi ip_{\text{lab}}}{M}[F_{S0}\rho _{S}+\gamma
_{0}F_{V0}\rho _{V}],  \label{trhor}
\end{eqnarray}%
where $\rho _{S}$ and $\rho _{V}$ are, respectively, the spatial
scalar and vector densities of an infinite nuclear matter.

The Dirac optical potential in Eq. (\ref{trhor}) is valid for
nucleons at high energies. With decreasing nucleon energy, medium
modification due to the Pauli blocking effect becomes important. As
described in detail in Ref. \cite{Mur87}, the Dirac optical
potential including the Pauli blocking effect can be written as
\begin{eqnarray}
U_{\text{opt}}=\left[ 1-a_{i}(E_{\mathrm{kin}})\left( \frac{\rho
_{B}}{\rho _{0}}\right) ^{2/3}\right] U,  \label{noisopb}
\end{eqnarray}
where $\rho _{B}$ is the nuclear baryon density and $\rho
_{0}=0.1934$ fm$^{-3}$. The parameter $a_{i}(E_{\mathrm{kin}})$
denotes the Pauli blocking factor for a nucleon with kinetic energy
$E_{\mathrm{kin}}$, and its value is given in Table II of Ref.
\cite{Mur87}. Although there are still many open questions on the
role of medium modification in the Dirac optical potential
\cite{Mur87}, the $\rho _{B}^{2/3}$ density dependence of the Pauli
blocking factor is consistent with the phase-space consideration for
isotropic scattering \cite{Che01}. For nucleon scattering in isospin
asymmetric nuclear matter, the Pauli blocking effect becomes
different for protons and neutrons. In Ref. \cite{Che01}, an
isospin-dependent Pauli blocking factor is introduced, resulting in
the following different Dirac optical potentials for protons and
neutrons:
\begin{eqnarray}
U_{\text{opt}}^{n(p)} =\left\{ 1-a_{i}(E_{\mathrm{kin}})\left[
\frac{(2\rho _{n(p)})^{2/3}+0.4(2\rho _{p(n)})^{2/3}}{1.4\rho
_{0}^{2/3}}\right] \right\}U^{n(p)}.  \label{isopb}
\end{eqnarray}
Obviously, Eq. (\ref{isopb}) reduces to Eq. (\ref{noisopb}) in the
symmetric nuclear matter with $\rho _{n}=\rho _{p}$.

\subsubsection{Nuclear scalar densities}

To evaluate the Dirac optical potential for nucleons in RIA, one
also needs to know the nuclear scalar and vector densities. They can
be determined from the RMF model \cite{Ser97,Ser86}. Currently,
there are many different versions of the RMF model \cite{Che07}, and
they mainly include the non-linear models
\cite{Ser97,Ser86,Rei89,Rin96}, the models with density dependent
meson-nucleon couplings \cite{Fuc95,She97,Typ99,Hof01}, and the
point-coupling models \cite{Fin04,Bur02,Mad04,Bur04,Rus97}, as to be
discussed in Chapter~\ref{chapter_rmf}. In Ref. \cite{Che05c}, the
nuclear scalar densities are calculated using the non-linear RMF
model with a Lagrangian density that includes the nucleon field
$\psi $, the isoscalar-scalar meson field $\sigma $, the
isoscalar-vector meson field $\omega $, the isovector-vector meson
field $\rho $, and the isovector-scalar meson field $\delta $ with
three typical parameter sets, namely, the very successful NL3 model
\cite{Lal97}, the Z271v model, which has been used to study the
neutron skin of heavy nuclei and the properties of neutron stars
\cite{Hor01a}, and the HA model which includes the isovector-scalar
meson field $\delta $ and fits successfully some results calculated
with the more microscopic DBHF approach \cite{Bun03}.

\begin{figure}[th]
\centering
\includegraphics[scale=0.9]{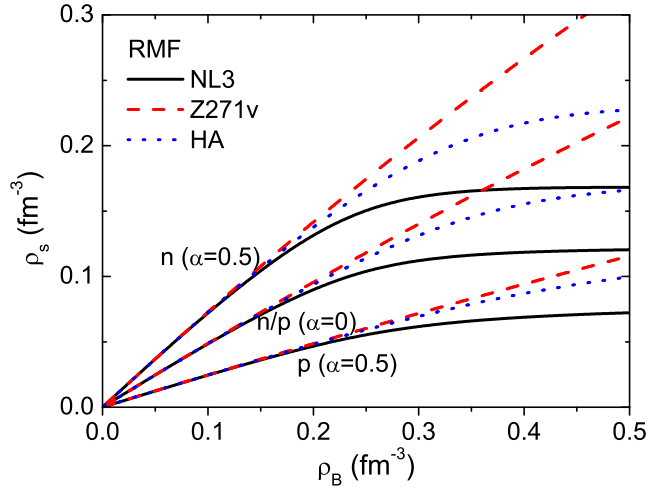}
\caption{(Color online) Neutron and proton scalar densities as
functions of baryon density in nuclear matter with isospin
asymmetry $\alpha =0$ and $0.5$ for the parameter sets NL3, Z271v,
and HA. Taken from Ref. \cite{Che05c}.} \label{rhos}
\end{figure}

Shown in Fig. \ref{rhos} are the neutron and proton scalar
densities $\rho _{\text{S}}$ as functions of the baryon density
$\rho _{\text{B}}$\ (vector density in the static infinite nuclear
matter) in nuclear matter with isospin asymmetry $\alpha =0$ and
$0.5$ for the parameter sets NL3, Z271v, and HA. It is seen that
the neutron scalar density is larger than that of the proton at a
fixed baryon density in the neutron-rich nuclear matter. While
results for different parameter sets are almost the same at lower
baryon densities, they become quite different when $\rho
_{\text{B}}\gtrsim 0.25$ fm$^{-3}$ with Z271v giving a larger and
NL3 a smaller $\rho _{\text{S}}$ than that from the parameter set
HA. For $\rho _{B}\lesssim 0.25$ fm$^{-3}$, the proton and neutron
scalar densities from these three RMF models are also consistent
with those from the RMF model with density-dependent meson-nucleon
couplings and the point-coupling models as shown in Ref.
\cite{Che07} and will be further discussed in the
Chapter~\ref{chapter_rmf}. The real and imaginary parts of the
scalar potential at higher densities ($\rho _{B}\gtrsim 0.25$
fm$^{-3}$) thus depend on the interactions used in evaluating the
nuclear scalar density and have, therefore, large uncertainties.
In Ref. \cite{Che05c}, only the HA parameter set is used, and the
focus is on nuclear densities smaller than $\rho _{B}\lesssim
0.25$ fm$^{-3}$ where the scalar densities of neutrons and protons
in asymmetric nuclear matter are essentially independent of the
model parameters \cite{Che07}.

\subsubsection{The nuclear symmetry potential}

In the Dirac spinor space of the projectile nucleon, the optical
potential $U_{\text{opt}}$ is a $4\times 4$ matrix and can be
expressed in terms of a scalar $U_{S}^{\mathrm{tot}}$ and a vector
$U_{0}^{\mathrm{tot}}$ piece:
\begin{eqnarray}
U_{\text{opt}}=U_{S}^{\mathrm{tot}}+\gamma
_{0}U_{0}^{\mathrm{tot}}.
\end{eqnarray}
Expressing $U_{S}^{\mathrm{tot}}$ and $U_{0}^{\mathrm{tot}}$ in
terms of their real and imaginary parts, i.e.,
\begin{eqnarray}
U_{S}^{\mathrm{tot}}=U_{S}+iW_{S},\text{ \ }U_{0}^{\mathrm{tot}%
}=U_{0}+iW_{0},
\end{eqnarray}
the following `Schr\"{o}dinger-equivalent potential' (SEP) can be
obtained from the Dirac optical potential \cite{Bro78,Jam80}:
\begin{eqnarray}
U_{\text{SEP}}=U_{S}^{\mathrm{tot}}+U_{0}^{\mathrm{tot}}
+\frac{1}{2M}(U_{S}^{\mathrm{tot2}}-U_{0}^{\mathrm{tot2}})
+\frac{U_{0}^{\mathrm{tot}}}{M} E_{\mathrm{kin}},  \label{SEP}
\end{eqnarray}
Solving the Schr\"{o}dinger equation with the SEP then gives the
same bound-state energy eigenvalues and elastic phase shifts as
the solution of the upper component of the Dirac spinor in the
Dirac equation using corresponding Dirac optical potential.

The real part of SEP is given by
\begin{eqnarray}
\text{Re}(U_{\text{SEP}})=U_{S}+U_{0}+\frac{1}{2M}
[U_{S}^{2}-W_{S}^{2}-(U_{0}^{2}-W_{0}^{2})]+\frac{U_{0}}{M}E_{\mathrm{kin}}.
\label{ReSEP}
\end{eqnarray}
The above equation corresponds to the nuclear mean-field potential
in non-relativistic models \cite{Fuc05,Jam89} and allows one to
obtain the following nuclear symmetry potential:
\begin{eqnarray}\label{dat}
U_{\text{sym}}=\frac{\text{Re}(U_{\text{SEP}})_{n}
-\text{Re}(U_{\text{SEP}})_{p}}{2\alpha},
\end{eqnarray}
where Re$(U_{\text{SEP}})_{n}$ and Re$(U_{\text{SEP}})_{p}$ are,
respectively, the real part of the SEP for neutrons and protons.

\subsection{The high-energy behavior of the nuclear symmetry potential}

In the following, we review the Dirac optical potential for neutrons
and protons in asymmetric nuclear matter based on the RIA using the
empirical NN scattering amplitude determined by McNeil, Ray, and
Wallace (MRW) \cite{Mcn83b}, which has been shown to be valid for
nucleons with kinetic energy greater than about $500$ MeV where
Pauli blocking and other medium effects can be neglected. The high
energy behavior of the nuclear symmetry potential from the resulting
Schr\"{o}dinger-equivalent potential can then be investigated
without adjustable parameters \cite{Che05c}.

\subsubsection{The relativistic Dirac optical potential}

\begin{figure}[th]
\centering
\includegraphics[scale=0.85]{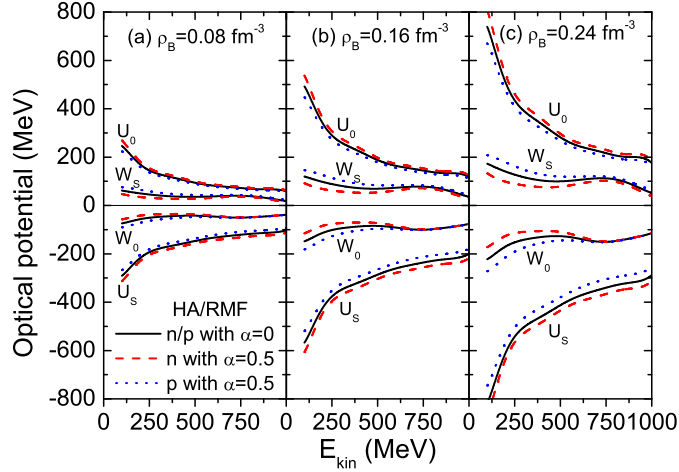}
\caption{(Color online) Energy dependence of real and imaginary
parts of the scalar and vector potentials for neutrons and protons
in nuclear matter with isospin asymmetry $\alpha =0$ and $0.5$ for
the parameter set HA. Taken from Ref. \cite{Che05c}.}
\label{OPReImEkin}
\end{figure}

With neutron and proton scalar densities obtained from the nonlinear
RMF theory with the parameter set HA and the empirical MRW NN
scattering amplitud, both the energy and density dependence of the
real and imaginary parts of the scalar and vector potentials for
neutrons and protons in nuclear matter with isospin asymmetry
$\alpha =0$ and $0.5$ have been studied \cite{Che05c}. In Fig.
\ref{OPReImEkin}, the resulting energy dependence of these
potentials is shown for the three nucleon densities $\rho
_{\text{B}}=0.08$ fm$^{-3}$ (panel (a)), $0.16$ fm$^{-3}$ (panel
(b)), and $0.24$ fm$^{-3}$ (panel (c)). For all densities, the
optical potential shows a strong energy dependence below $300$ MeV,
where it is known that the influences due to ambiguities in the
relativistic form of the NN interaction, the exchange contribution,
and the medium modification due to Pauli blocking are important. The
lower energy behavior of the optical potential can in principle be
studied in the generalized relativistic impulse approximation that
is based on the relativistic meson-exchange model of nuclear force
and the complete set of Lorentz-invariant NN amplitudes
\cite{Mur87,Hor85,Tjo85,Ott88,Tok01}. Many theoretical studies have
shown, however, that the experimental data on elastic
nucleon-nucleus scattering can be reproduced by using the above MRW
optical potential when the nucleon kinetic energy is greater than
about $500$ MeV and that this optical potential also agrees very
well with that extracted from the phenomenological analysis of the
nucleon-nucleus scattering data \cite{Mcn83a,She83,Cla83,Jin93}. As
shown in Fig. \ref{OPReImEkin}, for all three densities considered
here, there is a systematic difference or isospin splitting in the
optical potentials for protons and neutrons in asymmetric nuclear
matter. Specifically, the neutron exhibits a stronger real but
weaker imaginary scalar and vector potentials in neutron-rich
nuclear matter. Furthermore, both the proton and neutron optical
potentials become stronger with increasing density.

\begin{figure}[th]
\centering
\includegraphics[scale=0.85]{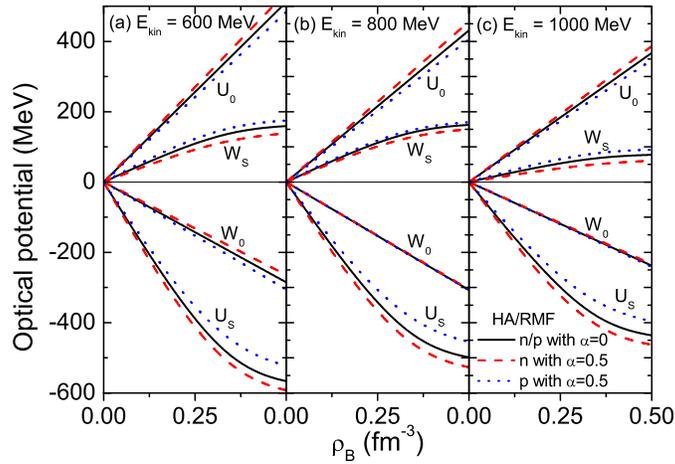}
\caption{(Color online) Density dependence of the real and
imaginary parts of the scalar and vector potentials for neutrons
and protons in nuclear matter with isospin asymmetry $\alpha =0$
and $0.5$ for the parameter set HA. Taken from Ref.
\cite{Che05c}.} \label{OPReImDen}
\end{figure}

The density dependence of the real and imaginary parts of the scalar
and vector potentials for neutrons and protons in nuclear matter
with isospin asymmetry $\alpha =0$ and $0.5$ obtained with the
parameter set HA is shown more explicitly in Fig. \ref{OPReImDen}
for the three nucleon kinetic energies of $E_{\mathrm{kin}}=600$ MeV
(panel (a)), $800$ MeV (panel (b), and $1000$ MeV (panel (c). An
isospin splitting of the nucleon optical potential in asymmetric
nuclear matter is again clearly seen.

\subsubsection{The nuclear symmetry potential}

\begin{figure}[th]
\centering
\includegraphics[scale=1.15]{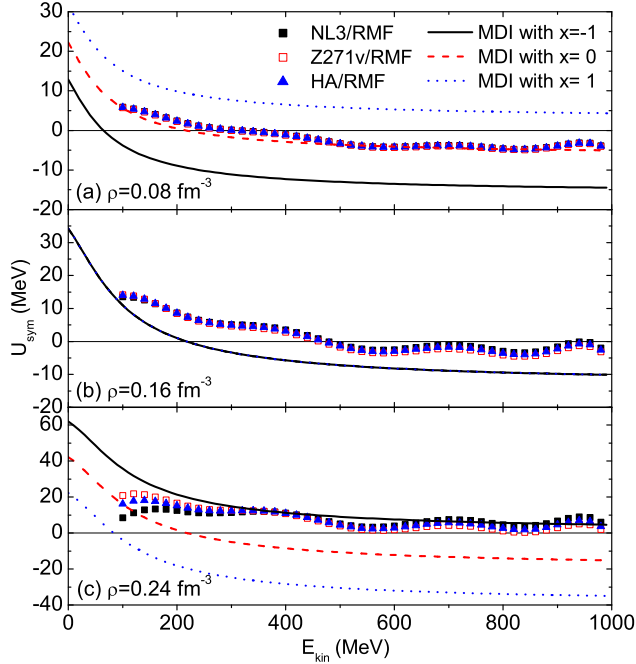}
\caption{(Color online) Energy dependence of the nuclear symmetry
potential using the parameter sets NL3, Z271v, and HA as well as
from the phenomenological interaction MDI with $x=-1,0,$ and $1$
at fixed baryon densities of $\rho _{\text{B}}=0.08$ fm$^{-3}$
(a), $0.16$ fm$^{-3}$ (b), and $0.24$ fm$^{-3}$ (c). Taken from
Ref. \cite{Che05c}.} \label{LaneEkin}
\end{figure}

The energy dependence of the nuclear symmetry potential for the
parameter sets NL3, Z271v, and HA at fixed baryon densities of
$\rho _{\text{B}}=0.08$ fm$^{-3}$ (panel (a)), $0.16$ fm$^{-3}$
(panel (b)), and $0.24$ fm$^{-3}$ (panel (c)) are shown in Fig.
\ref{LaneEkin}. All three parameter sets give similar nuclear
symmetry potential for nucleons with kinetic energy higher than
about $300$ MeV, i.e., it first decreases with nucleon kinetic
energy and then becomes essentially constant when the nucleon
kinetic energy is above about $500$ MeV. Specifically, the nuclear
symmetry potential starts from about $0$ MeV at lower density of
$\rho _{\text{B}}=0.08$ fm$^{-3}$ (about half of nuclear saturated
density), $4.8$ MeV at normal nuclear matter density ($\rho
_{\text{B}}=0.16$ fm$^{-3}$), and $12$ MeV at higher density of
$\rho _{\text{B}}=0.24$ fm$^{-3}$ (about $1.5$ time nuclear
saturated density) and then saturates to about $-3.8\pm 0.5 $ MeV,
$-1.8\pm 1.7$ MeV, and $5.3\pm 3.8$ MeV, respectively, when the
nucleon kinetic energy is greater than about $500$ MeV. The
uncertainties in the saturated values simply reflect the variation
in the energy dependence of the symmetry potential at high
energies.

For comparison, also shown in Fig. \ref{LaneEkin} are results from
the phenomenological parametrization of the momentum-dependent
nuclear mean-field potential, i.e., MDI interaction with $x=-1$,
$0$, and $1$. The energy dependence of the symmetry potential from
the MDI interaction is consistent with the empirical Lane potential
at normal nuclear matter density and low nucleon energies
\cite{LiBA04c} and has been used in the transport model for studying
isospin effects in intermediate-energy heavy ion collisions induced
by neutron-rich nuclei \cite{LiBA04a,Che04,Che05a}. It is seen from
Fig. \ref{LaneEkin} that results from RIA at lower density of $\rho
=0.08$ fm$^{-3}$ are comparable to those from the MDI interaction
with $x=0$, while at higher baryon density of $\rho
_{\text{B}}=0.24$ fm$^{-3}$ they are comparable to those from the
MDI interaction with $x=-1$. At normal nuclear matter density, the
MDI interaction, which gives same results for different $x$ values
by construction, leads to a smaller nuclear symmetry potential at
high nucleon kinetic energies compared with the results from the RIA
based on the empirical MRW NN scattering amplitude and the nuclear
scalar density from the relativistic mean-field theory.

\begin{figure}[th]
\centering
\includegraphics[scale=0.95]{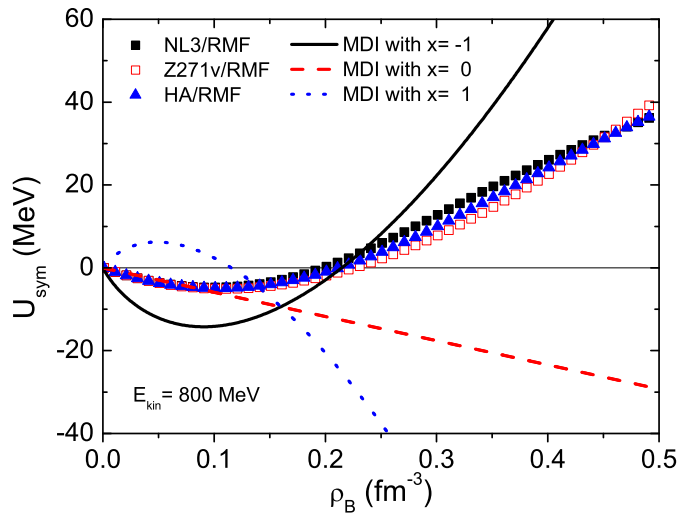}
\caption{(Color online) Density dependence of the nuclear symmetry
potential using the parameter sets NL3, Z271v, and HA as well as
from the MDI interaction with $x=-1$, $0$, and $1$ at a fixed
nucleon kinetic energy of 800 MeV. Taken from Ref. \cite{Che05c}.}
\label{LaneDen}
\end{figure}

For the density dependence of the nuclear symmetry potential using
the parameter sets NL3, Z271v, and HA at a fixed high nucleon
kinetic energy of $800$ MeV, it is shown in Fig. \ref{LaneDen}
together with corresponding results from the MDI interaction with
$x=-1$, $0$, and $1$. It is clearly seen that the nuclear symmetry
potential from all parameter sets NL3, Z271v, and HA changes from
negative to positive values at a fixed baryon density of about $\rho
_{\text{B}}=0.22$ fm$^{-3}$ and then increases almost linearly with
baryon density. Furthermore, the nuclear symmetry potential depends
not much on the choice of the parameter sets NL3, Z271v, and HA. At
such high nucleon kinetic energy, the nuclear symmetry potential
from the MDI interaction with $x=0$ reproduces nicely the results
from the RIA when $\rho _{\text{B}}\lesssim 0.1$ fm$^{-3}$ as in its
energy dependence at low densities shown in Fig. \ref{LaneEkin}. The
two differ strongly, however, at high densities. The MDI interaction
with both $x=-1$ and $1$, on the other hand, show very different
density dependence from the RIA results.

The nuclear symmetry potential derived from the Dirac optical
potential via its Schr\"{o}dinger-equivalent potential is thus not
very sensitive to the parameter sets used in the relativistic
mean-field calculation, particularly at low densities and high
nucleon energies where the RIA is an especially suitable approach,
although it gives very different nuclear scalar densities at high
baryon densities in both symmetric and asymmetric nuclear matters.
Furthermore, the nuclear symmetry potential at a fixed density
becomes almost constant when the nucleon kinetic energy is greater
than about $500$ MeV. For such high energy nucleon, the density
dependence of its nuclear symmetry potential is weakly attractive
at low densities but becomes increasingly repulsive as the nuclear
density increases. These results provide important constraints on
the high energy behavior of the nuclear symmetry potential in
asymmetric nuclear matter, which is an important input to the
isospin-dependent transport model \cite{LiBA04a,Bar05} for
studying heavy-ion collisions induced by radioactive nuclei at
intermediate and high energies. They are also useful in future
studies that extend the Lorentz-covariant transport model
\cite{Ko87,Ko88,Mar92,Ko96} to include explicitly the isospin
degrees of freedom.

\subsection{The intermediate-energy behavior of the nuclear
symmetry potential}

The empirical MRW \textsl{NN} scattering amplitude works well for
elastic nucleon-nucleus scattering at high energies (above about
$500$ MeV). However, the original RIA of MRW failed to describe spin
observables at laboratory energies lower than about $500$
MeV~\cite{Ray85}, and its predicted angular oscillations in the
analyzing power in proton-\textrm{Pb} scattering at large angles
were also in sharp disagreement with experimental data \cite{Dra85}.
These shortcomings are largely due to the implicit dynamical
assumptions about the relativistic \textsl{NN} interaction in the
form of the Lorentz covariance \cite{Ada84} and the somewhat awkward
behavior under the interchange of two particles \cite{Hor85} as well
as the omitted medium modification due to the Pauli blocking effect.
To overcome these theoretical limitations at lower energies, Murdock
and Horowitz (MH) \cite{Hor85,Mur87} extended the original RIA to
take into account following three improvements: i) an explicit
exchange contribution was introduced by fitting to the relativistic
\textsl{NN} scattering amplitude; ii) a pseudovector coupling rather
than a pseudoscalar coupling was used for the pion; and iii) medium
modification from the Pauli blocking was included. With these
improvements, the RIA with the free \textsl{NN} scattering amplitude
was then able to reproduce successfully measured analyzing power and
spin rotation function for all considered closed shell nuclei in
proton scattering near $200$ MeV. Particularly, the medium
modification due to the Pauli blocking effect was found to be
essential in describing the spin rotation function for $^{208}$Pb at
the proton energy of $290 $ MeV \cite{Mur87}.

The generalized RIA of MH has recently been used to study the
intermediate-energy ($100$ MeV$\leq E_{\mathrm{kin}}\leq 400$ MeV)
behavior of the nucleon Dirac optical potential, the
Schr\"{o}dinger-equivalent potential, and the nuclear symmetry
potential in isospin asymmetric nuclear matter \cite{LiZH06b}. In
the following, we review these results.

\subsubsection{The relativistic Love-Franey \textsl{NN} scattering amplitude}

\begin{figure}[h]
\centering
\includegraphics[scale=0.78]{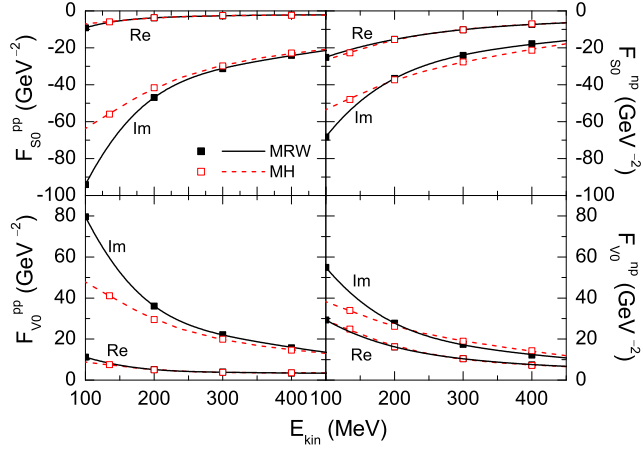}
\caption{(Color online) The scalar and vector parts of the free
\textsl{NN} forward scattering amplitudes $F_{S0}^{pp}$,
$F_{S0}^{np}$, $F_{V0}^{pp}$, and $F_{V0}^{np}$ at nucleon kinetic
energies $E_{\mathrm{kin}}=135$, 200, 300, and 400 MeV (open
squares) from the RIA of MH. Dashed lines are polynomial fits to the
energy dependence of the \textsl{NN} scattering amplitudes.
Corresponding results from the original RIA of MRW are shown by
solid squares and lines. Taken from Ref. \cite{LiZH06b}.}
\label{FFSV}
\end{figure}

Based on the generalized RIA of MH with the Love-Frany \textsl{NN}
scattering amplitudes \cite{Lov81}, one can evaluate the scalar and
vector parts of the \textsl{NN} forward scattering amplitudes
$F_{S0}^{pp}$, $F_{S0}^{np}$, $F_{V0}^{pp}$, and $F_{V0}^{np}$.
Their values at nucleon kinetic energies $E_{\mathrm{kin}}=135$,
$200$, $300$ and $400$ MeV can be found explicitly in Refs.
\cite{Hor85,Mur87}, and they are shown by open squares in Fig.
\ref{FFSV}. To obtain continuous and smooth results for the
\textsl{NN} scattering amplitude and other quantities in the
following, polynomial fits have been made to the energy dependence
of the \textsl{NN} scattering amplitudes, and the results are shown
by dashed lines in Fig.~\ref{FFSV}. For comparison, corresponding
results from the original RIA of MRW are also shown by solid squares
and lines in Fig.~\ref{FFSV}. It is seen that for both proton-proton
and proton-neutron scattering, the real parts of corresponding
amplitudes in the two approaches are in good agreement with each
other. However, for the imaginary parts of the amplitudes, the
strength of the scalar and vector amplitudes from the RIA of MH
displays a much weaker energy dependence in both proton-proton and
proton-neutron scattering at the energies $E_{\mathrm{kin}}\leq 300$
MeV. Since the imaginary part of the amplitude corresponds to the
real part of the Dirac optical potential as shown in
Eq.~(\ref{trhor}), the differences between the original RIA of MRW
and the generalized RIA of MH thus lead to different behaviors in
the Dirac optical potential at lower energies.

\subsubsection{The relativistic Dirac optical potential}

\begin{figure}[th]
\centering
\includegraphics[scale=0.8]{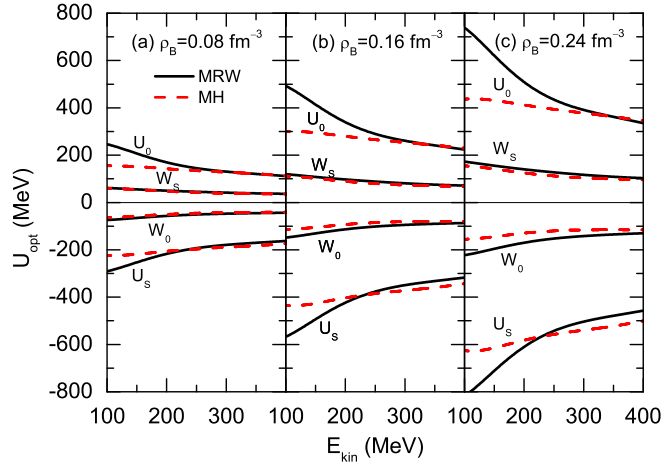}
\caption{(Color online) Energy dependence of the real and
imaginary parts of the scalar and vector optical potentials in
symmetric nuclear matter for different baryon densities
$\rho_{B}$, with MH and MRW scattering amplitudes. Taken from Ref.
\cite{LiZH06b}.} \label{OPden}
\end{figure}

With the free \textsl{NN} forward scattering amplitudes of MH and
MRW as well as the neutron and proton scalar and vector densities
obtained from the RMF theory with the parameter set HA, one can
also investigate the real and imaginary parts of the scalar and
vector Dirac optical potentials for nucleons in symmetric nuclear
matter as functions of nucleon energy. In Fig. \ref{OPden}, the
energy dependence of the Dirac optical potential is depicted at
three nucleon densities $\rho _{\text{B}}=0.08$ fm$^{-3}$ (panel
(a)), $0.16$ fm$^{-3}$ (panel (b)), and $0.24$ fm$^{-3}$ (panel
(c)). In each panel, both the scalar and vector optical potentials
based on the generalized amplitudes of MH and the original
amplitudes of MRW are shown. In calculating the Dirac optical
potential from the RIA of MH, the Pauli blocking effect as well as
the modifications from using the pseudovector coupling for pion
and the exchange term contribution have been included. For all
densities considered here, the energy dependence of the scalar and
vector optical potentials from the RIA of MH are significantly
reduced compared to those from the original RIA of MRW, especially
for the real part at low energies. Furthermore, their difference
becomes larger with increasing density. These results thus
demonstrate clearly the importance of the medium modifications
introduced in the RIA of MH for nucleons at lower energies. For
all three considered densities, the RIA of MH generates, on the
other hand, a similar systematic difference or isospin splitting
in the Dirac optical potentials for protons and neutrons in
asymmetric nuclear matter as in the original RIA of MRW
\cite{Che05c}. In particular, the neutron exhibits a stronger real
but weaker imaginary scalar and vector potentials than those of
the proton in neutron-rich nuclear matter.

\subsubsection{The Schr\"{o}dinger-equivalent optical potential}

\begin{figure}[th]
\centering
\includegraphics[scale=0.85]{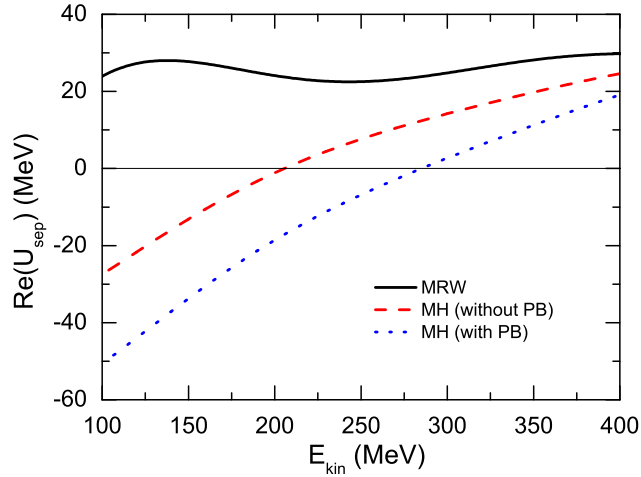}
\caption{(Color online) Energy dependence of the real part of the
nucleon Schr\"{o}dinger-equivalent potential at normal density in
symmetric nuclear matter from the original RIA of MRW and from the
RIA of MH with and without Pauli blocking correction. Taken from
Ref. \cite{LiZH06b}.} \label{Usep0}
\end{figure}

Fig. \ref{Usep0} shows the real part of the nucleon
Schr\"{o}dinger-equivalent potential in symmetric nuclear matter
at normal density obtained from above Dirac optical potential.
Because of uncertainties in the medium modification due to the
Pauli blocking effect at lower energies, results both with (dotted
line) and without Pauli blocking (dashed line) corrections based
on the MH free \textsl{NN} scattering amplitudes are shown. For
comparison, the real part of the nucleon
Schr\"{o}dinger-equivalent potential from the original RIA of MRW
(solid line) is also shown. The nucleon Schr\"{o}dinger-equivalent
potential from the original RIA of MRW is seen to be always
positive at considered energy range of $E_{\mathrm{kin}}=100\sim
400$ MeV. Including the pseudovector coupling and exchange term
corrections in the RIA of MH (dashed line), the behavior of the
resulting Schr\"{o}dinger-equivalent potential as a function of
energy is significantly improved, varying from $-27$\ MeV at
$E_{\mathrm{kin}}=100$ MeV to $0$\ MeV at
$E_{\mathrm{kin}}\thickapprox 200$ MeV and then continues to
increase monotonically as the nucleon energy increases. This
improvement is due to the fact that the pseudovector coupling and
exchange term corrections lead to a smaller strength for the
imaginary scalar and vector \textsl{NN} forward scattering
amplitudes while keep their sum roughly unchanged as shown in Fig.
\ref{FFSV}. From Eq. (\ref{trhor}), therefore, the term
$U_{S}+U_{0}$ does not change while the last two terms in Eq.
(\ref{ReSEP}) are reduced strongly and thus a smaller
Schr\"{o}dinger-equivalent potential is obtained. When the Pauli
blocking effect is further taken into account, the resulting
Schr\"{o}dinger-equivalent potential is seen to be more attractive
at the whole energy range considered here. At high enough energy,
the Schr\"{o}dinger-equivalent potentials from above three
approaches become similar as expected since the effects from both
Pauli blocking and exchange contribution play a minor role at high
energies.

With momentum/energy independent scalar and vector potentials from
the RMF calculation, the nucleon Schr\"{o}dinger-equivalent
potential in symmetric nuclear matter at normal nuclear density
exhibits already a linear energy dependence according to Eq.
(\ref{ReSEP}), with a change from negative to positive values
typically at kinetic energies between about $130$ MeV and $300$
MeV, depending on the model parameters \cite{Che07}. This behavior
is consistent with empirical results from the global relativistic
optical-model analysis of experimental data from proton-nucleus
scatterings, which also indicate that the nucleon
Schr\"{o}dinger-equivalent potential in symmetric nuclear matter
at normal nuclear density changes from negative to positive values
around $200$ MeV \cite{Che07}.

\subsubsection{The Nuclear symmetry potential}

\begin{figure}[th]
\centering
\includegraphics[scale=1.1]{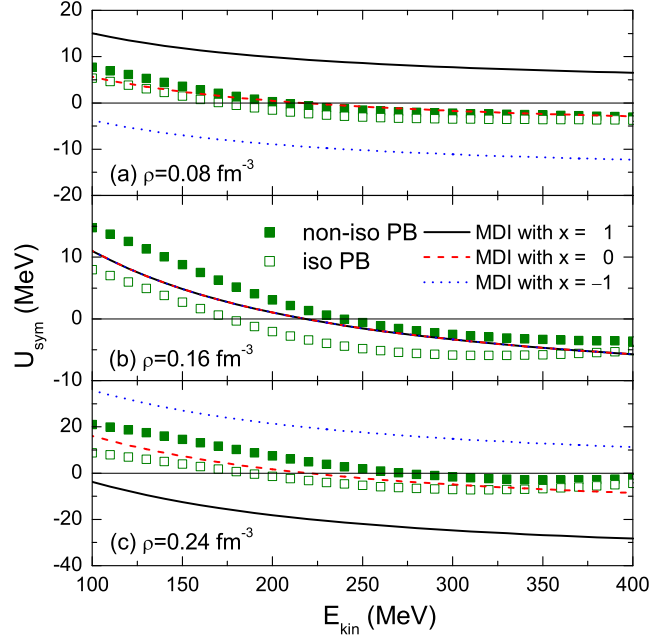}
\caption{{\protect\small (Color online) Energy dependence of the
nuclear symmetry potential from the RIA of MH with
isospin-dependent (open squares) and isospin-independent (solid
squares) Pauli blocking corrections, as well as the
phenomenological MDI interaction with }$x=1${\protect\small \
(solid line), }$0${\protect\small \ (dashed
line)}$,${\protect\small \ and }$-1${\protect\small \ (dotted
line) at fixed baryon densities of
}$\protect\rho_{B}=0.08${\protect\small \ \
fm}$^{-3}${\protect\small \ (a), }$0.16${\protect\small \
fm}$^{-3}${\protect\small \ (b), and }$0.24${\protect\small \
fm}$^{-3}${\protect\small \ (c). Taken from Ref. \cite{LiZH06b}.}}
\label{UsymEkin}
\end{figure}

For the nuclear symmetry potential based on the scattering
amplitudes of MH, its energy dependence is shown in Fig.
\ref{UsymEkin} for both cases of using isospin-dependent (Eq.
(\ref{isopb})) and isospin-independent Pauli blocking (Eq.
(\ref{noisopb})) corrections at fixed baryon densities of $\rho
_{B}=0.08$ fm$^{-3}$ (panel (a)), $0.16$ fm$^{-3}$ (panel (b)),
and $0.24$ fm$^{-3}$ (panel (c)). Also shown are results from the
phenomenological parametrization of the isospin- and
momentum-dependent nuclear mean-field potential, i.e., the MDI
interaction with $x=-1$, $0$, and $1$
\cite{LiBA04a,Che04,Che05a,Das03}. It is seen that at fixed baryon
density, the nuclear symmetry potential generally decreases with
increasing nucleon energy. At low nuclear density ($\rho
_{\text{B}}=0.08$ fm$^{-3}$), the symmetry potentials from the RIA
of MH with isospin-dependent and isospin-independent Pauli
blocking corrections are almost the same, especially at energies
higher than $E_{\mathrm{kin}}\geqslant 300$ MeV, where the Pauli
blocking correction is expected to be unimportant. The isospin
dependence of the Pauli blocking effect becomes, however, stronger
as the nuclear density increases, and an appreciable difference in
the resulting symmetry potentials is seen. The difference
disappears, however, for high energy nucleons when the Pauli
Blocking effect becomes negligible. At normal density ($\rho
_{\text{B}}=0.16$ fm$^{-3}$), the nuclear symmetry potential
changes from positive to negative values at nucleon kinetic energy
around $200$ MeV, with the one using the isospin-dependent Pauli
blocking correction at a somewhat lower energy than that using the
isospin-independent Pauli blocking correction. Comparing with
results from the MDI interaction, the one with $x=0$ is in
surprisingly good agreement with the results of the RIA by MH in
the region of nuclear densities and energies considered here.
Although the MDI interaction with different $x$ values give by
construction same symmetry potential at normal nuclear matter
density as shown in Fig.~\ref{UsymEkin}(b), the one with $x=0$ has
been found to give reasonable descriptions of the data on the
isospin diffusion in intermediate energy heavy ion collisions and
the neutron skin thickness of $^{208}$Pb
\cite{LiBA05c,Che05a,Che05b,Ste05b}.

\begin{figure}[th]
\centering
\includegraphics[scale=0.85]{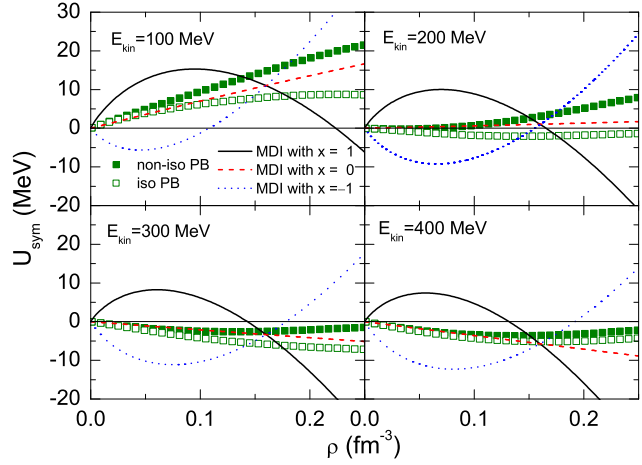}
\caption{{\protect\small (Color online) Density dependence of the
nuclear symmetry potential using RIA with isospin dependent and
independent Pauli blocking, as well as the results from the
phenomenological interaction MDI with} $x=-1,0,${\protect\small \
and }$1$ {\protect\small at nucleon kinetic energies of
}$E_{\mathrm{kin}}=100$ {\protect\small MeV, }$200$
{\protect\small MeV, }$300$ {\protect\small MeV and }$400$
{\protect\small MeV. Taken from Ref. \cite{LiZH06b}.}}
\label{UsymDen}
\end{figure}

The density dependence of the nuclear symmetry potential with
isospin-dependent and isospin-independent Pauli blocking corrections
at nucleon kinetic energies of $100$, $200$, $300$, and $400$ MeV
are shown in Fig.~\ref{UsymDen} together with corresponding results
from the MDI interaction with $x=-1$, $0$, and $1$. It is clearly
seen that the nuclear symmetry potentials are always positive at
lower nucleon kinetic energy of $E_{\mathrm{kin}}=100$ MeV while it
may become positive or negative at $E_{\mathrm{kin}}=200$ MeV
depending on if the Pauli blocking effect is isospin dependent or
not. At higher energies ($E_{\mathrm{kin}}=300$ and $400 $ MeV), the
nuclear symmetry potential is always negative in the density region
considered here. These features are consistent with the results
shown in Fig. \ref{UsymEkin}. Compared with results from the MDI
interaction, the nuclear symmetry potential from the generalized RIA
of MH reproduces nicely the results obtained from the MDI
interaction with $x=0$ when $\rho _{\text{B}}\lesssim 0.2$ fm$^{-3}$
even for nucleon kinetic energy as high as $400$ MeV. Moreover, in
the energy region of $E_{\mathrm{kin}}=100\sim 300$ MeV, the nuclear
symmetry potential from MDI interaction with $x=0$ always lies
between the results from the RIA of MH with isospin-dependent and
isospin-independent Pauli blocking corrections. On the other hand,
the MDI interaction with both $x=-1$ and $1$\ display a very
different density dependence from the results using the RIA of MH.

The above results indicate that in the relativistic impulse
approximation of MH, the low and intermediate energy behavior of
the Dirac optical potential has been significantly improved by
including the pseudovector coupling for pion, the exchange
contribution, and the medium modification due to the Pauli
blocking effect. Compared with results from the original RIA of
MRW, the generalized RIA of MH gives essentially identical real
parts of the scalar and vector amplitudes for both proton-proton
and neutron-proton scattering but significantly reduced strength
in their imaginary parts at low energies $E_{\mathrm{kin}}\leq
300$ MeV. These improvements in the RIA of MH modify the real
scalar and vector Dirac optical potentials at lower energies and
make the resulting energy dependence of the
Schr\"{o}dinger-equivalent potential and nuclear symmetry
potential more reasonable.

At saturation density, the nuclear symmetry potential has been found
to change from positive to negative values at nucleon kinetic energy
of about $200$ MeV. This is a very interesting result as it implies
that the proton (neutron) feels an attractive (repulsive) symmetry
potential at lower energies but repulsive (attractive) symmetry
potential at higher energies in asymmetric nuclear matter. Adding
also the repulsive Coulomb potential, a high energy proton in
asymmetric nuclear matter thus feels a very stronger repulsive
potential. This behavior of the nuclear symmetry potential can be
studied in intermediate and high energy heavy-ion collisions that
are induced by radioactive nuclei, e.g., by measuring two-nucleon
correlation functions \cite{Che03a} and light cluster production
\cite{Che03b} in these collisions.

Comparing the energy and density dependence of the nuclear
symmetry potential from the RIA of MH with that from the MDI
interaction indicates that results from the MDI interaction with
$x=0$ are in good agreement with those from the RIA of MH. For
baryon densities less than $0.25$ fm$^{-3}$ and nucleon energies
less than $400$ MeV as considered here, the nuclear symmetry
potential from the MDI interaction with $x=0$ lies approximately
between the two results from the RIA of MH with isospin-dependent
and isospin-independent Pauli blocking corrections. This provides
a strong evidence for the validity of the MDI interaction with
$x=0$ in describing both the isospin diffusion data in
intermediate energy heavy ion collisions and the neutron skin
thickness data for $^{208}$Pb.

\subsection{The low-energy behavior of the nuclear symmetry potential}

Compared with its high energy behavior, the low energy behavior of
the nuclear symmetry potential is much more involved since the
medium effects become much more important and complicated.
Empirically, a systematic analysis of a large number of
nucleon-nucleus scattering experiments and (p,n) charge-exchange
reactions at beam energies up to about $100$ MeV has shown that
the data can be very well described by the parametrization
\begin{eqnarray}\label{dat0}
U_{\rm sym}=a-bE
\end{eqnarray}
with $a\approx 22-34$ MeV and $b\approx 0.1-0.2$
\cite{Sat69,Hof72,Hod94,Kon03}. Although the uncertainties in both
parameters $a$ and $b$ are large, the nuclear symmetry potential
at nuclear matter saturation density, which is usually called the
Lane potential $U_{\mathrm{Lane}}$ \cite{Lan62}, clearly decreases
approximately linearly with increasing beam energy $E$. This
provides a stringent constraint on the low energy behavior of the
nuclear symmetry potential at saturation density. On the other
hand, the low energy behavior of the nuclear symmetry potential at
densities away from saturation density is presently not known
empirically. In the following, we review the present status of the
low-energy behavior of the nuclear symmetry potential from the
microscopic and phenomenological theoretical approaches.

\subsubsection{Microscopic approaches}

\begin{figure}[htb]
\centering
\includegraphics[scale=0.85]{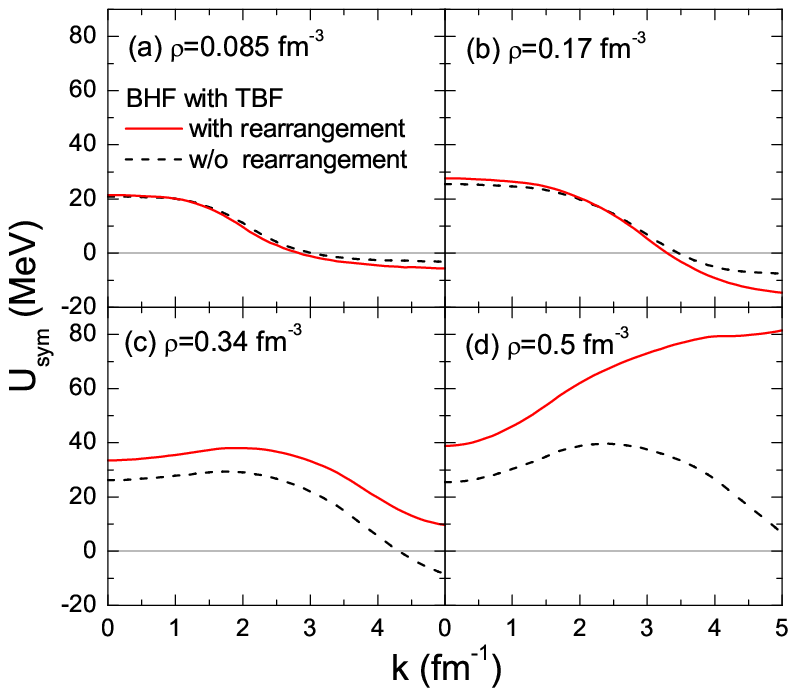}
\includegraphics[scale=0.8]{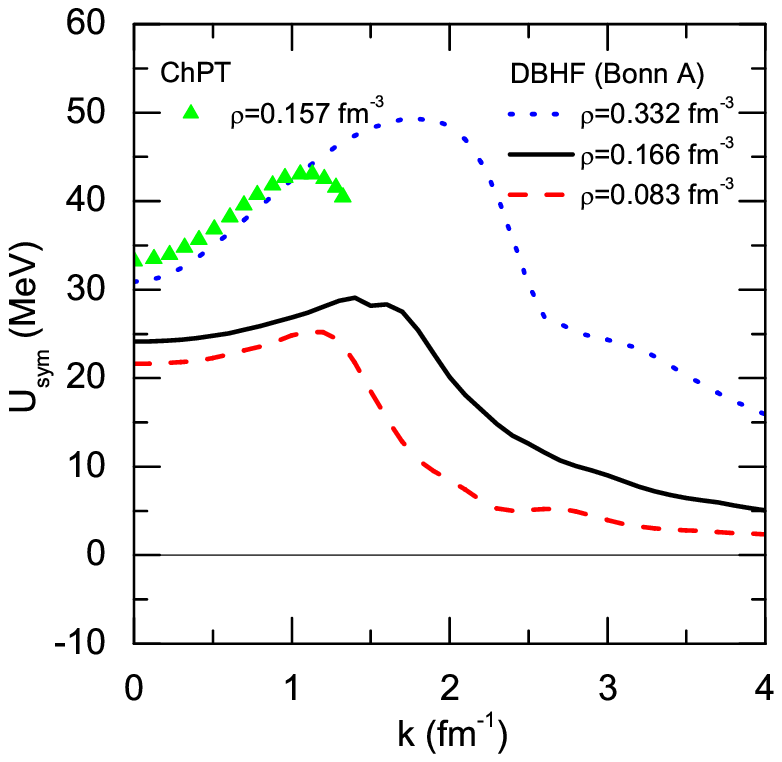}
\caption{Left window: Momentum dependence of the nuclear symmetry
potential at $\protect\rho= 0.085$, $0.17$, $0.34$ and $0.5$
fm$^{-3}$ in the BHF approach with (solid curves) and without
(dashed curves) TBF rearrangement modifications \cite{Zuo06}. Right
window: Same as left panel but for the DBHF approach at
$\protect\rho= 0.083$, $0.166$, and $0.332$ fm$^{-3}$ \cite{Fuc05b}
and the ChPT approach at $\protect\rho= 0.157$ fm$^{-3}$
\cite{Fri05}.} \label{UsymBHF}
\end{figure}

The nuclear symmetry potential has been extensively studied in
both non-relativistic and relativistic BHF approach as well as in
the ChPT approach. Fig. \ref{UsymBHF} shows the momentum
dependence of the nuclear symmetry symmetry potential at different
densities from the microscopic BHF and DBHF approaches. The left
window displays the momentum dependence of the symmetry potential
at four different densities obtained from recent BHF calculations
with and without the TBF rearrangement contribution \cite{Zuo06}.
The results indicate that the symmetry potential obtained without
the TBF rearrangement contribution stays as a constant or
increases slightly with momentum for nucleons with low momenta but
decreases when the momentum is higher and becomes negative at
sufficient high momenta. Including the TBF rearrangement
contribution does not change much the symmetry potential at lower
densities due to its small effect at lower densities and the
cancelation between its contributions to the neutron and proton
single-particle potentials \cite{Zuo06}. On the other hand, the
TBF rearrangement contribution enhances considerably the symmetry
potential at high densities. At sufficient high densities, the TBF
rearrangement contribution even modifies qualitatively the
momentum dependence of the symmetry potential. For example, at
density of $0.5$ fm$^{-3}$, the symmetry potential without the TBF
rearrangement contribution decreases as a function of momentum in
the relatively higher momentum region, whereas the one predicted
with the TBF rearrangement contribution increases monotonically in
the whole momentum region.

The right window of Fig.~\ref{UsymBHF} displays the momentum
dependence of the symmetry potential at $\rho= 0.083$, $0.166$,
and $0.332$ fm$^{-3}$ obtained from the DBHF calculations with the
Bonn A potential \cite{Fuc05b}. The result from the ChPT approach
including the effects from two-pion exchange with single and
double virtual $\Delta$(1232)-isobar excitation at $\rho= 0.157$
fm$^{-3}$ is also shown \cite{Fri05}. Similar to the results
obtained from the non-relativistic BHF approach, the symmetry
potentials obtained from the DBHF and ChPT approaches increase
slightly with momentum for nucleons with lower momenta but
decrease with momentum at higher momenta for the densities
considered here. However, the symmetry potential from the DBHF
approach seems not to change sign at momenta up to $4$ fm$^{-1}$
even at lower densities.

The above results indicate that all nonrelativistic BHF,
relativistic DBHF and ChPT approaches exhibit a common feature in
the momentum dependence of the symmetry potential, namely, it
stays as a constant or increases slightly with momentum for
nucleons with lower momenta but decreases with momentum at higher
momenta for densities up to about two times the normal nuclear
matter density. At higher densities up to about three times the
normal density, the BHF calculation indicates that the momentum
dependence of the symmetry potential strongly depends on the TBF
rearrangement contribution.

\subsubsection{Phenomenological approaches}

\begin{figure}[th]
\centering
\includegraphics[scale=1.2]{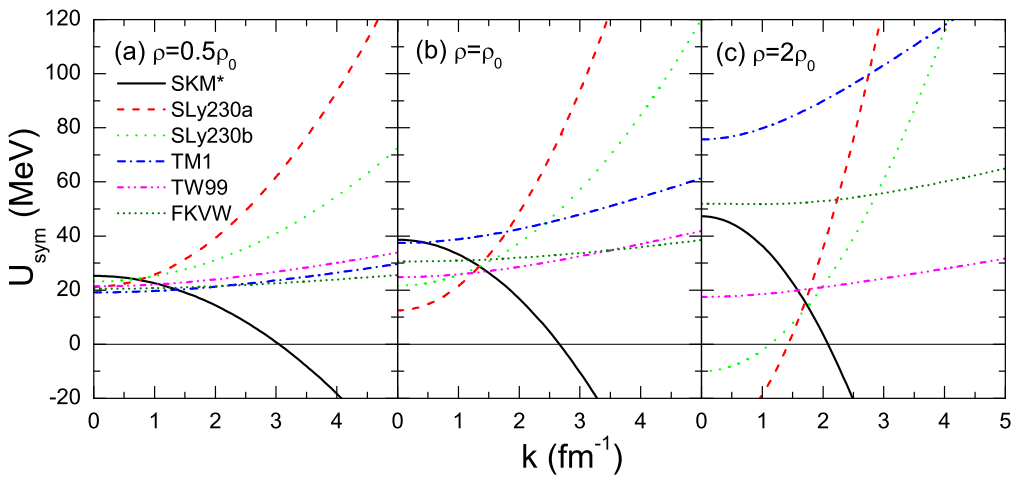}
\caption{(Color online) Same as Fig.~\protect\ref{UsymBHF} but for
SHF and RMF approaches at $\protect\rho= 0.5\protect\rho_{0}$,
$1.0\protect\rho _{0}$ and $2\protect\rho _{0}$.}
\label{UsymKSHFRMF}
\end{figure}

Besides the microscopic approaches, there are many predictions on
the momentum dependence of the symmetry potential based on
phenomenological approaches. Shown in Fig.~\ref{UsymKSHFRMF} is
the momentum dependence of the symmetry potential from the SHF and
RMF models using some typical interaction parameter sets. For the
SHF calculations, the symmetry potential is seen to decrease with
momentum with the old parameter sets (such as SKM$^*$ shown here)
while some new parameter sets from the Lyon group \cite{Cha97}
(SLy230a and SLy230a shown here) give the opposite results. For
RMF models \cite{Che07}, all the interactions TM1, TW99, and FKVW
shown here predict symmetry potentials that increase with
momentum. In addition, the density dependence of the symmetry
potential at a fixed momentum is strongly model dependent. Most
parameter sets from the SHF and RMF models display a weak momentum
dependence at low momenta. This feature is consistent with the
results from the microscopic approaches shown above. However, the
high momentum behavior is significantly different, especially
between the microscopic and phenomenological approaches.

\begin{figure}[th]
\centering
\includegraphics[scale=1.2]{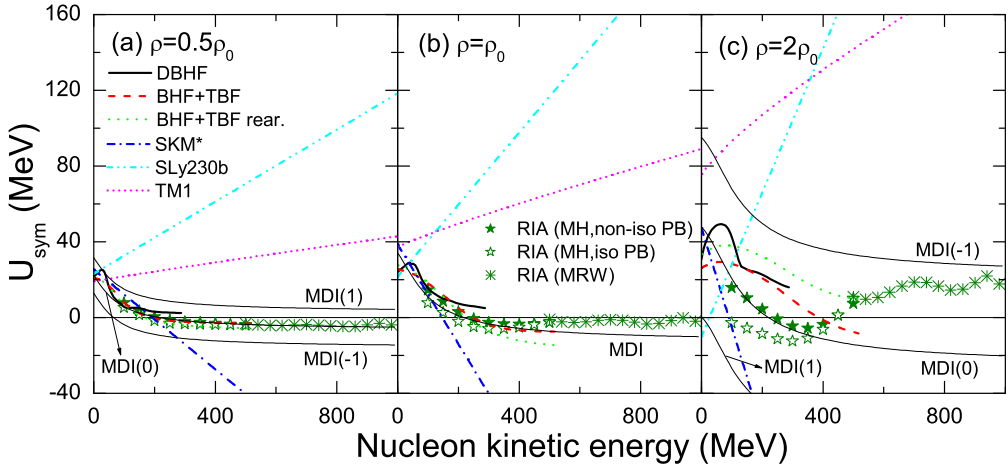}
\caption{(Color online) Nuclear symmetry potential as a function of
nucleon kinetic energy from different theoretical approaches.}
\label{UsymEkinAll}
\end{figure}

Fig.~\ref{UsymEkinAll} summarizes the results for the kinetic
energy dependence of the symmetry potential at $\rho=
0.5\rho_{0}$, $1.0\rho _{0}$ and $2\rho _{0}$ from different
theoretical approaches, including the microscopic DBHF and BHF
approaches with or without the TBF rearrangement contribution, the
phenomenological SHF approach with SKM$^*$ and SLy230a, the RMF
model with TM1, the RIA of MH and MRW as well as the MDI
interaction with $x=-1$, $0$, and $1$. At densities below the
saturation density, all models show a similar kinetic energy
dependence for the symmetry potential except the SLy230a and TM1
which give too large and the SKM$^*$ which gives too small values
for the symmetry potential at higher nucleon kinetic energies. At
higher densities (around two times the saturation density),
results from the microscopic DBHF and BHF approaches seem to be
consistent with each other while the RIA and the MDI interaction
with $x=0$ seem to give significantly smaller values for the
symmetry potential at lower nucleon kinetic energies compared with
the DBHF and BHF, although they predict a similar symmetry
potential at lower densities. For other interactions, their
results at higher densities exhibit very different behaviors for
the momentum dependence of the symmetry potential compared with
those from the DBHF, BHF, RIA approaches and the MDI interaction
with $x=0$. These results are of particular relevance for
understanding the dynamic of intermediate and high energy
heavy-ion collisions induced by radioactive nuclei.

\subsection{Isospin-splitting of the neutron and proton effective
masses in neutron-rich matter}

One of the important properties that characterize the propagation
of a nucleon in nuclear medium is its effective mass
\cite{Jeu76,Neg81,Dob99}. The latter describes the effects related
to the non-locality of the underlying nuclear effective
interactions and the Pauli exchange effects in many-fermion
systems. In neutron-rich matter, there arises an interesting new
question on whether the effective mass $m_n^*$ for neutrons is
higher or lower than that for protons $m_p^*$. Knowledge about
nucleon effective mass in neutron-rich matter is essential for
understanding a number of properties of neutron stars
\cite{Coo85,Bet90,Far01}. It is also important for the reaction
dynamics of nuclear collisions induced by radioactive nuclei, such
as the degree and rate of isospin diffusion, the neutron-proton
differential collective flow, and the isospin equilibration
\cite{LiBA04a,Riz04,Per02,Pan92}. Moreover, it influences the
magnitude of shell effects and the basic properties of nuclei far
from stability \cite{Dob99}. However, even the sign of the
neutron-proton effective mass splitting in asymmetric matter is
still a rather controversial theoretical issue.

\subsubsection{The nucleon effective mass}

In non-relativistic approaches, the effective mass $m_{\tau}^*$ of a
nucleon $\tau$ (n or p) measures the momentum (or equivalently
energy) dependence of the nucleon single-particle potential
$U_{\tau}$, and it can be defined via following three equivalent
expressions \cite{Jeu76}
\begin{eqnarray}\label{effmass}
\frac{m_{\tau }^{\ast }}{m_{\tau }}=1-\frac{dU_{\tau }(k,\epsilon
_{\tau}(k))}{d\epsilon _{\tau }}=\frac{k}{m_{\tau
}}\frac{dk}{d\epsilon _{\tau }} =\left[ 1+\frac{m_{\tau
}}{k}\frac{dU_{\tau }(k,\epsilon _{\tau }(k))}{dk} \right]
^{-1},\label{mstar}
\end{eqnarray}
where $\epsilon _{\tau }(k)$ is the nucleon single-particle energy
satisfying the following dispersion relation
\begin{eqnarray}
\epsilon _{\tau }(k)=\frac{k^{2}}{2m_{\tau }}+U_{\tau }(k,\epsilon
_{\tau }(k)).
\end{eqnarray}
The fact that $U_{\tau }(k,\epsilon _{\tau }(k))$ depends on $k$ and
$\epsilon _{\tau }$ leads, respectively, to the following so-called
$k$-mass $\widetilde{m}_{\tau }$ and $E$-mass $\overline{m}_{\tau
}$:
\begin{eqnarray}
\frac{\widetilde{m}_{\tau }}{m_{\tau }} =\left[ 1+\frac{m_{\tau
}}{k} \frac{\partial U_{\tau }(k,\epsilon _{\tau }(k))}{\partial
k}\right] ^{-1} \qquad {\rm and} \qquad \frac{\overline{m}_{\tau
}}{m_{\tau }} =1-\frac{\partial U_{\tau }(k,\epsilon _{\tau
}(k))}{\partial \epsilon _{\tau }}.
\end{eqnarray}
The $k$-mass $\widetilde{m}_{\tau }$ and $E$-mass
$\overline{m}_{\tau }$ reflect the non-locality in spatial
coordinates and in time, respectively. It can easily be checked that
the three masses $m_{\tau }^{\ast }$, $\widetilde{m}_{\tau }$ and
$\overline{m}_{\tau }$ satisfy the following relation
\begin{eqnarray}
\frac{m_{\tau }^{\ast }}{m_{\tau }}=\frac{\widetilde{m}_{\tau
}}{m_{\tau }}\cdot \frac{\overline{m}_{\tau }}{m_{\tau }}.
\end{eqnarray}
The effective mass is usually evaluated at the Fermi momentum
$k_{\tau }^{F}$ or corresponding Fermi energy $\epsilon _{\tau
}(k_{\tau }^{F})$, yielding the so-called Landau mass that is
related to the $f_1$ Landau parameter of a Fermi liquid.

In relativistic models, there exist many different definitions for
the nucleon effective mass in the literature. It has been argued
that it is the Lorentz mass $M_{\mathrm{Lorentz}}^{\ast }$, which
characterizes the energy dependence of the
Schr\"{o}dinger-equivalent potential $U_{\mathrm{SEP},\tau }$ in the
relativistic model, that should be compared with the usual
non-relativistic nucleon effective mass extracted from analyses
carried out in the framework of non-relativistic optical and shell
models \cite{Jam89}. In Ref. \cite{Fuc05}, a non-relativistic mass
$m_{NR,\tau }^{\ast }$ has been introduced via the momentum
dependence of the Schr\"{o}dinger-equivalent potential
$U_{\mathrm{SEP},\tau }$, i.e.,
\begin{eqnarray}
\frac{m_{NR,\tau }^{\ast }}{m_{\tau }}=\left[ 1+\frac{m_{\tau
}}{k}\frac{dU_{\mathrm{SEP},\tau }}{dk}\right] ^{-1}.
\end{eqnarray}
In standard relativistic mean-field models where the scalar and
vector nucleon self-energies are independent of momentum or energy,
the nonrelativistic mass $m_{NR,\tau }^{\ast }$ is the same as the
Lorentz mass $M_{\mathrm{Lorentz,\tau}}^{\ast }\equiv
m_\tau(1-dU_{\rm SEP,\tau}/d\epsilon_\tau)$ \cite{Jam89}, if one
neglects relativistic corrections to the kinetic energy in the
single-particle energy, which has been assumed in Ref. \cite{Fuc05}.

\subsubsection{Microscopic approaches}

\begin{figure}[htb]
\centering
\includegraphics[scale=0.3]{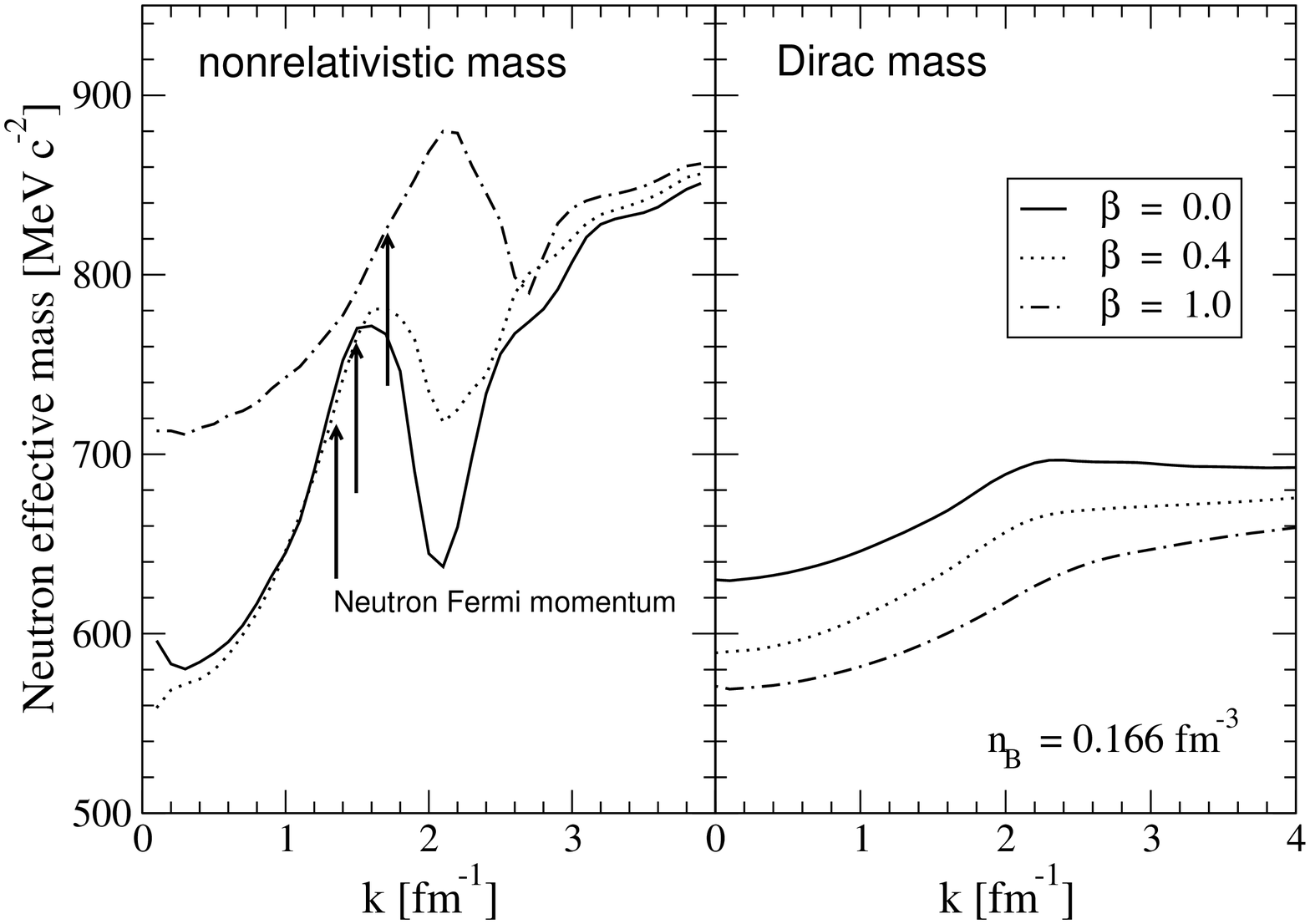}
\includegraphics[scale=0.3]{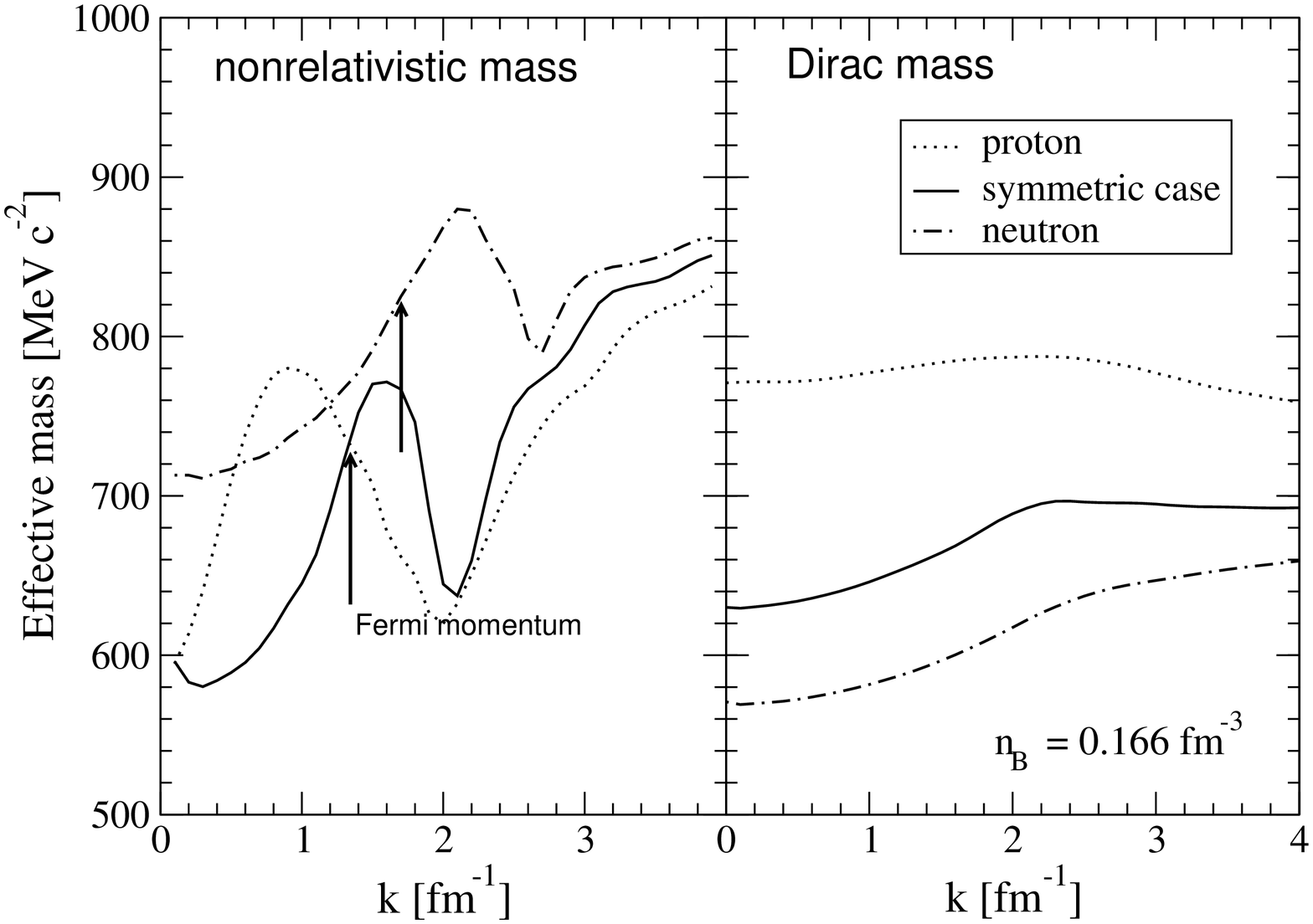}
\caption{Left window: Momentum dependence of the neutron effective
mass for various values of the asymmetry parameter $\beta$ at fixed
nuclear density at $\protect\rho= 0.166$ fm$^{-3}$ in the DBHF
approach. Right window: Same as left window but for neutrons and
protons at a fixed value of the asymmetry parameter $\beta = 1$.
Also shown is the effective mass in symmetric nuclear matter.
\protect Taken from Ref. \cite{Fuc05b}.} \label{IsoEffMassDBHF}
\end{figure}

In the following, we review the present status on the
isospin-splitting of neutron and proton effective masses in
neutron-rich nuclear matter from the microscopic DBHF and BHF
approaches. Shown in Fig.~\ref{IsoEffMassDBHF} is the momentum
dependence of the effective mass from recent DBHF calculations
\cite{Fuc05,Fuc05b}. In the left window of
Fig.~\ref{IsoEffMassDBHF}, the neutron nonrelativistic mass
($m_{NR}^{\ast }$) and Dirac mass are plotted for various values
of the asymmetry parameter $\beta$ previously called $\alpha$ at
nuclear density of $0.166$ fm$^{-3}$. Similar results for neutrons
and protons at a fixed value of the asymmetry parameter $\beta =
1$ (neutron matter) are shown in the right window of
Fig.~\ref{IsoEffMassDBHF}. One can see clearly that there exists a
pronounced peak in the nonrelativistic mass slightly above Fermi
momentum. The peak structure of the nonrelativistic mass is from
the nonlocalities in time which are generated by the Brueckner
ladder correlations due to the scattering to intermediate
off-shell states, inducing thus a strong momentum dependence with
a characteristic enhancement of the $E$-mass slightly above the
Fermi surface \cite{Fuc05,Fuc05b,Jam89,Mah85,Fri02,Has04}. This
peak structure reflects - as a model-independent result - the
increase of the level density due to the vanishing imaginary part
of the optical potential at the Fermi surface, which for example
is also seen in shell-model calculations \cite{Jeu76,Jam89,Mah85}.
As shown in the right window of Fig.~\ref{IsoEffMassDBHF}, the
nonrelativistic mass and the relativistic Dirac mass display
opposite isospin-splitting, i.e., in neutron-rich nuclear matter,
the neutron Dirac mass is smaller than the proton Dirac mass while
the nonrelativistic mass shows the opposite behavior, except
around the peak slightly above the proton Fermi momentum. This is
especially the case for the nonrelativistic mass at the Fermi
momentum. This opposite behavior of the nonrelativistic mass
deduced from the relativistic Dirac mass, i.e., $m_{NR,n}^{\ast }>
m_{NR,p}^{\ast }$, is in agreement with the results from
nonrelativistic BHF and most SHF calculations as will be discussed
in the following.

\begin{figure}[th]
\centering
\includegraphics[scale=1.0]{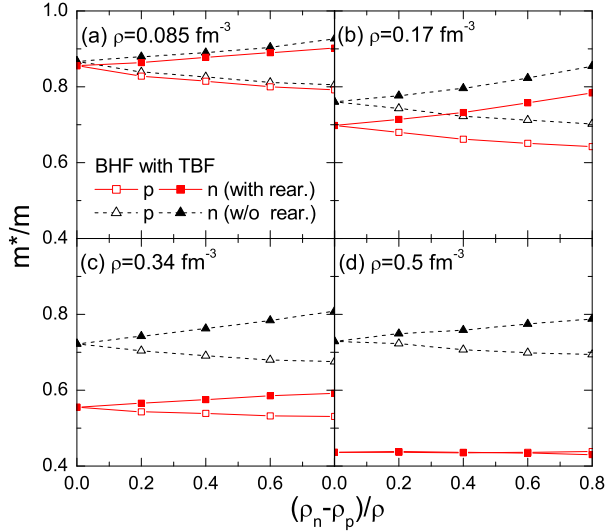}
\caption{(Color online) Isospin splitting of neutron and proton
effective masses in neutron-rich nuclear matter from the BHF
calculation with (solid lines) and without (dashed lines) the TBF
rearrangement contribution for different densities \cite{Zuo06}.}
\label{IsoEffMassBHF}
\end{figure}

Fig.~\ref{IsoEffMassBHF} shows recent nonrelativistic BHF
predictions \cite{Zuo06} for the neutron and proton effective
masses at their respective Fermi momenta, i.e., $m_{n}^{\ast }(k =
k^F_p)$ and $m_{p}^{\ast }(k = k^F_n)$ as functions of the isospin
asymmetry in neutron rich nuclear matter for the two cases with
(squares) and without (triangles) considering the TBF
rearrangement contribution. In both cases, the neutron effective
mass increases while that of a proton decreases with respect to
their common value in symmetric nuclear matter as the nuclear
matter becomes more neutron rich; i.e., the predicted isospin
splitting of the proton and neutron effective masses in
neutron-rich matter is such that $m_{n}^{\ast }(k = k^F_p)>
m_{p}^{\ast }(k = k^F_n)$, which is consistent with the above
relativistic DBHF predictions for the nonrelativistic mass.

For the BHF calculations without the TBF rearrangement contribution,
the absolute magnitude of the neutron-proton effective mass
splitting in neutron-rich matter is about the same for all four
densities considered here (i.e., $\rho = 0.085$, $0.17$, $0.34$, and
$0.5$ fm$^{-3}$), indicating a weak density dependence of the
isospin splitting. On the other hand, including the TBF
rearrangement contribution leads to a quite sensitive density
dependence for the magnitude of the effective mass isospin
splitting. At low densities around and below the normal nuclear
matter density, the magnitude of the isospin splitting is not
affected much by the TBF rearrangement contribution. However, the
rearrangement effect induced by the TBF gets increasingly larger as
the nuclear medium becomes denser, and it hinders the isospin
splitting of the neutron and proton effective masses in dense
neutron-rich matter. At high enough density (such as $\rho = 0.5$
fm$^{-3}$), the TBF rearrangement effect even suppresses almost
completely the isospin splitting. This disappearance of the isospin
splitting of the nucleon effective mass at high density neutron-rich
nuclear matter is due to the fact that the neutron and proton
single-particle potentials increase almost at the same rate as their
common one in symmetric nuclear matter as a function of momentum
when the TBF rearrangement effect is included \cite{Zuo06}.

The neutron-proton Dirac mass difference in asymmetric nuclear
matter was also studied recently in the framework of the
medium-modified Skyrme model that includes energy-dependent pion
optical potentials \cite{meissner07}. It was found that the
neutron-proton mass difference decreases strongly with increasing
density and isospin asymmetry of nuclear matter. This result is in
qualitative agreement with those from the relativistic mean-field
model as well as the nonrelativistic variation approach.

\subsubsection{Phenomenological approaches}

\begin{figure}[th]
\centering
\includegraphics[scale=1.2]{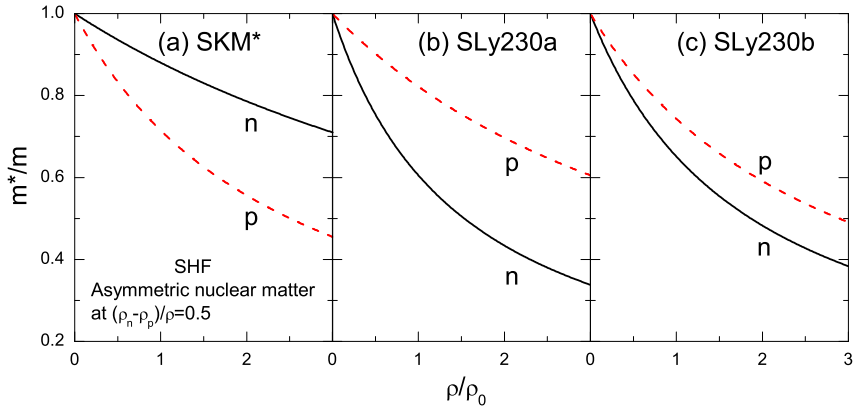}
\caption{(Color online) Same as Fig.~\protect\ref{UsymBHF} but for
SHF and RMF approaches at $\protect\rho= 0.5\protect\rho_{0}$,
$1.0\protect\rho _{0}$ and $1.5\protect\rho _{0}$.}
\label{IsoEffMassSHF}
\end{figure}

As discussed above, microscopic many-body theories, such as the
relativistic DBHF \cite{Fuc04,Ma04,Sam05a,Fuc05,Fuc05b} and
nonrelativistic BHF approaches \cite{Bom91,Zuo02,Zuo06,Zuo99},
predict that $m_n^*>m_p^*$ in neutron-rich matter. On the other
hand, opposite results are predicted by some effective
interactions within phenomenological approaches including the SHF
and potential models as well as all RMF models
\cite{Riz04,Che07,Beh05}. As an example, shown in
Fig.~\ref{IsoEffMassSHF} are the SHF predictions with SKM$^*$,
SLy230a and SLy230a on the density dependence of the neutron and
proton effective masses at their respective Fermi momenta at a
fixed isospin asymmetry of $0.5$. It is seen that the old
parameter set SKM$^*$ predicts an isospin-splitting of
$m_n^*>m_p^*$ while the new parameter sets from the Lyon group
\cite{Cha97} (SLy230a and SLy230a shown here) give opposite
results. Actually, almost all Skyrme forces predict an
isospin-splitting of $m_n^*>m_p^*$ except some new Lyon Skyrme
forces. Unfortunately, up to now almost nothing is known
experimentally about the neutron-proton effective mass splitting
$m_n^*-m_p^*$ in the neutron-rich medium.

Information on the isospin-splitting of the nucleon effective mass
can be obtained from the momentum dependence of the symmetry
potential. In a recent study by Rizzo {\it et al.} \cite{Riz04},
nuclear reactions with radioactive beams were proposed as a tool
to disentangle the sign of the neutron-proton effective mass
splitting in neutron-rich matter. On the other hand, Li has
recently shown that an effective mass splitting of $m_n^*<m_p^*$
leads to a symmetry potential that is inconsistent with the energy
dependence of the Lane potential constrained by existing
nucleon-nucleus scattering data \cite{LiBA04c}. In both Ref.
\cite{Riz04} and Ref. \cite{LiBA04c}, the single-nucleon potential
$U_{\tau}$ is taken from the phenomenological model of Bombaci
\cite{Bom01}, i.e.,
\begin{eqnarray}
U_\tau(k,u,\delta)&=&Au+Bu^\sigma-\frac{2}{3}(\sigma-1)\frac{B}{\sigma+1}
\umd{3}u^{\sigma}\delta^2\nonumber\\
&\pm& \qd{-\frac{2}{3}A \umd{0}u -
\frac{4}{3}\frac{B}{\sigma+1}\umd{3}u^{\sigma}\,}\delta\nonumber\\
&+&\frac{4}{5\rz} \qd{\frac{1}{2}(3C-4z_1) \inew{\tau} +
(C+2z_1)\inew{{\tau^{\prime}}}}+ \td{C \pm
\frac{C-8z_1}{5}\delta}u\cdot g(k), \label{ibob}
\end{eqnarray}
where $u\equiv \rho/\rho_0$ is the reduced density and $\pm$ is
for neutrons/protons. The $\delta$ is now the isospin asymmetry
previously denoted by $\alpha$ or $\beta$. In the above,
$\mathcal{I}_\tau=\itau$ with $g(k)\equiv 1/\qd{1+\td{k/\Lambda}^2
}$ being a momentum regulator, and $f_{\tau}(k)$ is the
phase-space distribution function. The parameter $\Lambda$ is
taken to be $\Lambda=1.5K_F^0$, where $K_F^0$ is the nucleon Fermi
wave number in symmetric nuclear matter at normal density
$\rho_0$. With $A=-144$ MeV, $B=203.3$ MeV, $C=-75$ MeV and
$\sigma=7/6$, the Bombaci model reproduces all ground state
properties including an incompressibility of $K_0$=210 MeV for
symmetric nuclear matter \cite{Bom01,Riz04}. The Bombaci model is
an extension of the well-known Gale-Bertsch-Das Gupta (GBD) model
\cite{Gal87} from symmetric to asymmetric nuclear matter. The
various terms in the nuclear potential are motivated by the HF
analysis using the Gogny effective interaction \cite{Gal90,Das03}.
This potential depends explicitly on the momentum but not the
total energy of the nucleons, leading thus to a k-mass which is
the same as the total effective mass. Since only the last term in
Eq.~(\ref{ibob}) is momentum dependent, the $\Lambda$ parameter
thus sets the scale for the momentum dependence of the nucleon
potential $U_{\tau}$ and also the scale for the effective mass,
and its value was determined by the ground state properties of
symmetric nuclear matter. As shown in the following, the Bombaci
model can also reproduce appropriately the momentum dependence of
the empirical isoscalar nuclear optical potential.

\begin{figure}[ht]
\centering
\includegraphics[scale=0.4,angle=-90]{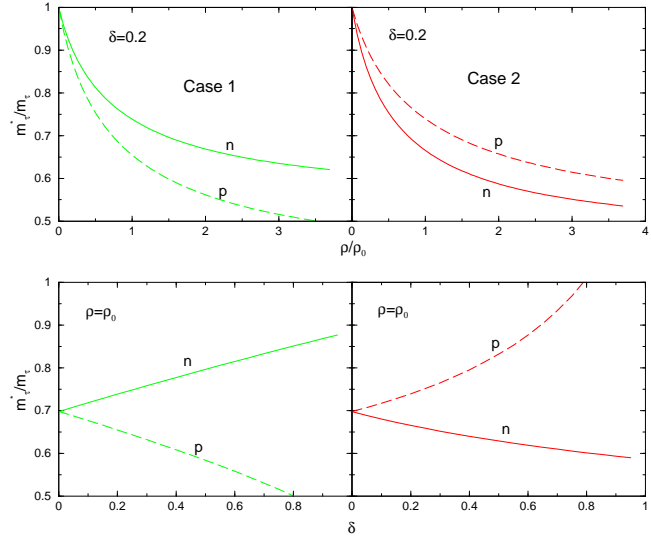}
\caption{(Color on line) Nucleon effective masses as functions of
density (upper window) and isospin asymmetry (lower window) from
the Bombaci model with two different parameter sets (see text).
Taken from Ref. \cite{LiBA04c}.} \label{EffMassRhoDel}
\end{figure}

Eq. (\ref{ibob}) leads to an effective mass \cite{Bom01,Riz04}
\begin{eqnarray}
\frac{m^*_\tau}{m_{\tau}}=\left\{1+\frac{-\frac{2m_{\tau}}{\hbar^2}
\frac{1}{\Lambda^2} \td{{C \pm \frac{C-8z_1}{5}\delta}}u}{ \left[ 1+
\left( \frac{k_{F0}}{\Lambda} \right)^{^2}   (1 \pm \delta)^{^{2/3}}
u^{^{2/3}}\right]^2}  \right\}^{-1},
\end{eqnarray}
where the $(1 \pm \delta)^{2/3}u^{2/3}$ term comes from the
nuclear Fermi wave number $k_{\tau}^F$ squared, and $\pm$ is for
$n/p$. The three parameters $x_0$, $x_3$ and $z_1$ can be adjusted
to mimic different behaviors of the density-dependent symmetry
energy and the neutron-proton effective mass splitting. Two sets
of parameters can be chosen to give two opposite nucleon effective
mass splittings, but almost the same symmetry energy $E_{\rm
sym}(\rho)$ \cite{Riz04}. The parameter set $z_1=-36.75$ MeV,
$x_0=-1.477$ and $x_3=-1.01$ (case 1) leads to $m_n^*>m_p^*$ while
the one with $z_1=50$ MeV, $x_0=1.589$ and $x_3=-0.195$ (case 2)
leads to $m_n^*<m_p^*$ at all non-zero densities and isospin
asymmetries. Shown in Fig.~\ref{EffMassRhoDel} are the nucleon
effective masses as functions of density and isospin asymmetry in
both cases. It is seen that although the neutron-proton effective
mass splittings have opposite signs in these two cases, they
increase in magnitude in both cases with increasing density and
isospin asymmetry. Thus, a large effective mass splitting can be
obtained in dense neutron-rich matter.

\begin{figure}[ht]
\centering
\includegraphics[scale=0.55,angle=-90]{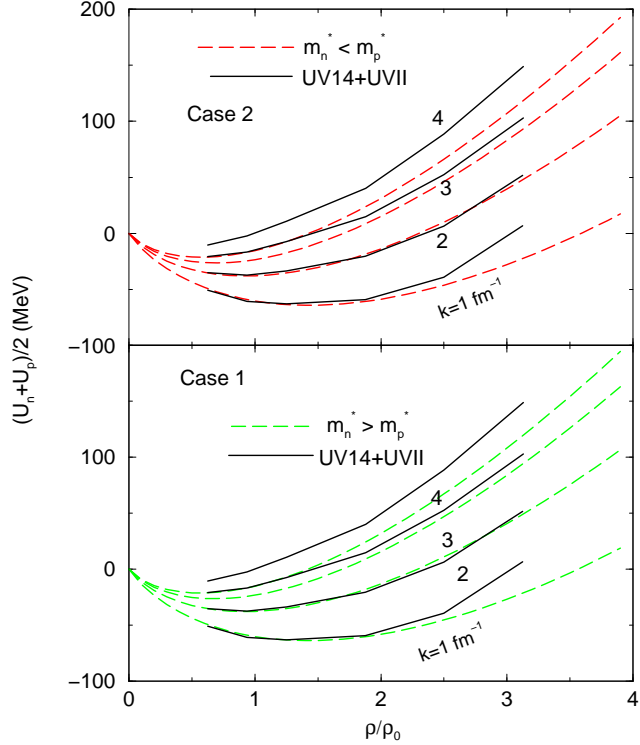}
\vspace{0.6cm} \caption{{\protect\small (Color on line) Strength
of the isoscalar potential as a function of density at four
different wave numbers for case 1 (lower panel) and case 2 (upper
panel) of the Bombaci model in comparison with results from the
variational many-body calculations. Taken from Ref.
\cite{LiBA04c}.}} \label{IsoScalarPotRho}
\end{figure}

Eq. (\ref{ibob}) allows one to calculate the isoscalar and isovector
parts of the single-nucleon potential, which must have asymptotic
values at $\rho_0$ in agreement with the real parts of the
corresponding nucleon optical potentials constrained by
nucleon-nucleus scattering experiments. For the isoscalar potential,
it is a good approximation to approximate it by $(U_n+U_p)/2$ as the
$\delta^2$ in Eq. (\ref{ibob}) is negligibly small. The resulting
isoscalar potential can be compared with that from the VMB
predictions by Wiringa \cite{Dan02a,Gal90,Pan93,Dan00,Dan91,Wir88b}.
In the VMB theory, the single-nucleon potential is obtained by using
a realistic Hamiltonian that fits the NN scattering data, few-body
nuclear binding energies and nuclear matter saturation properties.
It also reproduces the experimental nucleon optical potential
available mainly at low energies \cite{Cse92}. Shown in
Fig.~\ref{IsoScalarPotRho} is a comparison of isoscalar potentials
using Eq. (\ref{ibob}) with the VMB predictions using the $UV14$
two-body potential and the $UVII$ three-body potential
\cite{Wir88b}. It is seen that the isoscalar potentials for the two
sets of parameters are similar, indicating that they are almost
independent of the neutron-proton effective mass splittings as one
has expected. Furthermore, in both cases the isoscalar potentials
using Eq. (\ref{ibob}) are in good agreement with the VMB
predictions up to about $k=2.5~{\rm fm}^{-1}$. At higher momenta
where combinations of different two-body and three-body forces lead
to somewhat different predictions from the VMB approach, especially
at high densities \cite{Wir88b}, the Bombaci model leads to slightly
lower values. Nevertheless, the quality of agreement with the VMB
predictions shown here is compatible with those using other models
\cite{Gal90,Pan93,Dan00,Dan91}.

\begin{figure}[ht]
\centering
\includegraphics[scale=0.45,angle=-90]{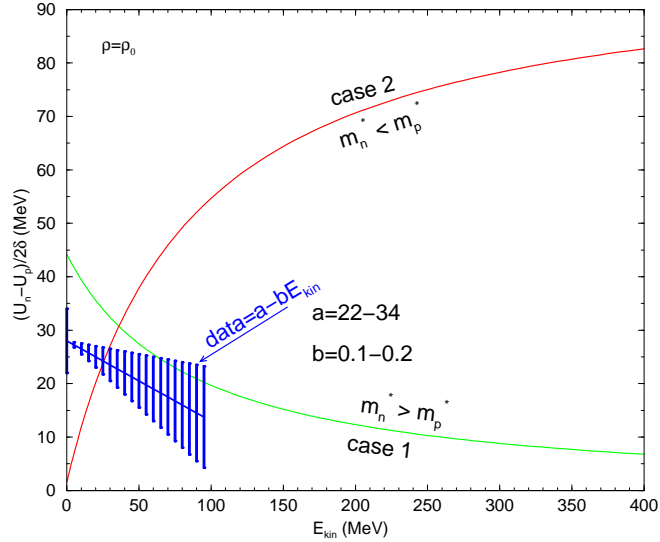}
\caption{{\protect\small (Color on line) Strength of the isovector
potential at normal density $\rho_0$ as a function of nucleon
kinetic energy. Taken from Ref. \cite{LiBA04c}.}}
\label{EffMassLanePot}
\end{figure}

For the isovector part of the nucleon potential, its strength at
the normal density, i.e., the Lane potential \cite{Lan62}, can be
extracted from $U_{\rm Lane}\equiv (U_n-U_p)/2\delta$ at $\rho_0$.
As mentioned before, systematic analyses of a large number of
nucleon-nucleus scattering experiments and (p,n) charge exchange
reactions at beam energies below about 100 MeV
\cite{Pat76,Kwi78,Rap79,Vit81,Leo81} have indicated undoubtedly
that the Lane potential (Eq.(\ref{dat0})) decreases approximately
linearly with increasing beam
energy\cite{Hod94,Roo73,Dab74,Jeu74}. Shown in
Fig.~\ref{EffMassLanePot} are the isovector or symmetry potentials
using above two parameter sets in comparison with the Lane
potential constrained by the experimental data. The vertical bars
are used to indicate the uncertainties of the coefficients $a$ and
$b$ in Eq. (\ref{dat0}). It is seen that with the effective mass
splitting $m_n^*>m_p^*$ (case 1) the strength of the symmetry
potential decreases with increasing energy. This trend is in
agreement with that extracted from the experimental data.
Moreover, the slope of the calculated symmetry potential with
respect to energy is also reasonable although the magnitude
obtained is slightly higher. In case 2, however, the most striking
feature is that the symmetry potential increases with increasing
beam energy. This is in sharp contrast with that indicated by the
experimental data. The incorrect energy dependence of the symmetry
potential in this case thus excludes the neutron-proton effective
mass splitting of $m_n^*<m_p^*$ in neutron-rich matter.

The results discussed above using the Bombaci model in the case $1$
is consistent with that of earlier work by Sj\"oberg in the
framework of the Landau-Fermi liquid theory \cite{Sjo76}. In the
latter, the nucleon effective mass splitting is given by
\cite{Sjo76}
\begin{eqnarray}
(m_n^* - m_p^*)/m = \frac{m_n^* k_n}{3 \pi^2} \left[ f_1^{nn} +
(k_p/k_n)^2 f_1^{np} \right]-\frac{ m_p^* k_p}{3\pi^2} \left[
f_1^{pp} + (k_n/k_p)^2 f_1^{np} \right], \label{sjo}
\end{eqnarray}
where $f_1^{nn}$, $f_1^{pp}$ and $f_1^{np}$ are the
neutron-neutron, proton-proton and neutron-proton quasiparticle
interactions projected on the $l=1$ Legendre polynomial, as for
the effective mass in a one-component Fermi liquid. Calculations
with microscopic NN interactions have predicted that all $f_1$'s
are negative in symmetric nuclear matter at normal density and
also in asymmetric matter at tree-level. It can then be seen from
Eq. (\ref{sjo}) that the proton effective mass is smaller than
that of neutrons ($m_n^*>m_p^*$) in neutron-rich matter as a
result of the coupling of protons to the denser neutron
background, i.e., the term $(k_n/k_p)^2 f_1^{np}$ is dominant in
Eq. (\ref{sjo}), as shown numerically in Fig.~\ref{EffMassLanePot}
of Ref. \cite{Sjo76}.


\section{Isovector nucleon potential and properties of asymmetric nuclear matter in
relativistic mean-field models}
\label{chapter_rmf}

In this Chapter, we review the isospin-dependent bulk and
single-particle properties of asymmetric nuclear matter based on
commonly used $23$ different parameter sets in three different
versions of the RMF model \cite{Che07}. In particular, we discuss
the density dependence of the nuclear symmetry energy from these
RMF models and compare them with the constraints recently
extracted from analyses of the isospin diffusion data from
heavy-ion collisions based on the isospin- and momentum-dependent
IBUU04 transport model with in-medium NN cross sections
\cite{Tsa04,Che05a,LiBA05c}, the isoscaling analyses of the
isotope ratios in intermediate energy heavy ion collisions
\cite{She07}, and measured isotopic dependence of the giant
monopole resonances (GMR) in even-A Sn isotopes \cite{Gar07}.
Moreover, as already mentioned in Chapter \ref{chapter_ria}, the
momentum dependence of the isovector potential and corresponding
neutron-proton effective mass splitting predicted by various RMF
models are very different from each other. They are also quite
different from those predicted by other approaches. We thus also
examine closely in this Chapter the momentum dependence of the
isovector potential using the $23$ different parameter sets of the
RMF models available in the literature.

\subsection{The nuclear symmetry potential in relativistic models}

The nuclear symmetry potential refers to the isovector part of the
nucleon mean-field potential in isospin asymmetric nuclear matter.
It was originally defined in non-relativistic models as discussed in
the previous Chapter. In relativistic models, it can be defined
similarly by using the non-relativistic reduction of the
relativistic single-nucleon potentials. The nuclear symmetry
potential in relativistic models therefore depends on the definition
of the real part of the non-relativistic optical potential or the
nucleon mean-field potential deduced from the relativistic effective
interactions, which are characterized by Lorentz covariant nucleon
self-energies. In the relativistic mean-field approximation, these
self-energies appear in the single-nucleon Dirac equation
\begin{eqnarray}
\lbrack \gamma _{\mu }(i\partial ^{\mu }-\Sigma _{\tau }^{\mu
})-(M_{\tau }+\Sigma _{\tau }^{S})]\psi _{\tau }=0,~~~\tau=n,p
\label{Dirac}
\end{eqnarray}
as the isospin-dependent nucleon vector self-energy $\Sigma _{\tau
}^{\mu }$ and scalar self-energy $\Sigma _{\tau }^{S}$. For the
Hartree approximation in the static limit, there are no currents
in a nucleus or nuclear matter, and the spatial vector components
vanish and only the time-like component of the vector self-energy
$\Sigma _{\tau }^{0}$ remains. Furthermore, the nucleon
self-energy is an energy-independent real, local quantity in the
standard RMF model.

There are different methods to derive the real part of the
non-relativistic optical potential based on the Dirac equation with
Lorentz covariant nucleon vector and scalar self-energies. The most
popular one is the `Schr\"{o}dinger-equivalent potential' (SEP).
From the nucleon scalar self-energy $\Sigma _{\tau }^{S}$ and the
time-like component of the vector self-energy $\Sigma _{\tau }^{0}$,
the `Schr\"{o}dinger-equivalent potential' is given by \cite{Jam80}
\begin{eqnarray}
U_{\mathrm{SEP},\tau } &=&\Sigma _{\tau }^{S}+\frac{1}{2M_{\tau
}}[(\Sigma _{\tau }^{S})^{2}-(\Sigma _{\tau }^{0})^{2}]+\frac{\Sigma
_{\tau }^{0}} {M_{\tau }}E_{\tau }\notag\\
&=&\Sigma _{\tau }^{S}+\Sigma _{\tau }^{0}+\frac{1}{2M_{\tau
}}[(\Sigma _{\tau }^{S})^{2}-(\Sigma _{\tau }^{0})^{2}]
+\frac{\Sigma _{\tau }^{0}}{M_{\tau }}E_{\mathrm{kin}},\notag\\
\label{Usep}
\end{eqnarray}%
where $E_{\mathrm{kin}}$ is the kinetic energy of a nucleon, i.e.,
$E_{\mathrm{kin}}=E_{\tau }-M_{\tau }$ with $E_{\tau }$ being its
total energy. Eq. (\ref{Usep}) shows that $U_{\mathrm{SEP},\tau }$
increases linearly with the nucleon energy $E_{\tau }$ or kinetic
energy $E_{\mathrm{kin}}$ if the nucleon self-energies are
independent of energy. By construction, solving the Schr\"{o}dinger
equation with above SEP gives same bound-state energy eigenvalues
and elastic phase shifts as the solution of the upper component of
the Dirac spinor in the Dirac equation with same nucleon scalar
self-energy and time-like component of the vector self-energy
\cite{Jam80}. The above SEP thus best represents the real part of
the nucleon optical potential in non-relativistic models
\cite{Fuc05,Jam89}. The corresponding nuclear symmetry potential is
given by
\begin{eqnarray}\label{UsymSEP}
U_{\mathrm{sym}}^{\mathrm{SEP}}=\frac{U_{\mathrm{SEP},n}-U_{\mathrm{SEP},p}}
{2\alpha},
\end{eqnarray}
with $\alpha $ being the isospin asymmetry, similar to
Eq.(\ref{dat}) in the previous chapter.

Another popular alternative for deriving the non-relativistic
nucleon optical potential in relativistic models is to take it as
the difference between the total energy $E_{\tau }$ of a nucleon
with momentum $\vec{p}$ in the nuclear medium and its energy at the
same momentum in free space \cite{Fe91}, i.e.,
\begin{eqnarray}
U_{\mathrm{OPT},\tau }=E_{\tau }-\sqrt{\mathbf{p}^{2}+M_{\tau
}^{2}}=E_{\tau }-\sqrt{(E_{\tau }-\Sigma _{\tau }^{0})^{2}-\Sigma
_{\tau }^{S}(2M_{\tau }+\Sigma _{\tau }^{S})}. \label{Uopt}
\end{eqnarray}
In obtaining the last step in above equation, the dispersion
relation
\begin{eqnarray}\label{dispersion}
E_{\tau }=\Sigma _{\tau }^{0}+\sqrt{\mathbf{p}^{2}+(M_{\tau }+\Sigma
_{\tau }^{S})^{2}}
\end{eqnarray}
has been used. This definition for the nucleon optical potential
has been extensively used in microscopic DBHF calculations
\cite{LiGQ93} and transport models for heavy-ion collisions
\cite{Dan00}. For energy-independent nucleon self-energies,
$U_{\mathrm{OPT},\tau }$ approaches the constant value $\Sigma
_{\tau }^{0}$ when $\left\vert \vec{p}\right\vert \rightarrow
\infty $, unlike the linear increase of $U_{\mathrm{SEP},\tau }$
with nucleon energy. For $\left\vert \vec{p}\right\vert =0$, one
has $U_{\mathrm{OPT},\tau }=\Sigma _{\tau }^{S}+\Sigma _{\tau
}^{0}$ while $U_{\mathrm{SEP},\tau }=\Sigma _{\tau }^{S}+\Sigma
_{\tau }^{0}+(\Sigma _{\tau }^{S}+\Sigma _{\tau
}^{0})^{2}/(2M_{\tau })$. Therefore, $U_{\mathrm{OPT},\tau }$
displays a more reasonable high energy behavior than
$U_{\mathrm{SEP},\tau }$. Unlike $U_{\mathrm{SEP},\tau }$,
$U_{\mathrm{OPT},\tau }$ does not give the same bound-state energy
eigenvalues and elastic phase shifts as the solution of the upper
component of the Dirac equation. As in the case of
$U_{\mathrm{SEP},\tau }$, the symmetry potential in this approach
is defined by
\begin{eqnarray}
U_{\mathrm{sym}}^{\mathrm{OPT}}=\frac{U_{\mathrm{OPT},n}-U_{\mathrm{OPT},p}}
{2\alpha}.  \label{UsymOPT}
\end{eqnarray}

In Ref. \cite{Ham90}, another optical potential was introduced
based on the second-order Dirac (SOD) equation, and it corresponds
to multiplying Eq.~(\ref{Usep}) by the factor $M_{\tau }/E_{\tau
}$, i.e.,
\begin{eqnarray}
U_{\mathrm{SOD},\tau } &=&[\Sigma _{\tau }^{S}+\frac{1}{2M_{\tau
}}[(\Sigma _{\tau }^{S})^{2}-(\Sigma _{\tau
}^{0})^{2}]+\frac{\Sigma _{\tau }^{0}}
{M_{\tau }}E_{\tau }]\frac{M_{\tau }}{E_{\tau }}  \notag \\
&=&\Sigma _{\tau }^{0}+\frac{M_{\tau }}{E_{\tau }}\Sigma _{\tau
}^{S}+\frac{1}{2E_{\tau }}[(\Sigma _{\tau }^{S})^{2}-(\Sigma
_{\tau }^{0})^{2}]. \label{Usd}
\end{eqnarray}
For energy-independent nucleon self-energies, $U_{\text{SOD,}\tau
}$ has the same asymptotical value of $\Sigma _{\tau }^{0}$ as
$U_{\mathrm{OPT},\tau }$ when $\left\vert \vec{p}\right\vert
\rightarrow \infty $. For $\left\vert \vec{p}\right\vert =0$, one
has $U_{\mathrm{SOD},\tau }=\Sigma _{\tau }^{0}+\frac{M_{\tau
}}{\Sigma _{\tau }^{S}+\Sigma _{\tau }^{0}+M_{\tau }}\Sigma _{\tau
}^{S}+\frac{1}{2(\Sigma _{\tau }^{S}+\Sigma _{\tau }^{0}+M_{\tau
})} [(\Sigma _{\tau }^{S})^{2}-(\Sigma _{\tau }^{0})^{2}] $. The
symmetry potential based on the optical potential of Eq.
(\ref{Usd}) is given by
\begin{eqnarray}
U_{\mathrm{sym}}^{\mathrm{SOD}}=\frac{U_{\mathrm{SOD},n}-U_{\mathrm{SOD},p}}
{2\alpha}.  \label{UsymSOD}
\end{eqnarray}

The above discussions thus show that the optical potentials
defined in Eqs. (\ref{Uopt}) and (\ref{Usd}) have similar high
energy behaviors, but they may be very different from that defined
in Eq. (\ref{Usep}). If one assumes that $\Sigma _{\tau
}^{S}+\Sigma _{\tau }^{0}\ll M_{\tau }$ and $\left\vert \Sigma
_{\tau }^{S}\right\vert \approx \left\vert \Sigma _{\tau
}^{0}\right\vert $, which have been shown to be generally valid in
the RMF model even at high baryon densities, one then has
$U_{\mathrm{SEP},\tau }\approx U_{\mathrm{SOD},\tau }\approx
U_{\mathrm{OPT},\tau }=\Sigma _{\tau }^{S}+\Sigma _{\tau }^{0}$ at
low momenta ($\left\vert \vec{p}\right\vert \approx 0$),
indicating that the above three definitions for the optical
potential in the RMF model behave similarly at low energies.
However, among the three optical potentials defined above, only
$U_{\mathrm{SEP},\tau }$ is obtained from a well-defined
theoretical procedure and is Schr\"{o}dinger-equivalent while
$U_{\mathrm{OPT},\tau }$ and $U_{\mathrm{SOD},\tau }$ are
introduced here for heuristic reasons as they are of practical
interest in microscopic DBHF calculations, transport models for
heavy-ion collisions, and the Dirac phenomenology study. As to be
discussed in the following, although the predicted energy
dependence of the nuclear symmetry potential at lower energies
from the RMF models does not agree with the empirical Lane
potential (Eq.(\ref{dat}), it is consistent with results at lower
momenta from the microscopic DBHF \cite{Fuc04}, the extended BHF
with TBF \cite{Zuo05}, and the chiral perturbation theory
calculations \cite{Fri05}.

\subsection{The nucleon effective mass in relativistic models}

Many different definitions for the nucleon effective mass can be
found in the literature \cite{Fuc05,Jam89}, and they are the Dirac
mass $M_{\mathrm{Dirac}}^{\ast }$ (also denoted as $M^{\ast }$ in
the following), the Landau mass $M_{\mathrm{Landau}}^{\ast }$, and
the Lorentz mass $M_{\mathrm{Lorentz}}^{\ast } $. The Dirac mass
$M_{\mathrm{Dirac}}^{\ast }$ is defined through the nucleon scalar
self-energy in the Dirac equation, i.e.,
\begin{eqnarray}
M_{\mathrm{Dirac},\tau }^{\ast }=M_{\tau }+\Sigma _{\tau }^{S}.
\end{eqnarray}
It is directly related to the spin-orbit potential in finite
nuclei and is thus a genuine relativistic quantity without
non-relativistic correspondence. The difference between the
nucleon vector and scalar self-energies determines the spin-orbit
potential, whereas their sum defines the effective single-nucleon
potential and is constrained by the nuclear matter binding energy
at saturation density. From the energy spacings between spin-orbit
partner states in finite nuclei, the constraint $0.55~M$ $\leq
M_{\mathrm{Dirac}}^{\ast }\leq 0.6~M$ has been obtained on the
value of the Dirac mass \cite{Typ05,Mar07}.

The Landau mass $M_{\mathrm{Landau}}^{\ast }$ is defined as
$M_{\mathrm{Landau},\tau }^{\ast } =p\frac{dp}{dE_{\tau }}$ in
terms of the single-particle density of state $dE_{\tau }/dp$ at
energy $E_{\tau }$ and thus characterizes the momentum dependence
of the single-particle potential. In the relativistic model, it is
given by \cite{Typ05}
\begin{eqnarray}
M_{\mathrm{Landau},\tau }^{\ast } =(E_{\tau }-\Sigma _{\tau
}^{0})(1- \frac{d\Sigma_{\tau}^{0}}{dE_{\tau }})-(M_{\tau }+\Sigma
_{\tau }^{S})\frac{d\Sigma _{\tau }^{S}}{dE_{\tau }}.
\label{MLandau}
\end{eqnarray}
Since $dp/dE_{\tau }$ is in principle measurable, the Landau mass
from the relativistic model should have a comparable value as that
in the non-relativistic model. Empirically, based on
non-relativistic effective interactions such as the Skyrme-type
interactions, calculations of the ground-state properties and the
excitation energies of quadrupole giant resonances have shown that
a realistic choice for the nucleon Landau mass is
$M_{\mathrm{Landau}}^{\ast }/M$ = $0.8\pm 0.1$
\cite{Cha97,Mar07,Cha98,Rei99}. The smaller Landau mass than that
of nucleon free mass leads to a smaller level density at the Fermi
energy and much spreaded single-particle levels in finite nuclei
\cite{Typ05}.

The Lorentz mass $M_{\mathrm{Lorentz}}^{\ast }$ characterizes the
energy dependence of the Schr\"{o}dinger-equivalent potential
$U_{\mathrm{SEP},\tau }$ in the relativistic model and is defined
as \cite{Jam89}
\begin{eqnarray}
M_{\mathrm{Lorentz},\tau }^{\ast } &=&M_{\tau
}(1-\frac{dU_{\mathrm{SEP},\tau }} {dE_{\tau }})\notag\\
&=&(E_{\tau }-\Sigma _{\tau }^{0})(1-\frac{d\Sigma _{\tau
}^{0}}{dE_{\tau }})-(M_{\tau }+\Sigma _{\tau }^{S})\frac{d\Sigma
_{\tau }^{S}}{dE_{\tau }} +M_{\tau }-E_{\tau}\notag\\
&=&M_{\mathrm{Landau},\tau }^{\ast }+M_{\tau }-E_{\tau }.
\label{MLorentz}
\end{eqnarray}
It has been argued in Ref. \cite{Jam89} that it is the Lorentz
mass $M_{\mathrm{Lorentz}}^{\ast }$ that should be compared with
the usual non-relativistic nucleon effective mass extracted from
analyses carried out in the framework of non-relativistic optical
and shell models. It can be easily seen that in the
non-relativistic approximation ($E_{\tau }\approx M_{\tau }$), the
Lorentz mass $M_{\mathrm{Lorentz}}^{\ast }$ reduces to the Landau
mass $M_{\mathrm{Landau}}^{\ast } $.

In relativistic models, the nucleon effective mass has sometimes
also been introduced via the energy dependence of the optical
potential in Eq. (\ref{Uopt}) and the second-order Dirac optical
potential in Eq. (\ref{Usd}), i.e.,
\begin{eqnarray}
M_{\mathrm{OPT},\tau }^{\ast } &=&M_{\tau
}\left(1-\frac{dU_{\mathrm{OPT},\tau }}{dE_{\tau }}\right)  \notag \\
&=&M_{\tau }\frac{(E_{\tau }-\Sigma _{\tau }^{0})(1-\frac{d\Sigma
_{\tau}^{0}} {dE_{\tau }})+(M_{\tau }-\Sigma _{\tau
}^{S})\frac{d\Sigma _{\tau }^{S}}{dE_{\tau }}}{\sqrt{(E_{\tau
}-\Sigma _{\tau }^{0})^{2}-\Sigma _{\tau }^{S}(2M_{\tau }
+\Sigma _{\tau }^{S})}}  \notag \\
&=&M_{\tau }\frac{M_{\mathrm{Landau},\tau }^{\ast
}}{\sqrt{(E_{\tau }-\Sigma _{\tau }^{0})^{2}-\Sigma _{\tau
}^{S}(2M_{\tau }+\Sigma _{\tau }^{S})}} \label{Mopt}
\end{eqnarray}
and
\begin{eqnarray}
M_{\mathrm{SOD},\tau }^{\ast }&=&M_{\tau }\left(1-
\frac{dU_{\mathrm{SOD},\tau }}{dE_{\tau }}\right)  \notag \\
&=&M_{\tau }[\frac{M_{\mathrm{Landau},\tau }^{\ast }}{E_{\tau }}
+\frac{(M_{\tau }+\Sigma _{\tau }^{S})^{2}-(E_{\tau }-\Sigma _{\tau
}^{0})^{2}+E_{\tau }^{2}-M_{\tau }^{2}}{2E_{\tau }^{2}}],
\label{Msod}
\end{eqnarray}
respectively.

The isospin-splitting of the nucleon effective mass in asymmetric
nuclear matter, i.e., the difference between the neutron and proton
effective masses, is currently not known empirically \cite{Lun03}.
Previous theoretical investigations have indicated that most RMF
calculations with the isovector $\delta $ meson predict
$M_{\mathrm{Dirac},n}^{\ast }<M_{\mathrm{Dirac},p}^{\ast }$ while in
the microscopic DBHF approach, $M_{\mathrm{Dirac},n}^{\ast }$ can be
larger or smaller than $M_{\mathrm{Dirac},p}^{\ast }$ depending on
the approximation schemes and methods used for determining the
Lorentz and isovector structure of the nucleon self-energy
\cite{Fuc05}. For the nucleon Lorentz mass, the microscopic DBHF or
BHF approach and most non-relativistic Skyrme-Hartree-Fock
calculations predict $ M_{\mathrm{Lorentz},n}^{\ast
}>M_{\mathrm{Lorentz},p}^{\ast }$, while most RMF and a few
Skyrme-Hartree-Fock calculations give opposite predictions.

\subsection{Relativistic mean-field models}

For completeness, we briefly review in the following the main
ingredients in the nonlinear RMF model, the density-dependent RMF
model, the nonlinear point-coupling RMF model, and the
density-dependent point-coupling RMF model. We neglect the
electromagnetic field in the following since we are interested in
the properties of the infinite nuclear matter. Furthermore, besides
the mean-field approximation in which operators of meson fields are
replaced by their expectation values (the fields are thus treated as
classical c-numbers), we also restrict the discussions to the
non-sea approximation which neglects the effect due to negative
energy states in the Dirac sea.

\subsubsection{The nonlinear RMF model}

The Lagrangian density in the nonlinear RMF model generally
includes the nucleon field $\psi $, the isoscalar-scalar meson
field $\sigma $, the isoscalar-vector meson field $\omega $, the
isovector-vector meson field $\vec{\rho}$, and the
isovector-scalar meson field $\vec{\delta}$, i.e.,
\begin{eqnarray}
\mathcal{L}_{\mathrm{NL}}&=&\bar{\psi}\left[ \gamma _{\mu
}(i\partial ^{\mu }-g_{\omega }\omega ^{\mu })-(M-g_{\sigma }\sigma
)\right] \psi +\frac{1}{2}(\partial _{\mu }\sigma \partial ^{\mu
}\sigma -m_{\sigma }^{2}\sigma ^{2})-\frac{1}{4}\omega _{\mu \nu
}\omega ^{\mu \nu } \notag\\
&+&\frac{1}{2} m_{\omega }^{2}\omega _{\mu }\omega ^{\mu }
-\frac{1}{3}b_{\sigma }M{(g_{\sigma }\sigma )}^{3}
-\frac{1}{4}c_{\sigma } {(g_{\sigma}\sigma)}^{4}
+\frac{1}{4}c_{\omega}{(g_{\omega }^{2}\omega _{\mu }\omega^{\mu })}^{2} \notag\\
&+&\frac{1}{2}(\partial _{\mu }\vec{\delta}\cdot
\partial ^{\mu }\vec{\delta} -m_{\delta
}^{2}\vec{\delta}^{2})+\frac{1}{2}m_{\rho }^{2}\vec{\rho}_{\mu
}\cdot \vec{\rho}^{\mu }-\frac{1}{4}\vec{\rho}_{\mu \nu }\cdot
\vec{\rho}^{\mu \nu }  \notag \\
&+&\frac{1}{2}(g_{\rho }^{2}\vec{\rho}_{\mu }\cdot \vec{\rho}^{\mu
})(\Lambda _{S}g_{\sigma }^{2}\sigma ^{2}+\Lambda _{V}g_{\omega
}^{2}{\omega _{\mu }\omega ^{\mu }})-g_{\rho }\vec{\rho}_{\mu }\cdot
\bar{\psi}\gamma ^{\mu }\vec{\tau}\psi +g_{\delta }\vec{\delta}\cdot
\bar{\psi}\vec{\tau}\psi \;, \label{lagNL}
\end{eqnarray}
where the antisymmetric field tensors $\omega _{\mu \nu }$ and
$\vec{\rho} _{\mu \nu }$ are given by $\omega _{\mu \nu }\equiv
\partial _{\nu }\omega _{\mu }-\partial _{\mu }\omega _{\nu }$ and
$\text{ }\vec{\rho}_{\mu \nu }\equiv \partial _{\nu
}\vec{\rho}_{\mu }-\partial _{\mu }\vec{\rho}_{\nu }$ ,
respectively, and other symbols have their usual meanings. Also,
vectors in isospin space are denoted by arrows. This model also
contains cross interactions between the isovector meson $\rho $
and isoscalar $\sigma $ and $\omega $ mesons through the
cross-coupling constants $\Lambda _{S}$ and $\Lambda _{V}$
\cite{Hor01a,Mul96}. Also included is the isovector-scalar channel
($\delta $ meson), which is important for the saturation of
asymmetric nuclear matter and has also been shown to be an
important degree of freedom in describing the properties of
asymmetric nuclear matter \cite{Kub97,Liu02}. The above Lagrangian
density is quite general and allows one to use most of presently
popular parameter sets in the nonlinear RMF model.

From the standard Euler-Lagrange formalism, one can deduce from the
Lagrangian density the equations of motion for the nucleon and meson
fields. The resulting Dirac equation for the nucleon field is
\begin{eqnarray}
\left[ \gamma _{\mu }(i\partial ^{\mu }-\Sigma _{\tau }^{\mu
})-(M+\Sigma _{\tau }^{S})\right] \psi =0\;,
\end{eqnarray}
with the following nucleon scalar and vector self-energies:
\begin{eqnarray}
\Sigma _{\tau }^{S}=-g_{\sigma }\sigma -g_{\delta }\vec{\delta}\cdot
\vec{\tau} \qquad {\rm and} \qquad \Sigma _{\tau }^{\mu }=g_{\omega
}\omega ^{\mu }+g_{\rho }\vec{\rho}^{\mu }\cdot \vec{\tau}.
\end{eqnarray}

For the isoscalar meson fields $\sigma $ and $\omega $, they are
described by the Klein-Gordon and Proca equations, respectively,
i.e.,
\begin{eqnarray}
(\partial _{\mu }\partial ^{\mu }+m_{\sigma }^{2})\sigma
&=&g_{\sigma }[\bar{\psi}\psi -b_{\sigma }M{(g_{\sigma }\sigma
)}^{2}-c_{\sigma }{(g_{\sigma}\sigma )}^{3}+\Lambda _{S}{(g_{\sigma
}\sigma )}g_{\rho }^{2}\vec{\rho}_{\mu }\cdot
\vec{\rho}^{\mu }]\;, \\
\partial _{\mu }\omega ^{\mu \nu }+m_{\omega }^{2}\omega ^{\nu }
&=&g_{\omega }[\bar{\psi}\gamma ^{\nu }\psi -c_{\omega }g_{\omega
}^{3}(\omega _{\mu }\omega ^{\mu }\omega ^{\nu })-\Lambda
_{V}g_{\rho }^{2}\vec{\rho}_{\mu }\cdot \vec{\rho}^{\mu }g_{\omega
}\omega ^{\nu }]\;.
\end{eqnarray}

Analogous equations for the isovector $\delta $ and $\rho $ meson
fields are
\begin{eqnarray}
(\partial _{\mu }\partial ^{\mu }+m_{\delta }^{2})\vec{\delta}
&=&g_{\delta
} \bar{\psi}\vec{\tau}\psi , \\
\partial _{\mu }\vec{\rho}^{\mu \nu }+m_{\rho }^{2}\vec{\rho}^{\nu }
&=&g_{\rho }[\bar{\psi}\gamma ^{\nu }\vec{\tau}\psi -\Lambda
_{S}(g_{\rho } \vec{\rho}^{\nu }){(g_{\sigma }\sigma )}^{2}-\Lambda
_{V}(g_{\rho }\vec{\rho}^{\nu })g_{\omega }^{2}\omega _{\mu }\omega
^{\mu }].
\end{eqnarray}

For a static, homogenous infinite nuclear matter, all derivative
terms drop out and the expectation values of space-like components
of vector fields vanish (only zero components $\vec{\rho}_{0}$ and
$\omega _{0}$ survive) due to translational invariance and
rotational symmetry of the nuclear matter. In addition, only the
third component of isovector fields ($\delta ^{(3)}$ and $\rho
^{(3)}$) needs to be taken into consideration due to the
rotational invariance around the third axis in the isospin space.
In the mean-field approximation, meson fields are replaced by
their expectation values, i.e., $\sigma \rightarrow \bar{\sigma}$,
$\omega _{\mu }\rightarrow \bar{\omega}_{0}$,
$\vec{\delta}\rightarrow \bar{\delta}^{(3)}$, and $\vec{
\rho}_{\mu }\rightarrow \bar{\rho}_{0}^{(3)}$, and the meson field
equations are reduced to
\begin{eqnarray}
m_{\sigma }^{2}\bar{\sigma} &=&g_{\sigma }[\rho _{S}-b_{\sigma
}M{(g_{\sigma }\bar{\sigma})}^{2}-c_{\sigma }{(g_{\sigma
}\bar{\sigma})}^{3}+\Lambda _{S}{(g_{\sigma }\bar{\sigma})(}g_{\rho
}\bar{\rho}_{0}^{(3)})^{2}], \\
m_{\omega }^{2}\bar{\omega}_{0} &=&g_{\omega }[\rho _{B}-c_{\omega }
{(g_{\omega }\bar{\omega}_{0})}^{3}-\Lambda {(g_{\omega
}\bar{\omega}_{0})(}g_{\rho }\bar{\rho}_{0}^{(3)})^{2}],  \label{OmgNL} \\
m_{\delta}^{2}{\bar{\delta}}^{(3)} &=&g_{\delta }(\rho _{S,p}-\rho
_{S,n}).\label{DelNL} \\
m_{\rho }^{2}\bar{\rho}_{0}^{(3)} &=&g_{\rho }[\rho _{B,p}-\rho
_{B,n}-\Lambda _{S}{(}g_{\rho }\bar{\rho}_{0}^{(3)}){(g_{\sigma
}\sigma )}^{2} -\Lambda _{V}{(}g_{\rho
}\bar{\rho}_{0}^{(3)}){(g_{\omega }\bar{\omega}_{0})}^{{2}}].
\label{RhoNL}
\end{eqnarray}
In the above, the nucleon scalar density $\rho _{S}$ is defined as
\begin{eqnarray}
\rho _{S}=\left\langle \bar{\psi}\psi \right\rangle =\rho
_{S,p}+\rho _{S,n}\;,  \label{RhoS}
\end{eqnarray}
with the proton ($p$) and neutron ($n$) scalar densities given by
\begin{eqnarray}
\rho _{S,i} &=&\frac{2}{{(2\pi
)}^{3}}\int_{0}^{k_{F}^{i}}d^{3}\!k\,\frac{M_{i}^{\ast
}}{\sqrt{\vec{k}^{2}+(M_{i}^{\ast })^{2}}} =\frac{M_{i}^{\ast
}}{2\pi ^{2}}\left[ k_{F}^{i}\tilde{E}_{F}^{i}-(M_{i}^{\ast
})^{2}\ln \frac{k_{F}^{i}+\tilde{E}_{F}^{i}}
{M_{i}^{\ast }}\right],~i=p,n \notag\\
\label{RhoSnp}
\end{eqnarray}
where
\begin{eqnarray}
\tilde{E}_{F}^{i}=\sqrt{(k_{F}^{i})^{2}+(M_{i}^{\ast })^{2}},
\label{Ef}
\end{eqnarray}
with $M_{p}^{\ast }$ and $M_{n}^{\ast }$ denoting the proton and
neutron Dirac masses, respectively, i.e.,
\begin{eqnarray}
M_{p}^{\ast }=M-g_{\sigma }\bar{\sigma}-g_{\delta
}{\bar{\delta}}^{(3)}, \text{ }M_{n}^{\ast }=M-g_{\sigma
}\bar{\sigma}+g_{\delta }{\bar{\delta}}^{(3)}.  \label{MDiracNL}
\end{eqnarray}
The nucleon scalar and vector self-energies are then given by
\begin{eqnarray}
\Sigma _{\tau }^{S} =-g_{\sigma }\bar{\sigma}-g_{\delta
}{\bar{\delta}} ^{(3)}\tau _{3} \qquad {\rm and} \qquad \Sigma
_{\tau }^{0} =g_{\omega }\bar{\omega}_{0}+g_{\rho }\bar{\rho}
_{0}^{(3)}\tau _{3}, \label{Sig0NL}
\end{eqnarray}
with $\tau _{3}=1$ and $-1$ for protons and neutrons,
respectively.

The set of coupled equations for the nucleon and meson fields can
be solved self-consistently using the iteration method, and the
properties of the nuclear matter can then be obtained from these
fields. From the resulting energy-momentum tensor, one can
calculate the energy density $\epsilon $ and pressure $P$ of
asymmetric nuclear matter, and the results are given by
\begin{eqnarray}
\epsilon &=&\epsilon _{\rm kin}^{n}+\epsilon _{\rm kin}^{p}
+\frac{1}{2}\left[ m_{\sigma }^{2}\bar{\sigma}^{2}+m_{\omega
}^{2}\bar{\omega}_{0}^{2}+m_{\delta }^{2}
{\bar{\delta}}^{(3)2}+m_{\rho }^{2}\bar{\rho}
_{0}^{(3)2}\right]  \notag \\
&+&\frac{1}{3}b_{\sigma }M{(g_{\sigma
}\bar{\sigma})}^{3}+\frac{1}{4}c_{\sigma }{(g_{\sigma
}\bar{\sigma})}^{4}+\frac{3}{4}c_{\omega }{(g_{\omega
}\bar{\omega}_{0})}^{4}+\frac{1}{2}(g_{\rho}\bar{\rho}_{0}^{(3)})^{2}
[\Lambda _{S}{(g_{\sigma }\bar{\sigma})}^{2}+3\Lambda
_{V}{(g_{\omega }\bar{\omega}_{0})}^{2}] \notag\\
\end{eqnarray}
and
\begin{eqnarray}
P &=&P_{\rm kin}^{n}+P_{\rm kin}^{p}-\frac{1}{2}\left[ m_{\sigma
}^{2}\bar{\sigma}^{2} -m_{\omega }^{2}
\bar{\omega}_{0}^{2}+m_{\delta }^{2}{\bar{\delta}}^{(3)2}
-m_{\rho }^{2}\bar{\rho}_{0}^{(3)2}\right]  \notag \\
&-&\frac{1}{3}b_{\sigma }M{(g_{\sigma
}\bar{\sigma})}^{3}-\frac{1}{4} c_{\sigma }{(g_{\sigma
}\bar{\sigma})}^{4}+\frac{1}{4}c_{\omega }{(g_{\omega
}\bar{\omega}_{0})}^{4}+\frac{1}{2}(g_{\rho
}\bar{\rho}_{0}^{(3)})^{2}[\Lambda _{S}{(g_{\sigma }
\bar{\sigma})}^{2}+\Lambda _{V}{(g_{\omega }\bar{\omega}_{0})}^{2}].
\notag\\
\end{eqnarray}
In the above, $\epsilon _{\rm kin}^{i}$ and $P_{\rm kin}^{i}$ are,
respectively, the kinetic contributions to the energy densities and
pressure of protons and neutrons in nuclear matter, and they are
given by
\begin{eqnarray}
\epsilon _{\rm kin}^{i}=\frac{2}{(2\pi )^{3}}\int_{0}^{k_{F}^{i}}d^{3}k \sqrt{%
\vec{k}^{2}+(M_{i}^{\ast })^{2}}=\frac{1}{4}[3\tilde{E}_{F}^{i}\rho
_{B,i}+M_{i}^{\ast }\rho _{S,i}],\quad i=p,n,
\end{eqnarray}
and
\begin{eqnarray}
P_{\rm kin}^{i}=\frac{2}{3(2\pi
)^{3}}\int_{0}^{k_{F}^{i}}d^{3}k\frac{\vec{k}
^{2}}{\sqrt{\vec{k}^{2}+(M_{i}^{\ast })^{2}}}
=\frac{1}{4}[\tilde{E}_{F}^{i}\rho _{B,i}-M_{i}^{\ast }\rho
_{S,i}],\quad i=p,n.
\end{eqnarray}

The binding energy per nucleon can be obtained from the energy
density via
\begin{equation*}
E=\frac{\epsilon }{\rho _{B}}-M,
\end{equation*}
while the symmetry energy is given by
\begin{eqnarray}
E_{\mathrm{sym}}(\rho _{B})
=\frac{k_{F}^{2}}{6\tilde{E}_{F}}+\frac{1}{2} \left( \frac{g_{\rho
}}{{m_{\rho }^{\ast }}}\right) ^{2}\rho _{B}-\frac{1}{2} \left(
\frac{g_{\delta }}{m_{\delta }}\right) ^{2}\times \frac{M^{\ast
2}\rho _{B}}{\tilde{E}_{F}^{2}[1+\left( \frac{ g_{\delta
}}{m_{\delta }}\right) ^{2}A(k_{F},M^{\ast })]}, \label{EsymNL}
\end{eqnarray}
with the effective $\rho $-meson mass given by \cite{Hor01a}
\begin{eqnarray}
{m_{\rho }^{\ast }}^{2}=m_{\rho }^{2}+g_{\rho }^{2}[\Lambda
_{S}{(g_{\sigma } \bar{\sigma})}^{2}+\Lambda _{V}{(g_{\omega
}\bar{\omega}_{0})}^{2}]
\end{eqnarray}
and
\begin{eqnarray}
A(k_{F},M^{\ast }) =\frac{4}{(2\pi
)^{3}}\int_{0}^{k_{F}}d^{3}k\frac{\vec{ k }^{2}}{\left(
\vec{k}^{2}+(M^{\ast })^{2}\right) ^{3/2}}=3\left( \frac{\rho
_{S}}{M^{\ast }}-\frac{\rho _{B}}{\tilde{E}_{F}} \right),
\end{eqnarray}
where $\tilde{E}_{F}=\sqrt{k_{F}^{2}+M^{\ast }{}^{2}}$ and
$M^{\ast } $ is the nucleon Dirac mass in symmetric nuclear
matter.

\subsubsection{The density-dependent RMF model}

In the density-dependent RMF model, instead of introducing terms
involving self-interactions of the scalar meson field and
cross-interactions of meson fields as in the nonlinear RMF model,
the coupling constants are density dependent. The Lagrangian
density in this model is generally written as
\begin{eqnarray}
\mathcal{L}_{\mathrm{DD}} &=&\bar{\psi}[\gamma _{\mu }(i\partial
^{\mu }-\Gamma _{\omega }\omega ^{\mu }-\Gamma _{\rho
}\vec{\rho}^{\mu }\cdot \vec{ \tau})-(M-\Gamma _{\sigma }\sigma
-\Gamma _{\delta }\vec{\delta}\cdot \vec{\tau}
)]\psi  \notag \\
&&+\frac{1}{2}(\partial _{\mu }\sigma \partial ^{\mu }\sigma
-m_{s}^{2}\sigma ^{2})+\frac{1}{2}(\partial _{\mu
}\vec{\delta}\cdot
\partial ^{\mu }\vec{\delta}-m_{\delta }^{2}\vec{\delta}^{2})  \notag \\
&&-\frac{1}{4}\omega _{\mu \nu }\omega ^{\mu \nu
}+\frac{1}{2}m_{\omega }^{2}\omega _{\mu }\omega ^{\mu }
-\frac{1}{4}\vec{\rho}_{\mu \nu }\cdot \vec{\rho}^{\mu \nu
}+\frac{1}{2} m_{\rho }^{2}\vec{\rho}_{\mu }\cdot \vec{\rho}^{\mu }
\label{lagDD}
\end{eqnarray}
The symbols used in above equation have their usual meanings as in
Eq.~(\ref{lagNL}) but the coupling constants $\Gamma _{\sigma }$,
$\Gamma _{\omega }$, $\Gamma _{\delta }$ and $\Gamma _{\rho }$ now
depend on the (baryon) density, which are usually parametrized as
\begin{eqnarray}
\Gamma _{i}(\rho )=\Gamma _{i}(\rho _{\rm sat})h_{i}(x),\quad x=\rho
/\rho _{\rm sat},
\end{eqnarray}
with
\begin{eqnarray}
h_{i}(x)=a_{i}\frac{1+b_{i}(x+d_{i})^{2}}{1+c_{i}(x+e_{i})^{2}},\quad
i=\sigma ,\omega ,\delta ,\rho ,
\end{eqnarray}
and $\rho _{\rm sat}$ being the saturation density of symmetric
nuclear matter. In some parameter sets,
\begin{eqnarray} h_{\rho }(x)=\exp
[-a_{\rho }(x-1)]
\end{eqnarray}
is used for the $\rho $ meson.

Since the coupling constants in the density-dependent RMF model
depend on the baryon fields $\bar{\psi}$ and $\psi $ through the
density, additional terms besides the usual ones in the nonlinear
RMF model appear in the field equations of motion when the partial
derivatives of $\mathcal{L}_{\text{DD}}$ are carried out with
respect to the fields $\bar{\psi}$ and $\psi $ in the
Euler-Lagrange equations. The resulting Dirac equation for the
nucleon field now reads:
\begin{eqnarray}
\left[ \gamma _{\mu }(i\partial ^{\mu }-\Sigma _{\tau }^{\mu
})-(M+\Sigma _{\tau }^{S})\right] \psi =0,
\end{eqnarray}
with the following nucleon scalar and vector self-energies:
\begin{eqnarray}
\Sigma _{\tau }^{S}=-\Gamma _{\sigma }\sigma -\Gamma _{\delta
}\vec{\delta} \cdot \vec{\tau} \qquad {\rm and} \qquad \Sigma _{\tau
}^{\mu }=\Gamma _{\omega }\omega ^{\mu }+\Gamma _{\rho }\vec{
\rho}^{\mu }\cdot \vec{\tau}+\Sigma ^{\mu (R)}.
\end{eqnarray}
The new term $\Sigma ^{\mu (R)}$ in the vector self-energy, which
is called the \textit{rearrangement} self-energy
\cite{Fuc95,Len95}, is given by
\begin{eqnarray}
\Sigma ^{\mu (R)} =\frac{j^{\mu }}{\rho }(\frac{\partial \Gamma
_{\omega } }{\partial \rho }\bar{\psi}\gamma _{\nu }\psi \omega
^{\nu }+\frac{\partial \Gamma _{\rho }}{\partial \rho
}\bar{\psi}\vec{\tau}\gamma ^{\nu }\psi \cdot \vec{\rho}_{\nu }
-\frac{\partial \Gamma _{\sigma }}{\partial \rho }\bar{\psi}\psi
\sigma - \frac{\partial \Gamma _{\delta }}{\partial \rho
}\bar{\psi}\vec{\tau}\psi \vec{\delta})~,
\end{eqnarray}
with $j^{\mu }=\bar{\psi}\gamma ^{\mu }\psi $. The rearrangement
self-energy plays an essential role in the applications of the
theory since it guarantees both the thermodynamical consistency
and the energy-momentum conservation \cite{Fuc95,Len95}.

For the meson fields, the equations of motion are
\begin{eqnarray}
(\partial _{\mu }\partial ^{\mu }+m_{\sigma }^{2})\sigma &=&\Gamma _{\sigma }%
\bar{\psi}\psi , \\
\partial _{\nu }\omega ^{\mu \nu }+m_{\omega }^{2}\omega ^{\mu } &=&\Gamma
_{\omega }\bar{\psi}\gamma ^{\mu }\psi , \\
(\partial _{\mu }\partial ^{\mu }+m_{\delta }^{2})\vec{\delta}
&=&\Gamma
_{\delta }\bar{\psi}\vec{\tau}\psi , \\
\partial _{\nu }\vec{\rho}^{\mu \nu }+m_{\rho }^{2}\vec{\rho}^{\mu }
&=&\Gamma _{\rho }\bar{\psi}\vec{\tau}\gamma ^{\mu }\psi .
\end{eqnarray}

In the static case for an infinite nuclear matter, the meson
equations of motion become
\begin{eqnarray}
m_{\sigma }^{2}\bar{\sigma} &=&\Gamma _{\sigma }\rho _{S}, \\
m_{\omega }^{2}\bar{\omega}_{0} &=&\Gamma _{\omega }\rho _{B}, \\
m_{\rho }^{2}\bar{\rho}_{0}^{(3)} &=&\Gamma _{\rho }(\rho _{p}-\rho _{n}), \\
m_{\delta }^{2}{\bar{\delta}}^{(3)} &=&\Gamma _{\delta }(\rho
_{S,p}-\rho _{S,n}),
\end{eqnarray}
so the nucleon scalar and vector self-energies are
\begin{eqnarray}
\Sigma _{\tau }^{S}=-\Gamma _{\sigma }\bar{\sigma}-\Gamma _{\delta }{\bar{%
\delta}}^{(3)}\tau _{3} \qquad {\rm and} \qquad \Sigma _{\tau }^{0}
=\Gamma _{\omega }\bar{\omega}_{0}+\Gamma _{\rho }\bar{
\rho}_{0}^{(3)}\tau _{3}+\Sigma ^{0(R)},
\end{eqnarray}
with
\begin{eqnarray}
\Sigma ^{0(R)} =\frac{\partial \Gamma _{\omega }}{\partial \rho
}\bar{ \omega}_{0}\rho _{B}+\frac{\partial \Gamma _{\rho }}{\partial
\rho } \bar{\rho}_{0}^{(3)}(\rho _{p}-\rho _{n})-\frac{\partial
\Gamma _{\sigma }}{\partial \rho }\bar{\sigma}\rho
_{S}-\frac{\partial \Gamma _{\delta }}{\partial \rho
}{\bar{\delta}}^{(3)}(\rho _{S,p}-\rho _{S,n}).
\end{eqnarray}

From the energy-momentum tensor, the energy density and pressure of
nuclear matter can be derived, and they are given by
\begin{eqnarray}
\epsilon =\epsilon _{\rm kin}^{n}+\epsilon _{\rm
kin}^{p}+\frac{1}{2}\left[ m_{\sigma }^{2}\bar{\sigma}^{2}+m_{\omega
}^{2}\bar{ \omega}_{0}^{2}+m_{\delta
}^{2}{\bar{\delta}}^{(3)2}+m_{\rho }^{2}\bar{\rho}
_{0}^{(3)2}\right]
\end{eqnarray}
and
\begin{eqnarray}
P =P_{\rm kin}^{n}+P_{\rm kin}^{p}+\rho _{B}\Sigma ^{0(R)}
-\frac{1}{2}\left[ m_{\sigma }^{2}\bar{\sigma}^{2}-m_{\omega
}^{2}\bar{ \omega}_{0}^{2}+m_{\delta
}^{2}{\bar{\delta}}^{(3)2}-m_{\rho }^{2}\bar{\rho}
_{0}^{(3)2}\right] .
\end{eqnarray}
It is seen that the rearrangement self-energy does not affect the
energy density but contributes explicitly to the pressure.
Furthermore, the symmetry energy can be written as
\begin{eqnarray}
E_{\mathrm{sym}}(\rho _{B})
=\frac{k_{F}^{2}}{6\tilde{E}_{F}}+\frac{1}{2} \left( \frac{\Gamma
_{\rho }}{m_{\rho }}\right) ^{2}\rho _{B}-\frac{1}{2} \left(
\frac{\Gamma _{\delta }}{m_{\delta }}\right) ^{2}\times
\frac{M^{\ast 2}\rho _{B}}{\tilde{E}_{F}^{2}[1+\left( \frac{\Gamma
_{\delta }}{m_{\delta }}\right) ^{2}A(k_{F},M^{\ast })]},
\label{EsymDD}
\end{eqnarray}
with notations similarly defined as in the nonlinear RMF model.

\subsubsection{The nonlinear point-coupling RMF model}

The point-coupling model is defined by a Lagrangian density that
consists of only nucleon fields. In Refs. \cite{Nik92,Bur02}, the
Lagrangian density of the nonlinear point-coupling model is given
by
\begin{eqnarray}
\mathcal{L}_{\mathrm{NLPC}}=\mathcal{L}^{\mathrm{free}}+\mathcal{L}^{\mathrm{\
4f}}+\mathcal{L}^{\mathrm{hot}}+\mathcal{L}^{\mathrm{der}},
\label{LagNLPC}
\end{eqnarray}
with
\begin{eqnarray}
\mathcal{L}^{\mathrm{free}} &=&\bar{\psi}(\mathrm{i}\gamma _{\mu
}\partial^{\mu }-M)\psi , \\
\mathcal{L}^{\mathrm{4f}}\hfill &=&-{{\frac{1}{2}}}\,\alpha
_{\mathrm{S}}( \bar{\psi}\psi )(\bar{\psi}\psi
)-{{\frac{1}{2}}}\,\alpha _{\mathrm{V}}(\bar{ \psi}\gamma _{\mu
}\psi )(\bar{\psi}\gamma ^{\mu }\psi ) -{{\frac{1}{2}}}\,\alpha
_{\mathrm{TS}}(\bar{\psi}\vec{\tau}\psi )\cdot
(\bar{\psi}\vec{\tau}\psi ) \notag\\
&-&{{\frac{1}{2}}}\,\alpha _{\mathrm{TV}}(\bar{\psi}\vec{\tau}\gamma
_{\mu}\psi )\cdot (\bar{\psi}\vec{\tau}\gamma ^{\mu }\psi ), \\
\mathcal{L}^{\mathrm{hot}} &=&-{{\frac{1}{3}}}\,\beta
_{\mathrm{S}}(\bar{\psi}\psi )^{3}-{{\frac{1}{4}}}\, \gamma
_{\mathrm{S}}(\bar{\psi}\psi )^{4} -{{\frac{1}{4}}}\,\gamma
_{\mathrm{V}}[(\bar{\psi}\gamma _{\mu }\psi ) (\bar{\psi}\gamma
^{\mu }\psi )]^{2} \notag\\
&-&{{\frac{1}{4}}}\,\gamma
_{\mathrm{TV}}[(\bar{\psi}\vec{\tau}\gamma _{\mu
}\psi )\cdot (\bar{\psi}\vec{\tau}\gamma ^{\mu }\psi )]^{2}, \\
\mathcal{L}^{\mathrm{der}} &=&-{{\frac{1}{2}}}\,\delta _{\mathrm{S}
}(\partial _{\nu }\bar{\psi}\psi )(\partial ^{\nu }\bar{\psi}\psi )
-{{\frac{1}{2}}}\,\delta _{\mathrm{V}}(\partial _{\nu
}\bar{\psi}\gamma_{\mu }\psi )(\partial ^{\nu }\bar{\psi}\gamma
^{\mu }\psi ) \notag\\
&-&{{\frac{1}{2}}}\,\delta _{\mathrm{TS}}(\partial _{\nu
}\bar{\psi}\vec{\tau} \psi )\cdot (\partial ^{\nu
}\bar{\psi}\vec{\tau}\psi ) -{{\frac{1}{2}}}\,\delta
_{\mathrm{TV}}(\partial _{\nu }\bar{\psi}\vec{\tau }\gamma _{\mu
}\psi )\cdot (\partial ^{\nu }\bar{\psi}\vec{\tau}\gamma ^{\mu }\psi
).
\end{eqnarray}
In the above, $\mathcal{L}^{\mathrm{free}}$ is the kinetic term of
nucleons and $\mathcal{L}^{ \mathrm{4f}}$ describes the four-fermion
interactions while $\mathcal{L}^{\mathrm{hot} } $ and
$\mathcal{L}^{\mathrm{der}}$ contain, respectively, higher-order
terms involving more than four fermions and derivatives in the
nucleon field. For the twelve coupling constants in the Lagrangian
density, $\alpha _{\mathrm{S}}$, $\alpha _{\mathrm{V}}$, $ \alpha
_{\mathrm{TS}}$, $\alpha _{\mathrm{TV}}$, $\beta _{\mathrm{S}}$, $
\gamma _{\mathrm{S}}$, $\gamma _{\mathrm{V}}$, $\gamma
_{\mathrm{TV}}$, $\delta _{\mathrm{S}}$, $\delta _{\mathrm{V}}$,
$\delta _{\mathrm{TS}}$, and $\delta _{\mathrm{TV}}$, the subscripts
denote the tensor structure of a coupling with `S', `V', and `T'
standing for scalar, vector, and isovector, respectively. The
symbols $\alpha _{\mathrm{i}}$, $\delta _{\mathrm{i}}$, $\beta
_{\mathrm{i }}$, and $\gamma _{\mathrm{i}} $ refer, respectively, to
four-fermion or second-order terms, derivative couplings, third- and
fourth order terms \cite{Nik92,Bur02}.

From the variation of the Lagrangian density Eq. (\ref{LagNLPC})
with respect to $\bar{\psi}$, one obtains the following Dirac
equation for the nucleon field:
\begin{eqnarray}
\lbrack \gamma _{\mu }(i\partial ^{\mu }-\Sigma ^{\mu })-(M+\Sigma
^{S})]\psi =0,
\end{eqnarray}
where the nucleon scalar ($\Sigma ^{S}$) and vector ($\Sigma ^{\mu
}$) self-energies are
\begin{eqnarray}
\Sigma ^{S}=V_{S}+\vec{V}_{TS}\cdot \vec{\tau} \qquad {\rm and}
\qquad \Sigma ^{\mu }=V^{\mu }+\vec{V}_{T}^{\mu }\cdot \vec{\tau},
\end{eqnarray}
respectively, with
\begin{eqnarray}
V_{S} &=&\alpha _{\mathrm{S}}(\bar{\psi}\psi )+\beta
_{\mathrm{S}}(\bar{\psi}\psi )^{2}+\gamma _{\mathrm{S}}(\bar{\psi}\psi )^{3}
-\delta _{\mathrm{S}}\square (\bar{\psi}\psi ), \\
\vec{V}_{TS} &=&\alpha _{\mathrm{TS}}(\bar{\psi}\vec{\tau}\psi )-
\delta _{\mathrm{TS}}\square (\bar{\psi}\vec{\tau}\psi ), \\
V^{\mu } &=&\alpha _{\mathrm{V}}(\bar{\psi}\gamma ^{\mu }\psi )
+\gamma _{\mathrm{V}}(\bar{\psi}\gamma ^{\mu }\psi)
(\bar{\psi}\gamma_{\mu }\psi ) (\bar{\psi}\gamma ^{\mu }\psi )
-\delta _{\mathrm{V}}\square (\bar{\psi}\gamma ^{\mu }\psi ), \\
\vec{V}_{T}^{\mu } &=&\alpha
_{\mathrm{TV}}(\bar{\psi}\vec{\tau}\gamma ^{\mu }\psi )+\gamma
_{\mathrm{TV}}(\bar{\psi}\vec{\tau}\gamma ^{\mu }\psi )\cdot (
\bar{\psi}\vec{\tau}\gamma _{\mu }\psi )(\bar{\psi}\vec{\tau}\gamma
^{\mu}\psi )-\delta _{\mathrm{TV}}\square
(\bar{\psi}\vec{\tau}\gamma ^{\mu }\psi ).
\end{eqnarray}
In the above, $\square =\partial ^{2}/(c^{2}\partial
t^{2}-\bigtriangleup )$ denotes the four-dimensional
d'Alembertian. In the translationally invariant infinite nuclear
matter, all terms involving derivative couplings drop out and the
spatial components of the four-currents also vanish. In terms of
the baryon density $\rho_B$ and scalar density $\rho_S$ as well as
the isospin baryon density $\rho_3=\rho_p-\rho_n$ and the isospin
scalar density $\rho_{S3}=\rho_{S,p}-\rho_{S,n}$, the nucleon
scalar and vector self-energies in asymmetric nuclear matter can
be rewritten as
\begin{eqnarray}
\Sigma _{\tau }^{S} &=&\alpha _{\mathrm{S}}\rho _{S}+\beta
_{\mathrm{S}}\rho _{S}^{2}+\gamma _{\mathrm{S}}\rho
_{S}^{3}+\alpha _{\mathrm{TS}}\rho
_{S3}\tau _{3}, \\
\Sigma _{\tau }^{0} &=&\alpha _{\mathrm{V}}\rho _{B}+\gamma
_{\mathrm{V}
}\rho _{B}^{3}+\alpha _{\mathrm{TV}}\rho _{3}\tau _{3}+\gamma _{\mathrm{TV}%
}\rho _{3}^{3}\tau _{3}.  \label{Sig0NLPC}
\end{eqnarray}

The energy density $\epsilon $ and the pressure $P$ derived from
the energy-momentum tensor in the nonlinear point-coupling RMF
model are given by
\begin{eqnarray}
\epsilon &=&\epsilon _{\rm kin}^{n}+\epsilon _{\rm
kin}^{p}-\frac{1}{2}\alpha _{ \mathrm{S}}\rho
_{S}^{2}-\frac{1}{2}\alpha _{\mathrm{TS}}\rho _{S3}^{2}
+\frac{1}{2}\alpha _{\mathrm{V}}\rho ^{2}+\frac{1}{2}\alpha
_{\mathrm{TV}}
\rho _{3}^{2}\;  \notag \\
&&-\frac{1}{3}\beta _{\mathrm{S}}\rho _{S}^{3}-\frac{3}{4}\gamma
_{\mathrm{S} }\rho _{S}^{4}+\frac{1}{4}\gamma _{\mathrm{V}}\rho
^{4}+\frac{1}{4}\gamma _{ \mathrm{TV}}\rho _{3}^{4},
\end{eqnarray}
\begin{eqnarray}
P &=&\tilde{E}_{F}^{p}\rho _{p}+\tilde{E}_{F}^{n}\rho
_{n}-\epsilon_{\rm kin}^{p}-\epsilon _{\rm kin}^{n}+\frac{1}{2}
\alpha _{\mathrm{S}}\rho _{s}^{2}+\frac{1}{2}\alpha _{\mathrm{\
TS}}\rho _{s3}^{2}+\frac{1}{2}\alpha _{\mathrm{V}}\rho
^{2}+\frac{1}{2}
\alpha _{\mathrm{TV}}\rho _{3}^{2}  \notag \\
&&+\frac{2}{3}\beta _{\mathrm{S}}\rho _{s}^{3}+\frac{3}{4}\gamma
_{\mathrm{S} }\rho _{s}^{4}+\frac{3}{4}\gamma _{\mathrm{V}}\rho
^{4}+\frac{3}{4}\gamma _{ \mathrm{TV}}\rho _{3}^{4},
\end{eqnarray}
where $\tilde{E}_{F}^{p}$ and $\tilde{E}_{F}^{n}$ are defined as in Eq. (\ref%
{Ef}) with the nucleon Dirac masses
\begin{eqnarray}
M_{p}^{\ast } &=&\alpha _{\mathrm{S}}\rho _{S}+\beta
_{\mathrm{S}}\rho _{S}^{2}+\gamma _{\mathrm{S}}\rho
_{S}^{3}+\alpha _{\mathrm{TS}}\rho _{S3},
\\
M_{n}^{\ast } &=&\alpha _{\mathrm{S}}\rho _{S}+\beta
_{\mathrm{S}}\rho _{S}^{2}+\gamma _{\mathrm{S}}\rho
_{S}^{3}-\alpha _{\mathrm{TS}}\rho _{S3}.
\end{eqnarray}
Furthermore, the symmetry energy in this model can be expressed as
\begin{eqnarray}
E_{\mathrm{sym}}(\rho _{B})
=\frac{k_{F}^{2}}{6\tilde{E}_{F}}+\frac{1}{2} \alpha
_{\mathrm{TV}}\rho _{B}+\frac{1}{2}\alpha_{\mathrm{TS}}\frac{M^{\ast
2}\rho _{B}}{\tilde{E} _{F}^{2}[1-\alpha
_{\mathrm{TS}}A(k_{F},M^{\ast })]},  \label{EsymNLPC}
\end{eqnarray}
with notations again similarly defined as in the nonlinear RMF
model.

\subsubsection{The density-dependent point-coupling RMF model}

For the density-dependent point-coupling RMF model, the Lagrangian
density can be written as \cite{Fin04,Fin06}
\begin{eqnarray}
\mathcal{L}_{\text{DDPC}}=\mathcal{L}_{\mathrm{free}}+\mathcal{L}_{\mathrm{%
4f }}+\mathcal{L}_{\mathrm{der}},  \label{LagDDPC}
\end{eqnarray}
with
\begin{eqnarray}
\mathcal{L}_{\mathrm{free}} &=&\bar{\psi}(i\gamma _{\mu }\partial
^{\mu}-M)\psi , \\
\mathcal{L}_{\mathrm{4f}}&=&-\frac{1}{2}~G_{S}(\hat{\rho})(\bar{\psi}\psi
)( \bar{\psi}\psi )-\frac{1}{2}~G_{V}(\hat{\rho})(\bar{\psi}
\gamma _{\mu }\psi )(\bar{\psi}\gamma ^{\mu }\psi )  \notag \\
&-&\frac{1}{2}~G_{TS}(\hat{\rho})(\bar{\psi}\vec{\tau}\psi )\cdot
(\bar{\psi}\vec{\tau}\psi)-\frac{1}{2}~G_{TV}(\hat{\rho})(\bar{\psi}\vec{\tau}\gamma
_{\mu }\psi)\cdot (\bar{\psi}\vec{\tau}\gamma ^{\mu }\psi ), \\
\mathcal{L}_{\mathrm{der}}
&=&-\frac{1}{2}~D_{S}(\hat{\rho})(\partial _{\nu } \bar{\psi}\psi
)(\partial ^{\nu }\bar{\psi}\psi ).
\end{eqnarray}
In the above, $\mathcal{L}^{\mathrm{free}}$ is the kinetic term of
nucleons and $\mathcal{L}^{\mathrm{4f}}$ is a four-fermion
interaction while $\mathcal{L}^{\mathrm{der}}$ represents
derivatives in the nucleon scalar densities. Unlike in the nonlinear
point-coupling RMF model, the density-dependent point-coupling RMF
model used here includes only second-order interaction terms with
density-dependent couplings $G_{i}(\hat{\rho})$ and
$D_{i}(\hat{\rho}) $ that are determined from finite-density QCD sum
rules and the in-medium chiral perturbation theory
\cite{Fin04,Fin06}.

Variation of the Lagrangian Eq. (\ref{LagDDPC}) with respect to
$\bar{\psi}$ leads to the single-nucleon Dirac equation
\begin{eqnarray}
\lbrack \gamma _{\mu }(i\partial ^{\mu }-\Sigma ^{\mu })-(M+\Sigma
^{S})]\psi =0,
\end{eqnarray}
with the nucleon scalar and vector self-energies given,
respectively, by
\begin{eqnarray}
\Sigma ^{S}=V_{S}+\vec{V}_{TS}\cdot \vec{\tau}+\Sigma _{rS} \qquad
{\rm and} \qquad \Sigma ^{\mu }=V^{\mu }+\vec{V}_{T}^{\mu }\cdot
\vec{\tau}+\Sigma _{r}^{\mu },
\end{eqnarray}
where
\begin{eqnarray}
V_{S} &=&G_{S}(\bar{\psi}\psi )-D_{S}\square (\bar{\psi}\psi ), \\
\vec{V}_{TS} &=&G_{TS}(\bar{\psi}\vec{\tau}\psi ), \\
V^{\mu } &=&G_{V}(\bar{\psi}\gamma ^{\mu }\psi ), \\
\vec{V}_{T}^{\mu } &=&G_{TV}(\bar{\psi}\vec{\tau}\gamma ^{\mu }\psi ), \\
\Sigma _{rS} &=&-\frac{\partial D_{S}}{\partial
\hat{\rho}}(\partial _{\nu }j^{\mu })u_{\mu }(\partial ^{\nu
}(\bar{\psi}\psi ))
\end{eqnarray}
and
\begin{eqnarray}
\Sigma _{r}^{\mu } &=&\frac{u^{\mu }}{2}\left( \frac{\partial
G_{S}}{\partial \hat{\rho}}(\bar{\psi}\psi )(\bar{\psi}\psi
)+\frac{\partial G_{TS} }{\partial
\hat{\rho}}(\bar{\psi}\vec{\tau}\psi )\cdot (\bar{\psi}\vec{\tau}
\psi )+\frac{\partial G_{V}}{\partial \hat{\rho}}(\bar{\psi}\gamma
^{\mu }\psi)(\bar{\psi}\gamma _{\mu }\psi ) \right. \notag \\
&+&\frac{\partial G_{TV}}{\partial \hat{\rho}}(
\bar{\psi}\vec{\tau}\gamma ^{\mu }\psi )\cdot
(\bar{\psi}\vec{\tau}\gamma _{\mu }\psi )\left. +\frac{\partial
D_{S}}{\partial \hat{\rho}}(\partial ^{\nu }(\bar{ \psi}\psi
))(\partial _{\nu }(\bar{\psi}\psi ))\right).
\end{eqnarray}
In the above, one has $\hat{\rho}u^{\mu }=\bar{\psi}\gamma ^{\mu
}\psi $, where the four-velocity $u^{\mu }$ is defined as
$(1-\mathbf{v}^{2})^{-1/2}(1,\mathbf{v})$ with $\mathbf{v}$ being
the three-velocity vector, and $\Sigma _{rS}$ and $\Sigma _{r}^{\mu
} $ represent the rearrangement contributions resulting from the
variation of the vertex functionals with respect to the nucleon
field in the density operator $\hat{\rho}$. The latter coincides
with the baryon density in the nuclear matter rest frame.

In the translationally invariant infinite asymmetric nuclear matter,
the nucleon scalar and vector self-energies become
\begin{eqnarray}
\Sigma _{\tau }^{S} =G_{S}\rho _{S}+G_{TS}\rho _{S3}\tau _{3} \qquad
{\rm and} \qquad \Sigma _{\tau }^{0} =G_{V}\rho _{B}+G_{TV}\rho
_{3}\tau _{3}+\Sigma ^{0(R)},
\end{eqnarray}
with the rearrangement contribution to the self-energy
\begin{eqnarray}
\Sigma ^{0(R)}=\frac{1}{2}[\frac{\partial G_{S}}{\partial \rho
}\rho _{S}^{2}+\frac{\partial G_{TS}}{\partial \rho }\rho
_{S3}^{2}+\frac{\partial G_{V}}{\partial \rho }\rho
^{2}+\frac{\partial G_{TV}}{\partial \rho }\rho _{3}^{2}].
\end{eqnarray}

For asymmetric nuclear matter, the energy density $\epsilon $ and
the pressure $P$ derived from the energy-momentum tensor in the
density-dependent point-coupling RMF model are
\begin{eqnarray}
\epsilon =\epsilon _{\rm kin}^{n}+\epsilon _{\rm
kin}^{p}-\frac{1}{2}G_{S}\rho _{S}^{2}-\frac{1}{2}G_{TS}\rho
_{S3}^{2}+\frac{1}{2}G_{V}\rho ^{2}+\frac{1}{2}G_{TV}\rho _{3}^{2},
\end{eqnarray}
and
\begin{eqnarray}
P &=&\tilde{E}_{F}^{p}\rho _{p}+\tilde{E}_{F}^{n}\rho _{n}-\epsilon
_{\rm kin}^{p}-\epsilon _{\rm kin}^{n}+\frac{1}{2}G_{V}\rho
^{2}+\frac{1}{2}G_{TV}\rho _{3}^{2}+\frac{1}{2}
G_{S}\rho _{S}^{2}+\frac{1}{2}G_{TS}\rho _{S3}^{2}  \notag \\
&+&\frac{1}{2}\frac{\partial G_{S}}{\partial \rho }\rho _{S}^{2}\rho
+\frac{ 1}{2}\frac{\partial G_{V}}{\partial \rho }\rho ^{3}
+\frac{1}{2}\frac{\partial G_{TV}}{\partial \rho }\rho _{3}^{2}\rho
+\frac{ 1}{2}\frac{\partial G_{TS}}{\partial \rho }\rho
_{S3}^{2}\rho,
\end{eqnarray}
where $\tilde{E}_{F}^{p}$ and $\tilde{E}_{F}^{n}$ are defined as
in Eq. (\ref{Ef}) with the effective nucleon masses
\begin{eqnarray}
M_{p}^{\ast } =M+G_{S}\rho _{S}+G_{TS}\rho _{S3} \qquad {\rm and}
\qquad M_{n}^{\ast } =M+G_{S}\rho _{S}-G_{TS}\rho _{S3}.
\end{eqnarray}
As in the density-dependent RMF model, the \textit{rearrangement}
contributions appear explicitly in the expression for the pressure.
Finally, the symmetry energy can be written as
\begin{eqnarray}
E_{\mathrm{sym}}(\rho _{B})
=\frac{k_{F}^{2}}{6\tilde{E}_{F}}+\frac{1}{2} G_{TV}\rho _{B}
+\frac{1}{2}G_{TS}\frac{M^{\ast 2}\rho _{B}}{\tilde{E}
_{F}^{2}[1-G_{TS}A(k_{F},M^{\ast })]},  \label{EsymDDPC}
\end{eqnarray}
with similar notations as in the nonlinear RMF model.

\subsection{RMF model predictions on the symmetry energy, symmetry
potential, and neutron-proton effective mass splitting}

All above models have been used to study the isospin-dependent
properties of asymmetric nuclear matter and the nuclear structure
properties of finite nuclei. In the following, we focus on results
regarding the nuclear symmetry energy, the nuclear symmetry
potential, the isospin-splitting of nucleon effective mass, and the
isospin-dependent nucleon scalar density in asymmetric nuclear
matter. For different versions of the RMF model considered in the
above, we mainly consider parameter sets commonly and successfully
used in nuclear structure studies. In particular, we select the
parameter sets NL1 \cite{Lee86}, NL2 \cite{Lee86}, NL3 \cite{Lal97},
NL-SH \cite{Sha93}, TM1 \cite{Sug94}, PK1 \cite{Lon04}, FSU-Gold
\cite{Tod05}, HA \cite{Bun03}, NL$\rho $ \cite{Liu02}, NL$\rho
\delta $ \cite{Liu02} for the nonlinear RMF model; TW99
\cite{Typ99}, DD-ME1 \cite{Nik02}, DD-ME2 \cite{Lal05}, PKDD
\cite{Lon04}, DD \cite{Typ05}, DD-F \cite{Kla06}, and DDRH-corr
\cite{Hof01} for the density-dependent RMF model; and PC-F1
\cite{Bur02}, PC-F2 \cite{Bur02}, PC-F3 \cite{Bur02}, PC-F4
\cite{Bur02}, PC-LA \cite{Bur02}, and FKVW \cite{Fin06} for the
point-coupling RMF model. There are totally $23$ parameter sets, and
most of them can describe reasonably well the binding energies and
charge radii of a large number of nuclei in the periodic table
except the parameter set HA, for which to our knowledge there are no
calculations for finite nuclei.

We note that all selected parameter sets include the
isovector-vector channel involving either the isovector-vector $\rho
$ meson or the isovector-vector interaction vertices in the
Lagrangian. The HA parameter set further includes the
isovector-scalar meson field $\vec{\delta}$ and fits successfully
some results obtained from the more microscopic DBHF approach
\cite{Bun03}. The parameter sets NL$\rho \delta $ and DDRH-corr also
include the isovector-scalar meson field $\vec{\delta}$, while
PC-F2, PC-F4, PC-LA, and FKVW include the isovector-scalar
interaction vertices. The parameter sets NL$\rho \delta $ as well as
NL$\rho $ are obtained from fitting the empirical properties of
asymmetric nuclear matter \cite{Liu02} and describe reasonably well
the binding energies and charge radii of a large number of nuclei
\cite{Gai04}. For the DDRH-corr, its parameters are determined from
the density-dependent meson-nucleon vertices extracted from the
self-energies of asymmetric nuclear matter calculated in the
microscopic DBHF approach with momentum corrections, and it
reproduces satisfactorily the properties of finite nuclei and the
EOS from the DBHF approach \cite{Hof01}. In the parameter sets
PC-F1, PC-F2, PC-F3, PC-F4 and PC-LA for the nonlinear
point-coupling model, their coupling constants are determined in a
self-consistent procedure that solves the model equations for
representative nuclei simultaneously in a generalized nonlinear
least-squares adjustment algorithm \cite{Bur02}. The parameter set
FKVW for the density-dependent point-coupling model are determined
by constraints derived from the finite-density QCD sum rules, the
in-medium chiral perturbation theory, and the experimental data on a
number of finite nuclei \cite{Fin06}.

\subsubsection{The density dependence of the nuclear symmetry energy}

\begin{figure}[th]
\centerline{\includegraphics[scale=1.2]{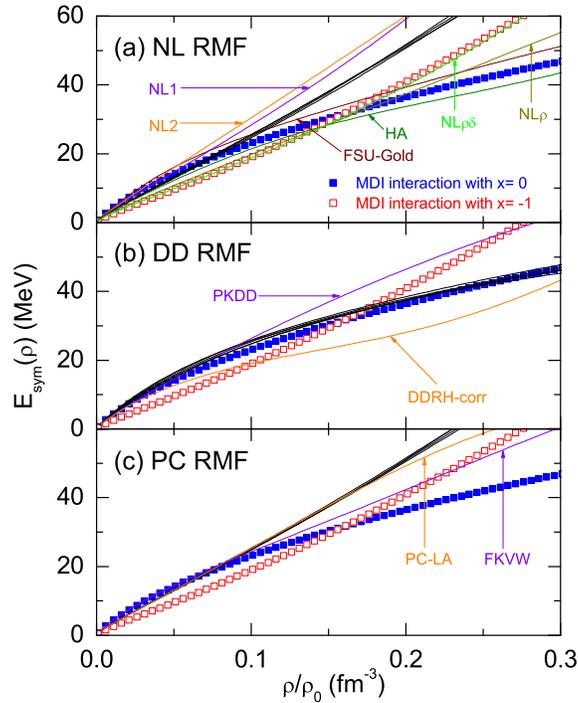}}
\caption{{\protect\small (Color online) Density dependence of the
nuclear symmetry energy }$E_{\mathrm{sym}}(\protect\rho
)${\protect\small \ for the parameter sets NL1, NL2, NL3, NL-SH,
TM1, PK1, FSU-Gold, HA, NL}$\protect \rho ${\protect\small , and
NL}$\protect\rho \protect\delta ${\protect\small \ in the
nonlinear RMF model (a); TW99, DD-ME1, DD-ME2, PKDD, DD, DD-F, and
DDRH-corr in the density-dependent RMF model (b); and PC-F1,
PC-F2, PC-F3, PC-F4, PC-LA, and FKVW in the point-coupling RMF
model (c). For comparison, results from the MDI interaction with
}$x=-1${\protect\small \ (open squares) and }$0${\protect\small \
(solid squares) are also shown. Taken from Ref. \cite{Che07}.}}
\label{EsymDen}
\end{figure}
Fig. \ref{EsymDen} displays the density dependence of the nuclear
symmetry energy $E_{\mathrm{sym}}(\rho )$ for the $23$ parameter
sets in the nonlinear, density-dependent, and point-coupling RMF
models. Also shown in Fig. \ref{EsymDen} for comparison are
results from the phenomenological parametrization of the
momentum-dependent nuclear mean-field potential based on the Gogny
effective interaction \cite{Das03}, i.e., the MDI interactions
with $x=-1$ (open squares) and $0$ (solid squares), where
different $x$ values correspond to different density dependence of
the nuclear symmetry energy but keep other properties of the
nuclear EOS the same \cite{Che05a}. From analyzing the isospin
diffusion data from NSCL/MSU using the IBUU04 transport model with
in-medium NN cross sections, it has been found that the MDI
interactions with $x=-1$ and $0$ give, respectively, the upper and
lower bounds for the stiffness of the nuclear symmetry energy at
densities up to about $1.2\rho_0$ \cite{LiBA05c,Che05a}.

It is seen from Fig. \ref{EsymDen} that the density dependence of
symmetry energy varies drastically among different interactions. In
the nonlinear RMF model, while the dependence on density is almost
linear for most parameter sets, it is much softer for the parameter
sets FSU-Gold and HA. The softening of the symmetry energy from the
latter two parameter sets is due to the mixed isoscalar-isovector
couplings $\Lambda _{S}$ and $\Lambda _{V}$ \cite{Hor01a,Mul96}
which modify the density dependence of symmetry energy as seen in
Eq. (\ref{EsymNL}). For the parameter set NL$\rho \delta $, it gives
a symmetry energy that depends linearly on density at low densities
but becomes stiffer at high densities due to inclusion of the
isovector-scalar $\delta $ meson. The approximate linear
density-dependent behavior of the symmetry energy for other
parameter sets in the nonlinear RMF model can also be understood
from Eq. (\ref{EsymNL}), which shows that the symmetry energy at
high densities is dominated by the potential energy that is
proportional to the baryon density if the mixed isoscalar-isovector
coupling and the isovector-scalar $\delta $ meson are not included
in the model.

The density dependence of the symmetry energy in the
density-dependent RMF model is essentially determined by the
density dependence of the coupling constants $\Gamma _{\rho }$ and
$\Gamma _{\delta }$ of isovector mesons. Most parameter sets in
this case give similar symmetry energies except the parameter sets
PKDD and DDRH-corr. Compared with other parameter sets in the
density-dependent RMF model, the PKDD gives a very large while the
DDRH-corr gives a very small value for the symmetry energy at
saturation density. For point-coupling models, all parameter sets
(PC-F1, PC-F2, PC-F3, PC-F4 and PC-LA) in the nonlinear
point-coupling RMF model predict almost linearly density-dependent
symmetry energies while the parameter set FKVW in the
density-dependent point-coupling RMF model gives a somewhat softer
symmetry energy.

\begin{table}[tbp]
\caption{{\protect\small Bulk properties of nuclear matter at the
saturation point: }$-B/A${\protect\small \ (MeV), }$\protect\rho
_{0}${\protect\small \ (fm}$^{-3}${\protect\small ),
}$K_{0}${\protect\small \ (MeV), }$E_{\text{sym }}(\protect\rho
_{0})${\protect\small \ (MeV), }$K_{\text{sym}}$ {\protect\small \
(MeV), }$L${\protect\small \ (MeV), and }$K_{\text{asy}}$
{\protect\small \ (MeV) using the }$23${\protect\small \ parameter
sets in the nonlinear, density-dependent, and point-coupling RMF
models. The last column gives the references for corresponding
parameter sets. Taken from Ref. \cite{Che07}.}}
\label{Bulk}%
\begin{center}
\begin{tabular}{ccccccccc}
\hline\hline
Model & $\quad -B/A$ & $\rho _{0}$ & $K_{0}$ & $E_{\text{sym}}$ & $L$ & $K_{%
\text{sym}}$ & $K_{\text{asy}}$ & Ref. \\ \hline
NL1 & $16.4$ & $0.152$ & $212$ & $43.5$ & $140$ & $143$ & $-697$ & \cite%
{Lee86} \\
NL2 & $17.0$ & $0.146$ & $401$ & $44.0$ & $130$ & $20$ & $-750$ & \cite%
{Lee86} \\
NL3 & $16.2$ & $0.148$ & $271$ & $37.3$ & $118$ & $100$ & $-608$ & \cite%
{Lal97} \\
NL-SH & $16.3$ & $0.146$ & $356$ & $36.1$ & $114$ & $80$ & $-604$ & \cite%
{Sha93} \\
TM1 & $16.3$ & $0.145$ & $281$ & $36.8$ & $111$ & $34$ & $-632$ & \cite%
{Sug94} \\
PK1 & $16.3$ & $0.148$ & $282$ & $37.6$ & $116$ & $55$ & $-641$ & \cite%
{Lon04} \\
FSUGold & $16.3$ & $0.148$ & $229$ & $32.5$ & $60$ & $-52$ & $-412$ & \cite%
{Tod05} \\
HA & $15.6$ & $0.170$ & $233$ & $30.7$ & $55$ & $-135$ & $-465$ & \cite%
{Bun03} \\
NL$\rho $ & $16.1$ & $0.160$ & $240$ & $30.3$ & $85$ & $3$ & $-507$ & \cite%
{Liu02} \\
NL$\rho \delta $ & $16.1$ & $0.160$ & $240$ & $30.7$ & $103$ &
$127$ & $-491$
& \cite{Liu02} \\
&  &  &  &  &  &  &  &  \\
TW99 & $16.2$ & $0.153$ & $241$ & $32.8$ & $55$ & $-124$ & $-454$ & \cite%
{Typ99} \\
DD-ME1 & $16.2$ & $0.152$ & $245$ & $33.1$ & $55$ & $-101$ & $-431$ & \cite%
{Nik02} \\
DD-ME2 & $16.1$ & $0.152$ & $251$ & $32.3$ & $51$ & $-87$ & $-393$ & \cite%
{Lal05} \\
PKDD & $16.3$ & $0.150$ & $263$ & $36.9$ & $90$ & $-80$ & $-620$ & \cite%
{Lon04} \\
DD & $16.0$ & $0.149$ & $241$ & $31.7$ & $56$ & $-95$ & $-431$ &
\cite{Typ05}
\\
DD-F & $16.0$ & $0.147$ & $223$ & $31.6$ & $56$ & $-140$ & $-476$ & \cite%
{Kla06} \\
DDRH-corr & $15.6$ & $0.180$ & $281$ & $26.1$ & $51$ & $155$ & $-151$ & \cite%
{Hof01} \\
&  &  &  &  &  &  &  &  \\
PC-F1 & $16.2$ & $0.151$ & $255$ & $37.8$ & $117$ & $75$ & $-627$ & \cite%
{Bur02} \\
PC-F2 & $16.2$ & $0.151$ & $256$ & $37.6$ & $116$ & $65$ & $-631$ & \cite%
{Bur02} \\
PC-F3 & $16.2$ & $0.151$ & $256$ & $38.3$ & $119$ & $74$ & $-640$ & \cite%
{Bur02} \\
PC-F4 & $16.2$ & $0.151$ & $255$ & $37.7$ & $119$ & $98$ & $-616$ & \cite%
{Bur02} \\
PC-LA & $16.1$ & $0.148$ & $263$ & $37.2$ & $108$ & $-61$ & $-709$ & \cite%
{Bur02} \\
FKVW & $16.2$ & $0.149$ & $379$ & $33.1$ & $80$ & $11$ & $-469$ & \cite%
{Fin06} \\ \hline\hline
\end{tabular}%
\end{center}
\end{table}

Fig. \ref{EsymDen} thus shows that only a few parameter sets can
give symmetry energies that are consistent with the constraint from
the isospin diffusion data in heavy-ion collisions, which is given
by results from the MDI interactions with $x=-1$ and $0$. The main
reason for this is that most parameter sets in the RMF model have
saturation densities and symmetry energies at their saturation
densities which are significantly different from the empirical
saturation density of $0.16$ fm$^{-3}$ and symmetry energy of $31.6$
MeV at this saturation density. This can be more clearly seen in
Table \ref{Bulk}, which gives the bulk properties of nuclear matter
at saturation density: the binding energy per nucleon $-B/A$\ (MeV),
the saturation density of symmetric nuclear matter $\rho _{0}$\
(fm$^{-3}$), the incompressibility of symmetric nuclear matter
$K_{0}$\ (MeV), the symmetry energy $E_{\mathrm{sym} }(\rho _{0})$\
(MeV), $K_{\mathrm{sym}}$\ (MeV), $L$\ (MeV), and $K_{\mathrm{\
asy}}$\ (MeV) using the $23$ parameter sets in the nonlinear,
density-dependent, and point-coupling RMF models. It is seen that
these parameter sets give saturation densities varying from $\rho
_{0}=0.145$ fm$^{-3}$ to $\rho _{0}=0.180$ fm$^{-3}$ and nuclear
symmetry energies $E_{\mathrm{sym}}(\rho _{0})$\ (MeV) ranging from
$26.1$ to $44.0$ MeV.

\begin{figure}[th]
\centerline{\includegraphics[scale=1.1]{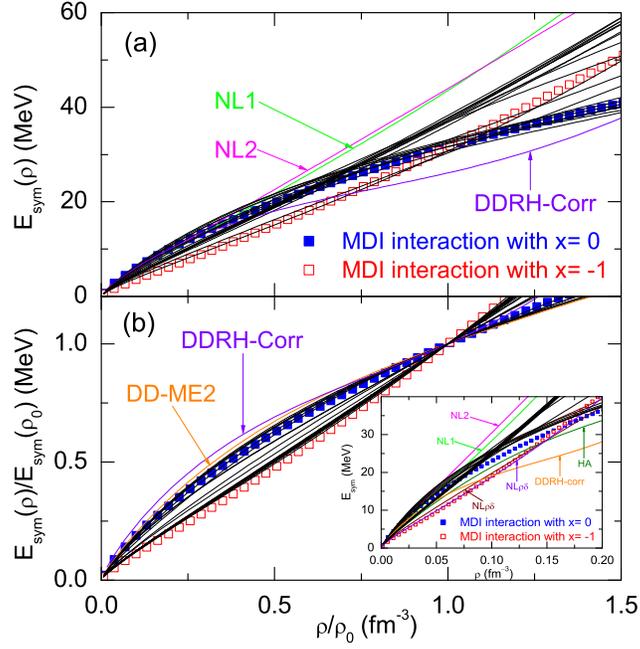}}
\caption{{\protect\small (Color online) The symmetry energy
}$E_{\mathrm{sym} }(\protect\rho )${\protect\small \ (a) and the
scaled symmetry energy }$E_{ \mathrm{sym}}(\protect\rho
)/E_{\mathrm{sym}}(\protect\rho _{0})$ {\protect\small \ (b) as
functions of the scaled baryon density }$\protect\rho
/\protect\rho _{0}${\protect\small \ for the 23 parameter sets in
the nonlinear, density-dependent, and point-coupling RMF models.
Results of the MDI interaction with }$x=-1${\protect\small \ (open
squares) and }$0$ {\protect\small \ (solid squares) are also
included for comparison. The inset in panel (b) shows the symmetry
energy }$E_{\mathrm{sym}}(\protect\rho )${\protect\small \ as a
function of the baryon density }$\protect\rho $ {\protect\small \
without scaling. Taken from Ref. \cite{Che07}.}}
\label{EsymDenSCL}
\end{figure}

\begin{figure*}[th]
\centerline{\includegraphics[scale=1.3]{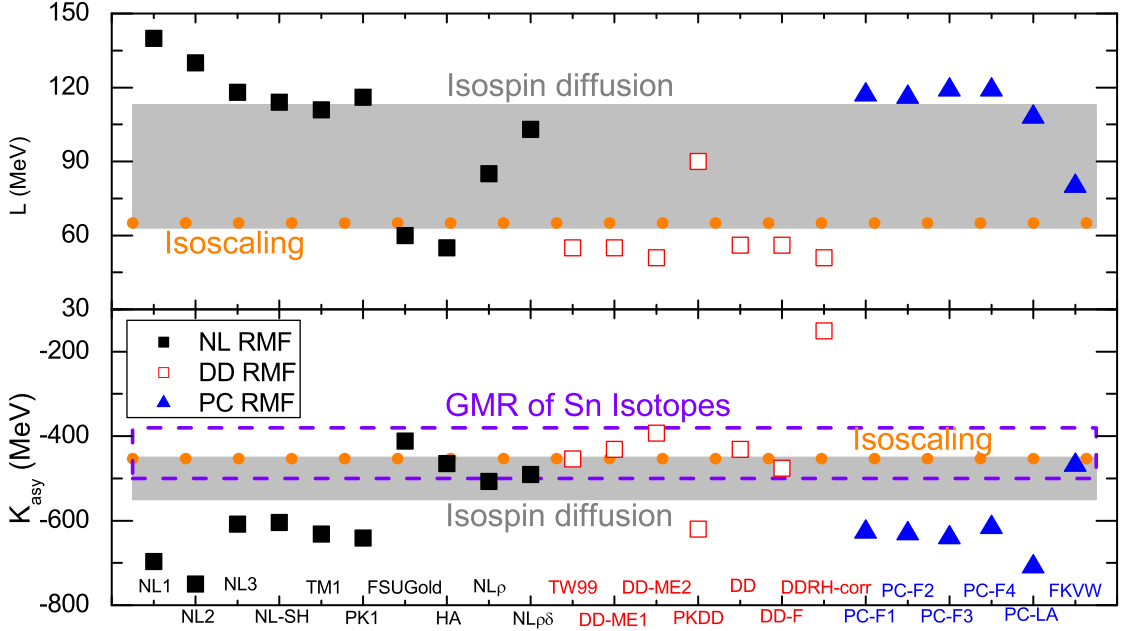}}
\caption{{\protect\small (Color online) Values of
}$L${\protect\small \ and } $K_{\mathrm{asy}}${\protect\small \
for the }$23${\protect\small \ parameter sets in the nonlinear
(solid squares), density-dependent (open squares), and
point-coupling (triangles) RMF models. The constraints from the
isospin diffusion data (shaded band), the isoscaling data (solid
circles), and the isotopic dependence of the GMR in even-A Sn
isotopes (dashed rectangle) are also included. Taken from Ref.
\cite{Che07}.}} \label{LKasy}
\end{figure*}

The effect due to differences in the saturation densities among
different parameter sets can be removed by considering both the
symmetry energy $E_{\mathrm{sym}}(\rho )$ and the symmetry energy
that is scaled by its value at corresponding saturation density,
i.e., $ E_{\mathrm{sym}}(\rho )/E_{\mathrm{sym}}(\rho _{0})$, as
functions of the scaled baryon density $\rho /\rho _{0}$, and this
is shown in Fig.~\ref{EsymDenSCL} for different parameter sets. For
comparison, the symmetry energy $E_{\mathrm{sym}}(\rho )$ as a
function of the baryon density $\rho $ without scaling is also shown
in the inset in panel (b) of Fig.~\ref{EsymDenSCL}. It is seen that
more parameter sets among the $23$ sets become consistent with the
constraint from the isospin diffusion data in heavy-ion collisions
after scaling the baryon density by the saturation density, and with
further scaling of the symmetry energy by its value at corresponding
saturation density, most of the parameter sets are in agreement with
the constraint from the isospin diffusion data. It is also
interesting to see from the inset in Fig.~\ref{EsymDenSCL} that most
of the parameter sets obtained from fitting the properties of finite
nuclei give roughly the same value of about $26$ MeV for the nuclear
symmetry energy at the same baryon density of $\rho =0.1$ fm$^{-3}$.
This interesting feature is very similar to that found with the
Skyrme interactions \cite{Che05b,Bro00}. It implies that the
constraint on the symmetry energy from fitting the properties of
finite nuclei is particularly sensitive to the nuclear properties at
lower densities, i.e., at densities slightly above half-saturation
density.

For the density dependence of the nuclear symmetry energy around
saturation density, a more reasonable and physically meaningful
comparison is through the values of $L$ and $K_{\mathrm{asy}}$
given by these parameter sets, since the $L$ parameter is
correlated linearly to the neutron-skin thickness of finite nuclei
while the $K_{\mathrm{asy}}$ parameter determines the isotopic
dependence of the GMR for a fixed element. From Table \ref{Bulk},
one can see that the values of $L$, $K_{\mathrm{sym}}$, and
$K_{\mathrm{asy}}$ vary drastically, and they are in the range of
$51\sim 140$ MeV, $-140\sim 143$ MeV and $-750\sim $ $-151$ MeV,
respectively. The extracted values of $L=88\pm 25$ MeV and $K_{
\mathrm{asy}}=-500\pm 50$ MeV from the isospin diffusion data,
$L\approx 65$ MeV and $K_{\mathrm{asy}}\approx -453$ MeV from the
isoscaling data, and $K_{ \mathrm{asy}}=-550\pm 100$ MeV from the
isotopic dependence of the GMR in even-A Sn isotopes give a rather
stringent constraint on the density dependence of the nuclear
symmetry energy and thus put strong constraints on the nuclear
effective interactions as well. This constraint can be more
clearly seen in Fig.~\ref{LKasy}, which shows the values of $L$
and $K_{\mathrm{asy }}$ obtained from the $23$ parameter sets in
the nonlinear, density-dependent, and point-coupling RMF models
together with the constraints from the isospin diffusion data,
isoscaling data, and the isotopic dependence of the GMR in even-A
Sn isotopes. From Fig.~\ref{LKasy} as well as Table \ref{Bulk},
one sees clearly that among the $23$ parameter sets considered
here, only six sets, i.e., TM1, NL$\rho $, NL$\rho \delta $, PKDD,
PC-LA, and FKVW, have nuclear symmetry energies that are
consistent with the extracted $L$ value of $88\pm 25$ MeV while
fifteen sets, i.e., NL3, NL-SH, TM1, PK1, HA, NL$\rho $, NL$\rho
\delta $, TW99, PKDD, DD-F, PC-F1, PC-F2, PC-F3, PC-F4, and FKVW,
have nuclear symmetry energies that are consistent with the
extracted $K_{\mathrm{asy}}$ value of $-500\pm 50$ MeV or $-550\pm
100$ MeV. Among the latter fifteen sets, only six sets, i.e., HA,
NL$\rho $, NL$\rho \delta $, TW99, DD-F, and FKVW are consistent
with $K_{\mathrm{asy}}=-500\pm 50$ MeV. One notes that most
parameter sets in the nonlinear and point-coupling RMF models
predict stiffer symmetry energies (i.e., larger values for the $L$
parameter and larger magnitudes for $K_{\mathrm{asy}} $) while
those in the density-dependent RMF model give softer symmetry
energies (i.e., smaller values for the $L$ parameter and smaller
magnitudes for $K_{\mathrm{asy}}$).

\begin{figure}[th]
\centerline{\includegraphics[scale=1.0]{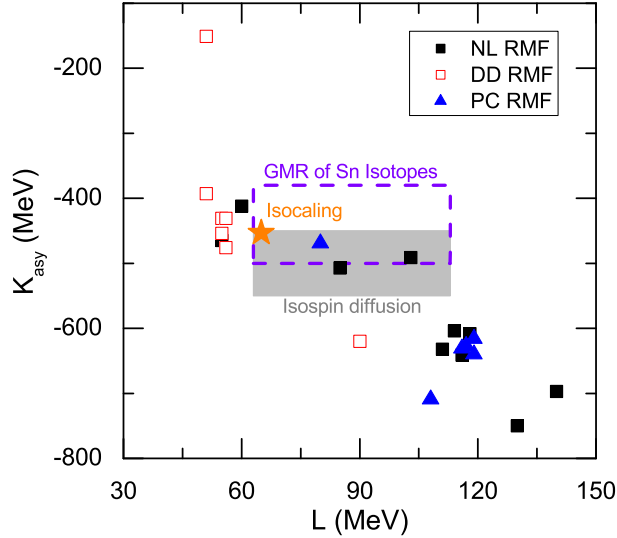}}
\caption{{\protect\small (Color online) Correlations between }$L$
{\protect\small \ and }$K_{\text{asy}}${\protect\small \ for the
23 parameter sets in the nonlinear (solid squares),
density-dependent (open squares), and point-coupling (triangles)
RMF models. The constraints from the isospin diffusion data
(shaded band), the isoscaling data (stars), and the isotopic
dependence of the GMR in even-A Sn isotopes (dashed rectangle with
$L$ constrained by the isospin diffusion data) are also included.
Taken from Ref. \cite{Che07}.}} \label{LKasyCorr}
\end{figure}

Table \ref{Bulk} also shows that only five parameter sets, i.e.,
TM1, NL$\rho $, NL$\rho \delta $, PKDD and FKVW, in the $23$
parameter sets have nuclear symmetry energies that are consistent
with the extracted values for both $L$ and $ K_{\mathrm{asy}}$
($-500\pm 50$ MeV or $-550\pm 100$ MeV). This can be seen more
clearly in Fig. \ref{LKasyCorr} where the correlation between $L$
and $K_{\mathrm{asy}}$ is displayed for the $23$ parameter sets
together with the constraints from the isospin diffusion data, the
isoscaling data, and the isotopic dependence of the GMR in even-A Sn
isotopes. Fig. \ref{LKasyCorr} further shows that there exists an
approximately linear correlation between $L$ and $
K_{\mathrm{asy}}$, i.e., a larger $L$ leads to a larger magnitude
for $K_{ \mathrm{asy}}$. A similar approximately linear correlation
between $L$ and $K_{ \mathrm{asy}}$ has also been observed in Ref.
\cite{Che05a} for the phenomenological MDI interactions, and this
correlation can be understood from the relation $K_{\rm asy}\approx
K_{\rm sys}-6L$, which shows that the value of $K_{\mathrm{asy}}$ is
more sensitive to the value of $L$ than to that of
$K_{\mathrm{sym}}$.

The above comparisons thus indicate that the extracted values of
$L=88\pm 25$ MeV and $K_{\mathrm{asy}}=-500\pm 50$ MeV from the
isospin diffusion data, $L\approx 65$ MeV and
$K_{\mathrm{asy}}\approx -453$ MeV from the isoscaling data, and
$K_{\mathrm{asy}}=-550\pm 100$ MeV from the isotopic dependence of
the GMR in even-A Sn isotopes indeed put a rather stringent
constraint on the values of the parameters in different RMF
models. The fact that most of the $23$ parameter sets considered
here give symmetry energies that are inconsistent with the
constraints of $L=88\pm 25$ MeV and $K_{\mathrm{asy}}=-500\pm 50$
MeV or $-550\pm 100$ MeV is probably related to the rather limited
flexibility in the parametrization of the isovector channel in all
RMF models. They are also probably connected to the fact that most
of the parameter sets are obtained from fitting properties of
finite nuclei, which are mostly near the $\beta $-stability line
and thus are not well constrained by the isospin-dependent
properties of nuclear EOS. Also, one is interested here in the
density-dependent behavior of the symmetry energy around
saturation density, as both $L$ and $K_{\mathrm{asy} } $ are
defined at saturation density, while the behavior of the nuclear
EOS at sub-subsaturation density may be more relevant when the
parameter sets are obtained from fitting the properties of finite
nuclei.

\subsubsection{The momentum dependence of the nuclear symmetry potential}

\begin{figure}[th]
\centerline{\includegraphics[scale=1.2]{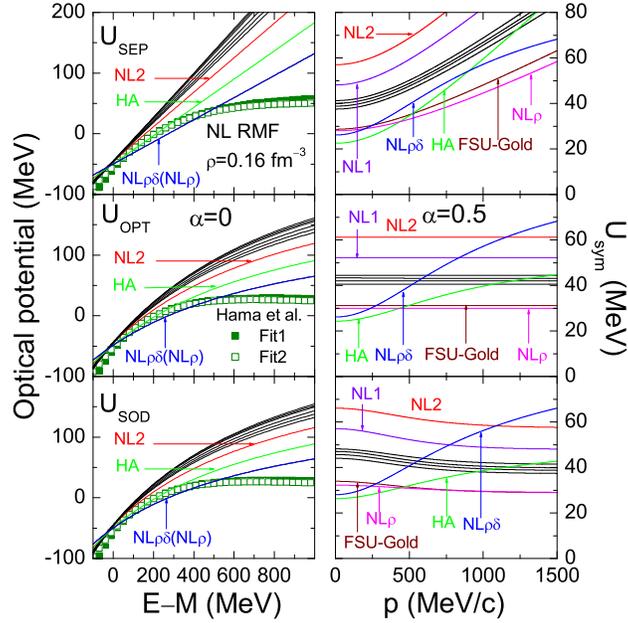}}
\caption{{\protect\small (Color online) Energy dependence of the
three different nucleon optical potentials, i.e.,\
}$U_{\text{SEP}}$ {\protect\small \ (Eq. (\protect\ref{Usep})),
}$U_{\text{OPT}}$ {\protect\small \ (Eq. (\protect\ref{Uopt})) and
}$U_{\text{SOD}}$ {\protect\small \ (Eq. (\protect\ref{Usd}))
(left panels) as well as their corresponding symmetry potentials
}$U_{\text{sym}}^{\text{SEP}}$ {\protect\small , }$U_{
\text{sym}}^{\text{OPT}}${\protect\small , and
}$U_{\text{sym}}^{\text{SOD}}${\protect\small as functions of
momentum (right panels), at a fixed baryon density }$\protect\rho
_{B}=0.16$ {\protect\small \ fm}$^{-3}${\protect\small \ for the
parameter sets NL1, NL2, NL3, NL-SH, TM1, PK1, FSU-Gold, HA,
NL}$\protect\rho ${\protect\small , and NL}$\protect\rho
\protect\delta ${\protect\small \ in the nonlinear RMF model. For
comparison, the energy dependence of the real part of the optical
potential in symmetric nuclear matter at saturation density
extracted from two different fits of the proton-nucleus scattering
data in the Dirac phenomenology are also included (left panels).
Taken from Ref. \cite{Che07}.}} \label{UpLabNL}
\end{figure}

\begin{figure}[th]
\centerline{\includegraphics[scale=1.2]{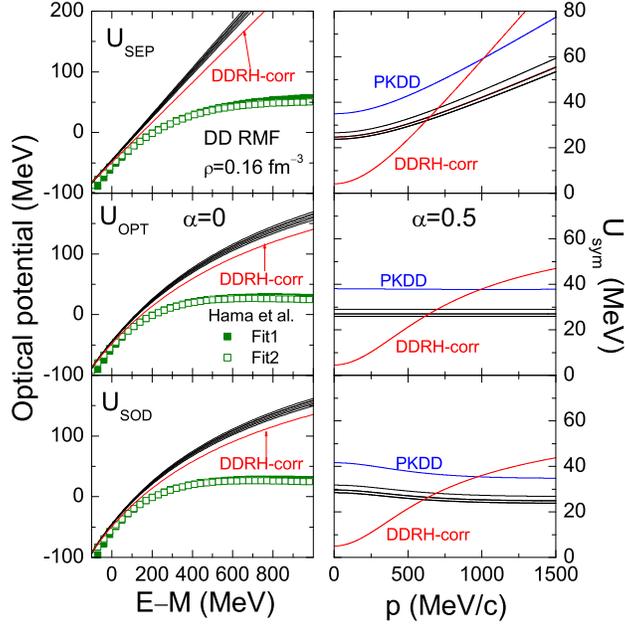}}
\caption{{\protect\small (Color online) Same as Fig.
\protect\ref{UpLabNL} for TW99, DD-ME1, DD-ME2, PKDD, DD, DD-F,
and DDRH-corr in the density-dependent RMF models. Taken from Ref.
\cite{Che07}.}} \label{UpLabDD}
\end{figure}

\begin{figure}[th]
\centerline{\includegraphics[scale=1.2]{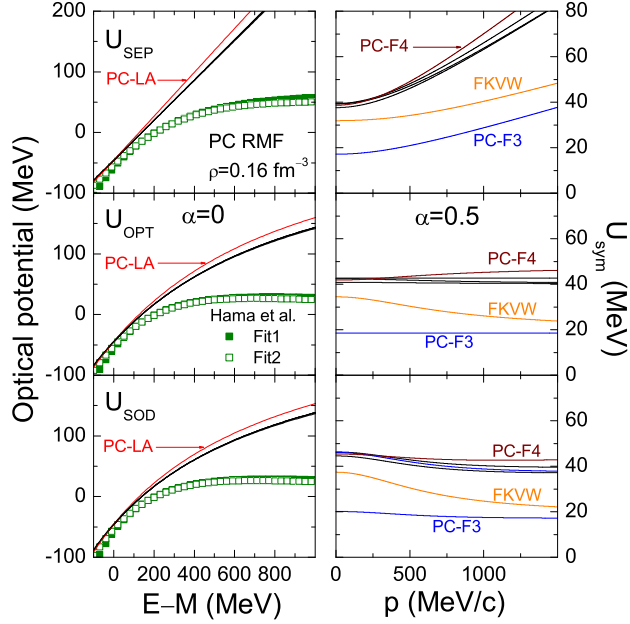}}
\caption{{\protect\small (Color online) Same as Fig.
\protect\ref{UpLabNL} for PC-F1, PC-F2, PC-F3, PC-F4, PC-LA, and
FKVW in the point-coupling RMF models. Taken from Ref.
\cite{Che07}.}} \label{UpLabPC}
\end{figure}

For the parameter sets NL1, NL2, NL3, NL-SH, TM1, PK1, FSU-Gold,
HA, NL$\rho $, and NL$\rho \delta $ \ in the nonlinear RMF model,
there are calculations \cite{Che07} for the energy dependence of
the three different nucleon optical potentials, i.e., the
\textquotedblleft Schr\"{o}dinger-equivalent
potential\textquotedblright\ $U_{\mathrm{SEP}}$ (Eq.
(\ref{Usep})), the optical potential from the difference between
the total energy of a nucleon in nuclear medium and its energy at
the same momentum in free space $U_{\mathrm{OPT}}$ (Eq.
(\ref{Uopt})), and the optical potential based on the second-order
Dirac equation $U_{\mathrm{SOD}}$ (Eq. (\ref{Usd} )), at a fixed
baryon density $\rho _{B}=0.16$ fm$^{-3}$ (roughly corresponding
to the saturation densities obtained from various RMF models). For
their corresponding symmetry potentials $U_{\mathrm{sym}}^{
\mathrm{SEP}}$, $U_{\mathrm{sym}}^{\mathrm{OPT}}$, and
$U_{\mathrm{sym}}^{ \mathrm{SOD}}$, their dependence on the
nucleon momentum in asymmetric nuclear matter at baryon density
$\rho _{B}=0.16~{\rm fm}^{-3}$ and with isospin asymmetry
$\alpha=0.5$ have also been studied \cite{Che07}. In contrast to
the energy dependence of the nuclear symmetry potential, the
momentum dependence of the nuclear symmetry potential is almost
independent of the isospin asymmetry of nuclear matter. These
results are shown in Fig. \ref{UpLabNL}. Corresponding results for
the parameter sets TW99, DD-ME1, DD-ME2, PKDD, DD, DD-F, and
DDRH-corr in the density-dependent RMF model and for PC-F1, PC-F2,
PC-F3, PC-F4, PC-LA, and FKVW in the point-coupling RMF model are
shown in Figs. \ref{UpLabDD} and \ref{UpLabPC}, respectively. Also
shown in these figures for comparison are results for the energy
dependence of the real part of the different optical potentials in
symmetric nuclear matter at saturation density that are extracted
from the proton-nucleus scattering data based on the Dirac
phenomenology \cite{Ham90,Coo93}.

It is seen that different optical potentials in symmetric nuclear
matter at $\rho _{B}=0.16$ fm$^{-3}$ exhibit similar energy
dependence at low energies but have different behaviors at high
energies. In particular, at high energies, $U_{\mathrm{SEP}}$
continues to increase linearly with energy while $U_{\mathrm{OPT}}$
and $U_{\mathrm{SOD}}$ seem to saturate at high energies and thus
display a more satisfactory high-energy limit, similar to what is
observed in the nuclear optical potential that is extracted from the
experimental data based on the Dirac phenomenology. The critical
energy at which the optical potential changes from negative to
positive values is between about $130$ MeV and $270$ MeV, depending
on the parameter sets used. These features are easy to understand
from the fact that the scalar and vector potentials are
momentum/energy-independent in the RMF models considered here.
Analysis of experimental data from the proton-nucleus scattering in
the Dirac phenomenology also indicates that the extracted different
nucleon optical potentials in symmetric nuclear matter at normal
nuclear density change from negative to positive values at nucleon
energy of about $208$ MeV. Furthermore, the different optical
potentials from all $23$ parameter sets are consistent with the
experimental data at lower energies, i.e.,
$E_{\mathrm{kin}}<100-200~{\rm MeV}$, but are generally too
repulsive at higher energies, especially for the \textquotedblleft
Schr\"{o}dinger-equivalent potential\textquotedblright\
$U_{\mathrm{SEP}}$. These features imply that the RMF models with
parameters fitted to the properties of finite nuclei can only give
reasonable description of the low energy behavior of the isoscalar
optical potentials. On the other hand, for optical potentials at
high energies, contributions from dispersive processes such as the
dynamical polarization by inelastic excitations, inelastic isobar
resonance excitation above the pion threshold, and particle
production become important \cite{Fuc06b,Are02}. Including such
continuum excitations is expected to improve significantly the high
energy behavior of the optical potential \cite{Are02}. Such studies
are, however, beyond the RMF model based on the Hartree level as
considered here.

For the momentum dependence of the symmetry potential, all $23$
parameter sets display similar behaviors in
$U_{\mathrm{sym}}^{\mathrm{SEP}}$, i.e., increasing with momentum,
albeit at different rates. This can be qualitatively understood as
follows. Expressing Eq. (\ref{Usep}) as
\begin{eqnarray}
U_{\mathrm{SEP},\tau }=\frac{1}{2M_{\tau }}[E_{\tau }^{2}-(M_{\tau
}^{2}+ \vec{p}^{2})],
\end{eqnarray}
and neglecting the difference in neutron and proton masses, one
can rewrite Eq. (\ref{UsymSEP}) as
\begin{eqnarray}
U_{\mathrm{sym}}^{\mathrm{SEP}}
&=&\frac{E_{n}^{2}-E_{p}^{2}}{4M_{\tau
}\alpha }  \notag \\
&=&\frac{1}{4M_{\tau }\alpha }[(\Sigma _{n}^{0})^{2}+2\Sigma
_{n}^{0}\sqrt{\vec{p}^{2}+(M_{n}+\Sigma _{n}^{S})^{2}}+(M_{n}+\Sigma
_{n}^{S})^{2}-(\Sigma _{p}^{0})^{2}  \notag \\
&&-2\Sigma _{p}^{0}\sqrt{\vec{p}^{2}+(M_{p}+\Sigma _{p}^{S})^{2}}
-(M_{p}+\Sigma _{p}^{S})^{2}]  \notag \\
&=&\frac{1}{4M_{\tau }\alpha }[(\Sigma
_{n}^{0})^{2}-(\Sigma_{p}^{0})^{2}+(M_{\mathrm{Dirac},n}^{\ast})^{2}
-(M_{\mathrm{Dirac},p}^{\ast})^{2}+2\Sigma_{n}^{0}\sqrt{\vec{p}^{2}
+(M_{\mathrm{Dirac},n}^{\ast})^{2}}  \notag \\
&&-2\Sigma_{p}^{0}\sqrt{\vec{p}^{2}+(M_{\mathrm{Dirac},p}^{\ast
})^{2}}].
\end{eqnarray}
In the simple case of the nonlinear RMF model without the
isovector-scalar $\delta $ meson, the neutron Dirac mass is the
same as that of proton. In this case, $U_{\mathrm{sym}}^{
\mathrm{SEP}}$ is reduced to
\begin{eqnarray}
U_{\mathrm{sym}}^{\mathrm{SEP}}=\frac{1}{4M_{\tau }\alpha }[(\Sigma
_{n}^{0})^{2}-(\Sigma _{p}^{0})^{2} +2(\Sigma _{n}^{0}-\Sigma
_{p}^{0})\sqrt{\vec{p}^{2}+(M_{\mathrm{Dirac}}^{\ast })^{2}}].
\end{eqnarray}
Since it can be shown from Eqs. (\ref{OmgNL}), ( \ref{RhoNL}), and
(\ref{Sig0NL}) that
\begin{eqnarray}
\Sigma _{n}^{0}-\Sigma _{p}^{0}=2\left( \frac{g_{\rho }}{m_{\rho
}}\right) ^{2}(\rho _{n}-\rho _{p}),
\end{eqnarray}
one thus has $\Sigma _{n}^{0}>\Sigma _{p}^{0}$ and an increase of
$U_{\mathrm{\ sym}}^{\mathrm{SEP}}$ with the momentum of a nucleon
in neutron-rich nuclear matter. The same argument applies to
density-dependent RMF models and point-coupling models if the
coupling constant $\alpha _{\mathrm{TV}}$ or $G_{TV}$ in the
point-coupling models is positive (at saturation density) so that
the potential energy part of the symmetry energy at saturation
density is also positive.

For $U_{\mathrm{sym}}^{\mathrm{OPT}}$, whether it increases or
deceases with nucleon momentum depends on the isospin splitting of
the nucleon scalar self energy (scalar potential) or Dirac mass in
neutron-rich nuclear matter. This can be seen from
Eq.~(\ref{UsymOPT}) if it is re-expressed as
\begin{eqnarray}
U_{\mathrm{sym}}^{\mathrm{OPT}} &=&\frac{E_{n}-E_{p}}{2\alpha }  \notag \\
&=&\frac{1}{2\alpha }(\Sigma _{n}^{0}-\Sigma
_{p}^{0}+\sqrt{\vec{p}^{2}+(M_{n}+\Sigma _{n}^{S})^{2}}
-\sqrt{\vec{p}^{2}+(M_{p}+\Sigma _{p}^{S})^{2}})  \notag \\
&=&\frac{1}{2\alpha }[\Sigma _{n}^{0}-\Sigma
_{p}^{0}+\sqrt{\vec{p}^{2}+(M_{ \mathrm{Dirac},n}^{\ast })^{2}}
-\sqrt{\vec{p}^{2}+(M_{\mathrm{Dirac},p}^{\ast })^{2}}].
\end{eqnarray}
One notes that $U_{\mathrm{sym}}^{\mathrm{OPT}}$ increases with
momentum for the parameter sets HA, NL$\rho \delta $, DDRH-corr, and
PC-F4 while the opposite behavior is observed for the parameter sets
PC-F2, PC-LA, and FKVW.

For the momentum dependence of $U_{\mathrm{sym}}^{\mathrm{SOD}}$,
it is similar to that of $U_{\mathrm{sym}}^{\mathrm{OPT}}$ if one
rewrites Eq. (\ref{UsymSOD}) as
\begin{eqnarray}
U_{\mathrm{sym}}^{\mathrm{SOD}} &=&\frac{E_{n}-E_{p}-(M_{\tau }^{2}+\vec{p}%
^{2})(\frac{1}{E_{n}}-\frac{1}{E_{n}})}{4\alpha }  \notag \\
&=&U_{\mathrm{sym}}^{\mathrm{OPT}}/2-\frac{(M_{\tau
}^{2}+\vec{p}^{2})(\frac{ 1}{E_{n}}-\frac{1}{E_{n}})}{4\alpha }.
\end{eqnarray}
In this case, $U_{\mathrm{sym}}^{\mathrm{SOD}}$ increases with
nucleon momentum for the parameter sets HA, NL$\rho \delta $, and
DDRH-corr while it decreases for other parameter sets considered
here.

In Ref. \cite{Jam89}, it has been argued that it is the
`Schr\"{o}dinger-equivalent potential' $U_{\mathrm{SEP}}$ (Eq.
(\ref{Usep})) and thus its corresponding symmetry potential
$U_{\mathrm{sym} }^{\mathrm{SEP}}$ that should be compared with the
results from non-relativistic models. As discussed before, the
experimental data indicate that the nuclear symmetry potential at
nuclear matter saturation density, i.e., the Lane potential
$U_{\mathrm{Lane}}$, clearly decreases at low energies (beam energy
$E_{\mathrm{kin}}$ up to about $100$ MeV and corresponding momentum
values ranging from about $300$ MeV/c to $470$ MeV/c), which is
obviously contradictory to the results for
$U_{\mathrm{sym}}^{\mathrm{SEP}}$ from all of the $23$ parameter
sets considered here. On the other hand, $U_{\mathrm{sym}
}^{\mathrm{OPT}}$ and $U_{\mathrm{sym}}^{\mathrm{SOD}}$ for some
parameter sets can decrease with nucleon momentum and are thus
qualitatively consistent with experimental results.

For nucleons with momenta less than about $250-300$ MeV/c or
$E_{\mathrm{kin}}<0$, although the observed increase of
$U_{\mathrm{sym} }^{\mathrm{SEP}}$ with momentum for all $23$
parameter sets, and $U_{\mathrm{sym} }^{\mathrm{OPT}}$ as well as
$U_{\mathrm{sym}}^{\mathrm{SOD}}$ with some parameter sets seem to
be consistent with the results from the microscopic DBHF
\cite{Fuc04}, the extended BHF with TBF \cite{Zuo05}, and chiral
perturbation theory calculations \cite{Fri05}, i.e., the symmetry
potential stays as a constant or slightly increases with momentum
before decreasing at high momenta, it fails to describe the high
momentum/energy behaviors of the nuclear symmetry potential
extracted from nucleon-nucleus scattering experiments and (p,n)
charge exchange reactions at beam energies up to about $100$ MeV.
This is in contrast with studies based on the relativistic impulse
approximation with empirical $NN$ scattering amplitude and the
nuclear scalar and vector densities from the RMF model, where the
Schr\"{o}dinger-equivalent nuclear symmetry potential at fixed
baryon density is found to decrease with increasing nucleon energy
in the range of $100\le E_{\mathrm{kin}}\le 400$ MeV \cite{LiZH06b}
and becomes essentially constant once the nucleon kinetic energy is
greater than about $500$ MeV \cite{Che05c}.

\subsubsection{Isospin-splitting of the nucleon effective mass}

\begin{table}[tbp]
\caption{{\protect\small Values of different nucleon effective
masses, i.e., }$M_{\mathrm{Dirac}}^{\ast }/M${\protect\small ,
}$M_{\mathrm{Landau}}^{\ast }/M${\protect\small ,
}$M_{\mathrm{Lorentz}}^{\ast }/M${\protect\small , }$
M_{\mathrm{OPT}}^{\ast }/M${\protect\small , and
}$M_{\mathrm{SOD}}^{\ast }/M ${\protect\small \ in symmetric
nuclear matter at saturation density using the
}$23${\protect\small \ parameter sets in the nonlinear,
density-dependent, and point-coupling RMF models. The last column
gives the references for corresponding parameter sets. Taken from
Ref. \cite{Che07}.}}
\label{EffMass0}%
\begin{center}
\begin{tabular}{ccccccc}
\hline\hline
Model & $\frac{M_{Dirac}^{\ast }}{M}$ & $\frac{M_{Landau}^{\ast }}{M}$ & $%
\frac{M_{Lorentz}^{\ast }}{M}$ & $\frac{M_{OPT}^{\ast }}{M}$ &
$\frac{ M_{SOD}^{\ast }}{M}$ & Ref. \\ \hline
NL1 & $0.57$ & $0.64$ & $0.65$ & $0.61$ & $0.59$ & \cite{Lee86} \\
NL2 & $0.67$ & $0.72$ & $0.74$ & $0.70$ & $0.68$ & \cite{Lee86} \\
NL3 & $0.60$ & $0.66$ & $0.67$ & $0.63$ & $0.61$ & \cite{Lal97} \\
NL-SH & $0.60$ & $0.66$ & $0.67$ & $0.63$ & $0.61$ & \cite{Sha93} \\
TM1 & $0.63$ & $0.69$ & $0.71$ & $0.67$ & $0.65$ & \cite{Sug94} \\
PK1 & $0.61$ & $0.66$ & $0.68$ & $0.64$ & $0.62$ & \cite{Lon04} \\
FSUGold & $0.61$ & $0.67$ & $0.69$ & $0.65$ & $0.62$ & \cite{Tod05} \\
HA & $0.68$ & $0.74$ & $0.75$ & $0.71$ & $0.69$ & \cite{Bun03} \\
NL$\rho $ & $0.75$ & $0.80$ & $0.82$ & $0.77$ & $0.76$ & \cite{Liu02} \\
NL$\rho \delta $ & $0.75$ & $0.80$ & $0.82$ & $0.77$ & $0.76$ &
\cite{Liu02}
\\
&  &  &  &  &  &  \\
TW99 & $0.55$ & $0.62$ & $0.64$ & $0.60$ & $0.57$ & \cite{Typ99} \\
DD-ME1 & $0.58$ & $0.64$ & $0.66$ & $0.62$ & $0.59$ & \cite{Nik02} \\
DD-ME2 & $0.57$ & $0.63$ & $0.65$ & $0.61$ & $0.59$ & \cite{Lal05} \\
PKDD & $0.57$ & $0.63$ & $0.65$ & $0.61$ & $0.59$ & \cite{Lon04} \\
DD & $0.56$ & $0.63$ & $0.64$ & $0.61$ & $0.58$ & \cite{Typ05} \\
DD-F & $0.56$ & $0.62$ & $0.64$ & $0.60$ & $0.57$ & \cite{Kla06} \\
DDRH-corr & $0.55$ & $0.63$ & $0.64$ & $0.60$ & $0.58$ & \cite{Hof01} \\
&  &  &  &  &  &  \\
PC-F1 & $0.61$ & $0.67$ & $0.69$ & $0.64$ & $0.62$ & \cite{Bur02} \\
PC-F2 & $0.61$ & $0.67$ & $0.69$ & $0.64$ & $0.62$ & \cite{Bur02} \\
PC-F3 & $0.61$ & $0.67$ & $0.69$ & $0.64$ & $0.62$ & \cite{Bur02} \\
PC-F4 & $0.61$ & $0.67$ & $0.69$ & $0.64$ & $0.62$ & \cite{Bur02} \\
PC-LA & $0.58$ & $0.64$ & $0.65$ & $0.61$ & $0.59$ & \cite{Bur02} \\
FKVW & $0.62$ & $0.68$ & $0.70$ & $0.65$ & $0.63$ & \cite{Fin06} \\
\hline\hline
\end{tabular}
\end{center}
\end{table}

For the different nucleon effective masses in symmetric nuclear
matter at saturation density, the results from the $23$ parameter
sets in the nonlinear, density-dependent, and point-coupling RMF
models are shown in Table \ref{EffMass0}. It is seen that the
values of $M_{\mathrm{Dirac} }^{\ast }/M$,
$M_{\mathrm{Landau}}^{\ast }/M$, $M_{\mathrm{Lorentz}}^{\ast }/M$,
$M_{\mathrm{OPT}}^{\ast }/M$, and $M_{\mathrm{SOD}}^{\ast }/M$ are
in the range of $0.55\sim 0.75$, $0.62\sim 0.80$, $0.64\sim 0.80$,
$0.60\sim 0.77$, and $0.57\sim 0.76$, respectively. The parameter
sets NL2, HA, NL$\rho $ and NL$\rho \delta $ seem to give too
large values, i.e., $0.67$, $0.68$, $0.75$, and $0.75$,
respectively, for the $M_{\mathrm{Dirac}}^{\ast }/M$, as values in
the range of $0.55\sim 0.60$ are needed to describe reasonably the
spin-orbit splitting in finite nuclei using the RMF models. On the
other hand, the larger Dirac masses leads to larger Landau masses
$M_{\mathrm{Landau}}^{\ast }/M$ of $0.72$, $0.74$, $0.80$, and $
0.80$, respectively, for the parameter sets NL2, HA, NL$\rho $ and
NL$\rho \delta $, which are consistent with the empirical
constraint of $M_{\mathrm{\ Landau}}^{\ast }/M$ = $0.8\pm 0.1$
\cite{Cha97,Mar07,Cha98,Rei99}.

\begin{figure}[th]
\centerline{\includegraphics[scale=0.8]{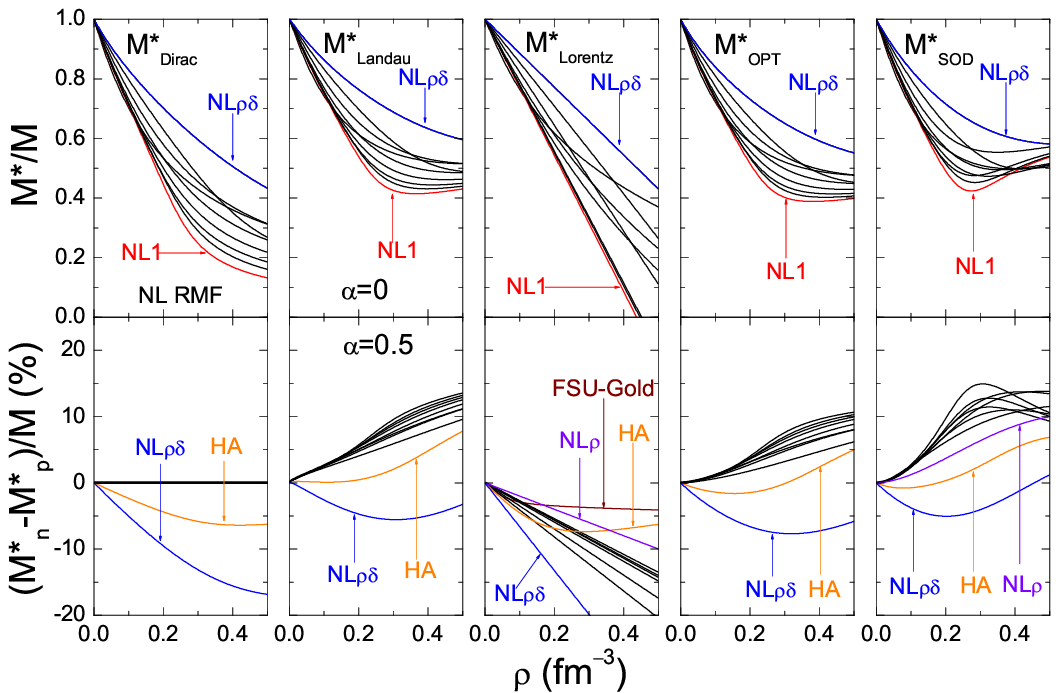}}
\caption{{\protect\small (Color online) Density dependence of
different nucleon effective masses, i.e.,
}$M_{\mathrm{Dirac}}^{\ast }/M$ {\protect\small ,
}$M_{\mathrm{Landau}}^{\ast }/M${\protect\small , }$M_{
\mathrm{Lorentz}}^{\ast }/M${\protect\small ,
}$M_{\mathrm{OPT}}^{\ast }/M$ {\protect\small , and
}$M_{\mathrm{SOD}}^{\ast }/M${\protect\small \ in symmetric
nuclear matter as well as their corresponding isospin splittings
in neutron-rich nuclear matter with isospin asymmetry
}$\protect\alpha =0.5$ {\protect\small \ for the parameter sets
NL1, NL2, NL3, NL-SH, TM1, PK1, FSU-Gold, HA, NL}$\protect\rho
${\protect\small , and NL}$\protect\rho \protect\delta
${\protect\small \ in the nonlinear RMF model.Taken from Ref.
\cite{Che07}.}} \label{MstarDenNL}
\end{figure}

\begin{figure}[th]
\centerline{\includegraphics[scale=0.8]{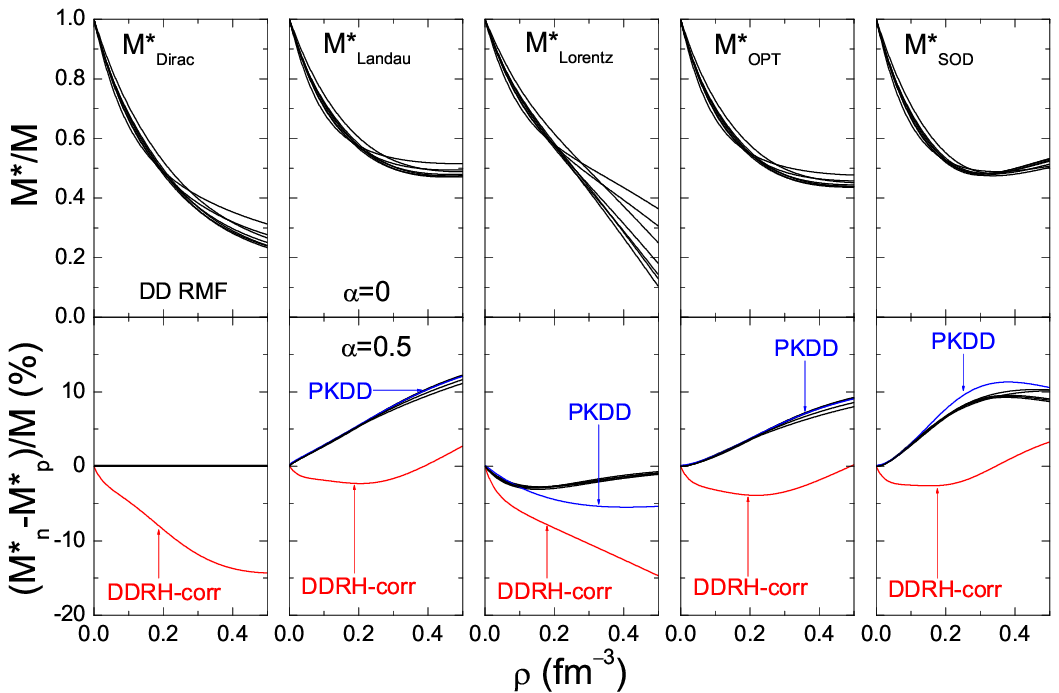}}
\caption{{\protect\small (Color online) Same as Fig.
\protect\ref{MstarDenNL} but for TW99, DD-ME1, DD-ME2, PKDD, DD,
DD-F, and DDRH-corr in the density-dependent RMF model. Taken from
Ref. \cite{Che07}.}} \label{MstarDenDD}
\end{figure}

\begin{figure}[th]
\centerline{\includegraphics[scale=0.8]{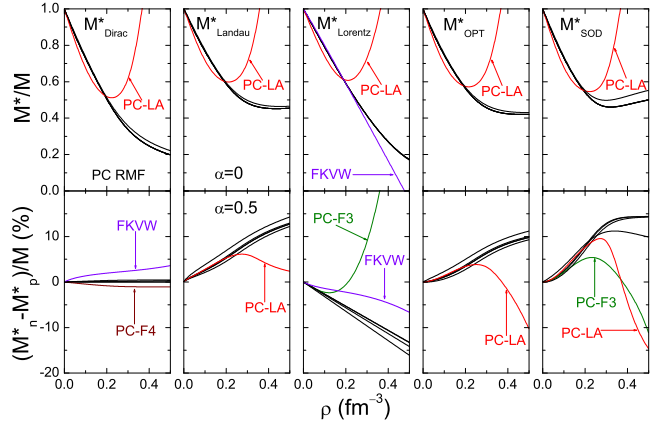}}
\caption{{\protect\small (Color online) Same as Fig.
\protect\ref{MstarDenNL} but for PC-F1, PC-F2, PC-F3, PC-F4,
PC-LA, and FKVW in the point-coupling RMF model. Taken from Ref.
\cite{Che07}.}} \label{MstarDenPC}
\end{figure}

The density dependence of the different nucleon effective masses
in symmetric nuclear matter and corresponding isospin splitting
$(M_{n}^{\ast }-M_{p}^{\ast })/M$ in asymmetric nuclear matter
with isospin asymmetry $\alpha =0.5$ are shown in Fig.
\ref{MstarDenNL} for the parameter sets NL1, NL2, NL3, NL-SH, TM1,
PK1, FSU-Gold, HA, NL$\rho $, and NL$\rho \delta $\ in the
nonlinear RMF model. Figs. \ref{MstarDenDD} and \ref{MstarDenPC}
display the same results as in Fig. \ref{MstarDenNL} but for the
parameter sets TW99, DD-ME1, DD-ME2, PKDD, DD, DD-F, and DDRH-corr
in the density-dependent RMF models and for PC-F1, PC-F2, PC-F3,
PC-F4, PC-LA, and FKVW in the point-coupling RMF model,
respectively. It is seen that different parameter sets in the
nonlinear RMF model give significantly different density
dependence for the nucleon effective masses while the different
parameter sets in the density-dependent and point-coupling RMF
models predict roughly the same density dependence for the nucleon
effective masses except that the parameter set PC-LA gives very
large values for the nucleon effective masses at high densities.
This unusual behavior for PC-LA was also observed in Ref.
\cite{Bur02}, and it is due to the fact that the coupling constant
$\gamma _{\mathrm{S}}$ for the higher-order interaction term in
PC-LA is positive \cite{Nik92} and dominates at high density,
leading thus to the very large nucleon effective mass.

For the Landau mass at a fixed baryon density, its value
$M_{\mathrm{Landau}}^{\ast }/M$ is generally larger than
$M_{\mathrm{Dirac}}^{\ast }/M$. This can be seen from Eq.
(\ref{MLandau}) if it is rewritten as
\begin{eqnarray}
M_{\mathrm{Landau},\tau }^{\ast } &=&(E_{\tau }-\Sigma _{\tau
}^{0})=\sqrt{ p_{F\text{,}\tau }^{2}+(M_{\tau }+\Sigma _{\tau
}^{S})^{2}} =\sqrt{p_{F,\tau }^{2}+M_{\mathrm{Dirac},\tau }^{\ast
2}} \label{MLanDir}
\end{eqnarray}
which shows that $M_{\mathrm{Landau},\tau }^{\ast }\geq
M_{\mathrm{Dirac},\tau }^{\ast }$ if nucleon self-energies are
independent of momentum/energy.

For the Lorentz mass $M_{\mathrm{Lorentz}}^{\ast }$, $M_{\mathrm{\
Lorentz}}^{\ast }/M$ depends almost linearly on density and thus
has a stronger density dependence than the Dirac and Landau
masses. One notes from Eqs. (\ref{dispersion}) and (\ref{MLanDir})
that Eq. (\ref{MLorentz}) can be reduced to
\begin{eqnarray}
M_{\mathrm{Lorentz},\tau }^{\ast }=M_{\tau }-\Sigma _{\tau }^{0},
\label{MLorentz2}
\end{eqnarray}
if nucleon self-energies are independent of momentum/energy.
Therefore, the density dependence of $M_{\mathrm{Lorentz}}^{\ast
}$ is determined uniquely by the density dependence of nucleon
vector self-energy. In the nonlinear RMF model, most of the
parameter sets, except for TM1, PK1 and FSU-Gold which include the
self-coupling of the $\omega $ meson field, give a linear density
dependence for $\Sigma _{\tau }^{0}$, leading thus to the observed
linear density dependence of $M_{\mathrm{Lorentz}}^{\ast }$. As to
the nonlinear density dependence of $M_{\mathrm{Lorentz}}^{\ast }$
in the density-dependent RMF model and point-coupling models, it
is due to the nonlinear density dependence of the coupling
constant or the inclusion of higher-order couplings.

For $M_{\mathrm{OPT}}^{\ast }/M$ and $M_{\mathrm{\ SOD}}^{\ast
}/M$, they are seen to have roughly same magnitude and also same
density dependence as $M_{\mathrm{Landau}}^{\ast }/M$. This
feature can be understood from the fact that with the dispersion
relation of Eq. (\ref{dispersion}), Eq. (\ref{Mopt}) and Eq.
(\ref{Msod}) can be re-expressed as
\begin{eqnarray}
M_{\mathrm{OPT},\tau }^{\ast }=\frac{M_{\tau }}{\sqrt{p_{F,\tau
}^{2}+M_{\tau }^{2}}}M_{\mathrm{Landau},\tau }^{\ast }
\label{Mopt2}
\end{eqnarray}
and
\begin{eqnarray}
M_{\mathrm{SOD},\tau }^{\ast }=M_{\tau
}\left[\frac{M_{\mathrm{Landau},\tau }^{\ast }}{E_{\tau
}}+\frac{E_{\tau }^{2}-(p_{F,\tau }^{2}+M_{\tau }^{2})}{ 2E_{\tau
}^{2}}\right], \label{Msod2}
\end{eqnarray}
respectively. Since $p_{F\text{,}\tau }^{2}\ll M_{\tau }^{2}$ (For
example, $p_{F}\approx 385$ MeV/c at $\rho _{B}=0.5$ fm$^{-3}$),
one has $M_{\tau }/\sqrt{p_{F,\tau }^{2}+M_{\tau }^{2}}\approx 1$
(with an error of a few percent) and thus $M_{\mathrm{OPT},\tau
}^{\ast }\approx M_{\mathrm{Landau},\tau }^{\ast }$. Furthermore,
the second term in Eq. (\ref{Msod2}) can be neglected compared
with the first term as $M_{\tau }/E_{\tau }\sim 1$ (it is a good
approximation at low densities and with an error of about $20\%$
at high densities, e.g., $\rho _{B}=0.5$ fm$^{-3}$). As a result,
one has $M_{\mathrm{SOD},\tau }^{\ast }\sim
M_{\mathrm{Landau},\tau }^{\ast }$.

From the Dirac equation, one sees that the condensed scalar fields
(scalar self-energies) lead to a shift of nucleon mass such that the
nuclear matter is described as a system of pseudo-nucleons with
masses $M^{\ast }$ (Dirac mass) moving in classical vector fields
with $\delta $ meson field or isovector-scalar potential further
generating the splitting of the proton and neutron Dirac masses in
asymmetric nuclear matter. For the isospin splitting of
$M_{\mathrm{Dirac}}^{\ast }$ in neutron-rich nuclear matter, it is
interesting to see that the parameter sets HA, NL$\rho \delta $,
DDRH-corr, and PC-F4 give $M_{\mathrm{Dirac},p}^{\ast
}>M_{\mathrm{Dirac},n}^{\ast }$ while PC-F2, PC-LA, and FKVW exhibit
the opposite behavior of $M_{\mathrm{\ Dirac},p}^{\ast
}<M_{\mathrm{Dirac},n}^{\ast }$. This feature implies that the
isospin-dependent scalar potential can be negative or positive
depending on the parameter sets used. In the nonlinear RMF model,
one obtains from Eqs. (\ref{DelNL}) and (\ref{MDiracNL})
\begin{eqnarray}
M_{\mathrm{Dirac},n}^{\ast }-M_{\mathrm{Dirac},p}^{\ast }=-2\left(
\frac{ g_{\delta }}{m_{\delta }}\right) ^{2}(\rho _{S,n}-\rho
_{S,p}),
\end{eqnarray}
which indicates that one always has $M_{\mathrm{Dirac},p}^{\ast
}>M_{\mathrm{\ Dirac},n}^{\ast }$ in the neutron-rich nuclear matter
where $\rho _{S,n}>\rho _{S,p}$. This argument is also applicable to
the density-dependent RMF model by replacing $g_{\delta }$ with the
density dependent $\Gamma _{\delta }$. For the nonlinear
point-coupling models, one has, on the other hand,
\begin{eqnarray}
M_{\mathrm{Dirac},n}^{\ast }-M_{\mathrm{Dirac},p}^{\ast }=2\alpha _{\mathrm{%
\ TS}}(\rho _{S,n}-\rho _{S,p}).
\end{eqnarray}
A similar equation can be obtained for the density-dependent
point-coupling models with the replacement of $\alpha
_{\mathrm{TS}}$ by the density dependent $G_{TS}$. Therefore, the
isospin splitting of $M_{\mathrm{Dirac}}^{\ast }$ in neutron-rich
nuclear matter depends on the sign of the isovector-scalar
coupling constant $\alpha _{\mathrm{TS}}$ and $G_{TS}$ in the
point-coupling models. Since the value of $\alpha _{ \mathrm{TS}}$
in PC-F2 and PC-LA as well as the value of $G_{TS}$ in FKVW are
positive, these parameter sets lead to the isospin-splitting
$M_{\mathrm{Dirac},n}^{\ast }>M_{ \mathrm{Dirac},p}^{\ast }$ in
neutron-rich nuclear matter, which is opposite to that in other
parameter sets considered here. The isospin splitting of
$M_{\mathrm{Dirac}}^{\ast }$ is directly related to the isovector
spin-orbit potential that determines the isospin-dependent
spin-orbit splitting in finite nuclei. Unfortunately, there are no
clear experimental indication about the isospin dependence of the
spin-orbit splitting in finite nuclei \cite{Hof01}, so detailed
experimental data on the single-particle energy levels in exotic
nuclei are needed to pin down the isospin splitting of
$M_{\mathrm{Dirac}}^{\ast }$ in asymmetric nuclear matter.

For the isospin splitting of $M_{\mathrm{Landau}}^{\ast }$ in
neutron-rich nuclear matter, most parameter sets give
$M_{\mathrm{Landau},n}^{\ast }>M_{\mathrm{Landau},p}^{\ast }$,
which is consistent with the usual results in non-relativistic
models. The parameter sets NL$\rho \delta $ and DDRH-corr give,
however, the opposite result due to the strong isospin-splitting
of $M_{\mathrm{Dirac}}^{\ast }$ with $M_{\mathrm{Dirac},n}^{\ast
}<M_{\mathrm{\ Dirac},p}^{\ast }$ for NL$\rho \delta $ and
DDRH-corr and the fact that $M_{\mathrm{Landau}}^{\ast }$ is
related to the Fermi momentum and $M_{\mathrm{Dirac}}^{\ast }$
according to Eq. (\ref{MLanDir}). The isospin-splitting
$M_{\mathrm{Landau},n}^{\ast }>M_{ \mathrm{Landau},p}^{\ast }$
implies that neutrons have a larger level density at the Fermi
energy and thus more compressed single-particle levels in finite
nuclei than protons.

For the isospin splitting of $M_{\mathrm{Lorentz}}^{\ast }$ in
neutron-rich nuclear matter, all parameter sets give
$M_{\mathrm{Lorentz},p}^{\ast }>M_{\mathrm{Lorentz},n}^{\ast }$
except that the PC-L3 gives $M_{\mathrm{Lorentz },p}^{\ast
}<M_{\mathrm{Lorentz},n}^{\ast }$ at high densities. From Eq.
(\ref{MLorentz2}), one has
\begin{eqnarray}
M_{\mathrm{Lorentz},n}^{\ast }-M_{\mathrm{Lorentz},p}^{\ast
}=-(\Sigma _{n}^{0}-\Sigma _{p}^{0}),
\end{eqnarray}
which leads to the observed isospin-splitting $M_{\mathrm{Lorentz}
,p}^{\ast }>M_{\mathrm{Lorentz},n}^{\ast }$ as one generally has
$\Sigma _{n}^{0}>\Sigma _{p}^{0}$ as discussed above. For the
parameter set PC-L3, it includes a higher-order isovector-vector
term through the parameter $\gamma _{\mathrm{TV}}$. Since the
latter has a negative value and dominates at high densities
according to Eq. (\ref{Sig0NLPC}), it leads to $\Sigma
_{n}^{0}<\Sigma _{p}^{0}$ and thus $M_{\mathrm{Lorentz},p}^{\ast
}<M_{\mathrm{Lorentz},n}^{\ast }$ at high densities. The isospin
splitting of $M_{\mathrm{OPT}}^{\ast }/M$ and
$M_{\mathrm{SOD}}^{\ast }/M$ in neutron-rich nuclear matter show a
similar behavior as $M_{\mathrm{Landau}}^{\ast }$ as expected from
the discussions below Eqs. (\ref{Mopt2}) and (\ref{Msod2}).

\subsubsection{The nucleon scalar density}

The nucleon scalar density as defined in Eq. (\ref{RhoS}) is the
source for the nucleon scalar self-energy (scalar potential). In
the RMF model, the isospin-dependent nucleon scalar density is
uniquely related to the nucleon Dirac mass as shown in Eq.
(\ref{RhoSnp}). The latter equation also shows that the scalar
density is less than the baryon density due to the factor
$M_{i}^{\ast}/\sqrt{\vec{k}^{2}+(M_{i}^{\ast })^{2}}$ which causes
a reduction of the contribution of rapidly moving nucleons to the
scalar source term. This mechanism is responsible for nuclear
matter saturation in the mean-field theory and essentially
distinguishes relativistic models from non-relativistic ones. In
practice, the isospin-dependent nucleon scalar density is also an
essential ingredient for evaluating the relativistic optical
potential for neutrons and protons in the relativistic impulse
approximation (See, e.g., Refs. \cite{Che05c,LiZH06b} and
references therein).

\begin{figure}[th]
\centerline{\includegraphics[scale=1.2]{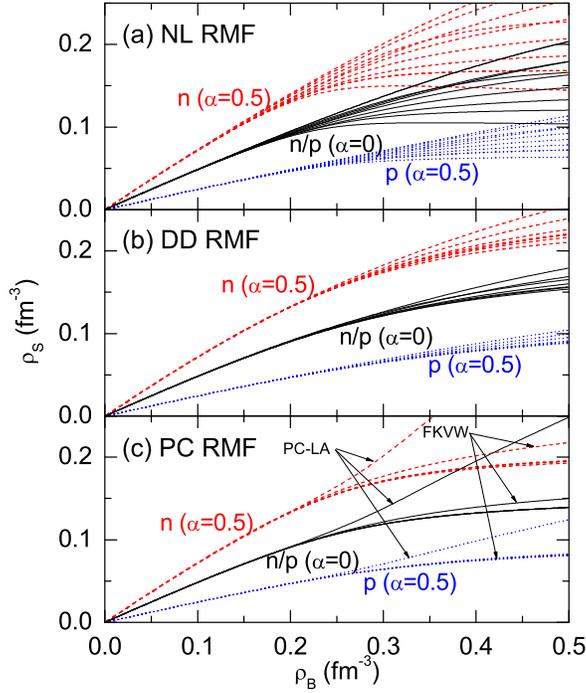}}
\caption{{\protect\small (Color online) Neutron and proton scalar
densities as functions of baryon density in nuclear matter with
isospin asymmetry }$\protect\alpha =0${\protect\small \ and
}$0.5${\protect\small \ for the parameter sets NL1, NL2, NL3,
NL-SH, TM1, PK1, FSU-Gold, HA, NL}$\protect \rho ${\protect\small
, and NL}$\protect\rho \protect\delta ${\protect\small \ of the
nonlinear RMF model (a); TW99, DD-ME1, DD-ME2, PKDD, DD, DD-F, and
DDRH-corr of the density-dependent RMF model (b); PC-F1, PC-F2,
PC-F3, PC-F4, PC-LA, and FKVW of the point-coupling RMF model (c).
Taken from Ref. \cite{Che07}.}} \label{RhoSRhoB}
\end{figure}

Fig. \ref{RhoSRhoB} shows the neutron and proton scalar densities
as functions of the baryon density $\rho _{B}$ in nuclear matter
with isospin asymmetry $\alpha =0$ and $0.5$ for the $23$
parameter sets from the nonlinear, density-dependent, and
point-coupling RMF models. It is seen that the neutron scalar
density is larger than that of protons in neutron-rich nuclear
matter at a fixed baryon density. Although results for different
parameter sets are almost the same at lower baryon densities, they
become different when $\rho _{B}\gtrsim 0.25$ fm$^{-3}$, and this
is consistent with the conclusions of Refs. \cite{Che05c,LiZH06b}.
In particular, different parameter sets in the nonlinear RMF model
predict a larger uncertainty for the value of the nucleon scalar
density at high baryon density while all the parameter sets
(except PC-LA) in the density-dependent RMF model and
point-coupling models give roughly same results for the nucleon
scalar density. These features are consistent with the results for
the density dependence of nucleon Dirac mass shown in Figs.
\ref{MstarDenNL}, \ref{MstarDenDD}, and \ref{MstarDenPC}. At low
baryon densities, neutron and proton scalar densities are seen to
increase roughly linearly with baryon density, and this can be
easily understood from Eq. (\ref{RhoSnp}), which is reduced to the
following expression at low densities ($|\vec{k}|\rightarrow 0$
due to $ k_{F}\rightarrow 0$):
\begin{eqnarray}
\rho _{S,i}\approx \frac{2}{{(2\pi
)}^{3}}\int_{0}^{k_{F}^{i}}d^{3}\!k\, \frac{M_{i}^{\ast
}}{M_{i}^{\ast }}=\frac{2}{{(2\pi
)}^{3}}\int_{0}^{k_{F}^{i}}d^{3}\!k\,=\rho _{B,i},\text{ } i=p,n.
\end{eqnarray}
Therefore, neutron and proton scalar densities generally approach
their respective baryon densities in asymmetric nuclear matter at
low baryon densities.

\subsection{Effects of charge symmetry breaking on the symmetry energy
in the RMF model with chiral symmetry restoration}

\begin{figure}[th]
\vspace{-2cm}
\centerline{\includegraphics[scale=0.5]{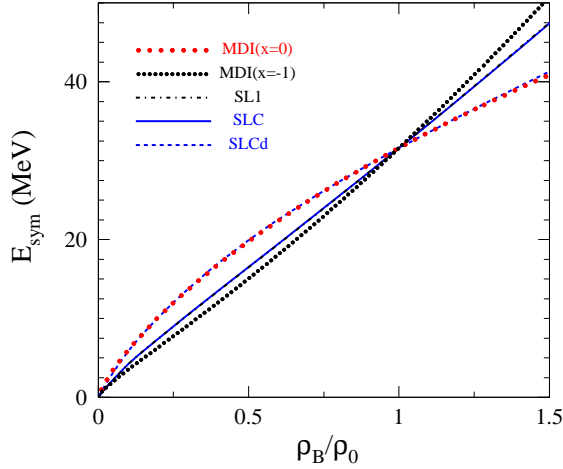}}
\vspace{-1cm} \caption{{\protect\small (Color online) The symmetry
energy as a function of density from the SLC and SLCd parameter
sets of the RMF model with in-medium hadron masses and coupling
constants in comparison with those from the MDI interactions with
x=0 and x=-1. Taken from Ref. \cite{Jia07b}.}} \label{Jia-esym}
\end{figure}

Recently, the standard RMF models have been extended in Refs.
\cite{Liu04,Ava06,Khv07,Jia07a,Jia07b} to include
density-dependent hadron masses and meson coupling constants via
the Brown-Rho (BR) scaling \cite{Bro91} to mimic the effects of
chiral symmetry restoration at high densities. As illustrated in
Fig.~\ref{Jia-esym}, the parameter sets SLC and SLCd constructed
in Ref. \cite{Jia07a,Jia07b} by Jiang {\it et al.} lead to a
symmetry energy that is consistent with that extracted from the
isospin diffusion data, i.e., the MDI interactions with x=0 and
x=-1. These parameter sets also give an EOS of symmetric nuclear
matter at supra-normal densities consistent with the experimental
constraints obtained from analyzing the nuclear collective flow in
relativistic heavy-ion collisions \cite{Dan02a} as well as a
fairly satisfactory description of the ground state properties of
both infinite nuclear matter and many finite nuclei, including
their binding energies, charge radii, and neutron skin thickness
\cite{Jia07a,Jia07b}.

Using the SLC parameter set, Jiang and Li investigated more
recently the effects of charge symmetry breaking on the density
dependence of the symmetry energy \cite{Jia08a}. Because of charge
symmetry breaking, which leads to a break down of the isospin
symmetry of nuclear interactions, neutron-neutron and
proton-proton interactions become different even after removing
the electromagnetic contributions \cite{mi90,mac}.  A lot of
efforts have been devoted to studying the charge symmetry breaking
(CSB) and its effect on few-body and bulk-matter observables in
nuclear systems, see, e.g., Refs. \cite{mi90,mac} for reviews.
The CSB is most explicitly displayed by the difference (about
10\%) between the neutron-neutron and proton-proton scattering
lengths in the $^1S_0$ state: $a_{nn}$ and $a_{pp}$. The CSB can
also be used to explain the well-known Nolen-Schiffer anomaly of
light mirror nuclei. One very successful approach to study the CSB
is to use many-body theories employing explicitly charge-dependent
nucleon-nucleon interactions that are adjusted to reproduce the
free-space nucleon-nucleon scattering data. However, the CSB
effects in free-space and/or symmetric nuclear matter are normally
very small. For example, the CSB-induced effects in symmetric
nuclear matter with the charge-dependent Bonn potential were shown
to be quite small \cite{ma01}. The results with the
charge-dependent Reid93 potential also showed that the CSB effect
on the equation of state (EOS) even in isospin-asymmetric nuclear
matter is negligible \cite{bord}.

Besides using many-body theories with interactions that are
explicitly charge-dependent, e.g., the Bonn \cite{ma01}, Reid93
\cite{bord} and V18 potentials \cite{wi95}, one can also explain
the difference between the scattering lengths $a_{nn}$ and
$a_{pp}$ successfully using approaches based on the $\rho-\omega$
meson mixing \cite{coon,mc,pi93,ki96,du97,du00} or the hadron mass
splitting \cite{ma01,gqli98}. Within the meson-mixing picture, the
two vector mesons may undergo a transition between each other
through the baryon-antibaryon loop (or polarization). Although the
contributions from the proton-antiproton and the
neutron-antineutron loop have opposite sign, as the
$\rho^0p\bar{p}$ vertex is opposite in sign to the
$\rho^0n\bar{n}$ vertex whereas the $\omega p\bar{p}$ and $\omega
n\bar{n}$ vertexes have the same sign, they do not cancel
completely as a result of the small neutron-proton mass difference
\cite{pi93}. In vacuum, the resulting small $\rho-\omega$ meson
mixing is sufficient within some models to explain the isospin
dependence of the nucleon-nucleon scattering lengths and the
Nolen-Schiffer anomaly of light mirror nuclei.  The CSB effect due
to the $\rho-\omega$ meson mixing gets, however, significantly
amplified in isospin-asymmetric nuclear matter, see, e.g., Refs.
\cite{ki97,is00}, because of the different neutron and proton
densities that lead to different stacking of protons and neutrons
in the Fermi sea.

\begin{figure}[tbh]
\vspace{-2cm}
\centerline{\includegraphics[scale=0.5]{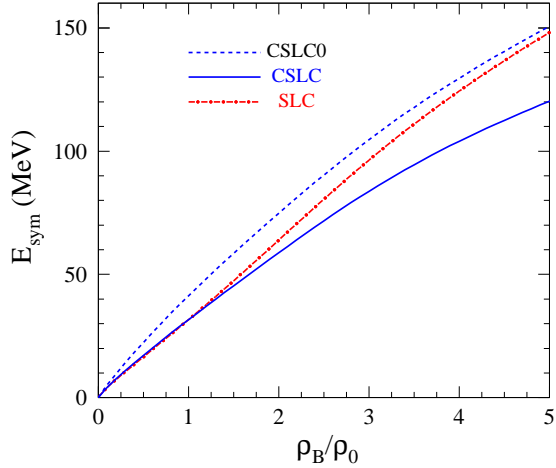}}
\vspace{-1cm} \caption{Effects of the charge symmetry breaking on
the density dependence of the symmetry energy. Taken from Ref.
\cite{Jia08a}.}\label{Jia-csb}
\end{figure}

The $\rho-\omega$ meson mixing in asymmetric nuclear matter also
results in significant modifications to the properties of the
isovector meson $\rho$ and its couplings with nucleons
\cite{ki97,is00,ki96,du97,du00}. Since the potential part of the
nuclear symmetry energy is dictated by the exchange of $\rho$
mesons, at least within the traditional RMF models, modifications
to the density dependence of the symmetry energy are thus expected
accordingly. While most studies \cite{du97,du00,mo03,ag04,mu07}
have focused on the meson spectra, effects of the CSB on the
symmetry energy were also examined in Refs.
\cite{Jia08a,ki97,is00}. It was found in Refs. \cite{ki97,is00}
that the symmetry energy was sharply stiffened. However, the
rearrangement term, which is crucial for the thermodynamic
consistency in deriving the matter pressure in the RMF models, was
neglected in these studies. Taking into account the rearrangement
term, Jiang and Li found recently that the symmetry energy is
actually significantly softened at high densities by the CSB
\cite{Jia08a}. Shown in Fig.~\ref{Jia-csb} are the symmetry energy
for three cases: the SLC, CSLC0 (SLC plus a CSB-induced energy
density), and the CSLC which includes the CSB but with the
parameters readjusted such that it has the same $E_{\rm
sym}(\rho_0)=31.6$ MeV at saturation density as the original SLC.
It is seen that the CSB effect on the symmetry energy is
significant at densities around $1-2\rho_0$. At higher densities,
the modification fades away because the vector meson coupling
constants become zero at high densities according to the BR
scaling \cite{Jia07b}. Comparing the results obtained with the SLC
and the CSLC, one can see clearly that the CSB has a large
softening effect at high densities.

\subsection{Outlooks}

In all standard RMF models, the nucleon self-energies are
independent of momentum/energy. As a result, the Dirac mass and
the Landau mass obtained from these models cannot be
simultaneously consistent with experimental data (see, e.g., Eq.
(\ref{MLanDir})). Also, the `Schr\"{o}dinger-equivalent potential'
$U_{\mathrm{SEP}}$ (Eq. (\ref{Usep})) in these models increases
linearly with nucleon energy even at high energies. Recently,
momentum-dependent nucleon self-energies have been introduced in
the RMF model by including in the Lagrangian density the couplings
of meson fields to the derivatives of nucleon densities
\cite{Typ03,Typ05}, and the results indicate that a reasonable
energy dependence of the `Schr\"{o}dinger-equivalent potential' in
symmetric nuclear matter at saturation density can be obtained,
and the Landau mass can also be increased to a more reasonable
value while keeping the Dirac mass unchanged, which further leads
to an improved description of $\beta $-decay half-lives of
neutron-rich nuclei in the $Z\approx 28$ and $Z\approx 50$ regions
\cite{Mar07}. In the framework of the density-functional theory,
including the couplings of meson fields to the derivatives of
nucleon densities in the Lagrangian density provides an effective
way to take into account higher-order effects.

Another way to introduce the momentum-dependence in nucleon
self-energies is to include the Fock exchange terms by means of
the relativistic Hartree-Fock (RHF) approximation, even though in
practice the inclusion of the Fock terms would increase
significantly the numerical complexity such that it is very
difficult to find appropriate effective Lagrangians for the RHF
model to give satisfactory quantitative description of the nuclear
structure properties compared with standard RMF models
\cite{Mil74,Bro78,Jam81,Hor83,Bou87,Lop88,Ber93,Blu87,Nie01,Mar04,Lop05}.
Recently, there have been some developments in the
density-dependent RHF approach \cite{Lon06a,Lon06b,Lon07} which
can describe the properties of both finite nuclei and nuclear
matter with results comparable to those from standard RMF models.
A more phenomenological way to improve the results of RMF models
is to introduce momentum- as well as isospin-dependent form
factors in the meson-nucleon coupling constants. It has been shown
in Refs. \cite{Mar94,Sah00,Web92} that such an approach can also
reproduce the empirically observed energy dependence of the
nuclear optical potential in symmetric nuclear matter at
saturation density.

Finally, to better understand the isospin-dependent properties of
asymmetric nuclear matter it is crucial to investigate the density
and momentum dependence of underlying isovector nuclear effective
interaction. To achieve this ultimate goal, we need not only more
advanced theoretical approaches but also more experimental data both
on finite nuclei, especially those far from $\beta $-stability line,
and from heavy-ion reactions induced by high energy neutron-rich
nuclei.


\section{Temperature dependence of the symmetry energy and the thermal properties
of hot neutron-rich nuclear matter}
\label{chapter_temperature}

Although significant efforts have been devoted to the study of the
properties of cold asymmetric nuclear matter during the last
decade, much less attention has so far been paid to those of hot
asymmetric nuclear matter, especially the temperature dependence
of the nuclear symmetry energy
\cite{Xu07,Xu07b,Che01b,Zuo03,LiBA06c,Mou07,Mos07}. For finite
nuclei at temperatures below about $3$ MeV, the shell structure
and pairing effects as well as vibrations of nuclear surfaces
remain important, and the symmetry energy was predicted to
increase only slightly with increasing temperature
\cite{Don94,Dea95,Dea02}. However, an increase by only about $8\%$
in the symmetry energy in the range of temperature $T$ from $0$ to
$1$ MeV was found to affect appreciably the physics of stellar
collapse, especially the neutralization processes \cite{Don94}. At
higher temperatures, one expects the symmetry energy to decrease
as the Pauli blocking becomes less important as a result of more
diffused nucleon Fermi surfaces
\cite{Xu07,Xu07b,Che01b,Zuo03,LiBA06c,Mou07}. The temperature
dependence of the nuclear symmetry energy also affects the nuclear
phase diagram. Due to the van der Waals-like behavior of the
nucleon-nucleon interaction, a liquid-gas (LG) like phase
transition is expected to also occur in nuclear matter. Since the
early work, see, e.g., Refs. \cite{Lam78,Fin82,Ber83,Jaq83}, many
investigations have been carried out to explore the properties of
the nuclear LG phase transition both experimentally and
theoretically over the last three decades. For a recent review,
see, e.g., Refs. \cite{Cho04,Das05,Cho06}. Most of these studies
have focused on investigating features of the LG phase transition
in symmetric nuclear matter. New features of the LG phase
transition in asymmetric nuclear matter are expected. In
particular, in a two-component asymmetric nuclear matter, there
are two conserved charges of baryon number and the third component
of isospin. The LG phase transition was suggested to be of second
order \cite{Mul95}. This suggestion together with the need to
understand better the properties of asymmetric nuclear matter have
recently stimulated a lot of work, see, e.g.,
Refs. \cite{LiBA97b,Xu07b,Su00,Wan00,Lee01,LiBA01c,Nat02,LiBA02b,Cho03,Sil04,Mek05,Duc06,Duc07,Leh08}%
.

While impressive progress has been made in the past decade, many
interesting questions about the properties of hot asymmetric
nuclear matter remain unanswered. Some of these questions can be
traced back to our poor understanding of the isovector nuclear
interaction and the density dependence of the nuclear symmetry
energy \cite{LiBA98,LiBA01b,Cho06}. With the recent progress in
constraining the nuclear symmetry energy from nuclear reactions
with radioactive beams, it is therefore of great interest to
investigate how these empirical constraints may allow one to
better understand the chemical, mechanical and thermal properties
of asymmetric nuclear matter. Moreover, both the isovector (i.e.,
the nuclear symmetry potential) and isoscalar parts of the
single-nucleon potential should be momentum dependent due to the
non-locality of the nucleon-nucleon interaction and the Pauli
exchange effects in many-fermion systems. However, effects of the
momentum-dependent interactions on the thermal properties of
asymmetric nuclear matter have received so far little theoretical
attention \cite{Xu07,Xu07b,Xu07c,Mou07}.

In this Chapter, we will mainly review recent progress on
understanding the chemical, mechanical and thermal properties of hot
neutron-rich nuclear matter. In particular, we will give special
emphasis on effects due to the momentum dependence of the isovector
nuclear interaction and the temperature dependence of the symmetry
energy.

\subsection{Thermal model with momentum-dependent interactions}

The effects of isospin and momentum-dependent interactions on the
thermal properties of asymmetric nuclear matter have been
investigated recently based on a self-consistent thermal model
using three different interactions \cite{Xu07c}. The first one is
the isospin and momentum-dependent MDI interaction discussed in
Chapter \ref{chapter_eos} of this review. The second one is a
momentum-independent interaction (MID) which leads to a fully
momentum-independent single-nucleon potential, and the third one
is an isoscalar momentum-dependent interaction (eMDYI) in which
the isoscalar part of the single-nucleon potential is momentum
dependent but the isovector part of the single-nucleon potential
is momentum independent by construction. Although the MID and
eMDYI interactions are not realistic compared to the MDI
interaction, they can be used to explore the effects of the
isospin and momentum dependence of nuclear interactions.

\subsubsection{The momentum-independent MID interaction}

In the momentum-independent MID interaction, the potential energy
density $V_{\text{MID}}(\rho ,\delta )$ of a thermally equilibrated
asymmetric nuclear matter at total density $\rho $ and isospin
asymmetry $\delta $ is written as
\begin{eqnarray}
V_{\text{MID}}(\rho ,\delta )=\frac{\alpha }{2}\frac{\rho ^{2}}{\rho
_{0}} +\frac{\beta }{1+\gamma }\frac{\rho ^{1+\gamma }}{{\rho
_{0}}^{\gamma }}+{\rho }E_{\rm sym}^{\rm pot}(\rho ,x){\delta }^{2}.
\label{MIDV}
\end{eqnarray}
The parameters $\alpha $, $\beta $ and $\gamma $ are determined by
the incompressibility $K_{0}$ of cold symmetric nuclear matter at
saturation density $\rho _{0}=0.16$ fm$^{-3}$ with the binding
energy per nucleon of $-16$ MeV \cite{LiBA97b}, and they are given
by
\begin{eqnarray}
\alpha &=&-29.81-46.90\frac{K_{0}+44.73}{K_{0}-166.32}~\text{(MeV)} \\
\beta &=&23.45\frac{K_{0}+255.78}{K_{0}-166.32}~\text{(MeV)} \\
\gamma &=&\frac{K_{0}+44.73}{211.05},
\end{eqnarray}
where $K_{0}$ is set to be $211$ MeV as in the MDI interaction. To
fit the MDI interaction at zero temperature,  the density
dependence of the potential part of the symmetry energy $E_{\rm
sym}^{\rm pot}(\rho ,x)$ is taken to be the same as that in the
MDI interaction, and it can be parameterized by \cite{Che05a}
\begin{eqnarray}
E_{\rm sym}^{\rm pot}(\rho ,x)=F(x)\frac{\rho }{\rho _{0}}+\left[
18.6-F(x)\right] \left(\frac{\rho }{\rho _{0}}\right)^{G(x)}
\label{epotsym}
\end{eqnarray}
with $F(x=0)=129.981$ MeV, $G(x=0)=1.059$, $F(x=-1)=3.673$ MeV,
and $G(x=-1)=1.569$. The MID interaction reproduces very well the
EOS of isospin-asymmetric nuclear matter at zero temperature
obtained from the MDI interaction with both $x=0$ and $x=-1$. The
single-nucleon potential in the MID interaction can be directly
obtained as
\begin{eqnarray}
U_{\text{MID}}(\rho ,\delta ,\tau )=\alpha \frac{\rho }{\rho
_{0}}+\beta \left(\frac{\rho }{\rho _{0}}\right)^{\gamma
}+U^{\text{asy}}(\rho ,\delta ,\tau ),
\end{eqnarray}%
with
\begin{eqnarray}
U^{\text{asy}}(\rho ,\delta ,\tau ) &=&\left[ 4F(x)\frac{\rho }{\rho
_{0}}+4(18.6-F(x))\left(\frac{\rho }{\rho _{0}}\right)^{G(x)}\right]
{\tau }{\delta }  \notag \\
&+&(18.6-F(x))(G(x)-1)\left(\frac{\rho }{\rho
_{0}}\right)^{G(x)}{\delta }^{2}. \label{Uasy}
\end{eqnarray}
The single-nucleon potential in the MID interaction is thus momentum
independent. As a result, the potential energy density and the
single-nucleon potential in the MID interaction are independent of
temperature as well.

\subsubsection{The extended MDYI (eMDYI) interaction}

To study the effect of the momentum dependence of the isovector
part of the single-nucleon potential (nuclear symmetry potential),
an isoscalar momentum-dependent interaction, called extended MDYI
(eMDYI) interaction, which has the same functional form as the
well-known MDYI interaction \cite{Gal90} for symmetric nuclear
matter has been constructed \cite{Xu07b}. In the eMDYI
interaction, the potential energy density $V_{\text{eMDYI}}(\rho
,T,\delta )$ of a thermally equilibrated asymmetric nuclear matter
at total density $\rho $, temperature $T$ and isospin asymmetry
$\delta $ is expressed as
\begin{eqnarray}
V_{\text{eMDYI}}(\rho ,T,\delta )&=&\frac{A}{2}\frac{\rho ^{2}}{\rho
_{0}}+ \frac{B}{1+\sigma }\frac{\rho ^{1+\sigma }}{{\rho
_{0}}^{\sigma }} +\frac{C}{\rho _{0}}\int \int d^{3}pd^{3}p^{\prime
}\frac{f_{0}(\vec{r},\vec{p})f_{0}(\vec{r},\vec{p}^{\prime
})}{1+(\vec{p}-\vec{p}^{\prime })^{2}/\Lambda ^{2}}\notag\\
&+&{\rho }E_{\rm sym}^{\rm pot}(\rho ,x){\delta }^{2}.\label{MDYIV}
\end{eqnarray}
Here $f_{0}(\vec{r},\vec{p})$ is the phase-space distribution
function of symmetric nuclear matter at total density $\rho $ and
temperature $T$, and $E_{\rm sym}^{\rm pot}(\rho ,x)$ has the same
expression as Eq.~(\ref{epotsym}). The parameters
$A=(A_{u}+A_{l})/{2}$ and $C=(C_{\tau ,-\tau }+C_{\tau ,\tau
})/{2}$, and $B$, $\sigma $ and $\Lambda $ have same values as in
the MDI interaction, so that the eMDYI interaction gives the same
EOS of asymmetric nuclear matter at zero temperature as the MDI
interaction with both $x=0$ and $x=-1$. The single-nucleon
potential in the eMDYI interaction can be obtained as
\begin{eqnarray}
U_{\text{eMDYI}}(\rho ,T,\delta ,\vec{p},\tau )=U^{0}(\rho
,T,\vec{p})+U^{\rm asy}(\rho ,\delta ,\tau ),
\end{eqnarray}
where
\begin{eqnarray}
U^{0}(\rho ,T,\vec{p}) &=&A\frac{\rho }{\rho _{0}}+B\left(\frac{\rho
}{\rho _{0}}\right)^{\sigma }+\frac{2C}{\rho _{0}}\int
d^{3}p^{\prime}\frac{f_{0}(\vec{r},\vec{p})}
{1+(\vec{p}-\vec{p}^{\prime})^{2}/\Lambda ^{2}}
\end{eqnarray}
and $U^{\text{asy}}(\rho ,\delta ,\tau )$ is the same as
Eq.~(\ref{Uasy}), which implies that the symmetry potential is
identical for the eMDYI and MID interactions. Therefore, in the
eMDYI interaction the isoscalar part of the single-nucleon
potential is momentum dependent but the nuclear symmetry potential
is not. For symmetric nuclear matter, the single-nucleon potential
in the eMDYI interaction is exactly the same as that in the MDI
interaction. A similar construction has been used in
Ref.~\cite{Che04} to study the momentum-dependent effects in
heavy-ion collisions.

\subsubsection{Thermodynamic properties of asymmetric nuclear matter}

At zero temperature, one has $f_{\tau }(\vec{r},\vec{p})$
$=\frac{2}{h^{3}}\Theta (p_{f}(\tau )-p)$ and all integrals in above
expressions can be calculated analytically \cite{Che07}.  At finite
temperature $T$, the phase-space distribution function becomes the
Fermi-Dirac distribution
\begin{eqnarray}
f_{\tau }(\vec{r},\vec{p})=\frac{2}{h^{3}}\frac{1}{\exp
\left(\frac{\frac{p^{2}}{2m_{_{\tau }}}+U_{\tau }-\mu _{\tau
}}{T}\right)+1}, \label{f}
\end{eqnarray}
where $\mu _{\tau }$ is the proton or neutron chemical potential and
can be determined from
\begin{eqnarray}
\rho _{\tau }=\int f_{\tau }(\vec{r},\vec{p})d^{3}p.
\end{eqnarray}%
In the above, $m_{_{\tau }}$ is the proton or neutron mass and
$U_{\tau }$ is the proton or neutron single-particle potential. For
fixed density $\rho $, temperature $T$, and isospin asymmetry
$\delta $, the chemical potential $\mu _{\tau }$ and the
distribution function $f_{\tau }(\vec{r},\vec{p})$ can be determined
numerically by a self-consistency iteration scheme
\cite{Gal90,Xu07}. One can then obtain the energy per nucleon
$E(\rho ,T,\delta )$ from
\begin{eqnarray}
E(\rho ,T,\delta )=\frac{1}{\rho }\left[{\sum_{\tau } \int
d^{3}p\frac{p^{2}}{2m_{\tau }}f_{\tau }(\vec{r},\vec{p})}+V(\rho
,T,\delta ) \right]  \label{E}
\end{eqnarray}
and the entropy per nucleon $S_{\tau }(\rho ,T,\delta )$ from
\begin{eqnarray}
S_{\tau }(\rho ,T,\delta )=-\frac{8\pi }{{\rho
}h^{3}}\int_{0}^{\infty }p^{2}[n_{\tau }\ln n_{\tau }+(1-n_{\tau
})\ln (1-n_{\tau })]dp,  \label{S}
\end{eqnarray}
where
\begin{eqnarray}
n_{\tau }=\frac{1}{\exp (\frac{\frac{p^{2}}{2m_{_{\tau }}}+U_{\tau
}-\mu _{\tau }}{T})+1}.
\end{eqnarray}
is the occupation number.

The above results then allow one to calculate the free energy per
nucleon $F(\rho ,T,\delta )$ and the pressure $P(\rho ,T,\delta )$
of a thermally equilibrated asymmetric nuclear matter according to
the thermal dynamic relations
\begin{eqnarray}
F(\rho ,T,\delta )=E(\rho ,T,\delta )-T{\sum_{\tau }}S_{\tau }(\rho
,T,\delta ) \label{F}
\end{eqnarray}
and
\begin{eqnarray} P(\rho ,T,\delta ) &=&\left[ T{\sum_{\tau
}}S_{\tau }(\rho ,T,\delta)-E(\rho ,T,\delta )\right]
\rho+\sum_{\tau }\mu _{\tau }\rho _{\tau }.  \label{P}
\end{eqnarray}

\subsection{Thermal effects on the isospin-dependent bulk and
single-particle properties of asymmetric nuclear matter}

\subsubsection{Nuclear symmetry energy at finite temperature}

As in the case of zero temperature, studies based on both
phenomenological and microscopic models
\cite{Xu07,Che01b,Zuo03,Mou07,Mos07,Sam07} have indicated that the
EOS of hot asymmetric nuclear matter at density $\rho $,
temperature $T$, and an isospin asymmetry $\delta $ can also be
written as a parabolic function of $\delta $, i.e.,
\begin{eqnarray}
E(\rho ,T,\delta )=E(\rho ,T,\delta =0)+E_{\rm sym}(\rho ,T)\delta
^{2}+\mathcal{O}(\delta ^{4}).  \label{eos}
\end{eqnarray}
This nice feature of the empirical parabolic law for the EOS of hot
asymmetric nuclear matter is very useful and convenient for
extracting the temperature and density dependence of the symmetry
energy $E_{\rm sym}(\rho ,T)$ in hot asymmetric nuclear matter,
i.e.,
\begin{eqnarray}
E_{\rm sym}(\rho ,T)\approx E(\rho ,T,\delta =1)-E(\rho ,T,\delta
=0).
\end{eqnarray}
Similar to the case of zero temperature, the symmetry energy at
finite temperature $E_{\rm sym}(\rho ,T)$ gives an estimation of
the energy cost to convert all protons in symmetric nuclear matter
to neutrons at fixed temperature $T$ and density $\rho $.

\begin{figure}[tbh]
\centering
\includegraphics[scale=0.8]{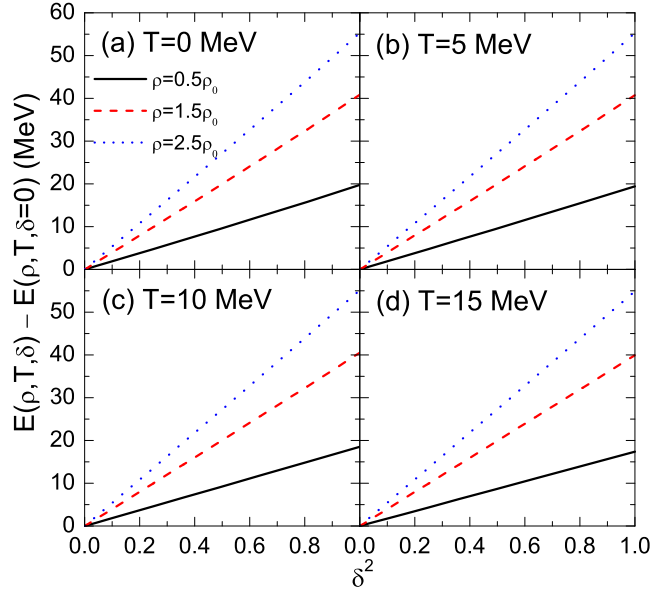}
\caption{{\protect\small (Color online) } Energy difference
between asymmetric and symmteric nuclear matter, $E(\protect\rho
,T,\protect\delta) -E(\protect\rho ,T,\protect\delta
=0)$,{\protect\small \ as a function of }$\protect\delta
^{2}${\protect\small \ at temperatures }$T=0${\protect\small \ MeV
(a), }$5${\protect\small \ MeV (b), }$10${\protect\small \ MeV (c)
and } $15${\protect\small \ MeV (d) for three different baryon
number densities }$\protect\rho =0.5\protect\rho _0$,
$1.5\protect\rho _{0}${\protect\small \ and }$2.5\protect\rho
_{0}${\protect\small \ from the MDI interaction with
}$x=0${\protect\small . Taken from Ref. \cite{Xu07}.}}
\label{EsymTParaMDI0}
\end{figure}

The parabolic approximation for the EOS of hot asymmetric nuclear
matter has been justified for the MDI interaction \cite{Xu07}. As
an example, Fig. \ref{EsymTParaMDI0} displays the quantity $E(\rho
,T,\delta )-E(\rho ,T,\delta =0)$ as a function of $\delta ^{2}$
at temperatures $T=0$, $5$, $10$ and $15$ MeV for three different
baryon number densities $\rho =0.5$, $1.5$ and $2.5~\rho _{0}$
using the MDI interaction with $x=0$. The clear linear relation
between $E(\rho ,T,\delta )-E(\rho ,T,\delta =0)$ and $\delta
^{2}$ shown in Fig.~\ref{EsymTParaMDI0} indicates that the
empirical parabolic law for the hot neutron-rich matter is valid.
As shown in Refs.~\cite{Che01b,Zuo03,Mou07}, the parabolic
approximation also holds very well for the MID and eMDYI
interactions as well as in other microscopic and phenomenological
calculations.

\begin{figure}[tbh]
\centering
\includegraphics[scale=0.8]{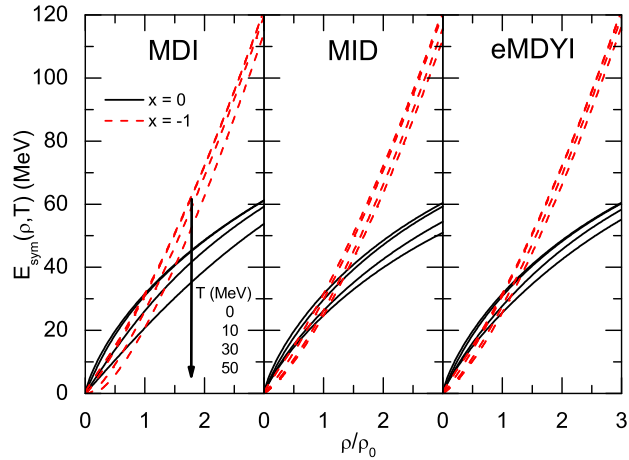}
\caption{{\protect\small (Color online) Density dependence of the
nuclear symmetry energy from the MDI, MID and eMDYI interactions
with $x=0$ and $x=-1$ at $T=0$, $10$, $30$ and $50$ MeV. Taken
from Ref. \cite{Xu07c}.}} \label{Esym}
\end{figure}

Fig.~\ref{Esym} shows the density dependence of the nuclear symmetry
energy at $T=0$, $10$, $30$ and $50$ MeV obtained from the MDI, MID
and eMDYI interactions with $x=0$ and $-1$. For different choices of
the parameter $x=0$ and $-1$, $E_{\rm sym}(\rho ,T)$ displays
different density dependence with $x=0$ ($-1$) giving a larger
(smaller) value for the symmetry energy at low densities while a
smaller (larger) value at high densities for a fixed temperature.
For all three interactions with both $x=0$ and $-1$, it is seen that
the symmetry energy decreases with increasing temperature. At higher
temperatures, one expects the symmetry energy $E_{\rm sym}(\rho ,T)$
to decrease as the Pauli blocking (a pure quantum effect) becomes
less important when the nucleon Fermi surfaces become more diffused
\cite{Xu07,Che01b,Zuo03,LiBA06c}.

\begin{figure}[tbh]
\centering
\includegraphics[scale=0.85]{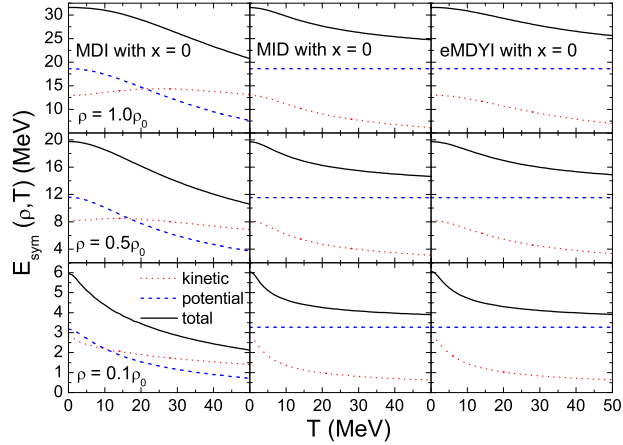}
\caption{{\protect\small (Color online) Temperature dependence of
the total symmetry energy and its kinetic and potential
contributions from the MDI, MID and eMDYI interactions with $x=0$
at $\protect\rho =0.1\protect\rho _{0}$, $0.5\protect\rho _{0}$
and $1.0\protect\rho_{0} $. Taken from Ref. \cite{Xu07c}.}}
\label{EsymT_x0}
\end{figure}

To study the temperature dependence of the potential and kinetic
parts of the symmetry energy $E_{\rm sym}(\rho ,T)$ is useful as
it reflect the effects due to the isospin and momentum dependence
of nuclear interactions. Because the self-consistent
single-particle potential derived from the MDI interaction for a
hot asymmetric nuclear matter is isospin and momentum dependent,
the potential part of the resulting symmetry energy is thus
temperature dependent as shown in Eq.~(\ref{MDIV}). On the other
hand, the potential part of the symmetry energy from the MID and
eMDYI interactions does not depend on the temperature by
construction as seen in Eq.~(\ref{MIDV}) and Eq.~(\ref{MDYIV}).
Fig.~\ref{EsymT_x0} displays the temperature dependence of the
symmetry energy $E_{\rm sym}(\rho ,T)$ as well as its potential
and kinetic energy parts from the MDI, MID and eMDYI interactions
with $x=0$ at $\rho =1.0\rho _{0}$, $0.5\rho _{0}$, and $0.1\rho
_{0}$. For the MDI interaction, both the total symmetry energy
$E_{\rm sym}(\rho ,T)$ and its potential energy part are seen to
decrease with increasing temperature at all three densities
considered. While the kinetic contribution increases slightly with
increasing temperature at low temperature and then decreases with
increasing temperature at high temperature for $\rho =1.0\rho
_{0}$ and $0.5\rho _{0}$, it decreases monotonically for $\rho
=0.1\rho _{0}$. These features observed within the self-consistent
thermal model for the MDI interaction are uniquely determined by
its isospin and momentum dependence. On the other hand, for the
MID and eMDYI interactions the kinetic part of the total symmetry
energy decreases with increasing temperature at all densities
while the potential contribution is independent of temperature and
has the same value for both interactions. These results indicate
that the temperature dependence of the total symmetry energy is
due to both the potential contribution and kinetic contribution
for the MDI interaction, but it is only due to the kinetic
contribution for the MID and eMDYI interactions. Because of the
balance between the kinetic and potential contributions as a
result of the self-consistent nucleon phase-space distribution
functions, the temperature dependence of the total symmetry energy
$E_{\rm sym}(\rho ,T)$ is quite similar for all three interactions
except that the MDI interaction exhibits a slightly stronger
temperature dependence at higher temperatures. Similar results are
obtained for these interactions with the parameter $x=-1$.

For the MDI interaction, the decrease of the kinetic energy part
of the symmetry energy with temperature at very low densities is
consistent with the predictions of the free Fermi gas model at
high temperatures and/or very low densities
\cite{LiBA06c,Mou07,Lee01,Mek05}. The temperature dependence of
nuclear symmetry energy has also been studied recently by
Moustakidis \cite{Mou07} using the isospin- and momentum-dependent
BGBD interaction developed by Bombaci \cite{Bom01} based on the
well known Gale-Bertsch-Das Gupta formalism \cite{Gal87}, and the
results indicate that both the potential and kinetic parts of the
symmetry energy can decrease with temperature for all the
densities considered there. The different temperature dependence
of the potential and kinetic parts of the symmetry energy between
the MDI and BGBD interactions is due to the different forms of the
energy density functional used in these two interactions, with the
MDI interaction leading to a more complicated momentum dependence
of the single-particle potential. This feature implies that the
temperature dependence of the potential and kinetic parts of the
symmetry energy depends on the isospin and momentum dependence of
the nuclear interactions. A similar conclusion was obtained in a
more recent study by Samaddar {\it et al.} \cite{Sam07} using the
density and momentum dependent Seyler-Blanchard interaction.

\subsubsection{Nuclear symmetry potential at finite temperature}

Besides the nuclear density, the symmetry potential of a nucleon
in nuclear matter also depends on the momentum or energy of the
nucleon. In hot asymmetric nuclear matter, the symmetry potential
of a nucleon can further depend on the temperature. The nuclear
symmetry potential is different from the nuclear symmetry energy
as the latter involves the integration of the isospin-dependent
mean-field potential of a nucleon over its momentum. Both the
nuclear symmetry potential and the nuclear symmetry energy are
essential for understanding many important questions in nuclear
physics and astrophysics. As we have already discussed in Chapter
\ref{chapter_ria} and Chapter \ref{chapter_rmf}, various
microscopic and phenomenological models have been used to study
the symmetry potential, and the predicted results vary widely as
in the case of the nuclear symmetry energy. In particular, whereas
most models predict a decreasing symmetry potential with
increasing nucleon momentum albeit at different rates, a few
nuclear effective interactions used in some models give an
opposite behavior. All these studies on the nuclear symmetry
potential are, however, for zero-temperature, and the temperature
dependence of the nuclear symmetry potential has received so far
little theoretical attention \cite{Xu07c}. The density,
temperature and momentum dependent nuclear symmetry potential can
be evaluated by generalizing Eq.(\ref{dat}) in
Chapter~\ref{chapter_ria} to include the temperature dependence,
i.e.,
\begin{eqnarray}
U_{\mathrm{sym}}(\rho ,\vec{p},T)=\frac{U_{n}(\rho
,\vec{p},T)-U_{p}(\rho ,\vec{p},T)}{2\delta }  \label{Usym}
\end{eqnarray}%
where $U_{n}(\rho ,\vec{p},T)$ and $U_{p}(\rho ,\vec{p},T)$
represent, respectively, the neutron and proton single-particle
potentials in hot asymmetric nuclear matter.

\begin{figure}[tbh]
\centering
\includegraphics[scale=0.8]{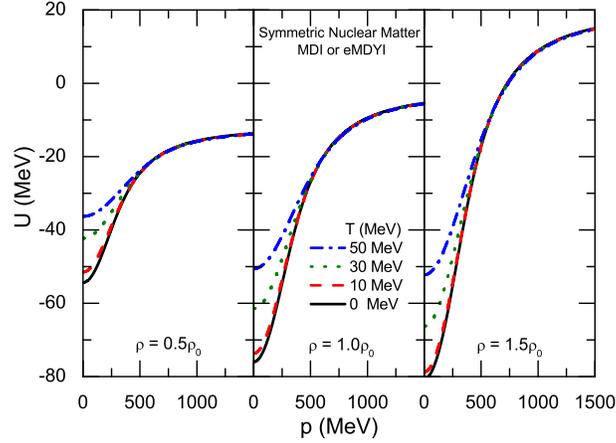}
\caption{{\protect\small (Color online) Momentum dependence of the
single particle-potential in symmetric nuclear matter at
$\protect\rho= 0.5\protect \rho _{0}$, $1.0\protect\rho_{0}$ and
$1.5\protect\rho _{0}$ and $T=0$, $10$, $30$ and $50$ MeV in the
MDI or eMDYI interaction. Taken from Ref. \cite{Xu07c}.}}
\label{SPPd0}
\end{figure}

To see the temperature effect on the nuclear symmetry potential,
it is worthwhile to first study the temperature dependence of the
nucleon single-particle potential in hot nuclear matter. Since the
MID interaction is momentum-independent, the single-particle
potential in symmetric nuclear matter from this interaction is
also temperature-independent. For the MDI and the eMDYI
interaction, they are exactly the same for symmetric nuclear
matter and thus give the same momentum-dependent single-particle
potential in symmetric nuclear matter, and this is shown in
Fig.~\ref{SPPd0} for symmetric nuclear matter at $T=0$, $10$, $30$
and $50$ MeV and $\rho =$ $0.5\rho _{0}$, $1.0\rho _{0}$ and
$1.5\rho _{0}$. It is seen that the single-particle potentials
increase with increasing momentum and saturate at high momenta.
The dependence of the single-particle potential on the nucleon
momentum also becomes stronger at higher densities. Furthermore,
only the low momentum part of the potential is affected by
temperature, becoming less attractive with increasing temperature.
For nucleons with high momenta, their potentials are essentially
independent of temperature as expected.

\begin{figure}[htb]
\centering
\includegraphics[scale=0.75]{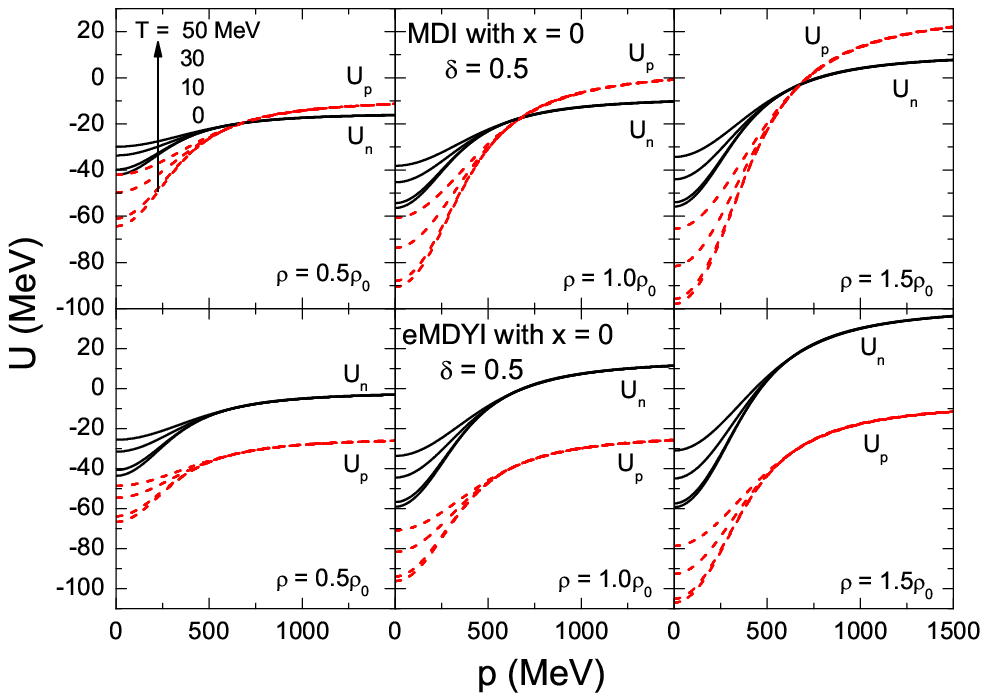}
\includegraphics[scale=0.75]{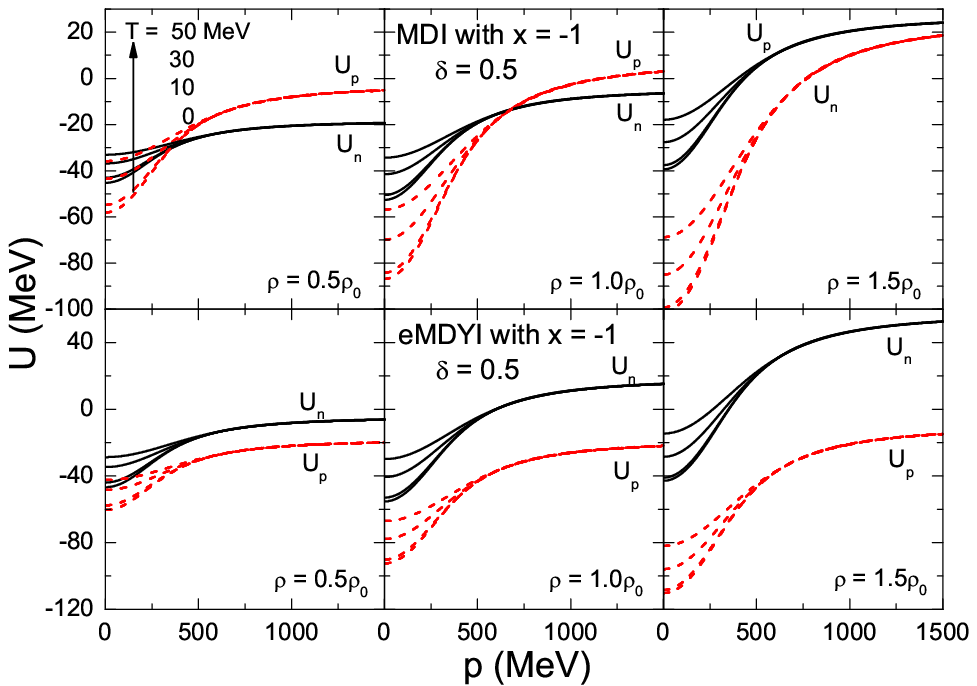}
\caption{Left window: Momentum dependence of the single-particle
potentials of protons and neutrons in asymmetric nuclear matter
with isospin asymmetry $\protect\delta =0.5$ at $\protect\rho=
0.5\protect\rho_{0}$, $1.0\protect\rho _{0}$ and $1.5\protect\rho
_{0}$ and $T=0$, $10$, $30 $ and $50$ MeV in the MDI and eMDYI
interactions with $x=0$. Right window: Same as right window but
for $x=-1$. Taken from Ref. \cite{Xu07c}.} \label{SPPd05}
\end{figure}

Shown in Fig.~\ref{SPPd05} is the momentum dependence of the
single-particle potentials of protons and neutrons in asymmetric
nuclear matter with the isospin asymmetry of $\delta =0.5$ at
$T=0$, $10$, $30$ and $50$ MeV and $\rho =0.5\rho _{0}$, $1.0\rho
_{0}$ and $1.5\rho _{0}$ for MDI (upper panels) and eMDYI (lower
panels) interactions with $x=0$ (left window) and $x=-1$ (right
window). The temperature and density effects are very similar to
those shown in Fig.~\ref{SPPd0} for symmetric nuclear matter,
i.e., only the low momentum part of the potential is affected by
temperature. In contrast to the results for symmetric nuclear
matter, the neutron and proton single-particle potentials in
asymmetric nuclear matter at a fixed temperature are, however,
different from each other. For the eMDYI interaction, finite
temperature causes a shift of the potential to a higher value for
neutrons and to a lower value for protons of any momentum, and the
shifted value is sensitive to the density and the EOS of the
asymmetric nuclear matter, i.e., the value of the $x$ parameter in
the interaction. For the MDI interaction, the isospin and momentum
dependence of the single-particle potentials is somewhat
complicated. In the case of MDI interaction with $x=0$ and at a
fixed temperature, the neutron potential is larger than the proton
potential at low momenta but is smaller at high momenta,
indicating that the neutron potential has a stronger momentum
dependence than that of protons. For other $x$ values, the
single-particle potentials from the MDI interaction are also
shifted at finite temperature, and the shifted value depends only
on the density as the term with $x$ in Eq.~( \ref{MDIU}) is
momentum-independent and depends only on the density.

\begin{figure}[tbh]
\centering
\includegraphics[scale=0.8]{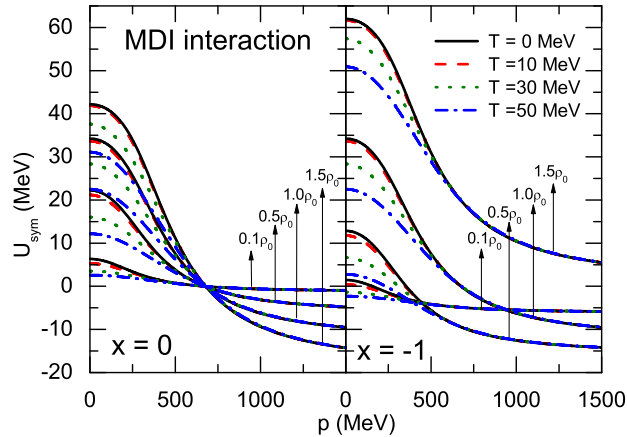}
\caption{{\protect\small (Color online) Momentum dependence of the
symmetry potential at $\protect\rho= 0.1\protect\rho _{0}$,
$0.5\protect\rho _{0}$, $1.0\protect\rho _{0}$ and
$1.5\protect\rho _{0}$ and $T=0$, $10$, $30$ and $50$ MeV in the
MDI interaction with $x=0$ and $x=-1$. Taken from Ref.
\cite{Xu07c}.}} \label{Usymmdi}
\end{figure}

For the nuclear symmetry potential, the one from the eMDYI
interaction is independent of momentum while that from the MDI
interaction is momentum-dependent as discussed previously. Shown in
Fig.~\ref{Usymmdi} is the momentum dependence of the nuclear
symmetry potential at $T=0$, $10$, $30$ and $50$ MeV and $\rho
=0.1\rho _{0}$, $0.5\rho _{0}$, $1.0\rho _{0}$ and $1.5\rho _{0}$
from the MDI interaction with $x=0$ and $x=-1$. It shows that the
symmetry potential decreases with increasing momentum for both $x=0$
and $x=-1$. As mentioned in Chapter~\ref{chapter_ria}, the symmetry
potential at saturation density given by the MDI interaction with
both $x=0$ and $x=-1$, agrees very well with the empirical Lane
potential at energy below about 100 MeV (Eq.(\ref{dat}) and is also
consistent with results from the relativistic impulse ($t$-$\rho $)
approximation based on the empirical NN scattering amplitude
\cite{Mcn83b} or the Love-Franey NN scattering amplitude developed
by Murdock and Horowitz \cite{Mur87,Hor85} for high energy nucleons.
From Fig.~\ref{Usymmdi}, one can also see clearly that the symmetry
potentials from different MDI interactions all decrease with
increasing temperature, especially at low momenta. At high momenta
(above about $500$ MeV/c), the temperature effect on the symmetry
potential is quite weak because nucleons with high momenta are
hardly affected by the temperature as mentioned before. Similar
conclusions are obtained for other values of isospin asymmetry
$\delta $ as corresponding results differ only slightly from the
ones for $\delta=0.5$ discussed in the above.

\subsubsection{Isospin-splitting of nucleon effective mass in hot
neutron-rich matter}

In non-relativistic models, as discussed in
Chapter~\ref{chapter_ria}, one of the important single-particle
properties of nuclear matter is the nucleon effective mass, which
characterizes the momentum dependence of the single-particle
potential of a nucleon. As defined in Eq.(\ref{effmass}), the
nucleon effective mass is related to the density of states
$m_{\tau }^{\ast }/(2\pi \hbar )^{3}$ in asymmetric nuclear
matter. By definition, the nucleon effective mass generally
depends on the density, isospin asymmetry of the medium, and the
momentum of the nucleon \cite{Fuc05,Jam89,Neg98}. In hot nuclear
medium, it depends on the temperature as well. At zero
temperature, when the nucleon effective mass is evaluated at the
Fermi momentum $p_{\tau }=p_{f}({\tau })$, Eq. (\ref{effmass})
yields the Landau mass which is related to the $f_{1}$ Landau
parameter of a Fermi liquid \cite{Fuc05,Jam89,Neg98}. We have
reviewed in Chapter~\ref{chapter_ria} and
Chapter~\ref{chapter_rmf} the different kinds of effective masses
in nuclear matter, and a more detailed discussion can be found in
Ref.~\cite{Jam89}.

Since the momentum-dependent part of the nuclear potential for the
MDI interaction is independent of the parameter $x$, same nucleon
effective mass is obtained for different values of $x$. In
asymmetric nuclear matter, the neutron and proton effective masses
are usually different due to differences in the momentum
dependence of the single-particle potential for neutrons and
protons. The isospin-splitting of nucleon effective mass in
asymmetric nuclear matter, i.e., the difference between the
neutron and proton effective masses, is currently not known
empirically \cite{Lun03}. Theoretical results on the
neutron-proton effective mass splitting are also highly
controversial among different approaches and/or different nuclear
effective interactions \cite{Riz04,Che07,LiBA04c,Beh05}. Being
phenomenological and non-relativistic in nature, the
neutron-proton effective mass splitting in the MDI interaction is
consistent with the predictions of all non-relativistic
microscopic models, see, e.g., Refs.~\cite{Bom91,Zuo05,Sjo76}, and
the non-relativistic limit of microscopic relativistic many-body
theories, see, e.g., Refs. \cite{Fuc04,Ma04,Sam05a,Fuc05}. Recent
transport model studies have indicated that the neutron/proton
ratio at high transverse momenta and/or rapidities is a
potentially useful probe of the neutron-proton effective mass
splitting in neutron-rich matter \cite{LiBA04a,Riz05}. Since the
momentum dependence of the single-particle potential is usually
also temperature dependent, it is thus of interest to study the
temperature effect on the nucleon effective mass.

\begin{figure}[tbh]
\centering
\includegraphics[scale=0.8]{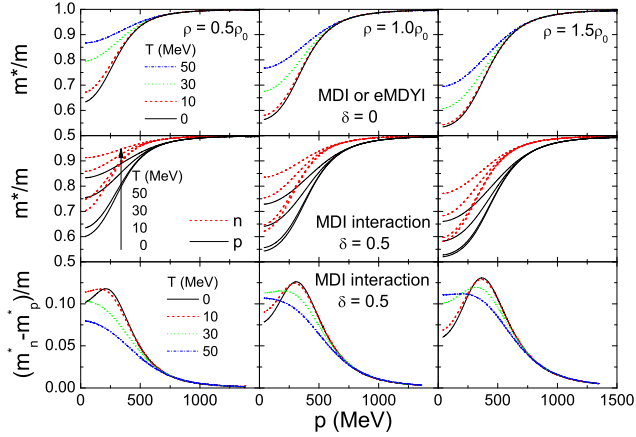}
\caption{{\protect\small (Color online) Momentum dependence of the
effective masses of protons and neutrons in symmetric nuclear
matter ($\protect\delta =0$, upper panels) and in asymmetric
nuclear matter ($\protect\delta =0.5$, middle panels) at
$\protect\rho= 0.5\protect\rho _{0}$, $1.0\protect\rho _{0}$ and
$1.5\protect\rho _{0}$ and $T=0$, $10$, $30$ and $50$ MeV in the
MDI or the eMDYI interaction. Corresponding results for the
reduced isospin-splitting of the nucleon effective mass, i.e.,
$(m_{n}^{\ast }-m_{p}^{\ast })/m$ in asymmetric nuclear matter
($\protect\delta =0.5$) are shown in the lower panels. Taken from
Ref. \cite{Xu07c}.}} \label{EffMass}
\end{figure}

The upper panels of Fig.~\ref{EffMass} show the momentum dependence
of nucleon effective mass in symmetric nuclear matter at $\rho
=0.5\rho _{0}$, $1.0\rho _{0}$ and $1.5\rho _{0}$ and $T=0$, $10$,
$30$ and $50$ MeV from the MDI or the eMDYI interaction as the two
give the same nucleon effective mass in symmetric nuclear matter.
These results are independent of the value of the $x$ parameter in
these interactions because the momentum dependence of the single
particle-potential does not depend on the $x$ parameter as mentioned
above. For the MID interaction, which gives a momentum-independent
and thus a temperature-independent single-particle potential, the
resulting nucleon effective mass is simply equal to the nucleon mass
in free space and is thus not shown here. Similar results for the
neutron and proton effective masses in neutron-rich nuclear matter
with isospin asymmetry $\delta =0.5$ are displayed in the middle
panels of Fig.~\ref{EffMass} for the MDI interaction. These results
show that for a fixed temperature, the nucleon effective mass
decreases with increasing density and decreasing momentum,
indicating that the momentum dependence of the single-particle
potential is stronger at higher densities and lower momenta. At high
momenta, the nucleon effective mass approaches the nucleon mass in
free space as the single-particle potential becomes saturated. For a
fixed momentum, the nucleon effective mass increases with
temperature, especially at lower momenta, implying that the
temperature effect weakens the momentum dependence of the nuclear
interaction at lower momenta. In asymmetric nuclear matter at a
fixed temperature, the neutron effective mass at a given momentum is
seen to be larger than the proton effective mass at same momentum,
leading thus to the isospin-splitting of the nucleon effective mass.

The temperature effect on the isospin-splitting of the nucleon
effective mass can be better seen in terms of the reduced
isospin-splitting of the nucleon effective mass, i.e.,
$(m_{n}^{\ast }-m_{p}^{\ast })/m$, as shown in the lower panels of
Fig.~\ref{EffMass}. It is seen that the temperature effect on the
isospin-splitting of the nucleon effective mass displays some
complicated behaviors. At lower densities, the temperature effect
seems to reduce the isospin-splitting of the nucleon effective
mass for a fixed momentum. At higher densities, it depends on the
momentum, i.e., the temperature effect reduces the
isospin-splitting of the nucleon effective mass at high momenta
but increases the isospin-splitting at lower momenta. These
features reflect the complexity of the temperature effect on the
momentum dependence of the neutron and proton single-particle
potentials in hot asymmetric nuclear matter for the MDI
interaction.

\subsection{Mechanical and chemical instabilities in hot neutron-rich nuclear
matter}

The mechanical and chemical instabilities of hot asymmetric
nuclear matter have been extensively studied based on various
theoretical models
\cite{Mul95,LiBA97b,Bar98,LiBA02b,Lat78,Bar80,Bar01,Cat01,Lee07}.
However, effects of the momentum-dependent interactions on the
mechanical and chemical instabilities have received so far not
much theoretical attention. In the following, we discuss the
mechanical and chemical instabilities using the MDI, MID, and
eMDYI interactions and focus on the effects due to the isospin and
momentum dependence of nuclear interactions.

\subsubsection{The mechanical instability}

The mechanical stability condition for a hot asymmetric nuclear
matter is
\begin{eqnarray}
\left( \frac{\partial P}{\partial \rho }\right) _{T,\delta }\geq
0. \label{Mstability}
\end{eqnarray}
If the above condition is not satisfied in certain part of the
system, any growth in its density would lead to a decrease of
pressure. As the pressure of this region is lower than other parts
of the system, the nuclear matter in this region would be
compressed, leading to further growth of the density. As a result,
any small density fluctuations in the matter can grow, and the
system would become mechanically unstable.

\begin{figure}[tbh]
\centering
\includegraphics[scale=0.8]{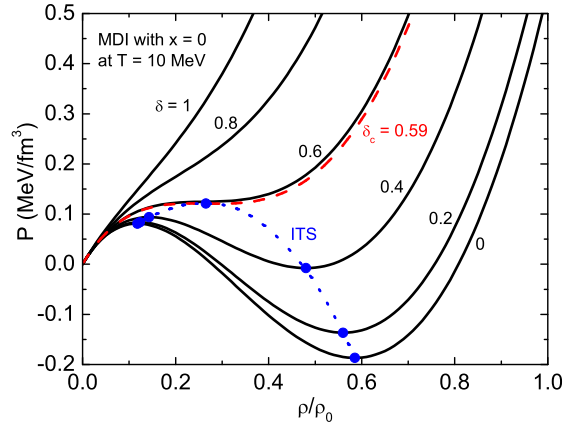}
\caption{{\protect\small (Color online) Pressure as a function of
density at fixed isospin asymmetry in the MDI interaction with
$x=0$ at $T=10$ MeV. The isothermal spinodals (ITS) and the case
for the critical isospin asymmetry are also indicated. Taken from
Ref. \cite{Xu07c}.}} \label{example_ITS1}
\end{figure}

An example of the boundary of mechanical instability in the $P-\rho$
plane of a hot asymmetric nuclear matter is shown in
Fig.~\ref{example_ITS1}. The isothermal lines for different values
of isospin asymmetry $\delta $ are obtained from the MDI interaction
with $x=0$  for asymmetric nuclear matter at $T=10$ MeV. One can see
that the mechanical stability condition can be violated for
isothermal lines with isospin asymmetries below the dashed line
corresponding to the critical asymmetry $\delta_c$, which is
determined by
\begin{eqnarray}
\left( \frac{\partial P}{\partial \rho }\right) _{T,\delta
_{c}}=\left( \frac{\partial ^{2}P}{\partial \rho ^{2}}\right)
_{T,\delta _{c}}=0 \label{inflection1}
\end{eqnarray}
and is about $0.59$ in this case. The extrema of the $P-\rho$
lines at different isospin asymmetries then form the boundary of
the mechanical instability region, namely, the isothermal spinodal
(ITS), and is shown by the dotted line. For isothermal lines above
the dashed one, corresponding to isospin asymmetry $\delta $
larger than the critical value, the pressure is seen to increase
monotonically with density and Eq.~(\ref{Mstability}) is thus
satisfied for all densities.

\begin{figure}[htb]
\centering
\includegraphics[scale=0.75]{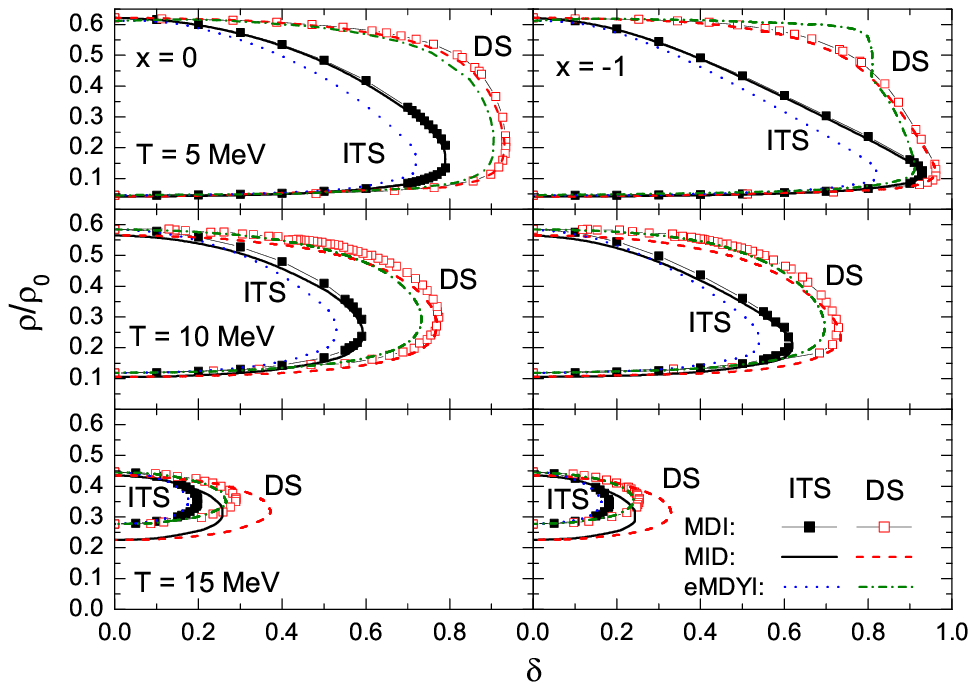}
\includegraphics[scale=0.75]{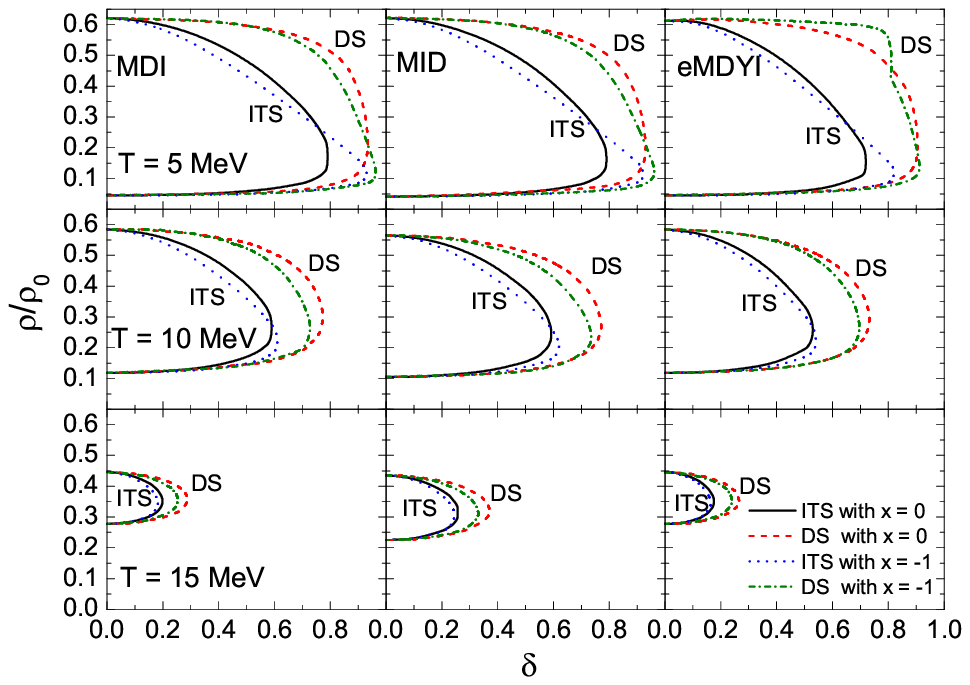}
\caption{Left window: Boundaries of mechanical (ITS) and chemical
(DS) instabilities in the $\protect\rho-\protect\delta $ plane at
$T=5$, $10$ and $15$ MeV for the MDI, MID and eMDYI interactions
with $x=0 $ and $x=-1$. Right window: Same as left window but
separately for the MDI, MID and eMDYI interactions to show the $x$
dependence. Taken from Ref. \cite{Xu07c}.} \label{rhodelta}
\end{figure}

\begin{figure}[tbh]
\centering
\includegraphics[scale=0.75]{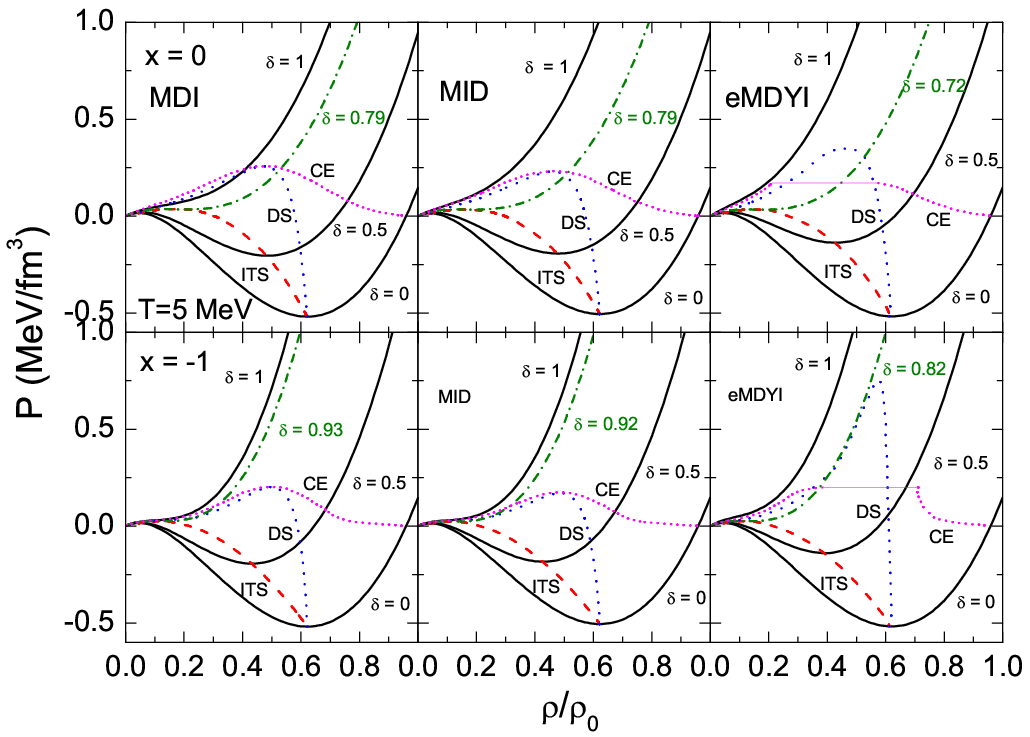}
\includegraphics[scale=0.75]{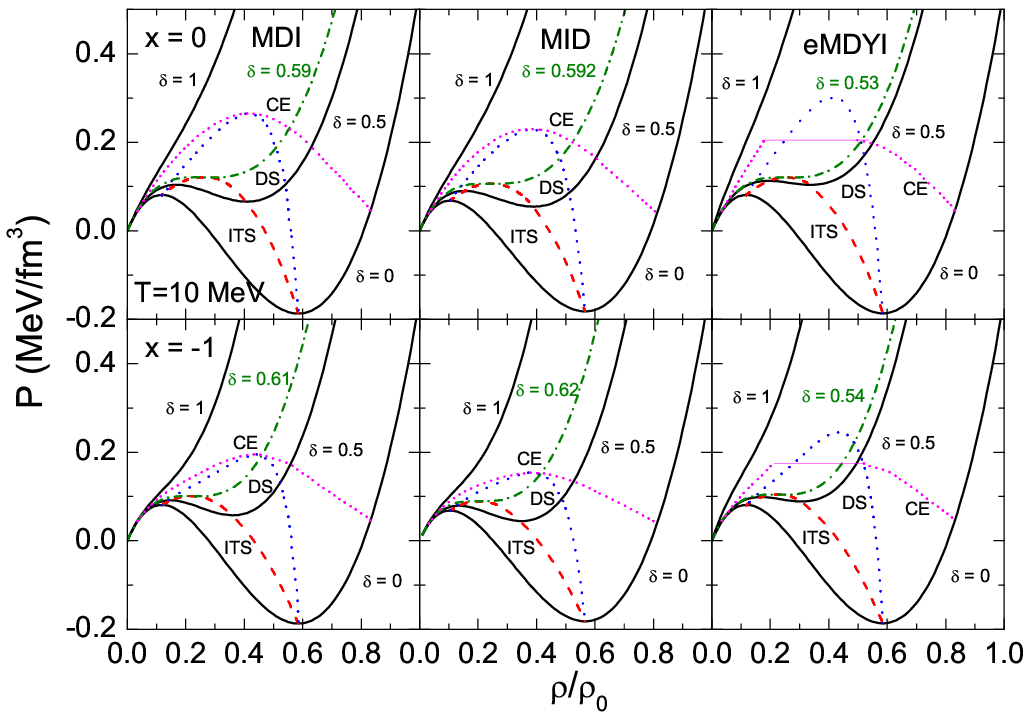}
\includegraphics[scale=0.75]{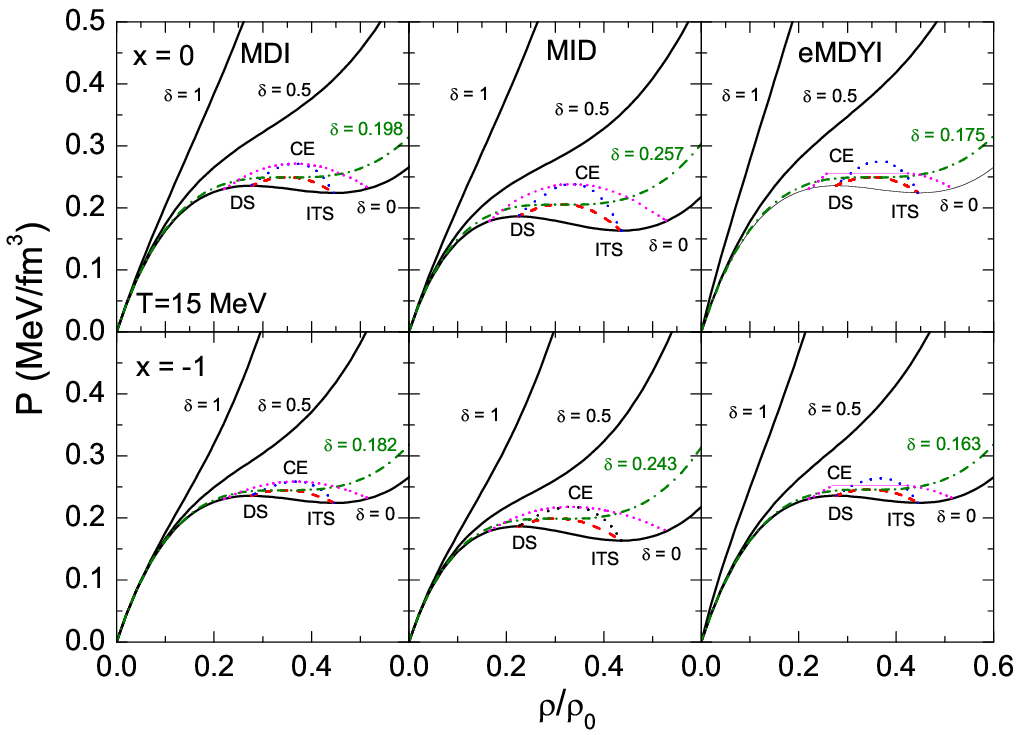}
\caption{{\protect\small (Color online) Boundaries of mechanical
(ITS) and chemical (DS) instabilities in the $P-\protect\rho $
plane at $T=5$ MeV (upper left window), $T=10$ MeV (upper right
window), and $T=10$ MeV (lower window) for the MDI, MID and eMDYI
interactions with $x=0$ and $x=-1$. Taken from Ref.
\cite{Xu07c}.}} \label{PrhoT5}
\end{figure}

Fig.~\ref{rhodelta} displays the boundary of mechanical
instability, i.e., ITS for the MDI, MID and eMDYI interactions at
$T=5,10$ and $15$ MeV with $x=0$ and $x=-1$ in the $\rho-\delta$
plane while Fig.~\ref{PrhoT5} displays the same curves in the
$P-\rho$ plane as well as the curves at constant isospin
asymmetries of $\delta =0$, $0.5$, $1$ and $\delta _{c}$.
Furthermore, the boundary of chemical instability as well as that
of the LG phase-coexistence region are also shown in
Fig.~\ref{PrhoT5}. From Fig.~\ref{rhodelta}, one can see that the
nuclear matter in the left region of the boundary of mechanical
instability indicated by ITS is mechanically unstable, and the
critical isospin asymmetry as well as the area of the mechanical
instability region decrease with increasing temperature. For each
interaction, the boundaries overlap at $\delta =0$ for different
values of $x$, since for symmetric nuclear matter the three
interactions are independent of the value of $x$. For the MDI and
eMDYI interactions, the ITS has the same value at $\delta =0 $, as
they are exactly the same model for symmetric nuclear matter as
mentioned above, while for the MID interaction it is shifted to
smaller densities at $\delta =0$. The values of critical isospin
asymmetry is sensitive to the density dependence of the symmetry
energy as shown by the dash-dotted lines in Fig.~\ref{PrhoT5}. At
$T=5$ and $10$ MeV the value of the critical isospin asymmetry is
larger for $x=-1$ than for $x=0$, while at $T=15$ MeV it is
smaller for $x=-1$ than for $x=0$. Therefore, the density
dependence of nuclear symmetry energy and the temperature are two
important factors in determining the value of critical isospin
asymmetry and the area of mechanical instability.
Fig.~\ref{rhodelta} further shows that both are also sensitive to
the isospin and momentum dependence of nuclear interactions,
especially at higher temperatures. Detailed comparisons indicate
that the critical isospin asymmetry from the MDI interaction is
very similar to that from the MID interaction at low and moderate
temperatures, while it is similar to that of the eMDYI interaction
at high temperatures. As to the area of the mechanically unstable
region, it is the largest for the MID interaction while the
smallest for the eMDYI interaction.

\begin{figure}[tbh]
\centering
\includegraphics[scale=0.8]{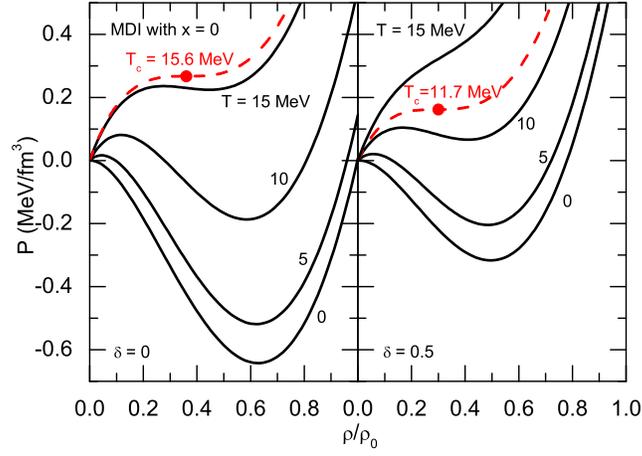}
\caption{{\protect\small (Color online) Pressure as a function of
density for different temperatures from the MDI interaction with
$x=0$ at $\protect\delta =0.0$ (left panel) and $\protect\delta
=0.5$ (right panel). Taken from Ref. \cite{Xu07c}.}}
\label{example_ITS2}
\end{figure}

The above results are for fixed temperature but different isospin
asymmetries. For fixed isospin asymmetries but different
temperatures, results from the the MDI interaction with $x=0$ are
shown in Fig.~\ref{example_ITS2} for isospin asymmetry $\delta =0$
and $\delta =0.5$ as an example. It is clearly seen that the
behavior of increasing the temperature at fixed isospin asymmetry
is similar to that of increasing the isospin asymmetry at fixed
temperature, and the mechanical stability condition
Eq.~(\ref{Mstability}) is satisfied at all densities once the
temperature is larger than the critical temperature $T_{c}$ (about
$15.6$ MeV at $\delta =0$ and $11.7$ MeV at $\delta =0.5$).

\begin{figure}[htb]
\centering
\includegraphics[scale=0.75]{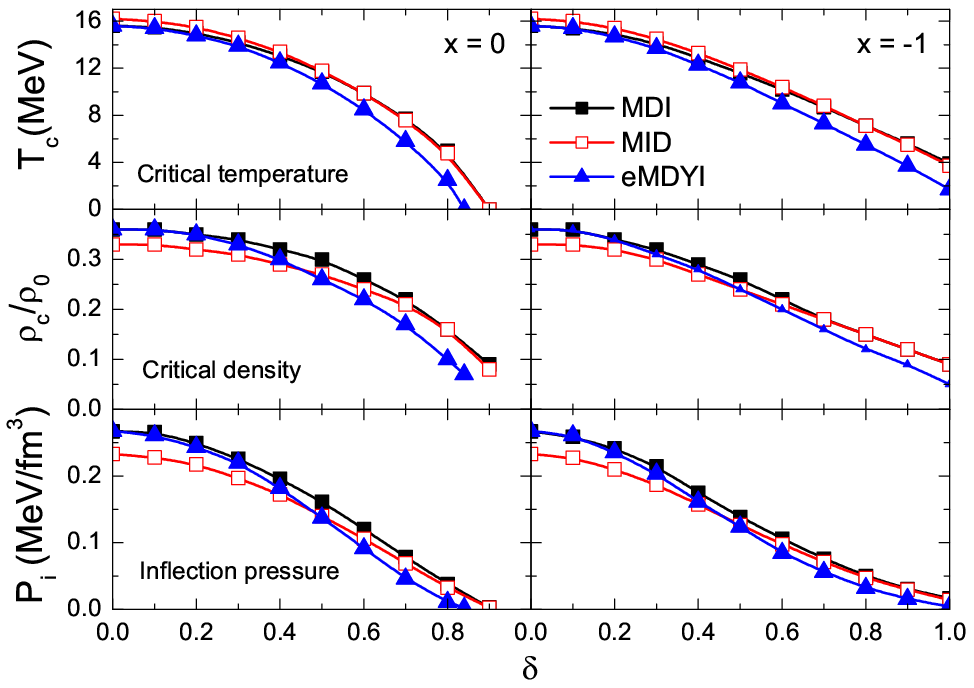}
\includegraphics[scale=0.75]{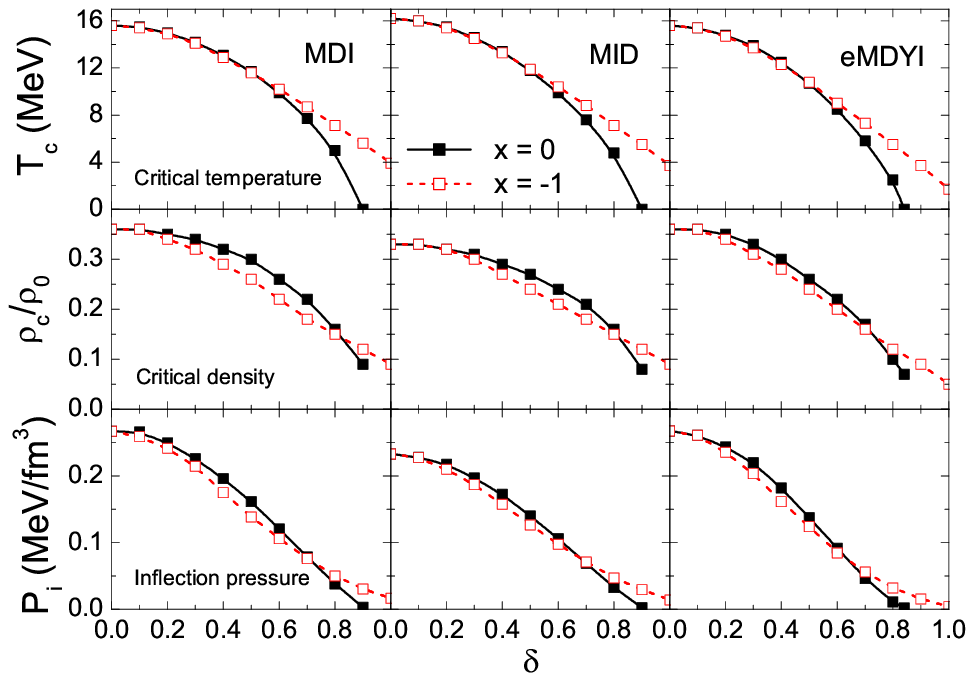}
\caption{(Color online) Left window: Isospin asymmetry dependence
of the critical temperature (upper panels), the critical density
(middle panels), and the inflection pressure (lower panels) for
the MDI, MID and eMDYI interactions with $x=0$ and $x=-1$. Right
window: Same as left window but separately for the MDI, MID and
eMDYI interactions with $ x=0$ and $x=-1$. Taken from Ref.
\cite{Xu07c}.} \label{deltaTcrhoc}
\end{figure}

The density at the inflection point, which satisfies
\begin{eqnarray}
\left( \frac{\partial P}{\partial \rho }\right) _{T_{c},\delta
}=\left( \frac{\partial ^{2}P}{\partial \rho ^{2}}\right)
_{T_{c},\delta }=0, \label{inflection}
\end{eqnarray}
is the critical density $\rho _{c}$, and the pressure at the
inflection point is named as the inflection pressure $P_{i}$. The
left window of Fig.~\ref{deltaTcrhoc} shows the isospin asymmetry
dependence of the critical temperature $T_{c}$, the critical
density $\rho _{c}$, the inflection pressure $P_{i}$ for the MDI,
MID and eMDYI interactions with $x=0$ and $x=-1 $. It is seen that
all these quantities decrease with increasing isospin asymmetry.
Below these curves the nuclear system can be mechanically
unstable. The critical temperature for symmetric nuclear matter is
$15.6$ MeV for the MDI and eMDYI interactions and $16.2$ MeV for
the MID interaction with both $x=0$ and $x=-1$. For $x=0$ the
system is stable above a certain high isospin asymmetry ($0.9$ for
MDI and MID model and $0.84$ for eMDYI model), but for $x=-1$ it
can be mechanically unstable even for pure neutron matter. These
features indicate again that the boundary of mechanical
instability is quit sensitive to the value of the parameter $x$ in
the interaction. The left window of Fig.~\ref{deltaTcrhoc} further
shows that results from the MDI interaction are similar to those
from the MID interaction at low temperatures, but are similar to
those from the eMDYI interaction at high temperatures.

The effect of the density dependence of symmetry energy on the
critical temperature, critical density and inflection pressure is
more clearly seen in the right window of Fig.~\ref{deltaTcrhoc},
where their isospin asymmetry dependence for each interaction with
$x=0$ and $x=-1$ is shown.  In each case the critical temperature
for the case of $x=0$ is a little higher than that for $x=-1$ at
smaller $\delta $, but is lower at larger $\delta $. For the
critical density and the inflection pressure, they are also larger
for $x=0$ than for $x=-1$ at low and moderate isospin asymmetries,
and become smaller at larger $\delta $.

\subsubsection{The chemical instability}

For a hot asymmetric nuclear matter, it becomes chemically unstable
if either of following inequalities are violated,
\begin{eqnarray}
\left( \frac{\partial {\mu }_{n}}{\partial {\delta }}\right)
_{P,T}>0\qquad\text{and}\qquad\left( \frac{\partial {\mu
}_{p}}{\partial {\delta }}\right) _{P,T}<0, \label{Cstability}
\end{eqnarray}
This is so as a small growth of the isospin asymmetry $\delta $ in
the region of chemical instability would grow further, since more
neutrons would move into the region from other part of the nuclear
system to lower the energy of the whole system as a result of the
low neutron chemical potential. This also holds true for the case of
protons. So any isospin fluctuations would make the system unstable
if either of the inequalities in Eq. (\ref{Cstability}) is not
satisfied.

\begin{figure}[tbh]
\centering
\includegraphics[scale=0.8]{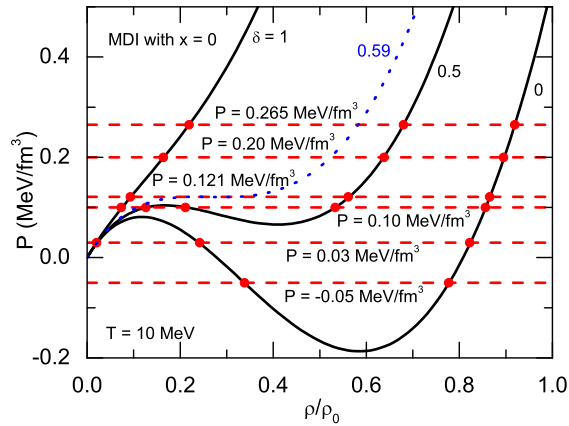}
\caption{{\protect\small (Color online) Pressure as a function of
density at fixed isospin asymmetry from the MDI interaction with
$x=0$ at $T=10$ MeV. See text for details on how to obtain the
chemical potential isobar. Taken from Ref. \cite{Xu07c}.}}
\label{example_DS1}
\end{figure}

To analyze the chemical instability in nuclear matter, one needs
information on the chemical potential isobar for neutrons and
protons at fixed temperature and pressure. This can be obtained
from searching for the cross point between the fixed pressure line
and the $P-\rho$ curves at fixed isospin asymmetry. This procedure
is illustrated in Fig.~\ref{example_DS1}, where the densities and
the chemical potentials of the cross points for one isospin
asymmetry at fixed pressure and temperature are shown. By changing
the isospin asymmetry from $0$ to $1$, one can then obtain the
whole chemical potential isobar at a fixed temperature and
pressure. It is seen that depending on the pressure, the number of
cross points can be one, two or three for a fixed isospin
asymmetry, which will be reflected in the shape of the resulting
chemical potential isobar. The critical isospin asymmetry of
mechanical instability is $0.59$ for the MDI interaction at $T=10$
MeV with $x=0$ and the corresponding curve in $P-\rho $ plane is
given by the dotted line in Fig.~\ref{example_DS1}. The pressure
of the inflection point is $0.121$ MeV, above which the mechanical
instability disappears and there only exists the chemical
instability, and the chemical potential isobar can only have one
branch for all values of $\delta $.

\begin{figure}[tbh]
\centering
\includegraphics[scale=0.8]{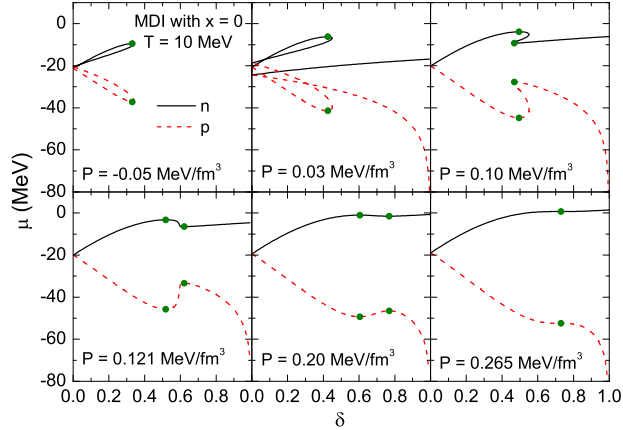}
\caption{{\protect\small (Color online) Chemical potential isobar
as a function of isospin asymmetry from the MDI interaction with
$x=0$ at $T=10$ MeV. The extrema of each curve are indicated by
solid circles. Taken from Ref. \cite{Xu07c}.}} \label{example_DS2}
\end{figure}

Fig.~\ref{example_DS2} displays the chemical potential isobar
calculated at the pressure of $P=-0.05$, $0.03$, $0.10$, $0.121$,
$0.20$ and $0.265$ MeV$/$fm$^{3}$. One sees that shapes of these
curves are different for different pressures. The extrema of these
curves, indicated by solid circles, are just the boundaries of
chemical instability, or diffusive spinodals (DS). The one for
$P=0.265$ MeV$/$fm$^{3}$ corresponds to the critical pressure
$P_{c}$ for the MDI interaction with $x=0$ at $T=10$ MeV, above
which the chemical potential of neutrons (protons) increases
(decreases) monotonically with $\delta $ and the chemical
instability disappears. The inflection point, which satisfies
\begin{eqnarray}
\left( \frac{\partial {\mu }}{\partial {\delta }}\right)
_{P_{c},T}=\left( \frac{\partial ^{2}{\mu }}{\partial {\delta
}^{2}}\right) _{P_{c},T}=0,\label{inflection2}
\end{eqnarray}
is also shown in the figure. For the MDI and MID interaction, the
extrema of $\mu _{n}$ and $\mu _{p}$ correspond to the same
$\delta $ value, so the critical pressure is the same for neutrons
and protons. For the eMDYI interaction the chemical potential
isobar shows an asynchronous behavior for neutrons and protons, as
will be shown in the following. This asynchronous behavior is also
different for different temperatures and values of $x$
\cite{Xu07b}.

The diffusive spinodals from the MDI, MID and eMDYI interactions
with $x=0$ and $x=-1$ at $T=5,10$ and $15 $ MeV as shown in the
$\rho-\delta $ plane in Fig.~\ref{rhodelta} and in the $P-\rho $
plane in Fig.~\ref{PrhoT5} clearly indicate that they envelope the
region of mechanical instability and extend further out into the
$\rho-\delta $ and $P-\rho $ planes. Furthermore, the area of
chemical instability region, which lies between the ITS and DS
curves, decreases with increasing temperature.  For $\delta =0$,
the DS and ITS from the MDI and eMDYI interactions coincide, and
the same happens to those from these interactions with $x=0$ and
$x=-1$. The boundary of chemical instability is sensitive to the
density dependence of the symmetry energy. At $T=5$ MeV the
maximum isospin asymmetry is larger for $x=-1$ than $x=0$, while
at $T=10$ and $15$ MeV it is smaller for $x=-1$ than $x=0$.
Fig.~\ref{rhodelta} further shows that both the critical isospin
asymmetry and the area of chemical instability are also sensitive
to the isospin and momentum dependence of the nuclear interaction,
especially at higher temperatures. The maximum $\delta $ value in
the MDI interaction is similar to that in the MID interaction at
low temperature, but it becomes similar to that in the eMDYI
interaction at high temperature. This also holds for the
mechanical instability as discussed before. The shape of the DS
curve in the eMDYI interaction with $x=-1$ at $T=5$ MeV shown in
Figs.~\ref{rhodelta} and \ref{PrhoT5} further exhibits some
unusual behaviors as a result of the asynchronous behavior of the
chemical potential isobar between neutrons and protons.

\subsection{The liquid-gas phase transition in hot neutron-rich nuclear matter}

\subsubsection{The chemical potential isobar}

The above theoretical models further allow one to study the LG
phase transition in hot asymmetric nuclear matter. The phase
coexistence is governed by the Gibbs conditions of equal pressures
and chemical potentials. For asymmetric nuclear matter with
different concentrations of protons and neutrons, the two-phase
coexistence conditions are
\begin{eqnarray}
P^{L}(T,\rho ^{L},\delta ^{L}) &=&P^{G}(T,\rho ^{G},\delta ^{G}),
\label{coexistenceP} \\
\mu _{n}^{L}(T,\rho ^{L},\delta ^{L}) &=&\mu _{n}^{G}(T,\rho
^{G},\delta
^{G}),  \label{coexistencemuN} \\
\mu _{p}^{L}(T,\rho ^{L},\delta ^{L}) &=&\mu _{p}^{G}(T,\rho
^{G},\delta ^{G}),  \label{coexistencemuP}
\end{eqnarray}%
where $L$ and $G$ stand for the liquid phase and the gas phase,
respectively.  For a fixed pressure, the solutions thus form the
edges of a rectangle in the proton and neutron chemical potential
isobars as functions of isospin asymmetry $\delta $ and can be found
by means of the geometrical construction method
\cite{Mul95,Xu07b,Su00}.

\begin{figure}[htb]
\centering
\includegraphics[scale=0.75]{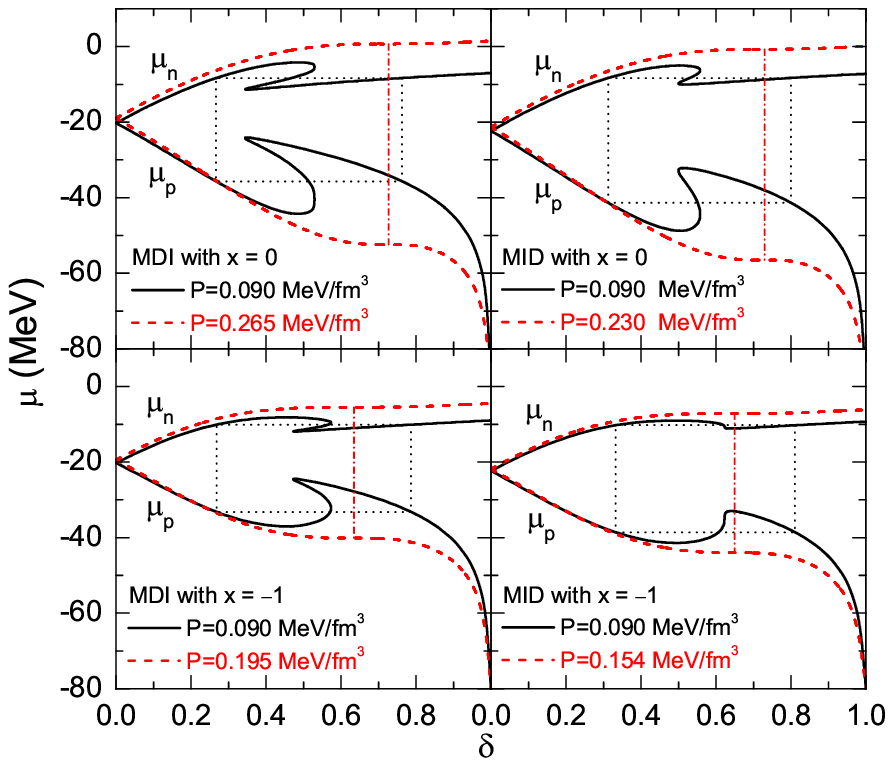}
\includegraphics[scale=0.75]{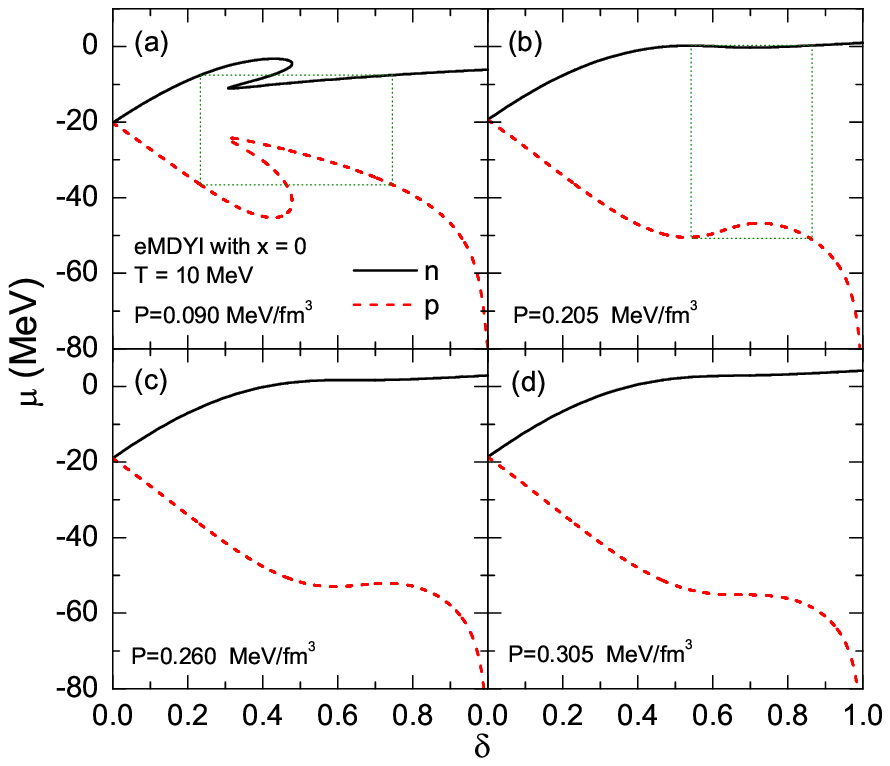}
\caption{(Color online) Left window: The chemical potential isobar
as a function of the isospin asymmetry $\protect\delta $ at $T=10$
MeV from the MDI and MID interactions with $x=0$. The geometrical
construction used to obtain the isospin asymmetries and chemical
potentials in the two coexisting phases is also shown
\cite{Xu07b}. Right window: Similar to left panel but for the
eMDYI interaction \cite{Xu07c}.} \label{mudelta}
\end{figure}

Shown in Fig.~\ref{mudelta} is an example for studying the LG
phase transition from the chemical potential isobars of an
aymmetric nuclear matter at $T=10$ MeV. The solid curves in the
left window are the proton and neutron chemical potential isobars
as functions of the isospin asymmetry $\delta $ at a fixed
pressure $P=0.090$ MeV$/$fm$^{3}$ by using the MDI and MID
interactions with $x=0$ and $x=-1$. The resulting rectangles from
the geometrical construction are shown by dotted lines, from which
one can see that different interactions lead to different shapes
for the chemical potential isobar. As the pressure increases and
approaches the critical pressure $P_{\mathrm{C}}$, an inflection
point defined by Eq.~(\ref{inflection2}) appears for both the
proton and the neutron chemical potential isobar. Above the
critical pressure, the chemical potential of neutrons (protons)
increases (decreases) monotonically with $\delta $ and the
chemical instability disappears. Also shown in Fig.~\ref{mudelta}
by dashed curves are the chemical potential isobars at the
critical pressure. In this case, corresponding rectangles in the
geometrical construction shrink to vertical lines perpendicular to
the $\delta $ axis as shown by the dash-dotted lines in
Fig.~\ref{mudelta}. As to the value of the critical pressure, it
depends on the interaction. For the MDI and MID interactions with
$x=0$, the values are $0.265$ and $0.230$ MeV$/$fm$^{3}$,
respectively. The critical pressure is also sensitive to the
density dependence of the nuclear symmetry energy with the stiffer
symmetry energy ($x=-1$) giving a smaller critical pressure
\cite{Xu07b}.

For the eMDYI interaction, its single-particle potential is
momentum-dependent but the momentum dependence is
isospin-independent, i.e., the same for protons and neutrons.
Comparing the results from the eMDYI interaction with those from
the MID interaction and the MDI interaction thus allows one to
study, respectively, the effects due to the momentum dependence of
the isoscalar and the isovector part of the single-nucleon
potential. As seen in the right window of Fig.~\ref{mudelta}, the
proton and neutron chemical potential isobars from the eMDYI
interaction with $x=0$ vary asynchronously with pressure, with the
chemical potential of neutrons increases more rapidly with
pressure than that of protons.  As a result, the left (and right)
extrema of $\mu _{n}$ and $\mu _{p}$ correspond to different
values of $\delta $, unlike that for the MDI and MID interactions
which have same extrema for $\mu_n$ and $\mu_p$ as shown in
Fig.~\ref{mudelta}. The relative behavior of neutron and proton
chemical potentials depends, however, on temperature.  For
example, for temperatures not shown here such as $T=5$ MeV, the
chemical potential of neutrons increases more rapidly than that of
protons for $x=0$ but they are reversed for $x=-1$, while at
$T=15$ MeV the asynchronous behavior seems not quite obvious. The
asynchronous variation of the neutron chemical potential relative
to that of protons is uniquely determined by the specific momentum
dependence in the eMDYI interaction within the present
self-consistent thermal model.

At lower pressures, such as $P=0.090$ MeV/fm$^{3}$ as shown in
Fig.~\ref{mudelta} (Panel (a) in the right window), the rectangle
can be accurately constructed and the Gibbs conditions
Eqs.~(\ref{coexistenceP}), (\ref{coexistencemuN}) and
(\ref{coexistencemuP}) thus have two solutions. Due to the
asynchronous variation of $\mu _{n}$ and $\mu _{p}$ with pressure,
there is a limiting pressure $P_{\lim }$ above which no rectangle
can be constructed, so there are no solutions to the coexistence
equation. Panel (b) in the right window of Fig.~\ref{mudelta}
shows the case at the limiting pressure of $P_{\lim }=0.205$
MeV/fm$^{3}$ for $x=0$. In this case, the left side of the
rectangle actually corresponds to the left extremum of $\mu _{p}$.
With increasing pressure, namely, at $P=0.260$, $\mu _{n}$ passes
through an inflection point while $\mu _{p} $ still has a
chemically unstable region, and this case is shown in panel (c) of
the right window in Fig.~\ref{mudelta}. When the pressure is
further increased to $P=0.305$ MeV/fm$^{3}$, as shown in panel (d)
of the left window in Fig.~\ref{mudelta}, $\mu _{p}$ passes
through an inflection point while $\mu _{n}$ increases
monotonically with $\delta $. As mentioned above, the asynchronous
variation of $\mu _{n}$ and $\mu _{p} $ with pressure also depends
on the value of $x$ \cite{Xu07b}.

\subsubsection{The binodal surface}

\begin{figure}[tbh]
\centering
\includegraphics[scale=0.8]{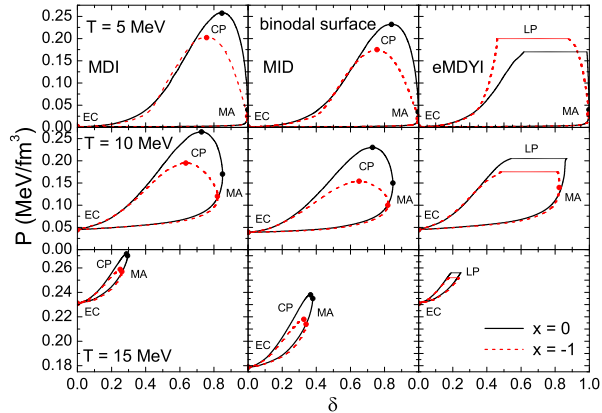}
\caption{{\protect\small (Color online) The binodal surface at
$T=5$, $10$ and $15$ MeV from the MDI, MID and eMDYI interactions
with $x=0$ and $x=-1$. The critical pressure (CP), the limiting
pressure (LP), and the points of equal concentration (EC) and
maximal asymmetry (MA) are also indicated. Taken from Ref.
\cite{Xu07c}.}} \label{Pdelta}
\end{figure}

For each interaction, the two different values of $\delta $
correspond to two different phases with different densities, with
the lower density phase (with larger $\delta $ value) being a gas
phase while the higher density phase (with smaller $\delta $
value) being a liquid phase. Collecting all such pairs of $\delta
(T,P)$ and $\delta ^{\prime }(T,P)$ forms the binodal surface.
Fig.~\ref{Pdelta} displays the binodal surface for the MDI, MID
and eMDYI interactions at $T=5$, $10$ and $15$ MeV with $x=0$ and
$x=-1$ in the $P-\delta $ plain. As expected, for the MDI and MID
interactions the binodal surface has a critical pressure, while
for the eMDYI interaction the binodal surface is cut off by a
limiting pressure. Above the critical pressure or below the
pressure of equal concentration (EC) point, no phase-coexistence
region can exist. The EC point indicates the special case that
symmetric nuclear matter with equal density coexists, which is
called `indifferent equilibrium' \cite{Mul95}. The maximal
asymmetry (MA) also plays an important role in LG phase
transition. The left side of the binodal surface is the region of
liquid phase and the right side the region of gas phase, and
within the surface is the phase-coexistence region.

The critical pressure is sensitive to the stiffness of the
symmetry energy, with a softer symmetry energy (with $x=0$) gives
a higher critical pressure and a larger area of phase-coexistence
region. This is the case for the eMDYI interaction, which has a
limiting pressure, at $T=10$ MeV and $T=15$ MeV. At $T=5$ MeV, the
behavior is reversed, so a softer symmetry energy gives a lower
critical pressure. For the eMDYI interaction, different values of
$x$ give the same EC point. The MDI interaction gives the same EC
point as the eMDYI interaction since the two are the same for
symmetric nuclear matter. For the MID interaction, the EC point
has a lower pressure which further decreases with increasing
temperature. Below the limit pressure, the binodal surface is
quite similar for the MDI and eMDYI interactions. Comparing the
results from the MDI and MID interactions, the isospin and
momentum dependence seems to increase the critical pressure
appreciably. At $T=5$ MeV and $T=10$ MeV, the area of
phase-coexistence region from the MDI interaction is larger than
that from the MID interaction, but at $T=15$ MeV the opposite
result is observed. Although the critical or limiting pressure
seems not to change monotonically with temperature, the area of
phase-coexistence region decreases with increasing temperature
while the pressure at the EC point increases with increasing
temperature. The feature that the gas phase is more neutron-rich
than the coexisting liquid phase leads to the so-called isospin
fractionation phenomenon that has been observed in heavy-ion
reaction experiments, see, e.g., Ref.~\cite{Xu00}.

Corresponding curves for the boundary of phase coexistence (CE)
region in $P-\rho $ plane together with the isothermal spinodals
(ITS) and diffusive spinodals (DS) have been shown in
Fig.~\ref{PrhoT5}. For the MDI and MID interactions, the critical
pressure is the same for the chemical instability and the binodal
surface. For the eMDYI interaction, the phase-coexistence region
can not extend beyond the region of chemical instability as the
binodal surface is cut off by the limit pressure.

\subsubsection{The Maxwell construction}

The binodal surface shown in the previous section provides rich
information about the LG phase transition. As discussed in Ref.
\cite{Mul95}, one can analyze the process of LG phase transition in
hot asymmetric nuclear matter by the Maxwell construction. This will
be discussed below using the MDI interaction with $x=-1$ at $T=10$
MeV as an example.

\begin{figure}[htb]
\centering
\includegraphics[scale=0.7]{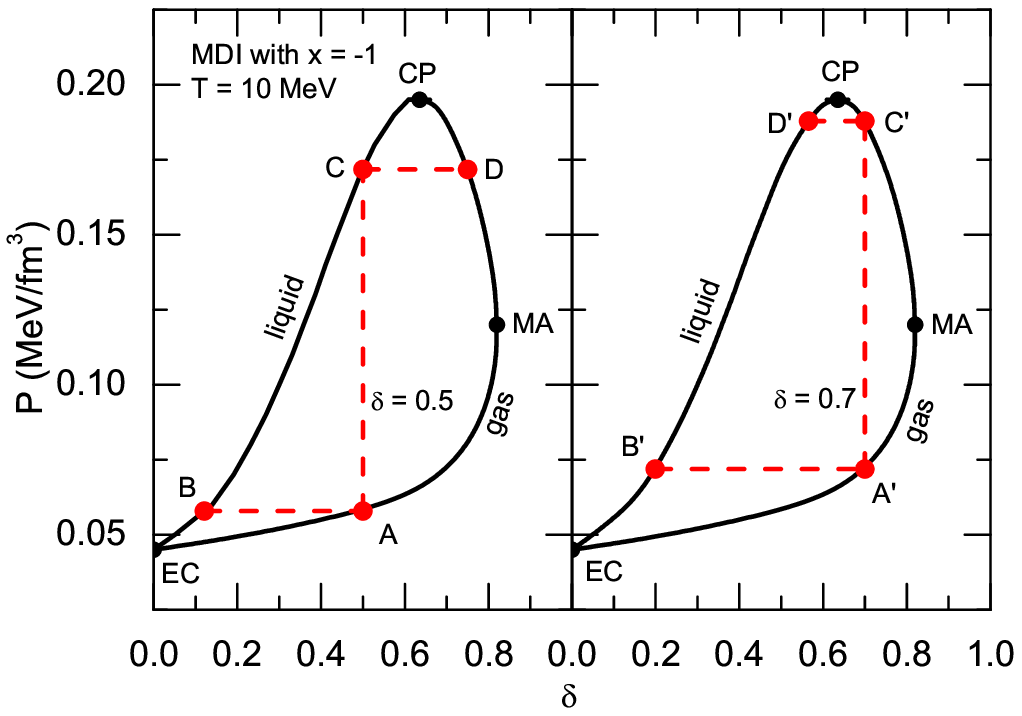}
\includegraphics[scale=0.7]{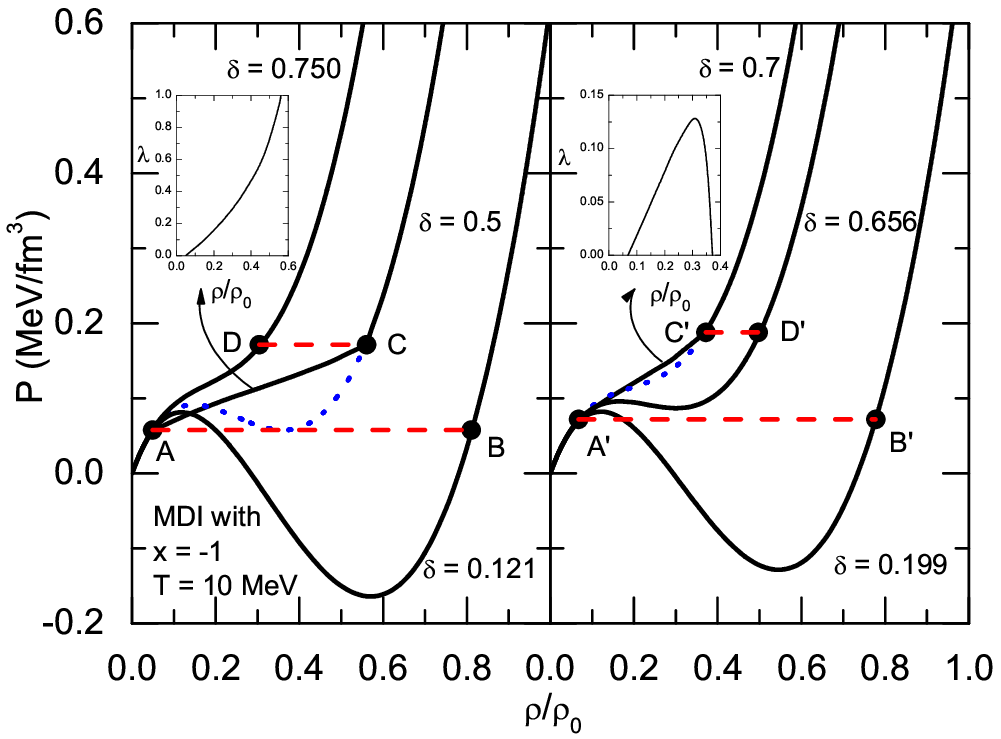}
\caption{(Color online) Left window: The binodal surface at $10$
MeV from the MDI interactions with $x=-1$. The points A
(A$^{\prime }$) through C (C$^{\prime }$) denote phases
participating in a phase transition. The critical pressure (CP) as
well as the points of equal concentration (EC) and maximal
asymmetry (MA) are also indicated. Right window: The LG phase
transition process in the $P-\protect\rho $ plain from the MDI
interaction with $x=-1$ at $T=10$ MeV. The system is initially
compressed at fixed total isospin asymmetry $\protect\delta =0.5$
(left panels) and $\protect\delta =0.7$ (right panels). The
Maxwell construction produces the curve AC (A$^{\prime
}$C$^{\prime }$). The inset displays the fraction of the liquid
phase $\protect\lambda $ from A(A$^{\prime }$) to C(C$^{\prime
}$). Taken from Ref. \cite{Xu07c}.} \label{MaxwellMDI}
\end{figure}

In the left panels of Fig.~\ref{MaxwellMDI}, the compression of
nuclear matter system at a fixed total isospin asymmetry $\delta
=0.5$ is shown. The system begins from the gas phase and enters the
two-phase region at the point A.  A liquid phase with higher density
then emerges from the point B with infinitesimal proportion. As the
system is compressed, the gas phase evolves from A to D, while the
liquid phase evolves from B to C. In this process the gas phase and
the liquid phase coexist and the proportion of each phase changes,
but the total isospin asymmetry is fixed. At the point C the system
totally changes from the gas phase to the liquid phase and leaves
the phase-coexistence region.

This process in the phase-coexistence region can be analyzed
through the isospin and baryon number conservation by solving
following equations
\begin{eqnarray}
\lambda \delta ^{L}\rho ^{L}+(1-\lambda )\delta ^{G}\rho ^{G}
&=&\delta \rho,
\label{maxwell1} \\
\lambda \rho ^{L}+(1-\lambda )\rho ^{G} &=&\rho,  \label{maxwell2}
\end{eqnarray}
where $\delta ^{L(G)}$ and $\rho ^{L(G)}$ are the isospin asymmetry
and density of liquid (gas) phase. The total isospin asymmetry
$\delta $ in this case is $0.5$. The fraction of the liquid phase
$\lambda $ and the total density $\rho $, from which the Maxwell
construction is produced, can be obtained by solving above
equations. The corresponding isotherms are drawn in the left panel
of the right window in Fig.~\ref{MaxwellMDI}. The dotted line
connecting A and C obtained by direct calculation is unphysical. The
nearly straight line connecting A and C is produced by the Maxwell
construction and corresponds to the realistic process. The fraction
of the liquid phase $\lambda $ from A to C is also shown in the
inset, and it changes monotonically from $0$ to $1$.

The geometry of the binodal surface offers a second possibility for
the LG phase transition process. The situation is displayed in the
right panels of Fig.~\ref{MaxwellMDI}, where the system is
compressed at fixed total isospin asymmetry $\delta =0.7$, which is
larger than the isospin asymmetry at the CP point. As in the
previous case, the system begins from the gas phase and enters the
two-phase region at the point A$^{\prime }$, so a liquid phase with
infinitesimal fraction emerges from the point B$^{\prime }$. As the
system is compressed, the gas phase evolves from A$^{\prime }$ to
C$^{\prime }$, while the liquid phase evolves from B$^{\prime }$ to
D$^{\prime }$. The system crosses the phase-coexistence region, but
at the point C$^{\prime }$ it remains in the gas phase and leaves
the binodal surface on the same branch. The corresponding isotherms
are shown in the right panel of the right window in
Fig.~\ref{MaxwellMDI}. The solid line rather than the dotted one
connecting A$^{\prime }$ and C$^{\prime }$ corresponds to the real
process of LG phase transition. In this case the fraction of the
liquid phase $\lambda $ increases from $0$ to $\lambda _{\rm max} $
(about $0.13$) and then drops to $0$ again as shown in the inset in
the right panel of the right window in Fig.~\ref{MaxwellMDI}.

\subsubsection{The order of liquid-gas phase transition in neutron-rich
nuclear matter}

In the following, we consider the order of LG phase transition and
focus on the realistic MDI interaction by observing the behavior of
thermodynamical quantities under a fixed pressure.  The pressure is
taken to be $P=0.05$ MeV$/$fm$^{3}$, but there are no qualitative
changes if other pressures below the critical pressure are used. A
relatively low pressure makes it easier to see more clearly the
effects of the phase transition on the thermodynamical quantities.

\begin{figure}[tbh]
\centering
\includegraphics[scale=0.8]{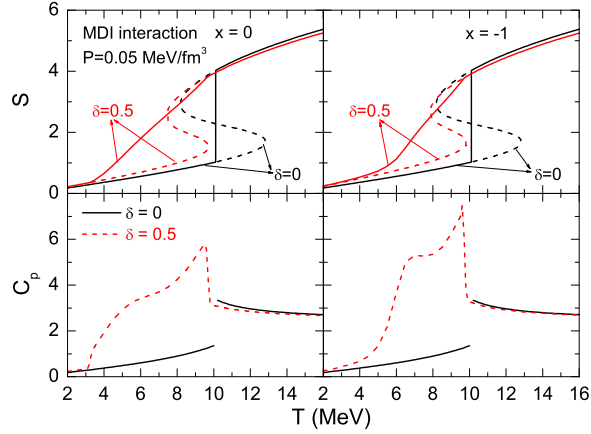}
\caption{{\protect\small (Color online) Evolution of the entropy
per nucleon and the specific heat per nucleon with temperature
under the fixed pressure $0.05$~MeV$/$fm$^{3}$ at $\protect\delta
=0$ and $0.5$ from the MDI interaction with $x=0$ (left panels)
and $x=-1$ (right panels). In the upper panels, the dashed line is
obtained by direct calculation while the solid line is from
Maxwell construction. Taken from Ref. \cite{Xu07c}.}}
\label{SCpTMDIP005}
\end{figure}

In the upper panels of Fig.~\ref{SCpTMDIP005}, the evolution of
entropy per nucleon with temperature at the fixed pressure $0.05$
MeV$/$fm$^{3}$ is shown for isospin asymmetries of $\delta =0$ and
$0.5$ using the MDI interaction with $x=0$ and $x=-1$. The method
of calculating the entropy isobar is similar to that of
calculating the chemical potential isobar discussed above. The
dashed line is obtained by direct calculations and is unphysical
as well known, while the solid line corresponds to the real
process and is obtained by the Maxwell construction. The chemical
potential isobar at every temperature of the phase-coexistence
region under fixed pressure can be calculated, and the densities
and isospin asymmetries of the coexistence phase can be found from
Gibbs conditions. Eqs. (\ref{maxwell1}) and (\ref{maxwell2}) then
allow one to obtain the fraction of each phase. Furthermore, the
total entropy per nucleon $S $ in the coexistence phase can be
calculated from
\begin{eqnarray}
S(\rho ,\delta ,T)=\lambda S^{L}(\rho ^{L},\delta
^{L},T)+(1-\lambda )S^{G}(\rho ^{G},\delta ^{G},T),
\label{entropy}
\end{eqnarray}
where $S^{L(G)}$ are the entropy per nucleon in the liquid or gas
phase and can be obtained from $\rho ^{L(G)}$ and $\delta ^{L(G)}$
by using Eq.~(\ref{S}). From the upper panels in
Fig.~\ref{SCpTMDIP005} one can see that at $\delta =0$ the entropy
jumps at $T=10.1$ MeV, which clearly indicates that the LG phase
transition for symmetric nuclear matter under the pressure of $0.05$
MeV$/$fm$^{3}$ (which is below the critical pressure) is of first
order. The transition temperature in this case is $T_{c}=10.1$ MeV,
and its value depends on the value of the fixed pressure. The curves
with $\delta =0.5$ is, on the other hand, continuous. From Eq.
(\ref{entropy}) for the entropy per nucleon, one can calculate the
heat capacity per nucleon under fixed pressure from
\begin{eqnarray}
C_{p}(T)=T\left( \frac{\partial S}{\partial T}\right) _{P,\delta
}.\label{heat capacity}
\end{eqnarray}
The lower panels in Fig.~\ref{SCpTMDIP005} display the heat
capacity per nucleon as a function of temperature under the fixed
pressure $0.05$ MeV$/$fm$^{3}$ for isospin asymmetries of $\delta
=0$ and $0.5$ using the MDI interaction with $x=0$ and $x=-1$,
respectively. For both cases of $x=0$ and $x=-1$ at $\delta =0.5$,
the heat capacity is continuous but not its first derivative,
which indicates that the LG phase transition for asymmetric
nuclear matter is of second order according to Ehrenfest's
definition of phase transitions \cite{Rei80}. Similar results are
obtained in Ref. \cite{Mul95} with a different model. Although the
discussions here are based on the MDI interaction, the order of
the LG phase transition will not depend on the isospin and
momentum dependence of the nuclear interaction.

\subsection{Evolution of the symmetry energy of hot neutron-rich nuclear matter
formed in heavy-ion reactions}

\subsubsection{Nuclear symmetry free energy at finite temperature}

\begin{figure}[tbh]
\centering
\includegraphics[scale=0.8]{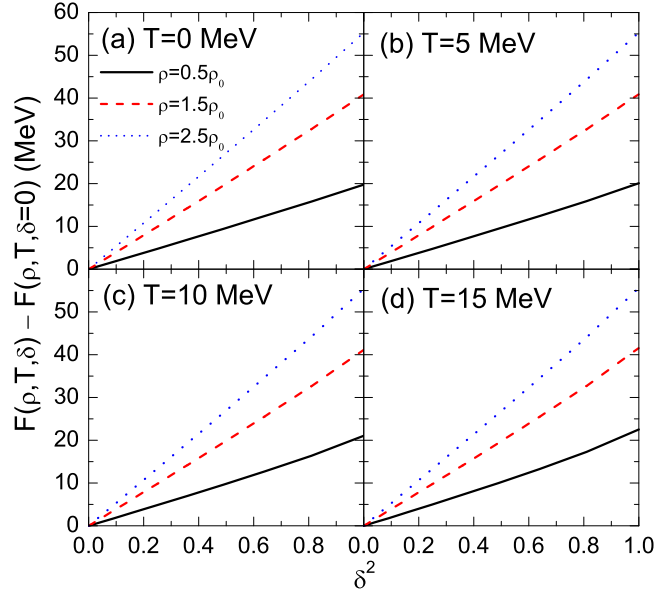}
\caption{{\protect\small (Color online) The free energy per
nucleon } $F(\protect\rho ,T,\protect\delta )$ as a function of
isospin asymmetry $\delta$ from the MDI interaction with $x=0$ for
different temperatures and densities. Taken from Ref.
\cite{Xu07}.} \label{FsymParaMDI0}
\end{figure}

Similar to the nuclear symmetry energy, the symmetry free energy
$F_{\rm sym}(\rho ,T)$ can be defined by the following parabolic
approximation to the free energy per nucleon~\cite{Xu07}
\begin{eqnarray}
F(\rho ,T,\delta )=F(\rho ,T,\delta =0)+F_{\rm sym}(\rho ,T)\delta
^{2}+\mathcal{O}(\delta ^{4}).  \label{eosF}
\end{eqnarray}
The temperature and density dependent symmetry free energy $F_{\rm
sym}(\rho ,T)$ for hot neutron-rich matter can thus be extracted
from $F_{\rm sym}(\rho ,T)\approx F(\rho ,T,\delta =1)-F(\rho
,T,\delta =0)$, which is just the free energy cost to convert all
protons in symmetry matter to neutrons at the fixed temperature $T$
and density $\rho $. The validity of the empirical parabolic law for
the free energy per nucleon of hot neutron-rich matter can be seen
from Fig. \ref{FsymParaMDI0} where $F(\rho ,T,\delta )-F(\rho
,T,\delta =0)$ is shown as a function of $\delta ^{2}$ at
temperature $T=0$ MeV, $5$ MeV, $10$ MeV and $15$ MeV for three
different baryon number densities $\rho =0.5\rho _{0},1.5\rho _{0}$
and $2.5\rho _{0}$ using the MDI interaction with $x=0$. One can see
that the parabolic law Eq. (\ref{eosF}) is approximately satisfied,
although it is slightly violated at low densities and high
temperatures. A similar conclusion has been obtained for the
parameter $x=-1$.

\begin{figure}[htb]
\centering
\includegraphics[scale=0.75]{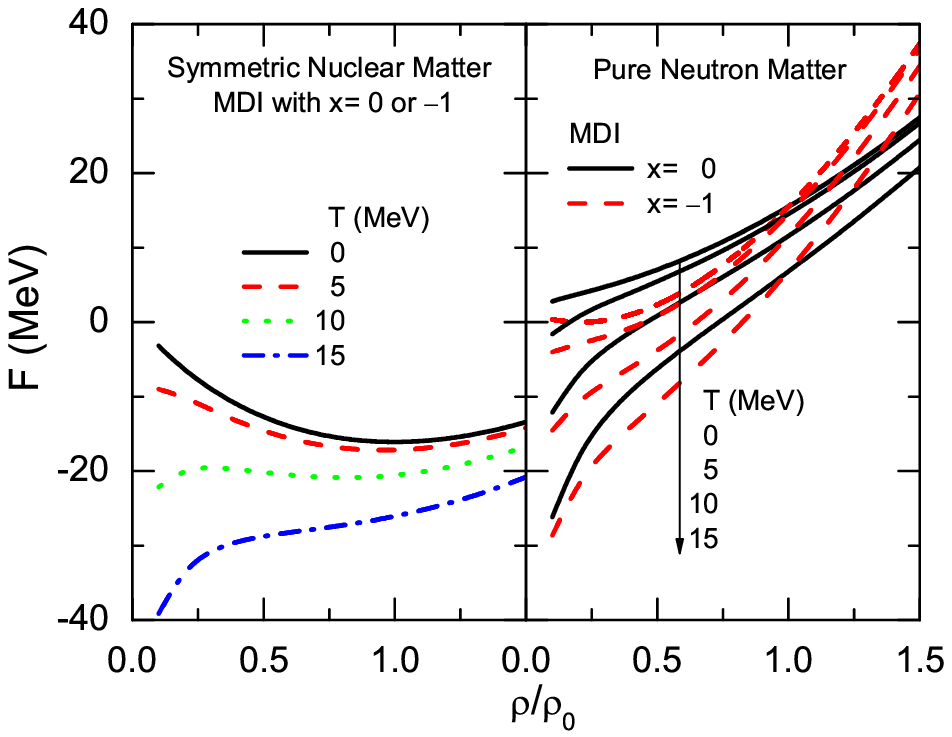}
\includegraphics[scale=0.75]{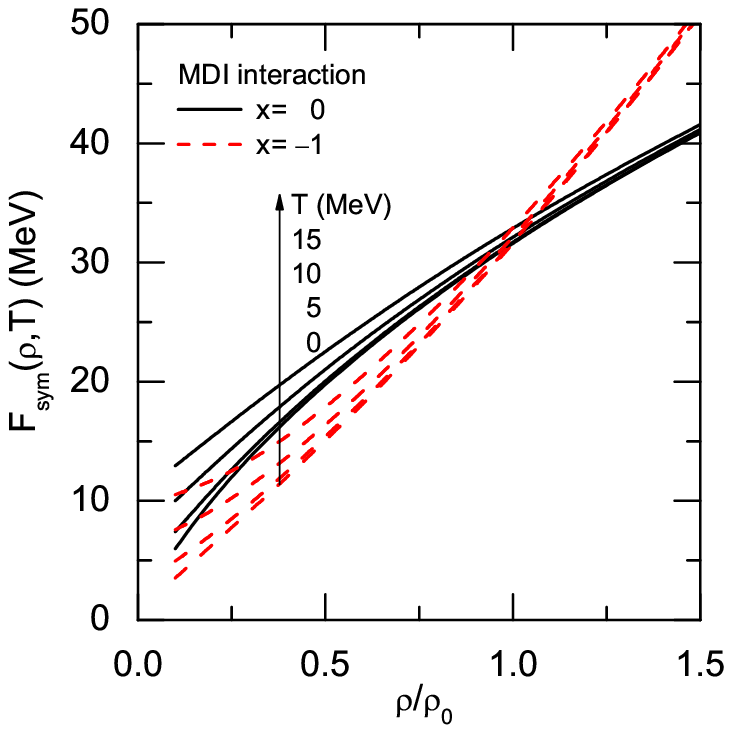}
\caption{Density dependence of the free energy per nucleon (left
window) and the symmetry free energy $F_{\rm sym}(\protect\rho
,T)$ (right window) for symmetric nuclear matter and pure neutron
matter at $T=0$, $5$, $10$ and $15$ MeV from the MDI interaction.
Data are taken from Ref. \cite{Xu07}.} \label{FsymDen}
\end{figure}

Shown in the left window of Fig. \ref{FsymDen} is the density
dependence of $ F(\rho ,T,\delta )$ for symmetric nuclear matter
and pure neutron matter at $T=0$ MeV, $5$ MeV, $10$ MeV and $15$
MeV using the MDI interaction with $x=0$ and $-1$. For symmetric
nuclear matter ($\delta =0$), the parameter $x=0$ gives same
results as the parameter $x=-1$ as discussed before and the curves
shown in the figure are thus the same for $x=0$ and $-1$. The free
energy per nucleon $F(\rho ,T,\delta )$ is seen to decrease with
increasing $T$, and this is opposite to the behavior of the energy
per nucleon $E(\rho ,T,\delta )$, which increases with temperature
as a result of the thermal excitation of the nuclear matter. The
decrease of the free energy per nucleon $F(\rho ,T,\delta )$ with
$T$ is mainly due to the increase of the entropy per nucleon with
increasing temperature. This feature also implies that the
increase of $TS(\rho ,T)$ with $T$ is larger than the increase of
$E(\rho ,T)$ with $T$. Furthermore, the temperature effects are
stronger at lower densities and become much weaker at higher
densities. At lower densities, the Fermi momentum $p_{f}(\tau )$
is smaller and thus temperature effects on the energy per nucleon
$E(\rho ,T,\delta )$ are expected to be stronger. On the other
hand, the entropy per nucleon becomes larger at lower densities
where the particles become more free in phase space and thus leads
to a smaller free energy per nucleon. For pure neutron matter, the
parameters $x=0$ and $-1$ give different density dependence for
the free energy per nucleon $F(\rho ,T,\delta )$, which just
reflects the fact that the symmetry free energy obtained from the
parameters have different density dependence. As shown in the
right window of Fig. \ref{FsymDen}, the symmetry free energy
$F_{\rm sym}(\rho ,T)$ displays different behaviors in the density
dependence with $x=0$ ($-1 $) giving larger (smaller) values for
the symmetry free energy at lower densities while smaller (larger)
ones at higher densities for a fixed temperature. Similar to the
$F(\rho ,T,\delta )$, the temperature effects on the symmetry free
energy $F_{\rm sym}(\rho ,T)$ are stronger at lower densities but
become much weaker at higher densities.

The temperature dependence of the symmetry free energy is, on the
other hand, different. As shown in the right window of Fig.
\ref{FsymDen}, the symmetry free energy $F_{\rm sym}(\rho ,T)$
increases with temperature, which is opposite to the case of
symmetry energy shown in the left window of Fig.~\ref{Esym}. This
also means that $F_{\rm sym}(\rho ,T)$ always has a larger value
than $E_{\rm sym}(\rho ,T) $ at fixed density and temperature
since they are identical at zero temperature. At higher
temperatures, one expects the symmetry energy $E_{\rm sym}(\rho
,T)$ to decrease as the Pauli blocking (a pure quantum effect)
becomes less important when the nucleon Fermi surfaces become more
diffused at increasingly higher temperatures
\cite{Xu07,Xu07b,Che01b,Zuo03,LiBA06c,Mou07}. Since the symmetry
free energy $F_{\rm sym}(\rho ,T)$ is related to the entropy per
nucleon of asymmetric nuclear matter, its increase with increasing
temperature can thus be understood from the following relation
\begin{eqnarray}
F_{\rm sym}(\rho ,T) &=&E_{\rm sym}(\rho ,T)+T\left[ S_{n}(\rho
,T,\delta =0)+S_{p}(\rho ,T,\delta =0)\right]-TS_{n}(\rho ,T,\delta
=1). \notag \\ \label{Fsym}
\end{eqnarray}
The first term of the right hand side in Eq. (\ref{Fsym}) is the
symmetry energy $E_{\rm sym}(\rho ,T)$, which decreases with
increasing temperature as discussed above. Since the total entropy
per nucleon of the symmetric nuclear matter is larger than that of
the pure neutron matter and their difference becomes larger with
increasing temperature, the difference between the last two terms of
the right hand side in Eq. (\ref{Fsym}) is positive. Therefore,
$F_{\rm sym}(\rho ,T)$ has a larger values than $E_{\rm sym}(\rho
,T)$ at fixed density and temperature. Furthermore, the increase of
$TS(\rho ,T)$ with $T$ is stronger than the increase of $E(\rho ,T)$
with $T$, so the combined effects cause the symmetry free energy
$F_{\rm sym}(\rho ,T)$ to increase with increasing temperature.

\subsubsection{Evolution of the symmetry energy observed in the
isoscaling analysis of heavy-ion collisions}

It has been observed experimentally and also theoretically in many
types of reactions that the ratio $R_{21}(N,Z)$ of the yields of a
fragment with proton number $Z$ and neutron number $N$ from two
reactions reaching about the same temperature $T$ satisfies an
exponential relation $R_{21}(N,Z)\propto \exp (\alpha N)$
\cite{Tsa01,Ono03,Ono04,Tia05,Tsa01b,She04,She05,She06,Sou06,Igl06,%
Fev05,Tra06,Kow07,Bot02,Dor06,Ma04b,Ma05,Rad07,Cha08}%
. In particular, several statistical and dynamical models
\cite{Ono03,Ono04,Tsa01b,Bot02} have shown, under some
assumptions, that the scaling coefficient $\alpha $ is related to
the symmetry energy $C_{\rm sym}(\rho ,T)$ via
\begin{eqnarray}
\alpha =\frac{4C_{\rm sym}(\rho ,T)}{T}\bigtriangleup \lbrack
(Z/A)^{2}], \label{scaling}
\end{eqnarray}
where $\bigtriangleup \lbrack (Z/A)^{2}]\equiv
(Z_{1}/A_{1})^{2}-(Z_{2}/A_{2})^{2}$ is the difference between the
$(Z/A)^{2} $ values of the two fragmenting sources created in the
two reactions.

As mentioned in Ref. \cite{LiBA06c}, because of the different
assumptions used in the various derivations, the validity of Eq.
(\ref{scaling}) is still disputable as to whether and when the
$C_{\rm sym}$ is actually the symmetry energy or the symmetry free
energy. Moreover, the physical interpretation of the $C_{\rm
sym}(\rho ,T)$ is also not clear, sometimes even contradictory, in
the literature. The main issue is whether the $C_{\rm sym}$
measures the symmetry (free) energy of the fragmenting source or
that of the fragments formed at freeze-out. This ambiguity is also
due to the fact that the derivation of Eq.~(\ref{scaling}) is not
unique. In particular, within the grand canonical statistical
model for multifragmentation \cite{Tsa01b,Bot02} the $C_{\rm sym}$
refers to the symmetry energy of primary fragments, while within
the sequential Weisskopf model in the grand canonical limit
\cite{Tsa01b} it refers to the symmetry energy of the emission
source. Very recently, referring the $C_{\rm sym}$ as the symmetry
energy of nuclear matter of the fragmenting source, Chaudhuri {\it
et al}. investigated the validity of Eq.~(\ref{scaling}) within
several models including a mean-field model and thermodynamic
models using both grand canonical and canonical ensembles
\cite{Cha08}. In particular, they have studied the conditions
leading to the observed deviations from the isoscaling behavior of
emitted fragments.

\begin{figure}[tbh]
\centering
\includegraphics[scale=0.85]{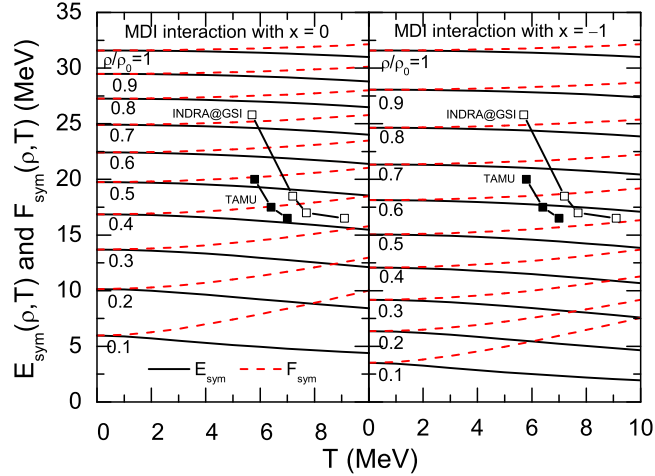}
\caption{{\protect\small (Color online) Temperature dependence of
the symmetry energy (solid lines) and symmetry free energy (dashed
lines) from the MDI interaction with }$x=0${\protect\small \ (left
panel) and }$-1${\protect\small \ (right panel) at different
densities from }$0.1\protect\rho _{0}${\protect\small \ to
}$\protect\rho _{0}${\protect\small . The experimental data from
Texas A}$\&${\protect\small M University (solid squares) and the
INDRA-ALADIN collaboration at GSI (open squares) are included for
comparison. Taken from Ref. \cite{Xu07}.}} \label{IsoScale}
\end{figure}

Fig.~\ref{IsoScale} shows the symmetry energy $E_{\rm sym}(\rho
,T)$ and the symmetry free energy $F_{\rm sym}(\rho ,T)$ as
functions of temperature using the MDI interaction with $x=0$ and
$-1$ at different densities from $0.1\rho _{0} $ to $\rho _{0}$.
It is seen that at a given density the symmetry energy does not
change much with temperature , especially around the saturation
density $\rho _{0}$. Furthermore, while the symmetry energy
$E_{\rm sym}(\rho ,T)$ at a given density deceases slightly with
increasing temperature, the symmetry free energy $F_{\rm sym}(\rho
,T)$ increases instead. Compared to their values at $T=0$ MeV, the
difference between the symmetry energy $E_{\rm sym}(\rho ,T)$ and
the symmetry free energy $F_{\rm sym}(\rho ,T)$ around the
saturation density $\rho _{0}$ is quite small, i.e., only a few
percent, even at temperatures up to $10$ MeV. This feature
confirms the assumption of identifying $C_{\rm sym}(\rho ,T)$ to
$E_{\rm sym}(\rho ,T)$ at low temperatures and not so low
densities \cite{She04,She05,She06,Sou06,Igl06}. At low densities,
on the other hand, the symmetry free energy $F_{\rm sym}(\rho ,T)$
exhibits a stronger temperature dependence and is significantly
larger than the symmetry energy $E_{\rm sym}(\rho ,T)$ at moderate
and high temperatures. This is due to the large entropy
contribution to the symmetry free energy $F_{\rm sym}(\rho ,T)$ at
low densities. The above results are based on the mean-field model
and thus neglect the clustering of nucleons in asymmetric nuclear
matter. At low densities, such clustering effects may affect
strongly the entropy of nuclear matter and thus the nuclear
symmetry energy as well as the symmetry free energy
\cite{Kow07,Hor06}, which will be discussed in the following.

Experimentally, the temperature $T$ and the scaling coefficient
$\alpha $ (thus the $C_{\rm sym}$) can be directly measured while
the determination of the freeze-out density of fragments usually
depends on the model used. Assuming the experimentally extracted
$C_{\rm sym}$ is the symmetry energy of the nuclear matter in the
fragmenting source, one can compare the calculations with the
experimental data in Fig.~\ref{IsoScale}. The solid squares are
the $C_{\rm sym}$ from Yennello's group at Texas A\&M University
(TAMU) \cite{She06,Sou06} and the open squares are from the
INDRA-ALADIN collaboration at GSI \cite{Fev05,Tra06}. It is
clearly seen that the experimentally observed evolution of the
symmetry energy is mainly due to the change in density rather than
temperature. This conclusion was first obtained in
Ref.~\cite{LiBA06c} using a simple Fermi gas model and was later
confirmed by more realistic models \cite{Xu07,Mou07,Sam07}.

One can estimate from Fig.~\ref{IsoScale} the average freeze-out
density of the fragment emission source from the measured
temperature-dependent symmetry energy based on the isotopic
scaling analysis in heavy-ion collisions. In particular, using the
symmetry energy $E_{\rm sym}(\rho ,T)$ from the MDI interaction
with $x=0$, one finds that the average freeze-out density of the
fragment emission source $\rho _{f}$ is between about $0.41\rho
_{0}$ and $0.52\rho _{0}$ for the TAMU data while about $0.42\rho
_{0}$ and $0.75\rho _{0}$ for the INDRA-ALADIN collaboration data.
These values are very similar to those extracted in Ref.
\cite{She06} using different models. On the other hand, using the
symmetry energy $E_{\rm sym}(\rho ,T)$ from the MDI interaction
with $x=-1$, the $\rho _{f}$ is found to be between about
$0.57\rho _{0}$ and $0.68\rho _{0}$ for the TAMU data while about
$0.58\rho _{0}$ and $0.84\rho _{0}$ for the INDRA-ALADIN
collaboration data. If the symmetry free energy $F_{\rm sym}(\rho
,T)$ from the MDI interaction with $x=0$ is used instead to
estimate the $\rho _{f}$, one finds that the $\rho _{f}$ would be
between about $0.36\rho _{0}$ and $0.49\rho _{0}$ for the TAMU
data and about $0.33\rho _{0}$ and $0.72\rho _{0}$ for the
INDRA-ALADIN collaboration data. Using the symmetry free energy
$F_{\rm sym}(\rho ,T)$ from the MDI interaction with $x=-1$, the
$\rho _{f}$ would, on the other hand, be between about $0.52\rho
_{0}$ and $0.66\rho _{0}$ for the TAMU data and about $0.51\rho
_{0}$ and $0.83\rho _{0}$ for the INDRA-ALADIN collaboration data.
Therefore, taking the measured $C_{\rm sym}(\rho ,T)$ as either
the symmetry energy or the symmetry free energy does not affect
much the extracted $\rho _{f}$ values. The extracted $\rho _{f}$
values are, however, sensitive to the $x$ parameter used in the
MDI interaction, namely, the density dependence of the symmetry
energy. Therefore, the isotopic scaling measurement in heavy-ion
collisions provides a potentially viable probe for the density
dependence of the nuclear matter symmetry energy.

If the $C_{\rm sym}$ is taken to be the symmetry energy of the hot
fragments at freeze-out, Fig.~\ref{IsoScale} then shows that the
experimentally measured values are smaller than that for a cold
nuclear matter at saturation density, which is about 30 MeV.
Several explanations have been given in the literature for this
small values of $C_{\rm sym}$. Among them are the finite size
effects on the symmetry energy and the temperature of the
fragments. However, from the very nature of the isoscaling
phenomenon that isotopes/isotones having very different masses
(sizes) fall on the same curve described by a single scaling
coefficient $\alpha$, one has to assume that the finite size
effects on both the $C_{\rm sym}$ and the temperature $T$
completely cancel with each other. Otherwise the isoscaling
phenomenon would not have been observed. Another possible reason
for the extracted small value of $C_{\rm sym}$ is that hot
fragments themselves at the so-called `{\it freeze-out}' are
dilute due to the strong and Coulomb interactions with surrounding
nuclei \cite{Buy08}. This picture, however, seems to contradict
the basic Fisher hypothesis that the only correlations inside a
dilute medium are those due to clusterization. Moreover, the
isoscaling phenomenon has actually been observed experimentally
for cold fragments. The sequential decay of hot primary fragments
thus may not affect much the isoscaling coefficient, although this
is still a question under debate as it depends on the model
calculations \cite{Ono05,she06a}. With this view the small value
of $C_{\rm sym}$ for cold fragments extracted in the isoscaling
experiments would therefore indicate that the fragments have
dilute internal density, and this would require re-considerations
of the statistical models from which Eq. (\ref{scaling}) was
derived. Furthermore, comparing the TAMU and the INDRA-ALADIN
data, one sees that they are actually parallel to each other in
the common density range. Since the evolution of the symmetry
energy is essentially independent of temperature for the
experiments considered, the two sets of data thus indicate the
same density dependence of the symmetry energy. On the other hand,
within the view that the extracted $C_{\rm sym}$ reflects the
symmetry energy of the fragments at freeze-out, the two sets of
data are incompatible.

\subsubsection{The symmetry energy of very dilute but hot $np\alpha$ matter}

\begin{figure}[h]
\centering
\includegraphics[scale=0.32,clip=]{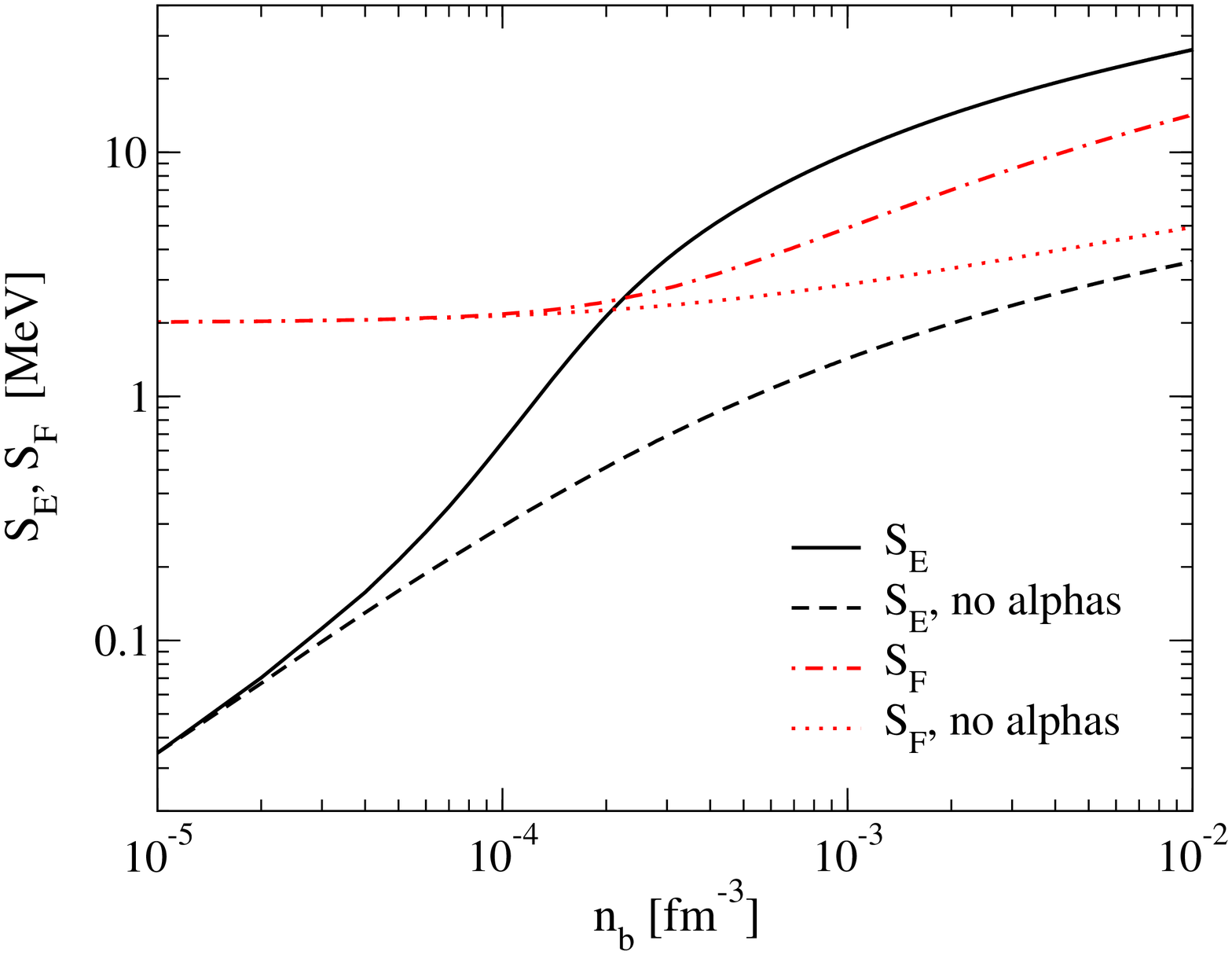}
\includegraphics[scale=0.32,clip=]{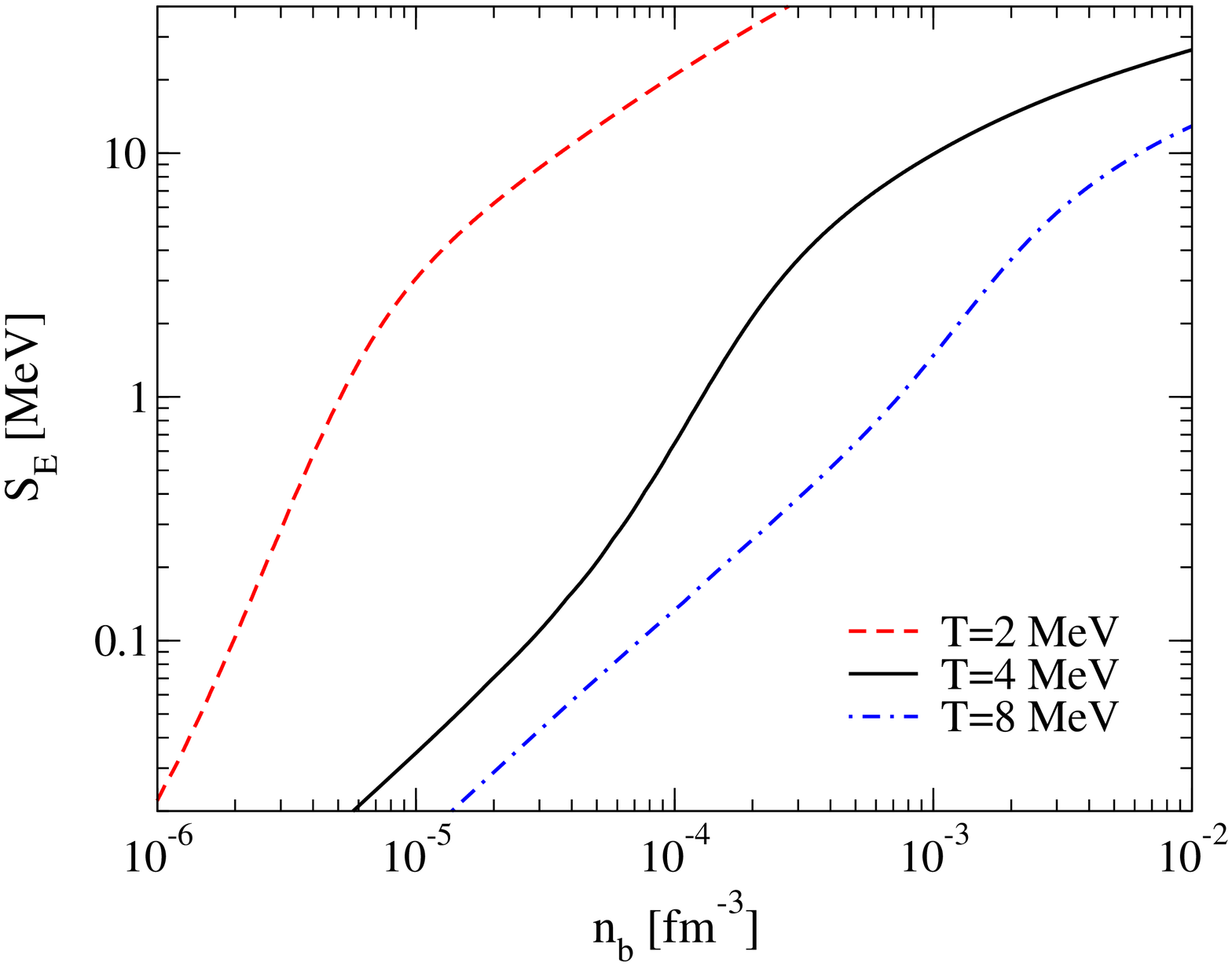}
\caption{(Color online) Left window: The virial symmetry energy
$S_E$ and the symmetry free energy $S_F$ versus baryon density
$n_b$ for $T=4$ MeV. Also shown are virial results without
$\alpha$ particles. Right window: The virial symmetry energy $S_E$
versus baryon density $n_b$ for $T=2$, $4$ and $8$ MeV. Taken from
Ref.~\cite{Hor06}.} \label{HorFig16}
\end{figure}

The above discussions are based on the properties of uniform
nucleonic matter. At very low densities, the ground state of nuclear
matter is expected to consist of light clusters. Using the virial
expansion and realistic $NN$, $N\alpha$ and $\alpha\alpha$ elastic
scattering phase shifts, Horowitz and Schwenk have recently
evaluated the symmetry energy and the symmetry free energy of very
dilute $np\alpha$ matter \cite{Hor06}. The left window of
Fig.~\ref{HorFig16} shows their calculations for the symmetry energy
$S_E$ and the symmetry free energy $S_F$ for $T=4$ MeV.  At very low
density, $S_E$ rises slowly with density. As $\alpha$ particles are
formed, both $S_E$ and $S_F$ rise much faster with density. As a
result of clustering, the symmetry energy is large even at a very
small fraction of saturation density. The symmetry energy remains to
decrease with increasing temperature as shown in the right window of
Fig.~\ref{HorFig16}, consistent with that observed for the uniform
nucleonic matter in other approaches as discussed previously.

\begin{figure}[h]
\centering
\includegraphics[scale=0.38]{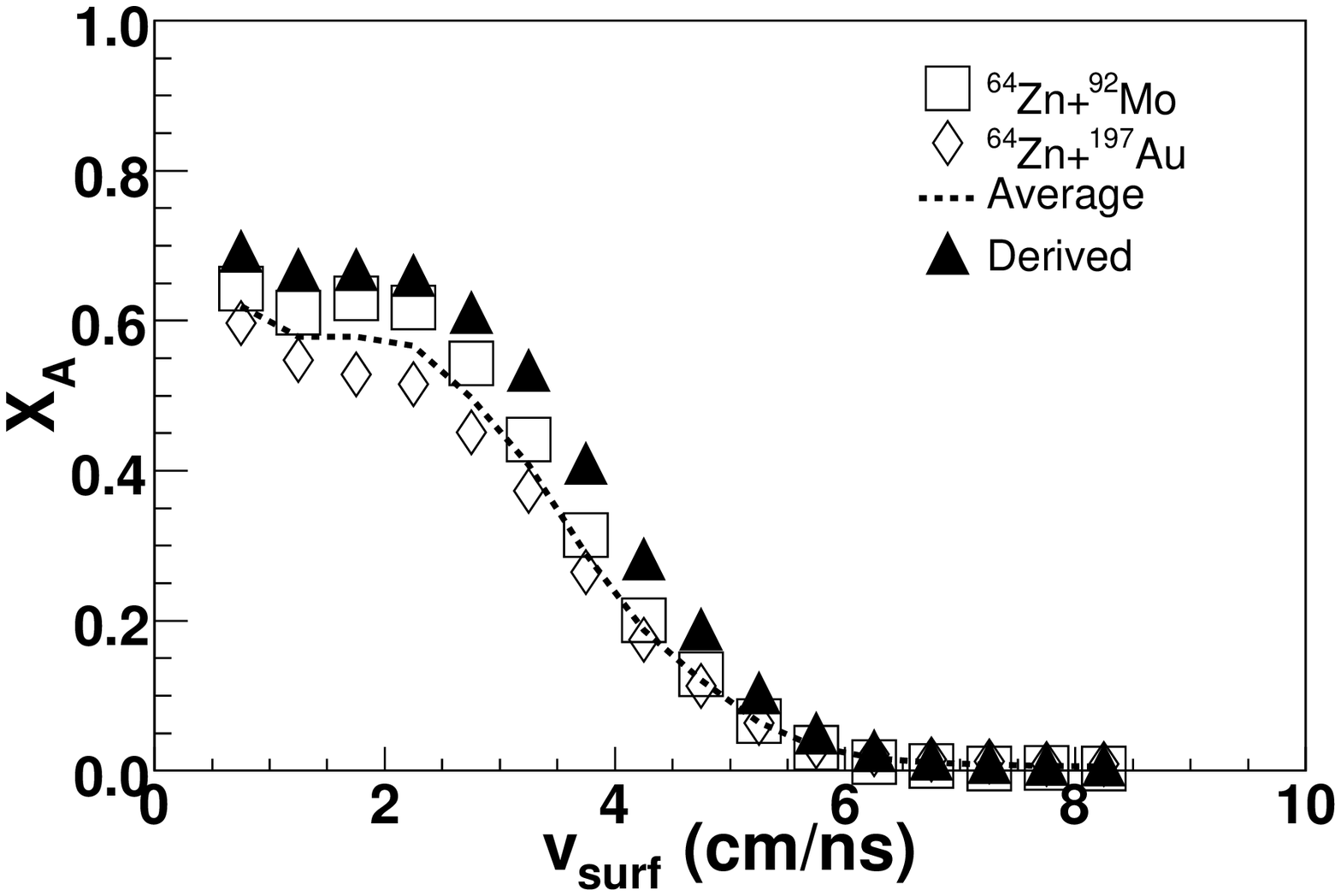}
\includegraphics[scale=0.38]{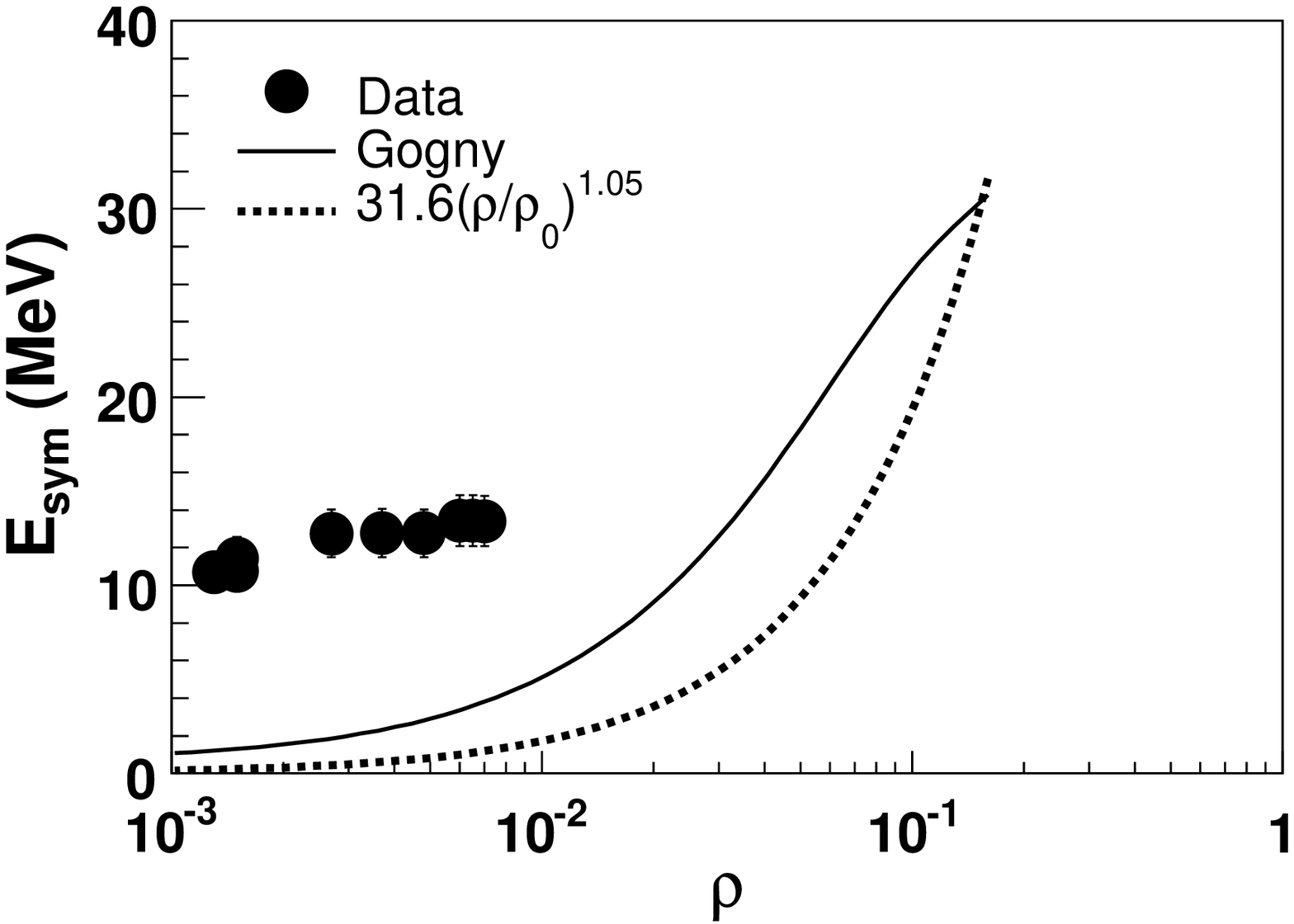}
\caption{(Color online) Left window: Alpha mass fractions,
X$_{A}$, as a function of surface velocity for 35 MeV/nucleon
$^{64}$Zn + $^{92}$Mo (squares) and $^{64}$Zn +$^{197}$Au
(diamonds) collisions. Also shown by the dashed line is the
average of the two. Right window: Derived symmetry energy
coefficients as a function of baryon density. Taken from
Ref.~\cite{Kow07}.} \label{Joefig}
\end{figure}

Very recently, Natowitz and his collabortors \cite{Kow07} have
extracted the symmetry energy at $0.01\rho_0$ to $0.05\rho_0$ from
analyzing the moderate temperature nuclear gases produced in the
violent collisions of 35 MeV/nucleon $^{64}$Zn projectiles with
$^{92}$Mo and $^{197}$Au target nuclei. Indeed, a large degree of
alpha particle clustering at these densities were revealed. Shown on
the left window of Fig.~\ref{Joefig} are the alpha mass fractions
X$_{A}$ for the intermediate velocity source ejectiles of both
systems.  For both colliding systems, X$_{A}$ evolves in a similar
fashion with surface velocity.  As the surface velocity decreases,
X$_{A}$ increases dramatically with a smaller X$_{A}$ for the more
neutron rich $^{64}$Zn +$^{197}$Au entrance channel. The symmetry
energy was also extracted from the isoscaling analyses. As shown by
the filled circles on the right window of Fig.~\ref{Joefig}, the
extracted symmetry energy at low densities are consistent with the
calculations by Horowitz and Schwenk \cite{Hor06} shown in
Fig.~\ref{HorFig16}. Shown in the same figure are comparisons with
the HF calculations for uniform nuclear matter with the Gogny
effective interaction. The function $31.6(\rho/\rho_0)^{1.05}$,
which corresponds to the lower boundary of the symmetry energy at
subsaturation density suggested by the analysis of the isospin
diffusion data \cite{Che05a}, is also shown. The derived values of
E$_{\rm sym}$ are much higher than those predicted by mean-field
calculations which ignore the cluster formation. While these
comparisons are useful, it is worth pointing out that the density
range explored by the isospin diffusion data is significantly higher
than $0.05\rho_0$.


\section{Isospin dependence of nucleon-nucleon cross sections in neutron-rich
medium}
\label{chapter_crosssection}

The isospin dependence of in-medium nuclear effective interactions
determines both the EOS, especially the nuclear symmetry energy,
and the transport properties of isospin asymmetric nuclear matter
\cite{DOE07}. All of these quantities are important, albeit at
different degrees, for determining the nature of nucleonic matter,
novel structures of radioactive nuclei, isospin-related phenomena
in heavy-ion reactions, properties of neutron stars, and the
mechanisms of supernova explosions
\cite{Lat04,Ste05a,Lat00,Lat01}. While much attention has been
given to finding experimental observables that can constrain the
EOS of isospin asymmetric nuclear matter, little effort has been
made so far to extract the isospin dependence (i.e., the ratio of
np to pp (nn) corss sections) of the in-medium NN cross sections.
The in-medium NN cross sections depend particularly on the
short-range part of nuclear effective interactions. They affect
the transport properties of isospin asymmetric nuclear matter
\cite{Che01} and are important for studying the structure of rare
isotopes. For instance, in the Glauber model used for extracting
information about the structure of radioactive nuclei, such as the
radii and the distributions of the constituent neutrons and
protons, the total interaction cross section in the optical limit
is determined by a transmission function, which is a convolution
of the NN cross section and the density distribution of nucleons
from both the target and projectile in the overlapping region.
Knowing the proton density distribution (which can be determined
from other means such as the electron scattering) and the NN cross
sections, the neutron density distribution can then be determined.
Usually, only the isospin averaged free NN cross section is used
as the input to the Glauber model. Effects of including the
isospin dependence of the in-medium NN cross section need to be
studied. Also, the isospin-dependent in-medium NN cross sections
are needed in transport models to extract more reliably the
density dependence of the nuclear symmetry energy. It is also
useful for understanding the isospin-dependent phenomena in
heavy-ion collisions.

In this Chapter, after recalling the isospin dependence of the
free-space NN cross sections, we review recent theoretical studies
on the isospin dependence of the in-medium NN cross sections and
discuss its determination from experiments.

\subsection{Isospin dependence of the free-space NN cross sections}

\begin{figure}[tbh]
\centering
\includegraphics[height=0.45\textheight,angle=90]{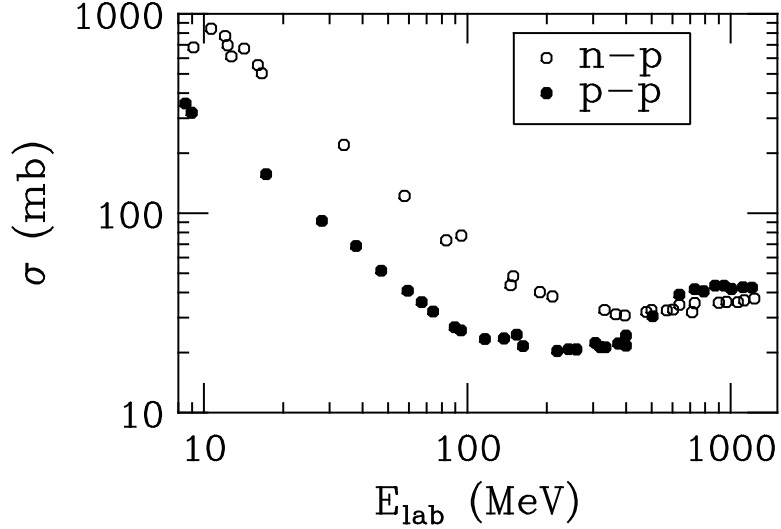}
\caption{Cross sections of neutron-proton and proton-proton
scatterings as functions of bombarding energy. Taken from Ref.\
\protect\cite{Alk77}.} \label{freesigma}
\end{figure}

It is well-known that the scattering cross section between two
nucleons depends on their isospin. Fig.\ \ref{freesigma} compares
the free-space cross sections for neutron-proton and proton-proton
or neutron-neutron scattering as functions of bombarding energy.
The data in the energy range of 10 MeV $\leq E_{\rm lab} \leq 1000
$ MeV can be parameterized by \cite{Cha90,Alk77}
\begin{eqnarray}
\sigma_{np}^{\rm free}&=&-70.67-18.18\beta^{-1}+25.26\beta^{-2}+113.85
\beta~({\rm mb}),\\
\sigma_{pp}^{\rm free}&=&13.73-15.04\beta^{-1}+8.76\beta^{-2}+68.67
\beta^4~({\rm mb}),
\end{eqnarray}
where $\beta\equiv v/c$ is the velocity of the projectile nucleon.

Because of the differences in the transition matrices of the isospin
$T=1$ and $T=0$ channeles, and the fact that both the iso-singlet
and iso-triplet channels contribute to neutron-proton (np)
scatterings, their cross sections ($\sigma_{np}^{\rm free}$) in free
space are higher than those for proton-proton (pp) or
neutron-neutron (nn) scatterings ($\sigma_{pp}^{\rm free}$) where
only iso-triplet channels are involved. It is seen that below about
500 MeV the neutron-proton cross section is about a factor of 2 to 3
larger than the proton-proton or neutron-neutron cross section.

\subsection{Theoretical predictions and experimental
information on NN cross sections in symmetric nuclear matter}

Theoretical studies of in-medium NN cross sections have been
carried out by many people, see, e.g.,
Refs.~\cite{Cug87,Har87,Ber88a,Boh89,Fas89,Koh91,Li93,Alm1,Alm2,Mao94,Gia96,Sch97,Koh98,Dic99,LiQF00,Fuc01,Gai05,Sam05b,Zhang07}.
Most of theses studies are carried out for symmetric nuclear
matter at zero temperature, and the results vary considerably. As
discussed in Chapter~\ref{chapter_eos},in microscopic models
medium effects appear in the Bethe-Goldstone equation mainly
through the Pauli blocking factor for intermediate states and the
self-energies of the two nucleons in the denominator of the
propagator. However, results from these studies differ
significantly, with some models predicting a decrease of the
in-medium NN cross sections compared to their free-space values
while others predict an increase. For instance, in the
Dirac-Brueckner approach of Refs.~\cite{Har87,Li93}, in which the
model parameters are fixed by fitting free-space NN scattering
data and deuteron properties, the NN cross sections in nuclear
medium at zero temperature have been predicted to decrease with
increasing density. For example, at the normal nuclear matter
density and a bombarding energy of 50 MeV, both $\sigma_{np}$ and
$\sigma_{pp}$ are reduced by about a factor of two. Results of the
calculations in Ref.~\cite{Li93} has been parameterized by
\begin{eqnarray}
\sigma_{np}^{\rm medium}&=&\left[31.5+0.092abs(20.2-E_{\rm
lab}^{0.53})^{2.9} \right]\cdot \frac{1.0+0.0034E_{\rm
lab}^{1.51}\rho^2}{1.0+21.55\rho^{1.34}}~({\rm mb}),\\
\sigma_{pp}^{\rm medium}&=&\left[23.5+0.0256(18.2-E_{\rm
lab}^{0.5})^{4} \right]\cdot \frac{1.0+0.1667E_{\rm
lab}^{1.05}\rho^3}{1.0+9.704\rho^{1.2}}~({\rm mb}).
\end{eqnarray}
In this study, respective effects of the Pauli blocking and the
self-energy corrections on the in-medium NN cross sections have not
been carried out \cite{Li93}. Opposite results have been found by
Bohnet {\it et al.}~\cite{Boh89} in studying the in-medium NN cross
sections during the collisions of two slabs of nuclear matter at
zero temperature. The Pauli blocking factor has been estimated using
two Fermi spheres separated by the beam momentum, and it is found
that the in-medium cross section generally increases with density.

\begin{figure}[tbh]
\centering
\includegraphics[height=0.45\textheight,angle=90]{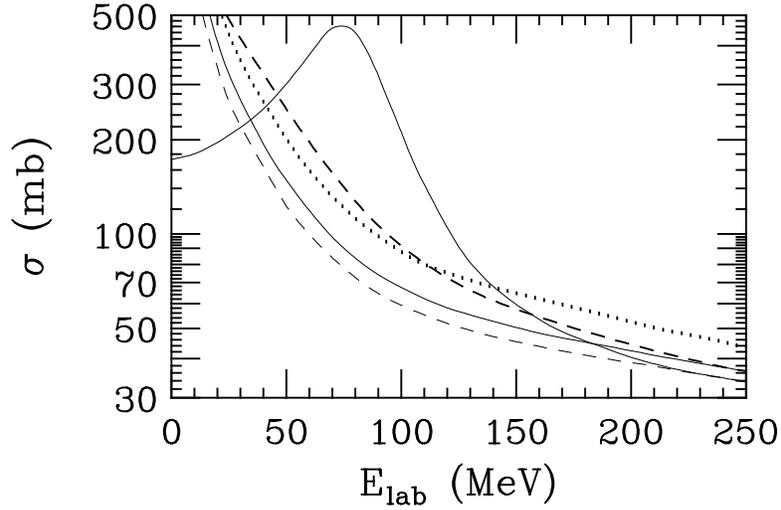}
\caption{Isospin averaged NN cross section as a function of
bombarding energy at a density of $0.5\rho_0$. The dotted line
represents the free cross section, the solid lines represent the
in-medium cross sections at temperatures of 10 (thin line) and 35
MeV (thick line). The dashed lines are the corresponding cases
without Pauli-blocking. Taken from Ref.\ \protect\cite{Alm2}.}
\label{alm}
\end{figure}

In the work by Alm {\it et al.} \cite{Alm1,Alm2}, the cross
section in a hot nuclear matter is evaluated by including also the
hole-hole collisions in the Pauli blocking operator. Fig.\
\ref{alm} shows their results for the isospin averaged NN
in-medium cross section at temperatures of 10 MeV and 35 MeV and a
density of $0.5\rho_0$ as a function of bombarding energy. Effects
of the Pauli blocking and self-energy corrections are separated by
comparing full calculations with those setting the Pauli blocking
operator to 1. First, it is seen that at both temperatures the
self-energy correction suppresses the cross section, while the
Pauli blocking operator for intermediate states enhances the cross
section. Second, at energies above about 200 MeV predictions for
different temperatures converge to values smaller than the
free-space cross section. It is also seen that at lower
temperatures the cross section has a strong peak above the
free-space cross section.  This has been interpreted as a
precursor effect of the superfluid phase transition in nuclear
matter \cite{Alm1,Alm2}.

Experimentally, strong evidences supporting reduced in-medium NN
cross sections have been found in heavy-ion collisions at
intermediate energies, see, e.g., Refs.~\cite{Wes93,Xu91,Dan02b}.
In particular, studies on collective flow, especially the balance
energy where the transverse flow disappears, have shown clearly
indications of reduced in-medium NN cross sections
\cite{Wes93,Kla93,Hun96}. An empirical relation \cite{Kla93}
\begin{eqnarray}\label{msigma}
\sigma_{NN}^{\rm
medium}=\left(1+\alpha\frac{\rho}{\rho_0}\right)\sigma_{NN}^{\rm
free}
\end{eqnarray}
with the parameter $\alpha\approx -0.2$ has been found to better
reproduce the flow data compared to transport model calculations
using the free-space NN cross sections. Very recently, in studying
the stopping power and collective flow in heavy-ion collisions at
SIS/GSI energies, there were indications that the in-medium NN
cross sections were reduced at low energies but enhanced at high
energies~\cite{Zha07}. However, all these analyses have been done
assuming simply some overall reduction of all NN scattering cross
sections without using in-medium cross sections that are evaluated
self-consistently at densities and temperatures determined by the
reaction dynamics. Furthermore, no information about the isospin
dependence of the in-medium NN cross sections has been extracted
from these experiments.

\subsection{Isospin dependence of NN cross sections in
neutron-rich matter}

\begin{figure}
\centering
\includegraphics[scale=0.6,angle=-90]{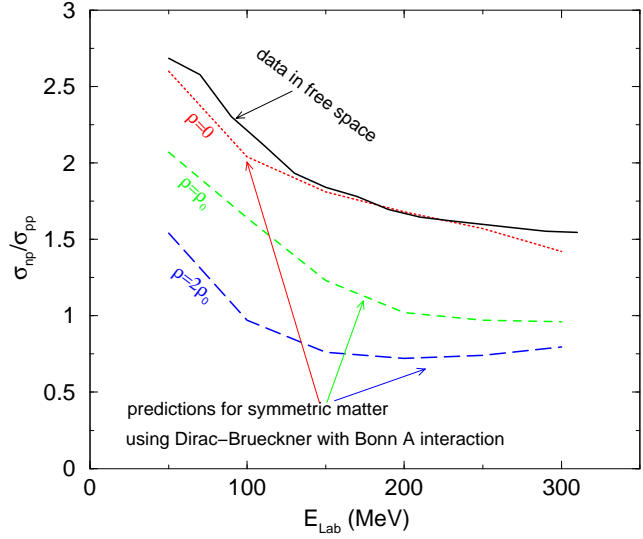}
\caption{(Color online) The ratio of np over pp scattering cross
sections as a function of incident nucleon energy. The solid line is
the one extracted from experimental data \protect
\cite{Cha90,Alk77,Che68,Bol85,Gru85,Lis82} while the dashed lines
are those extracted from calculations in symmetric matter using the
Bonn A potential within the Dirac-Bruckner approach \cite{Li93}.}
\label{XmedBonnA}
\end{figure}

The experimental free-space $\sigma_{np}/\sigma_{pp}$ ratio
\cite{Cha90,Alk77,Che68,Bol85,Gru85,Lis82} changes from about 2.7 at
$E_{\rm lab}=50$ MeV to 1.7 at $E_{\rm lab}=300$ MeV as shown by the
solid line in Fig.~\ref{XmedBonnA} . How the ratio
$\sigma_{np}/\sigma_{pp}$ changes with density and isospin asymmetry
in asymmetric medium encountered often in heavy-ion reactions and
astrophysical situations is an important question since its answer
may reveal directly useful information about the isospin dependence
of the in-medium nuclear effective interactions. However, very
little work has been done so far about the isospin dependence of the
in-medium NN cross sections in asymmetric nuclear matter although
extensive studies have been carried out in symmetric matter based on
various many-body theories and/or phenomenological approaches, see,
e.g., Refs.~\cite{Per02,Pan92,Li93,Sch97}. Therefore, only some
information about the density dependence of the
$\sigma_{np}/\sigma_{pp}$ ratio in symmetric nuclear matter can be
found in the literature.

As an example, shown in Fig.~\ref{XmedBonnA} with the dashed lines
are the $\sigma_{np}/\sigma_{pp}$ ratio in symmetric matter
extracted from predictions using the Bonn A potential within the
Dirac-Brueckner approach \cite{Li93}. In this approach not only
the in-medium NN cross sections are reduced compared to their
values in free-space, the ratio $\sigma_{np}/\sigma_{pp}$ is also
predicted to decrease with increasing density, becoming less than
1 for high energy nucleons at twice the normal density. Several
other microscopic studies have reached, however, opposite
conclusion, i.e., the $\sigma_{np}/\sigma_{pp}$ ratio increases in
symmetric medium, see, e.g., Refs.~\cite{Gia96,Koh98,LiQF00}. We
notice that there have also been some efforts to extend the above
Dirac-Brueckner calculations to isospin asymmetric matter
\cite{Sam05b}.

\begin{figure}[tbh]
\centering
\includegraphics[height=0.3\textheight,angle=-90]{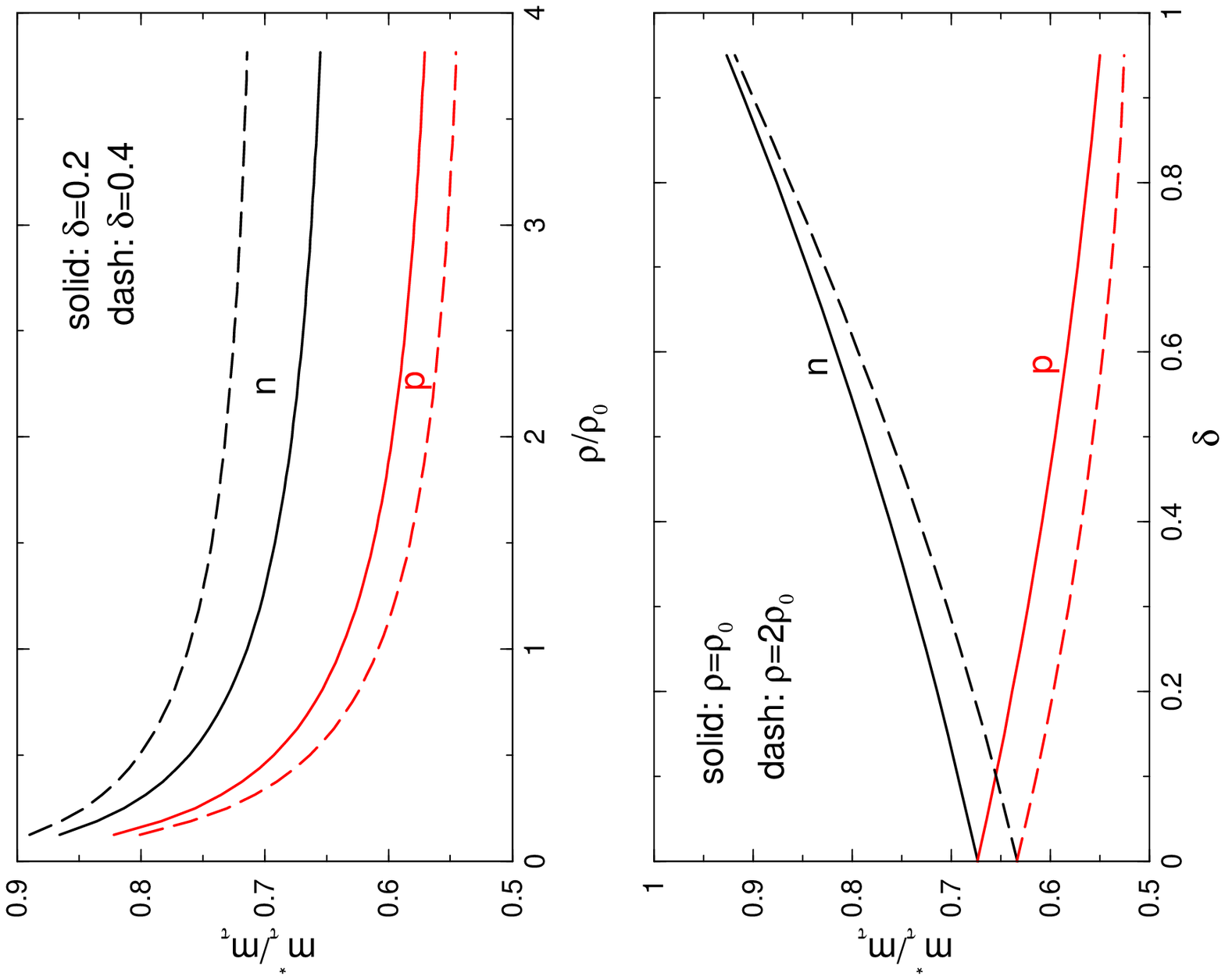}
\hspace{0.2cm}
\includegraphics[height=0.32\textheight,angle=-90]{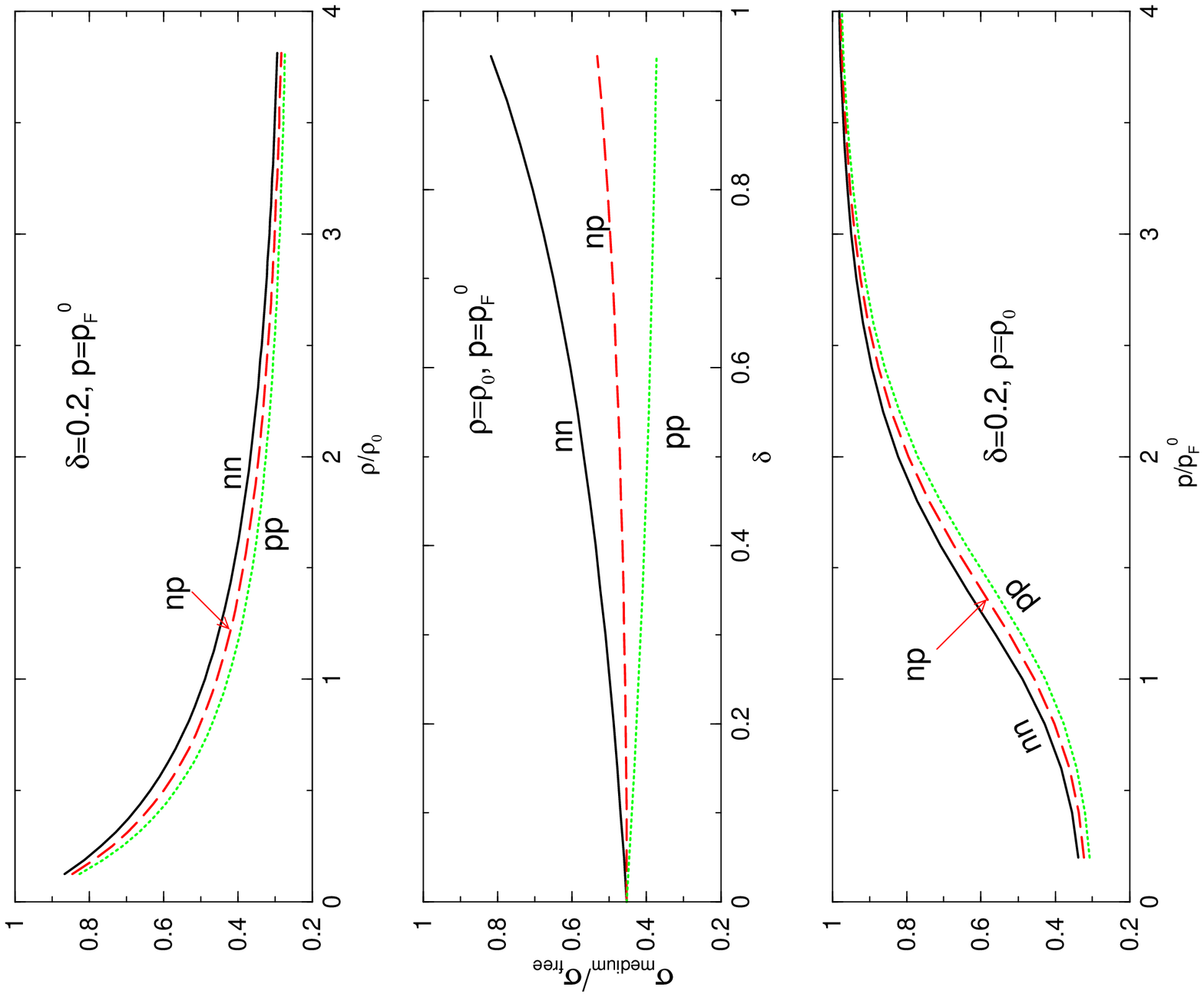}
\caption{{\protect\small (Color online) Left window: Nucleon
effective masses at the respective Fermi surface in asymmetric
matter as a function of density (upper panel) and isospin
asymmetry (lower panel). Right window: The reduction factor of the
in-medium NN cross sections compared to their free-space values as
a function of density (top panel), isospin asymmetry (middle
panel) and momentum (bottom panel). Taken from
Ref.~\cite{LiBA05c}.}} \label{ekmass-mdi}
\end{figure}

Based on an effective mass scaling model \cite{Per02,Neg81,Pan92},
the isospin dependence of NN cross sections in neutron-rich matter
was recently studied in Ref. \cite{LiBA05c}. In this model, the
matrix elements of the NN interactions are assumed to be the same
as those in free-space, so only medium effects due to nucleon
effective masses on the incoming current in the initial state and
the level density of the final state are included. The NN cross
sections in the medium $\sigma _{NN}^{\rm medium}$ are therefore
reduced in this model compared with their free-space values
$\sigma _{NN}^{\rm free}$ by a factor
\begin{eqnarray}
R_{\rm medium}\equiv \sigma _{NN}^{\rm medium}/\sigma _{NN}^{\rm
free}=(\mu _{NN}^{\ast}/\mu _{NN})^{2},  \label{xmedium}
\end{eqnarray}
where $\mu _{NN}$ and $\mu _{NN}^{\ast }$ are the reduced k-masses
of the colliding nucleon pairs in free-space and in the medium,
respectively. As an example, shown in the left window of
Fig.~\ref{ekmass-mdi} are the effective k-masses of nucleons at
their Fermi surfaces using the MDI interactions \cite{LiBA05c}. It
is seen that the effective mass of neutrons is higher than that of
protons and the splitting between them increases with both the
density and isospin asymmetry of the medium. As discussed in
Chapter~\ref{chapter_ria}, the momentum dependence of the symmetry
potential and the associated neutron-proton effective mass
splitting is still highly controversial within different
approaches and/or using different nuclear effective
interactions~\cite{LiBA04b,Riz04,Beh05}. Being phenomenological
and non-relativistic in nature, the neutron-proton effective mass
splitting with the MDI interaction is consistent with predictions
of all non-relativistic microscopic models, see, e.g.,
Refs.~\cite{Bom91,Zuo05,Sjo76}, and the non-relativistic limit of
microscopic relativistic many-body theories, see, e.g.,
Refs.~\cite{Fuc04,Ma04,Sam05a}.

As an illustration of a simplified case, shown in the right window
of Fig.~\ref{ekmass-mdi} is the in-medium reduction factor $R_{\rm
medium}$ of NN cross sections for two colliding nucleons having
the same magnitude of momentum $p$. The $R_{\rm medium}$ factor is
examined as a function of density (upper panel), isospin asymmetry
(middle panel) and the momentum (bottom panel). It is interesting
to see that the in-medium NN cross sections are not only reduced
compared to their free-space values, but the nn and pp cross
sections are also split while their free-space cross sections are
the same. Moreover, the difference between the nn and pp
scattering cross sections grows in more asymmetric matter. The
higher in-medium cross sections for nn than for pp are completely
due to the positive neutron-proton effective mass splitting
calculated with the MDI effective interaction.

\begin{figure}[tbh]
\centering
\includegraphics[height=0.35\textheight,angle=-90]{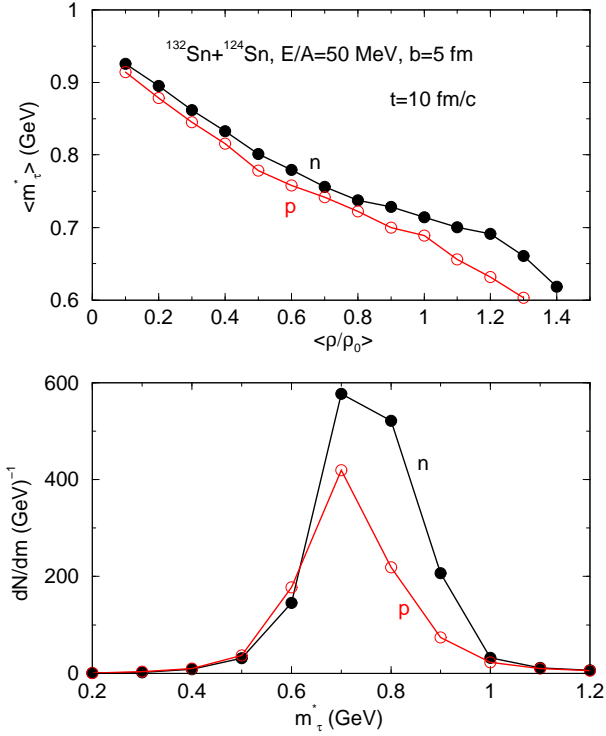}
\caption{{\protect\small (Color online) The correlation between
the average nucleon effective mass and the average nucleon density
(upper window), and the distribution of nucleon effective masses
(lower window) in the reaction of }$^{132} ${\protect\small
Sn+}$^{124}${\protect\small Sn at a beam energy of 50 MeV/A and an
impact parameter of 5 fm. Taken from Ref.~\cite{LiBA05c}.}}
\label{ixd}
\end{figure}

To include the in-medium effects in the transport model
description of heavy-ion reactions, both the effective masses and
the in-medium NN cross sections have to be calculated dynamically
in the created evolving environment. How the nucleon effective
masses and the NN cross sections are modified compared to their
free-space values in a typical heavy-ion reaction at intermediate
energies can be seen in Fig.\ \ref{ixd} from the correlation
between the average nucleon effective mass and average nucleon
density (top panel), and the distribution of nucleon effective
masses (bottom panel) at the instant of $10$ fm/c after the
contact of two $^{132}$Sn nuclei in the IBUU04 simulations of
their reactions at a beam energy of $50$ MeV/A and an impact
parameter of $5$ fm. It is seen that the nucleon effective masses
decrease with increasing density. The maximum density reached at
the instant considered, i.e., $10$ fm/c, is about $1.4\rho /\rho
_{0}$. Moreover, the neutron-proton effective mass splitting is
seen to increase slightly at supra-normal densities. However, the
increase is not large because the isospin asymmetry normally
decreases with increasing density, a phenomenon called the isospin
fractionation (distillation). These features are consistent with
our expectations discussed in previous sections. From the lower
panel of Fig.\ \ref{ixd} it is seen that the distribution of
nucleon effective masses peaks at about $0.7$ GeV, with a small
number of nucleons acquiring, however, effective masses above
their free masses. The latter happens, although rarely, when the
slope of the nucleon potential $dU_{\tau}/dp$, which is used in
calculating the effective mass with $\frac{m_{\tau }^{\ast
}}{m_{\tau }}= \left\{ 1+\frac{m_{\tau }}{p}\frac{dU_{\tau
}}{dp}\right\} $, becomes negative during heavy-ion reactions.

\begin{figure}[tbh]
\centering
\includegraphics[height=0.35\textheight,angle=-90]{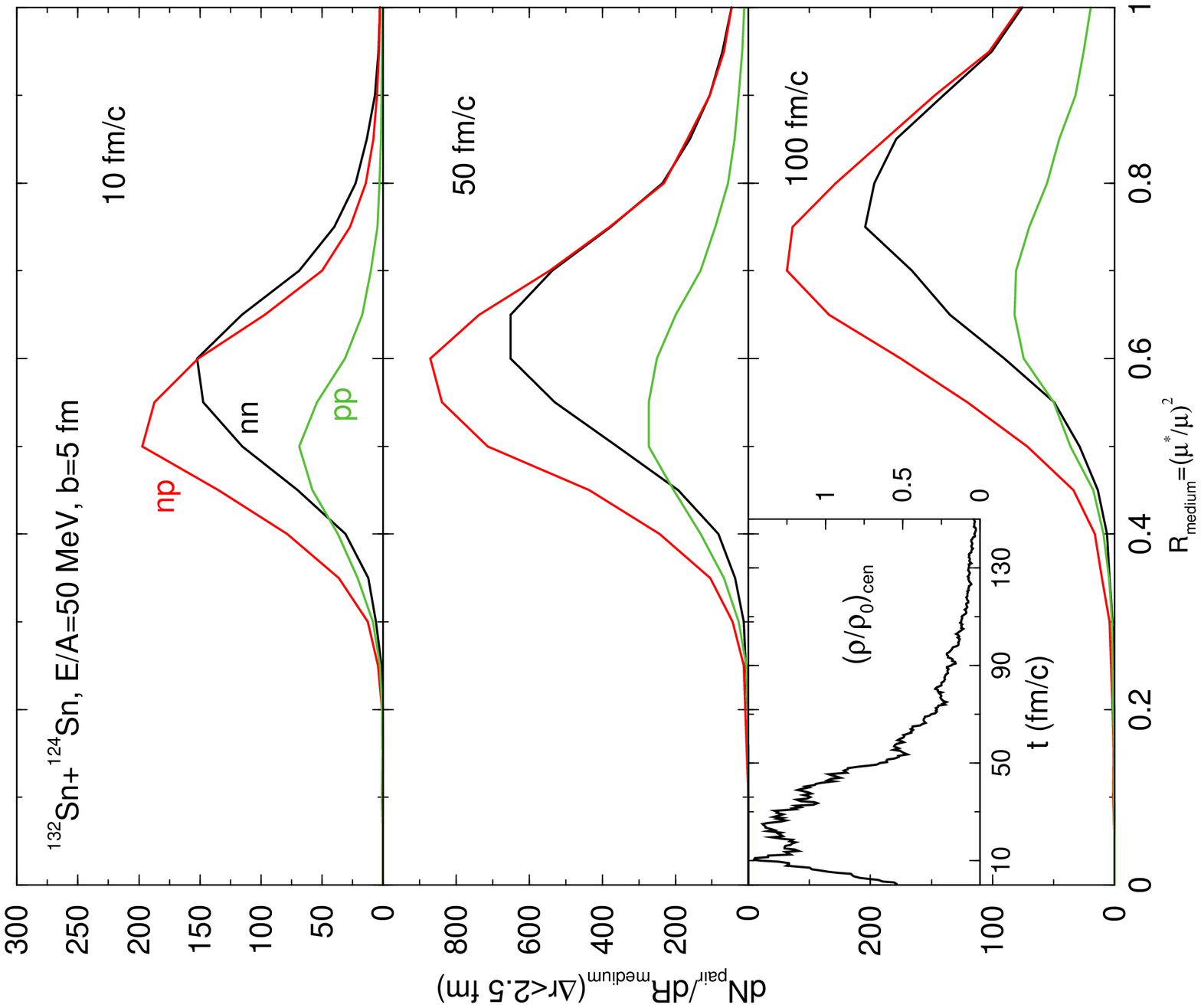}
\caption{{\protect\small (Color online) The distribution of the
reduction factor of in-medium NN cross sections in the reaction of
}$^{132}${\protect\small Sn+}$^{124}${\protect\small Sn at a beam
energy of 50 MeV/A and an impact parameter of 5fm at 10, 50 and
100 fm/c, respectively. The inset is the evolution of the central
density in the reaction. Taken from Ref.~\cite{LiBA05c}.}}
\label{rmedium}
\end{figure}

With the nucleon effective masses available, their effects on NN
scatterings during heavy-ion reactions can be examined. Shown in
Fig.\ \ref{rmedium} are the distributions of the reduction factor
$R_{\rm medium}$ in the reaction of $^{132}$Sn+$^{124}$Sn at a beam
energy of 50 MeV/A and an impact parameter of $5$ fm at $10$, $50$
and $100$ fm/c, respectively. The inset in the bottom panel shows
the evolution of the central density during the reaction. The three
instants represent the compression, expansion and freeze-out stages
of the reaction. The quantity $N_{\rm pair}(\Delta r<2.5~{\rm fm})$
is the number of nucleon pairs with spatial separations less than
$2.5$ fm. These are potential colliding nucleons whose scattering
cross section will be reduced by the factor $R_{\rm medium}$, i.e.,
$\sigma _{NN}^{\rm medium}=R_{\rm medium}\times \sigma _{NN}^{\rm
free}$. It is seen that on average as much as $50\%$ of the
reduction occurs for NN scatterings in the early stage of the
reaction. As the system expands, the average density decreases and
the reduction factor $R_{\rm medium}$ thus gradually shifts towards
$1$ in the later stage of the reaction.

\subsubsection{Global stopping power in heavy-ion
reactions as a probe of the isospin dependence of the in-medium NN
cross sections}

Several observables used to measure the global stopping power in
heavy-ion reactions are known to be sensitive to the in-medium NN
cross sections. These include the quadruple moment $Q_{zz}$ of the
nucleon momentum distribution, the linear momentum transfer (LMT)
and the ratio of the transverse to longitudinal energies (ERAT).
Unfortunately, these observables are sensitive only to the
magnitude but not to the isospin dependence of the in-medium NN
cross sections \cite{LiBA05d}. However, it has been claimed that
the quadruple moment $Q_{zz}$ my be a good measure of the isospin
dependence of the in-medium NN cross sections based on the IQMD
model calculations \cite{Liu01}. The origin of these seemingly
different conclusions is discussed in detail in Ref.
\cite{LiBA05d}. Here we recall some of the discussions.

\begin{figure}[htp]
\centering
\includegraphics[scale=0.7,angle=-90]{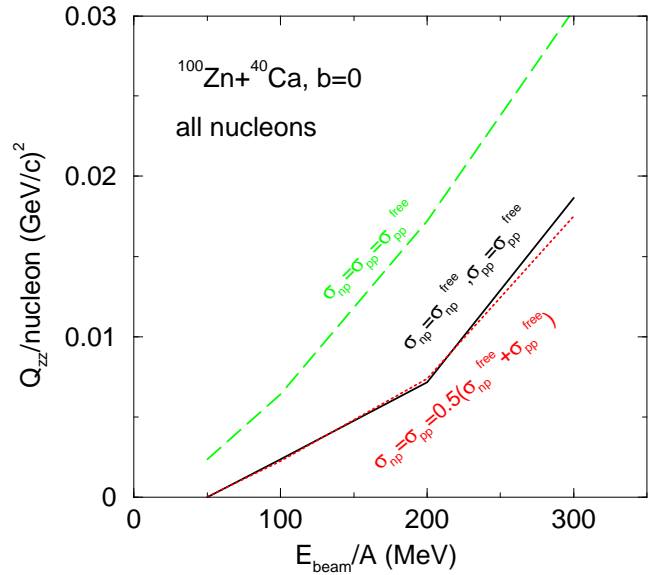}
\caption{(Color online) Quadruple moment as a function of beam
energy in head-on collisions of $^{100}$Zn+$^{40}$Ca with the
three choices of NN cross sections. Taken from
Ref.~\cite{LiBA05d}.} \label{XmedQzzEb}
\end{figure}

Shown in Fig.~\ref{XmedQzzEb} are the quadruple moment per nucleon
$Q_{zz}/A\equiv\frac{1}{A}\sum^A_{i=1}(2p_{iz}^2-p_{ix}^2-p_{iy}^2)$
as a function of beam energy for head-on collisions of $^{100}{\rm
Zn}+^{40}{\rm Ca}$ with three choices of in-medium NN cross
sections. Before discussing its dependence on NN in-medium cross
sections, we note that the $Q_{zz}$ is almost independent of the
symmetry energy simply because the isoscalar interaction
overwhelmingly dominates over the isovector interaction for the
globe thermalization of the nuclear system. In agreement with Ref.
\cite{Liu01}, setting artificially the cross section for
neutron-proton scatterings to be the same as that for proton-proton
scatterings in free-space (long dashed line), i.e., the ratio
$\sigma_{np}/\sigma_{pp}$ is one, the $Q_{zz}$ increases
significantly compared to calculations using the free-space np and
pp scattering cross sections $\sigma_{np}^{\rm free}$ and
$\sigma_{pp}^{\rm free}$ (solid line) with $\sigma_{np}^{\rm
free}>\sigma_{pp}^{\rm free}$ as shown in Fig.~\ref{freesigma}.
Based on this observation, it was proposed in Ref. \cite{Liu01} that
the stopping power measured by the $Q_{zz}$ could be used as a
sensitive probe of the isospin dependence of the in-medium NN cross
sections. However, as it was pointed out in Ref. \cite{LiBA05d} that
the observed increase of the $Q_{zz}$ was simply due to the
effective reduction of the np scattering cross sections although the
$\sigma_{np}/\sigma_{pp}$ ratio is indeed also changed. In fact, the
$Q_{zz}$ is insensitive to the $\sigma_{np}/\sigma_{pp}$ ratio if
one keeps the total number of NN collisions to be about the same.
This ambiguity of using the $Q_{zz}$ as a probe of the isospin
dependence of the in-medium NN cross sections can be demonstrated by
comparing the above calculations with the ones using
$\sigma_{np}=\sigma_{pp}=(\sigma_{np}^{\rm free}+\sigma_{pp}^{\rm
free})/2$. Although the ratio $\sigma_{np}/\sigma_{pp}$ in the
latter is one, the $Q_{zz}$ is, however, about the same as the
calculations using the free-space NN cross sections up to about
$E_{\rm beam}/A=220$ MeV. This observation can be understood
qualitatively from the total number of NN collisions $N_{\rm coll}$
that essentially determines the nuclear stopping power. Neglecting
the Pauli blocking, the $N_{\rm coll}$ scales according to $N_{\rm
coll}\propto N_{np}\sigma_{np}+(N_{pp}+N_{nn})\sigma_{pp}$, where
$N_{np}$ and $N_{pp}$ are the number of np and pp colliding pairs.
Assuming that only the first chance NN collisions contribute, one
then has the ratio $N_{np}/(N_{pp}+N_{pp}) \approx
(1-\delta_1\delta_2)/(1+\delta_1\delta_2)\approx
1-2\delta_1\delta_2$, where $\delta_1\equiv (N_1-Z_1)/A_1$ and
$\delta_2\equiv (N_2-Z_2)/A_2$ are the isospin asymmetries of the
two colliding nuclei.  To the second order in isospin asymmetry,
this ratio is about one even for very neutron-rich systems, and one
thus has $N_{coll}\propto N_{np}(\sigma_{np}+\sigma_{pp})$. With
either $\sigma_{np}=\sigma_{pp}=(\sigma_{np}^{\rm
free}+\sigma_{pp}^{\rm free})/2$ or $\sigma_{np}=\sigma_{np}^{\rm
free}$ and $\sigma_{pp}=\sigma_{pp}^{\rm free}$, the numbers of NN
collisions $N_{\rm coll}$ are then the same, leading thus to
approximately same $Q_{zz}$. At higher energies, however, secondary
collisions are expected to become gradually more important, and
above arguments become less valid.

The above discussions indicate clearly that the nuclear stopping
power is indeed sensitive to the in-medium NN cross sections.
However, the stopping power alone is insufficient to determine
simultaneously both the magnitude and the isospin dependence of
the in-medium NN cross sections. To determine both quantities, one
needs an additional observable besides the nuclear stopping power
that is sensitive to the ratio $\sigma_{np}/\sigma_{pp}$ as well.

\subsubsection{The backward neutron/proton ratio as a measure of
the isospin dependence of the in-medium NN cross sections}

Given the opportunities provided by the radioactive beams, it is of
great interest to find experimental observables that are sensitive
to the isospin dependence of the in-medium NN cross sections. In
Ref. \cite{LiBA05d}, it has been proposed that isospin tracers at
backward angles/rapidities in nuclear reactions induced by
radioactive beams in inverse kinematics are promising probes of the
isospin dependence of the in-medium NN cross sections
\cite{LiBA05d}. Several observables can be used as isospin tracers,
such as the neutron/proton ratio of free nucleons or the ratio of
mirror nuclei. The rapidity and angular distributions of the isospin
tracers measure directly the isospin transport in reactions
especially below the pion production threshold. These observables
were previously used also to study the momentum stopping power and
the nucleon translucency
\cite{LiBA95,Bas94,She94,Joh96,Joh97,LiBA98b,Hom99,Ram00} in
heavy-ion collisions, see, e.g., Ref.~\cite{LiBA98} for an earlier
review. In central collisions induced by highly asymmetric
projectiles on symmetric targets in inverse kinematics, the
deviation of the neutron/proton ratio from one at backward
rapidities/angles reflects the strength of net isospin transfer from
the projectile to the target. This proposal is based on the
consideration that only large angle and/or multiple np scatterings
are effective in transporting the isospin asymmetry from forward to
backward angles. With inverse kinematics, nucleons in the lighter
target moving backward with higher velocities in the center of mass
frame of the reaction are more likely to induce multiple np
scatterings.

The isospin tracers at backward rapidities/angles are also less
affected by the nuclear symmetry potential. Although the symmetry
potential is important for isospin transport in heavy-ion collisions
\cite{Bar05,LiBA97a,Shi03,Far91}, it is, however, unlikely for the
symmetry potential to change the directions of motion of nucleons.
Nevertheless, the relative importance and interplay of the symmetry
potential and the in-medium NN cross sections on the
rapidity/angular distributions of isospin tracers have to be studied
quantitatively. Ideally, one would like to identify observables in
special kinematic or geometrical regions where the sensitivity to
both the symmetry potential and the isospin dependence of the
in-medium NN cross sections is a minimum if it cannot be avoided
completely. Transport model simulations are useful for this purpose.
As a simple demonstration, one may use in transport model
calculations a symmetry energy of the form \cite{LiBA05d,Hei00b}
$E_{\rm sym}(\rho)=E_{\rm sym}(\rho_0)\cdot (\rho/\rho_0)^{\gamma}$,
where $E_{\rm sym}(\rho_0)\approx 30$ MeV is the symmetry energy at
normal nuclear matter density $\rho_0$ and $\gamma$ is a stiffness
parameter. By fitting earlier predictions of the variational
many-body calculations by Akmal {\it et al.} \cite{Akm98,Akm97}, one
obtains the values of $E_{\rm sym}(\rho_0)=$32 MeV and $\gamma=0.6$.
However, recent analyses of isospin diffusions in heavy-ion
collisions at intermediate energies favor strongly a $\gamma$ value
between 0.69 and 2 \cite{LiBA05c,Tsa04,Che05a} depending on whether
one includes the momentum dependence of the symmetry potential in
the analysis.

\begin{figure}[htp]
\centering
\includegraphics[scale=0.4,angle=-90]{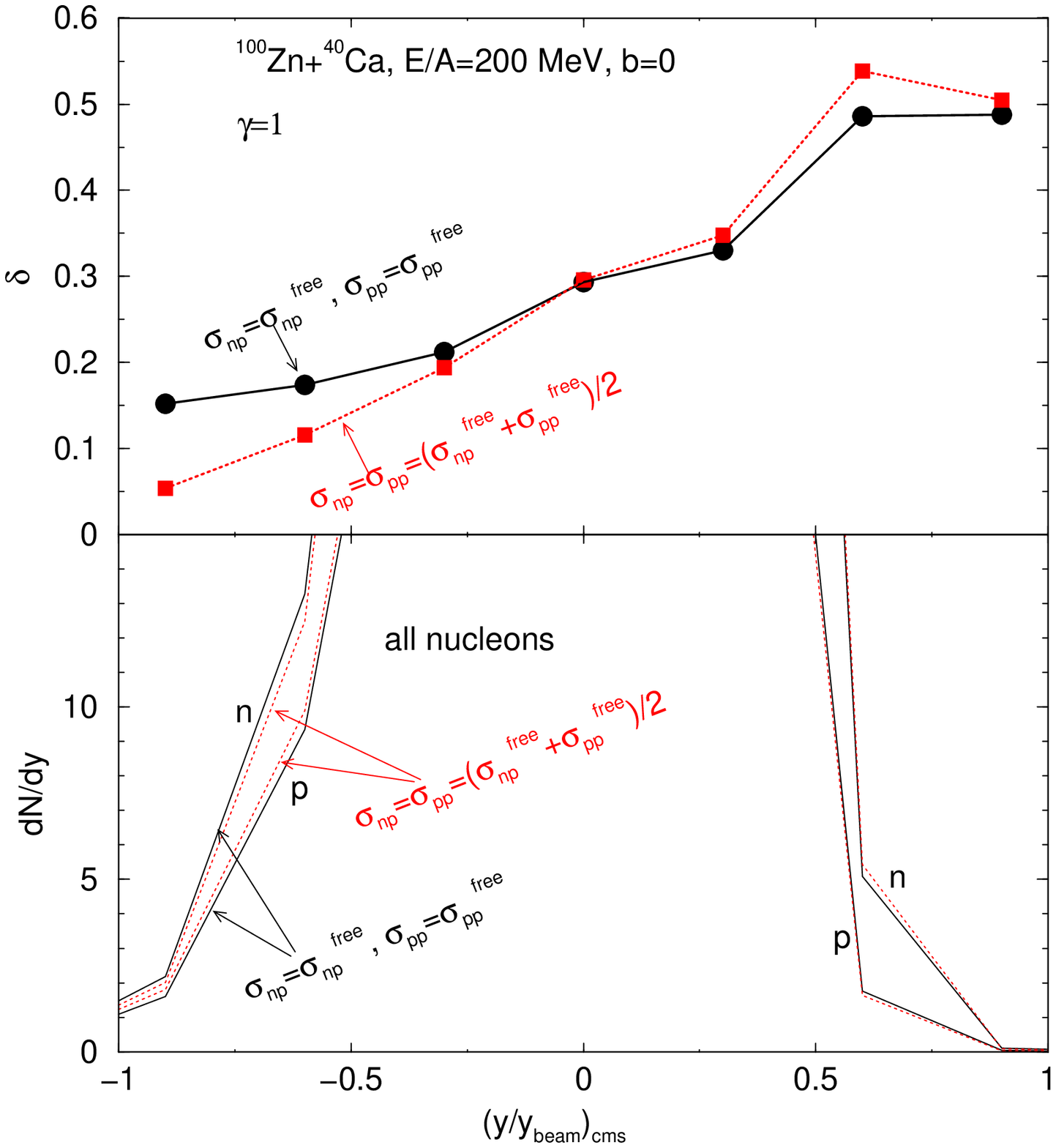}
\hspace{0.3cm}
\includegraphics[scale=0.4,angle=-90]{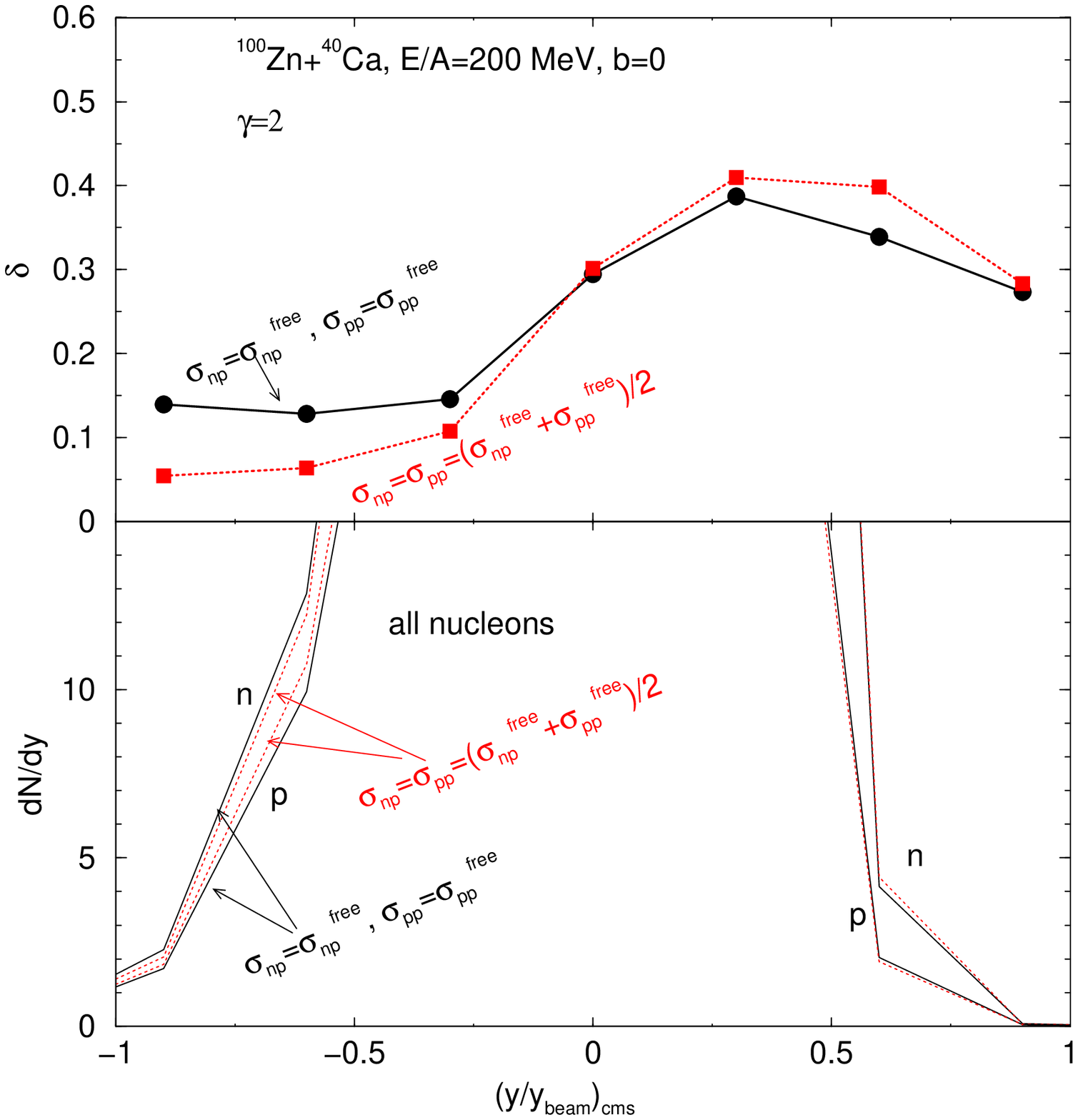}
\caption{(Color online) Rapidity distributions (lower panels) of
all nucleons and their isospin asymmetries (upper panels) in
head-on collisions of $^{100}{\rm Zn}+^{40}{\rm Ca}$ at a beam
energy of 200 MeV/A using a $\gamma$ parameter of 1 (left window)
or 2 (right window). Taken from Ref.~\cite{LiBA05d}.}
\label{XmedDeldNdY}
\end{figure}

Shown in Fig.~\ref{XmedDeldNdY} are the rapidity distributions of
all nucleons (lower panel) and their isospin asymmetries (upper
panel) at 100 fm/$c$ in head-on collisions of $^{100}{\rm
Zn}+^{40}{\rm Ca}$ at a beam energy of 200 MeV/A using $\gamma=1$
and $2$, respectively, for both the free-space NN cross sections and
the in-medium cross sections
$\sigma_{np}=\sigma_{pp}=(\sigma_{np}^{\rm free}+\sigma_{pp}^{\rm
free})/2$. These two cross sections have been shown in
Fig.~\ref{XmedQzzEb} to lead to identical quadruple moment $Q_{zz}$
at $E_{\rm beam}=200$ MeV/A. It is seen that the effects of the
in-medium NN cross sections on the overall nucleon rapidity
distributions are rather small for both values of the $\gamma$
parameter. Moreover, the symmetry energy also has very little effect
on the nucleon rapidity distributions. These observations are
consistent with those obtained from studying other global measures
of the nuclear stopping power. Concentrating on the forward and
backward nucleons, it is, however, clearly seen that the larger
$\sigma_{np}/\sigma_{pp}$ ratio in the case of using
$\sigma_{np}=\sigma_{np}^{\rm free}$ and
$\sigma_{pp}=\sigma_{pp}^{\rm free}$ leads to more (less) transfer
of neutrons (protons) from forward to backward rapidities. Since the
effect is opposite on neutrons and protons, it is much more
pronounced on the isospin asymmetry $\delta$ as shown in the upper
panels. In both cases, the isospin asymmetries are thus rather
sensitive to the isospin dependence of the in-medium NN cross
sections, especially at backward rapidities. Comparing the two upper
panels of Fig.~\ref{XmedDeldNdY}, one notices that the effects of
the symmetry potential are mostly at forward rapidities. At backward
rapidities, the influence of the isospin dependence of the in-medium
NN cross sections dominates, however, overwhelmingly over that due
to the symmetry potential. The effects on $\delta$ due to the
isospin dependence of the in-medium NN cross sections discussed
above are clearly measurable, especially at backward rapidities.

\begin{figure}[htp]
\centering
\includegraphics[scale=0.5,angle=-90]{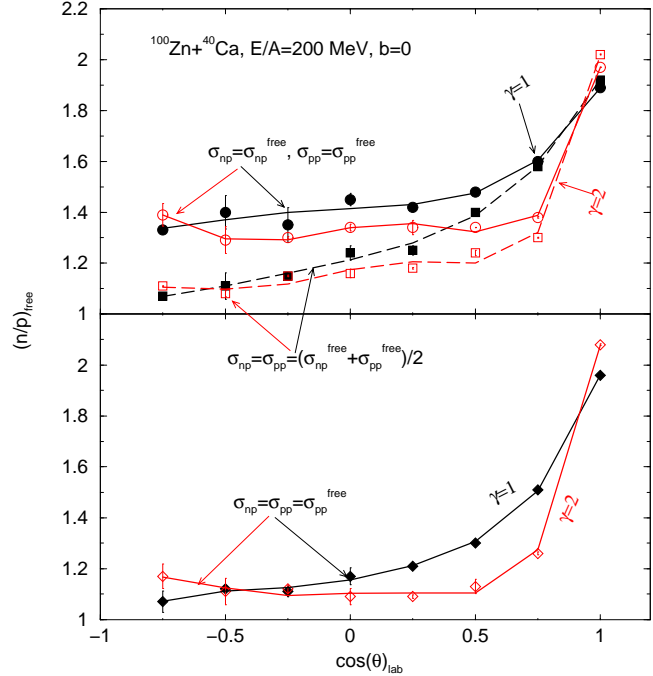}
\caption{(Color online) Angular distributions of the free neutron
to proton ratio $(n/p)_{free}$ in head-on collisions of
$^{100}{\rm Zn}+^{40}{\rm Ca}$ at a beam energy of 200 MeV/A.
Taken from Ref.~\cite{LiBA05d}.} \label{XmedNPth}
\end{figure}

The above discussed effect can be extracted experimentally, for
instance, by studying the free neutron/proton ratio, the ${\rm
t}/^3{\rm He}$ ratio or the isoscaling parameters. As an
illustration, one can study the polar angle distributions of the
neutron/proton ratio $(n/p)_{\rm free}$ of free nucleons identified
as those having local baryon densities less than $\rho_0/8$. These
are shown in Fig.~\ref{XmedNPth} for the three choices of the
in-medium NN cross sections and for both $\gamma=1$ and $2$. In the
upper panel, results obtained by using the free-space NN cross
sections and the choice $\sigma_{np}=\sigma_{pp}=(\sigma_{np}^{\rm
free}+\sigma_{pp}^{\rm free})/2$, the same choices as those in
Fig.~\ref{XmedDeldNdY}, are compared. It is clearly seen that the
$(n/p)_{\rm free}$ ratio at backward angles is rather insensitive to
the symmetry energy but very sensitive to the isospin dependence of
the in-medium NN cross sections. At forward angles an opposite
behavior is seen. Moreover, by comparing results using all three
choices considered for the in-medium NN cross sections, the choices
of $\sigma_{np}=\sigma_{pp}=(\sigma_{np}^{\rm free}+\sigma_{pp}^{\rm
free})/2$ and $\sigma_{np}=\sigma_{pp}=\sigma_{pp}^{\rm free}$ lead
to about the same $(n/p)_{\rm free}$ value at very backward angles.
The latter value is significantly less than the one obtained by
using the free np and pp cross sections. In other words, at these
very backward angles the $(n/p)_{\rm free}$ is sensitive only to the
$\sigma_{np}/\sigma_{pp}$ ratio but not the absolute values of the
individual nn and np cross sections nor the symmetry energies. Thus,
it would be very valuable to measure the $(n/p)_{\rm free}$ ratio at
large backward angles. On the other hand, at very forward angles the
$(n/p)_{\rm free}$ ratio is very sensitive to the symmetry potential
but not much to the in-medium NN cross sections.

\begin{figure}[htp]
\centering
\includegraphics[scale=0.5,angle=-90]{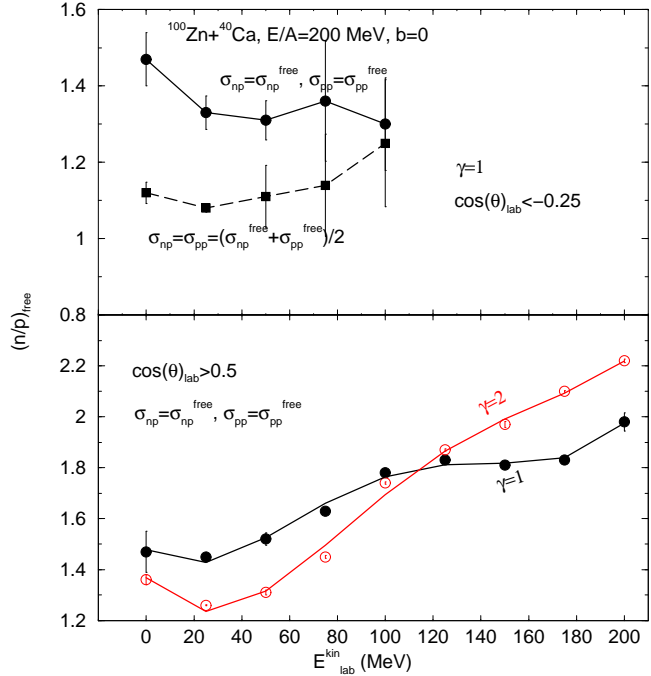}
\caption{(Color online) The $(n/p)_{\rm free}$ as a function of
nucleon kinetic energy at backward (upper panel) and forward
(lower panel) angles in head-on collisions of $^{100}{\rm
Zn}+^{40}{\rm Ca}$ at a beam energy of 200 MeV/A. Taken from
Ref.~\cite{LiBA05d}.} \label{XmedNPEkin}
\end{figure}

Further test of the above proposal can be seen in
Fig.~\ref{XmedNPEkin} which shows the $(n/p)_{\rm free}$ ratio as a
function of nucleon kinetic energy in the laboratory frame. The
limits of $\cos(\theta)\leq -0.25$ for backward (upper panel) and
$\cos(\theta)> 0.5$ for forward (lower panel) angles are used. Most
nucleons emitted to the backward angles have energies less than
about 100 MeV for the reaction considered. Only very few nucleons in
the backward regions have higher energies, and the calculations
using 12,000 events in each case do not have enough statistics to
show a meaningful $(n/p)_{\rm free}$ ratio. At backward angles, the
$(n/p)_{\rm free}$ ratio is significantly higher than one which is
the neutron/proton ratio of the target considered here. The value of
$(n/p)_{\rm free}$ is larger for the higher
$\sigma_{np}/\sigma_{pp}$ ratio, and the effect due to the isospin
dependence of the in-medium NN cross section is most pronounced at
very low energies. This is understandable because transferring
relatively more neutrons from the forward-going projectile to the
backward direction requires more np scatterings. Once neutrons move
backward through possibly multiple scatterings, they then have less
energies.

At the forward angles selected here, the $(n/p)_{\rm free}$ ratio
is, on the other hand, more affected by the symmetry energy. An
example is shown in the lower panel of Fig.~\ref{XmedNPEkin} which
is obtained by using the free NN cross sections and the still
relatively large angular range of $-60^0\leq \theta \leq 60^0$
selected by the cut $\cos(\theta)>0.5$. Although the in-medium NN
cross sections still have some effects on the $(n/p)_{\rm free}$
ratio at forward angles as indicated in Fig.~\ref{XmedNPth}, the
influence of the symmetry energy is seen, although it depends
strongly on the nucleon energy.  Since low energy nucleons are more
likely emitted at subnormal densities where the repulsive/attractive
symmetry potentials are stronger with the softer symmetry energy,
the $(n/p)_{\rm free}$ ratio is higher with $\gamma=1$ for low
energy nucleons. The high energy nucleons are mostly emitted
forwardly and more likely have gone through the supranormal density
region in the earlier stage of the reaction. The stiffer symmetry
energy with $\gamma=2$ thus results in higher values of $(n/p)_{\rm
free}$ for these nucleons. Qualitatively similar results are
obtained for other choices of the in-medium NN cross sections.

\begin{figure}[htp]
\centering
\includegraphics[scale=0.5,angle=-90]{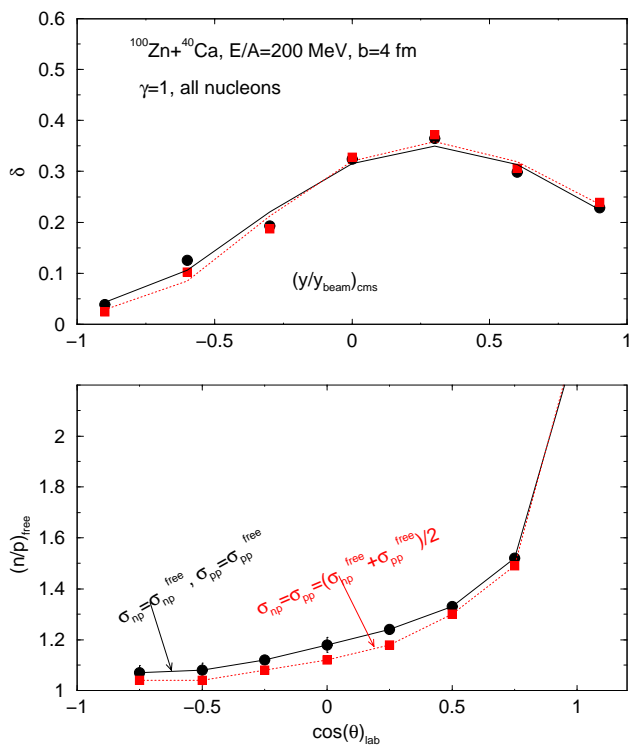}
\caption{(Color online) Rapidity (upper panel) and angular (lower
panel) dependence of isospin asymmetries of all nucleons in the
reaction of $^{100}{\rm Zn}+^{40}{\rm Ca}$ at a beam energy of 200
MeV/A and impact parameter of 4 fm using a $\gamma$ parameter of
1. Taken from Ref.~\cite{LiBA05d}.} \label{XmedDelNPth}
\end{figure}

The above discussions are all based on results from head-on
collisions. The conclusions remain qualitatively the same but with
reduced effects at finite impact parameters. As an example, shown in
Fig.~\ref{XmedDelNPth} are the rapidity and angular distributions of
the isospin asymmetry $\delta$ of all nucleons (upper window) and
the $(n/p)_{\rm free}$ of free ones at an impact parameter of 4 fm.
The effects of the in-medium NN cross sections are still clearly
observed but smaller than those in head-ion collisions. One also
notices that memories of the n/p ratios of the projectile and target
become now clearer as one expects.

\subsection{NN cross sections in neutron-rich matter
within the relativistic impulse approximation}

The nucleon-nucleus optical potential plays a central role in
determining the in-medium NN cross sections and the nucleon
mean-free-path (MFP). In particular, the imaginary part of the
optical potential allows one to extract easily the MFP and thus
also the in-medium NN cross sections. There are several possible
ways to derive the optical potential \cite{Hod94}. In the
microscopic Brueckner approach, it can be calculated from various
models either non-relativistically or relativistically, see, e.g.,
Refs. \cite{Ron06,Alm1,Fuc01,ha86,ja92}. Theoretically,
approximations have to be made to render the calculations
feasible. For instance, the evaluation of the imaginary part of
the optical potential depends on the treatment of the widths of
intermediate states that are rather unclear or the nucleon
polarization in the medium that is in principle coupled to the
vacuum. On the other hand, one can start from a physically
reasonable approximation for the optical potential and determine
its parameters using the experimental data \cite{kh06} as in the
RIA. As we have already discussed in Chapter \ref{chapter_ria},
the optical potential in the RIA is obtained in a form similar to
the non-relativistic $t\rho$ approximation. The basic ingredients
of the optical potential in this approach are the free Lorentz
invariant NN scattering amplitudes and the nuclear scalar and
vector densities in nuclear matter
\cite{Mcn83a,She83,Cla83,Mil83,Mcn83b}. An attractive feature of
the RIA is that the relativistic optical potentials are
experimentally constrained by the free-space NN scattering data.
The nuclear densities are calculated from the RMF models that
provide also a dynamical description for the spin-orbit coupling
\cite{Ser97,Ser86}. Along with the success of the RMF models in
describing the nuclear structure, the RIA was justified by nicely
reproducing the proton-nucleus elastic scattering data at high
energies. Of course, the limitation of the RIA is that it is valid
only at reasonably high energies and in not so dense matter. As
most of the existing microscopic calculations were devoted to the
low and intermediate energy regions, the in-medium NN cross
sections at high energies remain largely unknown. The recent work
by Jiang {\it et al.} \cite{Jiang2} on the nucleon MFP and
in-medium NN cross sections at high energies within the
relativistic impulsive approximation (RIA) is thus useful for the
ongoing and future studies on reactions with high energy
radioactive beams.

The MFP of a nucleon can be evaluated from its dispersion relation
in nuclear medium, which in the relativistic frame is written as
\begin{eqnarray}\label{disp}
(E_k-U^{\rm tot}_0)^2={\bf k}^2+({M+U_S^{\rm tot}})^2
\end{eqnarray}
where $E_k=E_{\rm kin}+M$, and $U_{S}^{\rm tot}$ and $U_{0}^{\rm
tot}$ denoting, respectively, the scalar and (0th-component)
vector parts of the relativistic optical potential in the RIA.
Equivalently, the nucleon dispersion relation can be written in
terms of the `Schr\"odinger equivalent potential' (SEP) $U^{\rm
tot}_{\rm sep}$ defined in Eq.(\ref{SEP}) of
Chapter~\ref{chapter_ria} \cite{Jam81}
\begin{eqnarray}\label{disp1}
\frac{k_\infty^2}{2M}=\frac{k^2}{2M}+U_{\rm sep}^{\rm tot}(E_{\rm
kin}),
\end{eqnarray}
with $k^2_\infty=E^2_{\rm kin}+2ME_{\rm kin}$. Since $U_{S}^{\rm
tot}$ and $U_{0}^{\rm tot}$ are generally complex, one can write
$U^{\rm tot}_{\rm sep}=U_{\rm sep}+iW_{\rm sep}$ and introduce a
complex momentum $k=k_R+ik_I$. The nucleon MFP $\ld$ is then given
exactly as \cite{ligq93}
\begin{eqnarray}\label{mfp0}
\ld=\frac{1}{2k_I}=\frac{1}{2}\left[-M\left(E_{\rm kin}+\frac{E_{\rm
kin}^2}{2M} -U_{\rm sep}\right) +M\left(\left(E_{\rm
kin}+\frac{E_{\rm kin}^2}{2M}-U_{\rm sep}\right)^2+W_{\rm sep}^2\right)^{1/2}
\right]^{-1/2}.\notag\\
\end{eqnarray}
Expanding the momentum in the vicinity of $k_R$ \cite{Neg81} one
can then approximate the real and imaginary parts of the momentum
as
\begin{eqnarray}\label{momentum}
k_R\approx(E_{\rm kin}^2+2ME_{\rm kin}-2MU_{\rm sep})^{1/2} \quad
{\rm and} \qquad k_I\approx-W_{\rm sep}\left(
\frac{k_R}{M}+\frac{\pp U_{\rm sep}}{\pp k_R}\right)^{-1}.
\end{eqnarray}
Since there is no explicit momentum dependence in the optical
potentials in RIA, the nucleon MFP can thus be approximated by
\begin{eqnarray}\label{mfp1}
\ld\approx\frac{1}{2k_I} =-\frac{k_R}{2MW_{\rm sep}}.
\end{eqnarray}

Since the nucleon MFP can also be measured as the length of the
unit volume defined by the matter density and the NN cross
section, i.e., it can be expressed as~\cite{Pan92}
\begin{eqnarray}\label{mfp2}
\ld_i=(\r_{p}\sg^*_{ip}+\r_n\sg^*_{in})^{-1}, \qquad i=p,n
\end{eqnarray}
where $\r_p$ and $\r_n$ are, respectively, the proton and neutron
densities. The in-medium NN cross sections can be obtained by
inverting the above equation. To write the results compactly, one
may define the following two quantities,
\begin{eqnarray}\label{mfp3}
\tilde{\Ld}^{-1}=\frac{1}{2}\left(\frac{1}{\ld_n}+\frac{1}{\ld_p}\right)\qquad
{\rm and} \qquad
\tilde{\ld}^{-1}=\frac{1}{2\delta}\left(\frac{1}{\ld_n}-\frac{1}{\ld_p}\right).
\end{eqnarray}
They can be further written in terms of the imaginary parts of the
symmetry potential and the isoscalar SEP, i.e.,
\begin{eqnarray}\label{mfp4}
\tilde{\Ld}^{-1}=\frac{2M}{k_R}\bar{W}_{\rm sep}, \hbox{ }
\tilde{\ld}^{-1}=\frac{2M}{k_R} W_{\rm sym},
\end{eqnarray}
where $W_{\rm sym}\equiv (W_{\rm sep}^n- W_{\rm sep}^p)/{2\delta}$
is the imaginary symmetry potential, similarly defined as the real
part of the symmetry potential $U_{\rm sym}\equiv (U_{\rm sep}^n-
U_{\rm sep}^p)/{2\delta}$, and $k_R^{n,p}$ have been approximated by
$k_R$. This is a very good approximation because at high energies
the $U_{\rm sym}$ is negligible compared to the kinetic energy and
the isoscalar SEPs are given by
\begin{eqnarray}\label{uwsep}
\bar{U}_{\rm sep}=({U}^n_{\rm sep}+{U}_{\rm sep}^p)/2\qquad {\rm
and} \qquad \bar{W}_{\rm sep}=({W}^n_{\rm sep}+{W}_{\rm sep}^p)/2.
\end{eqnarray}
As pointed out in Ref.~\cite{Che05c}, the $U_{\rm sym}$ itself is
isospin-independent because the difference between the neutron and
proton potentials is largely linear in isospin asymmetry. It has
been found that the $W_{\rm sym}$ also retains such an isospin
independence. The $\tilde{\Ld}$ and $\tilde{\ld}$ are thus
essentially independent of the isospin asymmetry of the medium at
high energies where the RIA is valid. Consequently, at these high
energies the in-medium NN cross sections are also independent of
the isospin asymmetry of the medium. Of course, there is still a
difference between the neutron-proton and proton-proton
(neutron-neutron) cross sections. In terms of these
isospin-independent quantities, the in-medium NN cross sections
are obtained as
\begin{eqnarray}\label{imcs}
\sg_{nn}^*=(\tilde{\Ld}^{-1}+\tilde{\ld}^{-1})/\r_B \qquad {\rm and}
\qquad \sg_{np}^*=(\tilde{\Ld}^{-1}-\tilde{\ld}^{-1})/\r_B.
\end{eqnarray}
Here it is assumed that $\sigma_{nn}^*=\sigma_{pp}^*$, thus
neglecting the small charge symmetry breaking effect \cite{li98}
and the isospin-dependent Pauli blocking effects in asymmetric
nuclear medium. The above results are only applicable to the high
energy region where the RIA is valid and the Pauli blocking
effects are negligible. At low energies, both $U_{\rm sym}$ and
Pauli blocking effects are not negligible, and the resulting NN
cross sections will depend on the isospin asymmetry of the medium.

\begin{figure}[tbh]
\centering \vspace{-1.7cm}
\includegraphics[width=8cm,height=10cm]{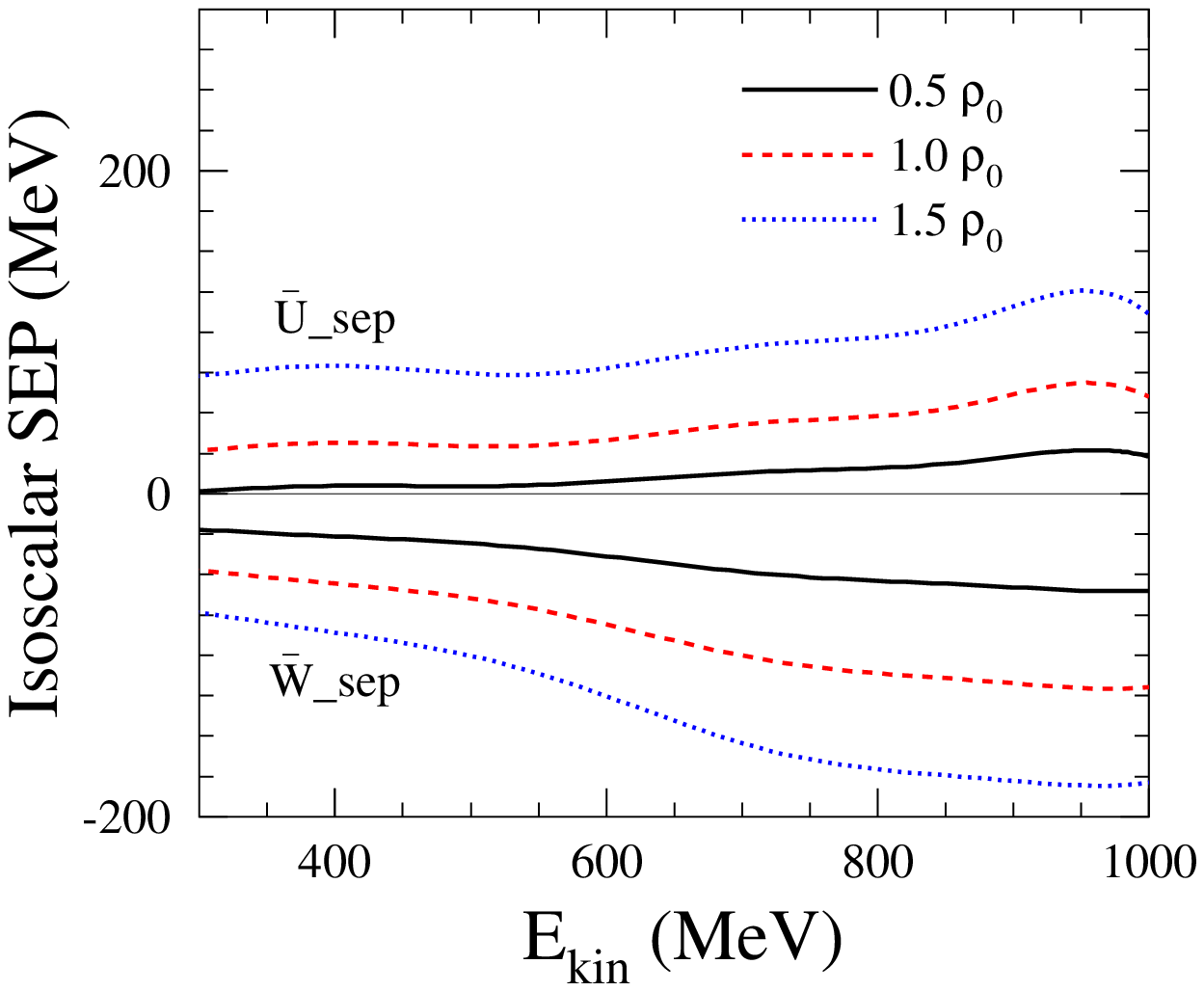}
\hspace{-1cm}\includegraphics[width=8cm,height=10cm]{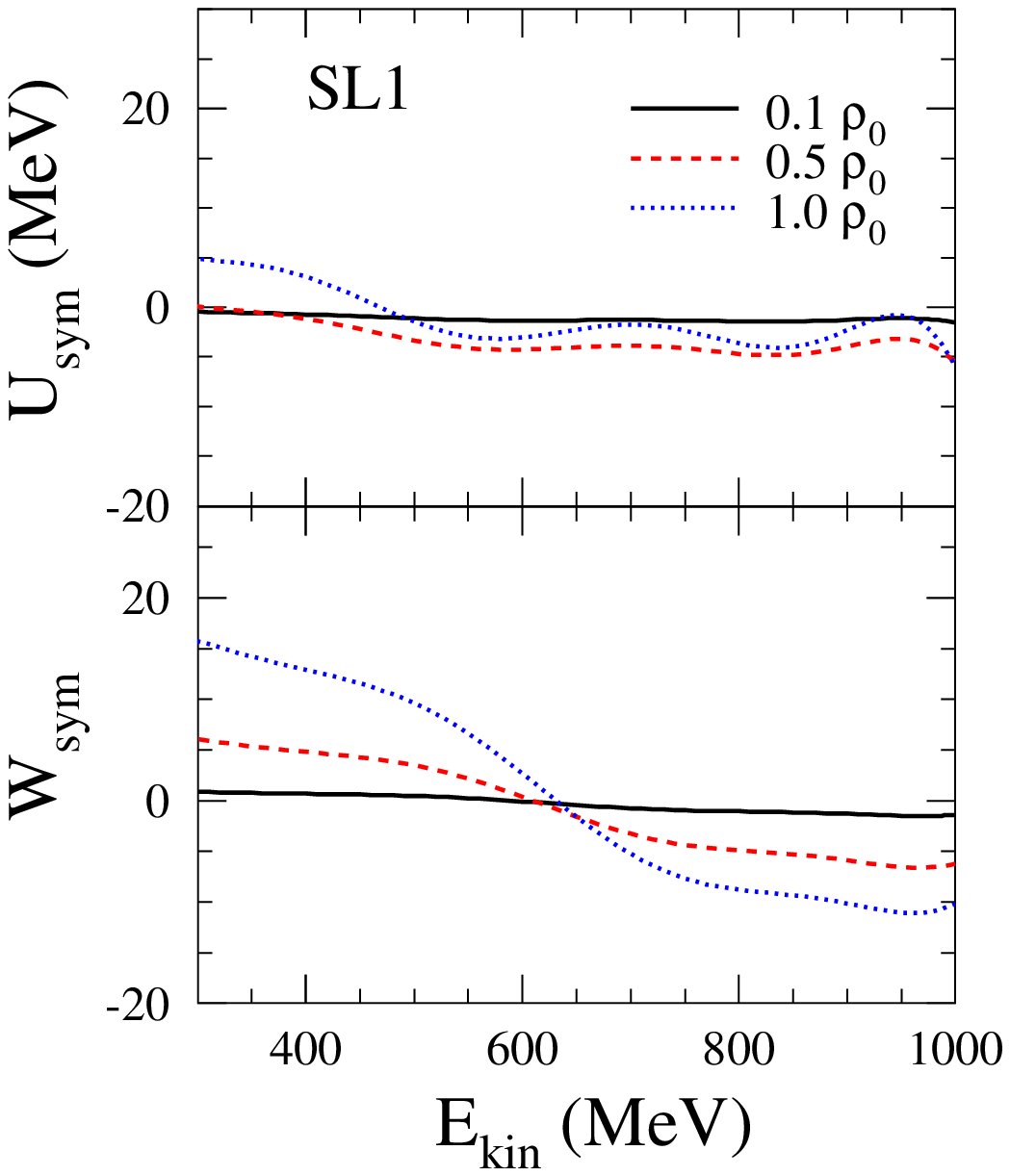}
\vspace{-1cm} \caption{The isoscalar (left window) and isovector
(right window) Schr\"odinger equivalent potential (SEP) as
functions of the nucleon kinetic energy for different densities.
Taken from Ref.~\cite{Jiang2}.}\label{f:f0}
\end{figure}

The work of Jiang {\it et al.} \cite{Jiang2}, mentioned previously,
was based on the above framework and the approximation that the NN
cross sections in neutron-rich matter are evaluated using a
relativistic model Lagrangian and taking into account the chiral
symmetry restoration \cite{Jia07a,Jia07b}. In the RIA approach, the
resulting proton and neutron SEP's serve as the basis for studying
the nucleon MFP and in-medium NN cross sections. The isoscalar
nucleon SEP, defined by the mean value of the neutron and proton
SEP's in Eq. (\ref{uwsep}), is displayed on the left window of
Fig.~\ref{f:f0}. It is seen that the imaginary SEP is stronger than
the real one for nucleons at high kinetic energies. As the kinetic
energy increases, the magnitude of the SEP increases very slowly.
Consequently, this results in a correspondingly small decrease of
the nucleon MFP, as given by Eq.~(\ref{mfp1}). Shown in the right
window of Fig.~\ref{f:f0} are the real and imaginary parts of the
symmetry potential. Indeed, these potentials are independent of the
isospin asymmetry of the medium. For the $U_{\rm sym}$, this was
also found previously \cite{Che05c}. The $U_{\rm sym}$ is negligibly
small, actually very close to zero, in the whole energy range
considered. This result is consistent with the energy dependence of
the Lane potential extracted from nucleon-nucleus scatterings, which
indicates that the Lane potential decreases with increasing incident
energy up to about 100 MeV, above which no data are available
\cite{Hod94}. The smallness of $U_{\rm sym}$ shown in the figure
justifies the approximation of using $k_R$ instead of the
$k_R^{n,p}$ in Eq.~(\ref{mfp4}). The imaginary part of the symmetry
potential, which relates to the splitting of proton and neutron
absorptions in nuclear medium, displays a different dependence on
the kinetic energy from that for $U_{\rm sym}$. It is particularly
interesting to note that its sign changes around $E_{\rm kin}\approx
630$ MeV from positive to negative. This change is reflected in the
energy dependence of the nucleon MFP's.

\begin{figure}[tbh]
\centering
\vspace{-1.7cm}
\includegraphics[width=10cm,height=10cm]{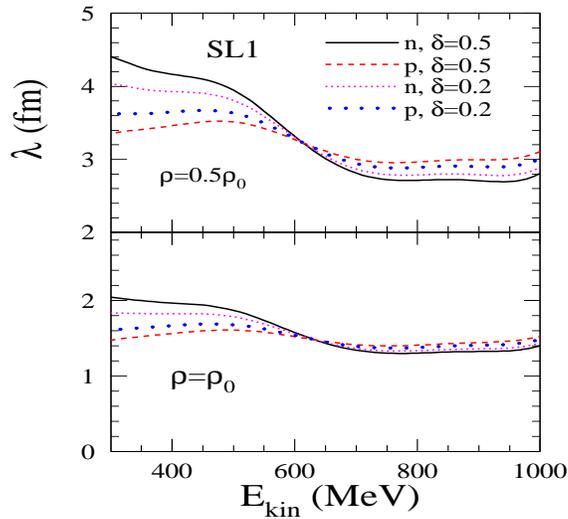}
\vspace{-1cm} \caption{(Color online) Nucleon mean free path as a
function of nucleon kinetic energy for different densities. Taken
from Ref.~\cite{Jiang2}.}\label{f:f2}
\end{figure}

The nucleon MFP's, calculated from Eq. (\ref{mfp0}) or
(\ref{mfp1}), are shown in Fig.~\ref{f:f2}. It is seen that at
higher kinetic energies the nucleon MFP changes very little with
the energy and is also less sensitive to the isospin asymmetry.
The rapid change of the nucleon MFP occurs around $E_{\rm
kin}\approx 630$ MeV, at which the isospin-splitting between the
neutron and proton MFP also changes appreciably. The nucleon MFP
is determined by both the real and imaginary parts of the SEP as
seen in Eq.~(\ref{mfp1}). The difference between the real parts of
the neutron and proton SEP's is small compared to the nucleon
momentum at high kinetic energies. The change of the
isospin-splitting between the neutron and proton MFP's can be
attributed to the sign change of the imaginary part of the
symmetry potential as shown in the lower panel of the right window
in Fig.~\ref{f:f0}. This change of the MFP's of neutrons and
protons around 600 MeV/A may have interesting experimental
consequences and certainly deserves further studies.

At $E_{\rm kin}\geq 700$ MeV, the nucleon MFP is insensitive to the
isospin asymmetry of the medium. This insensitivity is due to the
fact that at high nucleon kinetic energies $W_{\rm sym}$,
responsible for the difference between the proton and neutron MFP's,
is small compared to $\bar{W}_{\rm sep}$ that dominates the
contributions to the nucleon MFP's. At lower kinetic energies,
appreciable sensitivity to the isospin asymmetry, however, exists
because the imaginary part of the symmetry potential is not so small
in comparison. Comparing the two windows in Fig. \ref{f:f2}, it is
seen that the sensitivity of the nucleon MFP to the isospin
asymmetry is reduced with increasing density, as a result of the
decreasing nucleon MFP's and the drop of the $W_{\rm sym}$ to
$\bar{W}_{\rm sep}$ ratio at higher densities.

\begin{figure}[tbh]
\centering
\includegraphics[width=7cm,height=10cm]{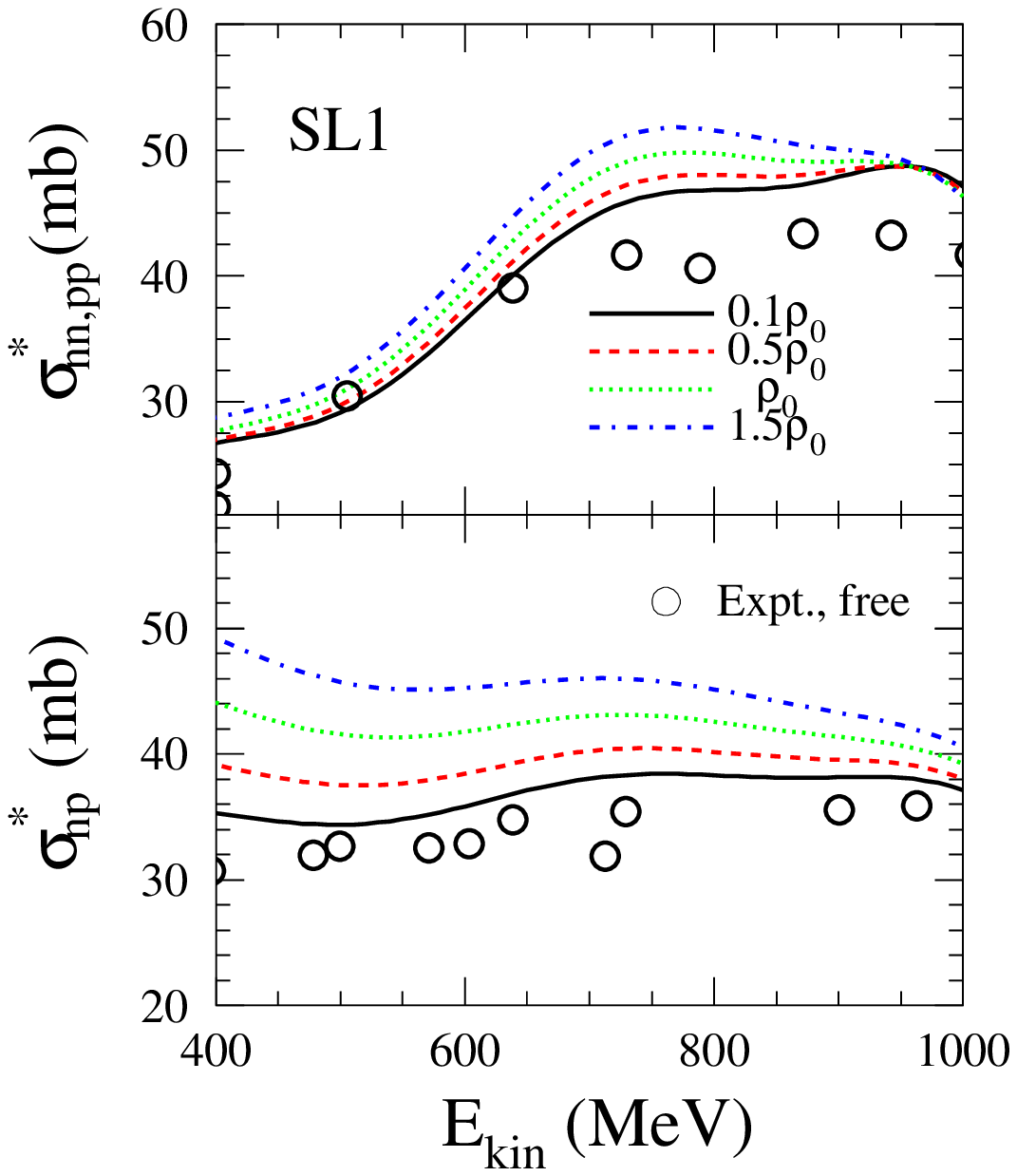}
\includegraphics[width=7cm,height=10cm]{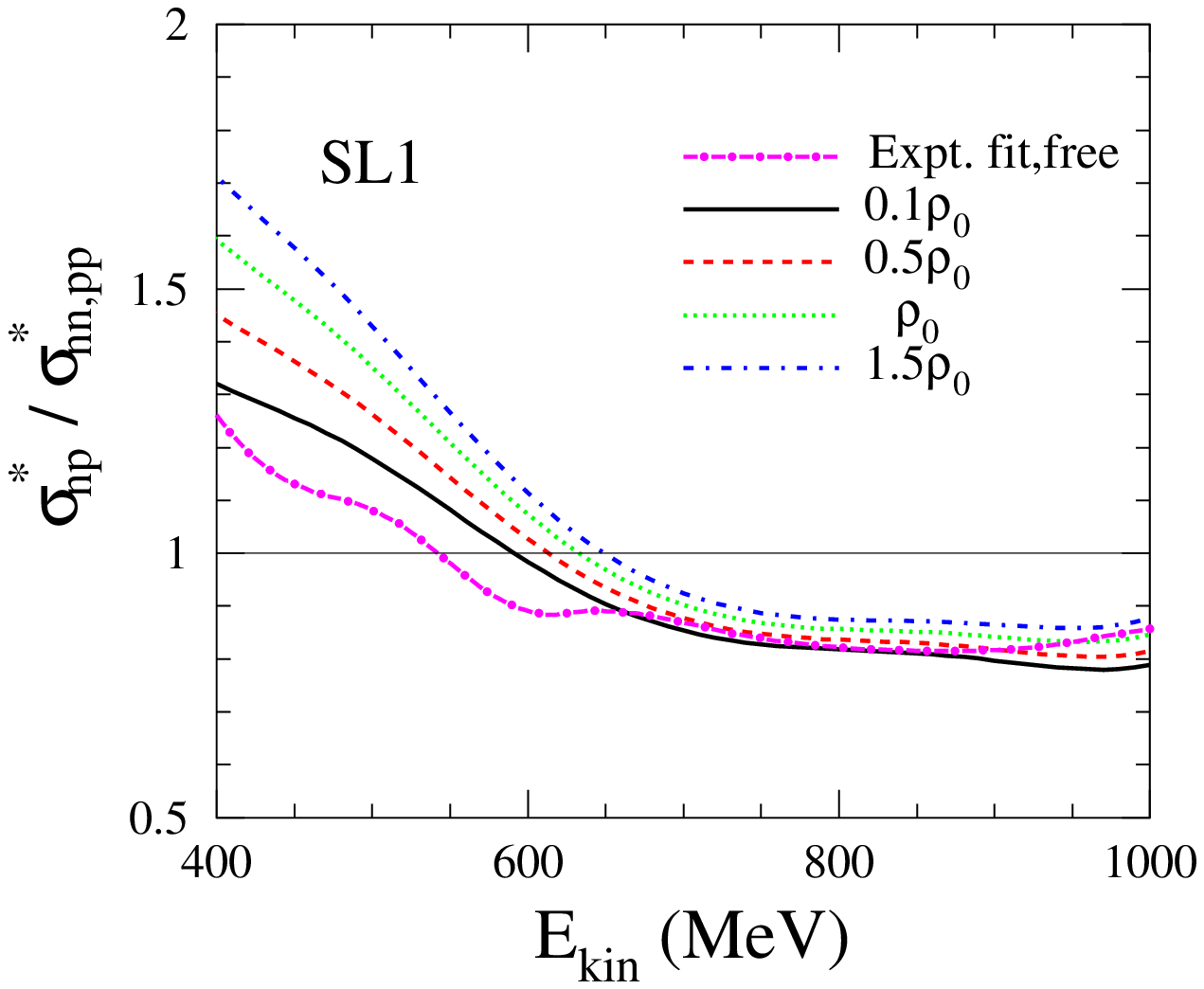}
\caption{The in-medium NN cross sections (left window) and their
ratios (right window) as a function of nucleon kinetic energy for
different densities. Taken from Ref.~\cite{Jiang2}.}\label{f:f3}
\end{figure}

In the energy range where the RIA is valid, the in-medium NN cross
sections $\sg_{np}^*$ and $\sg_{nn,pp}^*$ are independent of the
isospin asymmetry of the medium. They depend only on the density and
the nucleon kinetic energy. As shown in Fig.~\ref{f:f3}, the
in-medium NN cross sections increase with the density, in contrast
to that occurring at lower energies where the Pauli blocking plays
an important role in reducing the in-medium NN cross sections.
However, at higher kinetic energies the in-medium NN cross sections
are not necessarily a descending function of density. It is also
seen that the calculated NN cross sections at the very low density
of $0.1\r_0$ are very close to the free-space NN scattering data.
Moreover, the in-medium NN cross sections are shown to depend
linearly in density as seen at higher energies using some other
approaches \cite{Li93,Fuc01,Zha07}. For instance, in
Ref.~\cite{Zha07}, with the closed time path Green's function
approach it was found that the $\sg^*_{nn,pp}$ increases with
density at $E_{\rm kin}\ge 240$ MeV. Interestingly, a similar
observation was made more recently in the DBHF calculations using
the Bonn-B potential \cite{Sam08}.

The ratio $\sg_{np}^*/\sg_{nn,pp}^*$ is an ascending function of
density. This feature is also different from that at low and
intermediate energies. It is clearly shown that the magnitude of
$\sg_{nn,pp}^*$ exceeds $\sg_{np}^*$ around $E_{kin}\ge 600$ MeV
depending on the density. The experimental $\sg_{np}/\sg_{nn,pp}$
ratio in free-space becomes smaller than 1 above about 580 MeV. This
general trend is qualitatively reproduced by calculations in the
medium at the low density limit.

\subsection{General remarks on the NN cross sections in neutron-rich matter}

In summary of this Chapter, we note that a lot of theoretical
efforts have been devoted to calculating the NN cross sections in
symmetric nuclear matter based on various microscopic many-body
theories. There is, however, very little work on the isospin
dependence of the NN cross sections in isospin asymmetric nuclear
matter based on microscopic many-body theories. Phenomenological
models based on the nucleon effective mass scaling are useful but
need to be verified by more microscopic studies. Experimentally,
there are some convincing evidence for reduced in-medium NN cross
sections compared to their free-space values. However, so far
there is no experimental information about how the
$\sigma_{np}/\sigma_{pp}$ ratio may change in isospin asymmetric
medium. One of the challenging tasks is to identify experimental
observables that are mostly sensitive to the isospin dependence of
the in-medium NN cross sections, while simultaneously determining
also the density dependence of the symmetry energy. The usual
probes of the nuclear stopping power are sensitive to the
magnitude of the in-medium NN cross sections. They are, however,
ambiguous for determining the isospin dependence of the in-medium
NN cross sections. The isospin tracers, such as the neutron/proton
ratio of free nucleons, at backward rapidities/angles in nuclear
reactions induced by radioactive beams in inverse kinematics is a
sensitive probe of the isospin dependence of the in-medium NN
cross sections. At forward rapidities/angles, on the other hand,
the neutron/proton ratio is more sensitive to the density
dependence of the symmetry energy. It is thus very useful to
measure experimentally the rapidity and angular distributions of
isospin tracers to study the transport properties and the EOS of
isospin asymmetric matter. Indeed, a very recently experimental
proposal to investigate the isospin dependence of the in-medium NN
cross sections using the approach discussed here was approved at
the NSCL/MSU. Ultimately, these studies will enable us to better
understand the isospin dependence of the in-medium nuclear
effective interactions.


\section{Isospin effects in heavy-ion reactions as probes of
the nuclear symmetry energy and symmetry potential}
\label{chapter_observables}

\subsection{Overview}
In this Chapter, we review a number of interesting isospin
phenomena and effects in heavy-ion reactions. Many of them can be
used as effective probes of the density dependence of the nuclear
symmetry energy. Some of these probes are more promising than
others for investigating the symmetry energy at low densities
while others are more useful at supra-normal densities. Because
the symmetry potentials for neutrons and protons have opposite
signs and are generally smaller than the isoscalar potential at
same density, most of the observables proposed so far are based on
differences or ratios of isospin multiplets of baryons, mirror
nuclei and mesons, such as the neutron/proton ratio \cite{LiBA96},
the neutron-proton differential flow \cite{LiBA00}, the
neutron-proton correlation function \cite{Che03a}, the
$t$/$^{3}$He ratio \cite {Che03b,Zha05}, the isospin diffusion
\cite{Shi03,LiBA05c,Tsa04,Che05a,Far91}, the isoscaling
\cite{Tsa01,She07,She04,She05,She06,Sou06,Igl06,Fev05,Tra06,Kow07},
and the $\pi ^{-}/\pi ^{+}$
\cite{LiBA02,Gai04,LiBA05a,LiQF05b,LiBA03,LiQF05c}, $\Sigma
^{-}/\Sigma ^{+}$ \cite{Qli05b} and $K^{0}/K^{+}$ ratios
\cite{Fer05}, etc..

According to studies based on transport model calculations, among
the known observables sensitive to the density dependence of the
nuclear symmetry energy, the neutron/proton ratio of squeezed-out
nucleons with high transverse momenta perpendicular to the
reaction plane has probably the highest sensitivity to the
symmetry energy. This is because symmetry potentials act directly
on nucleons, which are abundantly emitted in heavy-ion reactions
at intermediate energies. Moreover, squeezed-out nucleons in the
direction perpendicular to the reaction plane are mostly from the
high density region formed  during the earlier stage of the
reaction, and they are thus not much affected by spectator
nucleons. However, it is very challenging to measure these
neutrons as their measurements, especially the low energy ones,
always suffer from low detection efficiencies even with the most
advanced neutron detectors. Therefore, observables involving
neutrons normally have large systematic errors. Also, for charged
particles the Coulomb potential plays an important role, and it
sometimes competes strongly with the symmetry potential that one
is interested in. One thus has to disentangle carefully effects of
the symmetry potentials from those due to the Coulomb potential.
Therefore, it is very desirable to find experimental observables
which are less sensitive to the influence of both the Coulomb
force and the systematic errors associated with neutrons. The
double neutron/proton ratio of emitted nucleons taken from two
reaction systems using four isotopes of the same element, namely,
the neutron/proton ratio in the neutron-rich system over that in
the more symmetric system, is such an observable
\cite{hils88,Fam06}. Theoretical studies have shown that the
double neutron/proton ratio has about same sensitivity to the
density dependence of symmetry energy as corresponding single
ratio in the respective neutron-rich system used in the study
\cite{LiBA06b}. For the cleanest observable that is free from
final-state strong interactions, one can use the hard photons
\cite{Ylc07}. The sensitivity of hard photons to the symmetry
energy is, however, modest. Furthermore, theoretical studies of
hard photons in heavy ion collisions is further hampered by our
poor knowledge on the cross section for the elementary
neutron-proton bremsstrahlung as its uncertainty is at present
larger than that of the symmetry energy. Fortunately, using the
ratio of hard photons from two reactions can reduce significantly
the effect due to uncertainty in the cross section of the
elementary neutron-proton bremsstrahlung \cite{Ylc07}.

Experimentally, only very limited data, mostly from reactions with
stable beams at intermediate energies, have so far been available
for comparisons with results from theoretical calculations.
Nevertheless, these comparisons have already provided valuable
information on the symmetry energy at sub-saturation densities.
Among the available experiments, the isospin diffusion
\cite{Tsa04} and the isoscaling coefficient
\cite{Tsa01,Tsa01b,Bot02} of fragments have been most extensively
studied. In particular, a quite stringent constraint on the
symmetry energy at sub-normal densities has been obtained from
comparisons of the isospin diffusion data with transport model
calculations \cite{LiBA05c,Che05a}. Interesting information has
also been obtained from studying the isoscaling coefficient
\cite{Tsa01,Tsa01b,She07,She04,She05,She06,Sou06,Igl06,Fev05,Tra06,Kow07,Bot02,Dor06,Ma04b,Ma05,Rad07},
particularly its variation with the impact parameter \cite{Fev05}
and excitation energy \cite{Sou06}. However, the relation between
the isoscaling parameter and the symmetry energy is rather model
dependent. Besides the difficulties of obtaining the freeze-out
density of fragments, physical interpretations of the isoscaling
parameter also have some serious ambiguities \cite{LiBA06c}.
Because of the problems associated with all known probes of the
density dependence of the nuclear symmetry energy, a more
definitive constraint on the symmetry energy can only be obtained
by studying the correlations of many observables, similar to the
strategy used in the search for the signatures of the Quark-Gluon
Plasma formed in relativistic heavy ion collisions.

Among the proposed probes of the high density behavior of the
symmetry energy, the $n/p$ ratio of squeeze-out nucleons, the
$\pi^-/\pi^+$ ratio, and the neutron-proton differential flow are,
in our opinion, most promising. Since there is little experimental
information on the symmetry energy at supra-normal densities,
mainly because of the lack of experimental data, it is thus
desirable to study reactions with high energy radioactive beams
from accelerators that are being planned or proposed at many
laboratories, e.g., IMP/Lanzhou \cite{Xu08}, RIKEN \cite{Sak07},
NSCL/MSU \cite{Bic08} and GSI \cite{Tra08}.

\subsection{Nuclear symmetry energy and symmetry potential}
\label{minsp}

Since most of the isospin effects observed in heavy-ion reactions
result from the competition between the Coulomb and the symmetry
potential, it is instructive to examine the major features of the
symmetry potential. We will first consider momentum-independent
symmetry potentials as they have been used widely in many of
available studies, especially earlier ones. To mimic the
predictions of microscopic many-body theories, the symmetry energy
in early studies of the cooling and structure of neutron stars
\cite{Pra88b,Lat91} was parameterized as
\begin{eqnarray}\label{esymfu}
e_{\rm sym}(\rho)=(2^{2/3}-1)\frac{3}{5}E_{F}^{0}[u^{2/3}-F(u)]
+e_{\rm sym}(\rho_{NM})F(u),
\end{eqnarray}
with $F(u)$ having one of following three forms
\begin{eqnarray}\label{fu}
F_1(u)=\frac{2u^2}{1+u},~F_2(u)=u,~F_3(u)=u^{1/2},
\end{eqnarray}
where $u\equiv \rho/\rho_0$ is the reduced baryon density and
$E_F^0$ is the Fermi energy.

Without momentum dependence, the symmetry potential $V_{\rm
asy}^{q}$ is given by
\begin{eqnarray}
V^{q}_{\rm asy}(\rho,\delta)=\partial w_a(\rho,\delta)/\partial
\rho_{q},
\end{eqnarray}\label{nomu}
where $w_a(\rho,\delta)$ is the contribution of nuclear
interactions to the symmetry energy density, i.e.,
\begin{eqnarray}
w_a(\rho,\delta)=e_a\cdot \rho F(u)\delta^2,
\end{eqnarray}
and
\begin{eqnarray}
e_a\equiv e_{\rm
sym}(\rho_{NM})-(2^{2/3}-1)\,{\textstyle\frac{3}{5}}E_F^0.
\end{eqnarray}

The symmetry potentials corresponding to the three forms of $F(u)$
are, respectively,
\begin{eqnarray}\label{vasy}
V_{\rm asy}^{n(p)}&=&\pm 2e_a u^2\delta+e_a u^2\delta^2,\\\
V_{\rm asy}^{n(p)}&=&\pm 2 e_a u\delta,
\end{eqnarray}
and
\begin{eqnarray}
V_{\rm asy}^{n(p)}=\pm 2e_a
u^{1/2}\delta-\frac{1}{2}e_au^{1/2}\delta^2.
\end{eqnarray}
For the simplest form of $F(u)$, i.e., $F(u)=F_2(u)=u$, one has
\begin{eqnarray}\label{vasy1}
V_{\rm asy}^{n(p)}=\pm 2e_a u\delta=\pm
2e_a\frac{\rho_n-\rho_p}{\rho_0}.
\end{eqnarray}
This is the asymmetric part of the nuclear mean-field potential
used in Refs.\ \cite{LiBA95,Xu91,Tsa89,Dan92,LiBA93}. Farine {\it
et al.} \cite{Far91} used instead the asymmetric energy density
\begin{eqnarray}\label{simple}
w_{\rm sym}=c\rho[(\rho_n-\rho_p)/\rho_0]^2,
\end{eqnarray}
where the coefficient $c=e_{\rm sym}(\rho_0)-\frac{1}{3}E_F^0$ is
the the symmetry energy at normal nuclear matter density due to
nuclear interactions. It leads to the same symmetry potential at
$\rho\approx \rho_0$ as Eq.\ (\ref{vasy1}).

\begin{figure}[tbh]
\centering
\includegraphics[scale=0.6]{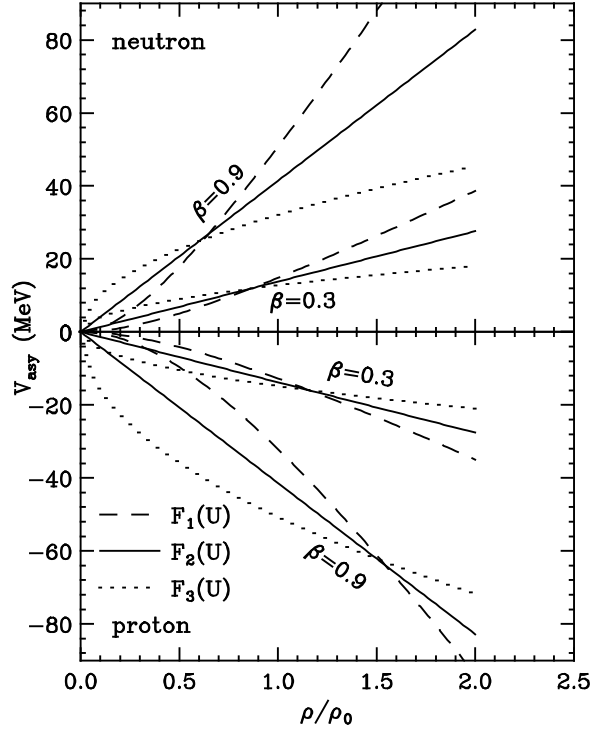}
\caption{Symmetry potentials for neutrons and protons
corresponding to the three forms of $F(u)$ (see text). Taken from
Ref.\ \protect\cite{LiBA97a}.} \label{spotential}
\end{figure}

The symmetry potentials $V_{\rm asy}^{n(p)}(\rho,\delta)$ at
different isospin asymmetries using the three forms of $F(u)$
given in Eq. (\ref{fu}) and $e_{\rm sym}(\rho_0)=32$ MeV are shown
in Fig.~\ref{spotential} as functions of density. It is seen that
the repulsive (attractive) mean-field potential for neutrons
(protons) depends sensitively on the form of $F(u)$, the neutron
excess $\delta$ (or $\beta$ used in Fig.~\ref{spotential}), and
the baryon density $\rho$. In collisions of neutron-rich nuclei at
intermediate energies, both $\delta$ and $\rho$ can be appreciable
in a large space-time volume where the isospin-dependent
mean-field potentials are strong. Since the symmetry potentials
have opposite signs for neutrons and protons, they affect
differently the reaction dynamics of neutrons and protons. For
protons, the nuclear mean-field potential also includes a Coulomb
term $V_C^p$. The competition between the Coulomb and the symmetry
potential then leads to possible differences in the yields and
energy spectra of protons and neutrons as well as on other isospin
effects. Because of the relatively small values of $V_{\rm
asy}^{n(p)}$, one needs to select observables that are sensitive
to the asymmetric part but not the symmetric part of the nuclear
{\sc eos}/potential in extracting information about the symmetry
energy/potential from the experimental data. In addition, these
observables should not depend strongly on other factors that
affect the reaction dynamics, such as the in-medium
nucleon-nucleon cross sections.

\begin{figure}[ht]
\centering
\includegraphics[scale=0.5,angle=-90]{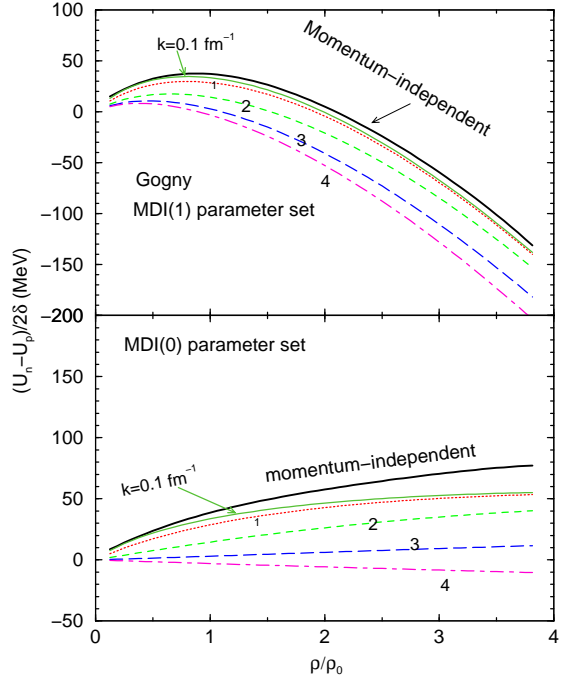}
\caption{(Color online) The strength of symmetry potential as a
function of density for the cases of with and without (solid line)
momentum dependence. Taken from Ref.~\cite{LiBA04a}.}
\label{usym2}
\end{figure}

As discussed in detail in Chapter~\ref{chapter_ria} and
Chapter~\ref{chapter_rmf}, significant progresses have been made
in recent years in understanding the momentum dependence of
symmetry potential. While there are still many uncertainties on
the latter, especially at high momenta, some transport models have
included the momentum dependence of symmetry potential. In some of
these models, the corresponding neutron-proton effective mass
splitting and the associated in-medium NN cross sections have also
been incorporated consistently too. Transport model calculations
with and without the momentum dependence of symmetry potential
generally give quite different predictions. Furthermore, the
various forms of the momentum dependence of symmetry potential
adopted in the different transport models also lead to different
predictions. It is therefore useful to compare the symmetry
potentials with and without the momentum dependence but
corresponding to the same density dependence of the symmetry
energy before we review the isospin effects in heavy-ion
reactions. For instance, it was shown in Ref.~\cite{LiBA04a} that
the following two momentum-independent symmetry potentials
\begin{eqnarray}
U^{MDI(1)}_{\rm
sym}(\rho,\delta,\tau)&=&4\tau\delta(3.08+39.6u-29.2u^2 +5.68u^3
\nonumber \\ && -0.52u^4) -\delta^2(3.08+29.2u^2-11.4u^3+1.57u^4)
\end{eqnarray}
and
\begin{eqnarray}
U^{MDI(0)}_{\rm
sym}(\rho,\delta,\tau)&=&4\tau\delta(1.27+25.4u-9.31u^2 +2.17u^3
\nonumber \\ && -0.21u^4) -\delta^2(1.27+9.31u^2-4.33u^3+0.63u^4),
\end{eqnarray}
which are obtained using Eq.~(\ref{nomu}) and the MDI symmetry
potential energy densities lead to the same density-dependent
symmetry energy shown in Fig.~\ref{MDIsymE} as the original MDI
symmetry potential shown in Fig.~\ref{MDIsymp}. Here, the symmetry
potential is denoted by $U$ instead of $V$. In Fig.\ \ref{usym2},
strengths of the symmetry potentials with and without the
momentum-dependence but corresponding to the same MDI $E_{\rm
sym}(\rho)$ with $x=0$ and $x=1$ are compared. It is clearly seen
that the symmetry potential without momentum dependence is
stronger than the one with momentum dependence. The difference
increases with increasing momentum because the strength of
momentum-dependent symmetry potential decreases with increasing
momentum. Different results are thus expected from calculation
with and without the momentum dependence of symmetry potential,
especially for high momentum nucleons. This expectation was
clearly verified in Refs.~\cite{LiBA04a,Che04,Riz05}. In the
following, we shall thus distinguish, whenever necessary
calculations carried out with or without the momentum dependence
of symmetry potential.

\subsection{Single and double neutron/proton ratios of pre-equilibrium nucleons}

The neutron/proton ratio of pre-equilibrium nucleons is among
first observables that were proposed as possible sensitive probes
of the symmetry energy \cite{LiBA97a}. The symmetry potential has
following effects on preequilibrium nucleons. First, it tends to
make more neutrons than protons unbound. One therefore expects
that a stronger symmetry potential leads to a larger ratio of free
neutrons to protons. Second, if both neutrons and protons are
already free, the symmetry potential makes neutrons more energetic
than protons. As an example, collisions of $^{112}{\rm
Sn}+^{112}{\rm Sn}$, $^{124}{\rm Sn}+^{124}{\rm Sn}$ and
$^{132}{\rm Sn}+^{132}{\rm Sn}$ at a beam energy of 40 MeV/nucleon
were studied using the BUU97 transport model in Ref.
\cite{LiBA97a}. To identify free nucleons, a phase-space
coalescence method was used, namely, a nucleon is considered free
if it is not correlated with another nucleon within a spatial
distance and a momentum distance of 3 fm and 300 MeV/c,
respectively, at a time of 200 fm/c after the initial contact of
the two reacting nuclei \cite{LiBA97a}.

\begin{figure}[tbh]
\centering
\includegraphics[scale=0.6]{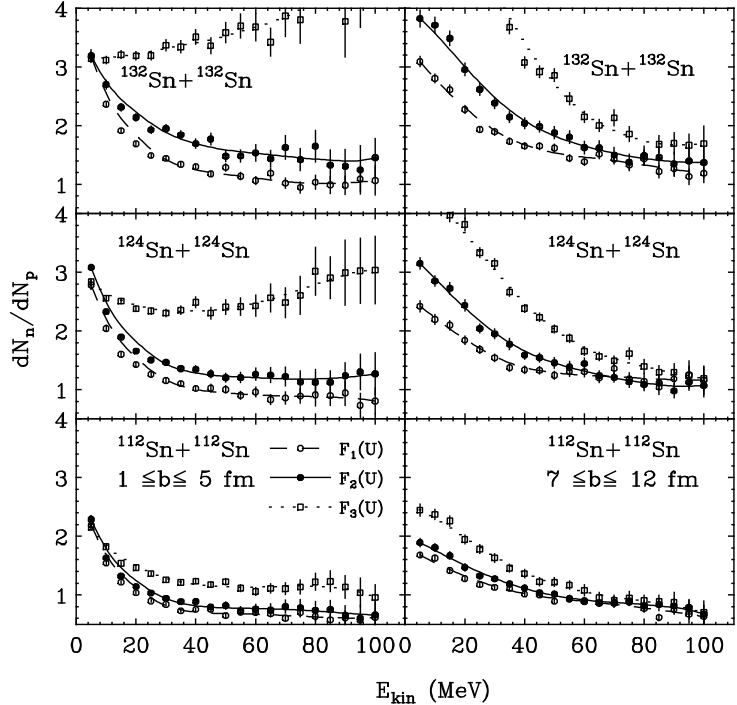}
\caption{The ratio of preequilibrium neutrons to protons as a
function of kinetic energy in central (left panels) and peripheral
(right panels) reactions using the three forms of $F(u)$. Taken
from Ref.~\cite{LiBA97a}.} \label{ratio1}
\end{figure}

The symmetry energy/potential effects are clearly demonstrated in
Fig.\ \ref{ratio1} where the ratios $R_{n/p}(E_{\rm kin})$ from
the collisions of $^{112}{\rm Sn}+^{112}{\rm Sn}$, $^{124}{\rm
Sn}+^{124}{\rm Sn}$, and $^{132}{\rm Sn}+^{132}{\rm Sn}$ are shown
as functions of kinetic energy. These results were obtained by
using the three forms of $F(u)$ for central (left panels) and
peripheral (right panels) collisions. The increase of these ratios
at lower kinetic energies in all cases is due to Coulomb repulsion
which shifts protons from lower to higher kinetic energies. The
different ratios calculated using different $F(u)$'s reflect
clearly the effect mentioned above, i.e., with a stronger symmetry
potential the ratio of preequilibrium neutrons to protons becomes
larger for more neutron rich systems.

It is interesting to note that effects due to different symmetry
potentials are seen in different kinetic energy regions for
central and peripheral collisions. In central collisions, effects
of the symmetry potential are most prominent at higher kinetic
energies. This is because most of finally observed free neutrons
and protons are already unbound in the early stage of the reaction
as a result of violent nucleon-nucleon collisions. The symmetry
potential thus mainly affects the nucleon energy spectra by
shifting more neutrons to higher kinetic energies with respect to
protons. In peripheral collisions, however, there are fewer
nucleon-nucleon collisions; whether a nucleon can become unbound
depends strongly on its potential energy. With a stronger symmetry
potential more neutrons (protons) become unbound (bound) as a
result of a stronger symmetry potential, but they generally have
smaller kinetic energies. Therefore, in peripheral collisions
effects of the symmetry potential show up chiefly at lower kinetic
energies. For the two systems with more neutrons the effects of
the symmetry potential are so strong that in central (peripheral)
collisions different forms of $F(u)$ can be clearly distinguished
from the ratio of preequilibrium neutrons to protons at higher
(lower) kinetic energies. However, because of energy thresholds in
detectors, it is difficult to measure low energy nucleons,
especially neutrons. Furthermore, the low energy spectrum also has
appreciable contribution from emissions at the later stage when
the system is in equilibrium. Therefore, the measurement of the
ratio $R_{n/p}(E_{\rm kin})$ in central heavy-ion collisions for
nucleons with energies higher than about 20 MeV is practically
more suitable for extracting the  {\sc eos} of asymmetric nuclear
matter.

The beam energy range where the symmetry potential is relevant for
heavy-ion collisions depends on the isospin asymmetry of the
reaction system and the observables to be studied. To observe
effects of a weak mean-field potential, such as the symmetry
potential, one needs to use relatively low beam energies so that
the dynamics is not dominated by nucleon-nucleon collisions. On
the other hand, to study the density dependence of the mean-field
potential and to reach a stronger mean-field potential the
reactions should be energetic enough to achieve sufficient
compression. Thus, it is necessary to study the beam energy
dependence of the isospin effects on preequilibrium nucleons.

Effects of the symmetry potential on the pre-equilibrium
neutron/proton ratio in heavy ion collisions have already been
seen in some experiments. For example, in early experiments of
heavy-ion collisions around the Fermi energy
\cite{hils88,hils87,hils92,hils95}, Hilscher {\it et al.} found
that the neutron/proton ratio, $((N/Z)_{\rm free})$, of
preequilibrium nucleons is consistently higher than that of the
projectile-target system, $(N_{P}+N_{t})/(Z_{p}+Z_{t})$, and
cannot be explained by the Coulomb effect alone. More
specifically, in the reaction of $^{12}{\rm C}+^{165}{\rm Ho}$ at
a beam energy of 32 MeV/nucleon, they have measured the neutron
and proton spectra at $14^{\circ}$ and energies between 70 and 100
MeV, and found that the multiplicity of neutrons is larger than
that of protons by a factor of $1.4\pm 0.2, 1.7\pm 0.3$ and
$2.4\pm 0.3$ for linear-momentum transfers of 52\%, 73\% and 93\%,
respectively. Therefore, the neutron to proton ratio is much
higher than that of the reaction system $(N/Z)_{\rm cs}$=1.42 in
central collisions corresponding to higher linear-momentum
transfer, This result cannot be explained by the standard Fermi
jet model for preequilibrium nucleon emission \cite{hils87}. On
the other hand, it is in agreement with the {\sc buu} predictions
discussed above.

\begin{figure}[tbh]
\centering
\includegraphics[scale=0.6]{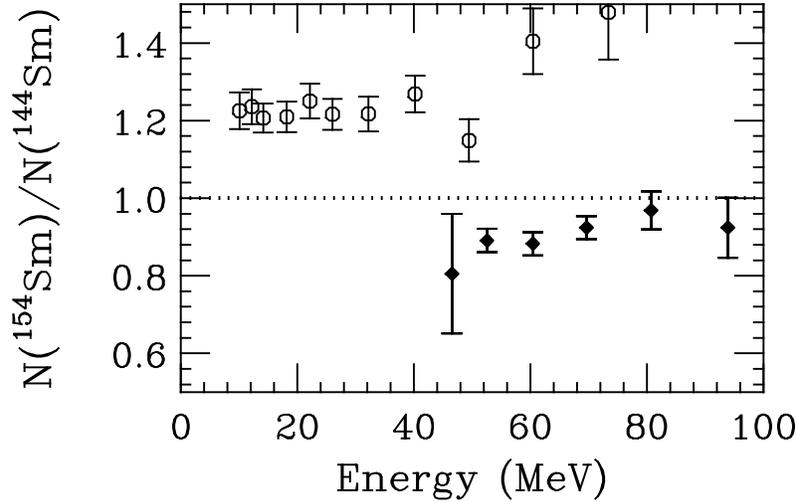}
\caption{The ratio of free neutrons (open circles) and protons
(solid diamond) as a function of the kinetic energy of the emitted
particle in central reactions of $^{32}{\rm S}+^{144, 154}{\rm
Sm}$ at a beam energy of 26 MeV/nucleon. Taken from Ref.
\protect\cite{hils88}.} \label{hilscher}
\end{figure}

Another interesting experimental observation is the ratio of free
neutrons and protons from two isotopic reaction systems $^{32}{\rm
S}+^{144,154}{\rm Sm}\rightarrow F,F+n,p$, where $F$ denotes
nuclear fragments, at $E_{\rm beam}/A=26$ MeV as shown in Fig.\
\ref{hilscher}. One notes that protons emitted at velocities
higher than the projectile velocity are mainly preequilibrium
particles. Several interesting observations can be made from these
data. First, the emission of protons in the neutron-richer system
$({\rm S}+^{154}{\rm Sm})$ is suppressed although both reaction
systems involve same number of protons. As discussed earlier,
protons feel an attractive symmetry potential, and the emission of
high energy protons is thus suppressed with respect to neutrons.
Correspondingly, the emission of high energy neutrons is enhanced,
and this is indeed consistent with the neutron data at higher
kinetic energies. Second, even for neutrons with lower kinetic
energies the ratio of free neutrons from the two systems is about
1.23 and is much larger than the ratio 1.12 of neutrons in the two
systems.

\begin{figure}[tbh]
\centering
\includegraphics[scale=0.6,angle=180]{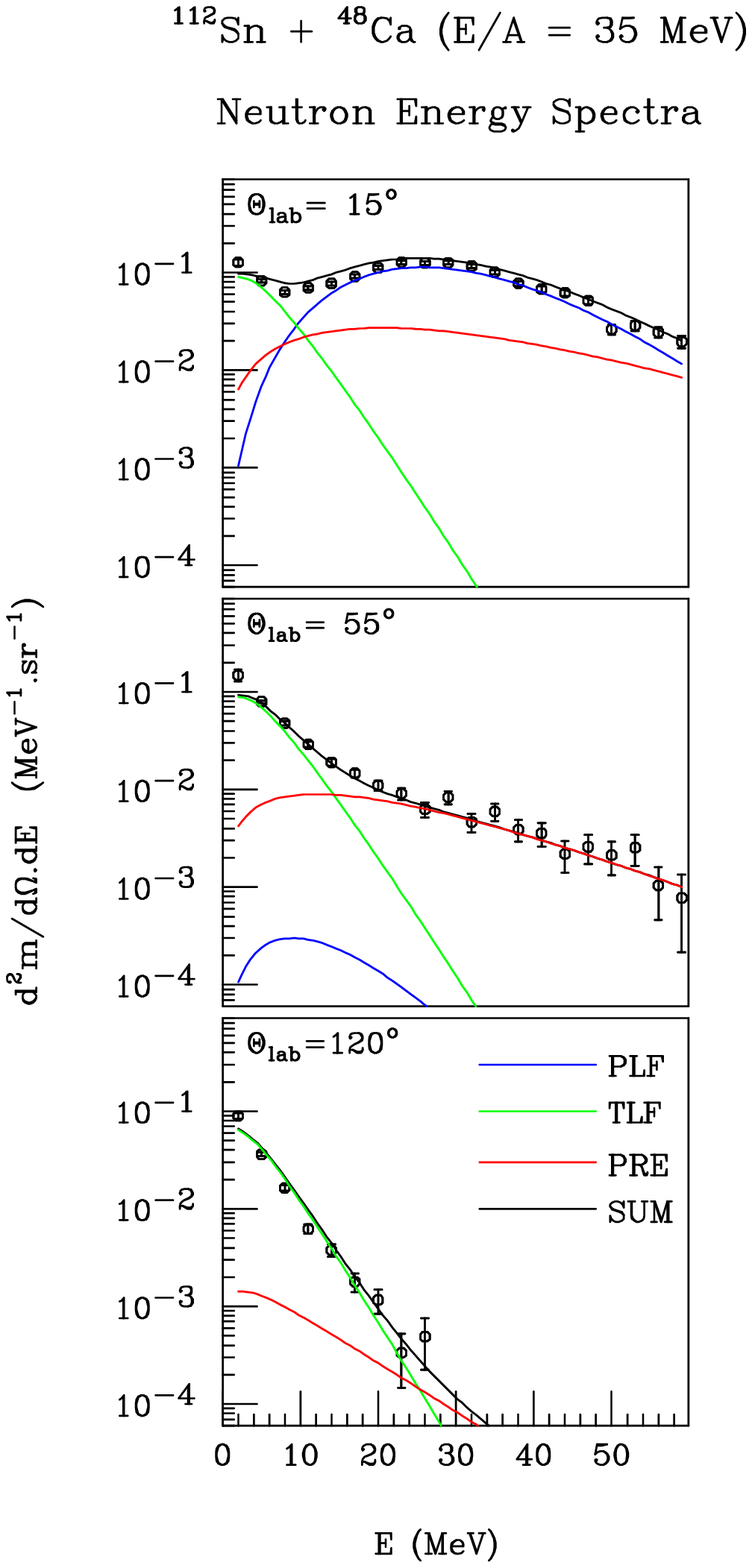}
\caption{(Color online) Neutron energy spectra for three lab
angles for the reaction $^{112}$Sn+$^{48}$Ca at 35 AMeV. Solid
lines are from the moving-source fits. Taken from
Refs.~\cite{udo1,udo2}.} \label{udo1}
\end{figure}

Schr\"oder {\it et al.} also studied systematically the spectra of
pre-equilibrium neutrons and protons in both isospin symmetric and
asymmetric collisions \cite{udo1,udo2}. Shown in Fig.\ \ref{udo1}
are the neutron spectra measured in coincidence with
projectile-like-fragments (PLFs) for the reaction
$^{112}$Sn+$^{48}$Ca. The solid lines in the figure indicate the
contributions assigned to the PLF, target-like-fragment (TLF), and
non-statistical (PRE) sources. Similar spectra were also obtained
for the reaction $^{112}$Sn+$^{40}$Ca, as well as for protons in
both reactions. From the information contained in the measured
post-evaporative properties of PLFs and the multiplicities and
energies of evaporated light particles, which are mostly neutrons,
the properties of the primary fragments were reconstructed for
different energy losses or reconstructed fragment excitation
energies.

\begin{figure}[tbh]
\centering
\includegraphics[scale=0.6,angle=180]{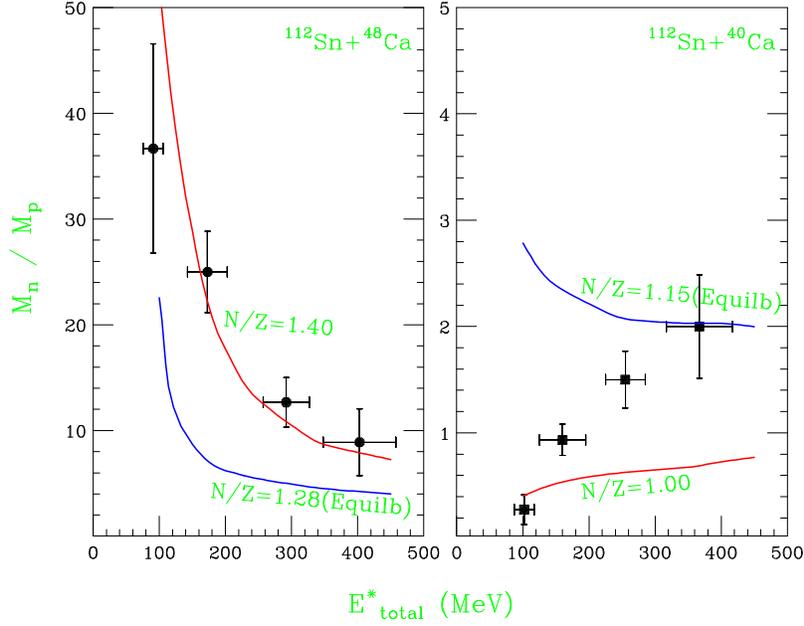}
\caption{(Color online) Neutron-to-proton multiplicity ratio vs.
total excitation energy for the two reactions $^{112}$Sn+$^{48}$Ca
(left panel) and $^{112}$Sn+$^{48}$Ca  (right panel) at 35 AMeV.
Taken from Refs.~\cite{udo1,udo2}.} \label{udof2}
\end{figure}

The relaxation of the isospin degree of freedom in the two
reactions $^{112}$Sn+$^{40,48}$Ca is illustrated in
Fig.~\ref{udof2}. Here, the neutron-to-proton multiplicity ratio
(circles with error bars) is plotted {\it {\it vs.}} excitation
energy. This ratio of multiplicities of evaporated particles,
combined with measured properties of the secondary
(post-evaporative) PLF, represents an observable sensitive to the
$N/Z$ ratio of the primary PLF. The sizable errors reflect overall
systematic uncertainties in the method, e.g., pertaining to the
level density parameters, etc. The abscissa scale in Fig.
\ref{udof2} can be thought of as an effective impact-parameter or
time scale. The sensitivity of the final neutron to proton
multiplicity ratio to the $N/Z$ ratio of the primary PLF can be
seen from the curves in the figure, which are obtained from
calculations assuming that the primary $N/Z$ ratio is that of
original projectile or of the combined system as labeled by
`Equilibrium'. It is seen that the $n/p$ ratio of pre-equilibrium
nucleons does not scale with the N/Z ratio of the combined system.
In particular, the double ratio $(n/p)_{^{48}{\rm Ca}+^{112}{\rm
Sn}}/(n/p)_{^{40}{\rm Ca}+^{112}{\rm Sn}}$ is about $4$ to $120$
depending on the impact parameter. This result is qualitatively
consistent with the observations made by Hilscher {\it et al.} and
both point towards the existence of a strong symmetry potential.
The explanations by Schr\"oder {\it et al.} about their
observations are as follows \cite{udo2}. The different behaviors
shown in Fig.~\ref{udof2} for the reactions $^{112}$Sn+$^{48}$Ca
and $^{112}$Sn+$^{40}$Ca indicates that the charge density
asymmetry is a dynamical variable, depending on the impact
parameter and evolving with the interaction time. The large
multiplicity ratios $M_n/M_p$ for $^{48}$Ca at low excitations
could be taken to reflect simply the large neutron excess of the
projectile and, perhaps to a lesser extent, the efficiency of the
Coulomb barrier in hindering the emission of protons. In such
picture, higher excitation energies would simply reduce the
Coulomb effect for proton emission. However, a similar
hypothetical scenario for the reaction $^{112}$Sn+$^{40}$Ca would
predict a trend opposite to that actually observed. One thus
concludes that the observed different evolution of the
neutron/proton multiplicity ratios must reflect differences in the
constitutions of the emitting PLFs, which change with dissipated
energy or impact parameter.

Since it is very difficult to measure neutrons accurately,
especially low energy ones, there have always been questions about
the experimental uncertainties in the measurements of the
neutron/proton ratio. As pointed out before, the ratio taken from
two reactions involving isotopes of same elements, as first done
by Hilsher {\it et al.}, can reduce the uncertainties. This
approach was also used in very recent experiments at the NSCL/MSU
by Famiano {\it et al.} \cite{Fam06} who have measured the double
neutron/proton ratio in central reactions of $^{124}$Sn$+^{124}$Sn
and $^{112}$Sn$+^{112}$Sn at a beam energy of $50$ MeV/nucleon.
The impact parameter for the selected data set was estimated to be
about $2$ fm.  Only neutrons and protons emitted between
$70^\circ$ and $110^\circ$ in the cms were measured to suppress
contributions from decays of the projectile-like fragment.
\begin{figure}[h]
\centering
\includegraphics[width=9.0cm]{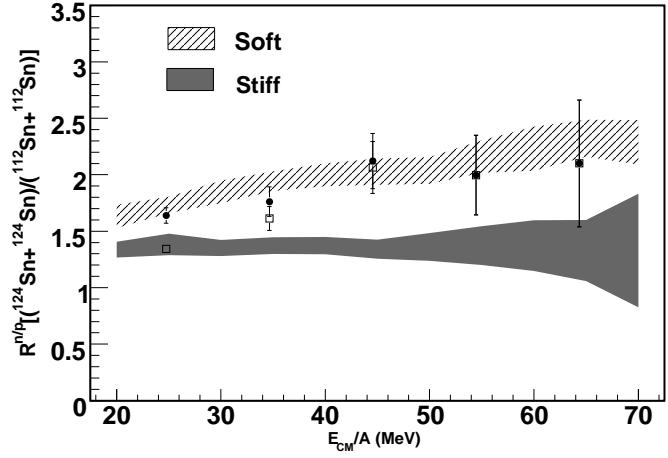}
\caption{Experimental double ratio $R_{124}$/$R_{112}$ at a beam
energy of 50 Mev/nucleon for free nucleons emitted in each
reaction compared to the double ratios of free nucleon yields
calculated earlier for the same systems but at 40 MeV/nucleon
using the BUU97 transport calculations with a momentum independent
interactions of Ref. \cite{LiBA97a}. The filled circles correspond
to double ratios of yields of transversely emitted free nucleons
and the open squares correspond to all nucleons including those
bound in clusters. Taken from Ref.~\cite{Fam06}.} \label{double}
\end{figure}
Shown by filled circles and open squares in Fig.~\ref{double} are
measured double ratios of free neutrons and protons from these two
reactions. Also shown in the figure are theoretical results from
the BUU97 calculations, obtained from the single $n/p$ spectra
from same reactions shown in Fig.~\ref{ratio1}. The `soft' and
`stiff' indicate results obtained using the functions $F_3(u)$ and
$F_1(u)$, respectively, in the symmetry energy/potential defined
in Section~\ref{minsp} and shown in Fig.~\ref{spotential}.
Although the data appears to favor the soft symmetry energy given
by $F_3(u)$, great cautions are required in drawing this
conclusion. First of all, the beam energy in the calculations is
40 MeV/nucleon, not the 50 MeV/nucleon as in experiments. As shown
in Ref.~\cite{LiBA00} (See also Fig.~\ref{ifragli} in the
following), the $n/p$ ratio of pre-equilibrium nucleons is higher
at lower beam energies. Secondly, the impact parameter used in the
calculations is between 1 and 5 fm, not exactly the 2 fm estimated
for the events selected in the data analyses. Thirdly, the data is
for transverse emissions only while the calculations are for
nucleons emitted in all directions, although calculations have
indicated a very weak angular dependence. Also, only nucleons
emitted with energies above 20 MeV/A are displayed as emission at
lower energies is significantly influenced by light cluster, which
is not modeled by the BUU97 calculations. Moreover, Coulomb
effects at low energies can adversely affect the comparison.

\begin{figure}[htp]
\centering
\includegraphics[scale=0.35,angle=-90]{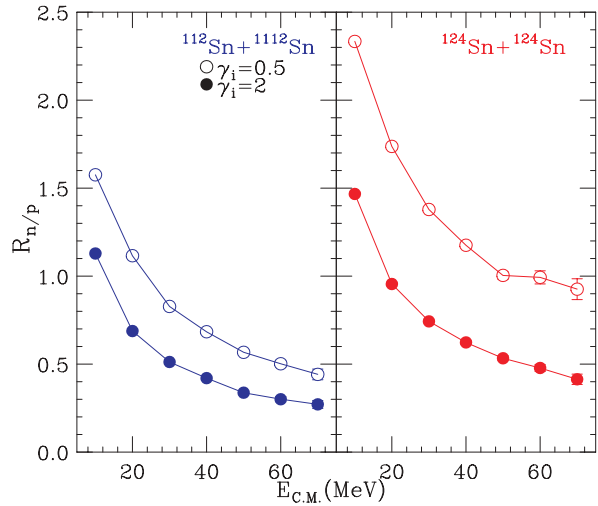}
\caption{(Color online) Ratios of neutron to proton yields from
the ImQMD model for the $^{112}$Sn+$^{112}$Sn reaction (left
panel) and the $^{124}$Sn+$^{124}$Sn reaction (right panel) as
functions of the kinetic energy of free nucleons emitted between
$70^{\circ}$ and $110^{\circ}$ in the center-of-mass system. Taken
from \cite{Zha08}.} \label{ZhangYX1}
\end{figure}

\begin{figure}[htp]
\centering
\includegraphics[scale=0.35,angle=-90]{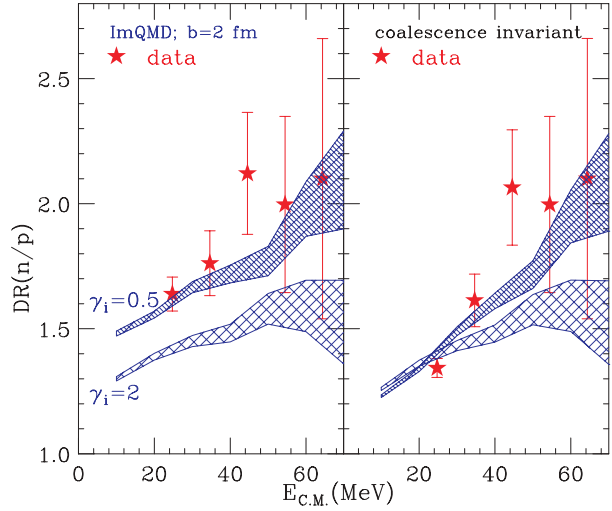}
\caption{(Color online) Free $n/p$ double ratio (left panel) and
coalescence invariant $n/p$ double ratios (right panel) as functions
of the kinetic energy of nucleons. The shaded regions are
calculations from the ImQMD simulations and the data (solid stars)
are from Ref.\cite{Fam06}. Taken from Ref.\cite{Zha08}.}
\label{ZhangYX2}
\end{figure}

Recent calculations by Zhang {\it et al.} \cite{Zha08} within the
Improved Quantum Molecular Dynamics (ImQMD) Model using a
momentum-dependent isoscalar but momentum-independent symmetry
potential give rather similar results as the BUU97 calculations with
momentum-independent potential and a similar density dependence in
the symmetry energy. Their results for the single and double $n/p$
ratios are shown in Figs.~\ref{ZhangYX1} and \ref{ZhangYX2},
respectively. To see the effects of cluster formation, both the
double $n/p$ ratio of free nucleons only (left panel) and that of
all nucleons including those bounded in clusters, i.e., the
so-called coalescence invariant $n/p$ double ratios (right panel),
are shown in Fig.~\ref{ZhangYX2}. From the comparison, it is seen
that the clustering effect on the double $n/p$ ratio is appreciable
mainly at lower kinetic energies in the case of the soft symmetry
energy with $\gamma=0.5$. At higher kinetic energies, the
coalescence invariant double $n/p$ ratio is less affected by
clusters and retains its sensitivity to the symmetry energy as in
the free double n/p ratio. It is thus more useful to measure
accurately the high energy single and/or double n/p ratio. However,
the above conclusions based on the momentum-independent symmetry
potential need to be taken with great cautions since the momentum
dependence of the symmetry potential, which would lead to very
different magnitude and density slope of the symmetry potential as
well as the effective masses of nucleons, is expected to affect
significantly the $n/p$ ratio of pre-equilibrium nucleons. Indeed,
as will be discussed in the following, new calculations within the
IBUU04 transport model using the momentum-dependent MDI interaction
introduced in Chapter~\ref{chapter_eos}, which leads to a decrease
of the strength of the symmetry potential with increasing nucleon
momentum, have indicated that the $n/p$ ratio of pre-equilibrium
nucleons is significantly reduced compared to earlier BUU97 results
using momentum-independent interactions \cite{LiBA06b}.

\begin{figure}[tbh]
\centering
\includegraphics[scale=0.9]{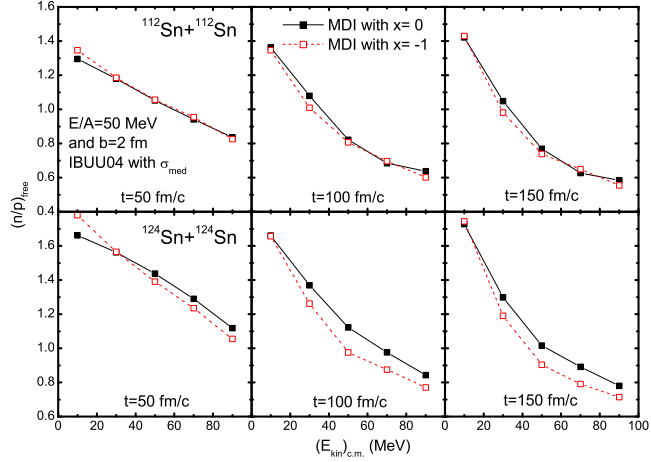}
\caption{(Color online) Time evolution of the neutron/proton ratio
of free nucleons as a function of kinetic energy obtained with the
MDI interaction of $x=0$ (filled square) and $x=-1$ (open square)
for the reaction of $^{124}$Sn$+^{124}$Sn (lower panels) and
$^{112}$Sn$+^{112}$Sn (upper panels) at $50$ MeV/nucleon and an
impact parameter of $2$ fm. Taken from Ref.~\cite{LiBA06b}.}
\label{dnpfigure2}
\end{figure}

Shown in Fig.\ \ref{dnpfigure2} are the IBUU04 model predictions
\cite{LiBA06b} on the time evolutions of the single neutron/proton
ratios versus the nucleon kinetic energy in the center-of-mass
frame of respective reaction. It is seen that the neutron/proton
ratio becomes stable after about $100$ fm/c. As one expects, the
neutron/proton ratio in the neutron-richer system is more
sensitive to the symmetry energy, especially for fast nucleons.
With the softer symmetry energy of $x=0$, the symmetry energy and
the magnitude of the symmetry potential are larger at
sub-saturation densities but smaller at supra-saturation densities
compared to the case with $x=-1$. Since the maximum density
reached in the two reactions studied here is about $1.2\rho _{0}$,
a higher neutron/proton ratio of free nucleons is expected for the
softer symmetry energy of $x=0$ due to the stronger repulsive
(attractive) symmetry potential for neutrons (protons). For the
more symmetric system $^{112}$Sn$+^{112}$Sn, effects of the
symmetry energy are negligible because of the small isospin
asymmetry in the system. The rise of the neutron/proton ratio at
low energies in both systems is due to the Coulomb force which
pushes protons away from the center of mass of the reaction. These
features are consistent with those found in an earlier study using
a momentum-independent transport model \cite{LiBA97a}. Unlike the
earlier results, however, the observed symmetry energy effect is
only about $10\%$ to $15\%$ even for the most energetic nucleons
in the $^{124}$Sn$+^{124}$Sn reaction. The larger symmetry energy
effect in the earlier study with the BUU97 model is due to a much
wider uncertainty range between approximately
$30(\rho/\rho_0)^{0.5}$ and $30(\rho/\rho_0)^{1.6}$ that was used
for the symmetry energy than severely constrained symmetry energy
from the isospin diffusion data that was used in the more recently
study using the IBUU04 model. Moreover, the MDI symmetry potential
shown in Fig.\ \ref{MDIsymp} and used in the IBUU04 model
decreases with increasing momentum, and its effects on the $n/p$
ratio of pre-equilibrium nucleons are thus much reduced compared
to calculations using the momentum-independent symmetry potential
shown in Fig.\ \ref{spotential} and used in the BUU97 model.

\begin{figure}[tbh]
\centering
\includegraphics[scale=1.]{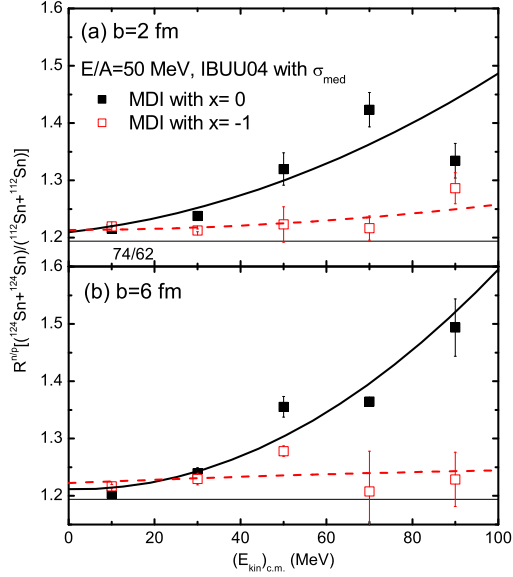}
\caption{{\protect\small (Color online) The double neutron/proton
ratio of free nucleons taken from the reactions of
}$^{124}${\protect\small Sn}$+^{124}${\protect\small Sn and
}$^{112}${\protect\small Sn}$+^{112}${\protect\small Sn at
}$50${\protect\small \ MeV/nucleon and impact parameters of
}$2${\protect\small \ fm (upper panel) and }$6${\protect\small \
fm (lower panel). Taken from Ref.~\cite{LiBA06b}.}}
\label{dnpfigure3}
\end{figure}

Fig.\ \ref{dnpfigure3} shows the double neutron/proton ratios
obtained from the IBUU04 model calculations \cite{LiBA06b} for the
$^{124}$Sn$+^{124}$Sn and $^{112}$Sn$+^{112}$Sn reactions at a
beam energy of $50$ MeV/nucleon and impact parameters of $2$ fm
(upper panel) and $6$ fm (lower panel) . As a reference, a
straight line at $74/62$ corresponding to the double
neutron/proton ratio of the entrance channel is also drawn. Below
the pion production threshold, the double neutron/proton ratio of
nucleon emissions is expected to be a constant close to this value
if one neglects effects due to both the Coulomb and symmetry
potentials. The observed double neutron/proton ratios, especially
at low kinetic energies with $x=-1$, at both impact parameters
indeed have almost a constant value. They are, however, slightly
above the straight line at $74/62$ as a result of the appreciable
repulsive/attractive symmetry potential on neutrons/protons in the
$^{124}$Sn$+^{124}$Sn reaction. Since the Coulomb effects are
largely cancelled out for the double ratios in the two reactions
involving isotopes of the same element, more energetic nucleons
are thus more affected by the symmetry potential as they go
through the denser regions of the reactions. The effect becomes
stronger as the $x$ parameter changes from $x=-1$ to $x=0$ because
the case with $x=0$ corresponds to a stronger symmetry potential
at sub-saturation densities compared to the case with $x=-1$. As a
result, the double neutron/proton ratios, especially for energetic
nucleons that are mostly from pre-equilibrium emissions, increase
when the $x$ parameter is changed from $x=-1$ to $x=0$. At both
impact parameters, the increase is about $10\%-15\%$, so the
expected sensitivity to the symmetry energy is about the same as
the single neutron/proton ratio. The calculated results with both
$x=0$ and $x=-1$ are, however, significantly lower than the
NSCL/MSU data from Famiano {\it et al.} \cite{Fam06}. It thus
remains a serious puzzle why the same transport model using the
same density dependence of the nuclear symmetry energy can not
reproduce both the isospin diffusion and the double neutron/proton
ratios data simultaneously, assuming that these data are
consistent. It should also be mentioned that since the
neutron/proton ratio at kinetic energies less than about $50$ MeV
is rather insensitive to the symmetry energy in reactions at a
beam energy of $50$ MeV/nucleon, neutron detectors with a
threshold energy of $50$ MeV is sufficient for the study discussed
here. However, as we shall discuss in the following, for reactions
at beam energies above the pion production threshold even the low
energy neutrons are useful.

\begin{figure}[tbh]
\centering
\includegraphics[scale=1.]{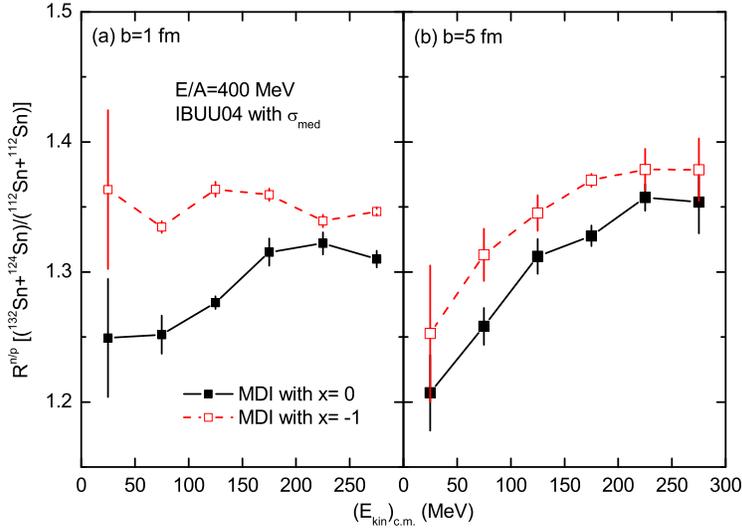}
\caption{(Color online) The double neutron/proton ratio of free
nucleons taken from the reactions of $^{132}$Sn$+^{124}$Sn and
$^{112}$Sn$+^{112}$Sn at 400 MeV/nucleon and impact parameters of
$1$ fm (left panel) and $5$ fm (right panel). Taken from
Ref.~\cite{LiBA06b}.} \label{dnpfigure4}
\end{figure}

For beam energies above the pion production threshold, the
reference line at $74/62$ is no longer useful. Instead, the $\pi
^{-}/\pi ^{+}$ ratio itself is a promising probe of the symmetry
energy at high densities. Shown in Fig.\ \ref{dnpfigure4} are the
double neutron/proton ratios from the reactions of
$^{132}$Sn$+^{124}$Sn and $^{112}$Sn$+^{112}$Sn at a beam energy
of $400$ MeV/nucleon and impact parameters of $1$ fm (left panel)
and $5$ fm (right panel). At both impact parameters, effects of
the symmetry energy are about $5\%-10\%$ when changing from the
case with $x=0$ to $x=-1$. One notices here that for such high
energy heavy-ion collisions the low energy nucleons have the
largest sensitivity to the variation of the symmetry energy. In
fact, the neutron/proton ratio of midrapidity nucleons, which have
gone through the high density phase of the reaction, are known to
be most sensitive to the symmetry energy \cite{LiBA05e}. Compared
to the results at the beam energy of $50$ MeV/nucleon, one can see
a clear inversion in the dependence of the double neutron/proton
ratio on the $x $ parameter, namely the double ratio is lower at
$50$ MeV/nucleon but higher at $400$ MeV/nucleon with $x=-1$ than
with $x=0$. The maximum density reached at the beam energies of
$50$ and $400$ MeV/nucleon are about $1.2\rho _{0}$ and $2\rho
_{0}$, respectively. The inversion clearly shows that the double
neutron/proton ratio reflects closely the density dependence of
the symmetry energy. This observation also indicates that
systematic studies of the double neutron/proton ratio over a broad
beam energy range will be very useful for mapping out the density
dependence of the symmetry energy.

\subsection{Light clusters and IMF production in intermediate-energy heavy
ion collisions}

While it is hard to measure the neutron/proton ratio, it is
relatively easier to measure ratios of charged particles. In
particular, ratios of light mirror nuclei are expected to provide
similar information as the neutron/proton ratio if the production
of these nuclei is based on the coalescence picture. Light cluster
production has been extensively studied in experiments involving
heavy-ion collisions at all energies, e.g., see Ref.
\cite{Hodgson03} for a review. A popular model for describing the
production of light clusters in these collisions is the
coalescence model, e.g., see Ref. \cite{Cse86}, which has been
used at both intermediate \cite{Gyu83,Koch90,Indra00} and high
energies \cite{Mattie95,Nagle96}. In this model, the probability
for producing a cluster is determined by the overlap of its Wigner
phase-space density with the nucleon phase-space distribution at
freeze out. Explicitly, the multiplicity of a $M$-nucleon cluster
in a heavy-ion collision is given by \cite{Mattie95}
\begin{eqnarray}
N_{M}=G\int d\mathbf{r}_{i_{1}}d\mathbf{q}_{i_{1}}\cdots d\mathbf{r}%
_{i_{M-1}}d\mathbf{q}_{i_{M-1}}\langle
\underset{i_{1}>i_{2}>...>i_{M}}{\sum
}\rho _{i}^{W}(\mathbf{r}_{i_{1}},\mathbf{q}_{i_{1}}\cdots \mathbf{r}%
_{i_{M-1}},\mathbf{q}_{i_{M-1}})\rangle .
\end{eqnarray}%
In the above, $\mathbf{r}_{i_{1}},\cdots ,\mathbf{r}_{i_{M-1}}$ and $\mathbf{%
q}_{i_{1}},\cdots ,\mathbf{q}_{i_{M-1}}$ are, respectively, the
$M-1$ relative coordinates and momenta taken at equal time in the
$M$-nucleon rest frame; $\rho _{i}^{W}$ is the Wigner phase-space
density of the $M$-nucleon cluster; and $\langle \cdots \rangle $
denotes event averaging. The spin-isospin statistical factor for
the cluster is given by $G$, and its value is $3/8$ for deuteron
and $1/3$ for triton or $^{3}$He, with the latter including the
possibility of coalescence of a deuteron with another nucleon to
form a triton or $^{3}$He \cite{Polleri99}.

\subsubsection{Ratios of light mirror nuclei}

Using the coalescence model, effects of the symmetry energy on the
$t/^{3}$He ratio was studied with the IBUU04 transport
model~\cite{Che03b,Che04}. In particular, effects of the momentum
dependence of the symmetry potential were examined. Besides the
MDI symmetry potential corresponding to the soft symmetry energy,
i.e., the MDI with $x=1$ and that for the hard symmetry energy,
i.e., the MDI with $x=-2$, two other potentials having the same
symmetry energy were used for comparisons. One of the latter is
taken to be $U_{\rm noms}(\rho ,\delta ,\mathbf{p},\tau )\equiv
U_{0}(\rho ,\mathbf{p})+U_{\rm sym}(\rho ,\delta ,\tau )$ with its
isoscalar part taken from the original MDYI interaction
\cite{Gal90}, i.e.,
\begin{eqnarray}
U_{0}(\rho ,\mathbf{p})=-110.44u+140.9u^{1.24}-\frac{130}{\rho
_{0}}\int d^{3}\mathbf{p}^{\prime
}\frac{f(\mathbf{r},\mathbf{p}^{\prime })}{1+(
\mathbf{p}-\mathbf{p}^{\prime })^{2}/(1.58p_{F}^{0})^{2}},
\end{eqnarray}%
which has a compressibility $K_{0}=215$ MeV and is almost the same
as the momentum-dependent isoscalar potential given by the MDI
interaction. For the momentum-independent symmetry potential
$U_{\rm sym}(\rho ,\delta ,\tau )$, it is obtained from $U_{\rm
sym}(\rho ,\delta ,\tau )=\partial W_{\rm sym}/\partial \rho
_{\tau }$ using the isospin-dependent part of the potential energy
density $W_{\rm sym}=E_{\rm sym}^{\rm pot}(\rho )\cdot \rho \cdot
\delta ^{2}$, where $E_{\rm sym}^{\rm pot}(\rho )$ is given by the
contributions of the MDI interactions to the symmetry energy,
i.e.,
\begin{eqnarray}
E_{\rm sym}^{\rm pot}(\rho
)=3.08+39.6u-29.2u^{2}+5.68u^{3}-0.523u^{4} \text{ (MeV) },\text{
} \label{mdi1}
\end{eqnarray}
for the soft symmetry energy, i.e., the MDI with $x=1$ and
\begin{eqnarray}
E_{\rm sym}^{\rm pot}(\rho
)=-1.83-5.45u+30.34u^{2}-5.04u^{3}+0.45u^{4}\text{ (MeV) }
\label{mdim2}
\end{eqnarray}
for the hard symmetry energy, i.e., the MDI with $x=-2$. In the
above, $u\equiv \rho /\rho _{0}$ is the reduced nucleon density.
The other potential considered is the usual momentum-independent
soft nuclear isoscalar potential with $K_{0}=200$ MeV (SBKD),
firstly introduced by Bertsch, Kruse and Das Gupta \cite{Ber84},
i.e.,
\begin{eqnarray}
U(\rho )=-356~u+303~u^{7/6}.
\end{eqnarray}
Comparing results from these potentials with those from the MDI
interaction then allows one to study the effects due to the
momentum dependence of the nuclear symmetry potential and the
momentum dependence of the isoscalar nuclear potential,
respectively.

\begin{figure}[th]
\centering
\includegraphics[scale=1.4]{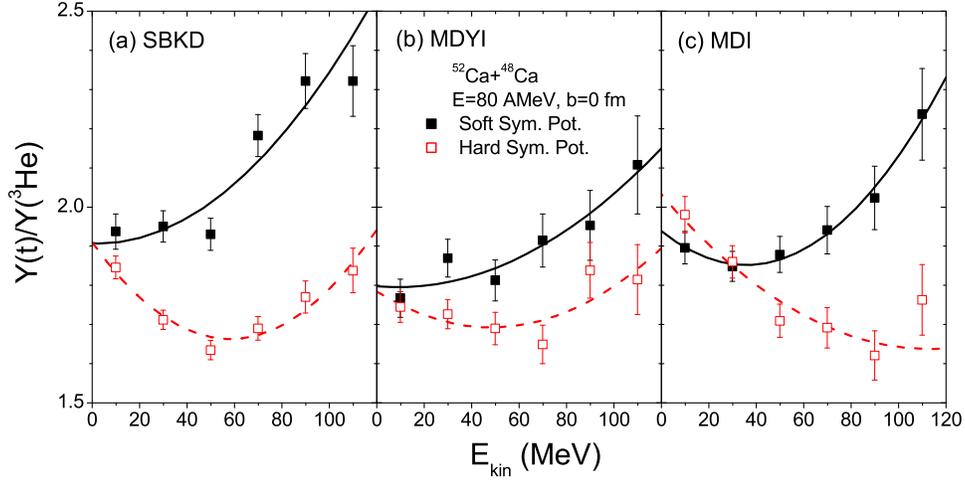}
\caption{(Color online) The t/$^{3}$He ratio as a function of the
cluster kinetic energy in the center-of-mass system for different
interactions (a) SBKD, (b) MDYI, and (c) MDI with the soft (solid
squares) and stiff (open squares) symmetry energies. The lines are
drawn to guide the eyes. Taken from Ref.~\cite{Che04}.}
\label{RtHe3}
\end{figure}

Shown in Fig. \ref{RtHe3} are the t/$^{3}$He ratios as functions
of the cluster kinetic energy in the center-of-mass system for the
SBKD, MDYI and MDI interactions with the soft (solid squares) and
stiff (open squares) symmetry energies. For all nuclear
potentials, the ratio t/$^{3}$He obtained with different symmetry
energies is seen to exhibit very different energy dependence.
While the t/$^{3}$He ratio increases with kinetic energy for the
soft symmetry energy, it decreases and/or increases weakly with
kinetic energy for the stiff symmetry energy. For both soft and
stiff symmetry energies, the ratio t/$^{3}$He is larger
than the neutron to proton ratio of the whole reaction system, i.e., \textsl{%
N/Z}$=1.5$. This is in agreement with results from both
experiments and the statistical model simulations for other
reaction systems and incident energies
\cite{Hagel00,Cibor00,Sobotka01,Vesel01,Chomaz99}. It is
interesting to note that the t/$^{3}$He ratio shows very different
energy dependence for the soft and hard symmetry potentials,
although the yield of light clusters is not so sensitive to the
density dependence of the symmetry potential for the MDI
interaction \cite{Che04}. This is related to the different
momentum dependence of the symmetry potential in the MDI
interaction, especially at low densities \cite{Che04}.

\begin{figure}[th]
\centering
\includegraphics[scale=1.0]{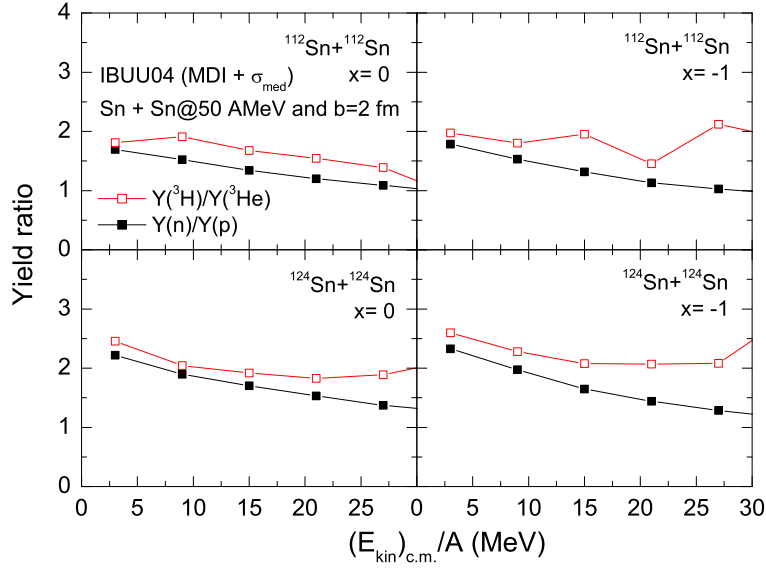}
\caption{{\protect\small (Color online) Ratios of t/$^3$He and
{\it n/p} for the $^{112}$Sn+$^{112}$Sn reaction (upper panels)
and the $^{124}$Sn+$^{124}$Sn reaction (lower panels) at 50
MeV/nucleon and an impact parameter of $2$ fm as functions of
energy per emitted nucleon from IBUU04 calculations using the MDI
interaction with $x=0$ and $x=-1$. Taken from Ref. \cite{Che08}.}}
\label{RtHe3np}
\end{figure}

As pointed out in Ref. \cite{Mattie95}, the validity of the
coalescence model introduced above is based on the assumption that
nucleon emissions are statistically independent and the binding
energies of formed clusters as well as the quantum dynamical
effect only play minor roles. Since the binding energies of triton
and $^{3}$He are $7.72$ \textrm{MeV} and $8.48$ \textrm{MeV},
respectively, the coalescence model for the production of these
light clusters in heavy-ion collisions is thus applicable if the
colliding system or the emission source has an excitation energy
per nucleon or a temperature above $\sim 9$ MeV. Furthermore, the
coalescence model is a perturbative approach and is valid only if
the number of clusters formed in the collisions is small. As shown
in Ref. \cite{Che03b,Che04}, this condition is indeed satisfied
for energetic tritons and $^3$He in the collisions considered
above. However, at lower incident energies, the coalescence model
based on the Wigner formulism introduced above may become invalid,
so other approaches have to be used. Shown in Fig. \ref{RtHe3np}
are the ratios of t/$^3$He and $n/p$ for the $^{112}$Sn+$^{112}$Sn
reaction and the $^{124}$Sn+$^{124}$Sn reaction from IBUU04
calculations based on the isospin-dependent phase-space
coalescence model, that has been used extensively in QMD-like
models \cite{Che98,Zha99}. Interestingly, one can see that the
ratios corresponding to the A=3 mirror nuclei indeed display a
similar energy dependence to that of free nucleons, especially for
$x=0$. It should be noted that besides the neglect of binging
energy effect, the effect of secondary decays is not included in
the above isospin-dependent phase-space coalescence analyses.

\subsubsection{The N/Z Ratio of intermediate mass fragments}

\begin{figure}[h]
\centering
\includegraphics[scale=0.33]{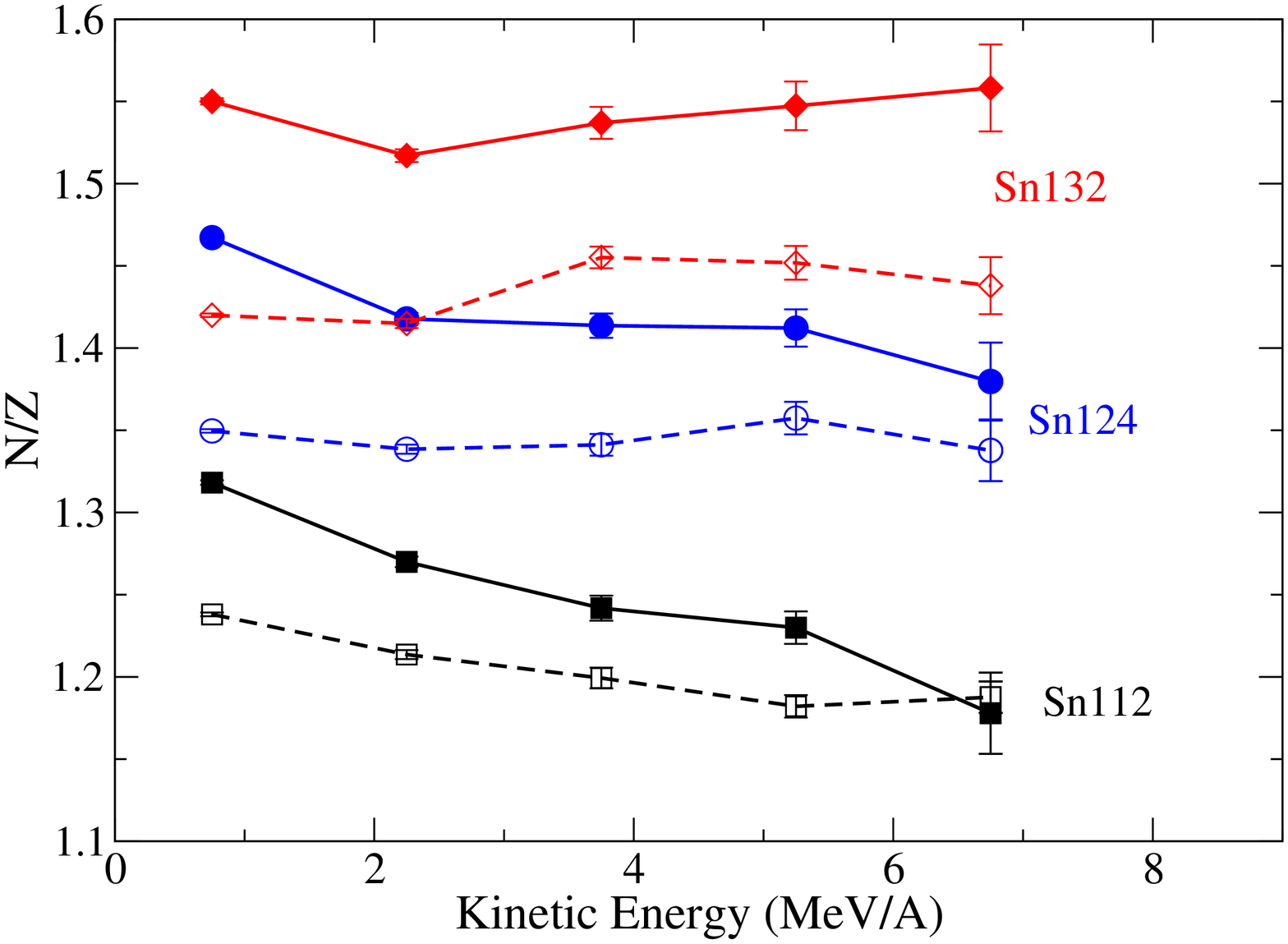}
\includegraphics[scale=0.78]{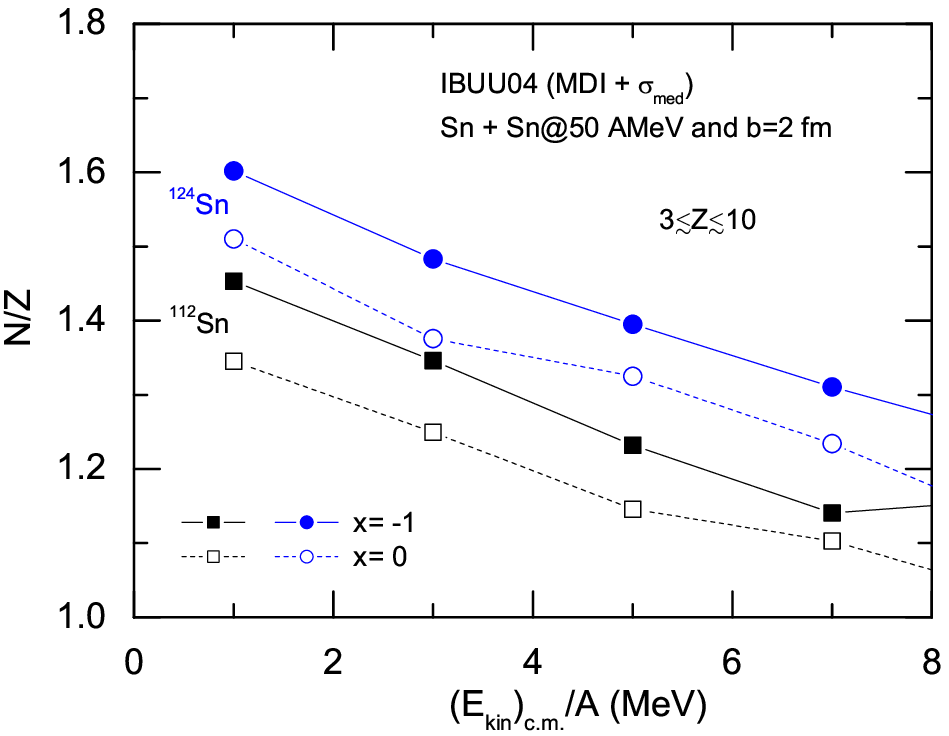}
\caption{(Color online) Left window: The fragment $N/Z$ as a
function of kinetic energy for reactions at 50 AMeV and b=2 fm from
the BNV calculations using momentum-independent potentials (see
text). Solid lines and full symbols are for 'asy-stiff' symmetry
energy; dashed lines and open symbols are for 'asy-soft' symmetry
energy \cite{Col07}. Right window: Same as left window but from the
IBUU04 using the MDI interaction with $x=0$ (open symbols) and
$x=-1$ (solid symbols) \cite{Che08}.} \label{colonna1}
\end{figure}

The $N/Z$ ratios of intermediate mass fragments can provide
information complementary to that extracted from the ratios of
neutron/proton and/or light mirror nuclei as a result of the
conservations of total charge and mass. This was demonstrated
nicely by the Catania group \cite{Col07} using the BNV model.
Unlike the IBU004 model, in which fluctuations are mainly due to
the finite number of test particles (200 here) used and the random
NN collisions in each run of the reaction simulation, the BNV
model includes explicitly the statistical fluctuations during the
collisions. Two types of symmetry energy were considered in this
study; one with a rapidly increasing behaviour in density, roughly
proportional to $\rho^2$ (asystiff), and the other with a
saturation above normal density (asysoft, $SKM^*$). The two
parameterizations obviously cross at normal density but the ranges
spanned by these two parameterizations is far beyond the available
constraints on the symmetry energy obtained from studying the
isospin diffusion \cite{LiBA05c} in intermediate-energy heavy ion
collisions. In their analysis, the $N/Z$ ratio of all fragments
with charge between $3$ and $10$ (intermediate mass fragments
(IMF)) was considered. As a measure of the isotopic composition of
the IMF's, the sums of neutrons, $N = \sum_i N_i$, and protons, $Z
= \sum_i Z_i$, of all IMF's in a given kinetic energy bin were
counted in each event. The ratio $N/Z$ averaged over the ensemble
of events was then studied as a function of the kinetic energy per
nucleon. The results are shown in the left window of Fig.\
\ref{colonna1} for the three reactions $^{112}$Sn+$^{112}$Sn,
$^{124}$Sn+$^{124}$Sn, and $^{132}$Sn+$^{132}$Sn, and the above
two symmetry energies. Indeed, the $N/Z$ ratio of the IMFs is seen
to be quite sensitive to the symmetry energy. One also sees that
the ratio decreases with the fragment kinetic energy, especially
in the asy-stiff case, for the neutron-poor system, but becomes an
increasing function of the fragment kinetic energy in systems with
larger initial asymmetry. The latter behavior is due to the larger
repulsive symmetry potential for neutrons in more neutron-rich
systems. However, this study was based on momentum-independent
symmetry potential and isoscalar potential. As in the case for the
neutron/proton ratio of pre-equilibrium nucleons, including the
momentum dependence in the symmetry potential and the isoscalar
potential would significantly influence the isospin effect on the
$N/Z$ ratio of IMFs. This is demonstrated in the right window of
Fig.\ \ref{colonna1}, which shows the results for
$^{124}$Sn$+^{124}$Sn and $^{112}$Sn$+^{112}$Sn based on the
IBUU04 model using the MDI interaction with $x=0$ and $x=-1$
\cite{Che08}. The clusters in this study were constructed by means
of an isospin-dependent phase-space coalescence model
\cite{Che98,Zha99}, in which a physical fragment is formed from a
cluster of particles with relative momenta smaller than
$P_{0}=263~{\rm fm}/c$ and relative distances smaller than
$R_{0}=3~{\rm fm}$ if the composition of the cluster can be
identified with an isotope in the nuclear data sheets and also if
its root-mean-square radius satisfies the condition $R_{\rm
rms}=1.14A^{1/3}$, where $A$ is the mass number of the cluster. It
is seen that the symmetry energy effects on the $N/Z$ ratio of the
IMFs are again observed, and the results seem to exhibit a
stronger energy dependence than those from the BNV calculations
with momentum-independence mean-field potentials.

\subsection{Isospin fractionation in heavy-ion reactions}

One of especially interesting new features of a dilute asymmetric
nuclear matter is the isospin-fractionation (IsoF) when it
undergoes the LG phase transition
\cite{Mul95,LiBA97b,Bar98,LiBA00}. The non-equal partition of the
system's isospin asymmetry with the gas phase being more
neutron-rich than the liquid phase has been found to be a general
phenomenon in essentially all thermodynamical models as well as in
simulations of heavy-ion reactions, for reviews see, e.g., Refs.
\cite{LiBA98,LiBA01b,Bar05,Cho04,Das05,Cho06}. Indications of the
IsoF in the nuclear system have been reported since early 1980's,
although their interpretations have not always been unique
\cite{Cho04,Ran81}. As discussed in the previous section, recent
experiments have confirmed unambiguously the IsoF phenomenon
\cite{Cho06}, particularly the the experiments and analyses by Xu
{\it et al.} \cite{Xu00} at the NSCL/MSU based on measured
isotope, isotone and isobar ratios. It was clearly found that the
gas phase was significantly enriched in neutrons relative to the
liquid phase that is represented by bound nuclei. However, in all
earlier studies in the literature, only the average neutron/proton
ratios integrated over the nucleon momentum in the liquid and gas
phases were studied, and they are referred in the above as the
integrated IsoF. The differential IsoF as a function of nucleon
momentum carries completely new and very interesting physics in
its fine structure \cite{LiBA07a}.

\begin{figure}[tbh]
\centering\includegraphics[scale=1]{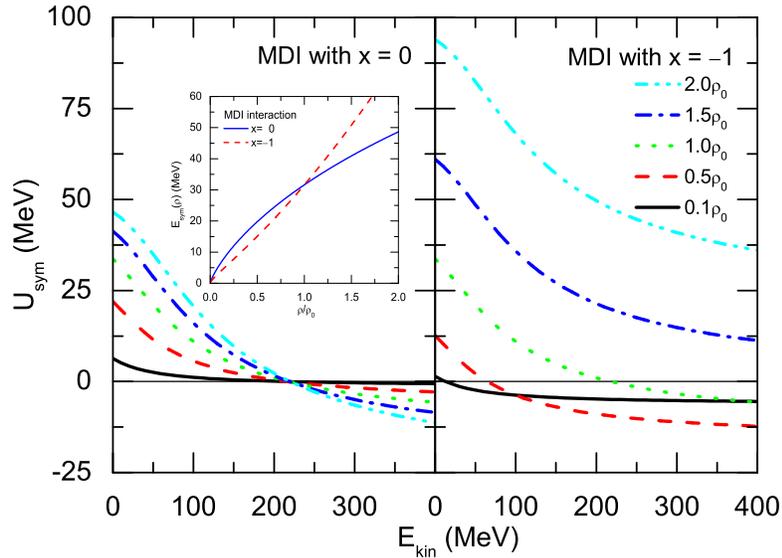}
\caption{{\protect\small (Color online) The symmetry potential and
energy (insert) in the MDI interaction with $x=0$ and $x=-1$.
Taken from Ref. \cite{LiBA07a}.}} \label{Difig1}
\end{figure}

As to be discussed in subsequent sections, many of the isospin
effects that have already been investigated experimentally,
especially the isospin diffusion and isoscaling data, have allowed
us to put some experimental constraints on the density dependence
of the nuclear symmetry energy at subsaturation densities.  Shown
in the inset of Fig. \ref{Difig1} is the experimentally
constrained range of symmetry energy $E_{\mathrm{sym}}(\rho )$
with $x=0$ and $x=-1$ using the MDI interaction. While this
constraint on $E_{\mathrm{sym}}(\rho )$ is the most stringent so
far in the field, the corresponding symmetry potential shown in
Fig.\ \ref{Difig1} still diverges widely with both momentum and
density. This is not surprising as the symmetry energy involves
the integration of the single-nucleon potential over its momentum.
To obtain information about the underlying momentum- and
density-dependence of the symmetry potential, which is more
fundamental than the $E_{\mathrm{sym}}(\rho )$ for many important
physics questions, one has to use differential probes. It was
recently demonstrated that the differential isospin fractionation
as a function of nucleon momentum is such an observable
\cite{LiBA07a}.

\subsubsection{Integrated isospin fractionation in heavy-ion
reactions}

\begin{figure}[th]
\centering
\includegraphics[scale=0.6]{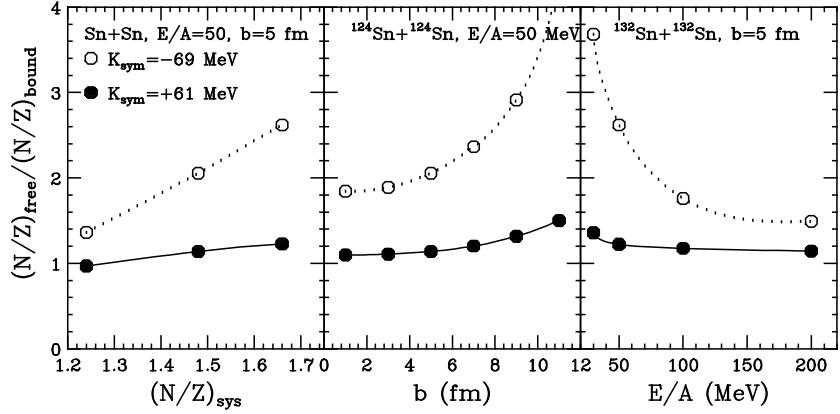}
\caption{{\protect\small Isospin fractionation as a function of
$N/Z$ of the reaction system (left panel), the impact parameter
(middle panel), and the beam energy (right panel) in {\rm Sn+Sn}
reactions. Taken from Ref.~\cite{LiBA00}.}} \label{ifragli}
\end{figure}

The neutron/proton ratio of pre-equilibrium nucleons, the t/$^3$He
ratio, and the $N/Z$ ratio of intermediate mass fragments are
related by charge and mass conservations. Denoting the neutron to
proton ratio of the reaction system and the light (gas) particles
by $(N/Z)_{\rm total}$ and $(N/Z)_{\rm free}$, respectively, then
the neutron to proton ratio of the heaver (liquid) ones is given
by
\begin{eqnarray}\label{nzratio}
(N/Z)_{\rm bound}=(N/Z)_{\rm total}+\frac{Z_g}{Z_l}[(N/Z)_{\rm
total}-(N/Z)_{\rm free}],
\end{eqnarray}
where $Z_g$ and $Z_l$ are the proton numbers of the light (gas) and
heavier (liquid) components, respectively. Given $(N/Z)_{\rm total}$
and $(N/Z)_{\rm free}$, the $(N/Z)_{\rm bound}$ is not unique but
depends on how the total charge is shared between the two phases,
i.e., the $Z_g/Z_l$ factor. As discussed in detail in
Chapter~\ref{chapter_temperature}, essentially all thermal and
dynamical models have predicted that the $(N/Z)_{\rm free}$ is
larger than the $(N/Z)_{\rm bound}$, and this phenomenon is
generally known as the isospin fractionation
\cite{Mul95,LiBA97b,Bar98,Ono03,LiBA02b,Cho03,Ran81,Shi00}. Since
$(N/Z)_{\rm free}$ and $(N/Z)_{\rm bound}$ are normally calculated
by integrating over momentum, the above phenomenon is thus referred
as the integrated isospin fractionation. This should be
distinguished from the differential isospin fractionation, which is
defined as the $(N/Z)_{\rm free}$ over $(N/Z)_{\rm bound}$ ratio as
a function of nucleon momentum \cite{LiBA07a} and will be discussed
in section \ref{difif}. In dynamical models, the degree of isospin
fractionation can be measured quantitatively by calculating the
ratio of $(N/Z)_{\rm free}$ to $(N/Z)_{\rm bound}$. This ratio is
shown in Fig.~\ref{ifragli} as a function of the neutron to proton
ratio $(N/Z)_{\rm sys}$ of the reaction system (left panel), the
impact parameter (middle panel), and the beam energy (right panel),
respectively, for reactions between several {\rm Sn} isotopes. It is
seen that the degree of isospin fractionation increases with both
$(N/Z)_{\rm sys}$ and impact parameter, but decreases with beam
energy. It is also rather sensitive to the $K_{\rm sym}$ parameter
of the asymmetric nuclear matter. The origin of isospin
fractionation and its dependence on the $K_{\rm sym}$ parameter can
be easily understood from the density dependence of the symmetry
energy. Since the repulsive symmetry potential for neutrons
increases with density, more neutrons are repelled from high density
regions to low density regions. The opposite is true for protons
because of their attractive symmetry potentials. Furthermore, the
magnitude of the symmetry potential is higher for $K_{\rm sym}=-69$
{\rm MeV} than for $K_{\rm sym}=+61$ {\rm MeV} for densities less
than about $\rho_0$. One thus expects to see a higher degree of
isospin fractionation with $K_{\rm sym}=-69$ {\rm MeV} as shown
here. Furthermore, the isospin fractionation is stronger at lower
energies, especially around the Fermi energy. As pointed out
previously, the strong incident energy dependence of the isospin
fractionation indicates that the comparison in Fig.~\ref{double} has
to be interpreted very carefully since the incident energy in the
calculations is lower than that in the experiments.

\begin{figure}[th]
\centering
\includegraphics[scale=0.6]{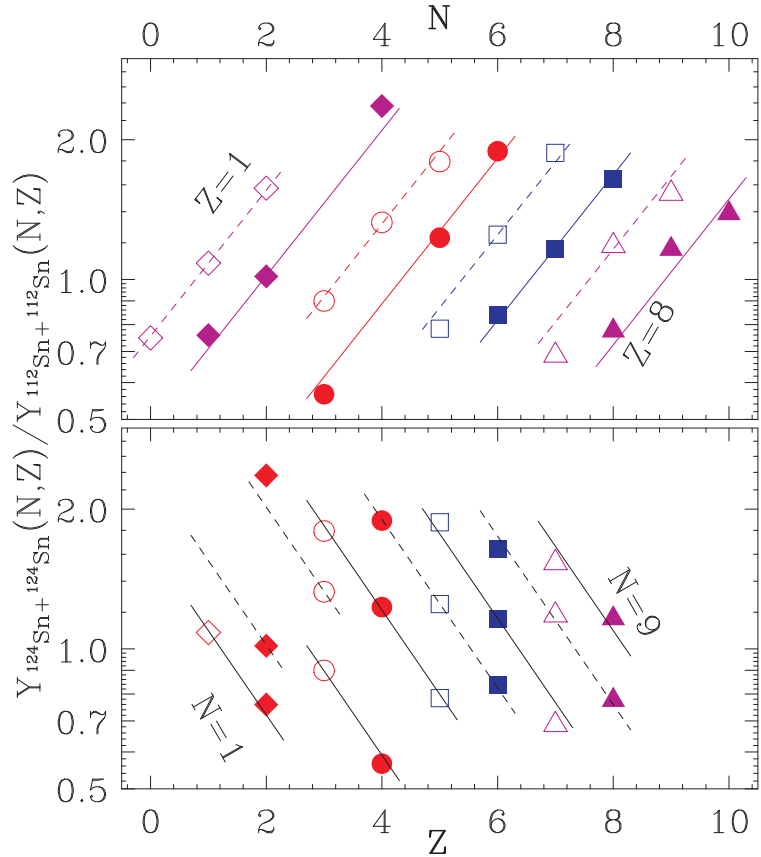}
\caption{(Color online) The ratio of fragments $R_{21}$ for
$^{124}$Sn+$^{124}$Sn and $^{112}$Sn+$^{112}$Sn systems as a
function of N (isotope data, upper panel) and as a function of Z
(isotone data, lower panel). Solid and dashed lines are best fits
to the data. Taken from Ref.~\cite{Tsa01}.} \label{Tsangfig}
\end{figure}

The isospin fractionation phenomenon has been experimentally
confirmed \cite{Cho06}, i.e., the gas phase was found to be
significantly enriched in neutrons relative to the liquid phase
represented by bound nuclei. In particular, in the above mentioned
experiments and analysis by Xu {\it et al.} \cite{Xu00} at the
NSCL/MSU, which are among the most interesting and detailed ones,
the isotope, isotone and isobar ratios were utilized to obtain an
estimate of the neutron/proton density ratio in the gas phase at
the breakup stage of the reaction. The analysis is based on the
isoscaling analysis \cite{Xu00,Tsa01} in the grand canonical
ensemble limit \cite{Ran81,Alb85}. To minimize the effect of
secondary decays, ratios from two reactions were used in their
analysis, as corrections to the primary yields due to secondary
decays appear to be similar in different reactions over a wide
range of bombarding energies. Specifically, the isotope yields of
two different systems with similar incident (excitation) energies
but different isospins are combined to construct ratios of the
form \cite{Xu00,Tsa01}
\begin{eqnarray}
R_{21}=\frac{Y_2(N,Z)}{Y_1(N,Z)}\approx
Ce^{N\alpha+Z\beta}\label{iscaling}
\end{eqnarray}
where $Y_1$ and $Y_2$ are the yields of fragments with proton
number $Z$ and neutron number $N$ from reactions $1$ and $2$,
respectively. The last approximation in the above equation follows
from the assumption that both chemical and thermal equilibriums
are reached in the reactions. The variables
$\alpha\equiv\Delta\mu_n/T$ and $\beta\equiv\Delta\mu_p/T$ reflect
the differences between the neutron and proton chemical potentials
for the two reactions, and $C$ is an overall normalization
constant. The $N$ and $Z$ dependence becomes most apparent if, for
each element $Z$, $R_{21}$ is plotted versus $N$ for all isotopes
on a semi-log plot. The resulting slopes would then be the same
for each $Z$. Similarly, plotting $R_{21}$ against $Z$ for all
isotones would provide a common slope for each $N$. This is
demonstrated in Fig.~\ref{Tsangfig} where the isotope yield ratios
from central collisions of $^{124}$Sn+$^{124}$Sn and
$^{112}$Sn+$^{112}$Sn at a beam energy of 50 MeV/nucleon are
plotted as a function of $N$ (upper panel) and as a function of
$Z$ (lower panel) for 24 isotopes spanning from Z =1 to Z=8
elements. The excellent agreement between the data and
Eq.~(\ref{iscaling}) can be seen more clearly by comparing the
experimental ratios with the best fitted straight lines with
$\alpha=0.36$ and $\beta=-0.41$. This almost perfect fit has been
known as the isoscaling \cite{Tsa01}.

\begin{figure}[th]
\centering
\includegraphics[scale=0.5,angle=90]{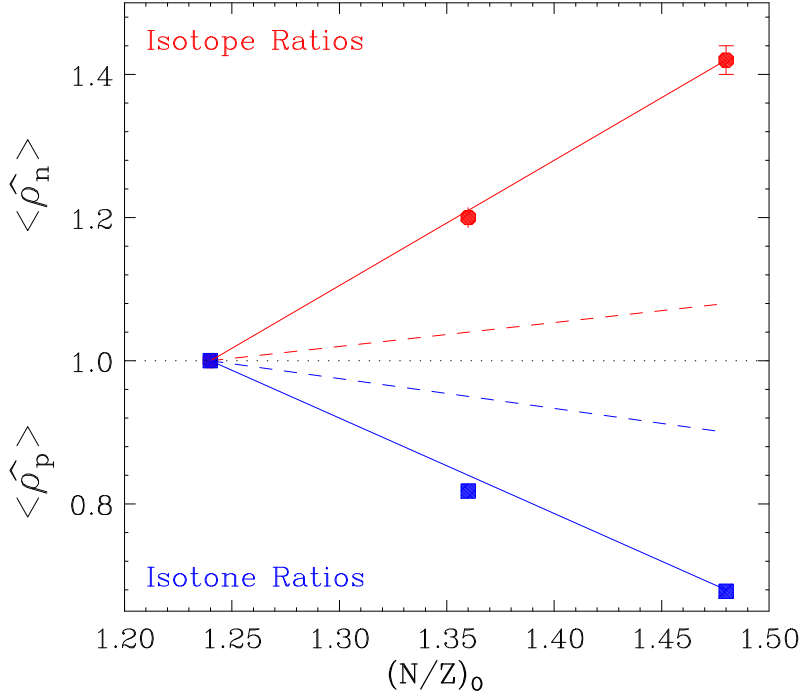}
\caption{The relative free neutron and free proton densities as
functions of $(N/Z)_O$. The solid lines are the best fit to data.
The dashed lines are the expected $n$-enrichment and $p$-depletion
with increase of the isospin of initial systems. Taken from
Ref.~\cite{Xu00}.}\label{Xufig}
\end{figure}

Since in the grand canonical ensemble limit of dilute
non-interacting gas, $\alpha$ and $\beta$ are related to the
relative nucleon density (with respect to the average matter
density in $^{112}$Sn) according to $\bar{\rho_n}=e^{\alpha}$ and
$\bar{\rho_p}=e^{\beta}$ \cite{Xu00}, the free neutron and proton
densities can thus be extracted from measured isotopic ratios as
first suggested in Refs.~\cite{Ran81,Alb85}. The resulting values
of $\bar{\rho_n}$ and $\bar{\rho_p}$, extracted from isotope
ratios from the three systems $^{112}$Sn+$^{112}$Sn,
$^{124}$Sn+$^{112}$Sn, and $^{124}$Sn+$^{124}$Sn, are shown in
Fig.~\ref{Xufig} by solid lines as functions of the $N/Z$ ratio of
the composite system, $(N/Z)_O$. The total neutron and proton
densities, assuming the same total matter density for the two
systems, are given by the dashed lines in Fig.~\ref{Xufig}. The
experimental data suggest that as the $(N/Z)_O$ increases, the
system responds by making the asymmetry of the gas (given by the
solid lines) much greater than the asymmetry of the total system
(given by the dashed lines). If the interpretation of these data
based on the equilibrium description is correct, the nucleon
density extracted from isotope, isotone, and isobar ratios is then
more enriched in neutrons than in the liquid phase represented by
bound nuclei, qualitatively consistent with the predicted isospin
fractionation. The neutron enrichment is much more enhanced in
collisions of neutron-rich systems as compared to collisions of
neutron-deficient systems. To compare the predictions from both
thermal and dynamical models on the isospin fractionation in
asymmetric nuclear matter at finite temperature, it is useful to
study isospin fractionation in future experiments at different
beam energies. Also, the study of $\bar{\rho_p}$ and
$\bar{\rho_p}$ as functions of nucleon momentum at freeze-out is
relevant for studying the differential isospin fractionation as
discussed in the next section.

\subsubsection{Thermodynamical approach to differential isospin
fractionation in asymmetric nuclear matter}\label{difif}

\begin{figure}[tbh]
\centering
\includegraphics[scale=1]{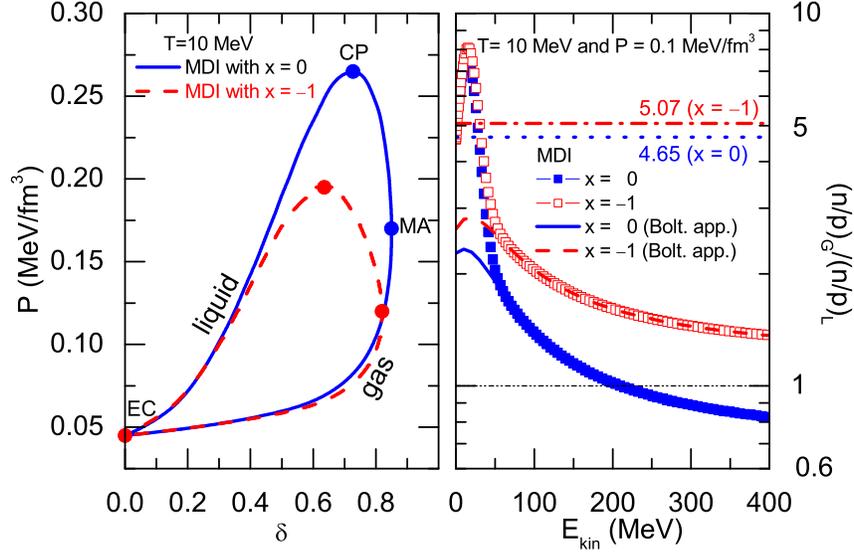}
\caption{{\protect\small Left window: the section of binodal
surface at $T=10$ MeV with $x=0$ and $x=-1$. The critical point
(CP), the points of equal concentration (EC) and maximal asymmetry
(MA) are also indicated. Right window: the double neutron/proton
ratio in the gas and liquid phases $(n/p)_{G}/(n/p)_{L}$ as a
function of the nucleon kinetic energy. Taken from Ref.
\cite{LiBA07a}.}} \label{Difig2}
\end{figure}

Shown in the left window of Fig.~\ref{Difig2} is a typical section
of the binodal surface at $T=10 $ MeV with $x=0$ and $x=-1$ within
the self-consistent thermal model~\cite{Xu07} using the isospin
and momentum-dependent MDI interaction.  The phenomenon of
integrated IsoF with the gas phase being more neutron-rich is
clearly seen. Also, the stiffer symmetry energy ($x=-1$)
significantly lowers the critical point (CP). However, below a
pressure of about $P=0.12\text{ MeV}$/fm$^{3}$, the magnitude of
the integrated IsoF becomes almost independent of the symmetry
energy used. The advantages of the differential IsoF analyses over
the integrated ones can be seen, for example, by selecting the gas
and liquid phases in equilibrium at T=10 MeV and $P=0.1\text{
MeV}$/fm$^{3}$. For $x=0$ the density and isospin asymmetry are,
respectively, $\rho _{G}=0.087\rho _{0}$ and $\delta _{G}=0.791$
for the gas phase, and $\rho _{L}=0.763\rho _{0}$ and $\delta
_{L}=0.296$ for the liquid phase. For $x=-1$ they are,
respectively, $\rho _{G}=0.114\rho _{0}$, $\delta _{G}=0.808$,
$\rho _{L}=0.714\rho _{0}$ and $\delta _{L}=0.30$. The
corresponding double neutron/proton ratio in the gas and liquid
phases $(n/p)_{G}/(n/p)_{L}(p)$ within the thermal model as a
function of nucleon momentum or kinetic energy, i.e., the
differential IsoF can be readily obtained \cite{LiBA07a}. Shown in
the right window of Fig.~\ref{Difig2} are the differential IsoFs
for both $x=0$ and $x=-1$. It is clearly seen that the
isospin-fractionation is strongly momentum dependent. Moreover,
while the integrated double neutron/proton ratios of $5.07$
($x=-1$) and $4.65$ ($x=0$) are very close to each other, the
differential IsoF for nucleons with kinetic energies high than
about $50$ MeV is very sensitive to the parameter $x$ used.
Surprisingly, a reversal of the normal IsoF is seen for $x=0$ for
nucleons with kinetic energies higher than about $220$ MeV. In
this case, there are more energetic neutrons than protons in the
liquid phase compared to the gas phase. At pressures higher than
$0.1$ $\text{MeV}$/fm$^{3}$, where the integrated IsoF is already
very sensitive to the $E_{\mathrm{sym}}(\rho )$, the differential
IsoF is much more sensitive to the $x$ parameter than that shown
in Fig.~\ref{Difig2}. For energetic nucleons where the
differential IsoF is very sensitive to the parameter $x$, their
momentum distribution $f_{\tau }$ can be well approximated by the
Boltzmann distribution as shown in Fig.~\ref{Difig2}. For these
nucleons in either the liquid ($L$) or gas ($G$) phase, the
neutron/proton ratio is
\begin{eqnarray}
(n/p)_{L/G}=\exp [-(E_{n}^{L/G}-E_{p}^{L/G}-\mu _{n}^{L/G}+\mu
_{p}^{L/G})/T].
\end{eqnarray}%
The energy difference of neutrons and protons having the same
kinetic energy and mass is then given by
\begin{eqnarray}
E_{n}^{L/G}-E_{p}^{L/G}=U_{n}^{L/G}-U_{p}^{L/G}\approx 2\delta
_{L/G}\cdot U_{\rm sym}(p,\rho _{L/G}),
\end{eqnarray}%
and is directly related to the symmetry potential $U_{\rm sym}$.
Because of the chemical equilibrium conditions, the chemical
potentials cancel out in the double neutron/proton ratio
\begin{eqnarray}
\frac{(n/p)_{G}}{(n/p)_{L}}(p)=\exp [-2(\delta _{G}\cdot U_{\rm
sym}(p,\rho _{G})-\delta _{L}\cdot U_{\rm sym}(p,\rho _{L}))/T].
\end{eqnarray}%
This general expression clearly demonstrates that the differential
IsoF for energetic nucleons carries direct information about the
momentum dependence of the symmetry potential. In the above
expressions, the weak temperature dependence of the symmetry
potential has been neglected \cite{Xu07c}.

For the liquid-gas phase transition, as for the hadron-QGP
(quark-gluon-plasma) phase transition, equilibrium model
calculations for infinite nuclear matter are very useful for
developing new concepts and predicting novel phenomena. However,
the experimental search/confirmation for the new
phenomena/concepts in real nuclear reactions is usually very
challenging. For example, the underlying nature and experimental
signatures of the LG phase transition, which was predicted first
for infinite nuclear matter based on thermodynamical
considerations, has been studied by the intermediate energy
heavy-ion reaction community for more than two decades, and they
are still far from well understood. The study of how nucleons
behave in the correlated momentum-and isospin-space may reveal
deeper insights into the nature of the LG phase transition.

\subsubsection{Dynamic approach to differential isospin
fractionation in asymmetric nuclear matter}

As for the structure functions of quarks and gluons in the initial
state of relativistic heavy-ion collisions, the momentum
distribution of the $n/p$ ratio in the liquid phase may not be
measured directly since what can be detected at the end of
heavy-ion reactions are free nucleons and bound nuclei in their
ground states. Nevertheless, precursors and/or residues of the
transition in the differential IsoF may still be detectable in
heavy-ion reactions, especially those induced by radioactive
beams.

\begin{figure}[tbh]
\centering
\includegraphics[scale=0.7]{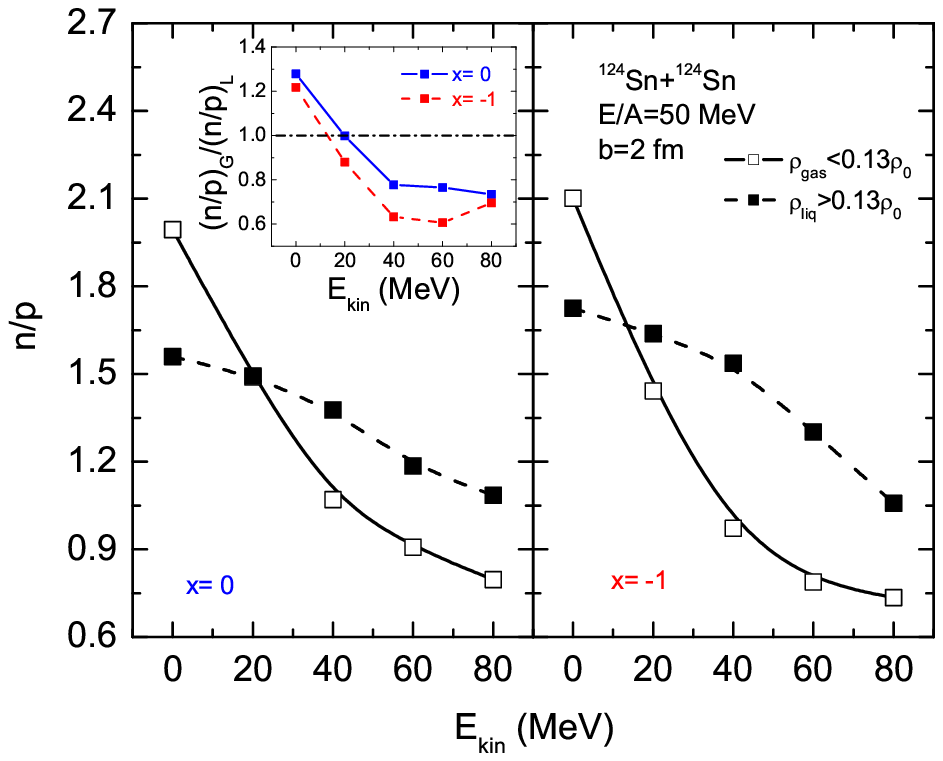}
\includegraphics[scale=0.7]{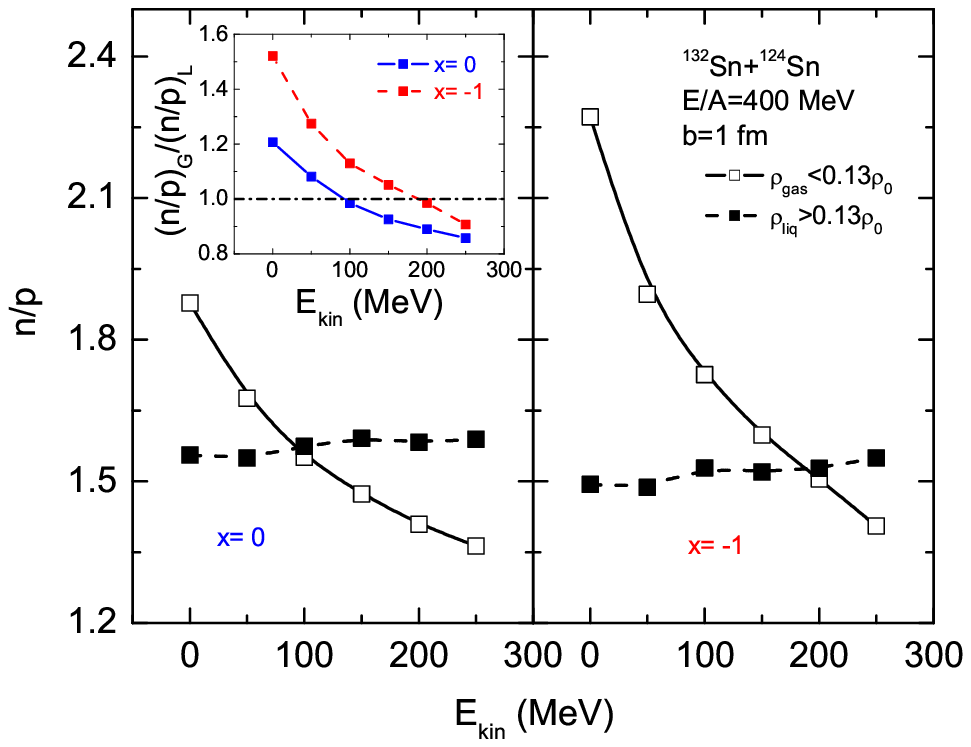}
\caption{{\protect\small The neutron/proton ratio in the `gas' and
`liquid' phases as a function of the nucleon kinetic energy for
the reaction of $^{124}$Sn$+^{124}$Sn at $E_{\rm beam}/A=50$ MeV
(left window) and $400$ MeV (right window), respectively. Taken
from Ref. \cite{LiBA07a}.}} \label{Difig3}
\end{figure}

Shown in Fig.~\ref{Difig3} are two typical examples for the
central reactions of $^{124}$Sn$+^{124}$Sn at $E_{\rm beam}/A=50$
MeV and $^{132}$Sn$+^{124}$Sn at $E_{\rm beam}/A=400$ MeV
calculated using the IBUU04 transport model with the same MDI
interaction. To separate approximately nucleons in the low density
`gas' region from those in the `liquid' region a density cut at
$0.13\rho _{0}$ is used. In both reactions there is indeed a
transition from the neutron-richer (poorer) `gas (liquid)' phase
normally known as the IsoF for low energy nucleons to the opposite
behavior (i.e., anti-IsoF) for more energetic ones. Moreover, the
transition nucleon energy from the normal IsoF to the anti-IsoF is
sensitive to the parameter $x$ used. This is more pronounced in
the reaction at $E_{\rm beam}/A=400$ MeV where effects of the
symmetry (Coulomb) potential are relatively stronger (weaker) for
more energetic nucleons consistent with predictions of the thermal
model. Comparing the thermal model predictions and the transport
model results, one sees that the two approaches predict
qualitatively the same phenomenon while there are quantitative
differences, especially for low energy nucleons. This is mainly
because in nuclear reactions the Coulomb repulsion shifts protons
in the gas phase from low to higher energies leading to the peak
in $(n/p)_G$ ratio at $E_{\rm kin}=0$, while it has little effects
on the protons in the liquid phase. The `gas' phase defined here
contains also the pre-equilibrium nucleons which are known to be
more neutron-rich than the reaction system. They are energetic and
thus affect mostly the high energy part of the $(n/p)_{G}$ ratio.
The subtraction of the pre-equilibrium nucleons from these
analyses thus mainly lowers the $(n/p)_{G}$ for high energy
nucleons, making the transition from the normal IsoF to the
anti-IsoF more obvious.

\begin{figure}[tbh]
\centering
\includegraphics[scale=1.0]{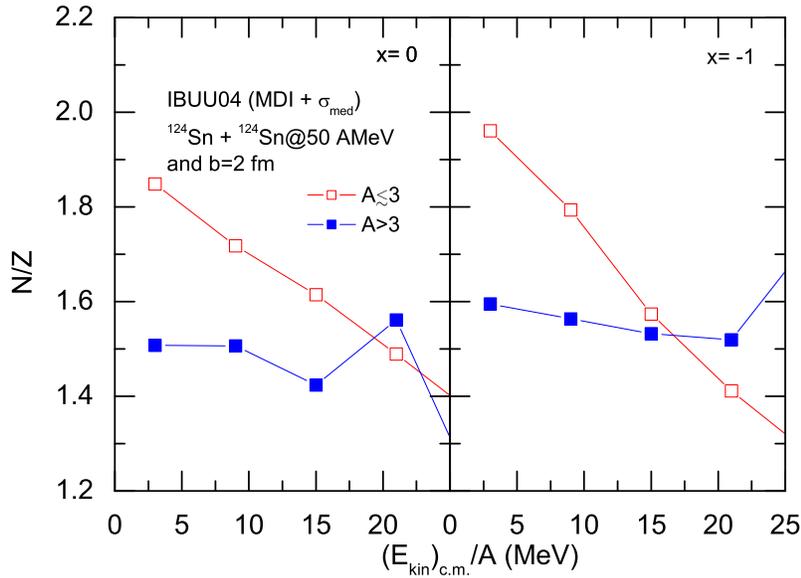}
\caption{{\protect\small Average N/Z ratio in `gas' (the free
nucleons and light clusters with A$\leq 3$) and `liquid'
(fragments with A$> 3$) phases as a function of the average
kinetic energy per nucleon for the reaction of
$^{124}$Sn$+^{124}$Sn at $E_{\rm beam}/A=50$ MeV and $b=2$ fm.
Taken from Ref. \cite{Che08}.}} \label{RatioGLSn124}
\end{figure}

In heavy ion collisions, there are free nucleons, light clusters,
and heavy fragments in the final state. The above analysis by
separating the `gas' phase from the `liquid' phase using a density
cut at $0.13\rho _{0}$ is a crude approach. To be more realistic,
the isospin-dependent phase-space coalescence model
\cite{Che98,Zha99} was used recently in analyzing the differential
IsoF \cite{Che08}. Considering the free nucleons and light
clusters of $A\leq 3$ as in the `gas' phase and the rest as in the
`liquid' phase, the differential IsoF was re-analyzed. Shown in
Fig.~\ref{RatioGLSn124} is a typical example for the central
reaction of $^{124}$Sn$+^{124}$Sn at $E_{\rm beam}/A=50$ MeV and
$b=2$ fm calculated using the IBUU04 transport model with the MDI
interaction. Indeed, the results clearly indicate again a
transition from the neutron-richer (poorer) `gas (liquid)' phase
(the IsoF) at low kinetic energies to the opposite behavior (i.e.,
anti-IsoF) at higher kinetic energies. Furthermore, the transition
energy from the normal IsoF to the anti-IsoF is sensitive to the
parameter $x$ used. These features nicely confirm qualitatively
the results shown in Fig.~\ref{Difig3}.

It is worthwhile to stress again that while the gas phase is
overall more neutron-rich than the liquid phase, the gas phase is
richer (poorer) only in low (high) energy neutrons than the liquid
phase. Clear indications of the differential IsoF consistent with
the thermodynamic model predictions are also seen in transport
model simulations of heavy-ion reactions. While the experimental
test of these predictions may be very challenging but can be done,
future comparisons between the experimental data and theoretical
calculations will allow one to extract critical information about
the momentum dependence of the isovector nuclear interaction.

\subsection{Neutron-proton correlation functions at low relative momenta}
\label{correlation}

The space-time properties of nucleon emission source, which are
important for understanding the reaction dynamics of heavy-ion
collisions, can be extracted from the two-particle correlation
functions; see, e.g., Refs. \cite{Boal90,Bau92,ardo97,Wied99} for
earlier reviews. In most studies, only the
two-proton correlation function is studied \cite%
{gong90,gong91,gong93,kunde93,handzy95,Verde02,Verde03}. Recently,
data on two-neutron and neutron-proton correlation functions have
also become available. The neutron-proton correlation function is
especially useful as it is free of correlations due to
wave-function anti-symmetrization and Coulomb interactions.
Indeed, Ghetti \textit{et al.} have deduced from measured
neutron-proton correlation function the emission sequence of
neutrons and protons in intermediate energy heavy-ion collisions \cite%
{Ghetti00,Ghetti01,Ghetti03} and have also studied the isospin
effects on two-nucleon correlation functions \cite{Ghetti04}.

In the standard Koonin-Pratt formalism
\cite{koonin77,pratt1,pratt2}, the two-particle correlation
function is obtained by convoluting the emission function
$g(\mathbf{p},x)$, i.e., the probability for emitting a particle
with momentum $\mathbf{p}$ from the space-time point
$x=(\mathbf{r},t)$, with the relative wave function of the two
particles, i.e.,
\begin{eqnarray}
C(\mathbf{P},\mathbf{q})=\frac{\int d^{4}x_{1}d^{4}x_{2}g(\mathbf{P}%
/2,x_{1})g(\mathbf{P}/2,x_{2})\left| \phi (\mathbf{q},\mathbf{r})\right| ^{2}%
}{\int d^{4}x_{1}g(\mathbf{P}/2,x_{1})\int
d^{4}x_{2}g(\mathbf{P}/2,x_{2})}. \label{CF}
\end{eqnarray}%
In the above, $\mathbf{P(=\mathbf{p}_{1}+\mathbf{p}_{2})}$ and $\mathbf{q(=}%
\frac{1}{2}(\mathbf{\mathbf{p}_{1}-\mathbf{p}_{2}))}$ are,
respectively, the
total and relative momenta of the particle pair; and $\phi (\mathbf{q},%
\mathbf{r})$ is the relative two-particle wave function with
$\mathbf{r}$
being their relative position, i.e., $\mathbf{r=(r}_{2}\mathbf{-r}_{1}%
\mathbf{)-}$ $\frac{1}{2}(\mathbf{\mathbf{v}_{1}+\mathbf{v}_{2})(}t_{2}-t_{1}%
\mathbf{)}$. This approach has been very useful in studying
effects of nuclear equation of state and nucleon-nucleon cross
sections on the reaction dynamics of intermediate energy heavy-ion
collisions \cite{Bau92}. In Ref.\ \cite{Che03a,Che04}, this
formalism was used to study effects of the momentum dependence of
nuclear mean-field potential and the density dependence of nuclear
symmetry energy on the nucleon-nucleon correlation functions.

\begin{figure}[th]
\includegraphics[scale=1.4]{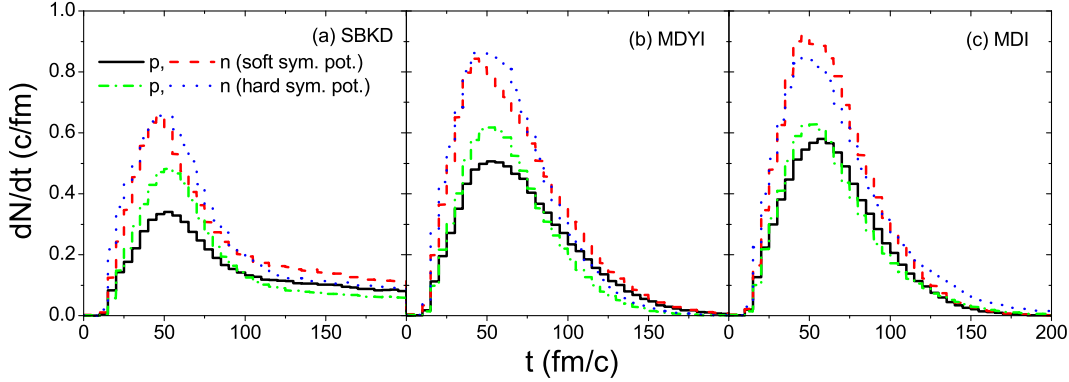}
\caption{{\protect\small (Color online) Emission rates of protons
and neutrons as functions of time for different nucleon effective
interactions. Taken from Ref. \cite{Che04}.}} \label{emRate}
\end{figure}

As an example, we quote here results for central collisions of
$^{52}$Ca + $^{48}$Ca at $E=80$ \textrm{MeV/nucleon}. This
particular reaction system with isospin asymmetry $\delta =0.2$
can be studied at future rare isotope facilities. Nucleons are
considered as emitted when their local densities are less than
$\rho _{0}/8$ and subsequent interactions do not cause their
recapture into regions of higher density. In Fig. \ref{emRate},
the emission rates of protons and neutrons are shown as functions
of time for the SBKD, MDYI, and MDI interactions with soft and
hard symmetry energies. It is clearly seen that there are two
stages of nucleon emissions: an early fast emission and a
subsequent slow emission. This is consistent with the long-lived
nucleon emission source observed in previous BUU calculations
\cite{handzy95}. For the momentum-independent nuclear potential
(SBKD), Fig. \ref{emRate} (a) shows that the hard symmetry energy
enhances the emission of early high momentum protons (dash-dotted
line) and neutrons (dotted line) but suppresses late slow emission
compared with results from the soft symmetry energy (protons and
neutrons are given by solid and dashed lines, respectively). The
difference between the emission rates of protons and neutrons is,
however, larger for the soft symmetry energy. Fig. \ref{emRate}
(b) shows results from the MDYI interaction which includes the
momentum-dependent isoscalar potential but the
momentum-independent symmetry potential. It is seen that the
momentum dependence of isoscalar potential enhances significantly
the nucleon emission rate due to the more repulsive
momentum-dependent nuclear potential at high momenta. As a result,
the relative effect due to the symmetry potential is reduced
compared with the results shown in Fig. \ref{emRate} (a). Fig.
\ref{emRate} (c) is obtained by using the MDI interaction which
includes momentum dependence in both isoscalar potential and
symmetry potential. The momentum dependence of symmetry potential
leads to a slightly faster nucleon emission but the symmetry
potential effects are reduced. The fraction of total number of
emitted nucleons, i.e., before $200$ fm/c in the IBUU04
simulations, in this study is about $80\%$ for the SBKD
interaction but almost $100\%$ for the MDYI and MDI interactions.

\begin{figure}[th]
\centering
\includegraphics[scale=0.72]{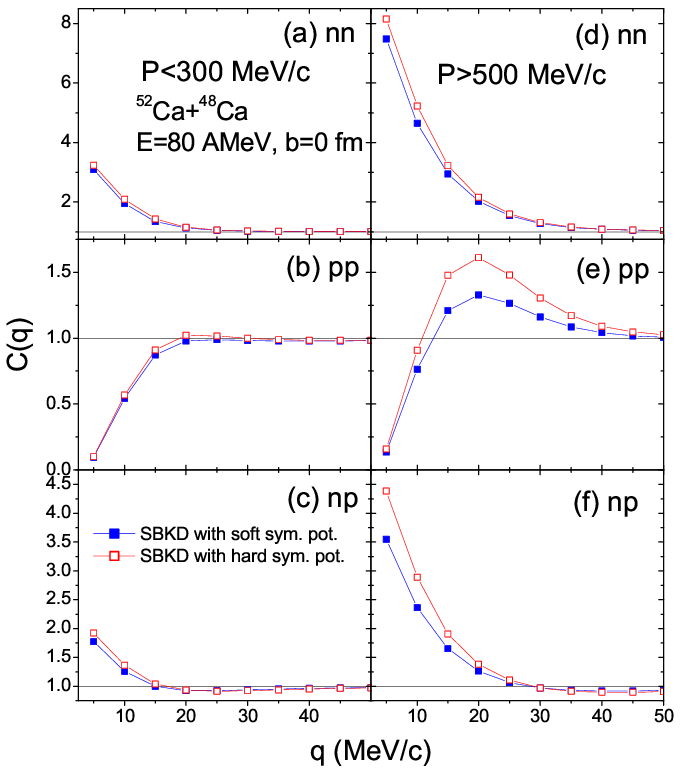}
\includegraphics[scale=0.72]{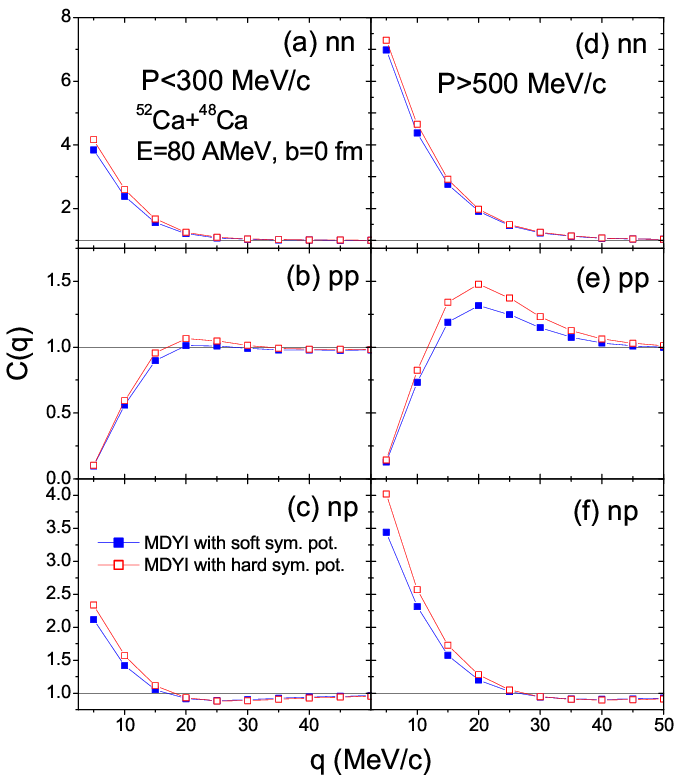}
\includegraphics[scale=0.72]{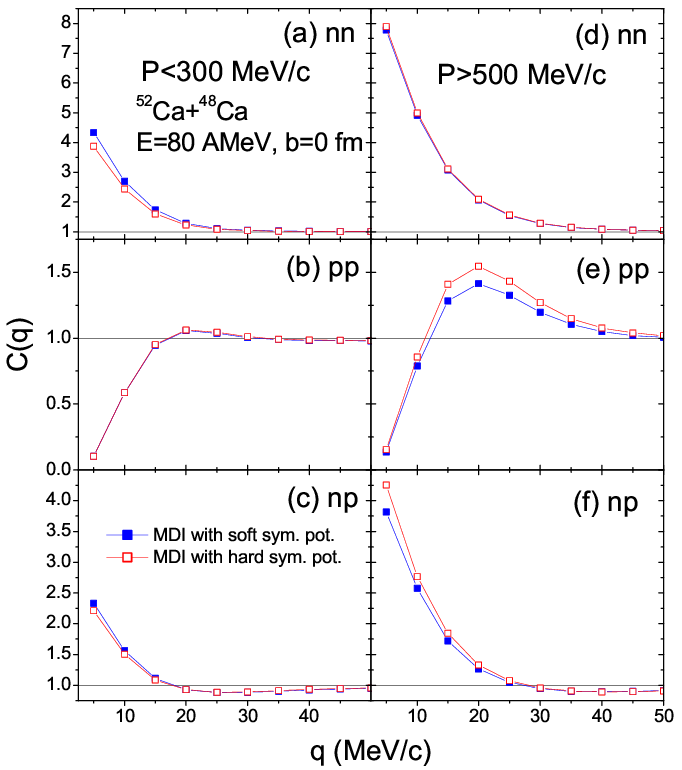}
\caption{{\protect\small (Color online) Two-nucleon correlation
functions gated on the total momentum $P$ of nucleon pairs using
the SBKD (left window), MDYI (middle window) and MDI (right
window) interactions, respectively, with the soft (filled squares)
or the stiff (open squares) symmetry energy. Taken from Ref.
\cite{Che04}.}} \label{CF35hSBKD}
\end{figure}

With the program Correlation After Burner \cite{hbt}, which takes
into account final-state nucleon-nucleon interactions, two-nucleon
correlation functions can be evaluated from the emission function
given by the IBUU04 model. Shown in Fig. \ref{CF35hSBKD} are
two-nucleon correlation functions gated on the total momentum $P$
of nucleon pairs from central collisions of $^{52}$Ca + $^{48}$Ca
at $E=80$ \textrm{MeV/nucleon} by using the SBKD (left window) ,
MDYI (middle window) and MDI (right window) interactions with the
soft and hard symmetry potentials. The left and right panels are
for $P<300$ \textrm{MeV/c} and $P>500$ \textrm{MeV/c},
respectively. Both neutron-neutron (upper panels) and
neutron-proton (lower panels) correlation functions peak at
$q\approx 0$ \textrm{MeV/c}, while the proton-proton correlation
function (middle panel) is peaked at about $q=20$ \textrm{MeV/c}
due to the strong final-state s-wave attraction. The latter is
suppressed at $q=0$ as a result of Coulomb repulsion and
anti-symmetrization of the two-proton wave function. These general
features are consistent with those observed in experimental data
from heavy-ion collisions \cite{Ghetti00}. For nucleon pairs with
high total momentum, their correlation function is stronger for
the hard symmetry energy than for the soft symmetry energy: about
$24\%$ and $9\%$ for neutron-proton pairs and neutron-neutron
pairs at low relative momentum $q=5$ MeV/c, respectively, and
$21\%$ for proton-proton pairs at $q=20$ \textrm{MeV/c}. The
neutron-proton correlation function thus exhibits the highest
sensitivity to the density dependence in nuclear symmetry energy
$E_{\mathrm{sym}}(\rho )$. For nucleon pairs with low total
momenta, the symmetry potential effects are weak.

The symmetry energy effect on two-nucleon correlation functions
after including the momentum-dependent isoscalar potential in the
IBUU04 model can be seen from the middle window. For nucleon pairs
with low total momentum, their correlation functions remain
insensitive to the nuclear symmetry energy. For nucleon pairs with
high total momentum, their correlation function is again stronger
for the hard symmetry energy than for the soft symmetry energy:
about $17\%$ and $4\% $ for neutron-proton pairs and
neutron-neutron pairs at low relative momentum $q=5$ MeV/c,
respectively, and $12\%$ for proton-proton pairs at $q=20$
\textrm{MeV/c}. Compared to the results from the SBKD interaction,
the momentum dependence of isoscalar potential thus reduces the
symmetry potential effects on two-nucleon correlation functions.
This is due to the fact that the repulsive momentum-dependent
potential enhances nucleon emissions and thus reduces the density
effect on nucleon emissions, leading to a weaker symmetry
potential effects on two-nucleon correlation functions. How the
momentum dependence of the nuclear symmetry potential affects the
two nucleon correlation functions can be seen from the right
window of Fig.~\ref{CF35hSBKD}, which shows the results from the
MDI interaction. Compared with results from the SBKD and MDYI
interactions, the two-nucleon correlation functions from the MDI
interaction are thus smaller, mainly due to the very small
difference between its neutron and proton potentials, especially
for higher momentum nucleons~\cite{Che04}. Experimentally, the
isospin effects on two-nucleon correlation functions has indeed
been observed \cite{Ghetti04}.

\subsection{Isospin transport in heavy-ion reactions}

Transport of the isospin degrees of freedom in heavy-ion
collisions can be used as a probe of the nuclear symmetry
potential and energy. It can also provide a measure of the nuclear
stopping power in these reactions. Many interesting phenomena have
been found in experiments that studied isospin transport. In
understanding the experimental results, nuclear transport models
have played a unique role. By comparing results from transport
model calculations with the experimental data, important
constraints on the nuclear symmetry energy at subsaturation
densities have been obtained. While it is important to carry out
transport model simulations, it is also very useful to study
analytically the isospin transport in isospin asymmetric and
nonuniform nuclear matter in order to have a better understanding
of the mechanisms for isospin transport and their relations to the
properties of symmetry potential and energy. For this reason, Shi
and Danielewicz \cite{Shi03} as well as the Catania group \cite
{Bar05b,Riz08} have analyzed the drift and diffusion terms for
isospin transport in some simplified special cases. Their results
are instructive for understanding why the isospin transport is a
useful tool for studying the symmetry energy and potential. It was
shown in Ref.~\cite{Shi03} that for a uniform system of protons
and neutrons at rest, but with the neutron and proton
concentrations changing in space, the isospin asymmetry $\delta $
satisfies the familiar diffusion equation
\begin{eqnarray}
\frac{\partial \delta }{\partial t}=D_{I}\,\nabla ^{2}\delta \,
\label{eq:diffusion}
\end{eqnarray}
where $D_{I}$ is the isospin diffusion coefficient. For systems
near thermal-chemical equilibrium, the mean-field contribution to
$D_{I}$ is proportional to the force due to isospin asymmetry
multiplied by the mean-free time \cite{Shi03}, i.e.,
\begin{eqnarray}
D_{I}\propto \frac{\Pi ^{\delta }}{\sigma _{np}}
\end{eqnarray}
where $\sigma _{np}$ is the neutron-proton scattering cross
section and the isospin asymmetry induced force $\Pi ^{\delta }$
is approximately given by \cite{Shi03}
\begin{eqnarray}
\Pi^{\delta }\approx \frac{\partial }{\partial \delta }\left(
\frac{\mu _{n} }{m_{n}}-\frac{\mu _{p}}{m_{p}}\right)
+\frac{\partial }{\partial \delta } \left(
\frac{U_{n}}{m_{n}}-\frac{U_{p}}{m_{p}}\right).
\end{eqnarray}
Neglecting the neutron-proton mass splitting, i.e.,
$m_{n}=m_{p}=m$, one then has
\begin{eqnarray}
\Pi ^{\delta }\approx \frac{1}{m}\left[ \frac{\partial \mu
_{np}}{\partial \delta }+\frac{\partial (U_{n}-U_{p})}{\partial
\delta }\right] .
\end{eqnarray}
In the above, $\mu _{np}=\mu _{n}-\mu _{p}=4\delta E_{\rm
sym}(\rho )$ is the difference  between the chemical potentials of
neutrons and protons. Replacing $U_{n}-U_{p}$ by $2\delta U_{\rm
sym}$ with $U_{\rm sym}$ being the strength of symmetry potential,
$\Pi ^{\delta }$ is then
\begin{eqnarray}
\Pi ^{\delta }\approx \frac{1}{m}\left[ 4E_{\rm sym}(\rho
)+2U_{\rm sym}\right] .
\end{eqnarray}
The isospin diffusion coefficient, even in this simplified case,
thus depends on the neutron-proton cross section, the symmetry
energy, and the symmetry potential.

In the work of the Catania group, both drift and diffusion
coefficients due to the gradients of both the density and isospin
asymmetry, i.e., $D^{I}$ and $D^{\rho }$, are considered
\cite{Bar05b,Riz08}. Since the nucleon current due to the
variation of its chemical potential with density and isospin
asymmetry can be expressed as
\begin{eqnarray}
\mathbf{j}_{N}=D_{N}^{\rho }\mathbf{\nabla }\rho -D_{N}^{\delta
}\mathbf{ \nabla }\delta,
\end{eqnarray}
the isovector current is then
\begin{eqnarray}
\mathbf{j}_{n}-\mathbf{j}_{p}=(D_{n}^{\rho }-D_{p}^{\rho
})\mathbf{\nabla }\rho -(D_{n}^{\delta }-D_{p}^{\delta })
\mathbf{\nabla }\delta.
\end{eqnarray}
It was argued in Ref.~\cite{Riz08} that
\begin{eqnarray}
D_{n}^{\rho }-D_{p}^{\rho }\propto 4\delta \frac{\partial E_{\rm
sym}}{\partial \rho },  \qquad D_{n}^{\delta }-D_{p}^{\delta
}\propto 4\rho E_{\rm sym}.
\end{eqnarray}
One thus sees that the isospin transport depends on both the slope
and magnitude of the symmetry energy. In more realistic situations
encountered in nuclear reactions, the neutron-proton cross section
and the isospin-dependent Pauli blocking also affect the isospin
transport. To take into account all these effects, one needs to
use the transport models. Using various techniques, one can then
suppress effects due to the in-medium NN cross sections to extract
more reliable information about the symmetry potential and energy,
or vice versa to learn more reliably the in-medium NN cross
sections.

In the following subsections, after a brief review of traditional
methods of measuring the nuclear stopping power, the new technique
of isospin tracing, i.e., using the degree of isospin
equilibration as a probe of nuclear stopping power is discussed.
Several examples of applying this method to heavy-ion collisions
from low to relativistic energies are then reviewed. We pay
special attention to extracting information about the symmetry
energy and potential from studying the isospin transport in
heavy-ion reactions.

\subsubsection{Traditional measures of the nuclear stopping power and
their limitations}

There has been considerable interest in studying the stopping
power of nuclei from low to ultra-relativistic energies. The
nuclear stopping power can be viewed as a measure of the degree to
which the energy of the initial relative motion of two colliding
nuclei is transformed into those in other degrees of freedom
\cite{Bus88,Won94,liwong93}. The degree of nuclear stopping power
determines parameters, such as, the energy density and volume of
the interaction region, which governs the reaction dynamics and
the possibility of reaching conditions capable of forming new
phases of nuclear matter. In heavy-ion collisions at intermediate
energies, nuclear stopping power is determined by both the nuclear
equation of state and the in-medium NN cross sections
\cite{Bau88}. Furthermore, a strong stopping is a necessary
condition to reach global thermal equilibrium in heavy-ion
collisions. Knowledge on the stopping power is also important for
developing theoretical models to understand and predict the
outcome of heavy-ion collisions at various energies. If thermal
equilibrium is established, a macroscopic statistical treatment of
the later stage in terms of temperature, volume and chemical
potential then becomes possible although the early stage of heavy
ion collisions still must be described by microscopic dynamical
models. Also, the interpretation of nuclear multifragmentation in
heavy ion collisions either as a dynamical or as a statistical
process depends on whether global or local chemical-thermal
equilibrium can be achieved in the collisions. Based on the
assumption that thermal equilibrium is established in subsystems
prior to fragment emission, statistical models have been quite
successful in describing heavy-ion reaction data
\cite{Bon95,Gro97}. However, a critical examination of whether
global or partial chemical-thermal equilibrium can be reached in a
model independent way is essential.

\begin{figure}[htp]
\centering
\includegraphics[scale=0.3]{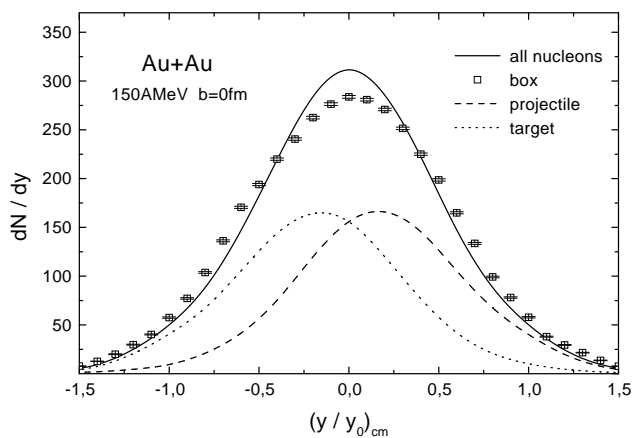}
\caption{Baryon rapidity distribution in a central Au+Au reaction
at a beam energy of 150 MeV/nucleon. The long (short) dashed line
is the contribution from the projectile (target) nucleus. The
squares are the results from simulating the thermalization by
putting all nucleons in a box, taken from Ref.
\protect\cite{Hom99}.} \label{mosel1}
\end{figure}

The stopping power is usually determined in experiments by
measuring: 1) final-state nucleon rapidity distributions; 2) the
energy remaining in forward-going baryons after the reaction; 3)
transverse energy distributions or 4) quadrupole moments of
momentum distributions or ratios of energies associated with the
transverse and longitudinal motions of nucleons and fragments,
see, e.g., Refs.~\cite{Bau88,Vid95,Har96}. All of these
measurements can provide information about the amount of energy
that is being transferred from initial longitudinal motion to
other directions and particle production, and they thus reflect
the stopping power from different aspects. Among these methods,
final-state proton rapidity distributions have been most
frequently used. However, these traditional measures have their
shortcomings. For instance, one major problem of studying the
stopping power using the rapidity distributions is well
illustrated in Fig.\ \ref{mosel1} based on the RBUU transport
model calculations \cite{Hom99}. Shown in Fig.\ \ref{mosel1} are
the rapidity distributions in the final state of a head-on Au+Au
reaction at a beam energy of 150 MeV/nucleon. The long (short)
dashed line is the contribution from the projectile (target)
nucleus. The sum of these two components is given by the solid
line. Although the total rapidity distribution is close to a
thermal distribution, which can be approximated by the squares
generated by putting all nucleons in a box with periodic boundary
conditions \cite{Hom99}, the rapidity distributions of target and
projectile nucleons are still clearly separated. The obvious
relative collective motion of the projectile and target nucleons
indicates that there is no complete stopping in the reaction.

\begin{figure}[htp]
\centering
\includegraphics[scale=0.3]{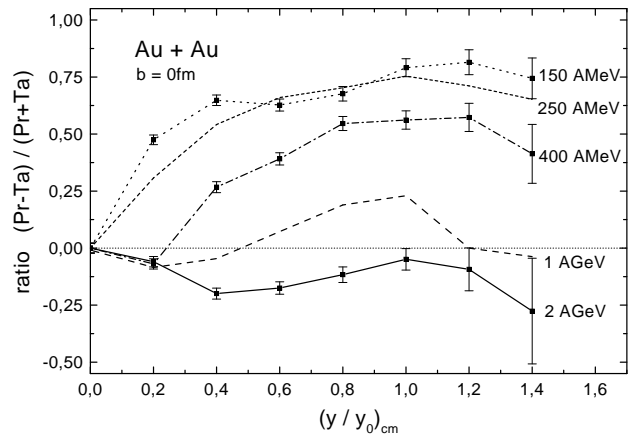}
\caption{Ratio of target to projectile nucleons as a function of
rapidity in a tube with 1 fm radius in head-on Au+Au reactions.
Taken from Ref.~\cite{Hom99}.} \label{mosel2}
\end{figure}

Even in head-on collisions there could still be some memory effect
due to nucleons close to the surfaces of the colliding nuclei. To
illustrate this problem, the stopping power of nucleons in a tube
with a radius of 1 fm along the beam direction was studied in the
same model. The ratio of target-nucleons over projectile-nucleons
in the tube is shown in Fig.\ \ref{mosel2} as a function of
rapidity for head-on Au+Au reactions at beam energies from 150
MeV/nucleon to 2 GeV/nucleon. Only at the mid-rapidity are there
equal numbers of nucleons from the projectile and target as one
expects from symmetry and geometry. At all other rapidities there
are unequal mixing of nucleons from the target and projectile.
These finding together with the relative motion seen in the
separated rapidity distributions of projectile and target nucleons
in Fig.\ \ref{mosel1} indicate that the complete stopping and
thermalization are not guaranteed even if the final baryon
rapidity distribution has a single peak and can be well described
by thermal models.

\subsubsection{The isospin tracing and transport as a measure of
nuclear stopping power}

If we were able to tag the nucleons from projectile and target in
experiments, the task of measuring the stopping power in a model
independent way would be much easier. By using nuclei with
different N/Z ratios, such a tag can be provided since final
nucleons can be attributed on average to either the projectile or
target. We stress that the indistinguishable nature of nucleons do
not allow one to separate unambiguously nucleons originally from
the target or projectile. The isospin tracing method only works in
an average sense.

\begin{figure}[htp]
\centering
\includegraphics[scale=0.6]{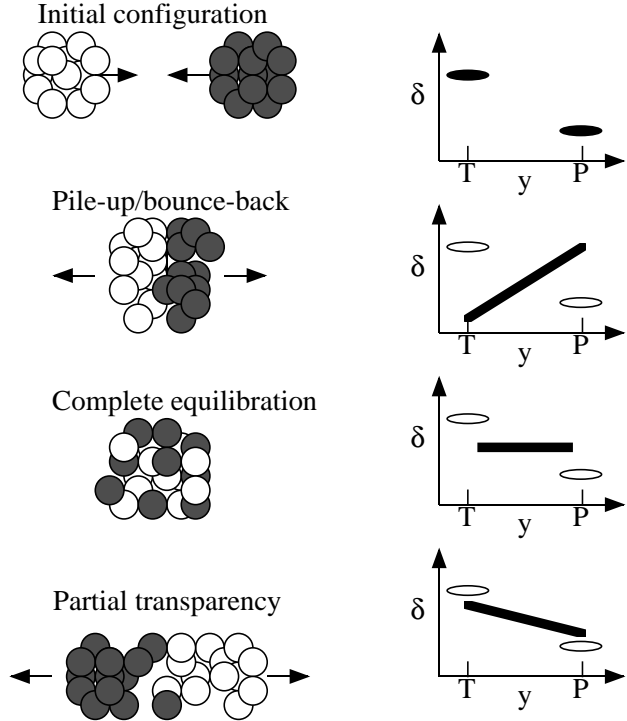}
\caption{Illustration of using the rapidity distribution of
neutron/proton ratio as a probe of nuclear stopping power. Taken
from Ref. \protect{\cite{LiBA98}}. } \label{lkb}
\end{figure}

The idea of using isospin transport as a measure of nuclear
stopping power is illustrated in Fig.\ \ref{lkb} where the isospin
tracer, neutron/proton ratio $N/Z$ or $\delta$, is shown in both
coordinate and momentum space. In practice, besides the $N/Z$
or$\delta$ of free nucleons one can also use ratios of mirror
nuclei as effective isospin tracers. The pictures on the left show
the nucleon distributions in coordinate space, while those on the
right are the rapidity distributions for the three scenarios of
pile-up and bounce-back, stopping and mixing (isospin
equilibrium), and translucency (partial transparency). It is seen
that for a projectile and a target with very different N/Z ratios,
a comparison of the rapidity distribution of N/Z before (shown in
top picture on the right) and after the collision can give direct
information about the degree of stopping between the target and
projectile \cite{LiBA98,Bas94}. This method thus allows one to
study whether there is a transition from full stopping to
translucency (partial transparency) as the beam energy increases.
Also, if isospin equilibrium can be reached in the collisions, it
is then possible to determine its time scale relative to that for
thermal equilibrium. Moreover, one can study the dependence of
isospin transport on the symmetry energy and the isospin
dependence of the in-medium nucleon-nucleon cross sections.

\subsubsection{Transition from isospin equilibrium at low energies to
translucency at intermediate energies}\label{imedium}

The isospin degree of freedom has been found to reach equilibrium
faster than all other degrees of freedom in deep inelastic heavy
ion collisions \cite{Far91,Gat75,Fed78,Udo84}. Recent experimental
studies \cite{She94,sherry2,sherry3} of the isotopic composition
of intermediate mass fragments (IMF) and their angular
distributions in heavy ion collisions have also shown that only at
low energies ($\sim 30$ MeV/nucleon) is isospin equilibrium
reached before fragment emission.

\begin{table}
\begin{center}
\caption{Values of N-Z for the quasitarget(QT) and quasiprojectile
(QP). Taken from Ref. \cite{Far91}.} \label{tableofc}
\medskip
\begin{tabular}{ccccccccccccccc}
\hline
\multicolumn{1}{c}{Reaction/N-Z} &\multicolumn{1}{c}{c=0}
&\multicolumn{1}{c}{c=20}&\multicolumn{1}{c}{c=28}\\
\hline \multicolumn{1}{c}{$^{56}{\rm Ca}+^{40}{\rm Ca}$~~QT}
&\multicolumn{1}{c}{3.74}
&\multicolumn{1}{c}{4.64}&\multicolumn{1}{c}{5.02}\\
\multicolumn{1}{c}{~~~~~~~~~~~~~~~~~~~QP}
&\multicolumn{1}{c}{8.76}
&\multicolumn{1}{c}{5.82}&\multicolumn{1}{c}{4.92}\\
\multicolumn{1}{c}{$^{48}{\rm Ca}+^{48}{\rm Ca}$~~QT}
&\multicolumn{1}{c}{6.52}
&\multicolumn{1}{c}{5.78}&\multicolumn{1}{c}{5.52}\\
\multicolumn{1}{c}{~~~~~~~~~~~~~~~~~~~QP}
&\multicolumn{1}{c}{6.96}
&\multicolumn{1}{c}{5.84}&\multicolumn{1}{c}{5.12}\\
\hline
\end{tabular}
\end{center}
\end{table}

Effects of the nuclear symmetry potential on isospin diffusion
towards isospin equilibrium at low beam energies was first studied
within a Landau-Vlasov transport model by Farine {\it et al.}
\cite{Far91}. To include the effects of the symmetry potential on
nuclear dynamics, they added to the Zamick potential energy
density for symmetric matter, which was widely used in early
transport model calculations \cite{Ber88b}, an asymmetric term in
Eq. (\ref{simple}). By varying the value of $c$, it has been found
that a stronger symmetry potential enhances the isospin diffusion
and thus the degree of isospin equilibrium. Shown in Table
\ref{tableofc} are the values of $N-Z$ in the quasitarget (QT) and
quasiprojectile (QP) formed in the reaction of $^{56}{\rm
Ca}+^{40}{\rm Ca}$ at an impact parameter of 7 fm and a beam
energy of 15 MeV/nucleon. For comparisons, results from reactions
of a symmetric system $^{48}{\rm Ca}+^{48}{\rm Ca}$ are also
listed in the table. In both reactions the total mass and charge
numbers are the same. The difference in the values of $N-Z$ for
the symmetric system is completely due to the numerical
fluctuations of the calculations. In the asymmetric reaction the
initial value of $N-Z$ is 16 and 0 for the projectile and target,
respectively. It is seen that significant mixing is achieved even
in the case of no symmetry potential ($c=0$). However, without
using the symmetry potential isospin equilibrium cannot be
reached. Moreover, it is seen that the degree of isospin mixing or
diffusion increases with the strength of the symmetry potential.

\begin{figure}[htp]
\centering
\includegraphics[scale=0.5]{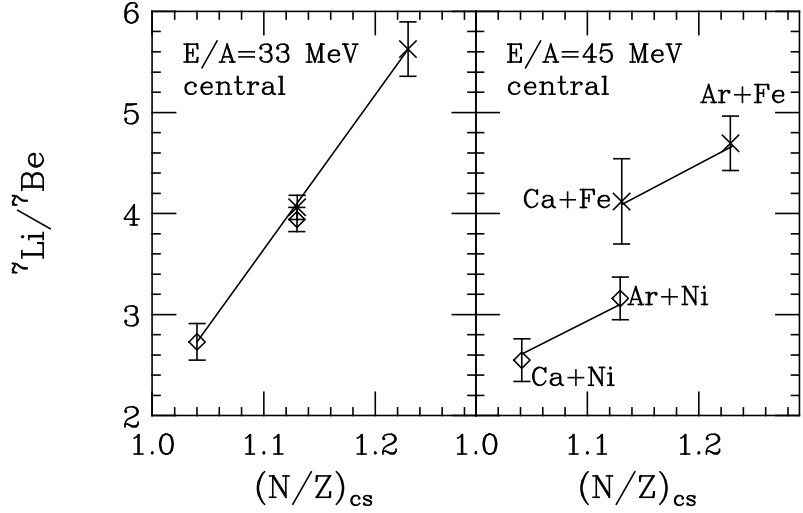}
\caption{Isobaric ratios from central collisions plotted as a
function of N/Z ratio of the combined target and projectile system
at $E_{\rm beam}/A$=35 and 45 MeV. Taken from Ref.
\protect\cite{sherry3}.} \label{sherry}
\end{figure}

The study of isospin transport using the isospin tracing method
has revealed a transition from isospin equilibrium to translucency
around the Fermi energy. In several early experiments by Yennello
{\it et al.} \cite{LiBA95,She94,sherry2,sherry3} both isotopic and
isobaric ratios of intermediate mass fragments from central
collisions of $^{40}{\rm Cl},~ ^{40}{\rm Ar}$ and $^{40}{\rm Ca}$
with $^{58}{\rm Fe}$ and $^{58}{\rm Ni}$ have been studied. A
consistent picture appears from the different analyses of these
data. For example, it was shown that at $E_{\rm beam}/A$=25 and 33
MeV the isotopic ratios $^9{\rm Be}/^7{\rm Be}, ~^{11}{\rm
B}/^{10}{\rm B}$ and $^{13}{\rm C}/^{12}{\rm C}$ increase linearly
with increasing $(N/Z)_{\rm cs}$ ratio of the combined target and
projectile system, but are independent of the $N/Z$ ratio of the
target or projectile. Shown in the left panel of Fig.\
\ref{sherry} are typical results of reactions at 33 MeV/nucleon,
which indicate that the isospin is equilibrated in the composite
system formed in these reactions before the emission of fragments.

The most striking and unexpected feature was observed from the
isobaric ratios in central collisions at $E_{\rm beam}/A=$ 45 and
53 MeV. A typical result at $E_{\rm beam}/A=$45 MeV is shown in
the right panel of Fig.\ \ref{sherry}. It is seen that the
isotopic ratios depend on the $N/Z$ ratio of the target and
projectile in reactions with target-projectile combinations having
the same $(N/Z)_{\rm cs}$ ratio, such as $^{40}{\rm Ca}+^{58}{\rm
Fe}$ and $^{40}{\rm Ar}+^{58}{\rm Ni}$. In Ref.\ \cite{sherry3},
similar results are also shown for other isotope ratios. Moreover,
data at very forward and backward angles show that the isotope
ratios do not simply depend on $(N/Z)_{\rm cs}$. Instead, light
fragments at backward angles are seen to have a much stronger
dependence on $(N/Z)_{\rm target}$, while at forward angles they
depend more on $(N/Z)_{\rm projectile}$. These results demonstrate
that the isospin degree of freedom in reactions at $E_{\rm
beam}/A=$ 45 and 53 MeV is not globally equilibrated prior to the
time when fragments are emitted. Therefore, a transition from
isospin equilibration to non-equilibration is observed, indicating
a change from complete mixing to translucency as the beam energy
increases from below to above the Fermi energy.

The above observation has profound implications on the reaction
mechanism leading to multifragmentation. It not only establishes
the relative time scale for multifragmentation in these reactions
but also indicates that the assumption of global isospin
equilibrium taken for granted in various statistical models for
nuclear multifragmentation at intermediate energies is not valid.
Indeed, a statistical model study was made and failed to show any
entrance channel effect \cite{She94}. Calculations using an
intranuclear cascade code {\sc isabel} \cite{Yar82} show that the
$N/Z$ ratio of the residue is very close to that of the initial
combined system \cite{She94} and thus also fails to reproduce
those features observed at $E_{\rm beam}/A=$ 45 and 53 MeV.
Although the exact origin of this failure is not clear, one
expects that the reaction dynamics at these relatively low
energies cannot be described by the nucleon-nucleon cascade alone,
and should include also nuclear mean-field potential, especially
the isovector one. The experimental observation discussed above
can, however, be well explained by using the IBUU transport model
\cite{LiBA95}. Calculations based on this model have been
performed over a range of impact parameters. For peripheral
collisions it shows a memory of the initial target and projectile,
which is, however, gradually lost as the collisions become more
central. A calculation at $b=0$ thus gives the most interesting
test of any non-equilibrium effect.

\begin{figure}[htp]
\centering
\includegraphics[scale=0.5]{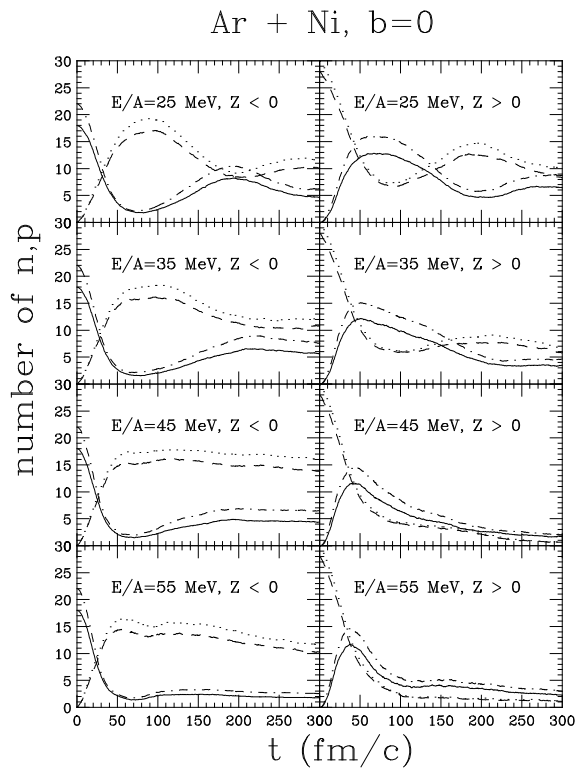}
\caption{The neutron and proton numbers in the residues on the
left ($Z < 0$) and right ($Z\geq 0$) side of the origin. Solid
lines are the proton number from the projectile, while dot-dashed
lines are the neutron number from the projectile which moves
towards the right. Dashed lines are the proton number from the
target, while dotted lines are the neutron number from the target
which moves towards the left. Taken from Ref.
\protect\cite{LiBA95}.} \label{lish2}
\end{figure}

To see if the heavy residues observed above are in isospin
equilibrium, we show in Fig.\ \ref{lish2} the neutron and proton
numbers in the residues on the left ($Z < 0$) and right ($Z\geq
0$) side of the origin in head-on reactions of Ar+Ni at $E_{\rm
beam}/A=$25, 35, 45 and 55 MeV. The solid and dot-dashed lines
are, respectively, the proton and neutron numbers from the
projectile, which is incident from the left. The dashed and dotted
lines are, respectively, the proton and neutron numbers from the
target, which moves from the right. It is seen that the neutron
and proton numbers on both sides not only decrease but also
fluctuate with time.  The decreases is mainly due to
nucleon-nucleon collisions and particle emissions, while the
fluctuation is due to both the restoring force from the mean-field
potential and nucleon-nucleon collisions. At $E_{\rm beam}/A=$ 25
MeV the neutron and proton numbers on the two sides become very
close to each other and the amplitude of oscillation is rather
small by the time of 300 fm/c. This indicates that the heavy
residue is very close to isospin equilibrium, i.e., the proton and
neutron distributions are independent of space-time. The damping
of the oscillation is faster at $E_{\rm beam}/A=35$ MeV so the
particle distribution also reaches isospin equilibrium sooner. The
isotopic composition of fragments emitted from the residues in
these low energy reactions after about 300 fm/c would therefore
essentially reflect the $(N/Z)_{\rm cs}$ ratio of the initial
composite system, and there is little forward-backward asymmetry.
These features are in good agreement with those found in the data
at $E_{\rm beam}/A=$25 and 35 MeV \cite{She94,sherry2,sherry3}.

At higher energies, such as $E_{\rm beam}/A=$ 45 and 55 MeV, there
is little oscillation in the overlapping region between target and
projectile. This is mainly because the incoming momenta of
projectile-nucleons and target-nucleons are very large so the
mean-field potential cannot reverse the directions of motion of
many nucleons during a relatively short reaction time. As a
result, there exists a large isospin asymmetry or
non-equilibration at these two energies. In particular, on the
left side of the origin the $N/Z$ ratio of the residue is more
affected by that of the target while on the right side it is more
affected by that of the projectile. However, the $N/Z$ ratios on
both sides are not simply those of the target and projectile but a
combination of the two, thus depending on the complicated reaction
dynamics. In the case of $E_{\rm beam}/A=$ 55 MeV, at the time of
about 200 fm/c the heavy residue has broken up into two pieces
with some longitudinal collectivity. The forward moving residue
has an excitation energy of about 8.6 MeV/nucleon, while the
backward moving residue has an excitation energy of about 6.8
MeV/nucleon. Both residues are found to be in approximate thermal
equilibrium in their own center of mass frame but not in thermal
equilibrium with each other \cite{LiBA95}. The relation between
the reaction mechanisms and the isospin equilibration in
intermediate energy heavy ion reactions has also been studied
using the isospin-dependent QMD model \cite{Che97}, and the
results indicate that the isospin equilibrium is reached if the
incomplete fusion mechanism is dominant but is not reached if the
fragmentation mechanism dominates. These results are consistent
with the conclusion obtained from the IBUU calculations.

\subsubsection{Nuclear translucency at high energies}\label{ihigh}

\begin{figure}[htp]
\centering \hspace{-3cm}
\includegraphics[scale=0.5]{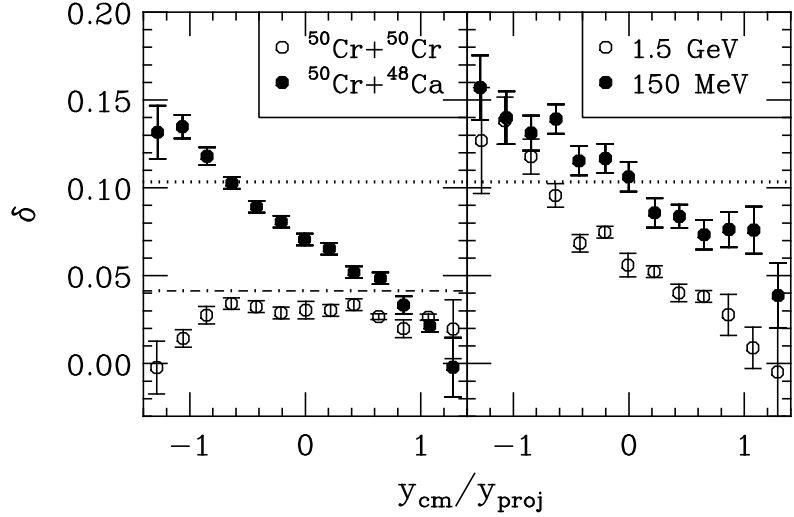}
\caption{The neutron to proton asymmetry versus rapidity predicted
by the {\sc qmd} model. Taken from Ref. \protect\cite{Bas94}.}
\label{qmd}
\end{figure}

From the above discussions, one expects nuclear translucency to be
more important at higher energies. This has indeed been observed in
both model calculations \cite{LiBA95,Bas94,Hom99} and experiments
\cite{Ram00}. In the left panel of Fig.\ \ref{qmd}, the neutron to
proton ratios in central collisions of $^{50}{\rm Cr}+^{48}{\rm Ca}$
and $^{50}{\rm Cr}+ ^{50}{\rm Cr}$ at $E_{\rm beam}/A=1.0$ GeV are
compared. For the symmetric system there is a significant stopping
as seen from the central-rapidity plateau. There is clearly a strong
translucency in the asymmetric system. In the right panel, results
from the asymmetric system at two different beam energies are shown,
and it is seen that even at 150 MeV/nucleon the asymmetric system
shows a strong translucency. Within the IQMD model it was further
shown that the signature for translucency seen in the $N/Z$ ratio is
not altered by cluster formations. On the other hand, the stopping
power is affected significantly by the magnitude of the in-medium
nucleon-nucleon cross sections. As expected, increasing the
in-medium nucleon-nucleon cross section by a factor of 5 results in
a transition from translucency to full stopping in the asymmetric
system \cite{Bas94}. However, the RBUU studies by Gaitanos {\it et
al.} from varying the in-medium neutron-proton cross sections by a
factor of two for the reactions of Ru(Zr)+Zr(Ru) at beam energies of
$0.4$ and $1.528~{\rm AGeV}$ indicate that the degree of isospin
translucency does not change much. Instead, it depends more on the
symmetry energy and a stiffer density dependence leads to a larger
transparency \cite{Gai04b}.

\begin{figure}[htp]
\centering
\includegraphics[scale=0.5]{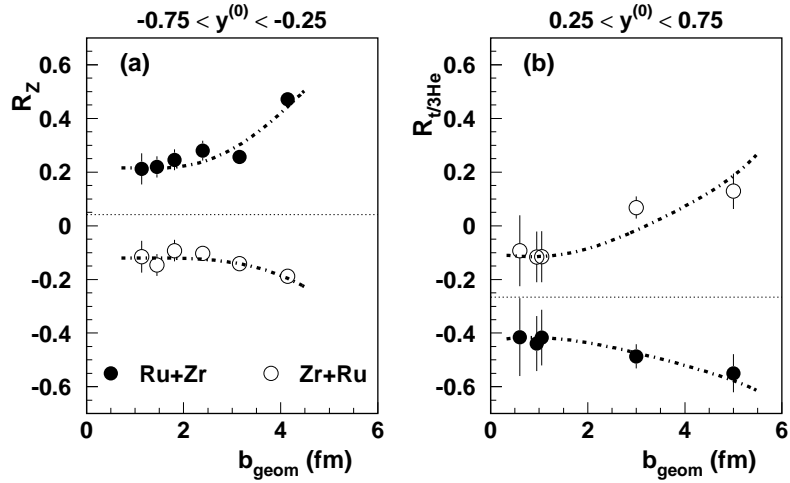}
\caption{Two isospin tracer observables as functions of impact
parameter in mixed reactions between Ru and Zr nuclei. Taken from
Ref. \protect\cite{Ram00}.} \label{ruzr1}
\end{figure}

Some very interesting results were obtained by the FOPI
collaboration at GSI \cite{Ram00}. Four different combinations of
$^{96}{\rm Ru}$ and $^{96}{\rm Zr}$, both as projectile and
target, were investigated at the same bombarding energy of 400
MeV/nucleon. The degree of isospin mixing between target and
projectile nucleons was mapped across a large portion of the phase
space using two different isospin-tracer observables, the number
of measured protons and the $^3{\rm H}/^{3}{\rm He}$ yield ratio.
In the first method, the relative abundance of the
projectile-target nucleons has been adopted:
\begin{eqnarray}
R_Z=\frac{2Z-Z^{\rm Zr}-Z^{\rm Ru}}{Z^{\rm Zr}-Z^{\rm Ru}}
\end{eqnarray}
where $Z$ is the final number of protons observed in a given cell
of the momentum space; $Z^{\rm Zr}$ and $Z^{\rm Ru}$ are the
values of $Z$ in ${\rm Zr+Zr}$ and ${\rm Ru+Ru}$ reactions,
respectively. Thus $R_z$ takes the value of $+1$ and $-1$ in the
${\rm Zr+Zr}$ and ${\rm Ru+Ru}$ reactions, respectively. In the
case of a mixed reaction, ${\rm Ru+Zr}$ or ${\rm Zr+Ru}$, the
measured proton yield $Z$ takes values intermediate between
$Z^{\rm Ru}$ and $Z^{\rm Zr}$. If $Z$ is close to $Z^{\rm Ru}$ in
a ${\rm Ru+Zr}$ reaction, it then indicates that the cell is
mainly populated by nucleons from the ${\rm Ru}$ projectile.
Similarly, an isospin tracer $R_{^{3}{\rm H}/^{3}{\rm He}}$ using
the $^{3}{\rm H}/^{3}{\rm He}$ yield ratio can be defined. Results
of the GSI measurements are shown as functions of centrality in
Fig.\ \ref{ruzr1}. Here the pseudo-proton yield includes both free
protons and protons still bound in deuterons. The two ratios $R_Z$
and $R_{^{3}{\rm H}/^{3}{\rm He}}$ are measured in the backward
and forward hemisphere, respectively. Except for an off-set, both
figures display the same trend, i.e., the global isospin
equilibrium is not reached even in the most central collisions.

\begin{figure}[htp]
\centering
\includegraphics[scale=0.5]{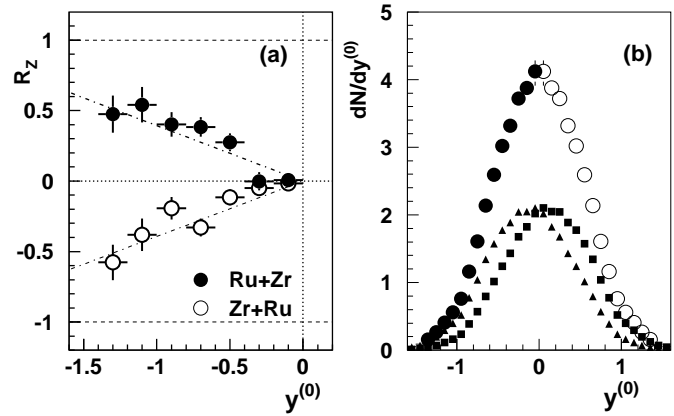}
\caption{Window (a) shows the isospin tracer observable $R_Z$ as a
function of rapidity in the mixed reactions between ${\rm Ru}$ and
${\rm Zr}$ nuclei. Window (b) displays the rapidity distributions
of total pseudo-protons (filled and open circles), deconvoluted
'projectile' (squares) and `target' (triangles) components,
respectively. Taken from Ref. \protect\cite{Ram00}.}\label{ruzr2}
\end{figure}

Furthermore, the rapidity dependence of the isospin mixing has
also been measured. Shown on the left window of Fig.\ \ref{ruzr2}
are the isospin tracer $R_z$ for the most central reactions
between ${\rm Ru}$ and ${\rm Zr}$ nuclei as a function of the
normalized center-of-mass rapidity $y^{(0)}$. It is seen that only
at mid-rapidity the isospin tracer $R_Z$ is about zero. The
measured variation of $R_Z$ can be described by a linear
dependence on the normalized rapidity in the form of $R_z\approx
\pm0.393 y^{(0)}$. This relation was used to deconvolute the total
rapidity distribution in central ${\rm Ru+Ru}$ reaction, shown by
filled and open circles in the right window of Fig.\ \ref{ruzr2},
into separated distributions for the projectile- and
target-nucleons, which are shown, respectively, by squares and
triangles in the same figure. Although the total rapidity
distribution peaks at the mid-rapidity, the `projectile' and
`target' rapidity distributions are clearly shifted relative to
each other, demonstrating that a memory of the initial
target/projectile relative motion survives throughout the central
collision. These results are in agreement with transport model
calculations as discussed earlier.

\subsection{Transport model analyses of the isospin diffusion data from NSCL/MSU}

\begin{figure}[tbh]
\centering
\includegraphics[scale=0.6]{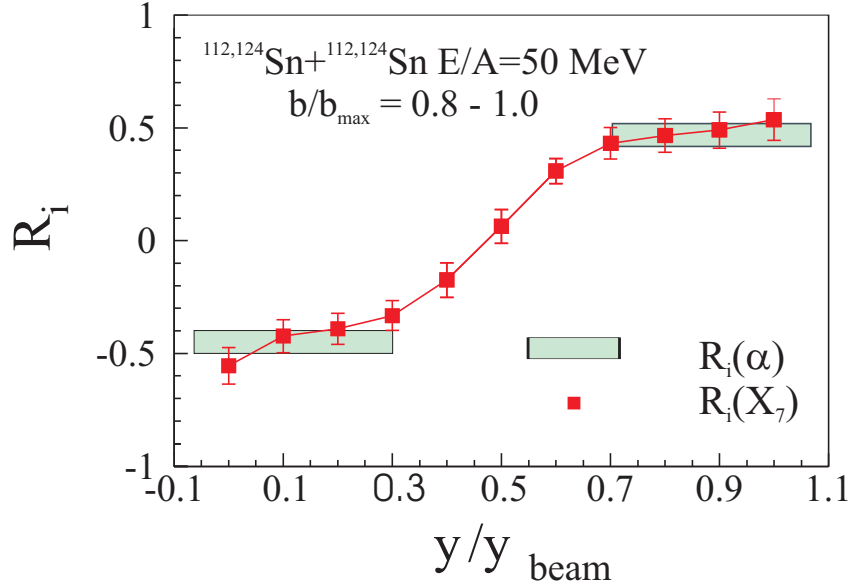}
\caption{(Color online) The rapidity dependence of the degree of
isospin diffusion $R_i$ in the reaction $^{124}$Sn + $^{112} $Sn
at $50$ MeV/nucleon. Taken from Ref.~\cite{Liu07}. }
\label{Liufig}
\end{figure}

In this subsection, we illustrate an example of extracting the
density dependence of nuclear symmetry energy using the isospin
transport/diffusion data in heavy-ion collisions taken at the
NSCL/MSU by Tsang \textsl{et al.} \cite{Tsa04,Liu07}. Isospin
transport/diffusion in heavy ion collisions can in principle be
studied by examining the isospin asymmetry of the projectile-like
residue $\delta_{\rm res}$ in the final state. Since reactions at
intermediate energies are complicated by pre-equilibrium particle
emission and production of neutron-rich fragments at mid-rapidity,
differences of isospin transport/diffusions in mixed and symmetric
systems are usually used to minimize these effects \cite{Tsa04}. To
study isospin transport/diffusion in $^{124}$Sn + $^{112}$Sn
reactions at $E=50$ \textrm{ MeV/nucleon}, reaction systems
$^{124}$Sn + $^{124}$S and $^{112}$Sn + $^{112}$Sn at the same
energy and impact parameter were also considered. The degree of
isospin transport/diffusion in the reaction of $^{124}$Sn + $^{112}
$Sn is then measured by \cite{Ram00}
\begin{eqnarray}
R_{i}(X)=\frac{2X_{^{124}\text{Sn}+^{112}\text{Sn}}-X_{^{124}\text{Sn}+^{124}
\text{Sn}}-X_{^{112}\text{Sn}+^{112}\text{Sn}}}{X_{^{124}\text{Sn}+^{124}
\text{Sn}}-X_{^{112}\text{Sn}+^{112}\text{Sn}}}  \label{Ri}
\end{eqnarray}
where $X$ is an isospin sensitive observable. In the NSCL/MSU
experiments, several experimental probes, such as the multiplicity
ratios of intermediate-mass mirror nuclei $X_7=^7{\rm Li}/^7{\rm
Be}$ and $X_{11}=^{11}{\rm B}/^{11}{\rm C}$ as well as the
isoscaling parameter $X=\alpha$ discussed in Chapter
\ref{chapter_temperature}, were used \cite{Tsa04,Liu07}. Lighter
mirror nuclei, such as $^3{\rm H}/^3{\rm He}$, are strongly
affected by pre-equilibrium emissions and are thus less useful.
While some probes can be more easily and accurately measured than
others, it was shown analytically in Ref.~\cite{Liu07} that as
long as the probe $X$ depends linearly on $\delta_{\rm res}$, one
has $R_{i}(X)=R_i(\delta_{\rm res})$. Namely, the measured degree
of isospin transport/diffusion is independent of probes used. This
nice feature makes comparisons with model calculations much
easier. As an example, shown in Fig.~\ref{Liufig} are the rapidity
dependence of the $R_i(X)$ obtained using $X_7=^{7}{\rm
Li}/^{7}{\rm Be}$ and $X=\alpha$ in peripheral reactions. While
the $R_i(\alpha)$ is only measured at projectile/target
rapidities, it is clearly seen that it indeed gives about the same
value as the $R_i(X_7)$. It is also worth noting that the measured
$R_i(\alpha)$ around projectile/target rapidities is almost a
constant. Moreover, the value of $R_i(X_7)$ is about zero at
mid-rapidity, which is consistent with the FOPI/GSI data shown in
Fig.~\ref{ruzr2}.

Because of the finding that $R_i(X) $ is approximately independent
of the probe $X$ used and of the fact that it is difficult for
most dynamical models to predict properly the formation of
intermediate mass fragments, most existing transport model
calculations have used the isospin asymmetry of projectile-like
residues \cite{Tsa04,Che05a,LiBA05c,Riz08}. In these calculations,
the average isospin asymmetry $\left\langle \delta \right\rangle $
of the projectile-like residue is normally defined as the
composition of nucleons with local densities higher than $\rho
_{0}/20$ and velocities larger than $1/2$ the beam velocity in the
c.m. frame. A density cut of $\rho_{0}/6$, $\rho_{0}/8$ or
$\rho_{0}/10$ was found to give almost same results
\cite{Tsa04,Che05a,LiBA05c,Riz08}. In the following, we review
some of the calculations and comparisons with the NSCL/MSU data.
First, it is worth mentioning that within a momentum-independent
transport model, in which the nuclear potential depends only on
local nuclear density, the isospin diffusion data from the
NSCL/MSU was found to favor a quadratic density dependence for the
interaction part of the nuclear symmetry energy \cite{Tsa04}. This
conclusion has stimulated much interest because of its
implications to nuclear many-body theories and nuclear
astrophysics. However, the nuclear potential acting on a nucleon
is known to depend also on its momentum. For the nuclear isoscalar
potential, its momentum dependence is well-known and is important
in extracting the information on the equation of state of
symmetric nuclear matter
\cite{Dan02a,Gal87,Wel88,Gal90,Pan93,Zha94,Gre99,Dan00,Per02}. The
momentum dependence of the isovector (symmetry) potential
\cite{Bom01,Das03,LiBA04c,Hod94} was also shown to be important
for understanding many isospin related phenomena in heavy-ion
reactions \cite{LiBA04a,Riz04,Che04}. It is thus necessary to
include the momentum dependence in both the isoscalar and
isovector potentials for studying the effect of nuclear symmetry
energy on isospin diffusion \cite{Che05a,LiBA05c,Riz08}.

\subsubsection{Effects of momentum-dependent interactions on
isospin diffusion}

\begin{figure*}[tbh]
\centering
\includegraphics[scale=1.5]{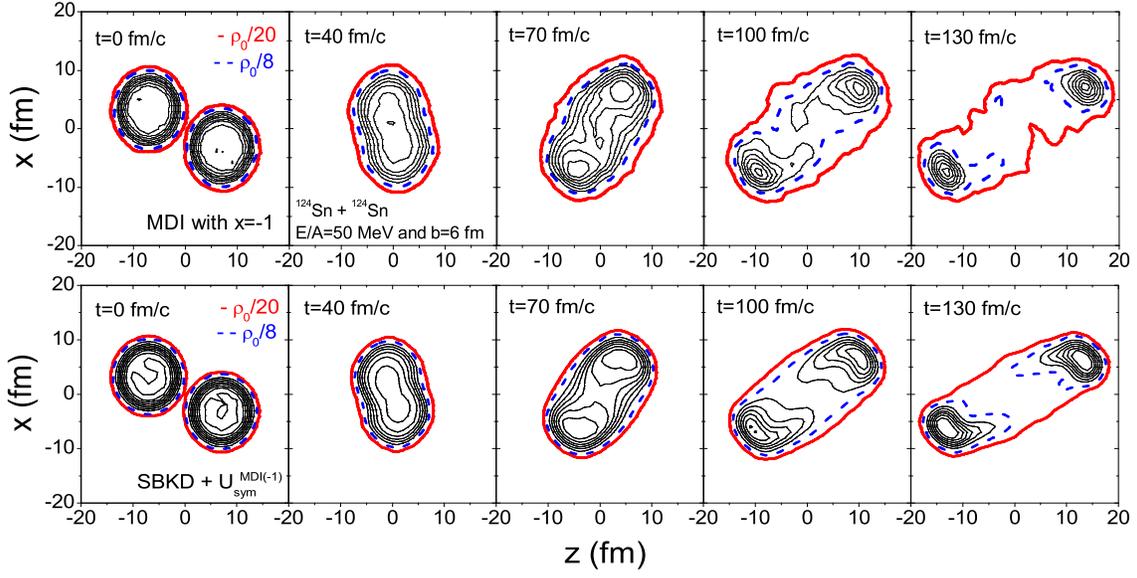}
\caption{{\protect\small (Color online) Density contour
}$\protect\rho (x,0,z)${\protect\small \ in the reaction plane at
different times for reaction }$^{124}${\protect\small Sn +
}$^{112}${\protect\small Sn at }$E/A=50${\protect\small \ MeV and
}$b=6${\protect\small \ fm by using momentum-dependent interaction
MDI with }$x=-1${\protect\small \ (upper panels) and
momentum-independent interaction SBKD with momentum-independent
symmetry potential }$U_{\text{sym}}^{\text{MDI}(-1)}(\protect\rho ,\protect%
\delta ,\protect\tau )${\protect\small \ (lower panels). The thick
solid lines represent }$\protect\rho _{{\protect\small
0}}${\protect\small /20
while dashed lines represent }$\protect\rho _{{\protect\small 0}}$%
{\protect\small /8. Taken from Ref.~\cite{Che07b}.}} \label{CouXZ}
\end{figure*}

Effects of the momentum-dependent interactions on the dynamical
evolution of heavy-ion collisions can be seen from the density
contour $\rho (x,0,z)$ in the reaction plane at different times as
shown in Fig.~\ref{CouXZ} for the reaction $^{124}$Sn + $^{112}$Sn
at $E/A=50$ MeV and $b=6 $ fm calculated with $x=-1$ using both
the MDI and the soft Bertsch-Das Gupta-Kruse (SBKD) interactions.
It should be noted that the former (MDI) interactions are momentum
dependent for both isoscalar and isovector nuclear potentials
while the later (SBKD) interactions do not include any momentum
dependence in either isoscalar or isovector nuclear potentials
though both interactions have similar incompressibility $K_{0}$
and the same density dependence of the symmetry energy. The
experimental free-space NN cross sections are adopted in these
calculations. It is seen from Fig.~\ref{CouXZ} that both the
momentum-dependent MDI interaction and momentum-independent SBKD
interaction give similar dynamic evolution in time, namely,
projectile-like residue and target-like residue can be separated
clearly after about $100$ fm/c. Detailed comparison indicates that
the momentum-dependent MDI interaction make the reaction system
expand more quickly and more nucleons are emitted~\cite{Che04}.
This feature was recently verified by the Catania
group~\cite{Riz08}.

\begin{figure}[tbh]
\centering
\includegraphics[scale=0.9]{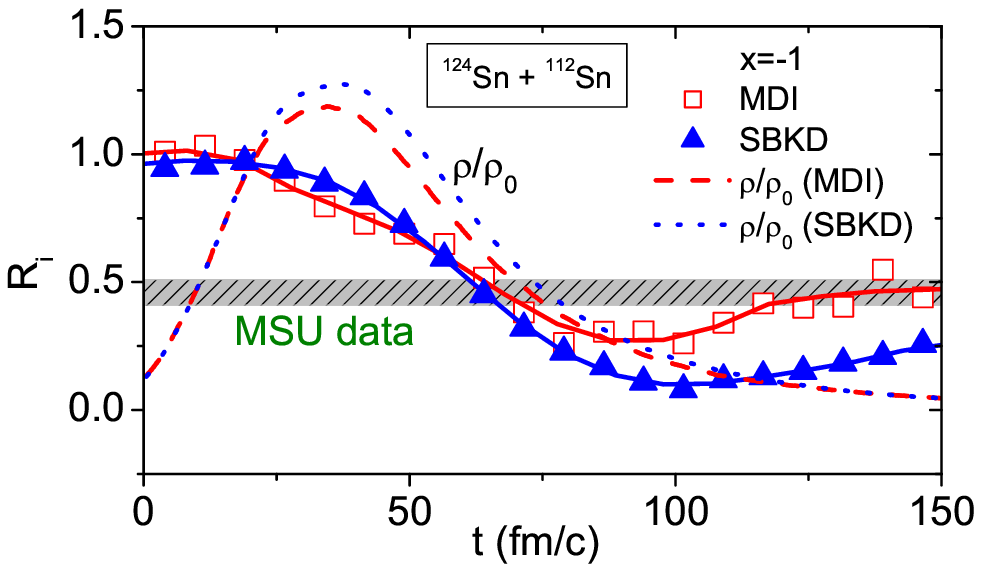}
\includegraphics[scale=0.85]{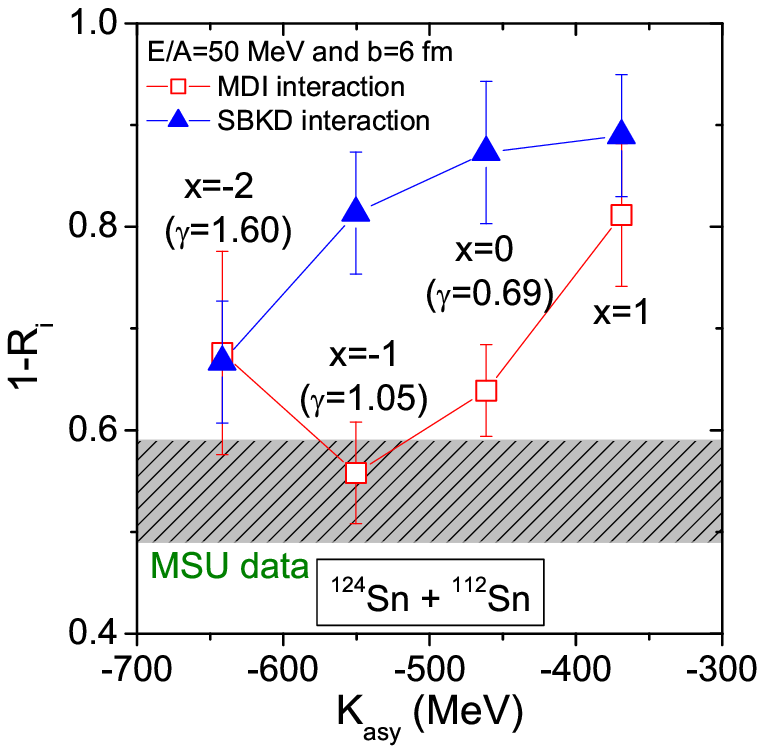}
\caption{The degree of isospin diffusion as a function of time (left
window) and $K_{\text{asy}}$ (right window) with the MDI and SBKD
interactions. The corresponding evolutions of central density are
also shown on the left. Taken from Ref. \protect\cite{Che05a}.}
\label{RiTime}
\end{figure}

Results from these studies on the isospin diffusion are shown in
Fig.~\ref{RiTime}. The left window shows the IBUU04 predictions with
$x=-1$ using both the MDI and the soft SBKD interactions and the
free-space NN cross sections. It is seen that the isospin diffusion
process occurs mainly from about $30$ fm/c to $80$ fm/c when the
average central density changes from about $1.2\rho _{0}$ to
$0.3\rho _{0}$. However, the value of $R_{i}$ still changes slightly
with time until after about $120$ fm/c when projectile-like and
target-like residues are well separated as shown in Fig.~
\ref{CouXZ}. This is partly due to the fact that the isovector
potential remains appreciable at low densities. Also, evaluating
isospin diffusion $R_i$ based on three reaction systems, that have
different time evolutions for the projectile residue as a result of
different total energies and numbers of nucleons, further
contributes to the change of $R_i$ at low density. For the two
interactions consider here, the main difference between the values
for $R_i$ appears in the expansion phase when densities in the
participant region are well below $\rho _{0}$. The experimental data
from MSU are seen to be reproduced nicely by the MDI interaction
with $x=-1$, while the SBKD interaction with $x=-1$ leads to a
significantly lower value for $R_{i}$ due to its stronger
momentum-independent potential, which has been shown to enhance the
isospin diffusion \cite{LiBA04b,Tsa04,Far91}.

Effects of the symmetry energy on isospin diffusion were also
studied by varying the parameter $x$ \cite{Che05a}. Show on the
right window of Fig.~\ref{RiTime} is the final saturated value for
$1-R_{i}$, which measures the degree of isospin diffusion, as a
function of $K_{\text{asy}}$ for both MDI and SBKD interactions.
It is obtained by averaging the value of $1-R_{i}$ after $120$
fm/c with error bars corresponding to its dispersion, whose
magnitude is similar to the error band shown in Ref.~\cite{Tsa04}
for the theoretical results from the BUU model. For the SBKD
interaction without momentum dependence, the isospin diffusion
decreases monotonically (i.e., increasing value for $R_i$) with
increasing strength of $K_{\mathrm{asy}}$ as the corresponding
isovector potential is mostly positive and decreases with
increasing stiffness of $E_{\mathrm{sym}}(\rho)$ in the whole
range of considered $x$ parameter. The isospin diffusion is
reduced when the momentum-dependent MDI interaction is used as the
momentum dependence weakens the strength of symmetry potential
except for $x=-2$.  For the symmetry energy in the MDI
interaction, besides the well-known contribution from nucleon
kinetic energies, i.e., $E_{\text{sym}}^{\text{kin}}(\rho
)=(2^{2/3}-1)\frac{3}{5}E_{F}^{0}(\rho /\rho _{0})^{2/3}\approx
13.0(\rho /\rho _{0})^{2/3}$, the interaction part of nuclear
symmetry energy can be well parameterized by \cite{Che05a}
\begin{eqnarray}
E_{\text{sym}}^{\mathrm{pot}}(\rho ) =F(x)\rho /\rho_{0}
+(18.6-F(x))(\rho /\rho _{0})^{G(x)},
\end{eqnarray}
with $F(x)$ and $G(x)$ given in Table \ref{MDIx} for $x=1$, $0$,
$-1$ and $-2$. Also shown in Table \ref{MDIx} are other
characteristics of the symmetry energy, including its slope
parameter $L$ and curvature parameter $K_{\text{sym}}$ at $\rho_0$,
as well as the isospin-dependent part $K_{\mathrm{asy}}$ of the
isobaric incompressibility of asymmetric nuclear matter.

\begin{table}[tbp]
\centering \caption{{\protect\small The parameters
}$F${\protect\small \ (MeV), }$G$
{\protect\small , }$K_{\text{sym}}${\protect\small \ (MeV), } $L$%
{\protect\small \ (MeV), and }$K_{\text{asy}}${\protect\small \
(MeV) for different values of} $x${\protect\small. Taken from Ref.
\protect\cite{Che05a}.}}
\label{MDIx}%
\begin{tabular}{ccccccc}
\hline\hline
$x$ & \quad $F$ & $G$ & $K_{\text{sym}}$ & $L$ & $K_{\text{asy}}$ &  \\
\hline
$1$ & $107.232$ & $1.246$ & $-270.4$ & $16.4$ & -368.8 &  \\
$0$ & $129.981$ & $1.059$ & $-88.6$ & $62.1$ & -461.2 &  \\
$-1$ & $3.673$ & $1.569$ & $94.1$ & $107.4$ & -550.3 &  \\
$-2$ & $-38.395$ & $1.416$ & $276.3$ & $153.0$ & -641.7 &  \\
\hline\hline
\end{tabular}%
\end{table}

\subsubsection{Effects of in-medium NN cross sections on isospin
diffusion}

In the above study on isospin diffusion, free-space NN cross
sections are used. However, the isospin degree of freedom plays an
important role in heavy-ion collisions through both the nuclear EOS
and the NN scatterings \cite{LiBA98,LiBA01b}. In particular, the
transport of isospin asymmetry between two colliding nuclei is
expected to depend on both the symmetry potential and the in-medium
NN cross sections. For instance, the drifting contribution to the
isospin transport in a nearly equilibrium system is proportional to
the product of the mean relaxation time $\tau _{np}$ and the isospin
asymmetric force $F_{np}$ \cite{Shi03}. While the $\tau _{np}$ is
inversely proportional to the neutron-proton (np) scattering cross
section $\sigma _{np}$ \cite{Shi03}, the $F_{np}$ is directly
related to the gradient of the symmetry potential. On the other
hand, the collisional contribution to the isospin transport in
non-equilibrium system is generally expected to be proportional to
the np scattering cross section. Thus the isospin transport in
heavy-ion reactions depends on both the long-range and the
short-range parts of the isospin-dependent in-medium nuclear
effective interactions, namely, the symmetry potential and the
in-medium np scatterings cross sections. The
former relates directly to the density dependence of the symmetry energy $E_{%
\text{sym}}(\rho )$.

\begin{figure}[tbh]
\centering
\includegraphics[scale=0.85]{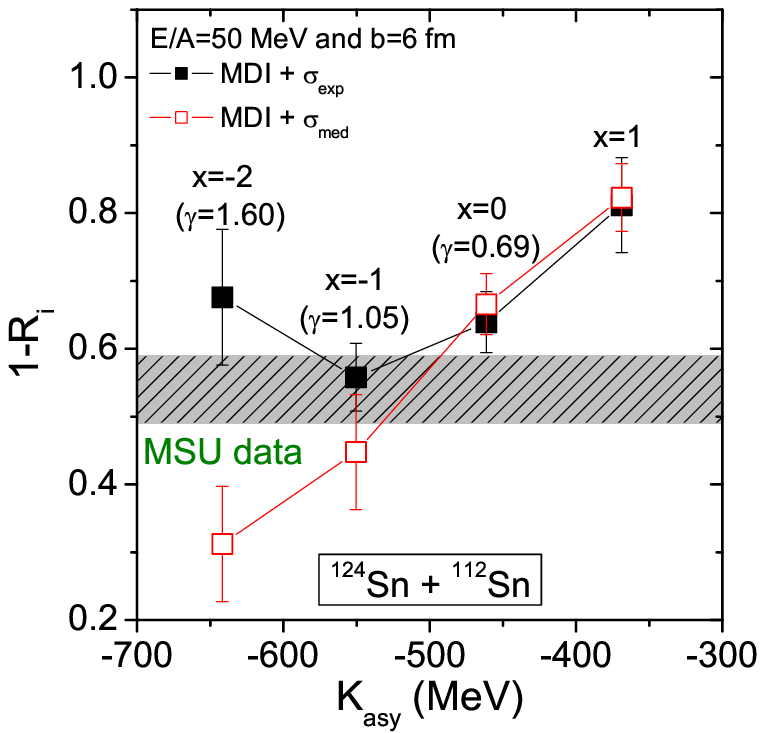}
\includegraphics[scale=0.95]{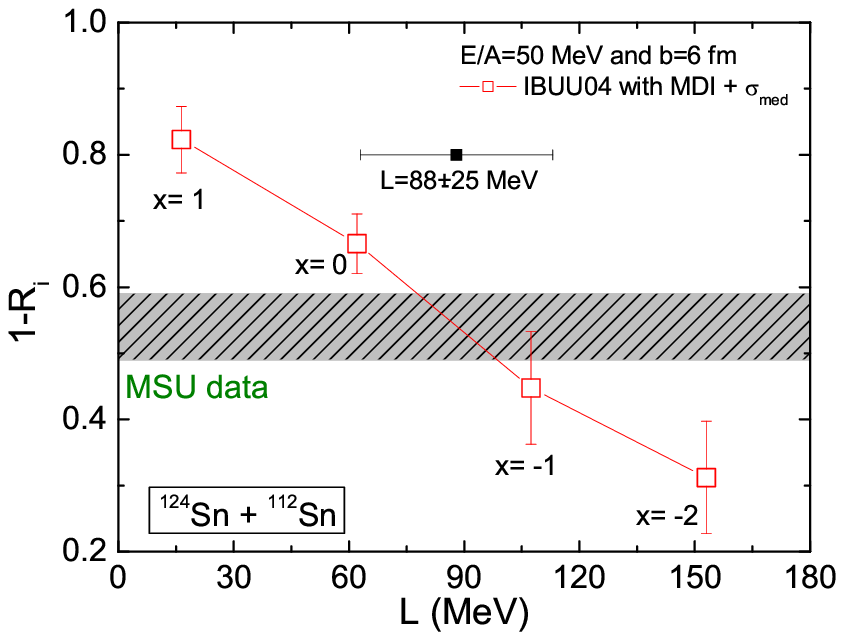}
\caption{The strength of isospin diffusion as a function of
$K_{\rm asy}$ (left window) and $L$ (right window) with the
in-medium nucleon-nucleon cross sections. Taken from Refs.
\protect\cite{LiBA05c} and \protect\cite{Che05b}.}
\label{RiTimeMed}
\end{figure}

Shown on the left window of Fig.~\ref{RiTimeMed} is a comparison
of the averaged strength of isospin transport $1-R_{i}$ obtained
with either the free or in-medium NN cross sections as a function
of the asymmetric part of the isobaric incompressibility of
nuclear matter at $\rho _{0}$, i.e., $K_{\rm asy}(\rho _{0})$. It
is seen that with the in-medium NN cross sections the strength of
isospin transport $1-R_{i}$ decreases monotonically with
decreasing value of $x$. With free-space NN cross sections, there
appears to be a minimum at around $x=-1$ and this minimum is the
point closest to the experimental data. One thus can extract a
value of $K_{\rm asy}(\rho _{0})=-550\pm 100$ MeV using the
free-space NN cross sections. With the in-medium NN cross
sections, one can further narrow down the value of $K_{\rm
asy}(\rho _{0})$ to about $-500\pm 50$ MeV. The left window of
Fig.~\ref{RiTimeMed} further shows that the difference in
$1-R_{i}$ obtained with the free-space and the in-medium NN cross
section is about the same for $x=1$ and $x=0$ but becomes
especially large between $x=-1$ and $x=-2$. The increasing effect
of the in-medium NN cross sections with decreasing $K_{\rm
asy}(\rho _{0})$ or $x$ parameter can be understood from
considering contributions from the symmetry potential and the np
scatterings. As mentioned above, both contributions to the isospin
transport depend on the np scattering cross section $\sigma
_{np}$. Schematically, the mean-field contribution is proportional
to the product of the isospin asymmetric force $F_{np}$ and the
inverse of the np scattering cross section $\sigma _{np}$. While
the collisional contribution is proportional to the $\sigma
_{np}$. The overall effect of the in-medium NN cross sections on
isospin transport is a result of a complicated combination of both
the mean field and the NN scatterings. Generally speaking, the
symmetry potential effects on the isospin transport become weaker
when the NN cross sections are larger while the symmetry potential
effects show up more clearly if smaller NN cross sections are
used. As the $x$ parameter decreases to $x=-1$ and $x=-2$,
however, the symmetry potential decreases and its density slope
can be even negative at low densities. In these cases, either the
collisional contribution dominates or the mean-field contribution
becomes negative. The reduced in-medium np scattering cross
section $\sigma _{np}$ leads then to a lower isospin transport
compared with the case with the free-space NN cross sections.
Shown also in the figure are the $\gamma $ values used in fitting
the symmetry energy with $E_{\text{sym}}(\rho )=31.6(\rho /\rho
_{0})^{\gamma }$ at subsaturation density ($\rho \leq \rho_0$).
The results with the in-medium NN cross sections constrain the
$\gamma $ parameter to be between $0.69$ and $1.05$ corresponding
to $x=0$ and $x=-1$.

Since the slope parameter $L$ of the nuclear symmetry energy gives
an important constraint on the density dependence of the nuclear
symmetry energy and is related to the neutron skin thickness of
heavy nuclei, as to be discussed in
Chapter~\ref{chapter_neutronskin}, it is of interest to see how
the isospin diffusion data constrain the value of $L$. This is
shown on the right window of Fig.~\ref{RiTimeMed}. It is seen that
the strength of isospin diffusion $1-R_{i}$ decreases
monotonically with decreasing value of $x$ or increasing value of
$L$. This is expected as the parameter $L$ reflects the difference
in the pressures on neutrons and protons. From comparison of the
theoretical results with the data, a value of $L=88\pm 25$ MeV as
shown by the solid square with error bar has been extracted.

\begin{figure}
\centering
\includegraphics[scale=0.4]{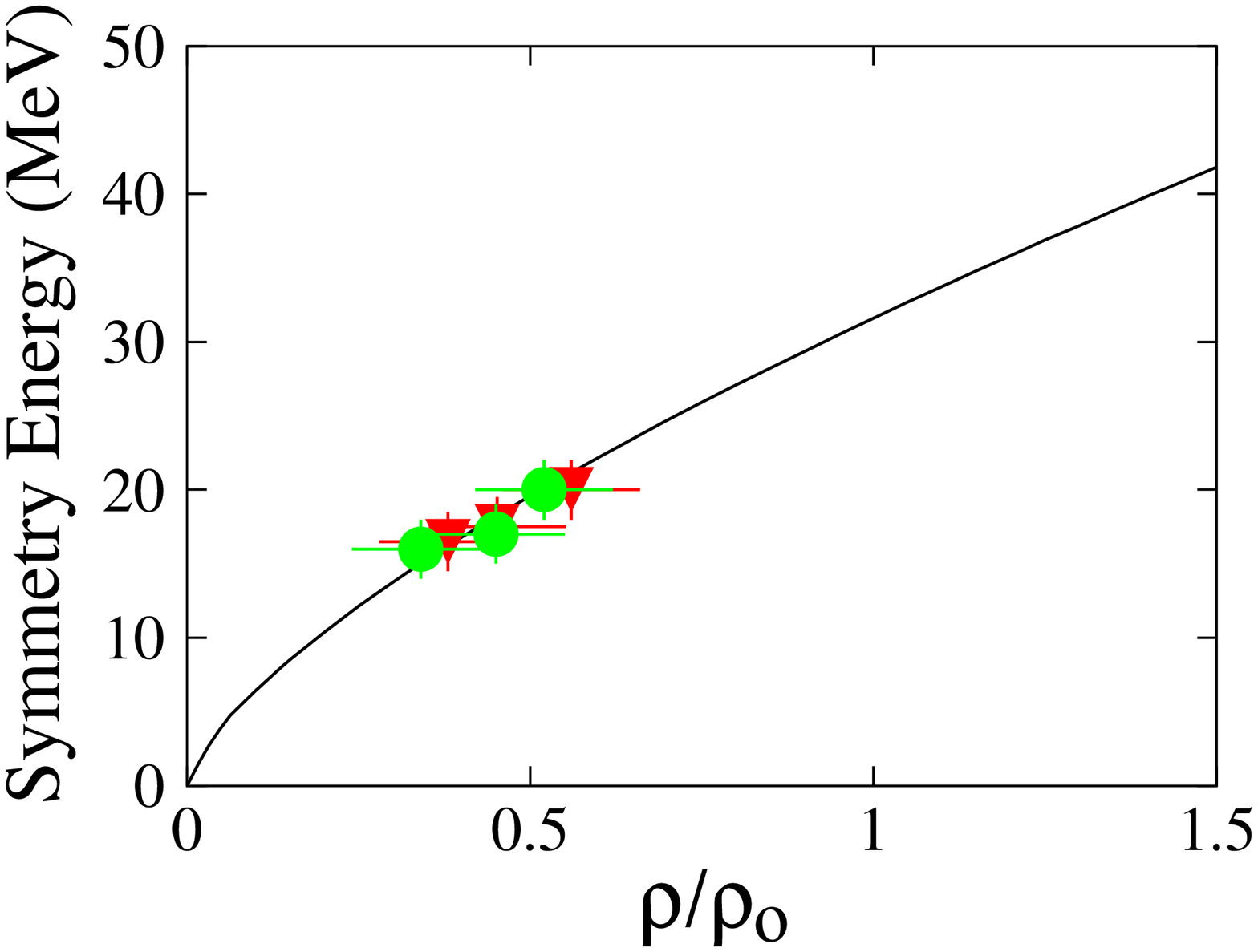}
\includegraphics[scale=0.4]{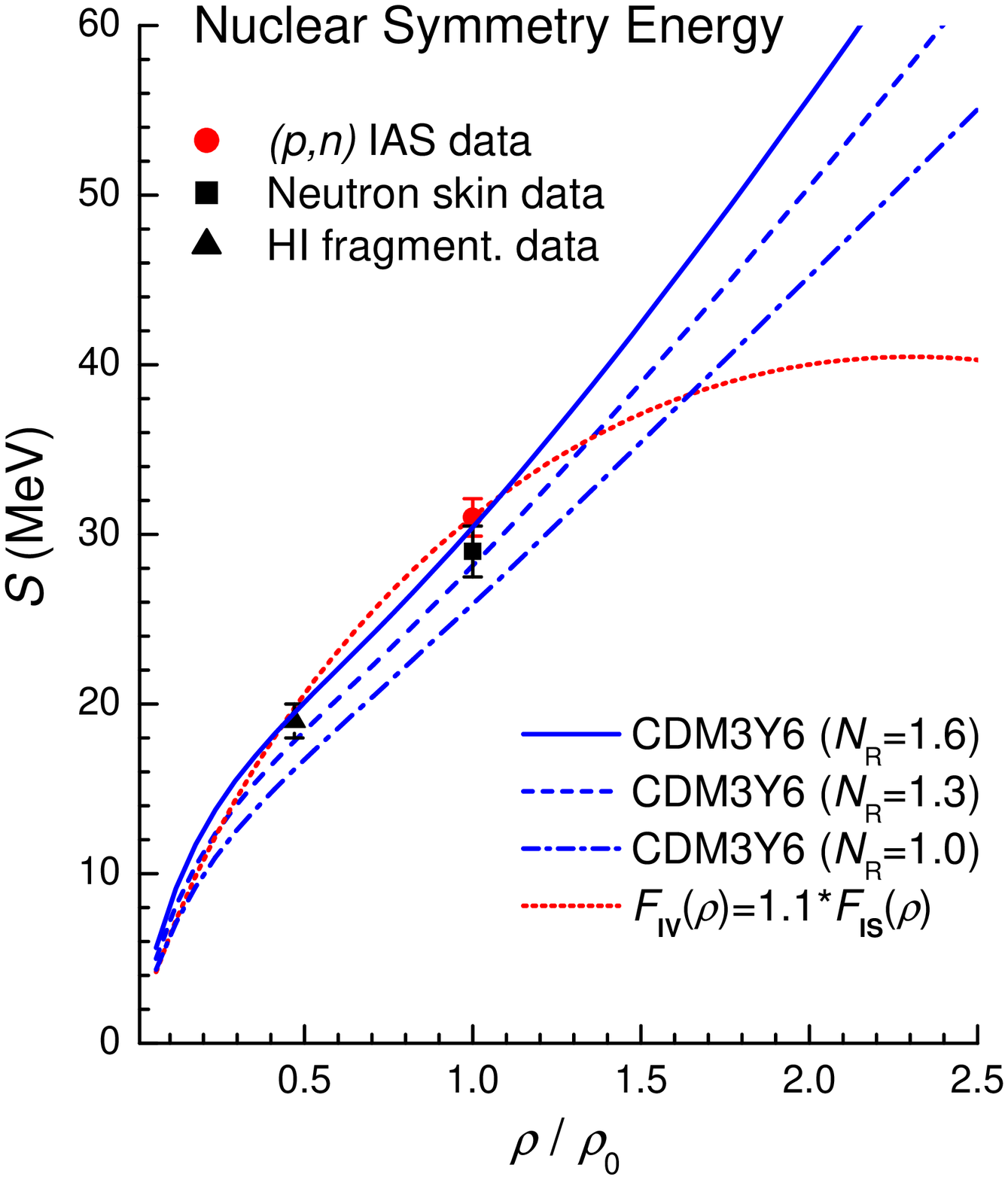}
\caption{Left window: Symmetry energy extracted from isosclaing
analyses for the Fe + Fe and Ni + Ni pair of reaction (inverted
triangles), and the Fe + Ni and Ni + Ni pair of reactions (solid
circles) at 30, 40 and 47 MeV/nucleon.  The solid curve is a fit
with $E_{\rm sym}(\rho)=31.6 (\rho/\rho_0)^{0.69}$. Taken from Ref.
\cite{She06}. Right window: The symmetry energy $S(\rho)$ given by
the HF calculations using different isovector density dependences of
the CDM3Y6 interaction and the empirical values deduced from the
analysis of the (p,n) charge-exchange reaction. Taken from Ref.
\cite{Kho07}.} \label{Shetty}
\end{figure}

It is also of interest to compare the range of symmetry energy
extracted from the analysis of the isospin diffusion data with the
results from other studies. The $E_{\text{sym}}(\rho )=31.6(\rho
/\rho _{0})^{1.05}$ extracted using the free-space NN cross
sections sets an upper limit. The lower bound of
$E_{\text{sym}}(\rho )=31.6(\rho /\rho _{0})^{0.69}$ is close to
the symmetry energy extracted from studying giant resonances
\cite{Pie05,Col05}. Moreover, the density dependence of the
symmetry energy with $x=0$ also fits very well the values
extracted by Shetty {\it et al.} \cite{She06} from the isoscaling
analyses as shown on the left window of Fig.~\ref{Shetty}. We
stress here that the symmetry energy extracted from comparing
transport model calculations with experimental data of heavy-ion
reactions, such as the isospin diffusion, is the symmetry energy
of neutron-rich matter at zero temperature constructed
analytically from the interaction used in the calculations. The
symmetry energy or symmetry free energy extracted directly from
analyzing the isoscaling data is at finite temperatures.
Therefore, the above comparison between the symmetry energies
extracted from the transport model calculations and the isoscaling
analyses is only meaningful if the temperature dependence of the
symmetry energy of hot matter is negligible. As first shown in
Ref.~\cite{LiBA06c} and later confirmed by several other studies,
the temperature dependence of the symmetry energy is rather weak
around the freeze-out temperature which is normally less than 10
MeV in heavy-ion reactions at intermediate energies \cite{Nat02a}.
The right window of Fig.~\ref{Shetty} further shows that the
symmetry energy extracted by Shetty {\it et al.} is consistent
with the HF calculation by Khoa {\it et al.} using the CDM3Y6
interaction \cite{Kho96}. Khoa also deduced a symmetry energy of
about 30 MeV at $\rho\approx \rho_0$ from the coupled-channel
analysis of the (p,n) charge-exchange reaction \cite{Kho05,Kho07}.
This result is consistent with earlier conclusions of several
other studies using different approaches.

\subsection{The isospin relaxation time in
heavy-ion collisions}

Although how fast the initial isospin asymmetry of a reaction
system approaches isospin equilibrium carries important
information on both the density dependence of the nuclear symmetry
energy and the in-medium nucleon-nucleon scattering cross
sections, only limited studies have so far been carried out
\cite{LiBA97b}. On the other hand, extensive efforts have been
devoted to investigate the rate of thermalization of nuclear
matter in intermediate energy heavy ion collisions. Using
transport models, the momentum relaxation both in collisions of
semi-infinite nuclear matter and in infinite nuclear matter have
been studied \cite{Ber78,Ran79,Won82}. It has also been studied
for nuclear reactions, see, e.g., Refs.
\cite{Cas87,Abg94,Had96,Bor97}. In many statistical and dynamical
models it is assumed that isospin equilibrium is reached either
instantaneously or as fast as momentum equilibrium. Our above
discussions show that this assumption is only approximately true
in deep inelastic heavy ion collisions at low energies, whereas
global isospin equilibrium is never achieved at higher energies.
It is thus of interest to compare the relaxation time of isospin
with other characteristic times of the reaction. In the following,
the relaxation times for isospin and for momentum from studies
based on the IBUU model are compared. In such model study, a heavy
residue can be identified as a collection of nucleons with
densities higher than 1/10 of normal nuclear matter density. To
characterize the degree of chemical equilibrium, one can introduce
the following quantity
\begin{eqnarray}
\lambda_I(t)\equiv \frac{(n/p)_{y>0}}{(n/p)_{y<0}},
\end{eqnarray}
where $(n/p)_{y>0}$ and $(n/p)_{y<0}$ are, respectively, the
neutron to proton ratio for positive and negative rapidity
nucleons in the rest frame of the residue. If chemical equilibrium
is established this quantity then has a value of one. Furthermore,
the isospin relaxation time $\tau_I$ can be defined as the time
when the quantity $(\lambda_I(t)-1)/(\lambda_I(0)-1)$ is 0.01,
i.e., it is one percent from its equilibrium value. This time is
an approximate measure of the rate at which the residue reaches
isospin equilibrium. This definition is somewhat different from
that one would usually use, i.e., $\lambda(\tau_I)=1/e$.

For describing thermal equilibrium of the heavy residue, the
quadrupole moment $Q_{zz}(t)$ defined earlier can be used.
Obviously $Q_{zz}=0$ is a necessary, although not sufficient,
condition for thermal equilibrium. Similar to the definition of
$\tau_I$, the momentum relaxation time $\tau_p$ can be defined as
the time when $Q_{zz}(t)/Q_{zz}(0)=1\%$. This quantity then
measures the rate at which the residue reaches thermal
equilibrium. Another important property of the heavy residue is
the possible existence of dynamical instabilities. To study this
phenomenon, the square of the adiabatic sound velocity has been
introduced as follows \cite{Bau92a,ligross}
\begin{eqnarray}
v_s^2 = \frac{1}{m}\left(\partial P\over\partial\rho\right)_S =
{1\over m} \left[ {10\over 9}\langle E_k \rangle + a
{\rho\over\rho_0} + b \sigma\left(\rho\over\rho_0\right)^{\sigma}
\right]\ ,
\end{eqnarray}
where $\langle E_k \rangle$ is the average kinetic energy per
nucleon, $a= -358.1$ MeV, $b= 304.8$ MeV and $\sigma=7/6$ are the
parameters corresponding to a soft nuclear equation of state. For
$v^2_s < 0$, a homogeneous nuclear matter is unstable against the
growth of fluctuation, leading to dynamical instability or spinodal
decomposition.

\begin{figure}[htp]
\centering
\includegraphics[scale=0.5]{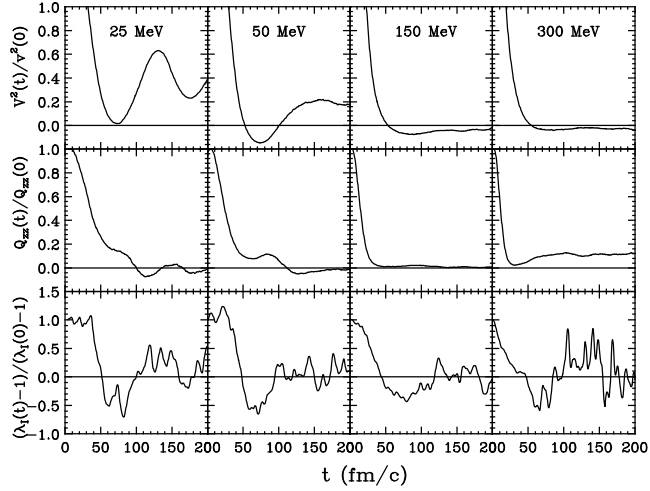}
\caption{The development of dynamical instability (upper panels),
thermal (middle panels) and chemical (lower panels) equilibrium in
$^{40}$Ca+$^{124}$Sn collisions at an impact parameter of 1 fm and
beam energies of 25, 50, 150 and 300 MeV/nucleon, respectively.
Taken from Ref. \cite{LiBA98b}. }\label{itime1}
\end{figure}

It is interesting to compare the time dependence of
$(\lambda_I(t)-1)/(\lambda_I(0)-1)$, $Q_{zz}(t)/Q_{zz}(0)$ and
$v_s^2(t)/v_s^2(0)$. They are shown in Fig.\ \ref{itime1} for
collisions of $^{40}$Ca+$^{124}$Sn at an impact parameter of 1 fm
and at beam energies of 25, 50, 150 and 300 MeV/nucleon. The
system considered here has an initial $\lambda_I(0)=1.48$. For
collisions at a beam energy of 25 MeV/nucleon the residue is found
to be dynamically stable up to 300 fm/c during the collision. This
time interval is long enough for both thermal and chemical
equilibrium to be fully established as shown in the middle and
lower windows of the first column. On the other hand, for
collisions at beam energies above 50 MeV/nucleon a significant
compression appears, and this is followed by expansion, leading
into the adiabatic spinodal region after about 50 fm/c. At this
time the heavy residue formed in the collision is still far from
thermal and chemical equilibrium. Both the momentum and isospin
asymmetries of the heavy residue are seen to oscillate with time.

It is seen from the middle window of the fourth column that
nuclear translucency occurs at E/A=300 MeV. After spinodal
decomposition the heavy residue quickly starts to break up into
fragments and nucleons \cite{ligross}. Although the isotopic
contents of these fragments and nucleons depend on the emission
angle, it is also strongly influenced by the neutron to proton
ratio of the target and projectile in the entrance channel instead
of the average neutron to proton ratio of the combined system.
This observation is consistent with recent experimental findings
\cite{She94,sherry2,sherry3}.

\begin{figure}[htp]
\centering
\includegraphics[scale=0.5]{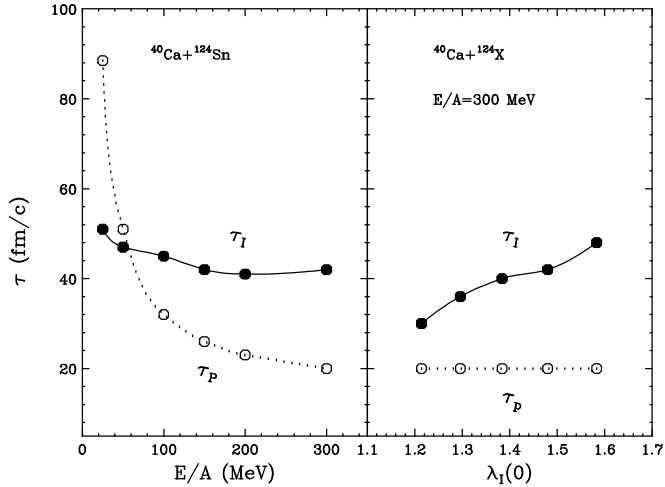}
\caption{Left panel: Isospin (open circles) and momentum (filled
circles) relaxation times as functions of beam energy. Right
panel: Relaxation times as functions of $\lambda_I(0)$ of the
reactions. Taken from Ref. \cite{LiBA98b}.}\label{relax}
\end{figure}

Although the isospin and thermal equilibrium are not completely
established at beam energies higher than the Fermi energy, it is
still interesting to compare their relaxation times. This is shown
in Fig.\ \ref{relax}. In the left panel, the comparison is made
for $^{40}$Ca+$^{124}$Sn collisions at beam energies from 25 to
300 MeV/nucleon and at an impact parameter of 1 fm. The momentum
relaxation time is seen to decrease with increasing beam energy.
This is in qualitative agreement with that found in Refs.
\cite{Ber78,Ran79,Won82,Cas87,Abg94,Bor97}. On the other hand, the
isospin relaxation time decreases slowly with the beam energy. The
shorter isospin relaxation time at incident energies below about
50 MeV/nucleon is in agreement with what was found in deep
inelastic heavy ion collisions \cite{Gat75,Fed78,Udo84}. At higher
incident energies the time for reaching momentum equilibrium is
shorter than that for isospin equilibrium. For example, a 20 fm/c
difference in the relaxation time is observed at E/A=300 MeV. In
the right panel, the isospin and momentum relaxation time as
functions of the initial isospin asymmetry $\lambda_I(0)\equiv
(n/p)_{\rm projectile}/(n/p)_{\rm target}$ are compared for
$^{40}$Ca induced reactions on several isobaric targets of mass
124 at a beam energy of 300 MeV and at an impact parameter of 1
fm. It is seen that although the momentum relaxation is almost
independent of the initial isospin asymmetry, the isospin
relaxation time increases with the initial isospin asymmetry.

The shorter relaxation time for isospin than momentum at low
incident energies can be understood as follows. First,
nucleon-nucleon collisions, which are responsible for momentum
relaxation, are more likely to be suppressed due to Pauli
blocking. Secondly, the repulsive symmetry potential for neutrons
and the attractive symmetry potential for protons make
pre-equilibrium emissions of neutrons more likely than protons in
low energy collisions, which thus enhances the isospin relaxation
rate in the residue. On the other hand, in high energy collisions,
Pauli blocking is less effective and the symmetry potential is
also less important, leading thus to a shorter momentum relaxation
time and a longer isospin relaxation time. Effects due to
different forms of symmetry potential and the charge exchange
reaction ($pn\rightarrow np$) on chemical and thermal equilibrium
have been studied in Ref.~\cite{LiBA98b}. They are found to have
no discernible effects on both the momentum and isospin relaxation
times. Only during the later stage of the collisions do they
affect slightly the momentum and isospin distributions.

Therefore, isospin and momentum relaxation times in the heavy
residues formed in heavy-ion collisions at intermediate energies can
be completely established only at beam energies below the Fermi
energy. At higher energies the dynamical instability sets in before
either chemical or thermal equilibrium is achieved. Moreover, the
isospin relaxation time is shorter (longer) than that for momentum
at beam energies lower (higher) than the Fermi energy.

\subsection{High density behavior of the nuclear symmetry energy and the isospin
asymmetry of the dense matter formed in high energy heavy-ion
reactions}\label{hdsymb}

\begin{figure}[tbh]
\centering
\includegraphics[scale=0.5]{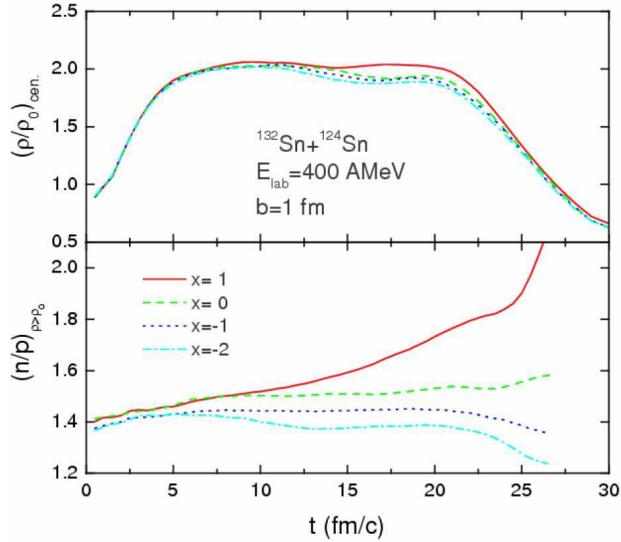}
\caption{(Color online) Central baryon density (upper panel) and
isospin asymmetry (lower panel) of high density region in the
reaction of $^{132}{\rm Sn}+^{124}{\rm Sn}$ at a beam energy of
400 MeV/nucleon and an impact parameter of 1 fm. Taken from Ref.
\protect\cite{LiBA05a}.} \label{CentDen}
\end{figure}

The maximum baryon density and isospin asymmetry achieved in
central heavy-ion collisions with high energy radioactive beams
can be appreciable, and this provides the opportunity to study the
EOS of asymmetric nuclear matter at future rare isotope
facilities. As an example, Fig.~\ref{CentDen} shows the central
baryon density (upper panel) and the average $(n/p)_{\rho \geq
\rho _{0}}$ ratio (lower panel) of regions with baryon densities
higher than $\rho _{0}$ in the reaction of $^{132}$Sn$+^{124}$Sn
at a beam energy of 400 MeV/nucleon and an impact parameter of 1
fm. These reactions will be available in future FAIR at GSI and
CSR at Lanzhou. It is seen that the maximum baryon density reached
in these reactions is about 2 times the normal nuclear matter
density, and it lasts for about 15 fm/c from 5 to 20 fm/c.
Although the compression is rather insensitive to the symmetry
energy because the latter is relatively small compared to the EOS
of symmetric matter around this density, the isospin asymmetry of
the high density region is affected appreciably by the symmetry
energy. The soft (e.g., $x=1$) symmetry energy leads to a
significantly higher value of $(n/p)_{\rho \geq \rho _{0}}$ than
the stiff one (e.g., $x=-2$). This is consistent with the
well-known isospin fractionation phenomenon. Because of the
$E_{\rm sym}(\rho )\delta ^{2}$ term in the EOS of asymmetric
nuclear matter, it is energetically more favorable to have a
higher isospin asymmetry $\delta $ in the high density region with
a softer symmetry energy functional $E_{\rm sym}(\rho )$. Since
the symmetry energy in the supranormal density region changes from
a soft one to a stiff one when the parameter $x$ varies from 1 to
$-2$, the value of $(n/p)_{\rho \geq \rho _{0}}$ becomes lower as
the parameter $x$ changes from 1 to $-2$. Because of the
neutron-skins of the colliding nuclei, especially that of the
neutron-rich projectile $^{132}{\rm Sn}$, the $n/p$ ratio on the
low-density surface is much higher than that in their interior. As
a result, the initial value of the quantity $(n/p)_{\rho \geq \rho
_{0}}$ is only about 1.4, which is less than the average $n/p$
ratio of 1.56 of the reaction system. In the dense region, the
matter can become, however, either more neutron-rich or more
neutron-poor with respect to the initial state depending on the
symmetry energy functional $E_{\rm sym}(\rho )$ used in the study.

At even higher densities above twice the normal nuclear matter
density, which are reachable at higher beam energies, the behavior
of the symmetry energy is probably among the most uncertain
properties of dense matter \cite{Kut94,Kut00}. Some predictions
show that the symmetry energy can decrease with increasing density
above certain density and may even finally becomes negative. This
extreme behavior was first predicted by some microscopic many-body
theories, see e.g., Refs.~\cite{Wir88a,Pan72,Kra06}. It has also
been shown that the symmetry energy can become negative at various
high densities within the Hartree-Fock approach using the original
Gogny force \cite{Ono03,Cha97}, the density-dependent M3Y
interaction \cite{Kho96,Bas07} and about 2/3 of the 87 Skyrme
interactions that have been widely used in the literature
\cite{Bom01,Mar02,Sto03,Sto05}. The mechanism and physical meaning
of a negative symmetry energy are still under debate and certainly
deserve more studies.

\begin{figure}[tbp]
\centering
\includegraphics[scale=0.6]{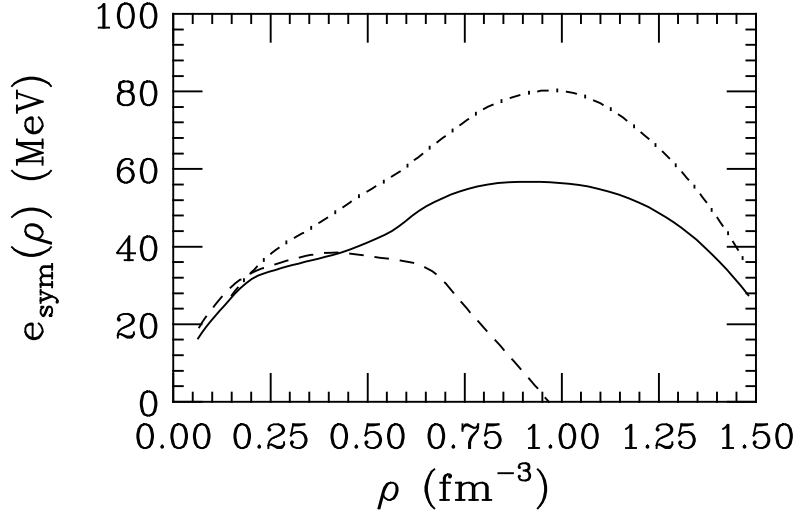}
\caption{Symmetry energy predicted by variational many-body
calculations: {\sc av14+uvii} (solid line), {\sc uv14+uvii}
(dot-dashed), and {\sc uv14+tni} (dashed). Results taken from Ref.\
\protect\cite{Wir88a}.} \label{wiringa}
\end{figure}

In the early studies by Wiringa {\it et al.} using the variational
many-body ({\sc vmb}) theory \cite{Wir88a}, the density dependence
of $e_{\rm sym}(\rho)$ was calculated using either the Argonne
two-body potential {\sc av14} or Urbana {\sc uv14} together with
either the three-body potential {\sc uvii} or {\sc tni}, and the
results differ appreciably as shown in Fig.\ \ref{wiringa}. Although
the symmetry energy remains positive at high densities for the first
two cases as in studies with many other interactions or models, it
vanishes and becomes negative at high densities for the case with
the three-body potential {\sc tni}. This prediction has important
consequences on the structure and magnetic properties of neutron
stars as stressed recently by Kutschera {\it et al.}
\cite{Kut94,Kut00}. A negative symmetry energy at high densities
implies that the pure neutron matter becomes the most stable state
leading to the onset of the isospin separation instability ({\rm
ISI}). Consequently, pure neutron domains or neutron bubbles
surrounding isolated protons may be formed in the cores of neutron
stars \cite{Kut94,Kut00}. Also, when the density is high enough the
chemical potential for neutrons can be higher than the rest mass of
$\Lambda$ hyperon, then a transition to the strange matter can
occur. Furthermore, with negative symmetry energy, the pressure of
the matter becomes negative and the matter would collapse
\cite{Sto05}. The high density behavior of the symmetry energy and
its astrophysical consequences are obviously interesting topics that
deserve further studies. We will return to this issue in Chapter
\ref{chapter_neutronstars} and discuss in more detail the effects of
the high density behavior of the nuclear symmetry energy on kaon
condensation and formation of a hadron-quark mixed phase in the
cores of neutron stars.

\begin{figure}[tbp]
\centering
\includegraphics[scale=0.5,angle=-90]{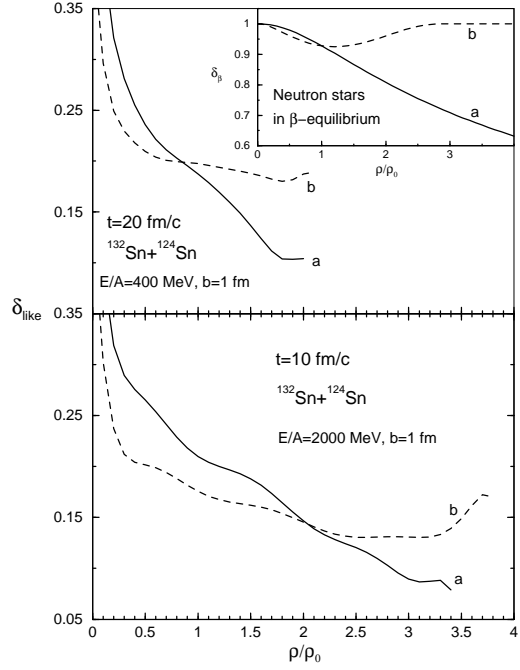}
\caption{Upper panel: Isospin asymmetry-density correlations at
$t=20$ fm/$c$ and $E_{\rm beam}/A=400$ MeV in central $^{132}{\rm
Sn}+^{124}{\rm Sn}$ reaction with the nuclear symmetry energy
$E^a_{\rm sym}$ and $E^b_{\rm sym}$, respectively. Lower panel: Same
correlation as in upper window but at 10 fm/$c$ and $E_{\rm
beam}/A=2$ GeV/nucleon. The corresponding correlation in neutron
stars is shown in the inset. Taken from Ref. \cite{LiBA02}.}
\label{delta-rho}
\end{figure}

Energetic nuclear reactions with rare isotopes provide an
opportunity to pin down the $E_{\rm sym}(\rho)$ at high densities.
They will also allow one to study the possible existence of ISI and
its consequences. It is thus of interest to see what will happen in
a heavy-ion reaction when the negative symmetry energy is reached at
high densities. As an example, one can compare results using the
following two examples of extreme density dependence for the
symmetry energy
\begin{eqnarray}\label{esyma}
E^a_{\rm sym}(\rho)\equiv E_{\rm sym}(\rho_0)u
\end{eqnarray}
and
\begin{eqnarray}\label{esymb}
E^b_{\rm sym}(\rho)\equiv E_{\rm sym}(\rho_0)u\cdot\frac{3-u}{2},
\end{eqnarray}
where $u\equiv\rho/\rho_0$. The $E^a_{\rm sym}(\rho)$ is a typical
RMF prediction while $E^b_{\rm sym}(\rho)$ mimics the prediction
of some VMB calculations and fits very well the VMB calculations
by Lagaris and Pandharipande up to about twice the normal nuclear
matter density \cite{Lag81,Coc04}. By construction, both symmetry
energies have the same value of $E_{\rm sym}(\rho_0)=30$ MeV at
the normal nuclear matter density $\rho_0$ and are very close to
each other at lower densities. At high densities they have
completely different behaviors reflecting the diverging
predictions of nuclear many-body theories. Shown in Fig.\
\ref{delta-rho} are the average (over all phase-space cells of the
same density) isospin asymmetry $\delta$ as a function of density
for central $^{132}{\rm Sn}+^{124}{\rm Sn}$ reactions at 400 and
2000 MeV/nucleon at the instants of approximately maximum
compression. The overall rise of $\delta$ at low densities is
mainly due to the neutron skins of the colliding nuclei and the
distilled neutrons from isospin fractionation. Effects due to
different symmetry energies are clearly revealed especially at
high densities. For a comparison with nuclear astrophysics, the
$\rho-\delta$ correlation in neutron stars at $\beta$ equilibrium
is shown in the inset of Fig.\ \ref{delta-rho}. With $E^{b}_{\rm
sym}(\rho)$, $\delta_{\beta}$ is $1$ for $\rho/\rho_0\geq 3$,
indicating that the neutron star becomes a pure neutron matter at
these high densities. To the contrary, with $E^{a}_{\rm
sym}(\rho)$, the neutron star becomes more proton-rich as the
density increases. An astonishing similarity is seen in the
resultant $\delta-\rho$ correlations for the neutron star and the
heavy-ion collisions. In both cases, the symmetry energy $E^b_{\rm
sym}(\rho)$ makes the high density nuclear matter more
neutron-rich than $E^a_{\rm sym}(\rho)$, and the effect grows with
increasing density. This is not surprising since the same
underlying nuclear ${\rm EOS}$ is at work in both cases. It is
particularly interesting to mention that the decreasing $E^b_{\rm
sym}(\rho)$ above $1.5\rho_0$ makes it more energetically
favorable to have the denser region more neutron-rich. One thus
sees an up turn of the $\delta$ at the high density end for
$E^{b}_{\rm sym}(\rho)$, especially in reactions at $E_{\rm
beam}/A=2$ GeV/A.

The above discussions clearly indicate that the neutron/proton
ratio of the high density phase in both neutron stars and
heavy-ion reactions are determined by the high density behavior of
the nuclear symmetry energy. By probing the neutron/proton ratio
of the high density matter formed in heavy-ion reactions, one can
thus obtain information about the symmetry energy at high
densities. This information has important ramifications for
nuclear astrophysics. We thus devote the next few subsections to
experimental probes of the high density behavior of the nuclear
symmetry energy. We first examine the neutron/proton ratio of
squeezed-out nucleons perpendicular to the reaction plane.
Compared to other potentially powerful probes of the symmetry
energy at supra-normal densities, such as the $\pi ^{-}/\pi ^{+}$
and $K^{0}/K^{+}$ ratios, the $n/p$ ratio of squeezed-out nucleons
carries most directly the information on the symmetry
potential/energy since the latter acts directly on nucleons. Pions
and kaons are mostly produced through nucleon-nucleon and
pion-nucleon inelastic scatterings, they thus carry indirectly and
often secondary or even higher order effects of the symmetry
energy \cite{LiBA06b}. Moreover, nucleonic observables such as the
$n/p$ ratio are essentially free of uncertainties associated with
the production mechanisms of pions and kaons. Of course, one
generally expects that all squeezed-out particles are more
sensitive to the properties of dense matter. Thus the
neutron/proton, $\pi ^{-}/\pi ^{+}$ and $K^{0}/K^{+}$ ratios
perpendicular to the reaction plane are all expected to be more
useful compared to their values in other directions.

\subsection{The neutron/proton ratio of squeezed-out nucleons}

It is well known that the squeeze-out of nuclear matter in the
participant region perpendicular to the reaction plane occurs in
noncentral heavy-ion collisions. In mid-central collisions, high
density nuclear matter in the participant region has larger
density gradient in the direction perpendicular to the reaction
plane. Moreover, in this direction nucleons emitted from the high
density participant region have a better chance to escape without
being hindered by the spectators. These nucleons thus carry more
direct information about the high density phase of the reaction.
They have been widely used in probing the EOS of dense matter,
see, e.g., Refs.
\cite{Dan02a,Zha00,Ber88b,Aic91,Sto86b,Cas90,Rei97} for a review.
Using the IBUU04 model, such study has been done recently to see
whether the squeeze-out nucleons can be used to constrain the high
density behavior of the nuclear symmetry energy \cite{Yon07}. An
example is shown here for the reaction of $^{132}$Sn+$^{124}$Sn at
a beam energy of $400$ MeV/nucleon and an impact parameter of $5$
fm. In this reaction the maximal baryon density reached is about
twice the normal nuclear matter density \cite{Yon06b}.

\begin{figure}[th]
\centering
\includegraphics[scale=0.7]{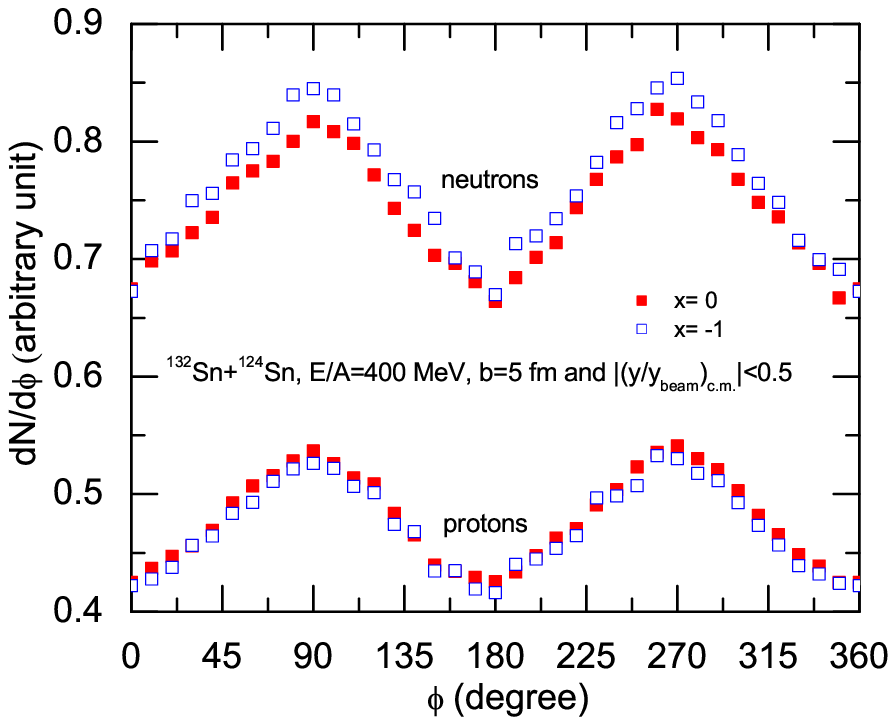}
\includegraphics[scale=0.7]{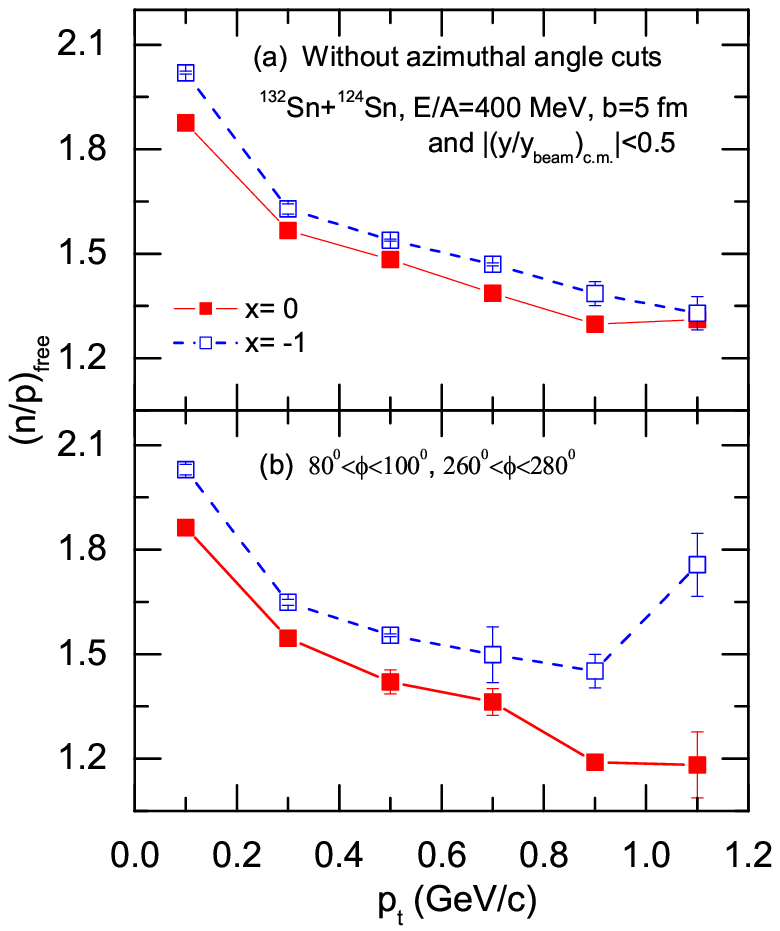}
\caption{Left window: Azimuthal distribution of midrapidity nucleons
emitted in the reaction of $^{132}$Sn+$^{124}$Sn at an incident beam
energy of $400$ MeV/nucleon and an impact parameter of $b=5$ fm.
Right window: Transverse momentum distribution of the ratio of
midrapidity neutrons to protons. For the lower panel, an azimuthal
angle cut of $80^{\circ }<\protect\phi <100^{\circ }$ and
$260^{\circ }<\protect\phi <280^{\circ }$ is used to select the free
nucleons that are from the direction perpendicular to the reaction
plane. Taken from Ref. \protect\cite{Yon07}.} \label{degree}
\end{figure}

Shown on the left window of Fig.\ \ref{degree} are the azimuthal
distributions of free nucleons in the midrapidity region
($|(y/y_{\rm beam})_{\rm c.m.}|<0.5$). A preferential emission of
nucleons perpendicular to the reaction plane is clearly observed
for both neutrons and protons as one expects. Most interestingly,
neutrons emitted perpendicular to the reaction plane show clearly
an appreciable sensitivity to the variation of the symmetry energy
compared to protons. This is mainly because the symmetry potential
is normally repulsive for neutrons and attractive for protons. For
the latter, the additional repulsive Coulomb potential works
against the attractive symmetry potential. Overall, one thus
expects the neutron emission to be more sensitive to the variation
of the symmetry energy. Since the symmetry potential is relatively
small compared to the isoscalar potential, it is always necessary
and challenging to find obervables that are sufficiently sensitive
to the symmetry energy to be useful for extracting information
about the symmetry potential/energy. Fortunately, the squeeze-out
neutrons appears to be a promising one. While it is very difficult
to measure experimentally the neutrons, both the transverse flow
and squeeze-out neutrons together with other charged particles
were measured accurately at both the BEVALAC \cite{Htu99} and
SIS/GSI \cite{Ven93,Lei93,Lam93}. These experiments and the
associated theoretical calculations, see, e.g., Refs.
\cite{Bas95,Lar00}, have all focused on extracting information
about the EOS of symmetric nuclear matter without paying much
attention to the effects due to the uncertainties in the symmetry
energy.

To probe the high density behavior of the symmetry energy, one
would like to avoid as much as possible all remaining
uncertainties associated with the EOS of symmetric nuclear matter.
It is known from previous studies \cite{LiBA98,LiBA97a} that the
$n/p$ ratio is determined mostly by the density dependence of the
symmetry energy and almost not affected by the EOS of symmetric
nuclear matter. As shown in the lower panel of the right window of
Fig.\ \ref{degree}, the symmetry energy effect on the $n/p$ ratio
of midrapidity nucleons emitted in the direction perpendicular to
the reaction plane is appreciable and increases with increasing
transverse momentum $p_{t}$. At a transverse momentum of $1$
GeV/c, the effect can be as high as $40\%$. The high $p_{t}$
particles most likely come from the high density region in the
early stage during heavy-ion collisions and they are thus more
sensitive to the high density behavior of the symmetry energy.
Without the cut on the azimuthal angle, the $n/p$ ratio of free
nucleons in the midrapidity region is much less sensitive to the
symmetry energy in the whole range of transverse momentum as shown
in the upper panel of the right window of Fig.\ \ref{degree}. In
fact, the neutron/proton ratio of squeezed-out nucleons
perpendicular to the reaction plane, especially at high transverse
momenta, is probably the most sensitive probe found so far among
all studied observables.

While it is very hard to measure neutrons, both the transverse
flow and the squeeze-out neutrons were measured at the BEVALAC by
Madey {\it et al.} \cite{Htu99,Mad93} and at the SIS/GSI by the
TAPS and the Land collaborations \cite{Ven93,Lei93,Lam94}. The
measurements were accurate enough to extract reliable information
about the EOS of symmetric nuclear matter and the reaction
dynamics. The analyses of the experimental data and the associated
theoretical calculations, see, e.g., Refs. \cite{Bas95,Lar00},
however, have all focused on extracting only information about the
EOS of symmetric nuclear matter without paying any attention to
the symmetry energy. In all of these experiments, it was essential
to measure simultaneously charged particles together with
neutrons. To study the symmetry energy at high densities using the
$n/p$ ratio of squeezed-out nucleons, similar experimental setups
are necessary, such as the $4\pi$ charged particle detectors for
constructing the reaction plane of the reaction and the neutron
walls or other neutron detectors for determining the momenta of
neutrons via the time of flight. The squeeze-out nucleons can then
be studied with respect to the reaction plane determined by using
the charged particles on the event-by-event basis. The symmetry
energy effects on the $n/p$ ratio of squeezed-out nucleons are
large enough to be measured even with some of the existing
detectors. This optimistic view and the past success in studying
neutron squeeze-out make us feel confident that the predicted
effects can be studied realistically.

\subsection{Isospin dependence of nucleon transverse, elliptical
and radial flow}

\begin{figure}[htp]
\begin{center}
\includegraphics[scale=0.50]{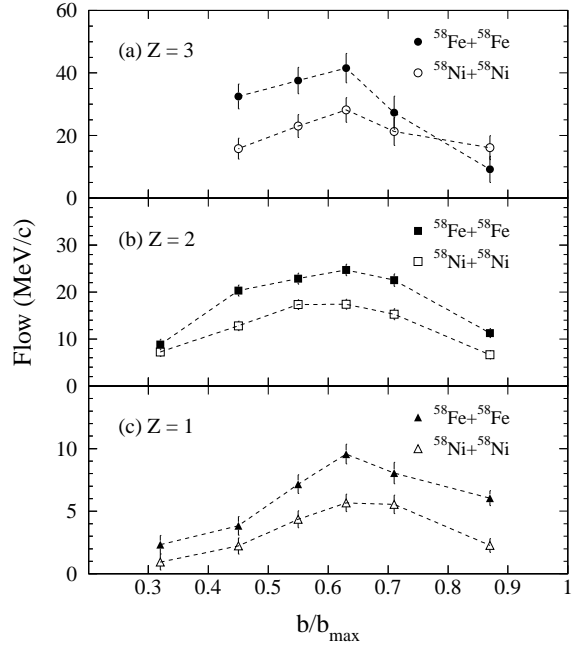}
\caption{Flow parameters for the reactions of $^{58}{\rm
Fe}+^{58}{\rm Fe}$ and $^{58}{\rm Ni}+^{58}{\rm Ni}$ as functions
of the reduced impact parameter at a beam energy of 55
MeV/nucleon. Taken from Ref. \protect\cite{Pak97a}.} \label{gary1}
\end{center}
\end{figure}

The transverse flow is a collective sidewise-deflection of forward
and backward moving particles within the reaction plane
\cite{Dan85}. The sideward flow is often represented in terms of the
average in-plane transverse momentum at a given rapidity
\begin{eqnarray}\label{dirfl}
\langle \frac{p_x}{A} \rangle\equiv \frac{1}{N(y)} \sum_{i=1}^{N(y)}
p_{x_{i}},
\end{eqnarray}
where $N(y)$ is the total number of nucleons at the rapidity $y$
and $p_{x_{i}}$ is the transverse momentum of particle $i$ in the
reaction plane. The transverse flow has been a major tool for
investigating the EOS of hot and dense matter \cite{Dan02a}. It
was first pointed out in Ref.~\cite{LiBA96} that the transverse
flow depends on the isospin asymmetry of the reaction system.
Furthermore, the so-called balance energy where the transverse
flow changes sign should also be sensitive to the isospin
asymmetry of the reaction system. These predictions were soon
verified by experiments carried out by Pak {\it et al.} at
NSCL/MSU \cite{Pak97a,Pak97b}. One normally measures the strength
of the transverse flow using the flow parameter $F$ defined as the
slope of the transverse momentum distribution at the center of
mass rapidity $y_{\rm cm}$. Since one cannot experimentally
identify the direction of flow using the transverse momentum
analysis, the absolute value of the flow parameter is usually
extracted. Shown in Fig.~\ref{gary1} are the flow parameters of
particles with charge $Z=1$, $Z=2$ and $Z=3$ as functions of
reduced impact parameter $b/b_{\rm max}$ for the reactions of
$^{58}{\rm Fe}+^{58}{\rm Fe}$ and $^{58}{\rm Ni}+^{58}{\rm Ni}$ at
a beam energy of 55 MeV/nucleon. Note that at this beam energy
flow is still dominated by the attractive mean-field potential and
is thus actually negative. It is seen that the flow parameter for
the more neutron-rich system is consistently higher and is in
agreement with the predictions in Ref.~\cite{LiBA96}.

\begin{figure}[htp]
\begin{center}
\includegraphics[scale=0.50]{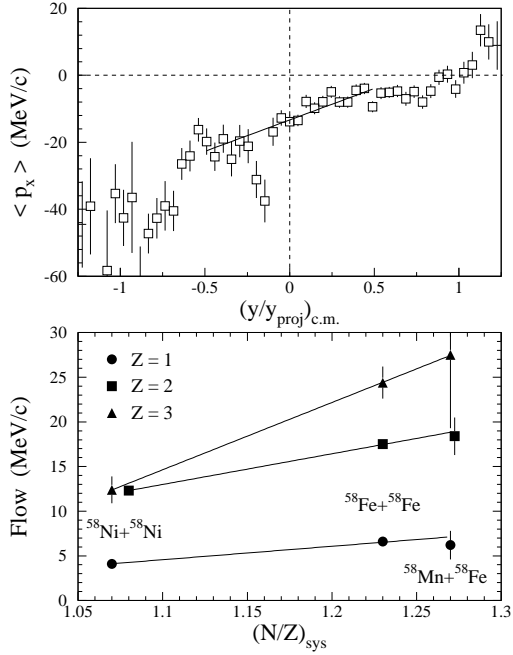}
\caption{Upper window: Mean transverse momentum in the reaction
plane versus the reduced c.m. rapidity for Z=2 fragments from
impact-parameter-inclusive $^{58}{\rm Fe}+^{58}{\rm Fe}$ collisions
at 55 MeV/nucleon. Lower window: Isospin dependence of the flow
parameter for inclusive collisions at a beam energy of 55
MeV/nucleon. Taken from Ref. \protect\cite{Pak97a}.}\label{gary2}
\end{center}
\end{figure}

Pak {\it et al.} also studied the flow parameter as a function of
the isotope ratio of the composite projectile plus target system
for three different fragment types from three isotopic entrance
channels. Shown in the upper window of Fig.\ \ref{gary2} is the
mean transverse momentum in the reaction plane versus the reduced
c.m. rapidity for Z=2 fragments from impact-parameter-inclusive
$^{58}{\rm Mn}+^{58}{\rm Fe}$ collisions at 55 MeV/nucleon. The
flow parameter extracted for the impact-parameter-inclusive events
is plotted in the lower window of Fig.\ \ref{gary2} as a function
of the ratio of neutrons to protons of the combined system
$(N/Z)_{\rm cs}$. The flow parameter is seen to increase linearly
with the ratio $(N/Z)_{\rm cs}$ for all three types of particles.

\begin{figure}[h]
\includegraphics[angle=-90,scale=0.4]{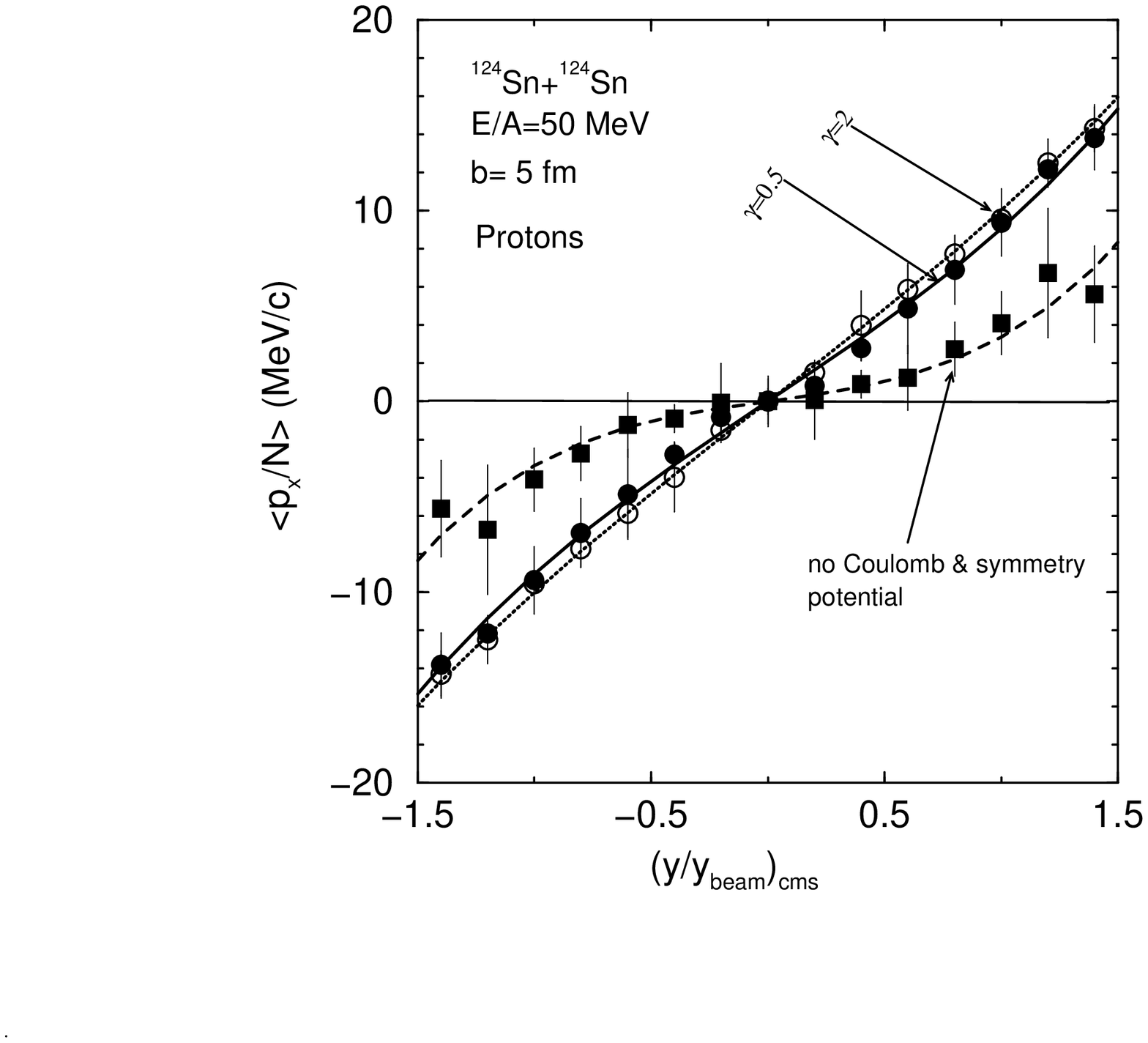}
\includegraphics[angle=-90,scale=0.42]{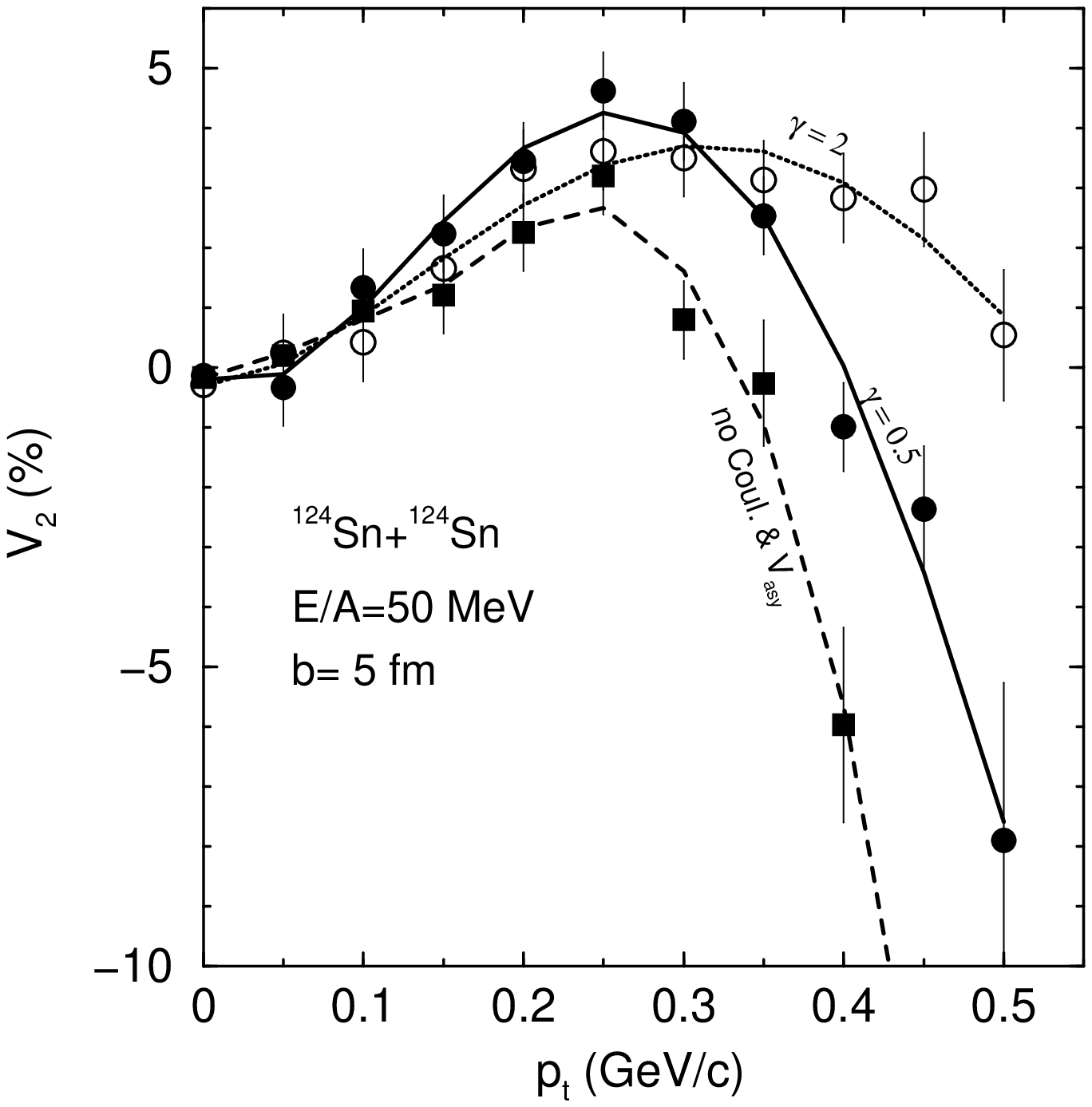}
\caption{Left window: Mean transverse momentum in the reaction plane
vs. reduced rapidity for protons in the $^{124}{\rm Sn}+^{124}{\rm
Sn}$ collisions at $50~{\rm AMeV}$ beam energy for semicentral
impact parameter, $b_{\rm red}=0.6$. Right window: Elliptic flow for
midrapidity protons as a function of transverse momentum. Taken from
Ref.~\cite{LiBA01a}.} \label{transv}
\end{figure}

While the isospin dependence of the transverse flow is
interesting, it has been very difficult to extract from it useful
information about the density dependence of the symmetry energy.
This is mainly because the isovector potential is much weaker than
the isoscalar potential in these reactions. This is actually why
the neutron-proton differential flow was introduced \cite{LiBA00}
as we shall discuss in detail in the section \ref{difflow}. To
illustrate this point more quantitatively, shown in the left
window of Fig.~\ref{transv} is the proton transverse flows for the
semicentral $^{124}$Sn+$^{124}$Sn reaction at a beam energy of
$50~{\rm AMeV}$. There is no appreciable difference in the
transverse flow with two quite different density dependencies of
the symmetry energy term in Eq.~(\ref{esymfu}), $F(u)=u^\gamma, u
\equiv \rho/\rho_0$, $\gamma=0.5$ (rather asy-soft) and $\gamma=2$
(asy-superstiff). On the other hand, the symmetry energy effect is
appreciable in the elliptic flow $V_2(p_t)$ of protons at high
transverse momentum as shown in the right window of
Fig.~\ref{transv}. The elliptic flow reflects the asymmetry of the
in-plane flow and the out-of-plane squeeze-out. It is normally
measured using the second coefficient of the Fourier expansion of
the azimuthal distribution \cite{Oll92,Vol97}
\begin{eqnarray}
\frac{dN}{d\phi}(y,p_t)=1+V_1(p_t)\cos(\phi)+2V_2(p_t)\cos(2\phi),
\end{eqnarray}
where $p_t=\sqrt{p_x^2+p_y^2}$ is the transverse momentum. The
directed flow $V_1$ and the elliptic flow $V_2$ can be expressed
as $V_1(y,p_t)=\langle p_x/p_t \rangle$ and $V_2(y,p_t)=\langle
(p_x^2-p_y^2)/p_t^2 \rangle$, respectively. Both $V_1(p_t)$ and
$V_2(p_t)$ have been found to be extremely useful in studying the
EOS and properties of dense matter at both intermediate
\cite{LiSu99,Zhe99} and ultra-relativistic energies, see, e.g.,
Ref.~\cite{BZh99,LHC}. The elliptic flow, especially at high
$p_t$, is expected to be more sensitive to the isospin dependence
of the nuclear EOS than the transverse flow. This is because all
three partial pressures lead approximately to a similar difference
$\delta p_{xy}^2\equiv p_x^2-p_y^2$ although their respective
contributions to the value of $<p_x>$ or $<p_y>$ is very different
\cite{LiBA01a}. Moreover, the pressure created in the participant
region during the early stage is revealed more clearly by the
value of $V_2(p_t)$ at high transverse momenta. This is due to the
fact that high $p_t$ particles can only be produced through the
most violent collisions in the early stage of the reaction and
these particles can only retain their high transverse momenta by
escaping from the reaction zone along the direction perpendicular
to the reaction plane without suffering much rescatterings. This
is an essentially universal phenomena in heavy ion collisions at
all energies.

The isospin dependence of radial flow at beam energies around 400
Mev/nucleon has also been investigated \cite{LiBA05b}. The
difference in the radial flow velocity for neutrons and protons is
the largest for the stiffest symmetry energy as one has expected.
As the symmetry energy becomes softer, the difference disappears
gradually. However, the overall effect of the symmetry energy on
the radial flow is small and is only about $4\%$ even for the
stiffest symmetry energy with $x=-2$. This is because the pressure
of the participant region is dominated by the kinetic
contribution. Moreover, the compressional contribution to the
pressure is overwhelmingly dominated by the isoscalar
interactions. For protons, the radial flow is affected much more
by the Coulomb potential than the symmetry potential. In fact, the
Coulomb potential almost cancels out the effect of the symmetry
potential at $x=-2$. As the symmetry energy becomes softer, the
radial flow for protons becomes higher than that for neutrons. The
radial flow thus seems to be less useful for studying the EOS of
neutron-rich matter \cite{LiBA05b}.

\subsection{Single and double neutron-proton differential
transverse flow}\label{difflow}

The concept of the neutron-proton differential flow was first
introduced several years ago \cite{LiBA00}. It was argued that the
neutron-proton differential flow minimizes the influences of the
isoscalar potential but maximizes the effects of the symmetry
potential. It can also reduce the effects of other dynamical
effects in intermediate energy heavy-ion reactions. It is
therefore among the most promising probes of the high density
behavior of the nuclear symmetry energy.

The neutron-proton differential transverse flow is defined as
\cite{LiBA00,LiBA02,LiBA05a}
\begin{eqnarray}
F_{n-p}^{x}(y)
\equiv\frac{1}{N(y)}\sum_{i=1}^{N(y)}p_{i}^{x}(y)w_{i}
=\frac{N_{n}(y)}{N(y)}\langle p_{n}^{x}(y)\rangle
-\frac{N_{p}(y)}{N(y)} \langle p_{p}^{x}(y)\rangle,  \label{npflow}
\end{eqnarray}
where $N(y)$, $N_{n}(y)$ and $N_{p}(y)$ are the numbers of free
nucleons, neutrons and protons, respectively, at rapidity $y$;
$p_{i}^{x}(y)$ is the transverse momentum of the free nucleon at
rapidity $y$; $w_{i}=1$ $(-1)$ for neutrons (protons); and
$\langle p_{n}^{x}(y)\rangle $ and $\langle p_{p}^{x}(y)\rangle $
are, respectively, the average transverse momenta of neutrons and
protons at rapidity $y$. Eq. (\ref{npflow}) shows that the
constructed neutron-proton differential transverse flow depends
not only on proton and neutron transverse momenta but also on
their relative multiplicities. The neutron-proton differential
flow thus combines effects due to both the isospin fractionation
and the different transverse flows of neutrons and protons. This
can be more clearly seen by considering two special cases. If
neutrons and protons have the same average transverse momentum in
the reaction plane but different multiplicities in each rapidity
bin, i.e., $\langle p_{n}^{x}(y)\rangle =\langle
p_{p}^{x}(y)\rangle =\langle p^{x}(y)\rangle $, and $N_{n}(y)\neq
N_{p}(y)$, then Eq. (\ref{npflow}) is reduced to
\begin{eqnarray}
F_{n-p}^{x}(y)=\frac{N_{n}(y)-N_{p}(y)}{N(y)}\langle
p^{x}(y)\rangle =\delta (y)\cdot \langle p^{x}(y)\rangle,
\end{eqnarray}
reflecting the effects of isospin fractionation. On the other
hand, if neutrons and protons have the same multiplicity but
different average transverse momenta, i.e., $N_{n}(y)=N_{p}(y)$
but $\langle p_{n}^{x}(y)\rangle \neq \langle p_{p}^{x}(y)\rangle
$, then Eq. (\ref{npflow}) reduces to
\begin{eqnarray}
F_{n-p}^{x}(y)=\frac{1}{2}(\langle p_{n}^{x}(y)\rangle -\langle
p_{p}^{x}(y)\rangle ).
\end{eqnarray}
In this case it reflects directly the difference of the neutron
and proton transverse flows. In heavy-ion collisions at higher
energies and for free nucleons in a given rapidity bin, one
expects that a stiffer symmetry potential generally leads to a
higher isospin fractionation and also contributes more positively
to the transverse momenta of neutrons compared to protons
\cite{LiBA06b,Yon06a}. The neutron-proton differential flow thus
combines constructively effects of the symmetry potentials for
neutrons and protons.

\begin{figure}[tbh]
\centering
\includegraphics[width=0.48\textwidth]{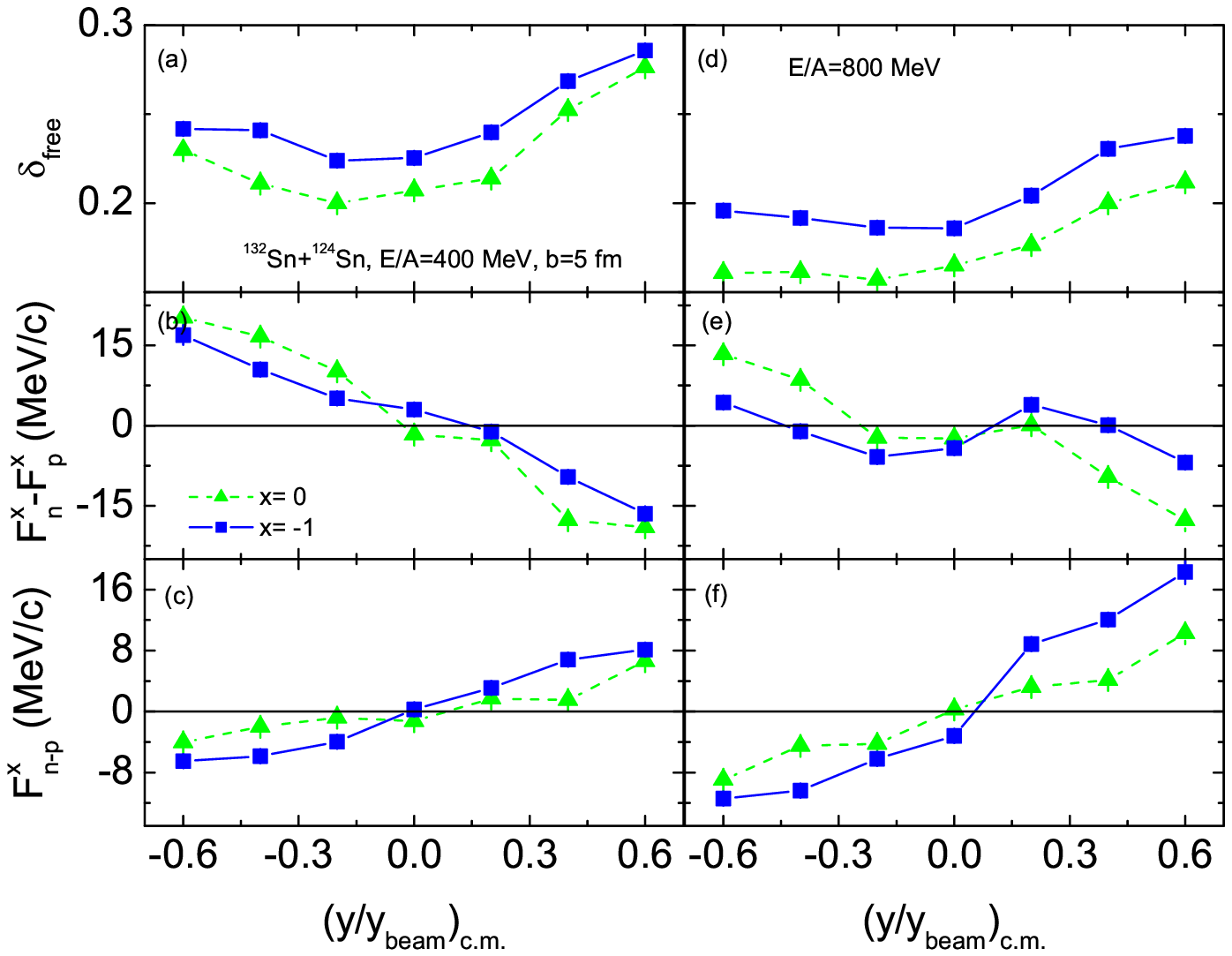}
\includegraphics[width=0.5\textwidth]{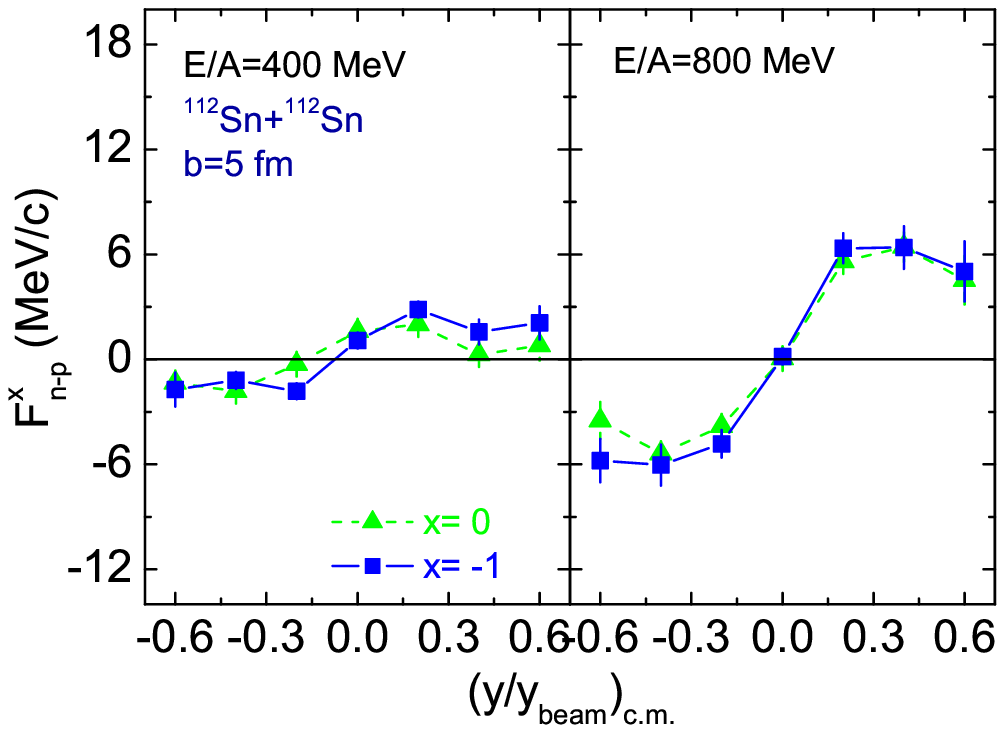}
\caption{(Color online) Left window: Rapidity distribution of the
isospin asymmetry of free nucleons (upper panels), the difference of
the average nucleon transverse flows (middle panels) and the
neutron-proton differential transverse flow (lower panels) from
$^{132}$Sn+$^{124}$Sn reaction at the incident beam energies of
$400$, $800$ MeV/nucleon and $b=5$ fm with two symmetry energies of
$x=0$ and $x=-1$. Right window: Same as the lowest two panels (c)
and (f) of the left, but for the reaction system of
$^{112}$Sn+$^{112}$Sn. Taken from Ref. \protect\cite{Yon06b}.}
\label{isoflow1}
\end{figure}

Shown in the left window of Fig.~\ref{isoflow1} are the rapidity
distribution of the isospin asymmetry of free nucleons (upper
panels), the difference of the average nucleon transverse flows
(middle panels) and the neutron-proton differential transverse
flow (lower panels) from the $^{132}$Sn+$^{124}$Sn reaction at
incident beam energies of $400$ and $800$ MeV/nucleon and an
impact parameter of $b=5$ fm with the two symmetry energies of
$x=0$ and $x=-1$. It is seen from the upper panels that a larger
isospin asymmetry of free nucleons (stronger isospin
fractionation) is obtained for the stiffer symmetry energy
($x=-1$), which thus leads to a stronger neutron-proton
differential transverse flow than the softer symmetry energy
($x=0$) as shown in the lower panels. Furthermore, the
neutron-proton differential transverse flow exhibits a stronger
sensitivity to the symmetry energy than the difference of the
average nucleon flows as shown in the middle panels. Since the
Coulomb potential normally dominates over the symmetry potential
for protons, protons thus have higher average transverse momenta
than neutrons, leading to the negative (positive) values of the
$F_n^x-F_p^x$ at forward (backward) rapidities.

The beam energy dependence of the neutron-proton differential
transverse flow is shown in the lowest two panels (c) and (f) in
the left window of Fig.\ \ref{isoflow1}. As one expects, with the
same symmetry energy, the slope of the neutron-proton differential
transverse flow around the mid-rapidity is larger for the higher
incident beam energy. This is mainly because a denser nuclear
matter is formed at higher incident beam energy. It then leads to
a stronger symmetry potential and thus higher transverse momenta
for neutrons compared to protons. The much larger neutron-proton
differential transverse flow at $800$ MeV/nucleon than that at
$400$ MeV/nucleon makes it easier to be measured experimentally,
although the net effect of the symmetry potential on the
neutron-proton differential transverse flow is not much larger
than that at $400$ MeV/nucleon. The right side of Fig.\
\ref{isoflow1} shows the rapidity distribution of the
neutron-proton differential transverse flow in the semi-central
reaction of $^{112}$Sn+$^{112}$Sn at incident beam energies of
$400$ and $800$ MeV/nucleon. Comparing with the case of
$^{132}$Sn+$^{124}$Sn, one can see that the slope of the
neutron-proton differential transverse flow around mid-rapidity
and effects of the symmetry energy become much smaller due to the
smaller isospin asymmetry in the reaction of
$^{112}$Sn+$^{112}$Sn.

\begin{figure}[tbh]
\centering
\includegraphics[scale=0.5]{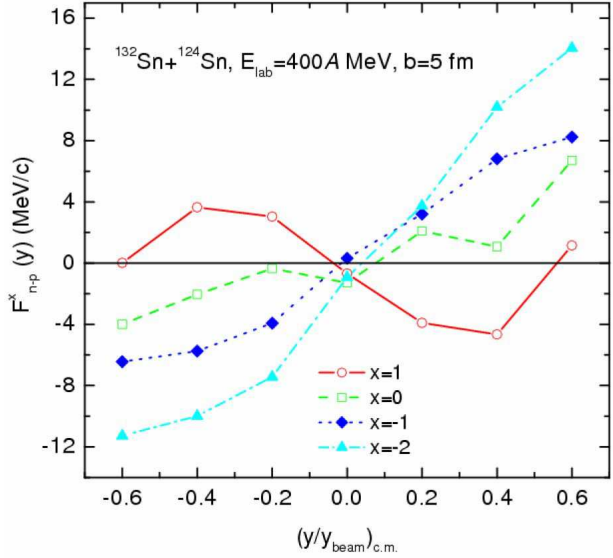}
\caption{(Color online) Neutron-proton differential flow for the
reaction of $^{132} $Sn$+^{124}$Sn at a beam energy of 400
MeV/nucleon and an impact parameter of 5 fm for different nuclear
symmetry energies. Taken from Ref. \protect\cite{LiBA05a}.}
\label{npdiff}
\end{figure}

Effects of the symmetry energy are clearly revealed by changing the
parameter $x$. As an example, shown in Fig.~\ref{npdiff} is the
$n-p$ differential flow for the reaction of $^{132} $Sn$+^{124}$Sn
at a beam energy of 400 MeV/nucleon and an impact parameter of 5 fm
using the four values of $x$.

As discussed previously, in order to extract accurately
information about the symmetry energy, one has to reduce as much
as possible the systematic errors involved in the experimental
observables. Moreover, the long range Coulomb force on charged
particles may play an important role in these observables. If all
possible, one would like to disentangle effects of the symmetry
energy from those due to the Coulomb force. Since this is often
impossible, one would thus like to construct observables that can
reduce the Coulomb effects as much as possible. Ratios and/or
differences of two observables from a pair of reactions using
different isotopes of the same element are among the promising
candidates to reduce both the systematic errors and the Coulomb
effects. Whether to use the ratio or the difference to construct
the desired observable depends on the nature of the observables
involved. For the neutron/proton ratio of pre-equilibrium nucleons
and the $\pi ^{-}/\pi ^{+}$ ratio, for instance, it is natural to
construct their double ratios as discussed above. However, the
neutron-proton differential flow is additive, it is thus more
useful to construct the double differences instead of ratios.

\begin{figure}[tbh]
\centering
\includegraphics[scale=0.7]{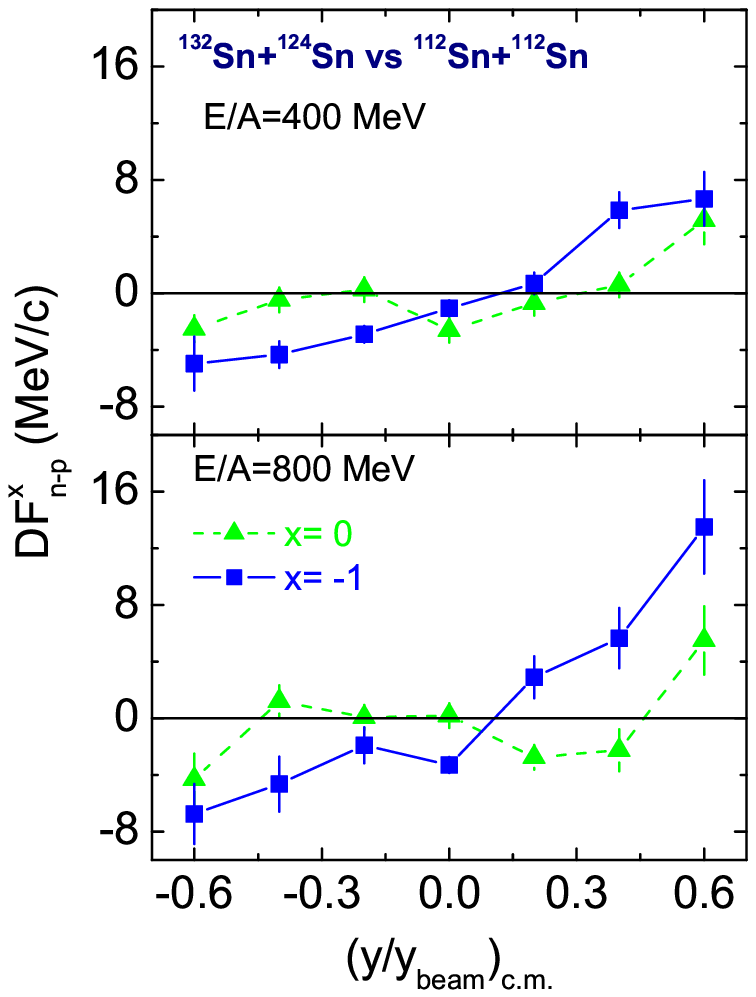}
\includegraphics[scale=0.7]{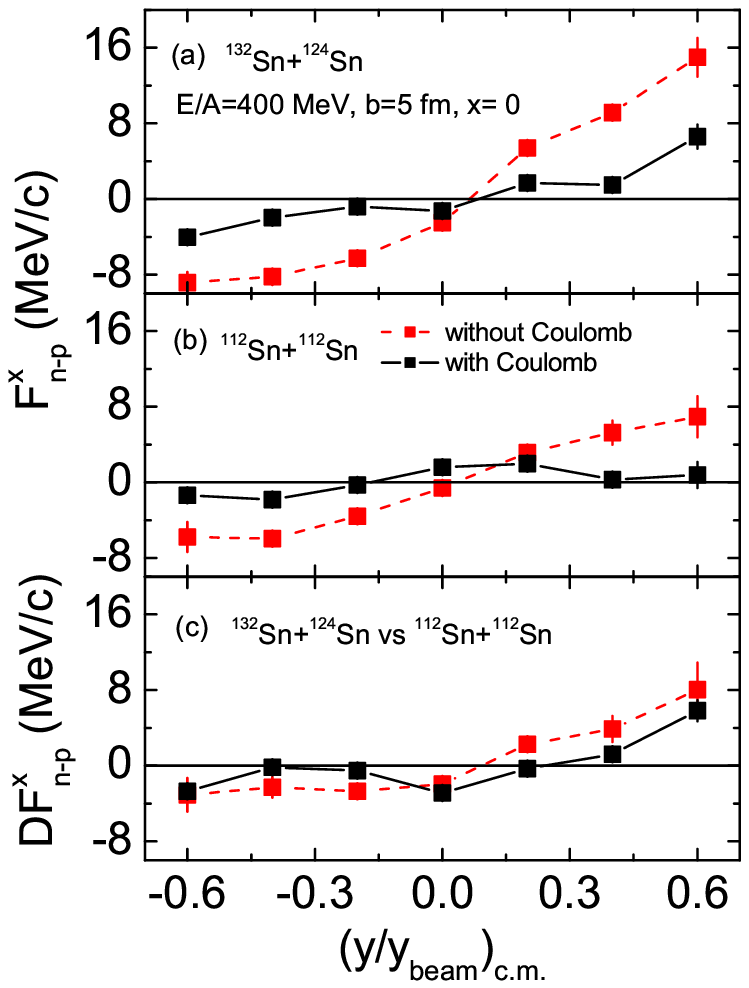}
\caption{Left window: Rapidity distribution of the double
neutron-proton differential transverse flow in the semi-central
reactions of Sn+Sn isotopes at the incident beam energies of $400$
and $800$ MeV/nucleon with two symmetry energies of $x=0$ and
$x=-1$. Right window: Coulomb effects on the neutron-proton
differential transverse flow (upper two panels) and the double
neutron-proton differential transverse flow (lowest panel) in the
semi-central reactions of Sn+Sn isotopes at the incident beam
energy of $400$ MeV/nucleon with the symmetry energy of $x=0$.
Taken from Ref. \protect\cite{Yon06b}. } \label{dflow}
\end{figure}

The left window of Fig.\ \ref{dflow} shows the rapidity
distribution of the double neutron-proton differential transverse
flow in the semi-central reactions of Sn+Sn isotopes. At both
incident beam energies of $400$ and $800$ MeV/nucleon, the double
neutron-proton differential transverse flow around mid-rapidity is
essentially zero for the soft symmetry energy of $x=0$. However,
it displays a clear slope with respect to the rapidity for the
stiffer symmetry energy of $x=-1$. Moreover, the double
neutron-proton differential transverse flow at the higher incident
energy indeed exhibits a stronger symmetry energy effect as
expected. Furthermore, the symmetry effect on the double
neutron-proton differential transverse flow is similar as in the
$^{132}$Sn+$^{124}$Sn reaction.

Also, the Coulomb effect, which competes strongly with the
symmetry potentials, is less important in the double
neutron-proton differential transverse flow than in the
neutron-proton differential transverse flow. This can be seen in
the right window of Fig.\ \ref{dflow} which shows the
neutron-proton differential transverse flow (upper two panels) and
the double neutron-proton differential transverse flow (lowest
panel) in the semi-central reactions of Sn+Sn isotopes at the
incident beam energy of $400$ MeV/nucleon with the symmetry energy
of $x=0$ for the two cases of with and without the Coulomb
potential. One also sees that the Coulomb effect reduces the
strength of the neutron-proton differential transverse flow as it
makes more protons unbound and to have large transverse momenta in
the reaction-plane. The Coulomb effect is, however, largely
reduced in the double neutron-proton differential transverse flow.

\subsection{Pions as a probe of the high density behavior of the nuclear
symmetry energy}

At beam energies above 300 MeV/nucleon, pion production become
significant. Pions carry interesting information about the high
density behavior of the symmetry energy
\cite{LiBA02,Gai04,LiBA05a,LiBA03}. In this subsection, we first
discuss why the $\pi ^{-}/\pi ^{+}$ ratio may be a sensitive probe
of the high density behavior of the symmetry energy based on two
idealized models for pion productions, i.e., the resonance model
and the thermal model. We then show results of transport model
calculations. Finally, we discuss some recent data from the FOPI
collaboration.

\subsubsection{The $\protect\pi ^{-}/\protect\pi ^{+}$ ratio}

It is well known that the $\pi^-/\pi^+$ ratio in heavy-ion
collisions depends strongly on the isospin asymmetry of the
reaction system, see, e.g.,
Refs.~\cite{Ben79,Nag81,Har85,Sto86,Li95}. It is also easy to
understand qualitatively why this dependence can be used to
extract crucial information about the {\rm EOS} of neutron-rich
matter and even about the structure of rare isotopes
\cite{Lom88,Libhu,Tel87}. In the $\Delta$ resonance model for pion
production from first-chance independent nucleon-nucleon
collisions \cite{Sto86}, the primordial $\pi^-/\pi^+$ ratio is
\begin{eqnarray}\label{pres}
(\pi^-/\pi^+)_{\rm res}\equiv (5N^2+NZ)/(5Z^2+NZ)\approx (N/Z)^2,
\end{eqnarray}
where $N$ and $Z$ are neutron and proton numbers in the
participant region of the reaction. It is thus a direct measure of
the isospin asymmetry $(N/Z)_{\rm dense}$ of the dense matter in
the participant region of heavy-ion collisions. As we have
discussed earlier, the $(N/Z)_{\rm dense}$ is uniquely determined
by the high density behaviour of the nuclear symmetry energy
\cite{LiBA02}. Therefore, the $\pi^-/\pi^+$ ratio can be used to
probe sensitively the {\rm EOS} of neutron-rich matter. On the
other hand, the $\pi^-/\pi^+$ ratio in the statistical model for
pion production \cite{Ber80} is proportional to ${\rm
exp}\left[(\mu_n-\mu_p)/T\right]$, where $T$ is the temperature,
and $\mu_n$ and $\mu_p$ are the chemical potentials of neutrons
and protons, respectively. At modestly high temperatures ($T\geq
4$ MeV), the difference in the neutron and proton chemical
potentials can be given by \cite{Jaq83}
\begin{eqnarray}
\mu_n-\mu_p=V^n_{\rm asy}-V^p_{\rm asy}-V_{\rm Coulomb}+T\left[{\rm
ln}\frac{\rho_n}{\rho_p}+\sum_m\frac{m+1}{m}b_m(\frac{\lambda_T^3}{2})^m
(\rho^m_n-\rho^m_p)\right],
\end{eqnarray}
where $V_{\rm Coulomb}$ is the Coulomb potential for protons,
$\lambda_T$ is the thermal wavelength of a nucleon and $b'_m$s are
the inversion coefficients of the Fermi distribution
function~\cite{Jaq83}. The difference in the neutron and proton
mean-field potentials is $V^n_{\rm asy}-V^p_{\rm asy}=2v_{\rm
asy}(\rho)\delta$, where $v_{asy}(\rho)$ is the symmetry
potential. Since the kinetic part of the difference $\mu_n-\mu_p$
relates directly to the isospin asymmetry $\rho_n/\rho_p$ or
$\rho_n-\rho_p$, the $\pi^-/\pi^+$ ratio in the statistical model
is also sensitive to the ratio $(N/Z)_{\rm dense}$. Moreover, the
value of the $\pi^-/\pi^+$ ratio is affected by the competition
between the symmetry potential and the Coulomb potential which all
depend on the isospin asymmetry of the reaction system.

The above expectations based on two idealized, extreme models
illustrate qualitatively the usefulness of the $\pi^-/\pi^+$ ratio
for investigating the {\rm EOS} of neutron-rich matter. For more
quantitative studies, however, advanced transport model
calculations are necessary. In heavy-ion collisions at beam
energies below about 1 GeV/nucleon, most pions are produced
through the decay of $\Delta (1232)$ resonances, see, e.g.,
Refs.~\cite{libauer1,libauer2}. The mean-field potentials for
$\Delta (1232)$ resonances in nuclear matter are still largely
unknown. Normally made is the minimum assumption that the
isoscalar part of the $\Delta$ potential is the same as that for
nucleons. To be consistent with the modelling of the isovector
potential for nucleons, one normally assumes that the isovector
potential for $\Delta$ resonances is an average of that for
neutrons and protons. The weighting factor depends on the charge
state of the resonance and is given by the square of the
Clebsch-Gordon coefficients for the isospin couplings in the
processes $\Delta\leftrightarrow \pi N$. In terms of the the
neutron and proton isoscalar potentials, the $\Delta$ isosclar
potentials are thus given by
\begin{eqnarray}
v_{\rm asy}(\Delta^-)&=&v_{\rm asy}(n),\\\
v_{\rm asy}(\Delta^0)&=&\frac{2}{3}v_{\rm asy}(n)+\frac{1}{3}v_{\rm asy}(p),\\\
v_{\rm asy}(\Delta^+)&=&\frac{1}{3}v_{\rm asy}(n)+\frac{2}{3}v_{\rm asy}(p),\\\
v_{\rm asy}(\Delta^{++})&=&v_{\rm asy}(p).
\end{eqnarray}
Similarly, the effective isospin asymmetry $\delta_{\rm like}$ for
excited baryonic matter is defined as
\begin{eqnarray}
\delta_{\rm like}\equiv \frac{(\rho_n)_{\rm like}-(\rho_p)_{\rm
like}} {(\rho_n)_{\rm like}+(\rho_p)_{\rm like}},
\end{eqnarray}
where
\begin{eqnarray}
(\rho_n)_{\rm
like}&=&\rho_n+\frac{2}{3}\rho_{\Delta^0}+\frac{1}{3}\rho_{\Delta^+}
+\rho_{\Delta^-},\\\
(\rho_p)_{\rm
like}&=&\rho_p+\frac{2}{3}\rho_{\Delta^+}+\frac{1}{3}\rho_{\Delta^0}
+\rho_{\Delta^{++}}.
\end{eqnarray}
It is evident that the $\delta_{\rm like}$ reduces naturally to
$\delta$ as the beam energy becomes smaller than the pion
production threshold. Moreover, for the hadronic matter produced
in heavy-ion reactions one can define the $(\pi^-/\pi^+)_{\rm
like}$ ratio as
\begin{eqnarray}
(\pi^-/\pi^+)_{\rm like}\equiv
\frac{\pi^-+\Delta^-+\frac{1}{3}\Delta^0}
{\pi^++\Delta^{++}+\frac{1}{3}\Delta^+}.
\end{eqnarray}
This ratio naturally goes to the $\pi^-/\pi^+$ at the end of the
reaction after all resonances have decayed.

\begin{figure}[htp]
\centering
\includegraphics[scale=0.6,angle=-90]{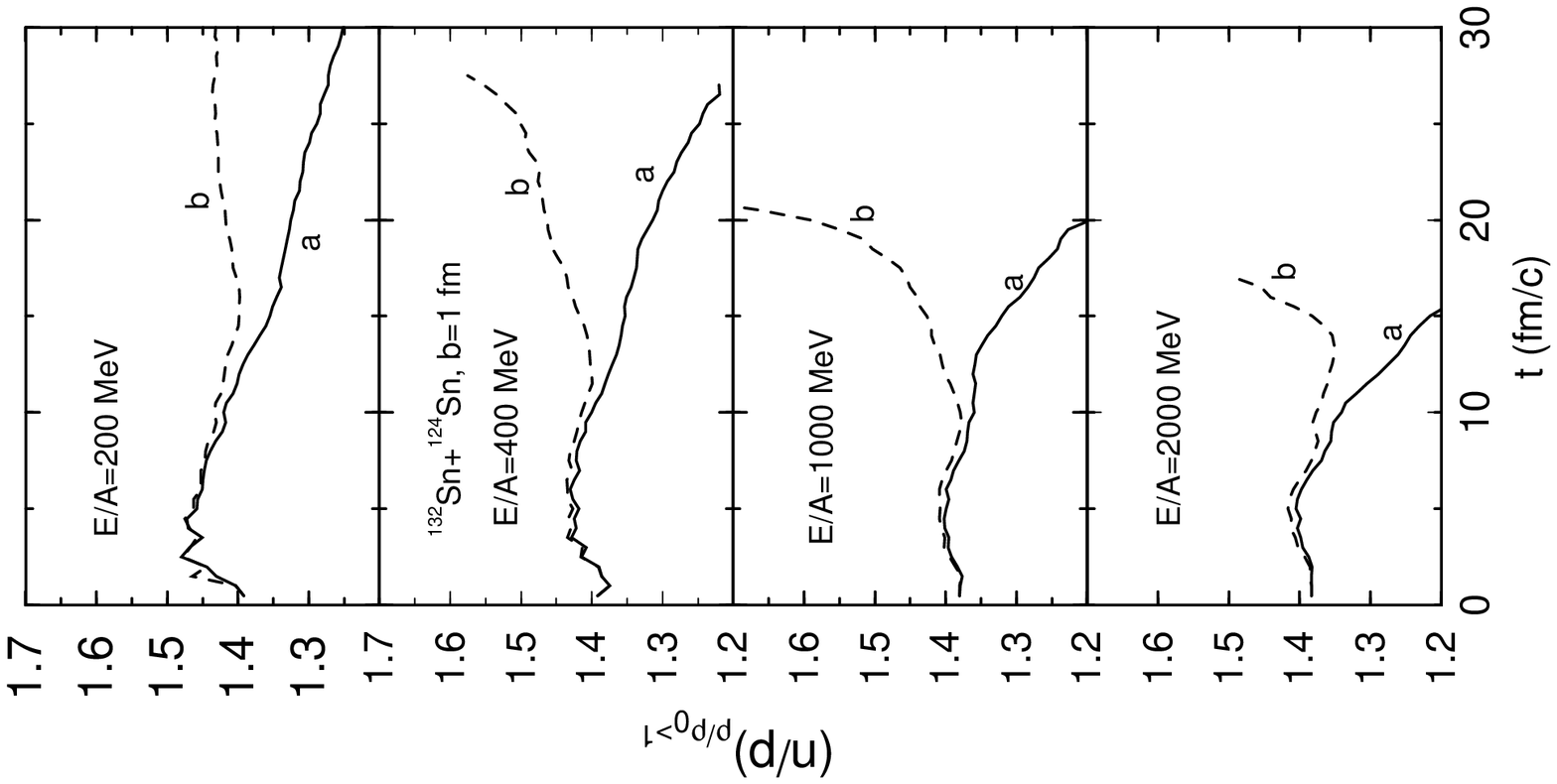}
\includegraphics[scale=0.6,angle=-90]{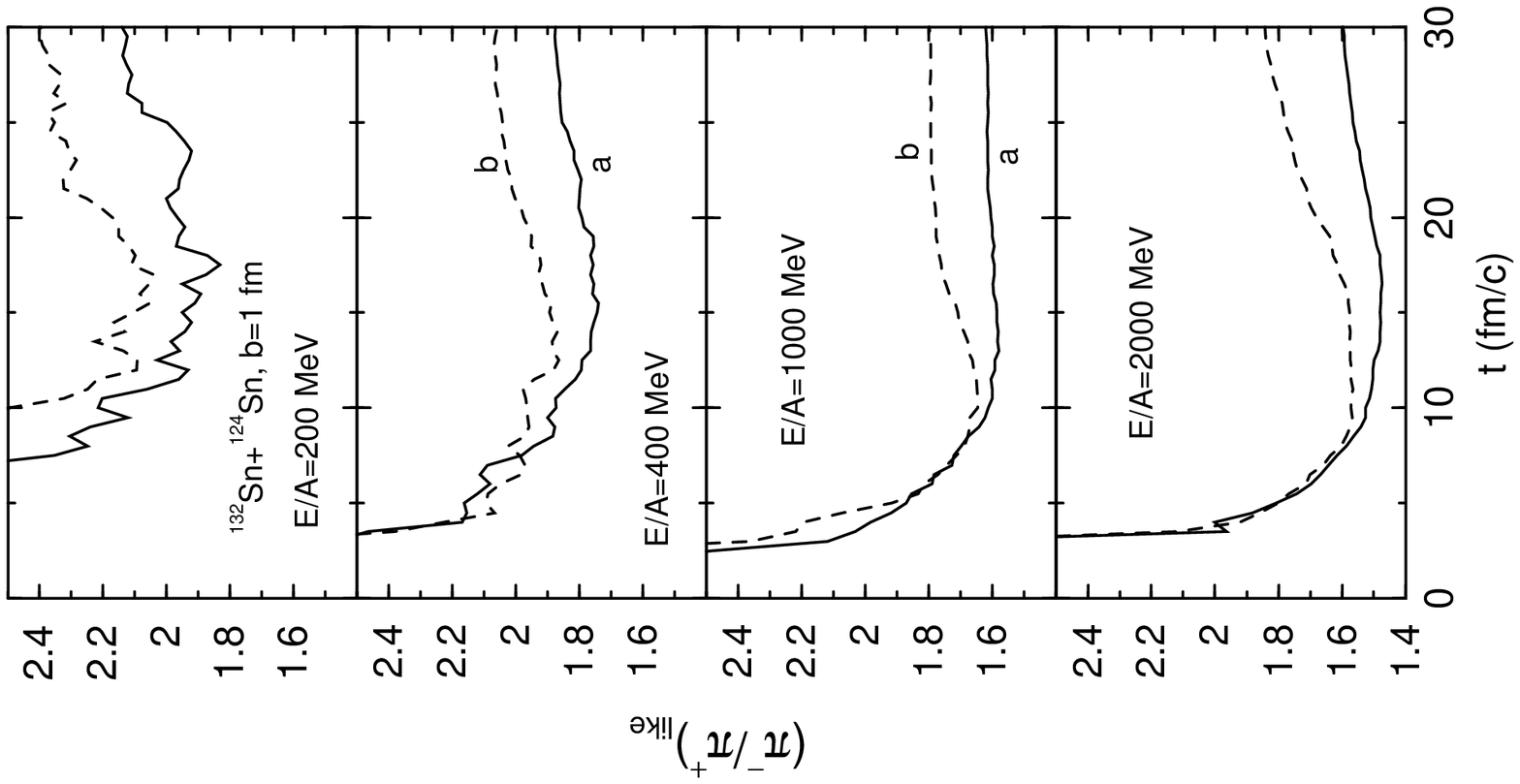}
\caption{Left window: The neutron/proton ratio of nuclear matter
with density higher than the normal nuclear matter density as a
function of time with the nuclear symmetry energy $E^a_{\rm sym}$
and $E^b_{\rm sym}$, respectively. Right window: The $\pi^-/\pi^+$
ratio as a function of time in the same reaction. Taken from Ref.
\cite{LiBA02}.} \label{npd1}
\end{figure}

There have been so far several studies on the $\pi^-/\pi^+$ ratio
using different transport models with various parameterizations
for the density dependence of the symmetry energy
\cite{Bar05,LiBA02,Gai04,LiBA05a,LiQF05b,LiBA03,Rei07}. While
qualitatively consistent, their predictions are quantitatively
different. We select here a few results to illustrate the major
points. Based on the symmetry energy given in Eqs.~(\ref{esyma})
and (\ref{esymb}) and the corresponding momentum-independent
potentials it is found in the IBUU approach \cite{LiBA02,LiBA03}
that the $(\pi^-/\pi^+)_{\rm like}$ and the average $n/p$ ratio of
the HD region are highly correlated. Shown in the left window of
Fig.\ \ref{npd1} are the $\rho/\rho_0\geq 1$ as a function of time
and beam energy. The effect on $(n/p)_{\rho/\rho_0\geq 1}$ due to
the different $E_{\rm sym}(\rho)$ is seen to grow with the
reaction time until the expansion has led the system to densities
below $\rho_0$, especially at higher beam energies. Although the
compression starts at about the same time, the expansion starts on
a faster time scale at higher beam energies as one expects. As in
collisions below the pion production threshold, whether the HD
region is neutron-rich or -poor depends critically on the HD
behavior of nuclear symmetry energy.

Shown in the right window of Fig.\ \ref{npd1} is the
$(\pi^-/\pi^+)_{\rm like}$ ratio as a function of time. This ratio
naturally becomes the final $\pi^-/\pi^+$ ratio at the freeze-out
when the reaction time $t$ is much longer than the lifetime of the
delta resonance $\tau_{\Delta}$. The $(\pi^-/\pi^+)_{\rm like}$
ratio is rather high in the early stage of the reaction because of
the large number of neutron-neutron scatterings near the surfaces
where the neutron skins of the colliding nuclei overlap. By
comparing the two windows of Fig.\ \ref{npd1}, it is seen that a
variation of about 30\% in the $(n/p)_{\rho/\rho_0\geq 1}$ ratio
due to the different $E_{\rm sym}(\rho)$ results in about 15\%
change in the final $\pi^-/\pi^+$ ratio. It thus has an
appreciable response factor of about 0.5 to the variation of the
HD $n/p$ ratio and is approximately independent of the beam
energy. Therefore, one can conclude that the $(\pi^-/\pi^+)_{\rm
like}$ ratio is a direct probe of the HD $n/p$ ratio, and thus an
indirect probe of the HD behavior of the nuclear symmetry energy.

\begin{figure}[htp]
\centering
\includegraphics[scale=0.5,angle=-90]{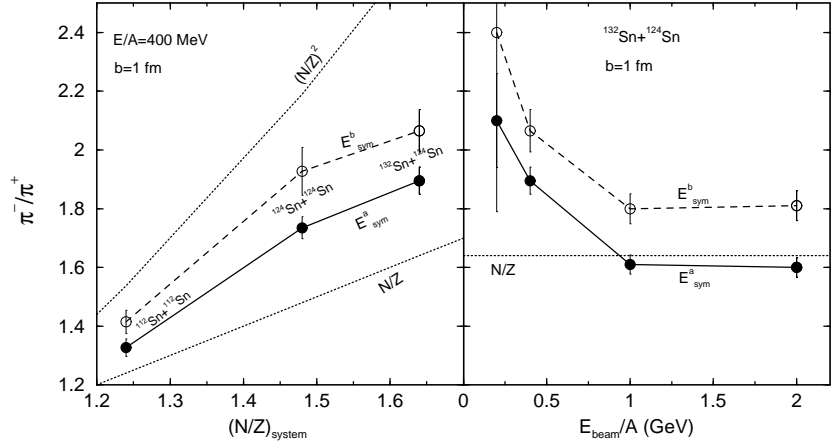}
\caption{The $(\pi^-/\pi^+)$ ratio as a function of the isospin
asymmetry (left panel) and beam energy (right panel) of the
reaction system. Taken from Ref. \cite{LiBA03}.} \label{pionsys}
\end{figure}

The final $\pi^-/\pi^+$ ratio is shown in Fig.\ \ref{pionsys} as a
function of $(N/Z)_{\rm system}$ (left panel) and beam energy
(right panel). Also plotted in the left panel for reference are
the ratios $(N/Z)$ and $(N/Z)^2$. It is seen that the
$\pi^-/\pi^+$ ratio falls far below the first-chance $\Delta$
resonance model prediction $(N/Z)^2$. This is because pion
reabsorption and rescattering ($\pi+N\leftrightarrow \Delta$ and
$N+\Delta\leftrightarrow N+N$) reduces the sensitivity of the
$\pi^-/\pi^+$ ratio to $(N/Z)_{\rm system}$, so what is more
important for the $\pi^-/\pi^+$ ratio is the local, changing $n/p$
ratio, particularly during the compression phase of the reaction.
The effect of the symmetry energy on the $\pi^-/\pi^+$ ratio thus
increases as one goes from $^{112}{\rm Sn}+^{112}{\rm Sn}$ to
$^{124}{\rm Sn}+^{124}{\rm Sn}$ but remains at about $15\%$ for
the $^{132}{\rm Sn}+^{124}{\rm Sn}$ system. Therefore,
neutron-rich stable beams, such as $^{124}{\rm Sn}$, seem to be
sufficient for probing the symmetry energy with the $\pi^-/\pi^+$
ratio. While the $\pi^-/\pi^+$ ratio decreases with increasing
beam energy, its sensitivity to the symmetry energy remains about
the same. Similar results were also found for other two reaction
systems. The decreasing $\pi^-/\pi^+$ ratio is mainly due to the
increasingly important contribution of pions from second-chance
nucleon-nucleon collisions as the beam energy increases. If a
first chance nucleon-nucleon collision converts a neutron to a
proton by producing a $\pi^-$, subsequent collisions of the still
energetic proton can convert itself back to a neutron by producing
a $\pi^+$. Eventually, at very high energies the sequential
multiple nucleon-nucleon collisions will lead to
$\pi^-/\pi^+\approx 1$.

\begin{figure}[tbh]
\centering
\includegraphics[scale=0.45]{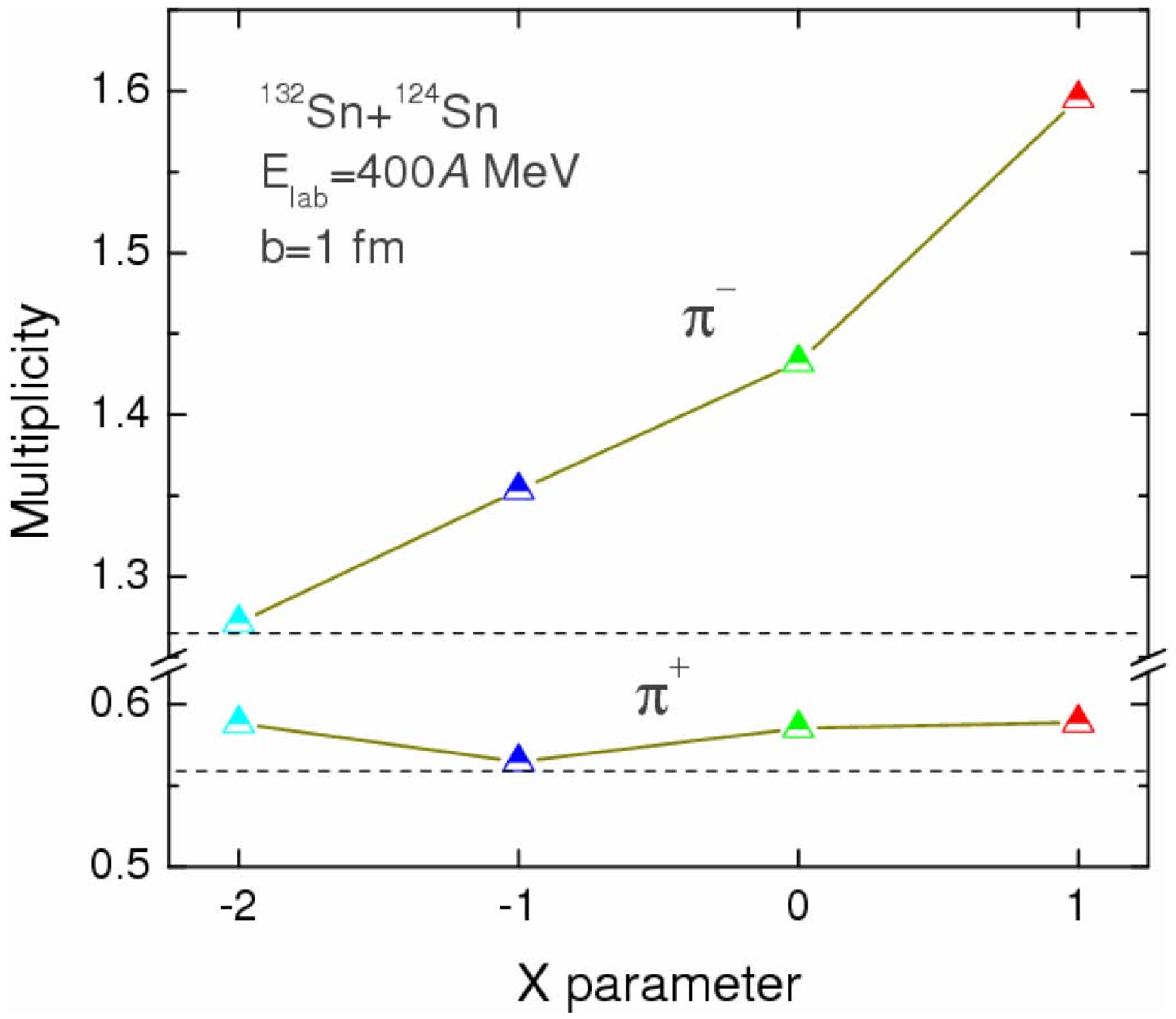}
\includegraphics[scale=0.45]{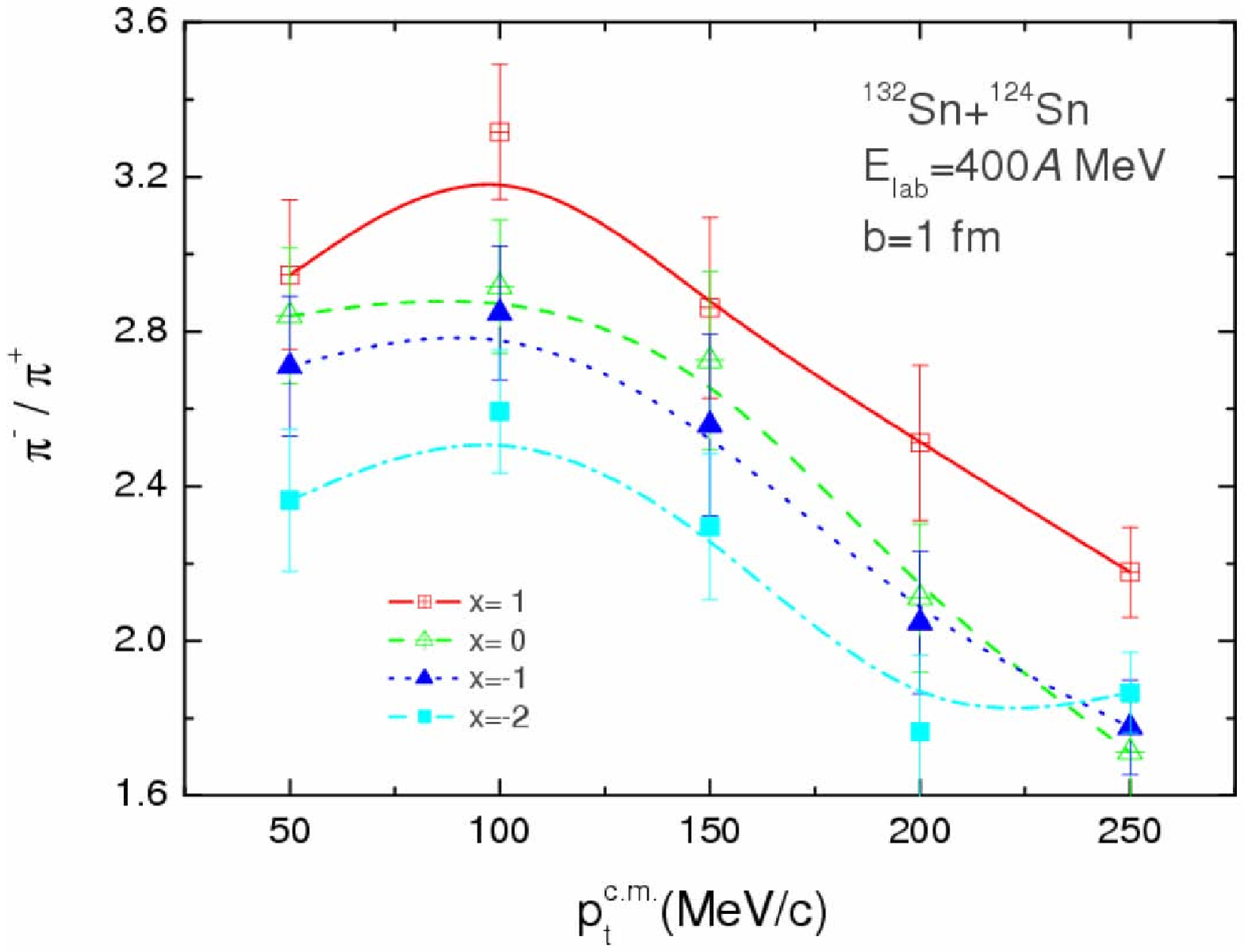}
\caption{The $\protect\pi ^{-}$ and $\protect\pi ^{+}$ yields as
functions of the $x$ parameter (left window) and transverse momentum
(right window). Taken from Ref. \protect\cite{LiBA05a}.}
\label{pionyield}
\end{figure}

Effects of the symmetry energy on pion production in high energy
heavy ion collisions can be studied in more detail in the IBUU04
transport model with momentum-dependent potentials \cite{LiBA05a}.
Shown in the left window of Fig.~\ref{pionyield} are the $\pi
^{-}$ and $\pi ^{+}$ yields as functions of the $x$ parameter. The
$\pi ^{-}$ multiplicity is seen to depend more sensitively on the
symmetry energy, as it increases by about $20\%$ while the $\pi
^{+}$ multiplicity remains about the same when the $x$ parameter
is changed from $-2$ to 1. Also, the multiplicity of $\pi ^{-}$ is
about 2 to 3 times that of $\pi ^{+}$ and this is because the $\pi
^{-}$ mesons are mostly produced from neutron-neutron collisions,
and with the softer symmetry energy the high density region is
more neutron-rich due to isospin fractionation \cite{LiBA05a}. The
$\pi ^{-}$ mesons are thus more sensitive to the isospin asymmetry
of the reaction system and the symmetry energy than the $\pi^+$
mesons. However, the pion yields are also sensitive to the
symmetric part of the nuclear EOS, so it is hard to obtain
reliable information about the symmetry energy from $\pi ^{-}$
yields alone. Fortunately, the $\pi ^{-}/\pi ^{+}$ ratio is a
better probe since according to the statistical model this ratio
is only sensitive to the difference in the chemical potentials of
neutrons and protons \cite{Ber80}. This expectation is well also
demonstrated in the transport model study as shown in the right
window of Fig.~\ref{pionyield}, where it is seen that the
$\pi^-/\pi^+$ ratio is quite sensitive to the symmetry energy,
especially at low transverse momenta. The $\pi ^{-}/\pi ^{+}$
ratio is thus a promising probe for the high density behavior of
the nuclear symmetry energy $E_{\rm sym}(\rho )$.

\begin{figure}
\centering
\includegraphics[scale=0.3]{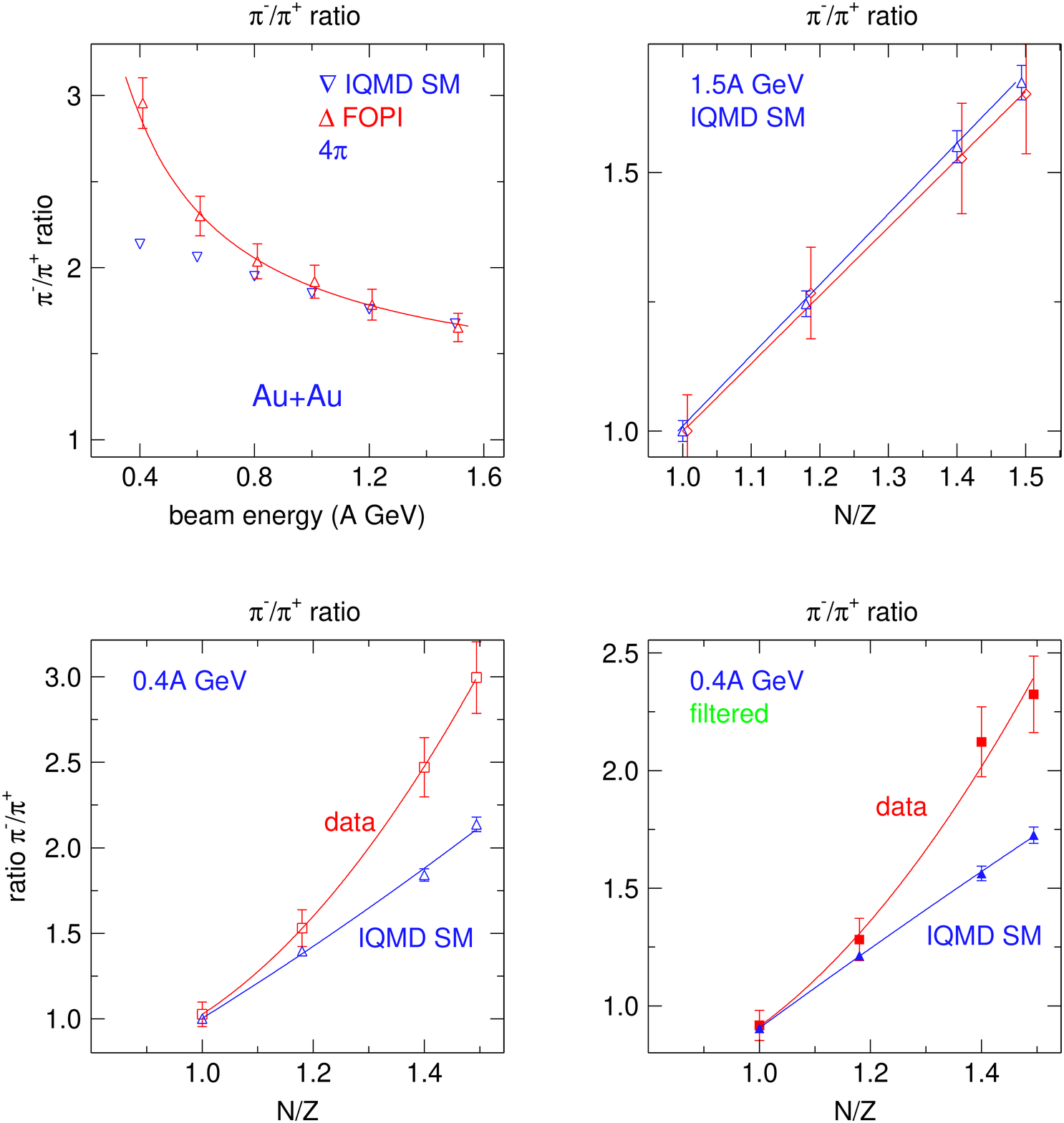}
\caption{ {\small Upper left window: Excitation function of the
$4\pi$-integrated ratio of $\pi^-/\pi^+$ yields in central Au+Au
collisions. The experimental data are joined by a least squares
fit of the function $c_0  + c_{-1}(E/A)^{-1}$ excluding the lowest
energy point. The IQMD SM prediction (triangles) is also given.
Upper right and lower left windows: The $N/Z$ dependence of the
$\pi^-/\pi^+$ ratio in reactions at $1.5$A and $0.4$A GeV,
respectively. The solid lines are least squares fits of linear or
quadratic $N/Z$ dependence. Lower right window: Same as lower left
window but for filtered data. Taken from Ref.
\protect\cite{Rei07}.} } \label{reisdorf1}
\end{figure}

\begin{figure}[h]
\centering
\includegraphics[scale=0.35]{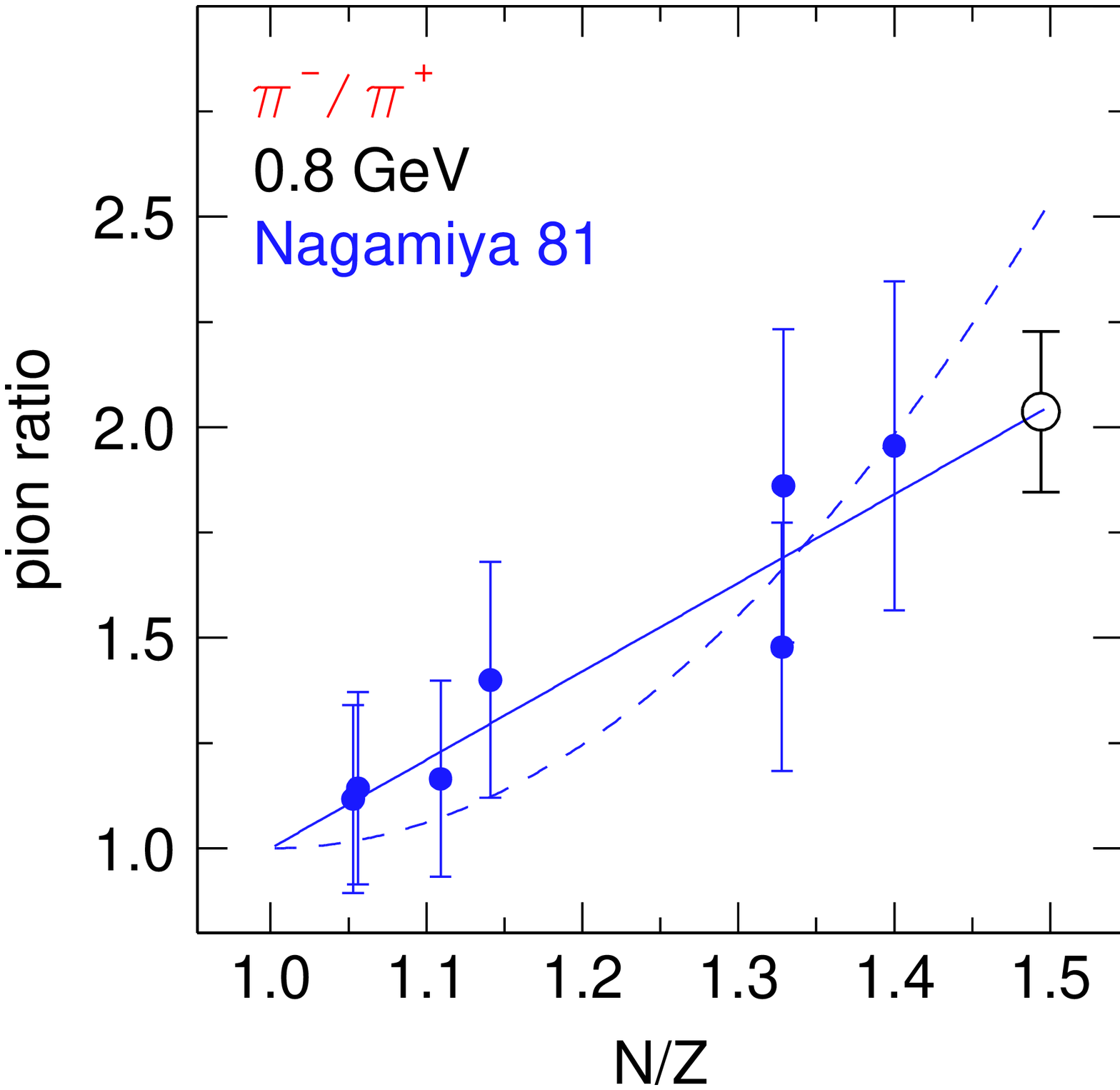}
\includegraphics[scale=0.35]{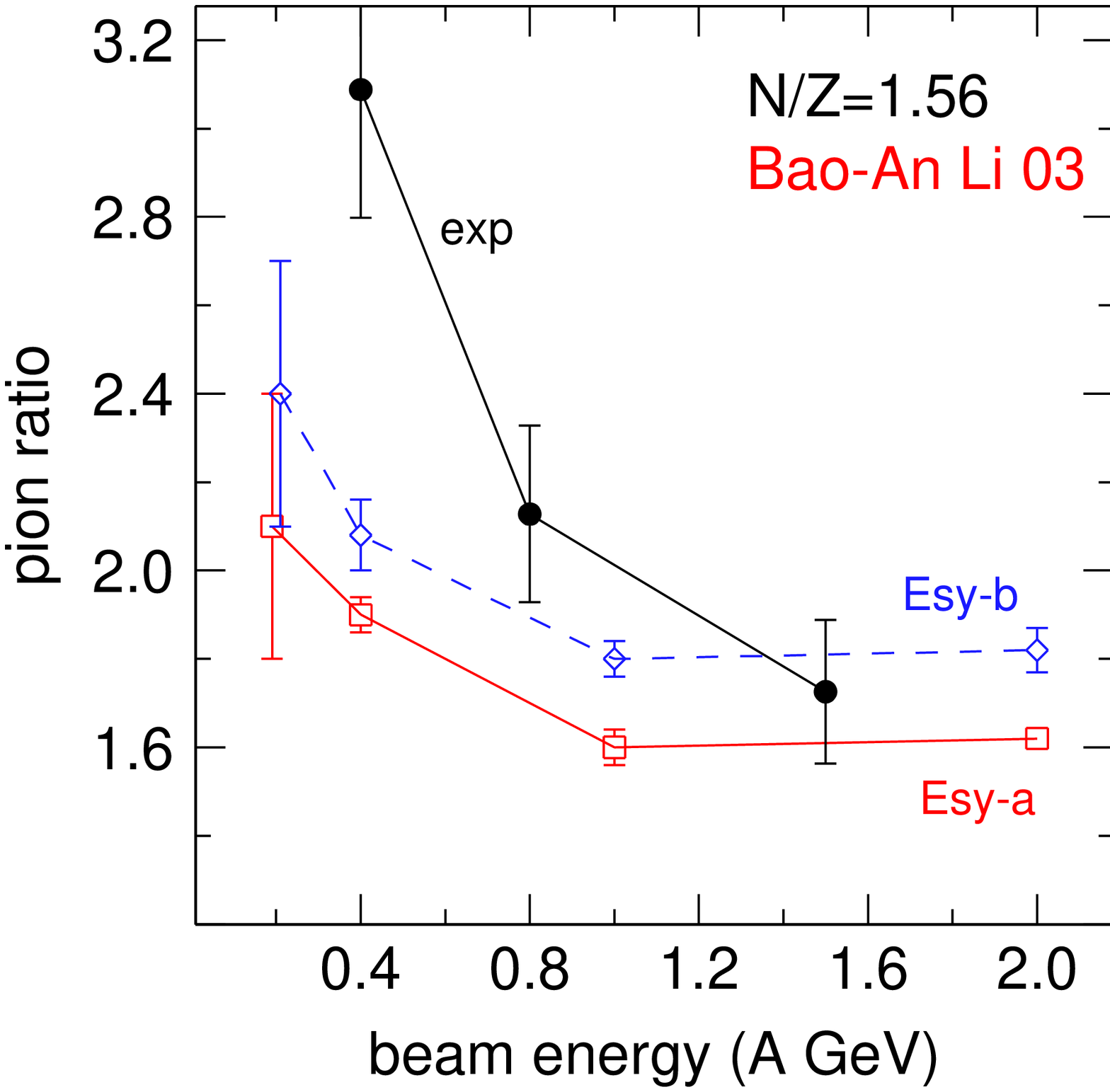}
\caption{ {\small Left window: The $\pi^-/\pi^+$ ratio versus
$N/Z$ of the 'fireball' measured~\cite{Nag81} in various inclusive
heavy ion reactions at $0.8A$ GeV (full circles). The solid
(dashed) curve is a linear (quadratic) least squares fit to the
data constrained to be one at $N/Z=1$. The data point denoted by
the open circle is for Au+Au and was not included in the fit.
Right window: The $\pi^-/\pi^+$ ratios versus beam energy obtained
in transport calculations~\cite{LiBA03} for the system
$^{132}$Sn+$^{124}$Sn ($N/Z=1.56$), using two options for the
symmetry energy, Esy-b (dashed line) and Esy-a (solid line). The
ratios obtained from the present Au+Au data (solid circles) by
linear extrapolation (from $N/Z=1.494$) are shown for comparison.
Taken from Ref. \protect\cite{Rei07}.} } \label{reisdorf2}
\end{figure}

Reisdorf {\it et al.} has recently made a very extensive review of
both old and new experimental data on the $\pi ^{-}/\pi ^{+}$
ratio from heavy-ion reactions at Bevalac and SIS/GSI energies
\cite{Rei07}. They also compared the data with several transport
model calculations. Shown in Fig.\ \ref{reisdorf1} is a summary of
measured $\pi^-/\pi^+$ ratios by the FOPI Collaboration and their
comparisons with the IQMD predictions. One observes a decrease of
the $\pi^-/\pi^+$ ratio with incident energy (upper left window)
as predicted by the IQMD. However, as shown in this window as well
as in other three windows, while the IQMD describes very well the
data at $1.5A$ GeV, including the dependence on $N/Z$, it clearly
underestimates the pion ratio at $0.4A$ GeV. Same conclusion is
obtained when the {\em filtered} data at $0.4A$ GeV is used as
shown in the right lower window of Fig.\ \ref{reisdorf1}.
Comparing the experimental data with the IBUU \cite{LiBA03} and
the RBUU \cite{Gai04} calculations leads to a similar conclusion
\cite{Rei07}. This is shown in the right window of Fig.\
\ref{reisdorf2}, which is a re-plot of Fig.\ \ref{pionsys} by
extrapolating linearly to the N/Z ratio of Au+Au \cite{Rei07}. It
is seen that the theoretical results are significantly below the
data at 400 Mev/A while come very close to the data at 1.5 GeV/A.
As to the effects of the symmetry energy, the difference predicted
from calculations using $E^a_{\rm sym}(\rho)$ and $E^b_{\rm
sym}(\rho)$ of Eqs. (\ref{esyma}) and (\ref{esymb}) is on the
$10-15\%$ level and hence in the order of present experimental
accuracy, and neither prediction follows the data. Similar
conclusions are obtained if one uses the results from the
calculations in Ref.~\cite{Gai04}. At this time, one can only
speculate several possible reasons for this discrepancy between
the data and the calculations. A more systematic comparison is
definitely needed before we can learn anything about the symmetry
energy from the $\pi^{-}/\pi ^{+}$ ratio.

The systematics of $\pi^-/\pi^+$ ratios was first established for
inclusive reactions \cite{Nag81} at $0.8A$ GeV beam energy using
various asymmetric systems. In the left window of
Fig.~\ref{reisdorf2} these older data and the new FOPI data are
plotted as functions of an estimated \cite{Nag81} 'fireball'
$(N/Z)$ composition. Due to limited accuracy,  both linear and
quadratic $(N/Z)$ dependences are compatible with these inclusive
data. The FOPI data point (open circle) for Au+Au at same energy
but for a central collision selection is perfectly compatible with
the linear extrapolation.

\subsubsection{Double $\protect\pi ^{-}/\protect\pi ^{+}$ ratio near
the Coulomb peak}

\begin{figure}[tbh]
\includegraphics[scale=0.7]{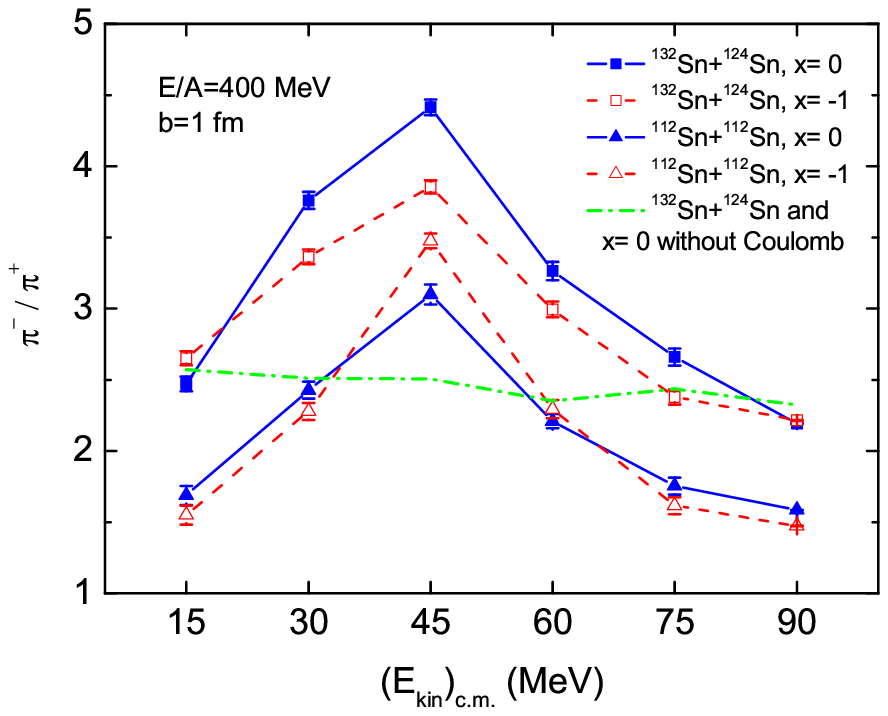}
\includegraphics[scale=0.7]{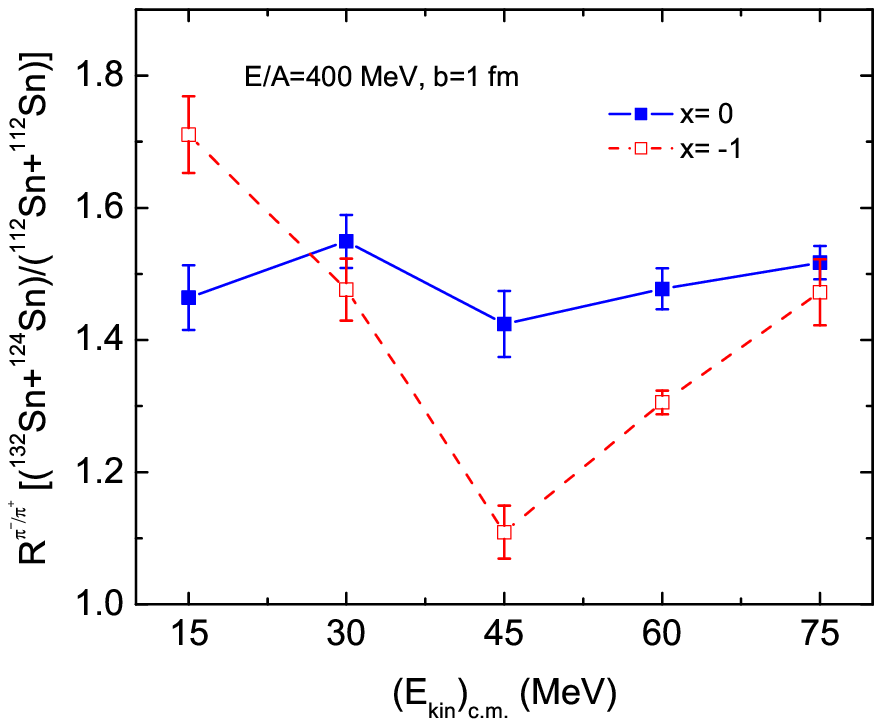}
\caption{(Color online) Left window: Kinetic energy distribution of
the single $\protect \pi ^{-}/\protect\pi ^{+}$ ratio for $^{132}$
Sn+$^{124}$ Sn and $^{112}$ Sn+$^{112}$ Sn at a beam energy of $400$
MeV/nucleon and an impact parameter of $b=1$ fm with the stiff
($x=-1$) and soft ($x=0 $) symmetry energies. The dash-dotted line
is the single $\protect\pi ^{-}/\protect\pi ^{+}$ ratio obtained by
turning off the Coulomb potentials in the $^{132}$ Sn+$^{124}$ Sn
reaction. Right window: Kinetic energy dependence of the double
$\protect\pi ^{-}/\protect\pi ^{+}$ ratio of $^{132}$ Sn+$^{124}$ Sn
over $^{112}$ Sn+$^{112}$ Sn. Taken from Ref.
\protect\cite{Yon06a}.} \label{Rpion}
\end{figure}

The double neutron/proton ratio of nucleon emissions taken from
two reaction systems using four isotopes of same element, namely,
the neutron/proton ratio in the neutron-rich system over that in
the more symmetric system, was found useful for reducing both
experimental uncertainties and the effects of the Coulomb force as
we have discussed earlier. Similarly, one can also take advantages
of the double $\pi ^{-}/\pi ^{+}$ ratios in these reactions. In
transport model calculations, the systematic errors are mostly
related to the physical uncertainties of in-medium NN cross
sections, techniques of treating collisions, sizes of the lattices
in calculating the phase space distributions, techniques in
handling the Pauli blocking, etc.. Since the double ratio is a
relative observable from two similar reaction systems, systematic
errors are thus expected to be reduced.  This is demonstrated in
Fig.~\ref{Rpion}, which shows the kinetic energy distributions of
the single (left window) and double (right window) $\pi ^{-}/\pi
^{+}$ ratios for the reactions of $^{132}$Sn+$^{124}$Sn and
$^{112}$Sn+$^{112}$Sn at a beam energy of $400$ MeV/nucleon and an
impact parameter of $b=1$ fm with the stiff ($x=-1$) and soft
($x=0$) symmetry energy, respectively.The results were obtained
with $12000$ events for each reaction.

For the overall magnitude of the single $\pi ^{-}/\pi ^{+}$ ratio,
it is larger for the neutron-rich system $^{132}$Sn+$^{124}$Sn
than for the neutron-deficient system $^{112}$Sn+$^{112}$Sn as
expected. The single $\pi ^{-}/\pi ^{+}$ ratio for the reaction
$^{112}$Sn+$^{112}$Sn is not so sensitive to the symmetry energy
due to the small isospin asymmetry. However, it becomes sensitive
to the symmetry energy for the neutron-rich system
$^{132}$Sn+$^{124}$Sn. These results are consistent with those
from previous studies \cite{LiBA02,LiBA05a,Gai04,LiQF05a,LiQF05b}.
It is further seen that the soft symmetry energy ($x=0$) leads to
a larger single $\pi ^{-}/\pi ^{+}$ ratio than the stiff one
($x=-1$). This is mainly because the high density region, where
most pions are produced, are more neutron-rich with the use of the
softer symmetry energy as a result of isospin fractionation
\cite{LiBA02,LiBA05a}.

For all cases considered here, the single $\pi ^{-}/\pi ^{+}$
ratio further exhibits a peak at a pion kinetic energy of about
$45$ MeV. In order to understand the origin of this peak,
calculations have also been done for the single $\pi ^{-}/\pi
^{+}$ ratios in both reactions by turning off the Coulomb
potentials for all charged particles. As an example, shown in
Fig.\ \ref{Rpion} with the dash-dotted line is the single $\pi
^{-}/\pi ^{+}$ ratio obtained by turning off the Coulomb
potentials in the $^{132}$Sn+$^{124}$Sn reaction. It is seen that
the single $\pi ^{-}/\pi ^{+}$ ratio now becomes approximately a
constant of about $2.4$, which is what one expects based on the
$\Delta $ resonance model. For central $^{132}$Sn+$^{124}$Sn
reactions, according to Eq.~(\ref{pres}) the $\pi ^{-}/\pi ^{+}$
ratio from the resonance model is approximately $2.43$. The
comparison of calculations with and without the Coulomb potentials
clearly indicates that the peak observed in the single $\pi
^{-}/\pi ^{+}$ ratio is indeed due to the Coulomb effects. The
$\pi ^{-}/\pi ^{+}$ ratio carries some information about the
symmetry energy mainly because it is sensitive to the isospin
asymmetry of the nucleonic matter where pions are produced. This
information might be distorted but is not completely destroyed by
the Coulomb interactions of pions with other particles. It is thus
natural to look for signals of the symmetry energy in kinematic
regions where the $\pi ^{-}/\pi ^{+}$ ratio reaches its maximum.
In this regard, the Coulomb peak is actually very useful for
studying the effects of the symmetry energy. Since the Coulomb
peak could appear at zero instead of a finite kinetic energy, one
needs to concentrate on the $\pi ^{-}/\pi ^{+}$ ratio of low
energy pions. Although most pions are produced in the high density
nucleonic matter (about $2\rho _{0}$) through $\Delta $
resonances, thus carrying important information about the high
density behavior of the symmetry energy, pions at lower kinetic
energies around the Coulomb peak experience many rescatterings
with nucleons at both high and low densities, with charged pions
further affected by the Coulomb potential from protons at
different densities. The information on high density symmetry
energy, that is carried by lower energy pions, may thus be
partially distorted by the low density behavior of the symmetry
energy \cite{LiQF05b}. Since the soft ($x=0$) and stiff ($x=-1$)
symmetry energies differ slightly at low densities but appreciably
at high densities (about $2\rho _{0}$), one thus expects the
observed symmetry energy effects on the energy dependence of the
$\pi ^{-}/\pi ^{+}$ ratio to mainly reflect (though not
completely) the information on the high density behavior of the
symmetry energy.

For the double $\pi ^{-}/\pi ^{+}$ ratio in the reactions of
$^{132}$Sn+$^{124}$Sn and $^{112}$Sn+$^{112}$Sn, the results are
shown in the right window of Fig.~\ref{Rpion}. It is seen that the
kinetic energy dependence of the double $\pi ^{-}/\pi ^{+}$ ratio
is rather different for the stiff ($x=-1 $) and soft ($x=0$)
symmetry energies. While it is quite flat for $x=0$, there is a
concave structure around the Coulomb peak for $x=-1$. These
different behaviors can be understood from corresponding single
$\pi ^{-}/\pi ^{+}$ ratios in the two reactions shown in the left
window of Fig. \ref{Rpion}. Although the double $\pi ^{-}/\pi
^{+}$ ratio has a weaker dependence on the pion kinetic energy
than the single $\pi ^{-}/\pi ^{+}$ ratio, its value around the
Coulomb peak is still sensitive to the symmetry energy. This is
because effects of the Coulomb potentials are reduced in the
double $\pi ^{-}/\pi ^{+}$ ratio. Compared with the double $n/p$
ratio for free nucleons shown in Fig.~\ref{dnpfigure4}, the double
$\pi ^{-}/\pi^{+}$ ratio displays an opposite symmetry energy
dependence. This is understandable since the soft symmetry energy
leads to a more neutron-rich dense matter in heavy-ion collisions
induced by neutron-rich nuclei, more $\pi ^{-}$'s are thus
produced due to more neutron-neutron inelastic scatterings.
Because of charge conservation, the $n/p$ ratio for free nucleons
is, on the other hand, expected to be smaller.

\subsection{The $K^{0}/K^{+}$ and $\Sigma ^{-}/\Sigma ^{+}$ ratios}

Since the proposal of Aichelin and Ko that subthreshold kaon yield
may be a sensitive probe of the EOS of nuclear matter at high
densities \cite{Aic85}, a lot of works have been done to
investigate the subthreshold kaon (and anti-kaon) production in
heavy-ion collisions both theoretically and experimentally
\cite{Fuc06a,Ko96,Cas90,Ko97,Cas99,Kol05}. The kaon is an
iso-doublet meson with the quark content of $d\overline{s}$ for
$K^{0}$ and $u\overline{s}$ for $K^{+}$, so the $K^{0}/K^{+}$
ratio provides a potentially good probe of the nuclear symmetry
energy, especially its high density behavior since kaons are
produced mainly from the high density region during the early
stage of the reaction and are essentially free of subsequent
reabsorption effects.

\begin{figure}[tbp]
\centering
\includegraphics[scale=0.7]{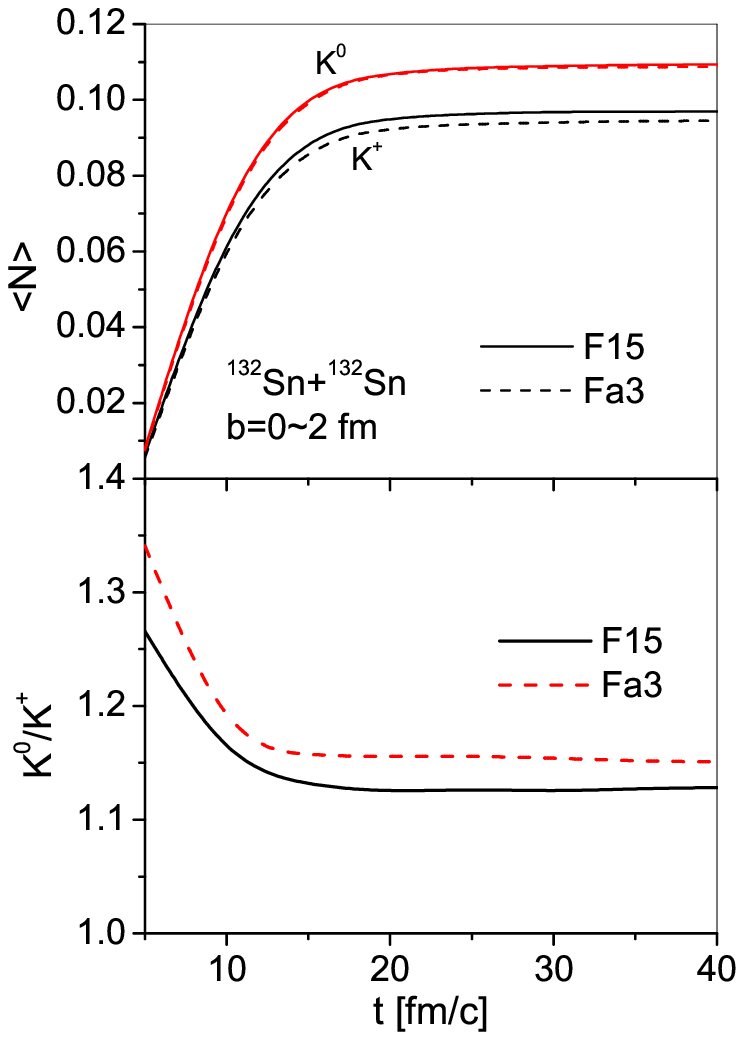}
\includegraphics[scale=0.7]{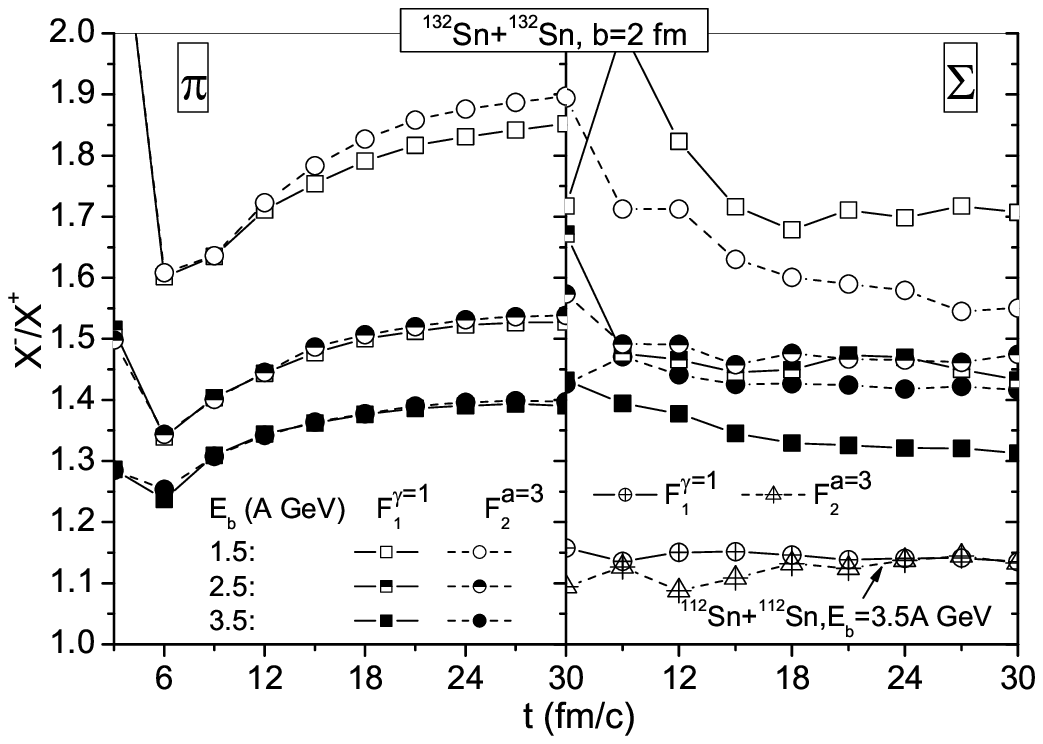}
\caption{Left window: Time evolution of the $K^{0}$ and $K^{+}$
abundances (upper panel) and their ratio (lower panel) for central
$^{132}{\rm Sn}+^{132}{\rm Sn}$ collisions at a beam energy $1.5A $
GeV and with the symmetry potentials F15 and Fa3. Right window: The
ratios $\protect\pi ^{-}/\protect\pi ^{+}$ (left panel) and $\Sigma
^{-}/\Sigma ^{+}$ (right panel) for the collisions $^{132}{\rm
Sn}+^{132}{\rm Sn}$ ($E_{\rm beam}=1.5A $, $2.5A$, and $3.5A$ GeV;
$b=2$ fm) and $^{112}{\rm Sn}+^{112}{\rm Sn}$ ($E_{\rm beam}=3.5A$
GeV; $b=2$ fm), calculated with the different symmetry potentials
$F_{1}^{\protect\gamma =1}$ and $F_{2}^{a=3}$. Taken from Ref.
\protect\cite{LiQF05c}.} \label{RatioKaonLi}
\end{figure}

Using the UrQMD model (version 1.3), Li \textit{et al.} have
investigated the symmetry energy effects on the $K^{0}/K^{+}$
ratio by studying the $K^{0}$ and $K^{+}$ production from the
central $^{132}$Sn$+^{132}$Sn collisions at a beam energy $1.5A$
GeV with two different forms of the symmetry energy, namely, a
still F15 and a soft Fa3 with its potential energy vanishing at
$3\rho _{0}$. The results are shown in the left window of Fig.\
\ref{RatioKaonLi}~\cite{LiQF05c}. It is seen that the
$K^{0}/K^{+}$ ratio displays only small symmetry energy effects at
an incident energy close to the kaon production threshold, which
is about $1.58$ GeV in nucleon-nucleon interaction in free space.
For energies far less than the kaon production threshold, they
have also calculated the kaon yields from the reaction
$^{208}$Pb$+^{208}$Pb at $E_{\mathrm{beam}}=0.8$ $A$ GeV and
$b=7\sim 9\ \mathrm{fm}$ with the symmetry energy forms of F15 and
Fa3, and the results indicate that the $K^{0}/K^{+}$ ratio for F15
is about $1.25$, whereas it is about $1.4$ for Fa3. These results
were obtained without including any nuclear in-medium effects on
kaon production in the UrQMD model simulations.

Besides the $K^{0}/K^{+}$ ratio, the $\Sigma ^{-}/\Sigma ^{+}$
ratio has also been proposed as a probe of the high density
behavior of the nuclear symmetry based on the UrQMD model (version
1.3) calculations \cite{LiQF05a}. Shown in the right window of
Fig.\ \ref{RatioKaonLi} is the time evolution of the $\pi ^{-}/\pi
^{+}$ ratios (left panel) and the $\Sigma ^{-}/\Sigma ^{+}$ ratios
(right panel) calculated with a stiff symmetry energy
$F_{1}^{\gamma =1}$ and a soft symmetry energy $F_{2}^{a=3}$ for
the reaction $^{132}$Sn$+^{132}$Sn at $E_{\rm beam}=1.5A$, $2.5A$,
$3.5A$ GeV and $b=2$ fm, and the reaction $^{112}$Sn$+^{112}$Sn at
$E_{\rm beam}=3.5A$ GeV and $b=2$ fm. It is seen that the $\Sigma
^{-}/\Sigma ^{+}$ ratio is sensitive to the density dependence of
the symmetry energy for neutron-rich $^{132}$Sn$+^{132}$Sn
collisions, but insensitive to that for the nearly symmetric
$^{112}$Sn$+^{112}$Sn collisions. For $^{132}$Sn$+^{132}$Sn at
$E_{\rm beam}=1.5A$ GeV, the $\Sigma ^{-}/\Sigma ^{+}$ ratio
calculated with the stiff symmetry energy ($F_{1}^{\gamma =1}$) is
higher than the one with the soft symmetry energy ($F_{2}^{a=3}$).
As the beam energy increases, the $\Sigma ^{-}/\Sigma ^{+}$ ratio
falls and the difference between the $\Sigma ^{-}/\Sigma ^{+}$
ratios calculated with $F_{1}^{\gamma =1}$ and $F_{2}^{a=3}$
decreases significantly. As the beam energy increases further to
$E_{\rm beam}=3.5A$ GeV, the $\Sigma ^{-}/\Sigma ^{+}$ ratio
continues to fall but the difference between the $\Sigma
^{-}/\Sigma ^{+}$ ratios calculated with $F_{1}^{\gamma =1}$ and
$F_{2}^{a=3} $ appears again, with the $\Sigma ^{-}/\Sigma ^{+}$
ratio with soft symmetry energy now becoming higher than that with
the stiff one. For pions, the results indicate that the ratio $\pi
^{-}/\pi ^{+}$ at high energies (as in the case with $E_{\rm
beam}=3.5A$ GeV) becomes insensitive to the symmetry energy. The
difference between the $\Sigma ^{-}/\Sigma ^{+}$ ratio and the
$\pi ^{-}/\pi ^{+}$ ratio can be understood from the fact that,
like nucleons, $\Sigma ^{\pm }$ hyperons are under the influence
of the mean field produced by surrounding nucleons. The symmetry
potential of hyperons thus play an important dynamic role and
results in a strong effect on the ratio of the negatively to
positively charged $\Sigma $ hyperons.

\begin{figure}[h]
\centering
\includegraphics[scale=0.3]{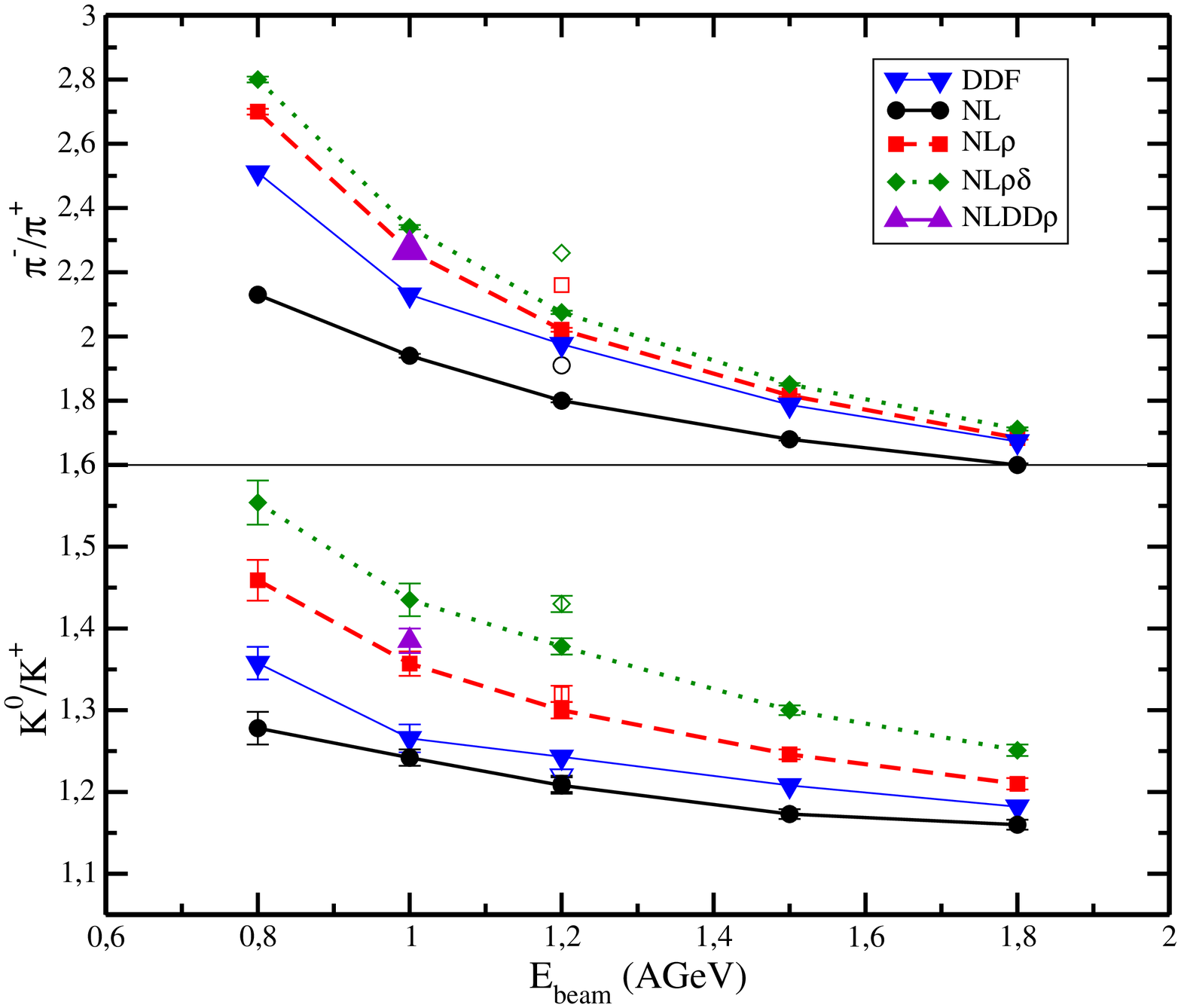}
\includegraphics[scale=0.4]{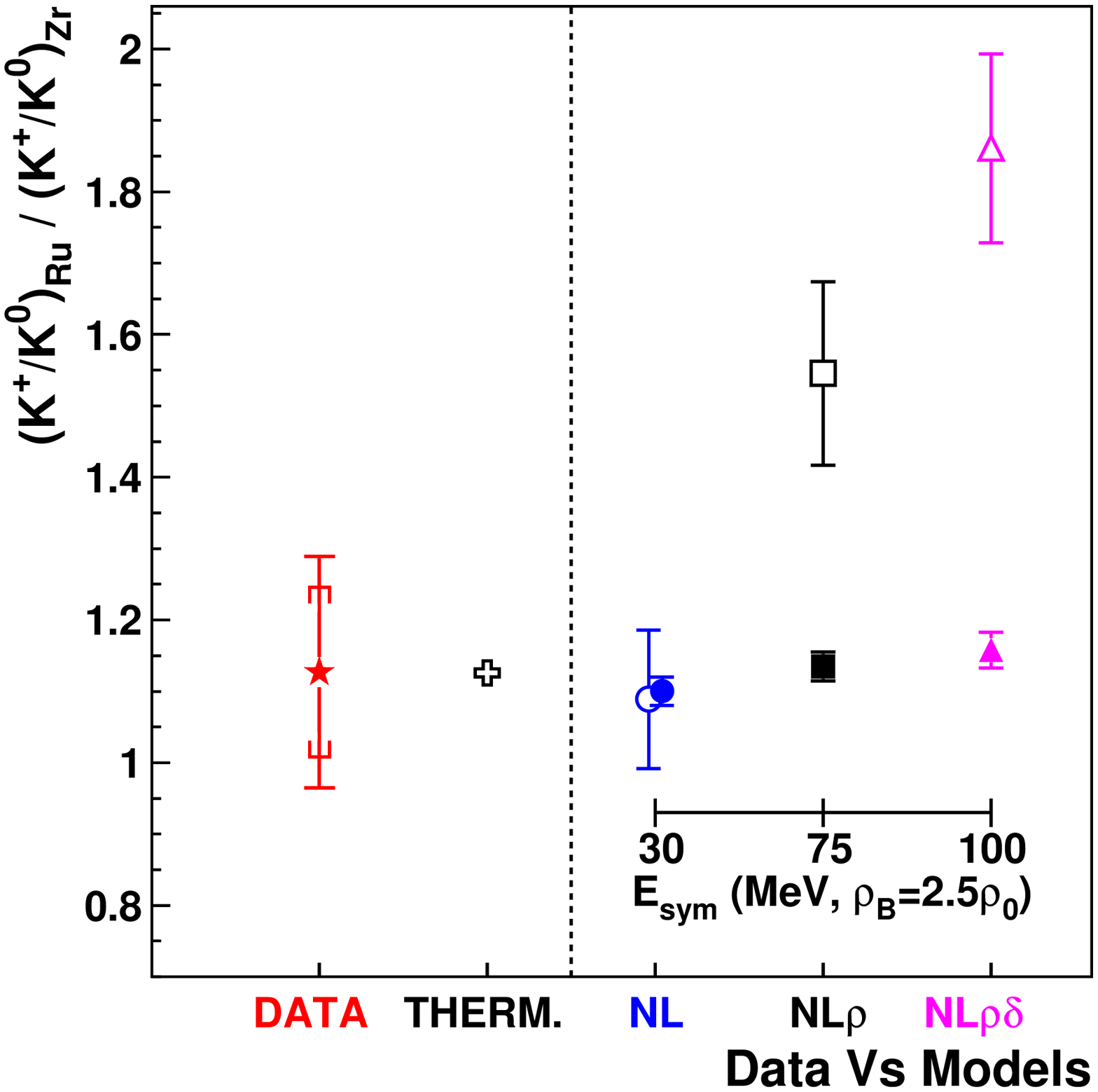}
\caption{Left window: $\protect\pi ^{-}/\protect\pi ^{+}$ (upper)
and $K^{+}/K^{0}$ (lower) ratios as a function of the incident
energy for central ($b=0$ fm impact parameter) Au+Au collisions with
the RBUU model . In addition, for $E_{\rm beam}=1~A$GeV,
$NL\protect\rho $ results with a density-dependent $\protect\rho
$-coupling (triangles) are also presented. The open symbols at
$1.2~{\rm AGeV}$ show the corresponding results for a
$^{132}$Sn+$^{124}$Sn collision with more neutron-rich isotopes.
Note the different scale for the $\protect\pi ^{-}/\protect\pi ^{+}$
ratios. Taken from Ref. \protect\cite{Fer05}. Right window:
Experimental ratio ($K^{+}/K^{0}$)$_{\rm Ru}$/($K^{+}/K^{0}$)$_{\rm
Zr}$ (star) and theoretical predictions of the thermal model (cross)
and the transport model with 3 different assumptions on the symmetry
energy: NL (circles), NL$\protect\rho $ (squares) and
NL$\protect\rho \protect\delta $ (triangles). The INM and HIC
calculations are represented by open and full symbols, respectively
(see text for more details). The statistic and systematic errors are
represented by vertical bars and brackets, respectively. Taken from
Ref. \protect\cite{Lop07}.} \label{RatioKaonFe}
\end{figure}

Recently, Ferini {\it et al.} also studied the symmetry energy
effects on the $K^{0}/K^{+}$ ratio in central ($b=0$ fm impact
parameter) Au+Au collisions using a relativistic hadronic
transport model of Boltzmann-Uehling-Uhlenbeck type (RBUU) with
different forms of symmetry energies \cite{Fer05}, and the results
are shown in Fig.\ \ref{RatioKaonFe}. Their results show that at
beam energies below and around the kinematical threshold of kaon
production, the $K^{0}/K^{+}$ inclusive yield ratio is more
sensitive to the symmetry energy than the $\pi ^{-}/\pi ^{+}$
ratio, thus indicating that sub-threshold kaon production could
provide a promising tool to extract information on the density
dependence of the nuclear symmetry energy.

Most recently, the FOPI collaboration has reported the results on
$K^{+}$ and $K^{0}$ meson production in $_{44}^{96}$Ru +
$_{44}^{96}$Ru and $_{40}^{96}$Zr + $_{40}^{96}$Zr collisions at a
beam kinetic energy of $1.528$ $A$ GeV. The measured double ratio
($K^{+}/K^{0}$)$_{\rm Ru}$/($K^{+}/K^{0}$)$_{\rm Zr}$ is compared
in the right window of Fig.\ \ref{RatioKaonFe} to the predictions
of a thermal model and the RBUU transport model using two
different collision scenarios and under different assumptions on
the stiffness of the symmetry energy. One can see a good agreement
with the thermal model prediction and the assumption of a soft
symmetry energy for infinite nuclear matter. While more realistic
transport simulations of the collisions show a similar agreement
with the data, they also exhibit a significantly reduced
sensitivity to the symmetry energy. In the present RBUU
calculations, the isospin dependence of the $K^{+}$- and
$K^{0}$-nucleon potentials in the asymmetric nuclear medium has,
however, been neglected. Recent studies by Mishra \textit{et al.}
based on the chiral SU(3) model have shown that the isospin
dependence of the kaon and antikaon optical potentials in dense
hadronic matter is appreciable. Also, results from the transport
model depend on the details how kaon production is implemented in
the model \cite{Kol05,Fuc06a}. To extract useful information on
the high density behavior of the nuclear symmetry energy from
subthreshold kaon production in heavy-ion collisions induced by
neutron-rich nuclei, further experimental and theoretical studies
are thus needed.

\subsection{Hard photon production as a probe of the symmetry
energy}

Recently, Yong {\it et al.} \cite{Ylc07} has studied the
neutron-proton bremsstrahlung from intermediate energy heavy-ion
reactions as a probe of the nuclear symmetry energy. Although the
results are promising, the experiments involved are very
challenging. Also, there exists the theoretical uncertainty on the
elementary neutron-proton bremsstrahlung cross section.

Hard photon production in heavy-ion reactions at beam energies
between about 10 and 200 MeV/A have been extensively studied both
experimentally and theoretically, especially in the mid 1980's, see,
e.g., Refs.~\cite{Ber88b,Cas90,nif90} for a comprehensive review.
Interesting physics has been obtained from the experimental data
taken by many collaborations. For instance, the TAPS collaboration
carried out a series of comprehensive measurements at various
experimental facilities, such as GSI, GANIL, KVI, to study in detail
the properties of hard photons, such as their energy spectra,
angular distributions, total multiplicities, and the di-photon
correlation functions, etc., from a large variety of nucleus-nucleus
systems in the energy range of $E_{\rm lab}\approx 20-200$
MeV/nucleon. They had used the bremsstrahlung photons as a tool to
study the nuclear caloric curve, the dynamics of nucleon-nucleon
interactions, as well as the time-evolution of the reaction process
before break-up \cite{TAPS}. From theoretical studies based on
various models, it has been concluded that the neutron-proton
bremsstrahlungs in the early stage of the reaction are the main
source of high energy $\gamma$ rays. In particular, the cascade and
BUU transport models have clearly demonstrated that the hard photons
can be used to probe the reaction dynamics leading to the formation
of the dense matter \cite{bertsch86,ko85,cassing86,bau86,stev86}.
However, the effects of the nuclear EOS on hard photon production
was found small \cite{ko87}. While these reaction models were able
to reproduce the qualitative features of experimental data, the
quantitative agreement was normally within about a factor of 2. One
of the major uncertainties is the input elementary $pn\rightarrow
pn\gamma$ probability $p_{\gamma}$ which is still rather model
dependent \cite{nif85,nak86,sch89,gan94,Bro73,Her91,tim06}. Early
model studies usually could only describe within a factor of 2 the
few existing data for the $pn\rightarrow pn\gamma$ process
\cite{Cas90}. However, very recent systematic measurements of the
$pn\rightarrow pn\gamma$ cross sections with neutron beams up to 700
MeV at Los Alamos are expected to improve the situation
significantly in the near future \cite{saf07}.

Since the photon production probability is small, a perturbative
approach has been used in all dynamical calculations of photon
production in heavy-ion reactions at intermediate energies
\cite{Ber88b,Cas90}. In this approach, one calculates the photon
production probability at each proton-neutron collision and then
sum over all such collisions for the entire history of the
reaction. As discussed in detail in Ref. \cite{Cas90}, the cross
section for neutron-proton bremsstrahlung in the long-wavelength
limit separates into a product of the elastic $np$ scattering
cross section and a $\gamma$-production probability. The
probability is often taken from the semiclassical hard sphere
collision model \cite{Ber88b,Cas90,nif90}. The double differential
probability, ignoring the Pauli exclusion in the final state, is
given by
\begin{eqnarray}\label{coss}
\frac{d^{2}N}{d\varepsilon_{\gamma}d\Omega_{\gamma}}=\frac{e^{2}}{12\pi^{2}\hbar
c}\times\frac{1}{\varepsilon_{\gamma}}(3\sin^{2}\theta_{\gamma}\beta_{i}^{2}
+2\beta_{f}^{2})=6.16\times10^{-5}\times
\frac{1}{\varepsilon_{\gamma}}(3\sin^{2}\theta_{\gamma}\beta_{i}^{2}
+2\beta_{f}^{2}),\notag\\
\end{eqnarray}
where $\theta_{\gamma}$ is the angle between the incident proton
direction and the emission direction of photon; and $\beta_i$ and
$\beta_f$ are the initial and final velocities of proton in the
proton-neutron center of mass frame. The above equation was
obtained from modifying the original semi-classical formula
\cite{jackson} to allow for energy conservation in the
$\gamma$-production process \cite{cassing86,bau86}. Integrating
Eq.\ (\ref{coss}) over the photon emission angle, one obtains the
single differential probability
\begin{eqnarray}\label{intc}
p^a_{\gamma}\equiv\frac{dN}{d\varepsilon_{\gamma}}
=1.55\times10^{-3}\times\frac{1}{\varepsilon_{\gamma}}
(\beta_{i}^{2}+\beta_{f}^{2}).
\end{eqnarray}
Other expressions involving the quantum-mechanical effects exist
in the literature, see, e.g.,
Refs.~\cite{nif85,nak86,sch89,gan94,tim06}. For example, Gan {\it
et al.} used the following \cite{gan94}
\begin{eqnarray}\label{QFT}
p^b_{\gamma}\equiv\frac{dN}{d\varepsilon_{\gamma}}
=2.1\times10^{-6}\frac{(1-y^{2})^{\alpha}}{y},
\end{eqnarray}
where $y=\varepsilon_{\gamma}/E_{\rm max}$, $
\alpha=0.7319-0.5898\beta_i$, and $E_{\rm max}$ is the energy
available in the center of mass of the colliding proton-neutron
pairs.

\begin{figure}[th]
\centering
\includegraphics[scale=0.7]{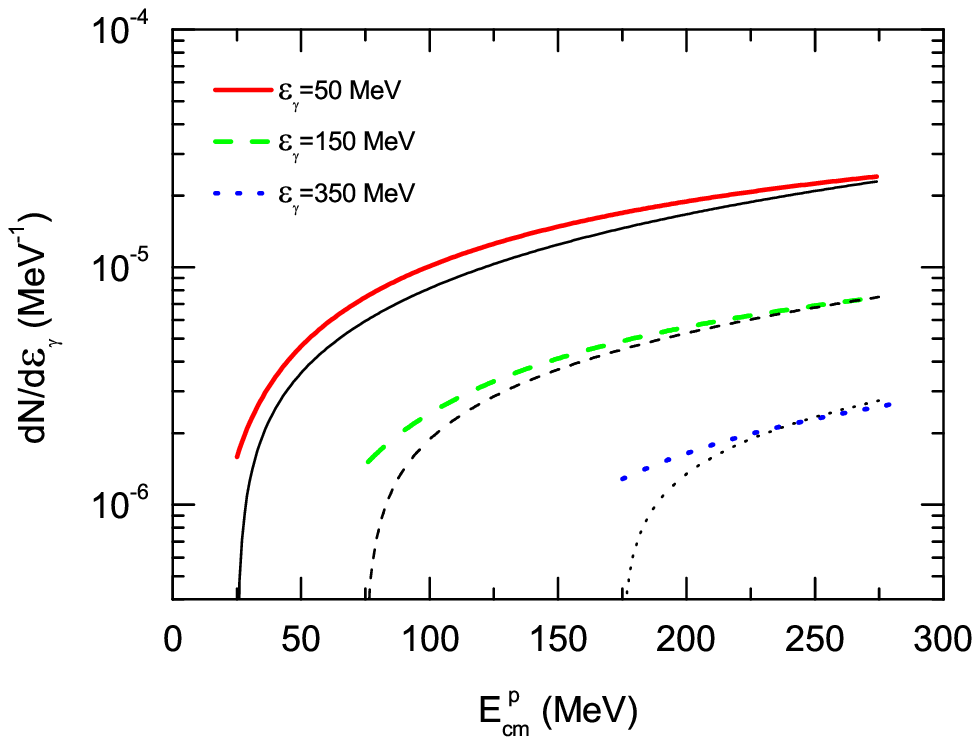}
\includegraphics[scale=0.7]{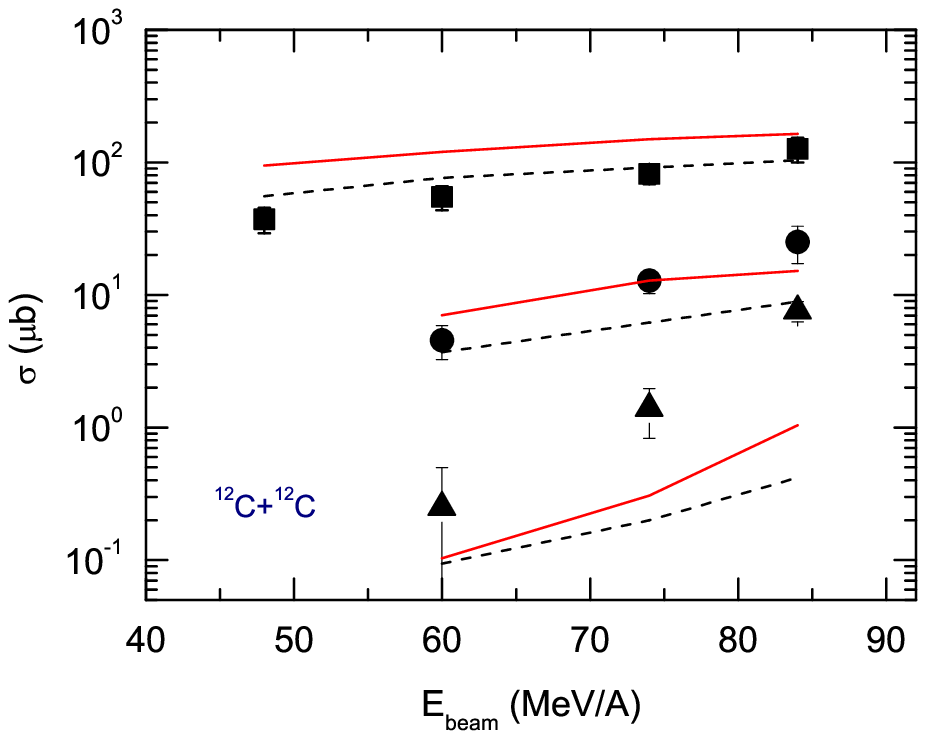}
\caption{Left window: The single differential probability as a
function of proton kinetic energy in the proton-neutron center of
mass frame for the production of photons at energies of $50$, $150$
and $350$ MeV. The lines with higher values are results calculated
with the semi-classical Eq. (\ref{intc}) while the ones with lower
values are obtained by using the quantum-mechanical Eq. (\ref{QFT}).
Right window: Beam energy dependence of the inclusive photon
production cross sections in $^{12}$C+$^{12}$C collisions. The solid
symbols stand for experimental data \cite{Cas90,grosse86}. (The
squares are for 50 MeV $\leq \varepsilon_{\gamma} <$ 100 MeV,
circles for 100 MeV $\leq \varepsilon_{\gamma} <$ 150 MeV and
triangles for $\varepsilon_{\gamma}\geq$ 150 MeV). The solid lines
are calculated using the $p^a_{\gamma}$ and the dashed ones using
the $p^b_{\gamma}$. Taken from Ref. \cite{Ylc07}.}\label{cross}
\end{figure}

The single differential probabilities $p^a_{\gamma}$ and
$p^b_{\gamma}$ from the two models are shown in the left window of
Fig.\ \ref{cross} as functions of proton kinetic energy in the
proton-neutron center of mass frame for the production of photons
at energies of $50$, $150$ and $350$ MeV. It is seen that the two
models give quite similar but quantitatively different results
especially near the kinematic limit where the $p^a_{\gamma}$ is
significantly higher than the $p^b_{\gamma}$, as noticed already
in Ref.~\cite{gan94}. Shown in the right window of Fig.\
\ref{cross} are the calculations with both $p^a_{\gamma}$ and
$p^b_{\gamma}$ within the IBUU04 transport model using
isospin-dependent in-medium NN cross sections \cite{Ylc07}. The
experimental data for the inclusive cross section of hard photon
production in the reaction of $^{12}$C+$^{12}$C
\cite{Cas90,grosse86} are also shown for comparison. The
calculated results are obtained with $x=0$. It is seen that both
results are in reasonable agreement with experimental data except
for very energetic photons. Quantitatively, the agreement is at
about the same level as previous calculations by others in the
literature \cite{ko85,bau86,gan94}. The uncertainty in the
elementary $pn\rightarrow pn\gamma$ probability leads to an
appreciable effect on the inclusive $\gamma$-production in
heavy-ion reactions. The effect is larger than that obtained by
varying the $x$ parameter from $x=0$ to $x=-1$ or $x=1$. It is
thus a very challenging task to extract useful information about
the symmetry energy from the total yield of photons from heavy ion
reactions. However, as in many experiments that search for minute
but interesting effects, ratios of two reactions can often reduce
not only the systematic errors but also some `unwanted' effects.
Within the perturbative approach, the uncertainty due to the
$\gamma$-production probability is thus expected to be removed in
the ratio of photons from two reactions. Depending on the relative
number of neutron-proton scatterings in the two reactions,
uncertainties due to the NN cross sections can also get
significantly reduced. It is thus better to measure experimentally
the spectra ratio $R_{1/2}(\gamma)$ of hard photons from two
reaction systems.

\begin{figure}[th]
\centering
\includegraphics[scale=1]{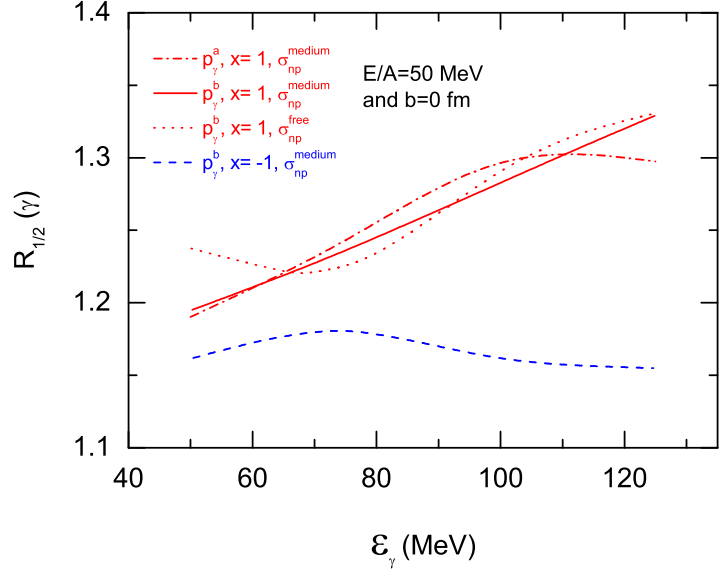}
\caption{(Color online) The spectra ratio of hard photons in the
reactions of $^{132}{\rm Sn}+^{124}{\rm Sn}$ and $^{112}{\rm
Sn}+^{112}{\rm Sn}$ at a beam energy of $50$ MeV/A with the
symmetry energies of $x=1$ and $x=-1$. Taken from Ref.
\cite{Ylc07}.} \label{relat}
\end{figure}

Shown in Fig.\ \ref{relat} is the $R_{1/2}(\gamma)$ for head-on
reactions of $^{132}$Sn+$^{124}$Sn and $^{112}$Sn+$^{112}$Sn, i.e.,
\begin{eqnarray}
R_{1/2}(\gamma)\equiv\frac{\frac{dN}{d\varepsilon_{\gamma}}(^{132}{\rm
Sn}+^{124}{\rm Sn})} {\frac{dN}{d\varepsilon_{\gamma}}(^{112}{\rm
Sn}+^{112}{\rm Sn})},
\end{eqnarray}
calculated for four different cases using both $p^a_{\gamma}$ and
$p^b_{\gamma}$. It is seen clearly that the full calculations with
$p^a_{\gamma}$ and $p^b_{\gamma}$ and the in-medium NN cross
sections indeed lead to about the same $R_{1/2}(\gamma)$ within
statistical errors as expected. It is also clearly seen that
effects of the in-medium NN cross sections are also essentially
cancelled out. These results thus demonstrated the advantage of
using $R_{1/2}(\gamma)$ as a robust probe of the symmetry energy
that is essentially free of the uncertainties associated with both
the elementary photon production and the NN cross sections.
Moreover, the spectra ratio $R_{1/2}(\gamma)$ remains sensitivity
to the symmetry energy especially for very energetic photons.
Since the symmetry energy is varied by at most 20\% in the
reaction considered when varying the parameter $x$ from $1$ ro
$-1$, the approximately 15\% maximum change in the spectra ratio
represents a relatively significant sensitivity, which is at about
the same level as most hadronic probes including the $K^0/K^+$
ratio. The latter is considered as among the most clean hadronic
probes of the symmetry energy, and it shows about a 15\% change
when the symmetry energy changes by at least 50\% at the density
reached in heavy-ion reactions near the kaon production threshold.
Compared to the $K^0/K^+$ ratio the hard photon production is an
even more sensitive and clean observable. However, while photons
are completely free from final-state strong interactions, one
needs to take consideration of photons from $\pi^{0}$ and fragment
decays in the data analysis \cite{grosse86}.

For hard photons in reactions at higher beam energies, which would
lead to higher densities and thus make it possible to explore the
behaviors of the symmetry energy there, other sources for hard
photon production may become important. Moreover, the reaction
dynamics at higher energies are dominated by nucleon-nucleon
collisions rather than the nuclear mean-field. Effects of the
symmetry energy on photons are then expected to become smaller as
hard photons are affected by the symmetry potential only
indirectly through the momentum distributions and the densities of
the colliding proton-neutron pairs. This is also the reason that
the hard photons were found not to be so sensitive to the nuclear
equation of state in an early study \cite{ko87}. Only at
intermediate energies both the mean-filed and the NN collisions
play about equally important roles in the reaction dynamics.


\section{Constraining the Skyrme effective interactions and the neutron skin
thickness of heavy nuclei using terrestrial nuclear laboratory
data}
\label{chapter_neutronskin}

Information on the density dependence of the nuclear symmetry energy
can also be obtained from the thickness of the neutron skin in heavy
nuclei. As first found by B. A. Brown \cite{Bro00}, there is a
correlation between the root-mean-square radius for neutrons in
nuclei and the equation of state for neutron matter. Subsequent
studies \cite{Die03,Hor01a,Typ01,Fur02,Kar02} further showed that a
particular strong correlation exists between the thickness of the
neutron skin in heavy nuclei and the slope parameter $L$ of the
nuclear symmetry energy at saturation density. A precise measurement
of the neutron radius and thus the thickness of neutron-skin in
heavy nuclei, such as $^{208}$Pb, thus would place an important
constraint on the equation of state for neutron matter. Because of
the large uncertainties in measured neutron skin thickness of heavy
nuclei, this has, however, not been possible. Instead, studies have
been carried out to use the extracted nuclear symmetry energy from
the isospin diffusion data to constrain the neutron skin thickness
of heavy nuclei \cite{LiBA05c,Che05b,Ste05b}. In the Hartree-Fock
approximation with parameters fitted to the phenomenological EOS
that was used in the IBUU04 transport model to describe the isospin
diffusion data from the NSCL/MSU, it was found that a neutron skin
thickness of less than $0.15$ fm \cite{LiBA05c,Che05b,Ste05b} for
$^{208}$Pb was incompatible with the isospin diffusion data.

In this Chapter, we discuss the correlation between the density
dependence of the nuclear symmetry energy and the thickness of the
neutron skin in a number of nuclei within the framework of the
Skyrme Hartree-Fock model. Using the extracted values of $L$ from
the isospin diffusion data in heavy-ion collisions, stringent
constraints on the neutron skin thickness of the nuclei $^{208}$Pb,
$^{132}$Sn, and $^{124}$Sn have been obtained. The extracted value
of $L$ also limits the allowed parameter sets for the Skyrme
interaction.

\subsection{Constraining the Skyrme effective interactions}

In the standard Skyrme Hartree-Fock model, the interaction is
taken to have a zero-range, density- and momentum-dependent form
\cite{Bra85,Sto03,Fri86,Bro98,Che99b}, i.e.,
\begin{eqnarray}
V_{12}(\mathbf{R},\mathbf{r}) &=&t_{0}(1+x_{0}P_{\sigma })\delta
(\mathbf{r})
+\frac{1}{6}t_{3}(1+x_{3}P_{\sigma })\rho ^{\sigma }(\mathbf{R})\delta
(\mathbf{r})  \notag \\
&+&\frac{1}{2}t_{1}(1+x_{1}P_{\sigma })(K^{^{\prime }2}\delta (\mathbf{r})
+\delta (\mathbf{r})K^{2})+t_{2}(1+x_{2}P_{\sigma })
\mathbf{K}^{^{\prime }}\cdot \delta (\mathbf{r})\mathbf{K}  \notag \\
&\mathbf{+}&iW_{0}\mathbf{K}^{^{\prime }}\cdot \delta
(\mathbf{r})[(\mathbf{\sigma }_{1}+\mathbf{\sigma }_{2})\times
\mathbf{K]},  \label{Sky}
\end{eqnarray}
with $\mathbf{r}=\mathbf{r}_{1}-\mathbf{r}_{2}$ and $\mathbf{R}=(\mathbf{r}%
_{1}+\mathbf{r}_{2})/2$. In the above, the relative momentum operators $%
\mathbf{K}=(\mathbf{\nabla }_{1}-\mathbf{\nabla }_{2})/2i$ and $\mathbf{K}%
^{\prime }=-(\mathbf{\nabla }_{1}-\mathbf{\nabla }_{2})/2i$ act on
the wave function on the right and left, respectively. The
quantities $P_{\sigma }$ and $\sigma _{i}$ denote, respectively,
the spin exchange operator and Pauli spin matrices. The $\sigma$,
$t_{0}-t_{3}$, $x_{0}-x_{3}$, and $W_{0}$ are Skyrme interaction
parameters that are chosen to fit the binding energies and charge
radii of a large number of nuclei in the periodic table. For
infinite nuclear matter, the symmetry energy from the Skyrme
interaction can be expressed as \cite{Sto03,Che99b}
\begin{eqnarray}
E_{\text{sym}}(\rho ) &=&\frac{1}{3}\frac{\hbar ^{2}}{2m}\left(
\frac{3\pi ^{2}}{2}\right) ^{2/3}\rho^{2/3}-\frac{1}{8}t_{0}(2x_{0}+1)\rho-
\frac{1}{48}t_{3}(2x_{3}+1)\rho ^{\sigma+1}  \notag \\
&+&\frac{1}{24}\left(\frac{3\pi ^{2}}{2}\right) ^{2/3}\left[
-3t_{1}x_{1}+\left( 4+5x_{2}\right) t_{2}\right] \rho ^{5/3}.
\label{EsymSky}
\end{eqnarray}

\begin{figure}[h]
\centering
\includegraphics[scale=0.9]{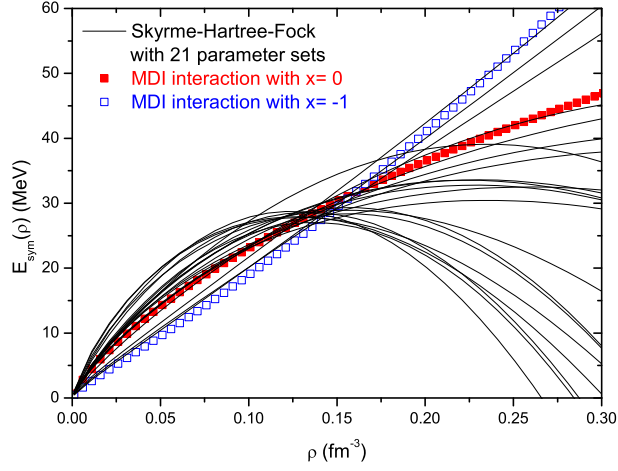}
\caption{(Color online) Density dependence of the nuclear symmetry energy $%
E_{\text{sym}}(\protect\rho )$ for 21 sets of Skyrme interaction
parameters. The results from the MDI interaction with $x=-1$ (open
squares) and $0$ (solid squares) are also shown. Taken from Ref.
\protect\cite{Che05b}.} \label{SymDen}
\end{figure}

Fig. \ref{SymDen} displays the density dependence of
$E_{\text{sym}}(\rho )$ for $21$ sets of Skyrme interaction
parameters, i.e., $SKM$, $SKM^{\ast }$, $RATP$, $SI$, $SII$, $SIII$,
$SIV$, $SV$, $SVI$, $E$, $E_{\sigma }$, $G_{\sigma }$, $R_{\sigma
}$, $Z$, $Z_{\sigma }$, $Z_{\sigma }^{\ast }$, $T$, $T3$, $SkX$,
$SkXce$, and $SkXm$. The values of the parameters in these Skyrme
interactions can be found in Refs. \cite{Bra85,Fri86,Bro98}. For
comparison, we also show in Fig. \ref{SymDen} results from the
phenomenological MDI interactions with $x=-1$ (open squares) and $0$
(solid squares). As we have discussed previously, from comparing the
isospin diffusion data from NSCL/MSU using the IBUU04 with in-medium
NN cross sections, these interactions are recently shown to give,
respectively, the upper and lower bounds for the stiffness of the
symmetry energy \cite{LiBA05c}. It is seen from Fig. \ref{SymDen}
that the density dependence of the symmetry energy varies
drastically among different interactions. Although the values of
$E_{\text{sym}}(\rho _{0})$ are all in the range of $26$-$35$ MeV,
the values of $L$ and $K_{\text{sym}}$ are in the range of
$-50$-$100$ MeV and $-700$-$50$ MeV, respectively.

The extracted value of $L=88\pm 25$ MeV from the isospin diffusion
data gives a rather stringent constraint on the density dependence
of the nuclear symmetry energy and thus puts strong constraints on
the nuclear effective interactions as well. For the Skyrme effective
interactions shown in Fig. \ref{SymDen}, for instance, all of those
lie beyond $x=0$ and $x=-1$ in the sub-saturation region are not
consistent with the extracted value of $L$. Actually, we note that
only $4$ sets of Skyrme interactions, i.e., $\mathrm{SIV}$,
$\mathrm{SV}$, $\mathrm{G}_{\sigma }$, and $\mathrm{R}_{\sigma }$,
in the $21$ sets of Skyrme interactions considered here have nuclear
symmetry energies that are consistent with the extracted $L$ value.

\subsection{Constraining the neutron skin thickness of heavy
nuclei}

\begin{figure}[h]
\centering
\includegraphics[scale=1.2]{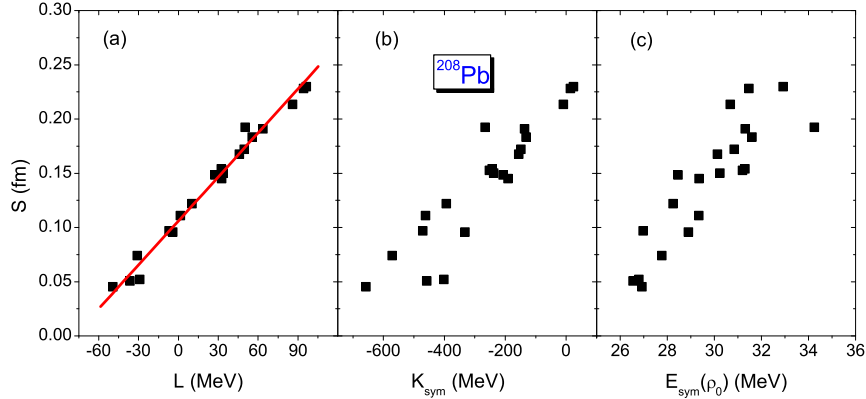}
\caption{(Color online) {Neutron skin thickness $S$ of $^{208}$Pb as
a function of (a) $L$, (b) $K_{\text{sym}}$, and (c)
$E_{\text{sym}}(\protect\rho _{0})$ for 21 sets of Skyrme
interaction parameters. }The line in panel (a) represents a linear
fit. Taken from Ref. \protect\cite{Che05b}.} \label{SPb208}
\end{figure}

The neutron skin thickness $S$ of a nucleus is defined as the
difference between the root-mean-square radii $\sqrt{\left\langle
r_{n}\right\rangle }$ of neutrons and $\sqrt{\left\langle
r_{p}\right\rangle }$ of protons, i.e.,
\begin{equation}
S=\sqrt{\left\langle r_{n}^{2}\right\rangle }-\sqrt{\left\langle
r_{p}^{2}\right\rangle }.
\end{equation}
It has been known that $S$ is sensitive to the density dependence
of the nuclear symmetry energy, particularly the slope parameter
$L$ at the normal nuclear matter density
\cite{Die03,Che05b,Bro00,Hor01a,Typ01,Fur02,Kar02}. The neutron
skin thickness of several nuclei have been evaluated using above
21 sets of Skyrme interaction parameteres. In Figs.
\ref{SPb208}(a), (b) and (c), we show, respectively, the
correlations between the neutron skin thickness of $^{208}$Pb with
$L$, $K_{\text{sym}}$, and $E_{\text{sym}}(\rho _{0})$. It is seen
from Fig. \ref{SPb208}(a) that there exists an approximate linear
correlation between $S$ and $L$. The correlations of $S$ with
$K_{\text{sym}}$ and $E_{\text{sym}}(\rho _{0})$ are less strong
and even exhibit some irregular behavior. The solid line in Fig.
\ref{SPb208}(a) is a linear fit to the correlation between $S$ and
$L$ and is given by the following expression:
\begin{equation}
S(^{\text{208}}\text{Pb)}=(0.1066\pm 0.0019)+(0.00133\pm 3.76\times
10^{-5})\times L,  \label{SLPb208a}
\end{equation}
or
\begin{equation}
L=(-78.5\pm 3.2)+(740.4\pm 20.9)\times S(^{\text{208}}\text{Pb)},
\label{SLPb208b}
\end{equation}
where the units of $L$ and $S$ are \textrm{MeV} and \textrm{fm},
respectively. Therefore, if the value for either
$S(^{\text{208}}$Pb) or $L$ is known, the value for the other can
be determined.

\begin{figure}[tbp]
\centering
\includegraphics[scale=1.2]{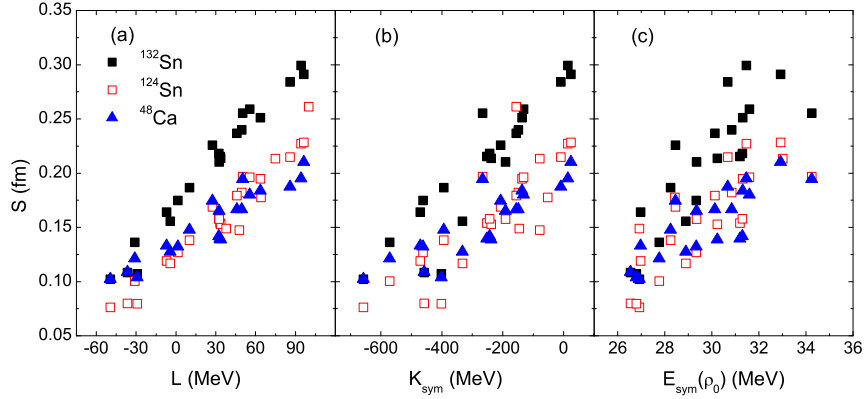}
\caption{(Color online) {Same as Fig. 2 but for nuclei $^{132}$Sn
(Solid squares), $^{124}$Sn (Open squares) and $^{48}$Ca
(Triangles).} Taken from Ref. \protect\cite{Che05b}.} \label{SSnCa}
\end{figure}

It is of interest to see if there are also correlations between the
neutron skin thickness of other neutron-rich nuclei and the nuclear
symmetry energy. Fig. \ref{SSnCa} shows the same correlations as in
Fig. \ref{SPb208} but for the neutron-rich nuclei $^{132}$Sn,
$^{124}$Sn, and $^{48}$Ca. For the heavy $^{132}$Sn and $^{124}$Sn,
there is a similar conclusion as for $^{208}$Pb, namely, $S$
exhibits an approximate linear correlation with $L$ but weaker
correlations with $K_{\text{sym}}$ and $E_{\text{sym}}(\rho _{0})$.
For the lighter $^{48}$Ca, on the other hand, all the correlations
become weaker than those of heavier nuclei. Therefore, the neutron
skin thickness of heavy nuclei is better correlated with the density
dependence of the nuclear symmetry energy. As in Eqs.
(\ref{SLPb208a}) and (\ref{SLPb208b}), a linear fit to the
correlation between $S$ and $L$ can also be obtained for $^{132}$Sn
and $^{124}$Sn, and the corresponding expressions are
\begin{equation}
S(^{\text{132}}\text{Sn)} =(0.1694\pm 0.0025)+(0.0014\pm 5.12\times
10^{-5})\times L,  \label{SLSn132a}
\end{equation}
\begin{equation}
L =(-117.1\pm 5.4)+(695.1\pm 25.3)\times S(^{\text{132}}\text{Sn)},
\label{SLSn132b}
\end{equation}
and
\begin{equation}
S(^{\text{124}}\text{Sn)} =(0.1255\pm 0.0020)+(0.0011\pm 4.05\times
10^{-5})\times L, \label{SLSn124a}
\end{equation}
\begin{equation}
L =(-110.1\pm 5.2)+(882.6\pm 32.3)\times S(^{\text{124}}\text{Sn)}.
\label{SLSn124b}
\end{equation}

\begin{table}[tbp]
\caption{{\protect\small {Linear correlation coefficients $C_{l}$
of $S$
with $L$, $K_{\text{sym}}$ and $E_{\text{sym}}(\protect\rho _{0})$ for $%
^{208}$Pb,\ $^{132}$Sn, $^{124}$Sn, and $^{48}$Ca from 21 sets of
Skyrme interaction parameters.} Taken from Ref.
\protect\cite{Che05b}.}} \label{Corr}
\centering
\begin{tabular}{ccccc}
\hline\hline $C_{l}$ $(\%)$ & \quad $^{208}$Pb\quad & $\quad
^{132}$Sn \quad & $^{124}$Sn & $^{48}$Ca \\ \hline
$S$-$L$ & $99.25$ & $98.76$ & $98.75$ & $93.66$ \\
$S$-$K_{\text{sym}}$ & $92.26$ & $92.06$ & $92.22$ & $86.99$ \\
$S$-$E_{\text{sym}}$ & $87.89$ & $85.74$ & $85.77$ & $81.01$ \\
\hline\hline
\end{tabular}%
\end{table}

To give a quantitative estimate of above discussed correlations,
we define the following linear correlation coefficient $C_{l}$:
\begin{equation}
C_{l}=\sqrt{1-q/t},
\end{equation}%
where%
\begin{eqnarray}
q =\underset{i=1}{\overset{n}{\sum }}[y_{i}-(A+Bx_{i})]^{2}, \qquad
t =\underset{i=1}{\overset{n}{\sum }}(y_{i}-\overline{y}), \qquad
\overline{y}=\underset{i=1}{\overset{n}{\sum }}y_{i}/n.
\end{eqnarray}
In the above, $A$ and $B$ are the linear regression coefficients, $(x_{i}$, $%
y_{i})$ are the sample points, and $n$ is the number of sample
points. The linear correlation coefficient $C_{l}$ measures the
degree of linear
correlation, and $C_{l}=1$ corresponds to an ideal linear correlation. Table %
\ref{Corr} gives the linear correlation coefficient $C_{l}$ for
the correlation of $S$ with $L$, $K_{\text{sym}}$ and
$E_{\text{sym}}(\rho _{0})$
for $^{208}$Pb, $^{132}$Sn, $^{124}$Sn, and $^{48}$Ca shown in Figs. \ref%
{SPb208} and \ref{SSnCa} for different Skyrme interactions. It is
seen that these correlations become weaker with decreasing nucleus
mass, and a strong linear correlation only exists between the $S$
and $L$ for the heavier nuclei $^{208}$Pb, $^{132}$Sn, and
$^{124}$Sn. Therefore, the neutron skin thickness of these nuclei
can be extracted once the slope parameter $L$ of the nuclear
symmetry energy at saturation density is known.

The extracted $L$ value from isospin diffusion data allows us to
determine from Eqs. (\ref{SLPb208a}), (\ref{SLSn132a}), and
(\ref{SLSn124a}), respectively, a neutron skin thickness of $0.22\pm
0.04$ fm for $^{208}$Pb, $0.29\pm 0.04$ fm for $^{132}$Sn, and
$0.22\pm 0.04$ fm for $^{124}$Sn. Experimentally, great efforts were
devoted to measure the thickness of the neutron skin in heavy nuclei
\cite{Sta94,Kra99,Trz01,Cla03}, and a recent review can be found in
Ref. \cite{Kra04}. The data for the neutron skin thickness of
$^{208}$Pb have large uncertainties, i.e., $0.1$-$0.28$ fm. Above
results for the neutron skin thickness of $^{208}$Pb are thus
consistent with present data but give a much stronger constraint. A
large uncertainty is also found experimentally in the neutron skin
thickness of $^{124}$Sn, i.e., its value varies from $0.1 $ fm to
$0.3$ fm depending on the experimental method. The proposed
experiment of parity-violating electron scattering from $^{208}$Pb
at the Jefferson Laboratory is expected to give another independent
and more accurate measurement of its neutron skin thickness (within
$0.05$ fm), thus providing improved constraints on the density
dependence of the nuclear symmetry energy \cite{Hor01b,Jef00}.

Recently, an accurately calibrated relativistic parametrization
based on the relativistic mean-field theory has been introduced to
study the neutron skin thickness of finite nuclei \cite{Tod05}.
This parametrization can describe simultaneously the ground state
properties of finite nuclei and their monopole and dipole
resonances. Using this parametrization, the authors
predicted a neutron skin thickness of $0.21$ fm in $^{208}$Pb, $0.27$ fm in $%
^{132}$Sn, and $0.19$ fm in $^{124}$Sn \cite{Tod05,Pie05}. These
predictions are in surprisingly good agreement with the results
constrained by the isospin diffusion data in heavy-ion collisions.

In addition, the neutron skin thickness of the nucleus $^{90}$Zr has
recently been determined to be $0.07\pm 0.04$ fm from the
model-independent spin-dipole sum rule value measured from the
charge-exchange spin-dipole excitations \cite{Yak06}. This value is
also reproduced by the symmetry energy with $L=88\pm 25$ MeV
extracted from the isospin diffusion data in heavy-ion collisions,
which predicts a neutron skin thickness of $0.088\pm 0.04$ fm for
$^{90}$Zr.

Most recently, there are new measurements of the neutron-sin
thickness of some heavy nuclei. Through analyzing the x-ray cascade
from antiprotonic atoms for $^{208}$Pb, Klos et al. \cite{Klo07}
recently deduced a value of $0.16 \pm 0.06$ fm for the neutron-sin
thickness of $^{208}$Pb. A neutron skin thickness of $0.18\pm 0.035$
fm for $^{208}$Pb and $0.24\pm 0.04$ fm for $^{132}$Sn, was derived
from pygmy dipole resonances by A. Klimkiewicz et al. \cite{Kli07}
and a value of $0.185\pm 0.017$ fm for $^{124}$Sn was obtained by S.
Terashima et al. \cite{Ter08} from the analysis of the proton
elastic scattering from Tin isotopes. These experimental results are
reasonably consistent with those constrained by the isospin
diffusion data discussed above.


\section{Astrophysical implications of the EOS of neutron-rich matter partially
constrained by terrestrial nuclear laboratory data}
\label{chapter_neutronstars}

\begin{figure}[htp]
\centering
\includegraphics[scale=0.8]{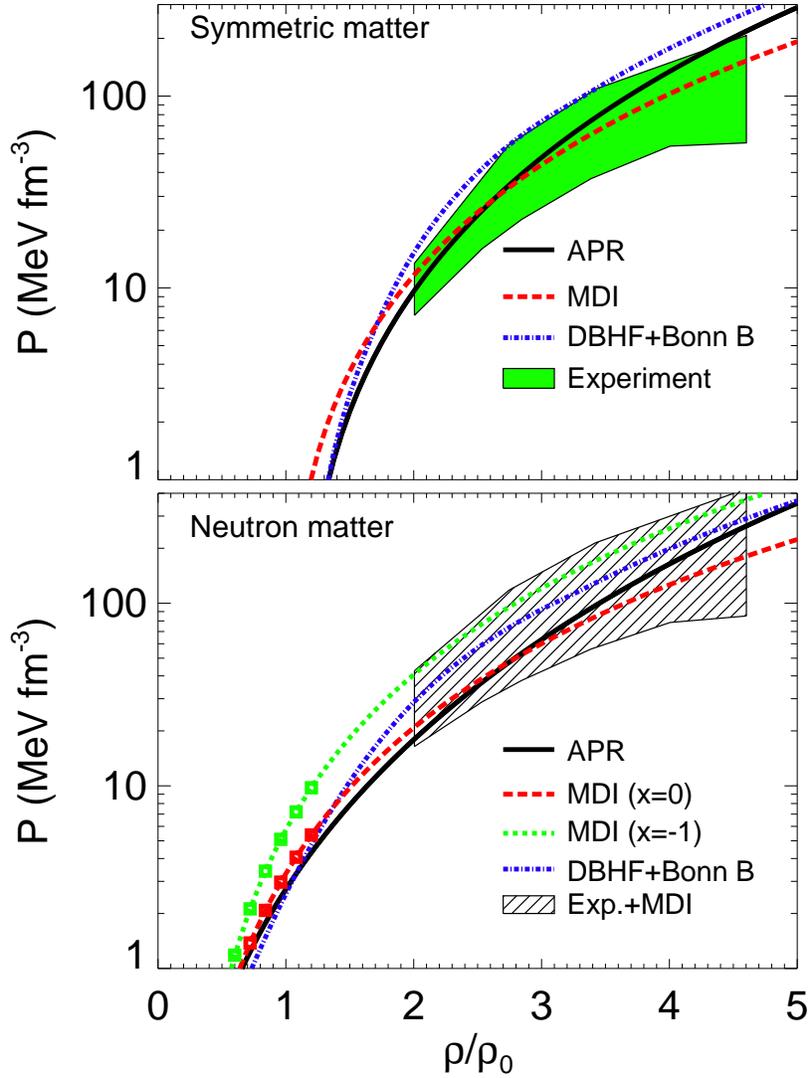}
\caption{Pressure as a function of density for symmetric (upper
panel ) and pure neutron (lower panel) matter. The green area in the
upper panel is the experimental constraint on symmetric matter
extracted by Danielewicz, Lacey and Lynch from analyzing the
collective flow in relativistic heavy-ion collisions. The
corresponding constraint on the pressure of pure neutron matter
obtained by combining the flow data and an extrapolation of the
symmetry energy functionals constrained below $1.2\rho_0$ by the
isospin diffusion data is the shaded black area in the lower panel.
Results taken from Ref.\ \protect\cite{Dan02a,Kra08b}.}
\label{Pressure-eos}
\end{figure}

Understanding the EOS of neutron-rich matter, especially the density
dependence of nuclear symmetry energy, is important not only for
nuclear physics, but also for many critical issues in astrophysics
\cite{Ste08}. Several recent reviews have dealt with extensively the
importance of the symmetry energy on various aspects of
astrophysics, see, e.g.,
Refs.~\cite{Lat04,Bom01,Lat00,Lat01,Hei00b,Pra01,Yak04}. For the
most recent reviews, we refer the reader to
Refs.~\cite{Ste05a,Dan06,Lat07}. All these articles are essentially
concerned with studies of astrophysical questions based on various
predictions on the EOS of neutron-rich matter using different
many-body theories. The present review differs significantly in that
we focus on the understanding of some global properties of neutron
stars using the EOS of neutron-rich matter that has been constrained
in certain density ranges by experimental data from terrestrial
nuclear laboratories.

For many astrophysical studies, it is more convenient to express the
EOS in terms of the pressure as a function of density and isospin
asymmetry. Shown in Fig.\ref{Pressure-eos} are the pressures for two
extreme cases: symmetric (upper panel) and pure neutron matter
(lower panel). The green area in the density range of $2-4.6\rho_0$
is the experimental constraint on the pressure $P_{0}$ of symmetric
nuclear matter extracted by Danielewicz, Lacey and Lynch from
analyzing the collective flow data from relativistic heavy-ion
collisions \cite{Dan02a}. It is seen that both the MDI and the APR
interaction are consistent with this constraint. For pure neutron
matter, its pressure is $P_{\rm PNM}=P_{0}+\rho^2dE_{\rm sym}/d\rho$
and depends on the density dependence of nuclear symmetry energy.
Since the constraints on the symmetry energy from terrestrial
laboratory experiments are only available for densities less than
about $1.2\rho_0$ as indicated by the green and red squares in the
lower panel, which is in contrast to the constraint on the symmetric
EOS that is only available at much higher densities, the most
reliable estimate of the EOS of neutron-rich matter can thus be
obtained by extrapolating the underlying model EOS for symmetric
matter and the symmetry energy in their respective density ranges to
all densities. Shown by the shaded black area in the lower panel is
the resulting best estimate of the pressure of high density pure
neutron matter based on the predictions from the MDI interaction
with x=0 and x=-1 as the lower and upper bounds on the symmetry
energy and the flow-constrained symmetric EOS. As one expects and
consistent with the estimate in Ref.\cite{Dan02a}, the estimated
error bars of the high density pure neutron matter EOS is much wider
than the uncertainty range of the symmetric EOS. For the four
interactions indicated in the figure, their predicted EOS's cannot
be distinguished by the estimated constraint on the high density
pure neutron matter. In the following, the astrophysical
consequences of this partially constrained EOS of neutron-rich
matter in the sense we discussed above are reviewed.

To give the reader a coherent and broad picture, we start by briefly
recalling those nuclear astrophysical phenomena which are strongly
affected by EOS of neutron-rich matter, especially the density
dependence of the nuclear symmetry energy. We will then give
specific examples. The mechanism for supernova explosions and the
properties of neutron stars have been the subjects of much interest
and extensive research. Various studies have indicated that the
symmetry energy affects mainly the chemical composition of neutron
stars \cite{Mut87,Pra87,Lat91,Tho94,Tok95,Mah97}. Other properties,
such as the cooling mechanisms of proto-neutron stars, the
possibility of kaon condensation in the cores of neutron stars,
lepton profiles and the neutrino flux, which all depend on the
chemical composition of stars, are therefore also affected. For
example, the prompt shock invoked to understand the explosion
mechanism of a type II supernova requires a relatively soft EOS
\cite{Bar85}. This can be understood in terms of the dependence of
the nuclear incompressibility on isospin as follows. In the model
for prompt explosion \cite{Kah89}, the electron-capture reaction
drives the star in the latest stage of collapse to an equilibrium
state where the proton concentration is about $1/3 $, which,
according to microscopic many-body calculations, reduces the nuclear
matter incompressibility by about $30\%$ compared to that for
symmetric nuclear matter. The exact magnitude of proton
concentration at $\beta $ equilibrium in a neutron star depends,
however, on the symmetry energy. Since the isobaric
incompressibility of a neutron-rich nuclear matter decreases with
its isospin asymmetry $\delta$ according to $K(\delta)=K_0+K_{\rm
asy}\delta^2$, a negative $K_{\rm asy}\approx -500\pm 50$ MeV as
discussed earlier thus leads to a smaller nuclear matter
incompressibility. The presence of protons in neutron stars affects
not only the stiffness of its EOS, including whether a kaon
condensation through the process $e^{-}\rightarrow K^{-}\nu _{e}$
can be formed \cite{Kap86,Sum94}, but also its cooling mechanisms
\cite{Lat91}. If the proton concentration is larger than a critical
value of about $15\%$, the direct URCA process $(n\rightarrow
p+e^{-}+\bar{\nu}_{e},~p+e^{-}\rightarrow n+\nu _{e})$ becomes
possible and would then enhance the emission of neutrinos, making it
a more important process in the cooling of a neutron star
\cite{Lat91}.

Besides those properties related to the proton fraction, there are
also properties of neutron stars that are directly related to the
magnitude and/or the density slope of the symmetry energy. Among
these properties, the most known example is probably the radius of a
neutron star. While many neutron star properties depend on both the
isospin symmetric and asymmetric parts of the equation of state, the
radius is primarily determined by the slope of the symmetry energy,
$E_{\mathrm{sym}}^{\prime}(\rho)$, in the density range of 1 to
2$\rho_0$~\cite{Lat04,Lat01,Pra01,Lat07}. In addition, the
transition density and pressure from the liquid core to the solid
crust and the fractional moment of inertia of the neutron star crust
are also directly related to the symmetry energy
\cite{Ste05a,Hor01a}.

\subsection{The symmetry energy and the proton fraction in neutron
stars at $\beta$-equilibrium}

In the most simple picture, a neutron star is composed of neutrons,
protons and electrons with a proton fraction of
\begin{eqnarray}
  x = {\textstyle\frac{1}{2}}(1-\delta).
\end{eqnarray}
The condition for $\beta$-equilibrium in terms of the chemical
potentials of electrons $(\mu_e)$, neutrons $(\mu_n)$ and protons
$(\mu_p)$ is
\begin{eqnarray}\label{xfraction}
\mu_e=\mu_n-\mu_p=-\frac{\partial e(\rho,\delta)}{\partial x}
=4e_{\rm sym}(\rho)(1-2x).
\end{eqnarray}
The last equality in the above equation is obtained by using the
parabolic approximation to the symmetry energy. For relativistic
degenerate electrons of density $\rho_e=\rho_p=x\rho$, charge
neutrality requires
\begin{eqnarray}
\mu_e=(m_e^2+p_{F_e}^2)^{1/2}\approx \hbar c(3\pi^2\rho x)^{1/3},
\end{eqnarray}
which together with Eq.~(\ref{xfraction}) determine an equilibrium
proton fraction $x$ given by
\begin{eqnarray}\label{fraction}
\hbar c(3\pi^2\rho x)^{1/3}=4e_{\rm sym}(\rho)(1-2x).
\end{eqnarray}
The equilibrium proton fraction $x$ is therefore determined solely
by the nuclear symmetry energy, $e_{\rm sym}(\rho)$. At high
densities such that $\mu_e\geq m_{\mu}$, where $m_\mu$ is the muon
mass, both electrons and muons are present at $\beta$-equilibrium
and should be included in determining the value of $x$. Since the
inclusion of muons mainly alters the value of the equilibrium
proton fraction $x$ but not its density dependence, the difference
in $x$ predicted by using different symmetry energies is about the
same with or without including muons~\cite{Wir88a,Lat91}.

\begin{figure}[htp]
\centering
\includegraphics[scale=0.5]{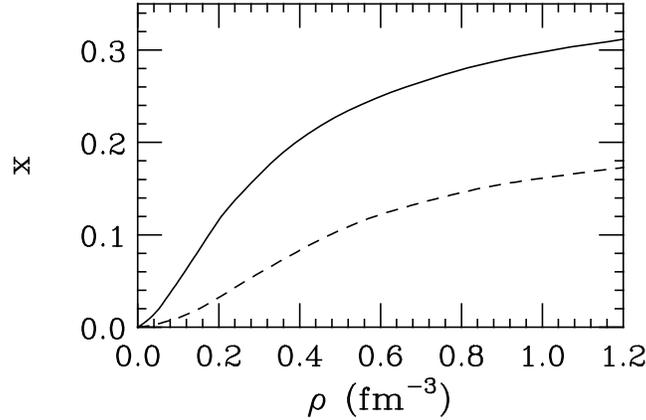}
\caption{Equilibrium fraction of protons as a function of density
obtained from the relativistic mean field theory with
$g_{\rho}=5.507$ (solid curve) or $g_{\rho}=2.78$ (dashed curve).
Results taken from Ref.\ \protect\cite{Sum92}.} \label{ProFracRMF}
\end{figure}

The dependence of the equilibrium proton fraction on the
underlying nuclear interaction was illustrated nicely in the
earlier work of Sumiyoshi {\it et al.} using the relativistic
mean-field ({\sc rmf}) theory \cite{Sum92,Tok95}. The symmetry
energy is found to vary almost linearly with density, and its
strength is related to the $\rho$ meson-nucleon coupling constant
$g_{\rho}$ via
\begin{eqnarray}\label{e0sym}
e_{\rm sym}(\rho_0)=\frac{k_F^2}{6\sqrt{M^{*2}+k_F^2}}
+g_{\rho}^2\rho_0/2m_{\rho}^2,
\end{eqnarray}
where $M^*$ is the nucleon effective mass, $m_{\rho}$ is the
$\rho$ meson mass, and $k_F$ is the Fermi momentum. The first and
second terms are the kinetic and potential contributions to the
symmetry energy, respectively. From the above expression and Eq.\
(\ref{fraction}), one obtains the $g_{\rho}$ dependence of the
proton fraction shown in Fig.\ \ref{ProFracRMF}.  As $g_{\rho}$
increases from 2.78 to 5.507, the proton fraction is seen to
increase by about a factor of two. More detailed discussions about
the density dependence of the symmetry energy within different
versions of the RMF models can be found in
Chapter~\ref{chapter_rmf}.

\begin{figure}[htp]
\centering
\includegraphics[scale=0.4]{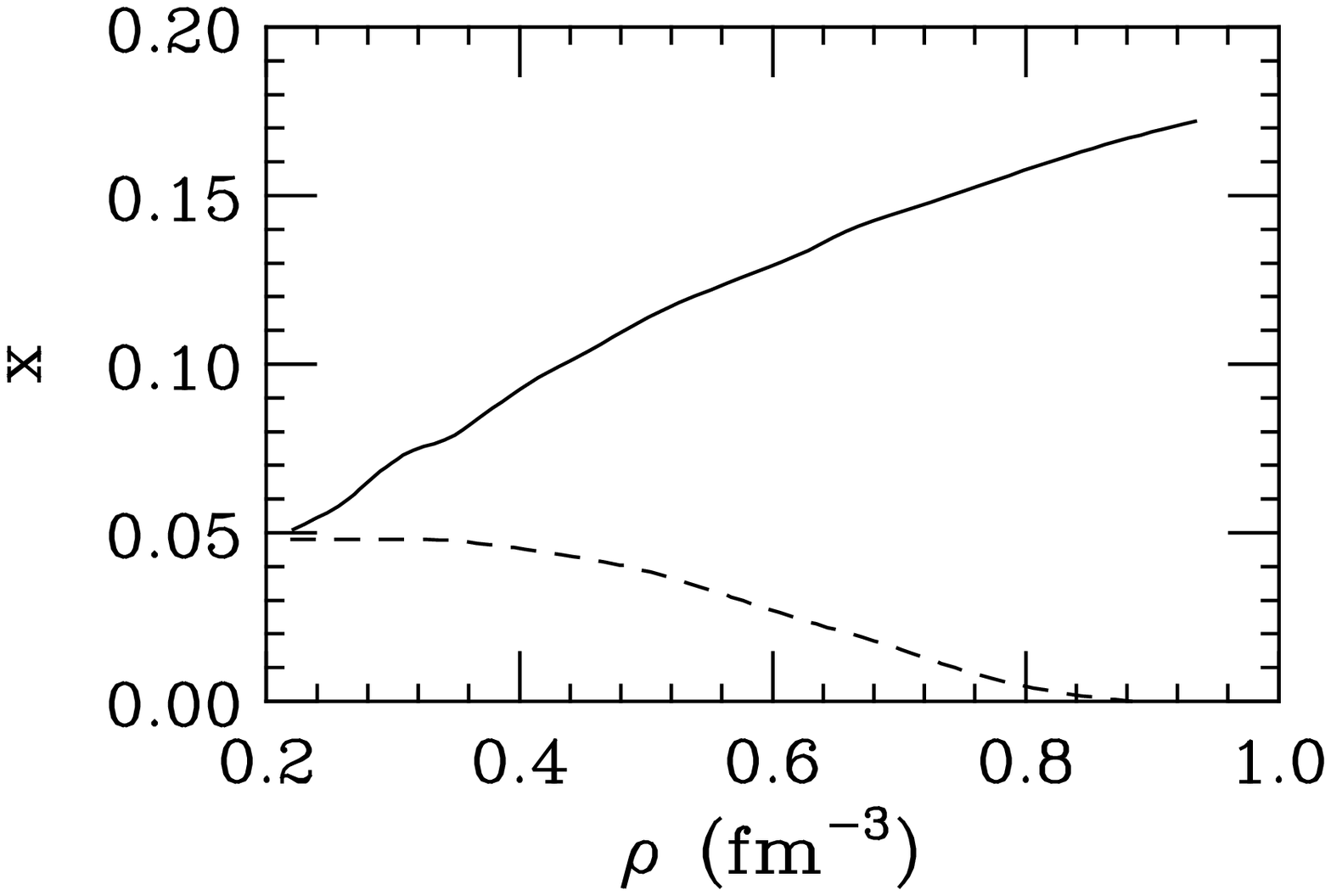}
\includegraphics[scale=0.4]{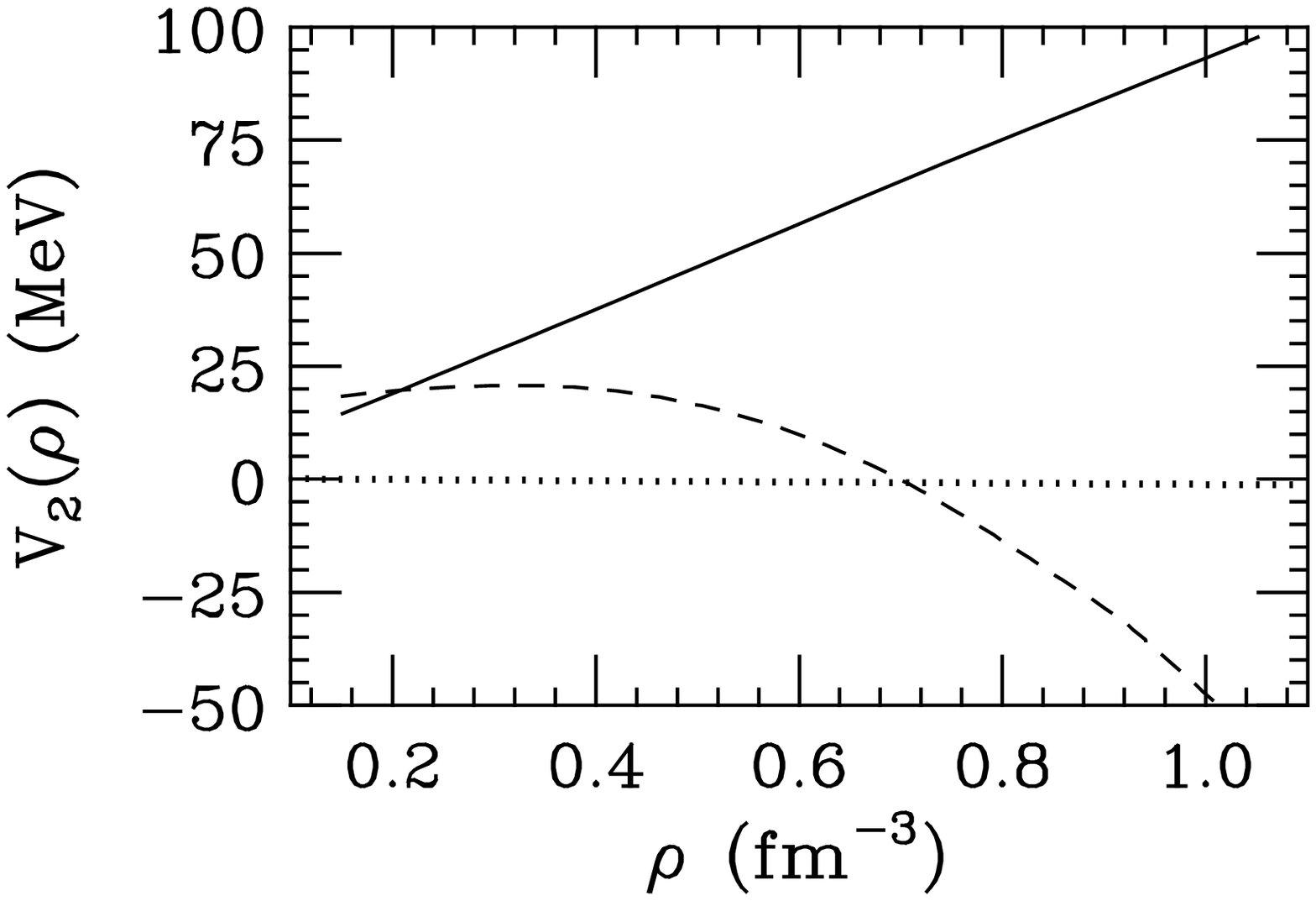}
\caption{Left window: Equilibrium fraction of protons in neutron
stars predicted by the relativistic mean-field ({\sc rmf}, solid
line) and variational many-body ({\sc vmb}, dashed line) theories.
Right window: Contribution from nuclear interactions to the
symmetry energy in the relativistic mean field ({\sc rmf}) theory
(solid line) and the variational many-body ({\sc vmb}) theory
(dashed line). Results taken from Ref.\ \protect\cite{Kut94}.}
\label{v2rmf}
\end{figure}

As two extreme examples, the {\sc rmf} and the {\sc vmb} theory
with the {\sc av14+tni} interaction differ most in their
predictions on the symmetry energy at high densities. Their
predicted equilibrium proton fractions at high densities are
therefore also very different as shown in the left window of Fig.\
\ref{v2rmf}. Their dramatically different behaviors in the
symmetry energy and proton fraction at high densities can be
traced back to the very different contributions from the potential
energy to the symmetry energy in these two models, as the
contribution from the kinetic energy to the symmetry energy is
about the same in all model calculations. Shown in the right
window of Fig.\ \ref{v2rmf} are the potential contributions
$V_2(\rho)$ to the symmetry energy from the {\sc rmf} and the {\sc
vmb} theory. This contribution is always repulsive and increases
linearly with density in the {\sc rmf} theory, while in the {\sc
vmb} theory it changes from repulsion to attraction as the density
increases. The variation of $V_2(\rho)$ with density in the {\sc
vmb} theory was first explained by Pandharipande in terms of the
behavior of nuclear interactions in dense nuclear matter
\cite{Pan72}. At high densities the short-range repulsion
dominates and is greater for a nucleon pair in an isospin singlet
($T=0$) than in an isospin triplet ($T=1$) state. A pure neutron
matter is therefore more stable. At moderate densities the strong
attractive isospin-singlet tensor potential and correlation keep
the isospin-singlet pairs more bound, and a symmetric nuclear
matter is thus more stable than a pure neutron matter. These
features do not exist in the {\sc rmf} theory where the symmetry
energy is due to the rho meson exchange, which leads to a
repulsive $V_2(\rho)$ at all densities.

It is further seen in the left window of Fig.~\ref{v2rmf} that the
{\sc rmf} theory predicts a linear increase of the proton fraction
with increasing density, while in the {\sc vmb} theory the proton
fraction in neutron stars gradually decreases as the density
increases. The disappearance of protons in neutron stars is a common
feature of the {\sc vmb} theory, although the critical density at
which this occurs depends on the interaction used in the
calculation. As we have mentioned earlier in section \ref{hdsymb}, a
decreasing symmetry energy above certain densities also appears in
many other models, such as the HF approach with density-dependent
M3Y (DDM3Y) interaction~\cite{Kho96,Bas07}, the DBHF using various
Bonn potentials \cite{Kra06} and the HF using many Skyrme and/or
Gogny effective interactions \cite{Ono03,Che05b,Das03,Sto03}. When
the symmetry energy becomes negative, the isospin separation
instability would occur and this could lead to the formation of
polarons, which are localized protons surrounded by neutron bubbles,
in neutron stars \cite{Kut94}. However, most of these models predict
that the density for the symmetry energy to become negative is above
the critical densities for the formation of the QGP and the hyperon
matter. Therefore,  a transition of the nuclear matter to these
exotic phases should happen before the symmetry energy becomes
negative. The effects of symmetry energy on the transition from the
hadron to the hadron-quark mixed phase were studied in
Ref.~\cite{Kut00} for neutron stars and Ref.~\cite{Tor06} for
heavy-ion reactions. In particular, using symmetry energies
predicted by the RMF and VMB models and a MIT bag model for the
quark phase, Kutschera and Niemiee found that the role of the
nuclear symmetry energy changes with the value of the bag constant
$B$. Although for lower values of $B$ the properties of the mixed
phase do not depend strongly on the symmetry energy, this changes
for larger $B$. In the latter case, the critical pressure for the
first quark droplets to form in the nucleon medium is strongly
dependent on the nuclear symmetry energy, while the pressure at
which last nucleons disappear is independent of it. Also, the
allowed range of surface tension for the mixed phase that is
energetically favorable depends strongly on the nuclear symmetry
energy \cite{Kut00}. Using a similar approach and a bag constant of
$B^{1/4}=150$ Mev, Di Toro {\it et al.} \cite{Tor06} found that the
transition density between the hadron and hadron-quark mixed phases
would decrease quickly with increasing neutron-excess, especially in
the proton-fraction region of $0.3\geq x \geq 0.5$. Moreover, for
these values of proton fraction the transition density is very
sensitive to the high density behavior of the symmetry energy.
Precursor effects due to the decreasing transition density in the
neutron-rich matter formed in heavy-ion reactions at several
GeV/nucleon beam energies were proposed although none of the
suggested signatures is unique. Given the fact that no convincing
evidence of a phase transition from the hadron to the quark phase
has been found in heavy-ion reactions in this energy range despite
of the great efforts at the AGS and other facilities for many years,
it is very challenging to extract information from these experiments
about the isospin dependence of the transition density.

\begin{figure}[htp]
\centering
\includegraphics[scale=0.5]{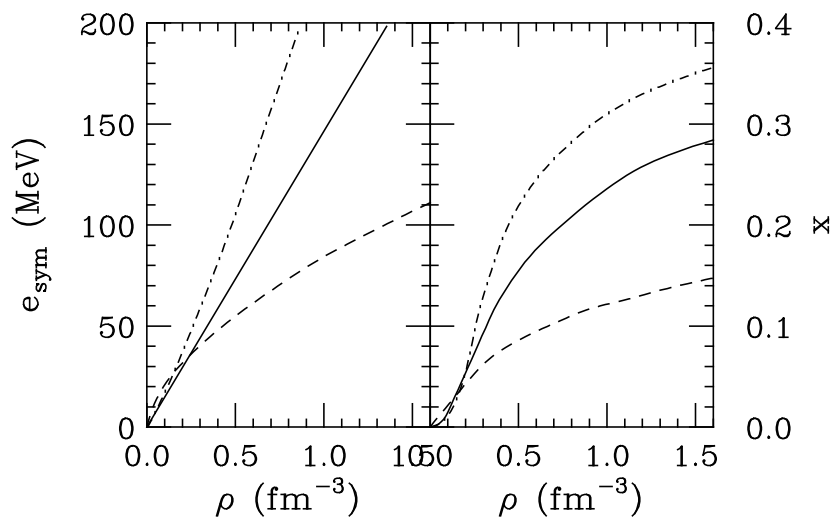}
\caption{Symmetry energy as a function of density (left window) and
corresponding equilibrium fraction of protons in neutron stars
(right window) for nuclear symmetry energies $F_1(u)$ (dash-dotted
lines), $F_2(u)$ (solid lines), and $F_3(u)$ (dashed lines). Results
taken from Ref. \protect\cite{Lat91}.}\label{ProFracLat}
\end{figure}

As illustrated by the above comparison between predictions of the
{\sc rmf} and {\sc vmb} theories, the uncertainty in the density
dependence of the nuclear symmetry energy is mainly due to our
poor understanding of the potential contribution to the symmetry
energy. To clearly and explicitly examine the effects of nuclear
interactions on the proton fraction, Prakash and Lattimer
parameterized the potential part of the symmetry energy using
$F_1(u)=\frac{2u^2}{1+u}$, $F_2(u)=u$, and $ F_3(u)=u^{1/2}$,
where $u\equiv \rho/\rho_0$ is the reduced baryon density, as
introduced in Chapter~\ref{minsp}. These forms of the symmetry
energy resemble closely three typical results from microscopic
many-body calculations. Fig.\ \ref{ProFracLat} shows the
equilibrium proton fractions (right window) corresponding to the
above three different symmetry energies using the $F_1(u), F_2(u)$
and $F_3(u)$ (left window). In all three cases the proton fraction
$x$ increases with density, and the differences among them are
appreciable, reflecting the effect of the potential contribution
to the symmetry energy as a function of density.

The proton fraction has significant effects on the cooling of
proto-neutron stars. In the so-called standard model for neutron
stars, their cooling is mainly due to the modified {\sc urca}
process
\begin{eqnarray}\label{murca}
(n,p)+p+e^-\rightarrow (n,p)+n+\nu_e,~~ (n,p)+n\rightarrow
(n,p)+p+e^-+\bar{\nu_e}.
\end{eqnarray}
The direct  {\sc urca} process
\begin{eqnarray}\label{durca}
n\rightarrow p+e^-+\bar{\nu_e},~~ p+e^-\rightarrow n+\nu_e
\end{eqnarray}
is usually forbidden by energy-momentum conservation. However,
Lattimer {\it et al.} \cite{Lat91} have shown that if the proton
fraction is higher than a critical value of about 0.14
\cite{Ste05a,Lat91}, the direct  {\sc urca} process can also
occur. This would then enhance the emission of neutrinos, thus
increasing significantly the neutron star cooling rate. As shown
in Fig.\ \ref{ProFracLat}, whether the proton fraction can exceeds
the critical value and at what density this happens are entirely
determined by the symmetry energy of the nuclear matter.

Another important effect of the symmetry energy on the properties
of neutron stars is the possible formation of kaon condensation in
their dense cores. This happens when the chemical potential of
electrons exceeds the kaon mass, so the process $e^{-}\rightarrow
K^{-}\nu_e$ can occur. It was shown in
Refs.~\cite{Lee96,Kub99,Kub03,Odr07} that the critical density for
forming the kaon condensation depends sensitively on the density
dependence of the symmetry energy. Using the RMF and the VMB
predictions shown in Fig.~\ref{v2rmf}, Kubis, Kutschera and
Odrzywolek \cite{Kub99,Kub03,Odr07} found that the high density
behavior of the nuclear symmetry energy plays an essential role in
determining the composition of the kaon-condensed neutron star
matter, and this in turn affects its cooling properties. In
particular, the symmetry energy which decreases at higher
densities (e.g., VMB) makes the kaon-condensed neutron star matter
fully protonized. This effect inhibits strongly direct URCA
processes and results in a slower cooling of neutron stars as only
kaon-induced URCA cycles are present. In contrast, for the
increasing symmetry energy (e.g., RMF) direct URCA processes are
allowed in almost the whole density range where the kaon
condensation exists \cite{Kub99,Kub03,Odr07}.

\subsection{Constraining the proton fraction in neutron stars}

\begin{figure}[tbh]
\centering
\includegraphics[scale=0.40,angle=0]{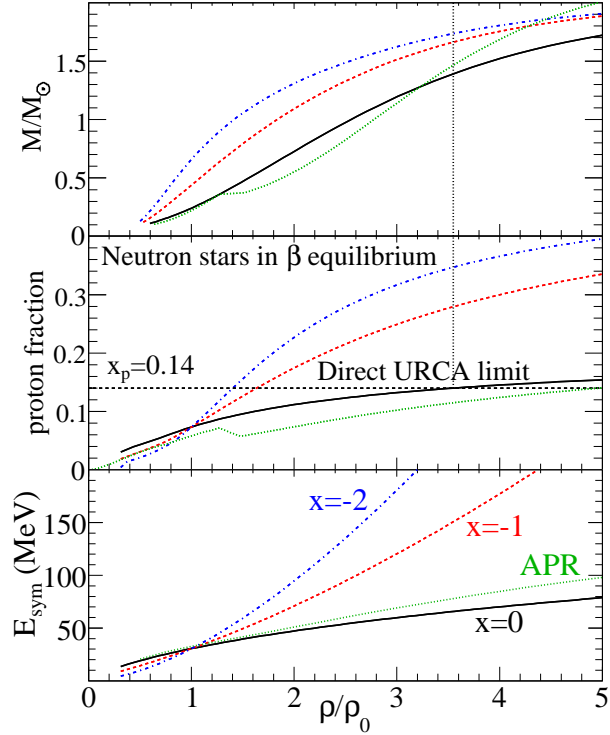}
\caption{{\protect\small Neutron star mass (upper panel), proton
fraction for beta-equilibrated matter (middle panel), and symmetry
energy (lower panel) as functions of density for the EOS with
$x=0$, $-1,$ and $-2$. The dotted lines give the corresponding
results for the APR EOS. Taken from Ref.~\cite{LiBA06a}.}}
\label{MnsProFracEsymRho}
\end{figure}

The structure of a non-rotating neutron star with an isotropic mass
distribution can be determined by solving the well-known
Tolman-Oppenheimer-Volkov (TOV) equation
\begin{eqnarray}
\frac{dP}{dr}=-\frac{[\rho(r)+P(r)][M(r)+4\pi,
r^3P(r)]}{r^2-2rM(r)},
\end{eqnarray}
where $P(r)$ is the pressure and $M(r)$ is the gravitational mass
inside a radius $r$.

Solving the TOV equation with the symmetry energy constrained by
the heavy-ion reaction data allows one to limit the critical
central density for the direct URCA process to happen. This study
was first carried out by Steiner and Li in Refs.
\cite{Ste05b,LiBA06a}. Here we recall their main results. Shown in
the lower panel of Fig.~\ref{MnsProFracEsymRho} are the symmetry
energies from the MDI interaction with the parameter $x$ equals to
$0$, $-1$, and $-2$, as already shown in in Fig. \ref{MDIsymE}.
The symbol $x$ used here should be distinguished from the proton
fraction used in the previous section. With $x=0$ the symmetry
energy agrees very well with the prediction from Akmal {\it et.
al.} (APR) \cite{Akm98} up to about $5\rho_0$. Around $\rho_0$,
the EOS from $x=0$ can be well approximated by
$E_{\mathrm{sym}}^{x=0}(\rho)\approx 32(\rho/\rho_0)^{0.7}$. With
$x=-1$, the parametrization $E_{\mathrm{sym}}^{x=-1}(\rho)\approx
32(\rho/\rho_0)^{1.1}$ is closer to the predictions of typical
relativistic mean-field models \cite{Ste05a}.

The middle panel shows the proton fraction $x_p$, as a function of
density, while the top panel gives the mass of a neutron star as a
function of its central density. It is seen that the proton
fraction at a given density varies appreciably among the three
symmetry energies, as it is sensitive to the slope of the symmetry
energy \cite{Lat04,Lat00,Lat01,Pra01}. Since the direct URCA
process occurs only for $x_p$ greater than 0.14 because of energy
and momentum conservation \cite{Ste05a,Lat91}, the condition for
direct URCA for the $x=-1$ and $x=-2$ EOSs is thus fulfilled for
nearly all neutron stars above 1 \Msun. For the $x=0$ EOS, the
minimum density for direct URCA is indicated by the vertical
dotted line in the middle panel, and the corresponding minimum
neutron star mass is indicated by the horizontal dotted line in
the top panel. For the $x=0$ EOS, neutron stars with masses above
1.39 \Msun thus have a central density above the threshold for the
direct URCA process. This constraint nearly matches the constraint
for the direct URCA process of 1.30 \Msun obtained in Ref.
\cite{Tod05}. It is, however, markedly different from the result
of APR, which gives a larger threshold for the direct URCA process
(even though the symmetry energy is very similar to the $x=0$
EOS).

\subsection{Constraining the pressure and radii of static neutron stars}

\begin{figure}[th]
\centering
\includegraphics[width=7cm,height=8.5cm,angle=-90]{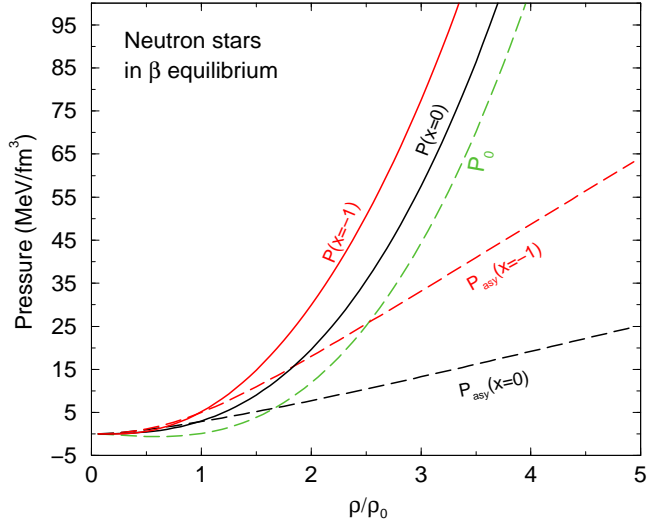}
\caption{{\protect\small The symmetric ($P_0$), asymmetric ($P_{\rm
asy}$), and total pressure in neutron stars at $\beta$-equilibrium
using the MDI interaction with $x=0$ and $x=-1$. Taken from Ref.
\cite{Li06}.}} \label{nspre}
\end{figure}

The radii of neutron stars are primarily determined by the isospin
asymmetric pressure that is proportional to the slope of the
symmetry energy $E_{\rm sym}^{\prime}(\rho)$~\cite{Lat01}.  For the
simplest case of a neutron-proton-electron ($npe$) matter in neutron
stars at $\beta$ equilibrium, the pressure is given by
\begin{eqnarray}\label{pre}
P(\rho,\delta)&=&P_0(\rho)+P_{\rm asy}(\rho,\delta)=
\rho^2\left(\frac{\partial E}{\partial \rho}
\right)_{\delta}+\frac{1}{4}\rho_e\mu_e\nonumber\\
&=&\rho^2\left[E'(\rho,\delta=0)+E'_{\rm sym}(\rho)\delta^2\right]
+\frac{1}{2}\delta(1-\delta)\rho E_{\rm sym}(\rho),
\end{eqnarray}
where $\rho_e=\frac{1}{2}(1-\delta)\rho$ and
$\mu_e=\mu_n-\mu_p=4\delta E_{\rm sym}(\rho)$ are, respectively,
the density and chemical potential of electrons. The value of the
isospin asymmetry $\delta$ at equilibrium is determined by the
chemical equilibrium and charge neutrality conditions, i.e.,
$\delta=1-2x_p$ with
\begin{eqnarray}
x_p\approx 0.048 \left[E_{\rm sym}(\rho)/E_{\rm sym}(\rho_0)\right]^3
(\rho/\rho_0)(1-2x_p)^3.
\end{eqnarray}
For pure neutron matter at the nuclear saturation density
$\rho_0$, the pressure $P$ in Eq. (\ref{pre}) reduces to
\begin{eqnarray}
P_{PNM}(\rho_0)=\rho_0^2E'_{\rm sym}(\rho_0)=\frac{1}{3}\rho_0L,
\end{eqnarray}
where $L$ is the slope of the symmetry energy at normal density
given in Eq.~(\ref{lsyme}). Because of the large isospin asymmetry
or value of $\delta$ in neutron stars, the electron degenerate
pressure is small. Also, the isospin symmetric contribution to the
pressure is also very small around normal nuclear matter density
as $E'(\rho_0,\delta=0)=0$. The latter can be seen from Fig.
\ref{nspre}, which gives the isospin symmetric ($P_0$) and
asymmetric ($P_{\rm asy}$) as well as the total pressure in
neutron stars at $\beta$-equilibrium calculated from
Eq.~(\ref{pre}) using the MDI interaction with $x=0$ and $x=-1$.
Up to about $2.5\rho_0$ for $x=-1$ and about $1.5\rho_0$ for $x=0$
the total pressure is indeed dominated by the isospin asymmetric
contribution. Because neutron star radii are determined by the
pressure at moderate densities where the proton content of matter
is small, they are thus very sensitive to the slope of the
symmetry energy near and just above $\rho_0$. In particular, a
stiffer symmetry energy is expected to lead to a larger neutron
star radius.

\begin{figure}[h]
\centerline{\includegraphics[scale=0.4,angle=0]{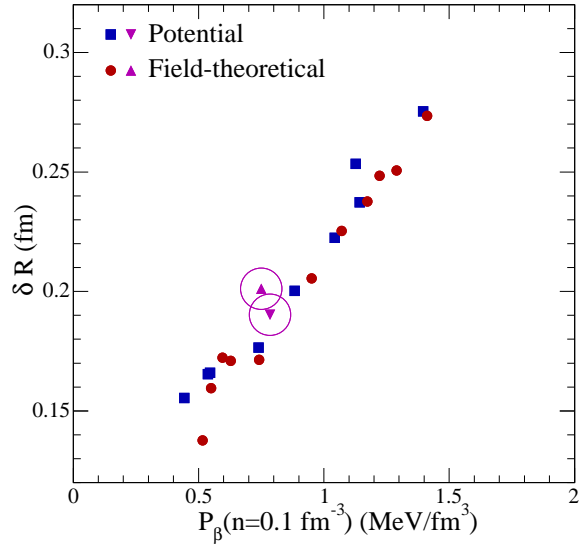}}
\caption{The neutron skin thickness $\delta R$ of finite nuclei
versus the pressure of $\beta$-equilibrated matter at a density of
0.1 fm$^{-3}$ for a variety of potential and field-theoretical
models. Taken from Ref.~\cite{Ste05a}.} \label{tplot}
\end{figure}

An empirical relation between the radii $R$ and the pressure $P$ has
been found by Lattimer and Prakash \cite{Ste05a,Lat01}, i.e.,
\begin{eqnarray}
R \simeq C(\rho,M)~[P(\rho)]^{0.23-0.26}\,,
\label{correl}
\end{eqnarray}
where $P(\rho)$ is the total pressure, including the leptonic
contributions, evaluated at a density $\rho$ in the range 1 to
2$\rho_0$, and $C(\rho,M)$ is a number that depends on the density
$\rho$ at which the pressure was evaluated and the stellar mass
$M$. It is then crucial to know the pressure in neutron stars
using information about the symmetry energy from terrestrial
nuclear laboratory experiments. One of the most interesting ideas
is to use the sizes of neutron skins in heavy nuclei
\cite{Bro00,Hor01a,Typ01,Fur02}. Since extra neutrons in
neutron-rich nuclei are pushed further out of an isospin symmetric
core of nearly normal density by the isospin asymmetry pressure,
the thickness of neutron-skins in these nuclei thus increases with
increasing slope of the symmetry energy. This idea is well
illustrated in Fig.\ \ref{tplot}, where the correlation between
the neutron-skin thickness in $^{208}$Pb and the pressure of pure
neutron matter at a density of $\rho=0.1$ fm$^{-3}$ is shown. As
it was pointed out by Steiner {\it et al.}, to the extent that
this correlation can be applied, a measurement of the neutron-skin
thickness $\delta R$ will help to establish an empirical
calibration point for the pressure of neutron star matter at
subnuclear densities \cite{Ste05a}.

\begin{figure}[h]
\centering
\includegraphics[scale=0.4,angle=-90]{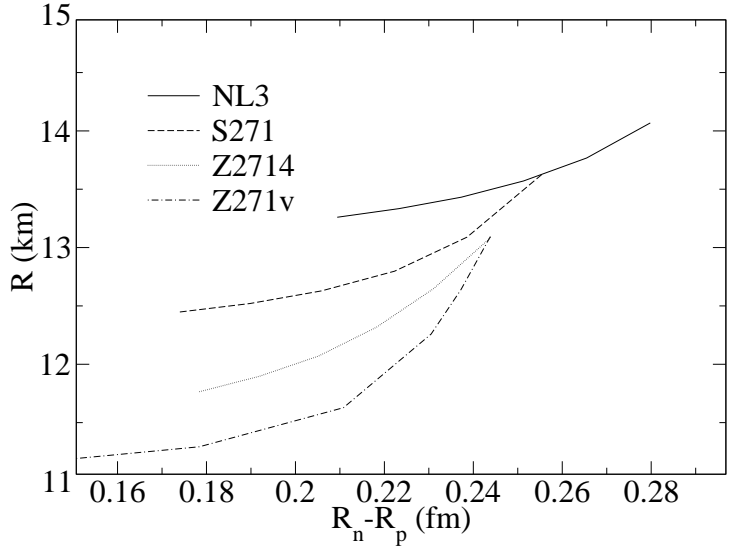}
\caption{Radius of a 1.4 solar mass neutron star versus
neutron-minus-proton radius in $^{208}$Pb for four parameter sets
used in the RMF model \cite{Hor01a}.} \label{horowitz1}
\end{figure}

Using the EOS of neutron matter constrained by the thickness of
neutron-skins in heavy nuclei, Horowitz and Piekarewicz further
established a correlation between the radius of a canonical
neutron star of $1.4$ solar mass and the size of neutron-skin in
$^{208}$Pb \cite{Hor01a}. This is shown in Fig.\ \ref{horowitz1}
for four parameter sets used in the RMF model \cite{Hor01a}. For a
given parameter set, the neutron star radius $R$ increases
monotonically with the neutron-skin thickness $R_{\rm
n}\!-\!R_{\rm p}$. Although the radius $R$ is not uniquely
constrained by a measurement of the neutron-skin thickness because
$R_{\rm n}\!-\!R_{\rm p}$ only depends on the equation of state at
normal and lower densities while $R$ is also sensitive to the
equation of state at higher densities, one can combine separate
measurements of $R_{\rm n}\!-\!R_{\rm p}$ and $R$ to obtain
information about the equation of state at both low and high
densities. For example, if $R_{\rm n}\!-\!R_{\rm p}$ is relatively
large while $R$ is small this could suggest a phase transition in
the equation of state. A large $R_{\rm n}\!-\!R_{\rm p}$ implies
that the low density equation of state is stiff while a small $R$
suggests that the high density equation of state is soft. The
transition from a stiff to a soft equation of state could be
accompanied by a phase transition \cite{Hor01a}.

\begin{figure}[th]
\centering
\includegraphics[width=7.2cm,height=7cm,angle=-90]{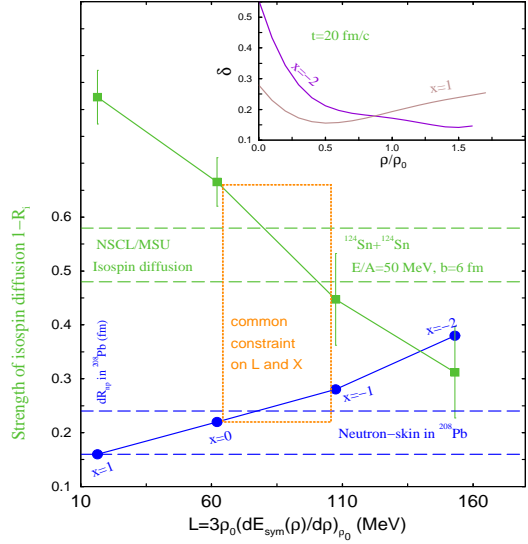}
\caption{{\protect\small The strength of isospin diffusion in
$^{124}$Sn$+^{112}$Sn reactions and the size of neutron skin in
$^{208}$Pb as functions of the slope of the symmetry energy. The
inset is the correlation between the average isospin asymmetry and
density of matters at the instant of 20 fm/c in the considered
reaction~\cite{LiBA06a}.}} \label{slope}
\end{figure}

More recently, Li and Steiner have examined the correlations among
the radii of neutron stars, the size of neutron-skin in
$^{208}$Pb, and the strength of isospin diffusion in heavy-ion
collisions at intermediate energies \cite{Ste05b,LiBA06a}. The
isospin diffusion in heavy-ion reactions is essentially a
re-distribution of isospin asymmetries that is initially carried
by the colliding nuclei. The degree and rate of this process
depend on the relative pressures of neutrons and protons, namely
the slope of $E_{\rm sym}(\rho)$. With a stiffer $E_{\rm
sym}(\rho)$, it is more difficult for neutrons and protons to mix,
leading thus to a smaller/slower isospin diffusion.  Because of
isospin asymmetric pressure, dilute neutron-rich clouds
surrounding a more symmetric dense region are dynamically
generated in heavy-ion reactions through isospin diffusion as
illustrated in the inset of Fig. \ref{slope}, where the
correlation between the isospin asymmetry and density of matters
at the instant of 20 fm/c in $^{124}$Sn$+^{112}$Sn reactions is
shown. One can also see from the inset that this dynamical isospin
fractionation depends sensitively on the symmetry energy.

From the above discussions, one thus expects that the radii of
neutron stars, the degree of isospin diffusion in heavy-ion
collisions, and the sizes of neutron-skins in heavy nuclei are all
correlated through the same underlying asymmetric pressure. This
expectation was confirmed as demonstrated in Fig. \ref{slope},
where the strength of the isospin diffusion $1-{\rm R_i}$,
calculated with the IBUU04 model with in-medium NN cross sections,
and the thickness of neutron skin $dR_{np}$ in $^{208}$Pb,
calculated using the Skyrme Hartree-Fock with interaction
parameters adjusted to give an EOS which is similar to the
effective interaction used in the IBUU04 model~\cite{Ste05b}, are
examined simultaneously as functions of the slope parameter $L$ of
the symmetry energy. It is seen that $1-{\rm R_i}$ decreases while
$dR_{np}$ increases with increasing $L$ as expected. Taken the
fiducial value $dR_{np}=0.2\pm 0.04$ fm, that is measured and
supported strongly by many theoretical studies \cite{Ste05a}, and
the NSCL/MSU data $1-{\rm R_i}=0.525\pm0.05$ \cite{Tsa04}, the $L$
parameter is constrained in a common range between 62.1 MeV
($x=0$) and 107.4 MeV ($x=-1$) \cite{LiBA06a}. For a comparison,
the RMF with the FSUGold and the NL3 parameter sets gives $L=60.5$
MeV and 118.4 MeV~\cite{Piek07}, respectively.

\begin{figure}[tbh]
\centering
\includegraphics[scale=0.40,angle=0]{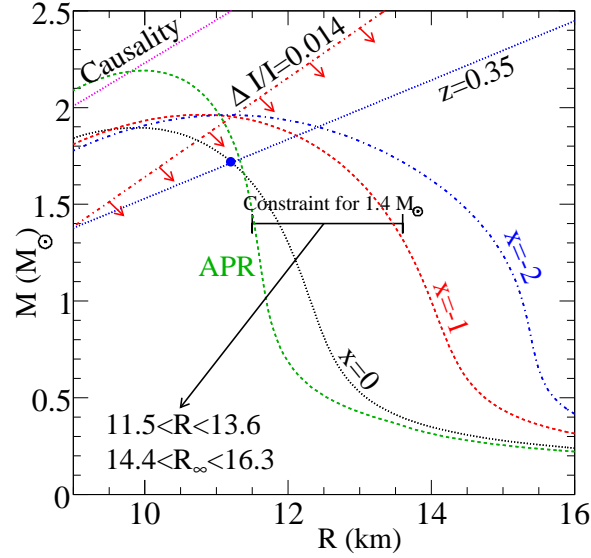}
\caption{The mass-radius curves for $x=0,-1,$ and $-2$ and the APR
EOS. The limit from causality, the Vela pulsar, and the redshift of
EXO0748 are all indicated. The inferred radius of a 1.4 solar mass
neutron star and the inferred value of $R_{\infty}$ are given. Taken
from Ref.~\cite{LiBA06a}.} \label{MRx0xm1}
\end{figure}

The corresponding mass versus radius curves for the MDI EOSs as well
as for APR (using the AV18+$\delta v$+UIX$^{*}$ interaction) are
given in Fig.~\ref{MRx0xm1} for non-rotating neutron stars. As we
shall discuss later, the rotation generally increases both the
masses and radii of neutron stars by up to about 20\% to 30\% for a
given EOS. In addition, the constraints of causality, the
mass-radius relation from estimates of the crustal fraction of the
moment of inertia ($\Delta I/I=0.014$) in the Vela
pulsar~\cite{Lin99}, and the mass-radius relation from the redshift
($z$) measurement from Ref. \cite{Cot02} are also given. Any
equation of state should be to the right of the causality line and
the $\Delta I/I$ line and should cross the $z=0.35$ line. The
horizontal bar indicates the inferred limits on the radius $R$ and
the radiation radius (the value of the radius which is observed by
an observer at infinity) defined as $R_{\infty}=R/\sqrt{1-2GM/Rc^2}$
for a 1.4 \Msun~neutron star. Since all three calculations with
$x=0, -1$ and $x=-2$ have the same compressibility ($K_0=211$ MeV)
but rather different radii, it is clear that the radius is indeed
rather sensitive to the symmetry energy while the maximum mass is
only slightly modified \cite{Lat04,Pra88b,Lat00,Lat01,Pra01}.  The
APR EOS has a compressibility of $K_0=269$ MeV but almost the same
symmetry energy as with $x=0$. It leads, however, to a 16\% higher
maximum mass ($1.9M_{\odot}$ to $2.2M_{\odot}$) but only a 5\%
decrease in radius (12.0 km to 11.5 km) as compared to the results
with $x=0$.

Since only EOSs with symmetry energies between $x=0$ and $x=-1$
are consistent with the isospin diffusion data and measurements of
the skin thickness of lead, the resulting neutron star structure
can be taken as representative of the possible variation that is
consistent with terrestrial data. The APR and the $x=0$ EOS have
nearly identical symmetry energies and slightly different radii.
Neutron star radii are sensitive to the symmetry energy but also
contain contributions from the isospin-symmetric part of the EOS,
especially at higher densities. Even though the compressibility of
the APR EOS is larger than that of the $x=0$ EOS, the pressure is
typically lower in the APR EOS at densities just above saturation,
giving the APR EOS a smaller radius by about 5\%. This 5\%
difference can be taken as representative of the remaining
uncertainty in the symmetric part of the EOS, so the minimum
neutron star radius can be extended to 11.5 km. Neutron stars with
radii larger than 13.6 km are difficult to make without a larger
symmetry energy or compressibility \cite{Ste05a}. It was concluded
that only radii between 11.5 and 13.6 km (or radiation radii
between 14.4 and 16.3 km) are consistent with the $x=0$ and $x=-1$
EOSs, and thus consistent with the laboratory data \cite{LiBA06a}.
It is interesting to note that a radius of R=12.66 km was recently
predicted for canonical neutron stars using the FSUGold
interaction \cite{Tod05}. This radius falls right in the range
favored by the isospin diffusion data. The constraints from
analyzing the isospin diffusion data are also consistent with the
extensive analysis of neutron star radii in Ref. \cite{Ste05a}
with only a few exceptions.

The observational determination of a neutron star radius from the
measured spectral fluxes relies on a numerical model of the
neutron star atmosphere and uses the composition of the
atmosphere, a measurement of the distance, the column density of
x-ray absorbing material, and the surface gravitational redshift
as inputs. Many of these quantities are difficult to measure,
which thus leads to the paucity of radius measurements. While
estimates of radii based on astrophysical observations are still
very challenging, it is useful to compare the above results with
recent Chandra/XMM-Newton observations. Assuming a mass of 1.4
\Msun, the inferred radiation radius, $R_{\infty}$, (in km) is
$13.5\pm 2.1$ \cite{Rut02a,Rut02b} or $13.6\pm 0.3$ \cite{Gen03a}
for the neutron star in $\omega$ Cen, $12.8\pm 0.4$ in M13
\cite{Gen03b}, $14.5^{+1.6}_{-1.4}$ for X7 in 47 Tuc \cite{Hei06}
and $14.5^{+6.9}_{-3.8}$ in M28 \cite{Bec03}, respectively. Except
the neutron star in M13 that has a slightly smaller radius, all
others fall into the constraints of 14.4 km $<R_{\infty}< 16.3$ km
within the observational error bars that are often larger than the
given range. It is also interesting to note that the upper limit
for the radius of a 1.4 \Msun neutron star shown in Fig.
\ref{MRx0xm1} also agrees with the lower limit inferred by
Tr\"umper {\it et al.} \cite{Tru04}. The maximum mass with the MDI
interaction is $1.95\pm 0.05$ \Msun. It is close to but lower than
the M(PSR J0751+1807)=$2.1\pm 0.2$ \Msun~originally reported by
Nice {\it et al.} \cite{Nic05}. The recently reported revision,
however, puts the mass at M(PSR J0751+1807)=$1.3\pm 0.2$ \Msun
\cite{Piek07}. While the Vela $\Delta I/I$ upper limit does not
provide any new information, the crossing of the $z=0.35$ line
with the $x=0$ and $x=-1$ curves are interesting. It implies a
mass larger than 1.4 \Msun~for EXO-0748~(the minimum mass would be
about 1.7 \Msun~corresponding to the dot in Fig.~\ref{MRx0xm1})
and a radius similar to that of a canonical neutron star. This is
not unreasonable since this object is accreting \cite{Cot02}.

\subsection{Constraining properties of rapidly rotating neutron stars}

To conserve the total angular momentum in supernova explosions
neutron stars normally spins. Because of their strong gravitational
binding neutron stars can rotate very fast without breaking apart
\cite{Bej07}. The first millisecond pulsar PSR1937+214, spinning at
$\nu=641 {\rm Hz}$ \cite{Bac82}, was discovered in 1982, and during
the next decade or so almost every year a new one was reported. In
recent years the situation changed considerably with the discovery
of an anomalously large population of millisecond pulsars in
globular clusters \cite{Web99a}. These are very favorable sites for
formation of rapidly rotating neutron stars which have been spun up
by means of mass accretion from a binary companion. Presently the
number of observed pulsars is close to 2000, and the detection rate
is rather high. In 2006 Hessels {\it et al.} \cite{Hes06} reported
the discovery of a very rapid pulsar J1748-2446ad, rotating at
$\nu=716~{\rm Hz}$ and thus breaking the previous record (of
$641~{\rm Hz}$). However, even this high rotational frequency is too
low to affect the structure of neutron stars with masses above
1\Msun~\cite{Bej07}. Such pulsars belong to the slow-rotation regime
since their frequencies are considerably lower than the Kepler
(mass-shedding) frequency
\begin{eqnarray}\label{kfre}
\nu_K=\frac{1}{2\pi}\left(\frac{GM}{R_{\rm eq}^3}\right)^{1/2},
\end{eqnarray}
where $R_{\rm eq}$ is the equatorial radius. $\nu_K$ is the highest
possible frequency for a star before it starts to shed mass at the
equator. Neutron stars with masses above 1\Msun~enter the
rapid-rotation regime if their rotational frequencies are higher
than $1000~{\rm Hz}$ \cite{Bej07}. A recent report by Kaaret {\it et
al.}~\cite{Kaa06} suggests that the X-ray transient XTE J1739-285
contains the most rapid pulsar ever detected rotating at
$\nu=1122~{\rm Hz}$. This discovery has reawaken the interest in
building models for rapidly rotating neutron stars \cite{Bej07}.

While global properties of spherically symmetric static
(non-rotating) neutron stars have been studied extensively,
properties of rapidly rotating neutron stars have been investigated
to lesser extent. Models of rapidly rotating neutron stars have been
constructed only by several research groups with various degree of
approximation, see e.g., Refs. \cite{Web99a,Ste03} for a review.
Using the MDI EOS constrained by data from heavy-ion reactions,
Krastev, Li and Worley recently studied properties of rapidly
rotating neutron stars \cite{Kra07b} by solving Einstein's field
equation using a code developed by Nikolaos Stergioulas and J.L.
Friedman \cite{Ste94,Ste96,Ste98}.

\begin{figure}[h]
\centering
\includegraphics[totalheight=3.5in]{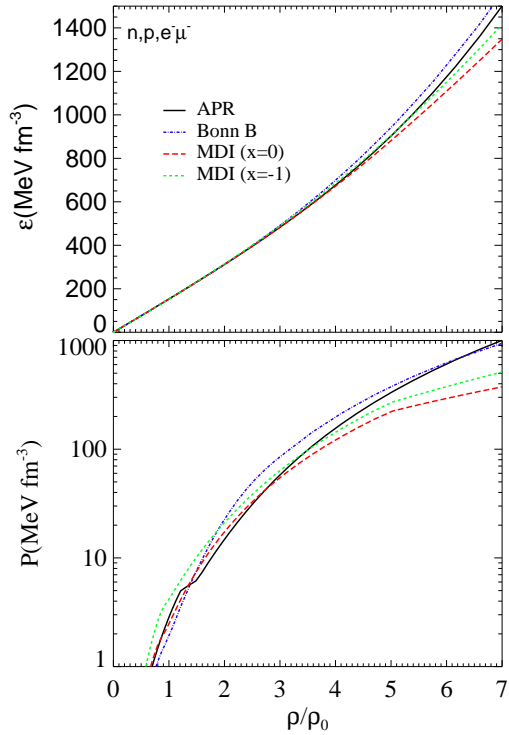}
\vspace{1cm} \caption{\protect\small Several typical equation of
state of stellar matter in $\beta$-equilibrium. The upper panel
shows the energy density and lower panel the pressure as functions
of the baryon number density (in units of $\rho_0$). Taken from Ref.
\cite{Kra07b}.} \label{GeDenPRho}
\end{figure}

\begin{figure}[h]
\centering
\includegraphics[totalheight=4.0in]{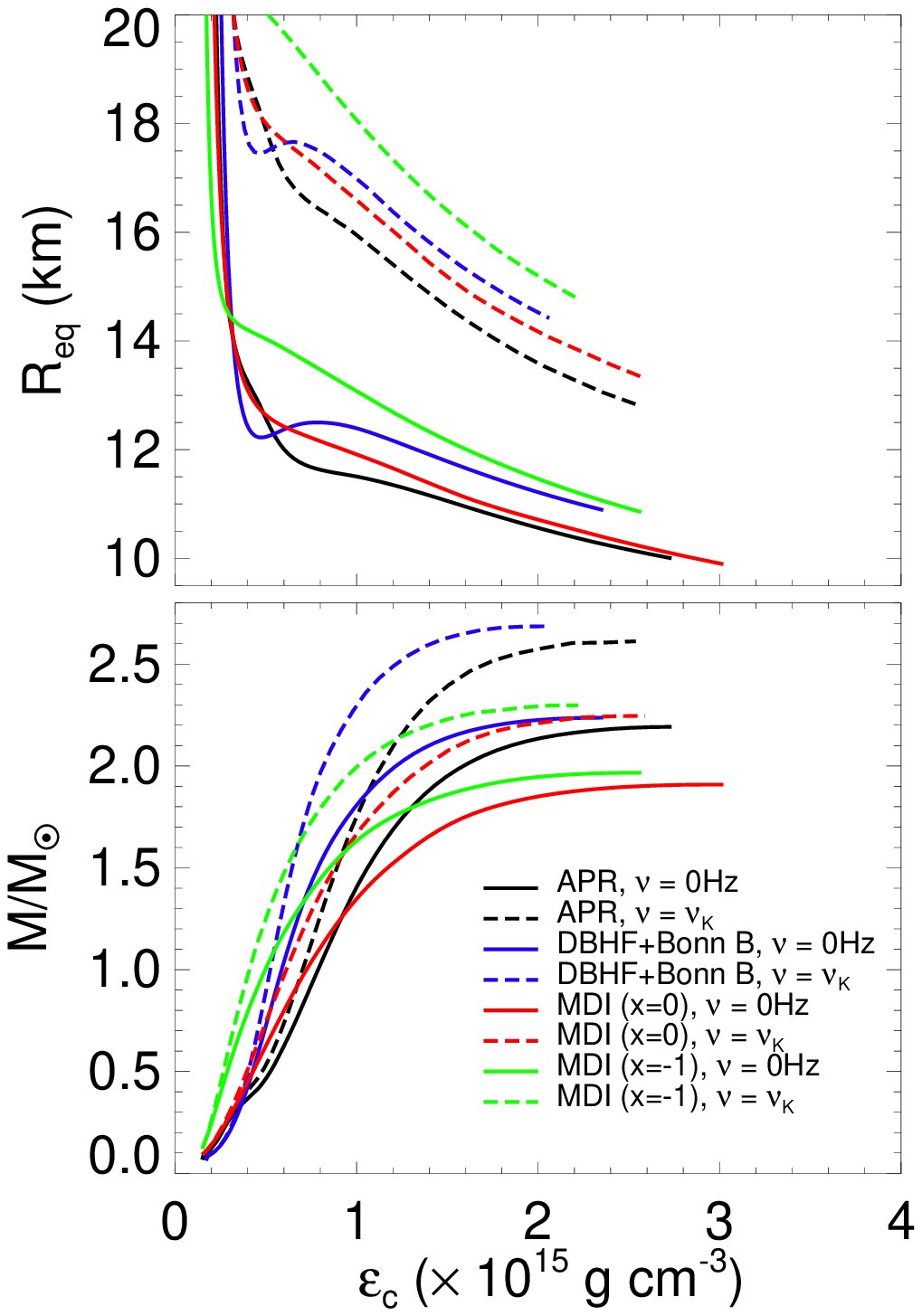}
\includegraphics[totalheight=2.50in]{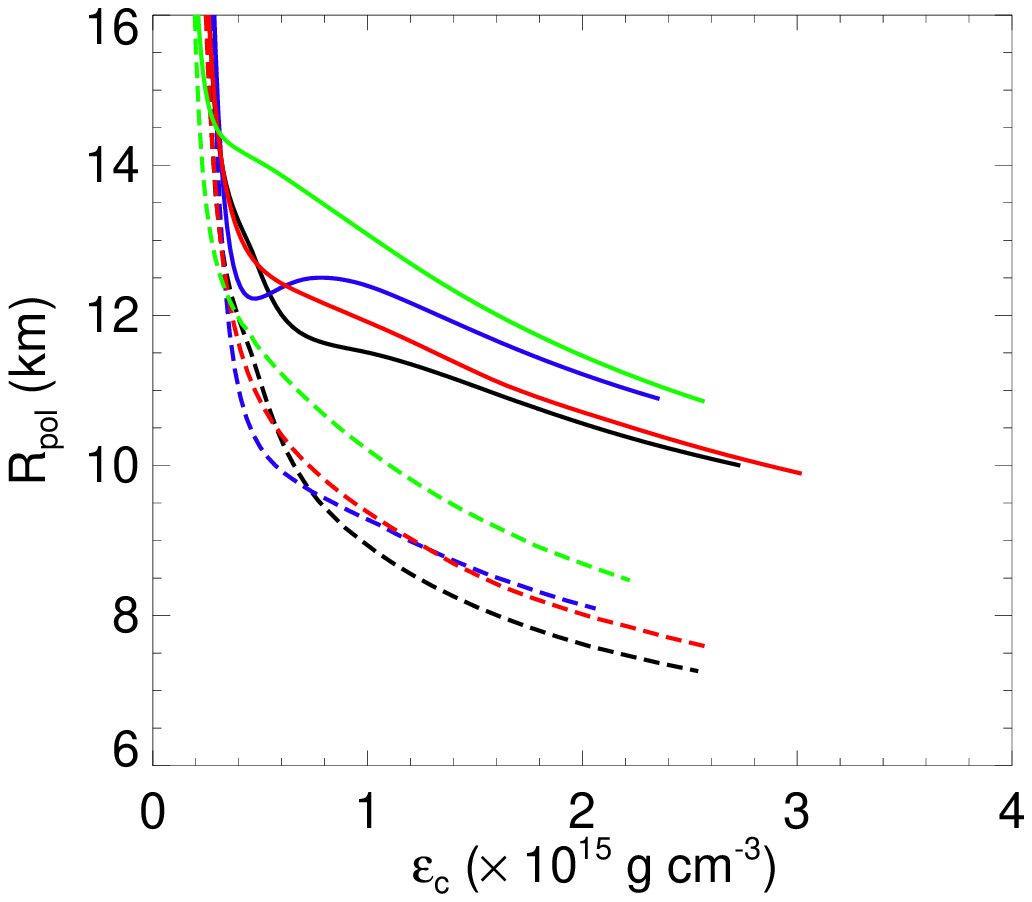}
\caption{Equatorial radii (upper left panel), total gravitational
mass (lower left panel) and the polar radii (right panel) versus
central energy density $\epsilon_c$. Both static (solid lines) and
Keplerian (broken lines) models are shown. Taken from Ref.
\cite{Kra07b}.}\label{rotatingns}
\end{figure}

Shown in Fig.~\ref{GeDenPRho} are several EOSs used by Krastev, Li
and Worley in studying the rotational effects on the mass and
geometry of neutron stars. The upper panel displays the total
energy density, $\epsilon$ (including leptons), as a function of
baryon number density and the lower panel shows the total
pressure. Besides the MDI EOS with $x=$ and $x=-1$, the Akmal EOS
with the $A18+\delta v+UIX^*$ interaction (APR) \cite{Akm98} and
the recent DBHF calculations (Bonn B) \cite{Kra06,Alo03} are also
used. Below the density of approximately $0.07~{\rm fm}^{-3}$ the
equations of state shown in Fig. \ref{GeDenPRho} are supplemented
with a crust EOS \cite{Pet95b,Hae94} which is more suitable for
the low density regime. Shown in Fig. \ref{rotatingns} are the
mass and radius versus the central energy density for static
(solid) and pulsars (dashed) rotating at the Kepler frequency
$\nu_K$.

\begin{figure}[h]
\centering
\includegraphics[totalheight=2.8in]{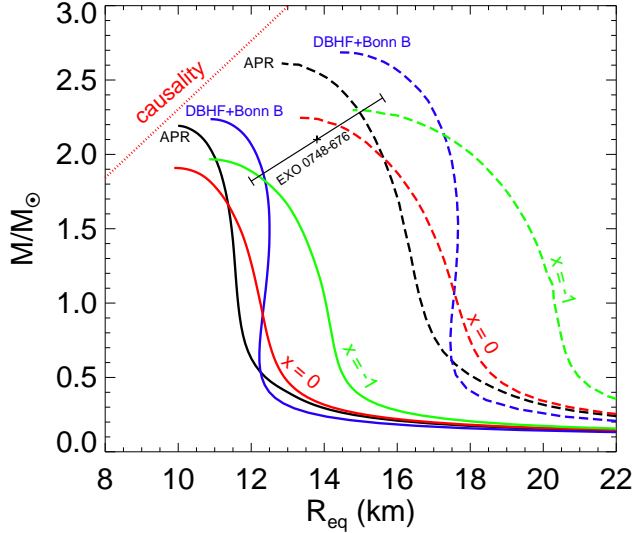}
\caption{Mass-radius correlation. Both static (solid lines) and
Keplerian (broken lines) sequences are shown. Taken from Ref.
\cite{Kra07b}.} \label{ms-rns}
\end{figure}

There exists a maximum gravitational mass that a given EOS can
support for a neutron star. This holds for both static and rapidly
rotating stars. The sequences shown in Fig.~\ref{rotatingns}
terminate at the `maximum mass' point. Comparing the results for
static and rotating stars, it is seen clearly that the rapid
rotation increases noticeably the mass that can be supported against
collapse while lowering the central density of the maximum-mass
configuration. This is what one should expect since rotation
stabilizes the star against the gravitational pull providing an
extra (centrifugal) repulsion. For rapid rotation at the Kepler
frequency, a mass increase up to $\sim 17\%$ is obtained, depending
on the EOS. The equatorial radius increases while the polar radius
decreases correspondingly by several kilometers, leading to an
overall oblate shape of the rotating star. In each case the upper
mass limit is attained for a model at the mass-shedding limit with
central density $\sim 15\%$ below that of the static model. The
rotational effect on the mass-radius correlation is more clearly
illustrated in Fig.~\ref{ms-rns} where the gravitational mass is
given as a function of the equatorial radius. The $1-\sigma$ error
bar corresponding to the mass and radius of EXO 0748-676 reported in
Ref. \cite{Oze06} is also shown for a comparison. Since this object
has a spinning frequency of only $47$~Hz, rotational effects on its
mass and radius are very small. It is seen that static calculations
using the $x=0$ and $x=-1$ lead to appreciably lower values for both
the radii and maximum masses compared to the reported observations.

\subsection{The pulsars at 716 and 1122 Hz}

\begin{figure}[h]
\centering
\includegraphics[totalheight=2.5in]{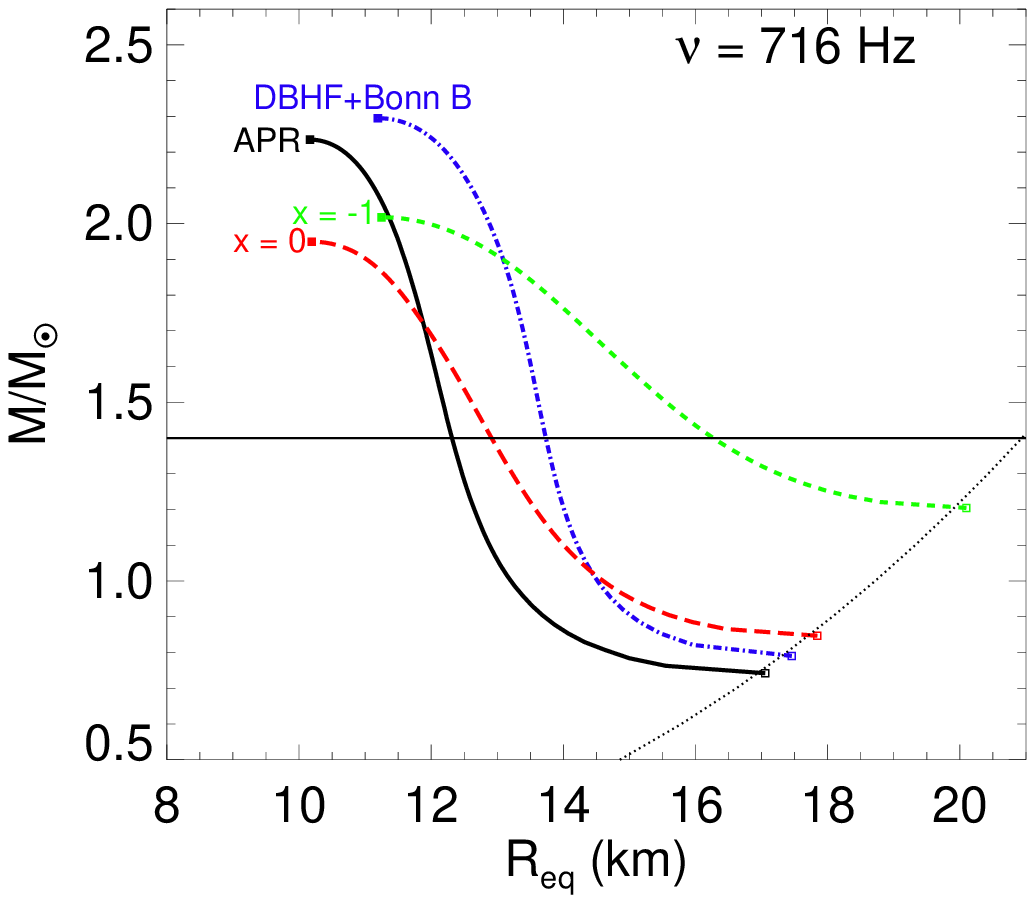}
\includegraphics[totalheight=2.5in]{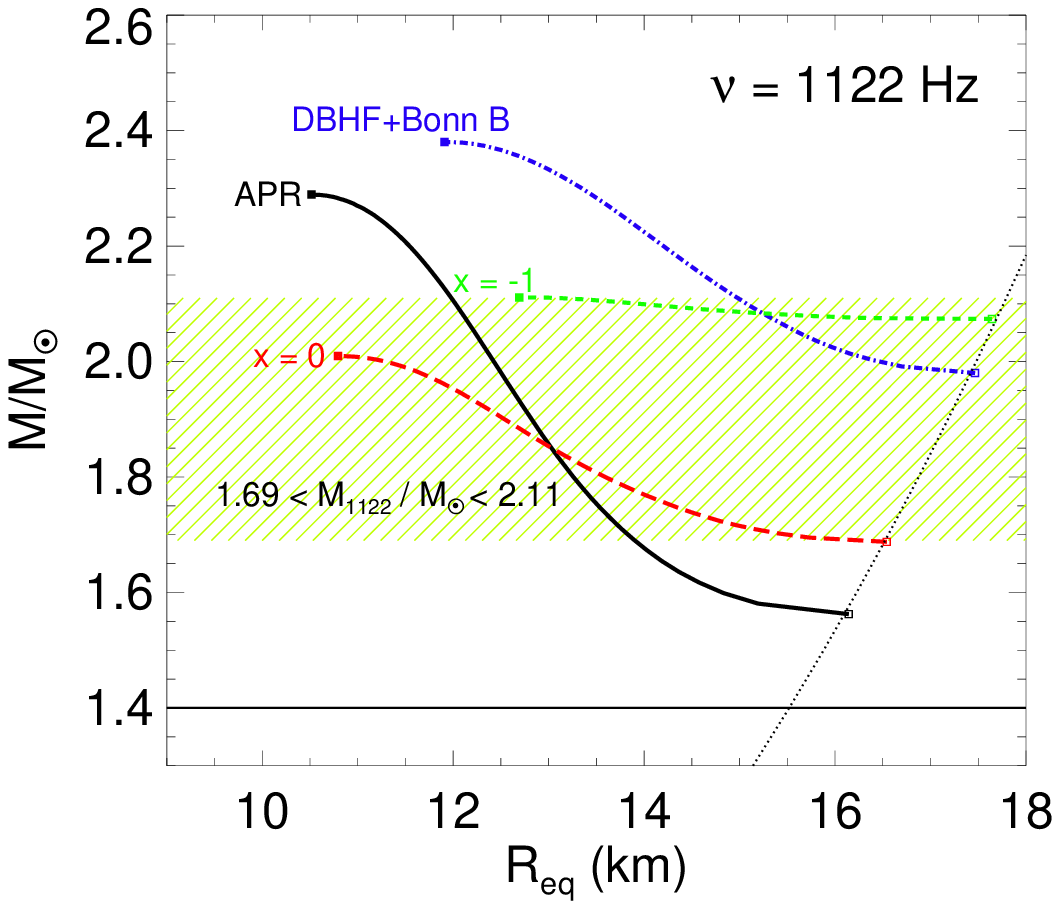}
\caption{(Color online) Gravitational mass versus equatorial radius
for neutron stars rotating at $\nu=716~{\rm Hz}$ and $\nu=1122~{\rm
Hz}$. Taken from Ref.~\cite{Kra07b}.}\label{radii}
\end{figure}

The two fastest pulsars discovered so far are spinning at 716
\cite{Hes06} and 1122 Hz \cite{Kaa06}, respectively. However, based
on the observational data available so far their properties have not
been fully understood yet. The analysis of their properties based on
the EOS and symmetry energy constrained by the terrestrial
laboratory data is thus especially interesting. Setting the observed
frequency of the pulsar as the Kepler frequency one can obtain an
estimate of its maximum radius as a function of mass $M$
\begin{eqnarray}\label{eq.17}
R_{\rm max}(M)=\chi\left(\frac{M}{1.4M_{\odot}}\right)^{1/3}~{\rm
km},
\end{eqnarray}
with $\chi=20.94$ for rotational frequency $\nu=716~{\rm Hz}$ and
$\chi=15.52$ for $\nu=1122~{\rm Hz}$. The maximum radii are shown
with the dotted lines in Fig.~\ref{radii}. It is seen that the range
of the allowed masses supported by a given EOS for rapidly rotating
neutron stars becomes narrower than the one of static
configurations. This effect becomes stronger with increasing
frequency and depends upon the EOS. For instance, for neutron stars
rotating at $1122~{\rm Hz}$ the allowed mass range is only $\sim
0.1$\Msun~for the x=-1 EOS. Since predictions from the $x=0$ and
$x=-1$ EOSs represents the limits of the neutron star models
consistent with the experimental data from terrestrial nuclear
laboratories, it was predicted by Krastev, Li and Worley that the
mass of the neutron star in XTE J1739-285 is between 1.7 and
$2.1$\Msun~\cite{Kra07b}.

\subsection{Rotational effects on the cooling mechanism of neutron stars}

\begin{figure}[h]
\centering
\includegraphics[totalheight=3.6in]{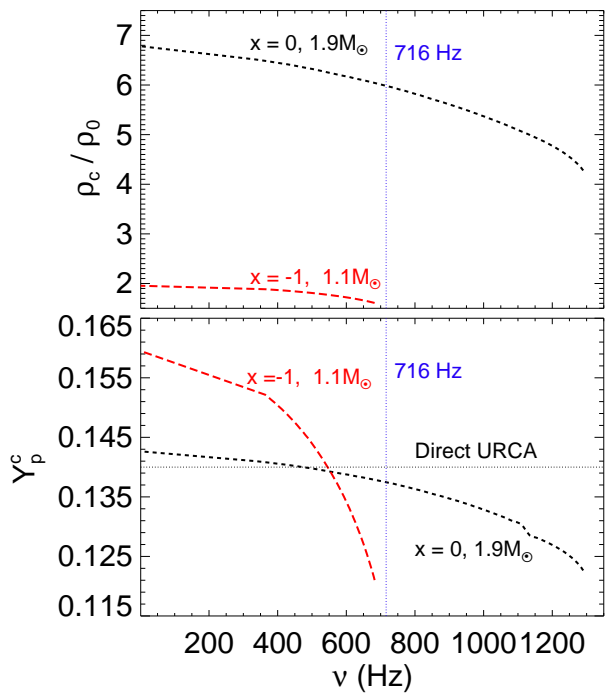}
\caption{Central density (upper panel) and central proton fraction
(lower panel) versus rotational frequency for fixed neutron star
mass. Taken from Ref. \cite{Kra07b}.}\label{rns-yp}
\end{figure}

As the neutron star rotates, because of the elongation along the
equator the central density drops. The proton fraction in the
rotating neutron star is thus different from that in a static one
having the same mass. The rotation may thus also affect the cooling
mechanism of neutron stars. Fig.~\ref{rns-yp} shows the central
baryon density (upper panel) and central proton fraction (lower
panel) as functions of the rotational frequency for fixed-mass
models. Predictions from both $x=0$ and $x=-1$ EOSs are shown. The
masses of the models are chosen so that the proton fraction in
stellar core is just above the direct URCA limit for the {\it
static} configurations. It is seen that the central density and the
proton fraction $Y_p^c$ decrease with increasing frequency. This
reduction is more pronounced in heavier neutron stars. We recall
that large proton fraction (above $\sim 0.14$ for $npe\mu$-stars)
leads to fast cooling of neutron stars through direct URCA
reactions. One sees here that depending on the stellar mass and
rotational frequency, the central proton fraction could, in
principle, drop below the threshold for the direct {\it nucleonic}
URCA channel and thus making the fast cooling in rotating neutron
stars impossible. The stellar sequences in Fig.~\ref{rns-yp} are
terminated at the respective Kepler (mass-shedding) frequency for
the given mass. In both cases of x=0 and x=-1, the central proton
fraction drops below the direct URCA limit at frequencies lower than
the sping rate of PSR J1748-244ad \cite{Hes06}. This implies that
the fast cooling can be effectively blocked in millisecond pulsars
depending on the exact mass and spin rate. It might also explain why
heavy neutron stars (could) exhibit slow instead of fast cooling.
For instance, with the $x=0$ EOS (with softer symmetry energy) for a
neutron star of mass approximately $1.9$\Msun, the Direct URCA
channel closes at $\nu\approx 470~{\rm Hz}$. On the other hand, with
the $x=-1$ EOS (with stiffer symmetry energy) the direct URCA
channel can close only for low mass neutron stars, in fact only for
masses well below the canonical mass of $1.4$\Msun. This is due to
the much stiffer symmetry energy (see Fig.~9 middle panel) because
of which the direct URCA threshold is reached at much lower
densities and stellar masses.

\subsection{The core-crust transition density and momenta of inertia
of neutron stars and their crusts}

\begin{figure}[h]
\centering
\includegraphics[scale=1,angle=0]{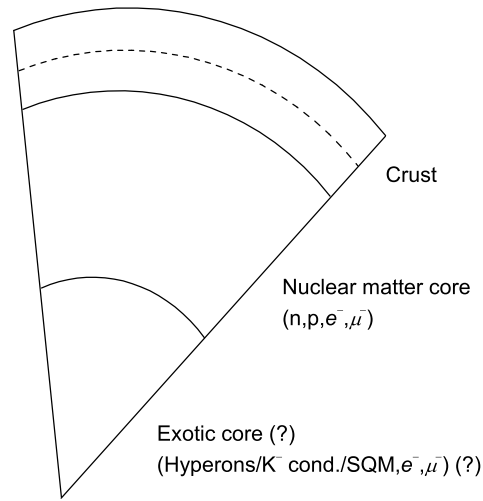}
\caption{Schematic cross section of a neutron star. The thicknesses
of the various layers are not drawn in scale. Taken from Ref.
\cite{Bom01}.} \label{nstar}
\end{figure}

\begin{figure}
\centering
\includegraphics[scale=0.4,angle=-90]{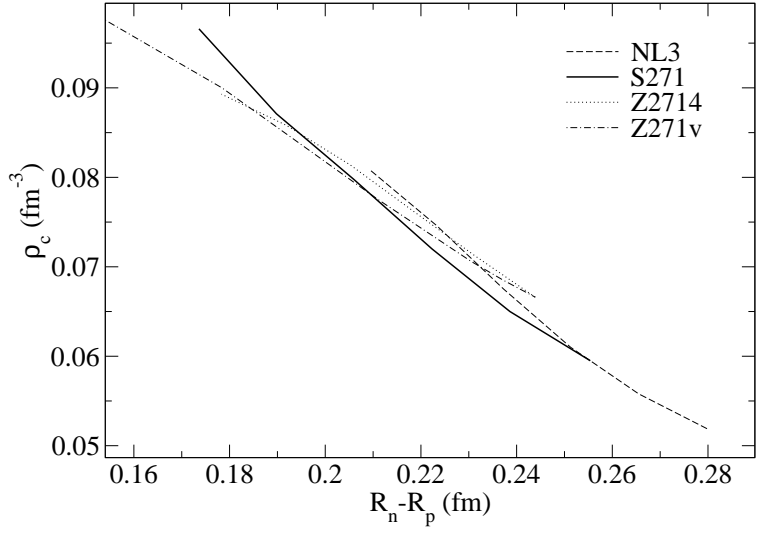}
\caption{Estimate of the transition density from nonuniform to
uniform neutron-rich matter versus neutron-minus-proton radius in
$^{208}$Pb. The curves are for four parameter sets used in the RMF
model. Taken from Ref. \cite{Hor01a}.} \label{horowitz2}
\end{figure}

Shown in Fig.\ \ref{nstar} is a schematic cross section of a
neutron star. Neutron stars are expected to have a solid crust of
nonuniform neutron-rich matter above a liquid mantle. The phase
transition between solid and liquid depends on the properties of
neutron-rich matter. Indeed, as discussed in detail by Horowitz
and Piekarewicz \cite{Hor01a}, high pressure implies a rapid rise
of the energy with density making it energetically unfavorable to
separate uniform matter into regions of high and low densities.
Thus a high pressure typically implies a low transition density
from a solid crust to a liquid mantle. This suggests an inverse
relationship: the thicker the neutron-rich skin of a heavy
nucleus, the thinner the solid crust of a neutron star. This
expectation is demonstrated in Fig.\ \ref{horowitz2} where the
transition density is shown as a function of the neutron-skin
thickness in $^{208}$Pb.

The estimate of the transition density by Horowitz and Piekarewicz
was obtained by using the random-phase approximation
\cite{Hor01a}. We notice here that the estimation of the
transition density itself is a very complicated problem. Different
approaches used often give quite different results, see, e.g.,
discussions in Refs.
\cite{Ste05a,Duc07,Pet95b,Lat07,Bay71,Kub07,Ste07}. Similar to
finding the critical density for the spinodal decomposition for
the liquid-gas phase transition in nuclear matter, for the uniform
$npe$ matter, Lattimer and Prakash \cite{Lat07} as well as Kubis
\cite{Kub07} have evaluated the crust transition density by
investigating when the incompressibility of the $npe$ matter
becomes negative. Using the parabolic approximation for the
symmetry energy, they obtained the following condition
\begin{eqnarray}\label{knpe}
\rho^2{d^2E_0\over d\rho^2}+2\rho{dE_0\over
d\rho}+\delta^2\left[\rho^2{d^2E_{\rm sym}\over
d\rho^2}+2\rho{dE_{\rm sym}\over d\rho}-2E_{\rm
sym}^{-1}\left(\rho{dE_{\rm sym}\over d\rho}\right)^2\right]<0
\end{eqnarray}
where the $E_0$ is the EOS of symmetric nuclear matter. Using this
approach and the MDI interaction, Kubis found the transition
density of $0.119, 0.092, 0.095$ and $0.160~{\rm fm}^{-3}$ for the
$x$ parameter of $1, 0, -1 $ and $-2$, respectively. Indeed, as
stressed by Lattimer and Prakash \cite{Lat07} the transition
density is very sensitive to the density dependence of the
symmetry energy. Using the same approach and the APR EOS, Worley,
Krastev and Li \cite{Wor07} found a transition density and
pressure of $\rho_t=0.097~{\rm fm}^{-3}$ and $P_t=0.6827~{\rm
MeV}\cdot{\rm fm}^{-3}$, respectively. With the DBHF+Bonn B EOS
they found a $\rho_t= 0.082~{\rm fm}^{-3}$ and $P_t= 0.3659~{\rm
MeV}\cdot{\rm fm}^{-3}$, respectively. A comparison is made in
Table \ref{tstable}. It is interesting to note that the transition
density is in the same density range as that explored by heavy-ion
collisions around the Fermi energy. The MDI interactions with
$x=0$ and $x=-1$ constrained by the isospin diffusion data thus
limits the transition density rather tightly at about
$0.092-0.095~{\rm fm}^{-3}$.

\begin{table}[!h]
\caption{Transition and saturation densities of several typical
nuclear EOSs used in Fig.~\ref{MI-static}. The first row
identifies the equation of state. The remaining rows exhibit the
saturation density and the transition density from the liquid core
to solid crust in neutron stars calculated using Eq.~(\ref{knpe}).
Taken from Ref. \cite{Wor07}.}

\begin{center}
\begin{tabular}{lcccccc}\label{tstable}
EOS &  MDI(x=1) & MDI(x=0) & MDI(x=-1) & MDI(x=-2) & APR & DBHF+Bonn B\\
\hline\hline
$\rho_0({\rm fm}^{-3})$  & 0.160 & 0.160 & 0.160 & 0.160 & 0.160 & 0.185\\
$\rho_t({\rm fm}^{-3})$  & 0.119 & 0.092 & 0.095 & 0.160 &
0.097 & 0.082\\
\hline
\end{tabular}
\end{center}
\end{table}

As it was discussed extensively by Lattimer and Prakash
\cite{Lat01,Lat07} and others, the neutron star crust thickness
might be measurable from observations of pulsar glitches, the
occasional disruptions of the otherwise extremely regular pulsations
from magnetized, rotating neutron stars. The canonical model of Link
{\it et al.} \cite{Lin99} suggests that glitches are due to the
transfer of angular momentum from superfluid neutrons to normal
matter in the neutron star crust, the region of the star containing
nuclei and nucleons that have dripped out of nuclei. This region is
bounded by the neutron drip density at which nuclei merge into
uniform nucleonic matter. Link {\it et al.} \cite{Lin99} concluded
from glitches of the Vela pulsar that at least 1.4\% of the total
moment of inertia resides in the crust of the Vela pulsar. For
slowly rotating neutron stars using realistic hadronic EOSs that
permit maximum masses greater than about 1.6 \Msun, Lattimer \&
Schutz \cite{LS05} found that the fractional moment of inertia,
$\Delta I/I$ can be expressed approximately as \cite{Lat01,Lat07}
\begin{eqnarray}
{\Delta I\over I}\simeq{28\pi P_t R^3\over3
Mc^2}{(1-1.67\beta-0.6\beta^2)\over\beta}\left[1+{2P_t(1+5\beta-14\beta^2)
\over n_tm_bc^2\beta^2}\right]^{-1}\,, \label{dii}
\end{eqnarray}
where $\beta = GM/Rc^2$ and $I$ is the star's total moment of
inertia
\begin{eqnarray}
I\simeq(0.237\pm0.008) MR^2(1+2.84\beta+ 18.9\beta^4)
{\rm~M}_\odot {\rm~km}^2.\label{momls}
\end{eqnarray}

\begin{figure}
\centering
\includegraphics[scale=0.6,angle=0]{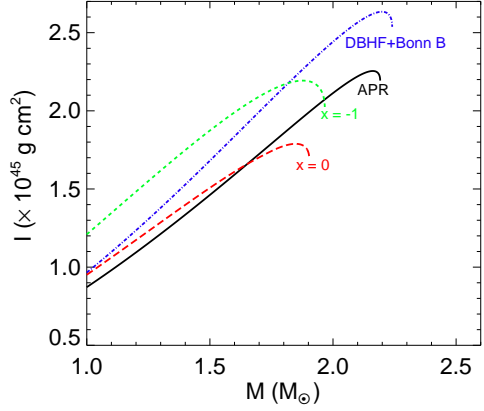}
\caption{The total moment of inertia of neutron stars estimated
using Eq.~(\ref{momls}). Taken from Ref. \cite{Wor07}.}
\label{MI-static}
\end{figure}

\begin{figure}
\centering
\includegraphics[scale=0.6,angle=0]{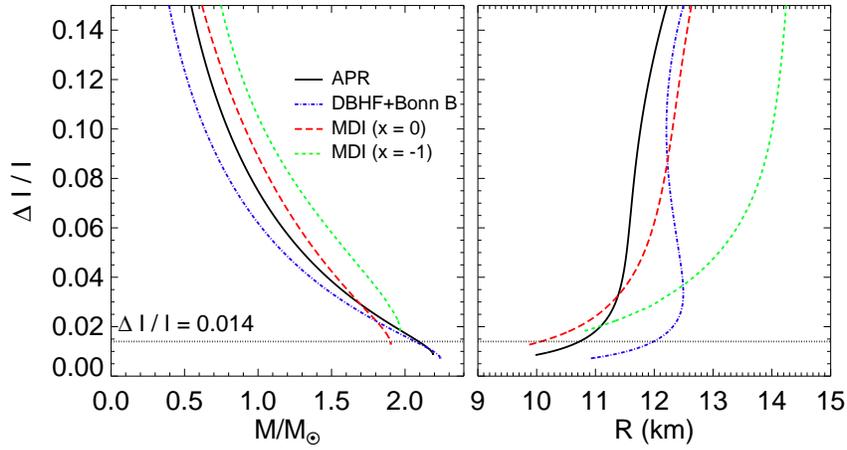}
\caption{The fractional moment of inertia of the neutron star
crusts estimated using Eq.~(\ref{dii}). Taken from Ref.
\cite{Wor07}.} \label{ifrac}
\end{figure}

Using the above formalism, Worley, Krastev and Li examined very
recently the total and fractional moment of inertia using the APR,
MDI and the DBHF+Bonn B EOSs \cite{Wor07}. Shown in
Fig.~\ref{MI-static} are the momenta of inertia versus mass. As
shown earlier in the mass-radius correlation of
Fig.~\ref{MRx0xm1}, above about 1.0\Msun the radii of neutron
stars remain about the same before reaching the maximum mass. The
moment of inertia thus increases almost linearly with mass. After
reaching the maximum mass the radius starts to decreases and thus
causes the drop of the moment of inertia. Since the latter is
proportional to the mass and the square of the radius, it is more
sensitive to the density dependence of the symmetry energy. For a
canonical neutron star of 1.4\Msun, for instance, the difference
in the moment of inertia predicted using the $x=0$ and $x=-1$ EOSs
is more than 30\%. The fractional momenta of inertia $\Delta I/I$
of the neutron star crusts are shown in Fig.\ \ref{ifrac}. It is
seen that the condition $\Delta I/I>0.014$ extracted from studying
the glitches of the Vela pulsar does put a strict lower limit on
the radius for a given EOS. It also limits the maximum mass to be
less than about 2\Msun for all of the EOSs considered. Similar to
the total momenta of inertia the $\Delta I/I$ changes more
significantly with the radius as the EOS is varied.

\begin{figure}[h]
\centering
\includegraphics[scale=0.8,angle=0]{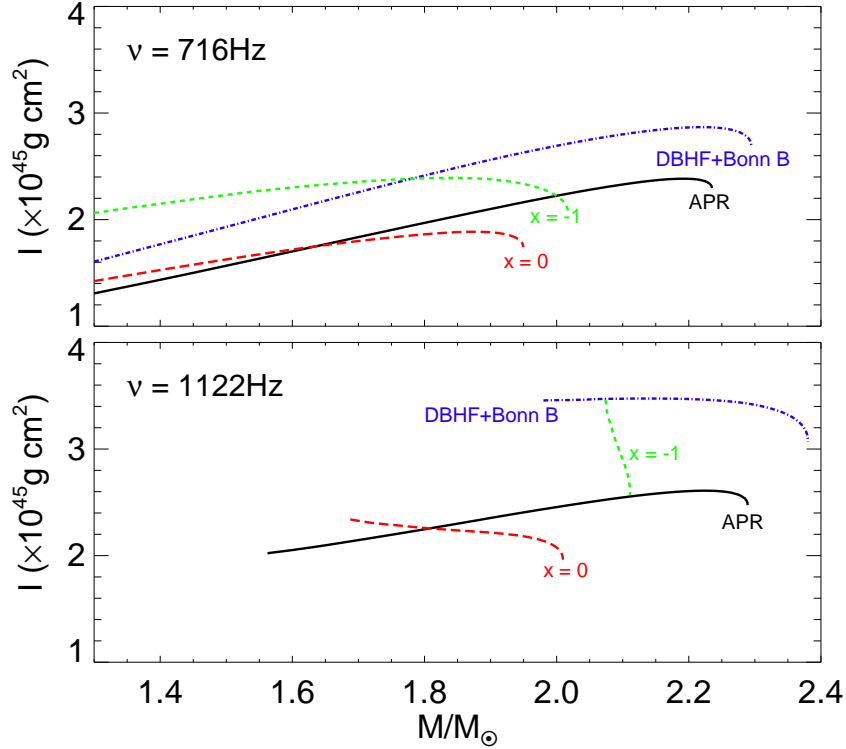}
\caption{The total moment of inertia of pulsars at 716 Hz and
1122Hz. Taken from Ref. \cite{Wor07}.} \label{Irotating}
\end{figure}

For rapidly rotating neutron stars the calculation of the moment
of inertia becomes much more complicated because of the
deformation. In principle, it has to be done within the framework
of general relativity. Using the RNS code of Nikolaos Stergioulas
and J.L. Friedman \cite{Ste94,Ste96,Ste98} for fast pulsars,
Worley, Krastev and Li \cite{Wor07} calculated the total moment of
inertia with respect to the rotational axis. Their results for
pulsars at 716Hz and 1122Hz are shown in Fig.~\ref{Irotating}.
Compared to the results for static stars, because of the
deformation the rotating ones have a significantly higher moment
of inertia. Comparing the momenta of inertia of the two fastest
pulsars at 716~Hz and 1122~Hz, one sees that the allowed ranges of
the mass-moment of inertia are quite different using different
EOSs, especially at 1122~Hz. This is mainly because the
instability of axial asymmetry and the mass-shedding limit are
strongly frequency dependent as we discussed earlier.

\subsection{Constraining a possible time variation of the
gravitational constant $G$}

The question whether or not the fundamental constants of nature
vary with time has been of considerable interest in physics. The
constancy of Newton's gravitational coupling parameter $G$ was
first addressed in 1937 by Dirac \cite{Dir37} who suggested that
the gravitational force might be weakening continuously due to the
ongoing expansion of the Universe. Although general relativity
assumes a strictly constant $G$, time variations of the Newton's
constant are predicted by some alternative theories of
gravity~\cite{Bra61} and a number of modern cosmological models
\cite{Kra95,Bon02}. Many theoretical approaches, such as models
with extra dimensions \cite{Ove97}, string theories
\cite{Hor96,Dam02a,Dam02b}, and scalar-tensor quintessence models
\cite{Zla99,Arm00,Arm01,Ste99,Heb00,Heb01}, have been proposed in
which the gravitational coupling parameter becomes a
time-dependent quantity.  More recently, the debate over the
constancy of $G$ has been revived by new astronomical observations
\cite{Per99,Rie00} of distant high-red-shift type Ia supernovae
suggesting that presently the Universe is in a state of
accelerated expansion \cite{Bon02}. This acceleration can be
interpreted in terms of a ``dark energy'' with negative pressure,
or alternatively by allowing a time variation of the gravitational
constant \cite{Gar06a}. Right after Dirac had published his
hypothesis \cite{Dir37}, Chandrasekhar \cite{Cha37} and Kothari
\cite{Kot38} pointed out that a decreasing G with time could have
some detectable astrophysical consequences. Since then many
attempts have been made to find astrophysical signs due to the
possible time variation of G. However, there is no firm conclusion
so far (see Ref. \cite{Uza03} for a review). Interestingly though,
as pointed by Uzan \cite{Uza03}, contrary to most of the other
fundamental constants, as the precision of the measurements
increased the discrepancy among the measured values of $G$ also
increased. This circumstance led the $CODATA$ (Committee on Data
for Science and Technology) to raise the relative uncertainty for
$G$ \cite{Uza03} by a factor of about 12 in 1998. Given the
current status of both theory and experiment, it is fair to say
that whether or not the gravitational constant G varies with time
is still an open question and therefore additional work is
necessary to investigate further this fundamental issue.

\renewcommand{\thetable}{\arabic{table}}
\begin{table}[!h]
\caption{Upper bounds on $\dot{G}/G$ published recently in the
literature}
\centering
\begin{tabular}{llll}\label{Gtable}
Method &  $(\dot{G}/G)_{max}$ & Time scale & reference\\
       &  $[10^{-12}~{\rm yr}^{-1}]$ & [yr] & \\
\hline\hline
Big Bang Nucleosynthesis  & 0.4    & $1.4\times 10^{10}$  & C. Copi {\it et al.},
PRL 92, 17 (2004).\\
\hline
Microwave Background & 0.7    & $1.4\times 10^{10}$ & R. Nagata {\it et al.},
PRD 69, 3512 (2004).\\
\hline Globular ClusterIsochrones & $10^{10}$ & 10.81  & S.
Degl'Innocenti {\it et al.},
A\&A 312, 345 (1996).\\
\hline
Binary Neutron Masses & 2.6    & $10^{10}$ & S.E. Thorsett, PRL 77, 1432 (1996).\\
\hline
Helioseismology & 1.6    & $4\times 10^{9}$ & D.B. Guenther {\it et al.}, AJ
498, 871 (1998).\\
\hline
Paleontology & 20   & $4\times 10^{9}$  & W. Eichendorf and M. Reinhardt (1977).\\
\hline
Lunar Laser Ranging & 1.3   & 24  & J.G. Williams {\it et al.}, PRL 93, 261101
(2004).\\
\hline
Binary Pulsar Orbits & 9    & 8 & V.M. Kaspi {\it et al.}, AJ 428, 713 (1994).\\
\hline
White Dwarf Oscillations & 250   & 25  & O. Benvenuto {\it et al.}, PRD 69,
2002 (2004).\\
\hline Gravitochemical heating of \\
n-stars & 4 to 200   & $10^8$  & P. Jofr\'e {\it et al.}, PRL 97, 131102 (2006).\\
\hline Gravitochemical heating of \\
n-stars with constrained EOS  & 4 to 21   & $10^8$ & Krastev and Li, PRC 76,
055804 (2007).\\
\hline\hline
\end{tabular}
\end{table}

Shown in Table \ref{Gtable} are estimates, often upper limits, of
the absolute values of the relative changing rate of G, i.e.,
$|\dot{G}/G|$, obtained using several different methods. Depending
on the approaches used, the estimates were made over different
time spans and they gave rather diverse results. Recently a new
method named {\it gravitochemical heating} was introduced to
constrain the $|\dot{G}/G|$~\cite{Jof06}. It is based on the
expectation that a variation of G would perturb the internal
composition of a neutron star, producing entropy which is
partially released through the emission of neutrinos and thermal
photons. A constraint on the changing rate of $G$ is achieved via
a comparison of the calculated and measured surface temperatures
of old neutron stars \cite{Kar04}. The gravitochemical heating
formalism is based on the results of Fern\'{a}ndez and Reisenegger
\cite{Fer05b} (see also Ref. \cite{Rei95}) who demonstrated that
internal heating could result also from spin-down compression in a
rotating neutron star ({\it rotochemical heating}). In both cases
(gravito- and rotochemical heatings) predictions rely heavily on
the equation of state of stellar matter used to calculate the
neutron star structure. Accordingly, detailed knowledge of the EOS
is crucial for setting a reliable constraint on the time variation
of $G$. Adopting the gravitochemical heating approach, Krastev and
Li \cite{Kra07a} recently evaluated the upper bound on the
$|\dot{G}/G|$ using the asymmetric EOS constrained by the
terrestrial laboratory data. In the following we first outline the
gravitochemical heating formalism and then summarize the main
results of Krastev and Li.

\subsubsection{The gravitochemical heating formalism}

In neutron stars, neutrons and protons can transform into each
other through direct and inverse $\beta$-reactions. The neutrinos
($\nu$) and antineutrinos ($\bar{\nu}$) produced in these
reactions leave the star without further interactions,
contributing to its cooling. At $\beta$-equilibrium the balance
between the rates of direct and inverse processes is reflected
through the following relation among the chemical potentials of
the particle species
\begin{eqnarray}
\mu_n-\mu_p=\mu_e=\mu_{\mu}
\end{eqnarray}
A time-variation of $G$ would cause continuously a perturbation in
the stellar density. Since the chemical potentials are
density-dependent, a variation of $G$ would thus cause neutron
stars to depart from their $\beta$-equilibrium. This departure can
be quantified by the chemical imbalances
\begin{eqnarray}
\eta_{npe}=\delta\mu_n-\delta\mu_p-\delta\mu_e \qquad {\rm and}
\qquad \eta_{np\mu}=\delta\mu_n-\delta\mu_p-\delta\mu_{\mu},
\end{eqnarray}
where $\delta\mu_i=\mu_i-\mu_i^{eq}$ is the deviation of the
chemical potential of particle species $i$ ($i=n,p,e,\mu$) from
its equilibrium value at a given pressure. The chemical imbalances
enhance the rates of reactions driving the star to a new
equilibrium state. If $G$ changes continuously with time the star
will always be out of equilibrium, storing an excess of energy
that is dissipated as internal heating and enhanced neutrino
emission~\cite{Jof06}.

The evolution of the internal temperature is given by the thermal
balance equation
\begin{eqnarray}\label{geq1}
\dot{T}^{\infty}=\frac{1}{C}[L_H^{\infty}-L_{\nu}^{\infty}-L_{\gamma}^{\infty}]
\end{eqnarray}
where $C$ is the total heat capacity of the star, $L_H^{\infty}$ is
the total power released by heating mechanisms, $L_{\nu}^{\infty}$
is the total neutrino luminosity, and $L_{\gamma}^{\infty}$ is the
photon luminosity ("$\infty"$ labels the quantities as measured by a
distant observer). The evolution of the red-shifted chemical
imbalances is governed by
\begin{eqnarray}
\dot{\eta}_{npe}^{\infty}=\delta\dot{\mu}_n^{\infty}-\delta\dot{\mu}_p^{\infty}
-\delta\dot{\mu}_e^{\infty} \qquad {\rm and} \qquad
\dot{\eta}_{np\mu}^{\infty}=\delta\dot{\mu}_n^{\infty}-\delta\dot{\mu}_p^{\infty}
-\delta\dot{\mu}_{\mu}^{\infty}.
\end{eqnarray}
These equations can be written as \cite{Jof06}
\begin{eqnarray}
\dot{\eta}_{npe}^{\infty}&=&-[A_{D,e}(\eta_{npe},T^{\infty})
+A_{M,e}(\eta_{npe},T^{\infty})]-[B_{D,e}(\eta_{np\mu},T^{\infty})
+B_{M,e}(\eta_{np\mu},T^{\infty})] \notag \\
\dot{\eta}_{np\mu}^{\infty}&=&-[A_{D,\mu}(\eta_{npe},T^{\infty})
+A_{M,\mu}(\eta_{npe},T^{\infty})]-[B_{D,\mu}(\eta_{np\mu},T^{\infty})
+B_{M,\mu}(\eta_{np\mu},T^{\infty})]\notag \\ \label{geq3}
\end{eqnarray}
The functions $A$ and $B$ quantify the effect of reactions toward
restoring chemical equilibrium, and thus have the same sign as
$\eta_{npl}$ ($l=e,\mu$) \cite{Fer05b}. The subscripts $M$ and $D$
refer to the modified (Eq.~(\ref{murca})) and direct
(Eq.~(\ref{durca})) URCA reactions.

The constants $C_{npe}$ and $C_{np\mu}$ that quantify the
departure from chemical equilibrium due to the time-variation of
$G$ can be written as \cite{Jof06}
\begin{eqnarray}
C_{npe}=(Z_{npe}-Z_{np})I_{G,e}+Z_{np}I_{G,p}\qquad {\rm and} \qquad
C_{np\mu}=(Z_{np\mu}-Z_{np})I_{G,\mu}+Z_{np}I_{G,p}.\nonumber\\
\end{eqnarray}
Here $I_{G,i}=(\partial N_i^{eq}/\partial G)_A$ is the change of the
equilibrium number of particles species $i$ ($i=n,p,e,\mu$),
$N_i^{eq}$, due to the variation of $G$, and $Z$ are constants
depending only on the stellar structure \cite{Jof06}. Eqs.
(\ref{geq1}) and (\ref{geq3}) determine completely the thermal
evolution of a neutron star with gravitochemical heating. The main
consequence of this mechanism is that eventually the star arrives at
a quasi-equilibrium state, with heating and cooling balancing each
other \cite{Jof06}. The properties of this stationary state can be
obtained by solving simultaneously Eqs.(\ref{geq1}) and (\ref{geq3})
by setting
$\dot{T}^{\infty}=\dot{\eta}_{npe}^{\infty}=\dot{\eta}_{np\mu}^{\infty}=0$.
The existence of a quasi-equilibrium state makes it possible to
compute the equilibrium temperature of an old neutron star without
knowing its complete evolution and exact age for a given value of
$|\dot{G}/G|$~\cite{Jof06}.

First, it is instructive to see analytically how the stationary
surface temperature of an old neutron star is related to the
changing rate of the gravitational constant $G$ by considering the
modified URCA process only. In this case, for a given stellar
model, it is possible to derive an analytic expression relating
the photon luminosity in the stationary state,
$L_{\gamma,eq}^{\infty}$, to $|\dot{G}/G|$. This is because the
longer time required to reach a stationary state when only the
modified URCA processes operate. In this case, the chemical
imbalances satisfy the condition $\eta_{npl}>>k_BT$ \cite{Jof06}.
Under these conditions the photon luminosity in the
quasi-equilibrium state is given by
\begin{eqnarray}\label{lum}
L_{\gamma,eq}^{\infty}=C_M\left(\frac{k_BG}{C_H}\right)^{8/7}
\left[\left(\frac{I_{G,e}^8}{\tilde{L}_{Me}}\right)^{1/7}+
\left(\frac{I_{G,\mu}^8}{\tilde{L}_{M\mu}}\right)^{1/7}\right]
\left|\frac{\dot{G}}{G}\right|^{8/7}.
\end{eqnarray}
The meaning of the constants $C_M$ and $C_H$, and the functions
$\tilde{L}_{M_i}$ ($i=e,\mu$) are explained in Ref. \cite{Fer05b}.
From $L_{\gamma,eq}^{\infty}$ the neutron-star surface temperature
can be calculated by assuming an isotropic blackbody spectrum
\begin{eqnarray}
L_{\gamma,eq}^{\infty}=4\pi\sigma R^2_{\infty}(T_s^{\infty})^4
\end{eqnarray}
with $\sigma$ the Stefan-Boltzmann constant and $R_{\infty}$ the
redshifted radius of the star. The stationary surface temperature
can then be written as \cite{Jof06}
\begin{eqnarray}
T_s^{\infty}=\tilde{D}\left|\frac{\dot{G}}{G}\right|^{2/7},
\end{eqnarray}
where the function $\tilde{D}$ is a quantity depending only on the
stellar model and the equation of state. The above formalism can
be applied to constrain the value of $|\dot{G}/G|$, provided one
knows (i) the surface temperature of a neutron star, and (ii) that
the star is certainly older than the time-scale necessary to reach
a quasi-stationary state. So far the only identified object
satisfying these conditions is the closest millisecond pulsar to
our solar system PSR J037-4715. Its surface temperature was
deduced from ultraviolet observations \cite{Kar04} while its mass
was measured to be in the range of $M=(1.40-1.76)M_{\odot}$
\cite{Str01}. To constrain the value of $|\dot{G}/G|$ one,
therefore, needs to consider neutron-star models in the above mass
range and calculate the surface temperature for each stellar
configuration. Clearly, predictions of the surface temperature
and, in turn, value of $|\dot{G}/G|$ depend heavily on the EOS of
neutron-star matter since the later is crucial for determining the
neutron-star structure. If both the direct and modified URCA
processes are allowed, then the calculations become much more
complicated and have to be carried out numerically as done by
Krastev and Li \cite{Kra07a}.

\subsubsection{Constraining the changing rate of the gravitational
constant G using terrestrial nuclear laboratory data}

\begin{figure}[h]
\centering
\includegraphics[totalheight=2.6in]{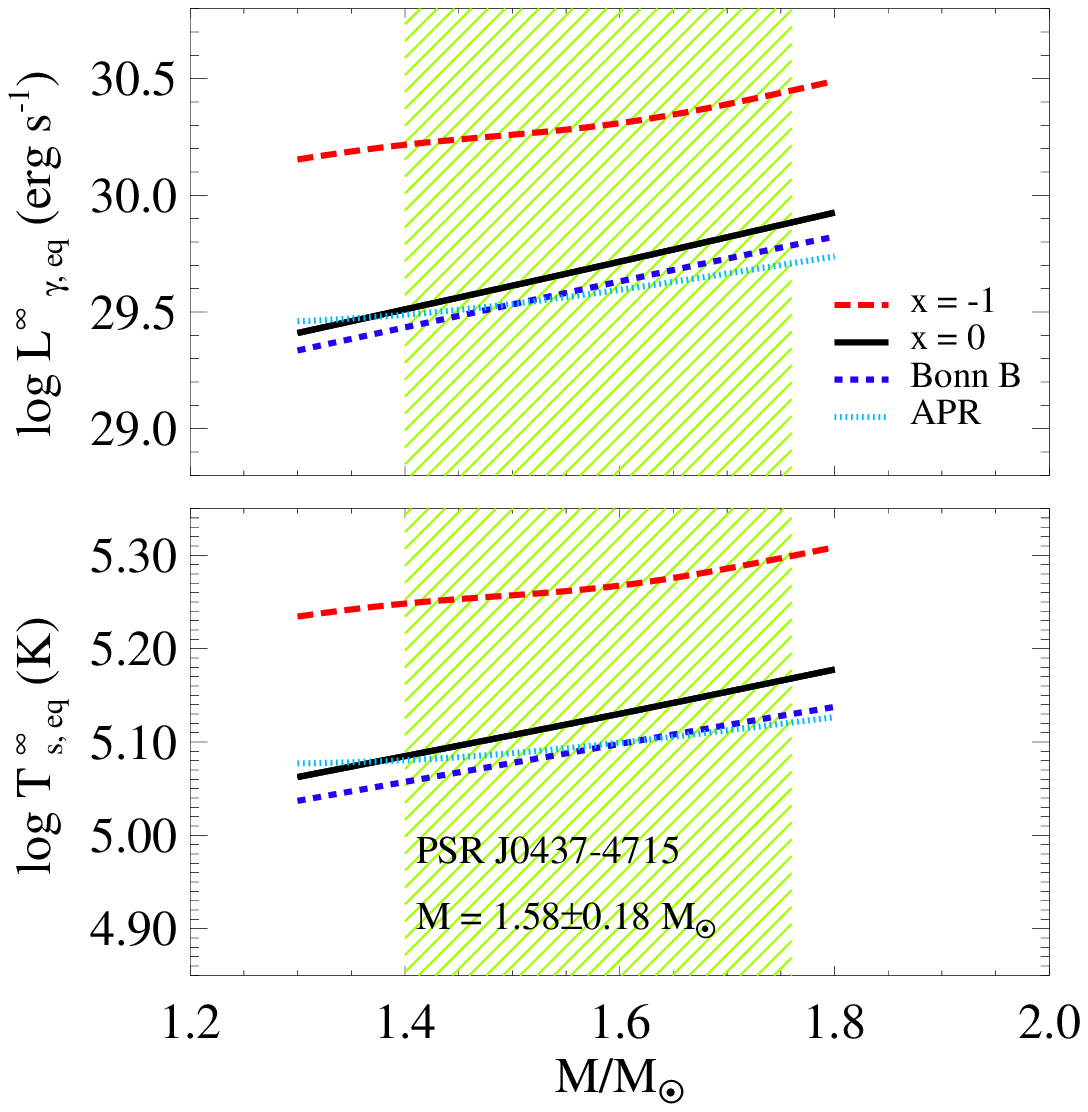}
\includegraphics[totalheight=2.6in]{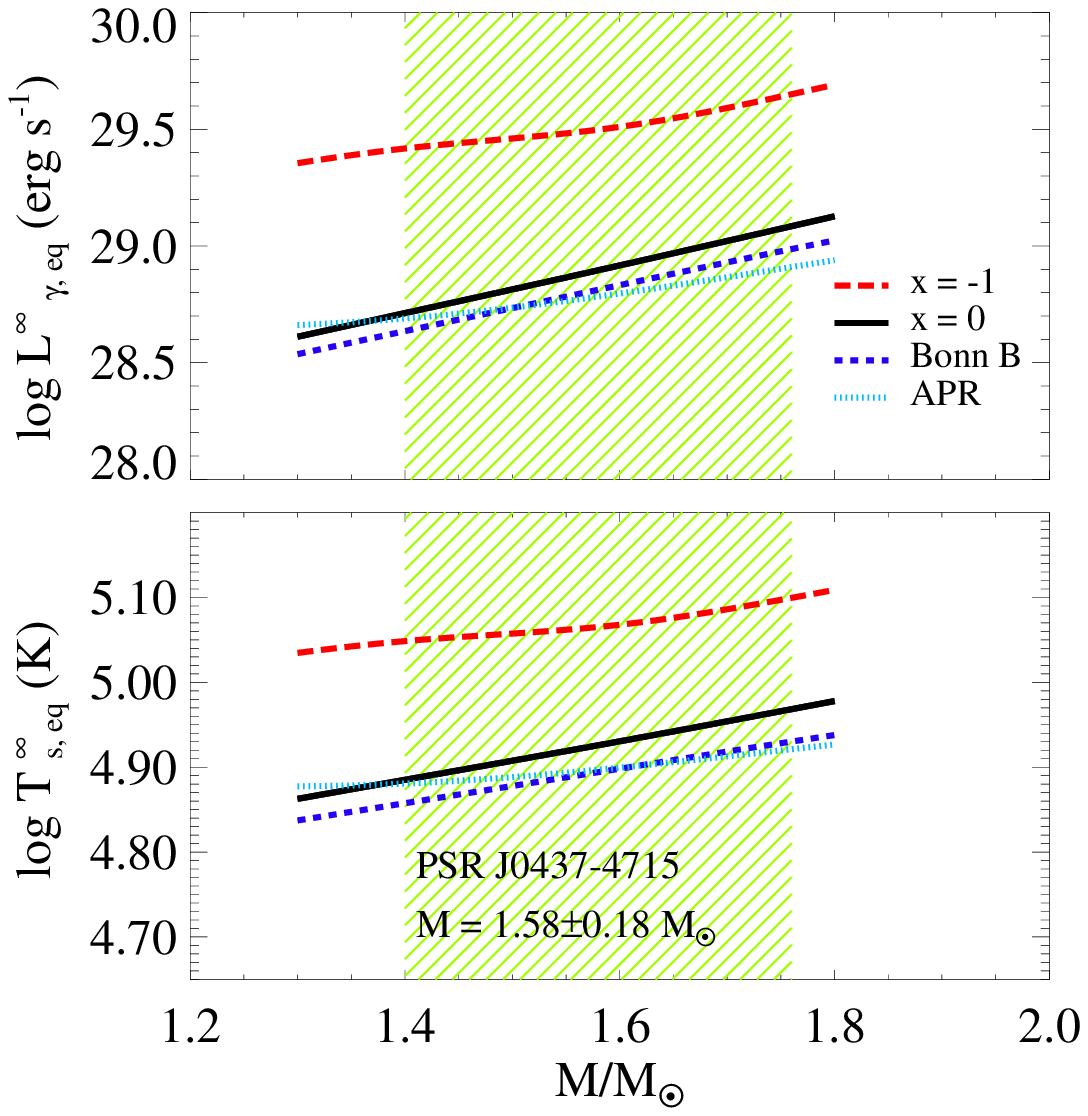}
\vspace{5mm} \caption{{\protect\small Stationary photon luminosity
(upper panel) and neutron star surface temperature (lower panel)
as functions of stellar mass, assuming $|\dot{G}/G|=4\times
10^{-12}~{\rm yr}^{-1}$ (left window) and $8\times 10^{-12}~{\rm
yr}^{-1}$ (right window). The shaded region corresponds to the
mass constraint form van Straten {\it et al.} \cite{Str01}. Taken
from Ref. \cite{Kra07a}.}} \label{GLT_mf}
\end{figure}

\begin{figure}[h]
\centering
\includegraphics[totalheight=1.9in]{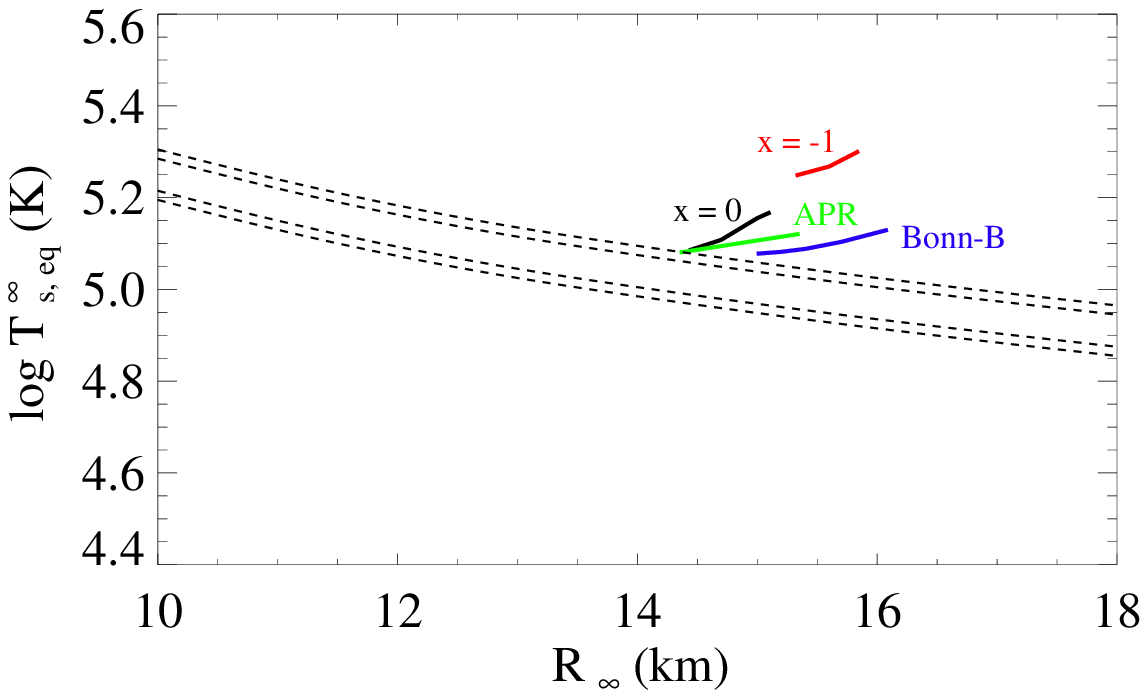}
\includegraphics[totalheight=1.9in]{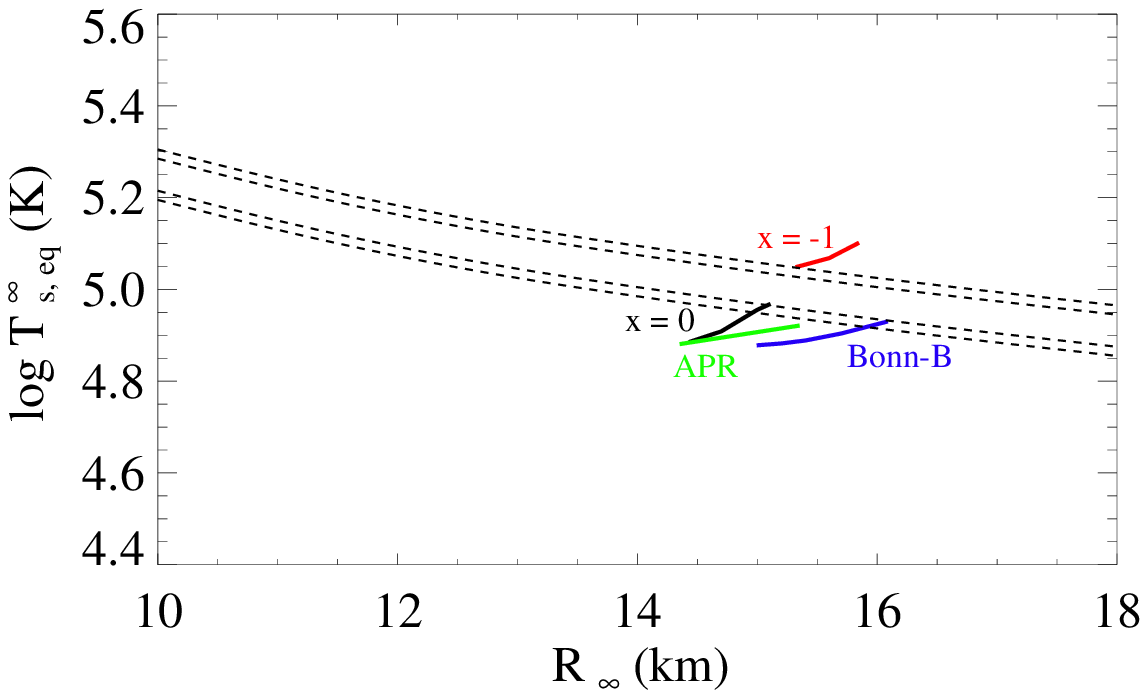}
\vspace{2mm} \caption{{\protect\small (Color online) Neutron star
stationary surface temperature for stellar models satisfying the
mass constraint by van Straten {\it et al.} \cite{Str01}. The
solid lines are the predictions versus the stellar radius for the
considered neutron star sequences. Dashed lines correspond to the
68\% and 90\% confidence contours of the black-body fit of
Kargaltsev {\it et al.} \cite{Kar04}. Values of $|\dot{G}/G|$ are
chosen to be $4\times 10^{-12}~{\rm yr}^{-1}$ (left window) and
$8\times 10^{-12}~{\rm yr}^{-1}$ (right window) so that
predictions from the $x=0$ EOS are just above the observational
constraints. Taken from Ref. \cite{Kra07a}.}} \label{GT_rf}
\end{figure}

Shown in the left window of Fig.~\ref{GLT_mf} are the neutron star
stationary photon luminosity (upper panel) and the steady surface
temperature (lower panel) versus the stellar mass, as computed from
Eq.~(\ref{lum}) assuming only the modified URCA processes are active
and using $|\dot{G}/G|=4\times 10^{-12}~{\rm yr}^{-1}$. The value of
$\dot{G}$ is chosen so that predictions from the $x=0$ EOS are just
above the 90\% confidence contour of Kargaltsev {\it et al.}
\cite{Kar04} (see left window of Fig.~\ref{GT_rf}). This upper limit
agrees exactly with the one by Jofr\'{e} {\it at al.} \cite{Jof06}
under the same assumptions. One notices that predictions from the
$x=0$, APR and Bonn B EOSs all lie just above this observational
constraint, with those from the $x=0$ and APR EOSs being very
similar to each other because they have very similar symmetry
energies up to several times normal nuclear matter density as shown
in Fig.~\ref{MnsProFracEsymRho}. The right windows of
Figs.~\ref{GLT_mf} and \ref{GT_rf} display predictions assuming
$|\dot{G}/G|=8\times 10^{-13}~{\rm yr}^{-1}$. In this case the value
of $\dot{G}$ is chosen so that predictions from the $x=-1$ EOS are
just above the observational constraints at the 90\% confidence
level. Although this value is among the most restrictive results
available in the literature \cite{Uza03}, the above analytic
expression used to calculate the stationary photon luminosity
$L_{\gamma,eq}^{\infty}$ (and in turn surface temperature
$T_s^{\infty}$) becomes a very poor approximation if the direct URCA
channels open in the neutron star core. In fact, the direct URCA
channels do happen easily for stellar models constructed from the
$x=-1$ EOS (see Fig.~\ref{MnsProFracEsymRho}).

\begin{figure}[h]
\centering
\includegraphics[totalheight=1.8in]{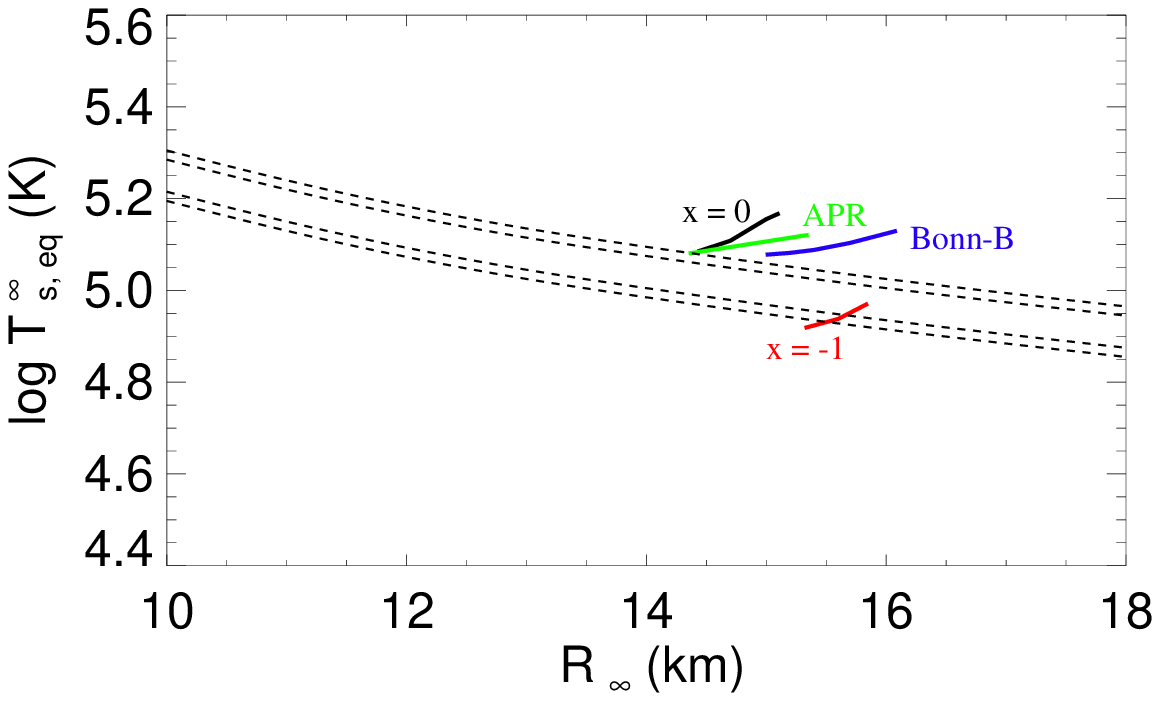}
\includegraphics[totalheight=1.8in]{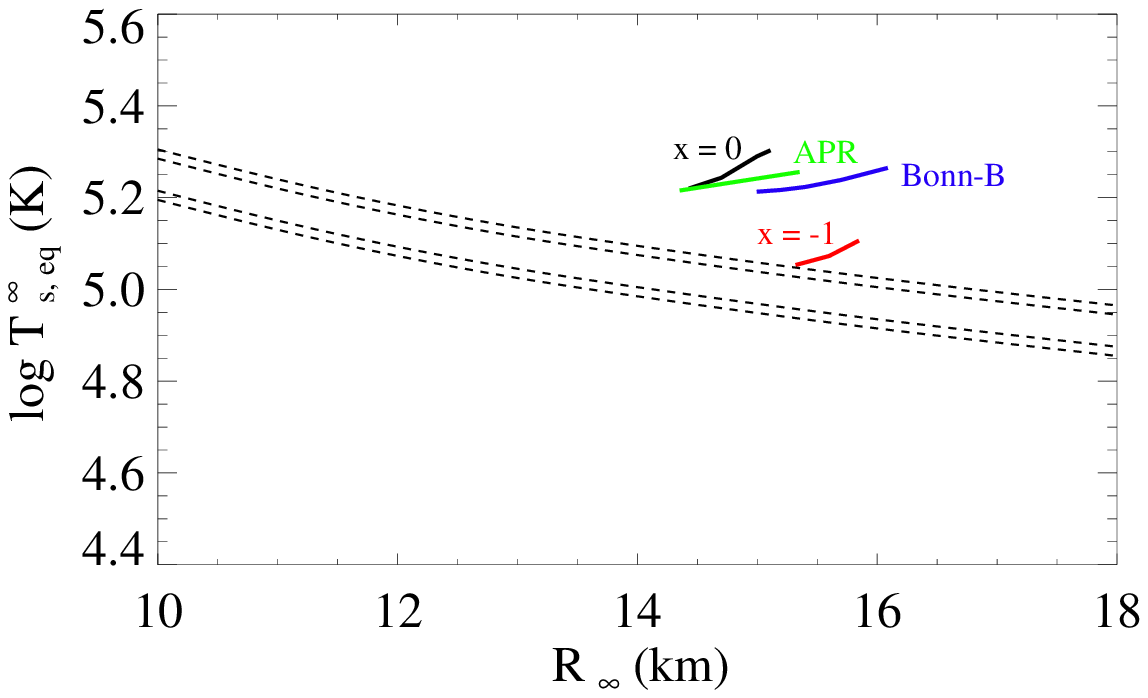}
\vspace{2mm} \caption{(Color online) Same as Fig.\ \ref{GT_rf} but
including the direct URCA channels for $x=-1$. Values of
$|\dot{G}/G|$ are chosen to be $4\times 10^{-12}~{\rm yr}^{-1}$
(left window) and $21\times 10^{-12}~{\rm yr}^{-1}$ (right
window). Taken from Ref. \cite{Kra07a}.} \label{T_r1}
\end{figure}
When the neutron star mass becomes large enough for the central
density to exceed the direct URCA threshold, the surface temperature
is expected to drop abruptly, due to the faster cooling. In this
case the thermal evolution Eqs. (\ref{geq1}) and (\ref{geq3}) have
to be solved numerically. Shown in Fig.\ \ref{T_r1} is the new
result including the direct URCA channels for $x=-1$ and
$|\dot{G}/G|=4\times 10^{-12}~{\rm yr}^{-1}$. Comparing with Fig.\
\ref{GT_rf}, it is seen that indeed the direct URCA channels for
$x=-1$ lead to a dramatic drop in the surface temperature. To set an
upper limit of $|\dot{G}/G|$ such that the new surface temperature
calculated with $x=-1$ including the direct URCA is just above the
90\% confidence level, a significantly higher value of
$|\dot{G}/G|=21\times 10^{-12}~{\rm yr}^{-1}$ has to be used as
shown in the right window of Fig.\ \ref{T_r1}. Based on the above
results and considering the fact that the parameter $x$ is
constrained between 0 and $-1$ by the terrestrial nuclear laboratory
data, one can conclude that an upper bound of the $|\dot{G}/G|$ is
between 4 to $21\times 10^{-12}~{\rm yr}^{-1}$. This constraint is
relatively tight compared to other estimates listed in Table
\ref{Gtable}.


\section{Summary and outlook}\label{chapter_summary}

The field of isospin physics with heavy-ion reactions has witnessed
many exciting new developments over the last few years. We have
reviewed the recent progress in several selected areas of isospin
physics that we are familiar with. Undoubtedly, because of the
limitation of our knowledge the review is incomplete and our
opinions may be considered biased by some experts. But we have made
our best efforts in good faith to minimize possible mistakes.
Naturally, we have concentrated mostly on our own relevant
contributions to this fast growing new field that result from the
hard work of many people in the community.

The heavy-ion reaction community has achieved a great deal in
isospin physics. In our opinion, the most important achievements
include but are not limited to

\begin{itemize}
\item Based on several complementary approaches to heavy-ion reactions, a
symmetry energy of $E_{\rm sym}(\rho )\approx 31.6(\rho /\rho
_{0})^{\gamma }$ with $\gamma =0.69-1.05$ was extracted for
densities between $0.1\rho_0$ and $1.2\rho_0$.

\item The isospin asymmetric part of the isobaric
incompressibility was determined to be $K_{\mathrm{asy}}=-500\pm 50$
MeV, while the slope parameter of the symmetry energy at normal
density was found to be $L=88\pm 25$ MeV.

\item At extremely low densities below $0.05\rho_0$, nuclear clustering
is important. The predicted symmetry energy for the $np\alpha$
matter using the virial expansion method was verified by heavy-ion
reaction experiments.

\item The evolution of the symmetry energy with excitation energy
and impact parameter was observed in the isoscaling analyses of
heavy-ion reactions. It was found that the evolution was mainly due
to the variation of the freeze-out density rather than temperature
of the fragmenting sources.

\item The conclusions about the symmetry energy have important
implications on nuclear effective interactions and the nuclear
many-body theories. For instance, it was found that a large number
of the Skyrme interactions and the RMF Lagrangians lead to symmetry
energies outside the experimental constraints.

\item Some important astrophysical implications of the above
conclusions about the symmetry energy have also been examined. For
instance, nuclear constraints on the mass-radius relationship of
neutron stars, properties of very fast pulsars and the changing rate
of the gravitational constant G were obtained for the first time.

\item Several isospin-related new phenomena were observed
in heavy-ion reaction experiments. These include the isoscaling of
nuclear fragments, isospin fractionation during the liquid-gas phase
transition in asymmetric matter as well as isospin non-equilibrium
and diffusion.

\item Several new probes of the symmetry energy/potential at both
sub-saturation and supra-saturation densities were predicted,
mostly, based on transport models. Moreover, some new phenomena in
heavy-ion reactions, such as the differential isospin fractionation,
were also predicted.

\item Chemical, mechanical and thermal properties of hot
neutron-rich matter were also studied in more detail. In particular,
the symmetry energy and the isovector potential at finite
temperatures were studied systematically using mostly
thermodynamical models. Several new features of the liquid-gas phase
transition in neutron-rich matter, especially their dependence on
the momentum dependence of the isovector potential, were also
predicted.

\item Some new features about the nuclear mean-free-path and
nucleon-nucleon cross sections in neutron-rich matter were also
predicted. Moreover, based on transport model simulations
proposals were also made on how to experimentally test these
predictions.

\end{itemize}

Although considerable progress has been made in isospin physics with
heavy-ion reactions, there are still many very challenging and
exciting problems to be solved. Among the most important theoretical
challenges, we notice the following:

\begin{itemize}

\item The high density behavior of the nuclear symmetry energy

\item The momentum dependence of the isovector potential and the
associated neutron-proton effective mass splitting in asymmetric matter

\item The isospin-dependence of the in-medium nucleon-nucleon cross
sections in asymmetric matter

\item The development of practically implementable quantum transport model
with dynamical formation of clusters for nuclear reactions involving
rare isotopes

\end{itemize}

As we have discussed in detail earlier, microscopic model
calculations are extremely important. However, the results so far
have been very model dependent. Besides the theoretical problems,
the progress on the topics listed above is hindered by the lack of
relevant experimental data. For instance, a number of probes of
the high density behavior of the symmetry energy have been
predicted. However, there is so far very little data available. On
the other hand, there are also some interesting isospin-related
phenomena that are not fully understood because of the lack of
appropriate theoretical tools. The isospin degree of freedom plays
important roles in the reaction dynamics. However, many features
of the reactions involving rare isotopes need to be better
understood theoretically. Only then, one can extract from the
isospin-related phenomena relevant physics that may help us solve
many existing problems in the field. Moreover, to make further
progress in isospin physics with heavy-ion reactions, we also need
significantly better knowledge on the isovector potential at
normal density, i.e., the Lane potential, especially its energy
dependence, that can be obtained from nucleon-nucleus scatterings
and/or (p,n) charge exchange reactions. Furthermore, more reliable
data on neutron-proton bremsstrahlung will allow us to use
confidently the hard photons in heavy-ion reactions as the most
clean probe of the symmetry energy at supra-normal densities. If
we were asked to identify the single most important theoretical
question to be solved urgently before major new progress can be
made, it would be the momentum and density dependence of the
isovector potential.

Given all the challenges mentioned above, there are great
opportunities. Especially, with the development of more advanced
radioactive beams up to a few GeV/nucleon incident energies, we are
hopeful that most of the predictions on the high density behavior of
the symmetry energy and the high energy behavior of the symmetry
potential will be tested soon. Moreover, progress is also being made
with other approaches/fields. Thus, combining measurements of the
neutron skin of $^{208}{\rm Pb}$ at the Jefferson National
Laboratory, more refined observations of neutron stars with advanced
x-ray satellites and heavy-ion reactions will ultimately allow us to
constrain consistently the isovector nuclear effective interaction
and the EOS of neutron-rich matter over a broad density range. We
stress that essentially all progress in isospin physics with
heavy-ion reaction was made as a result of the close collaborations
between experimentalists and theorists. To make further progress and
meet the new challenges, the continuation of this practice across
several sub-fields of nuclear physics and astrophysics are
essential.


\section{Acknowledgements}

We are very grateful to Wolfgang Bauer, Pawel Danielewicz, Champak
B. Das, Subal Das Gupta, Massimo Di Toro, Charles Gale, Vincenzo
Greco, Wei-Zhou Jiang, Gen-Ming Jin, Plamen Krastev, Zeng-Hua Li,
Umberto Lombardo, Bill Lynch, Hong-Ru Ma, Joe Natowitz, W. Udo
Schr\"oder, Andrew Steiner, Andy Sustich, Betty Tsang, Gary
Westfall, Aaron Worley, Jun Xu, Sherry J. Yennello, Gao-Chan Yong,
Bin Zhang, Feng-Shou Zhang, Zhi-Yuan Zhu and Wei Zuo for
collaborating with us on some of the topics discussed in this
review. The work was supported in part by the US National Science
Foundation under Grant No. PHY-0652548, PHY-0457265, the Research
Corporation under Award No. 7123, the Welch Foundation under Grant
No. A-1358, the National Natural Science Foundation of China under
Grant Nos. 10575071 and 10675082, MOE of China under project
NCET-05-0392, Shanghai Rising-Star Program under Grant No.
06QA14024, the SRF for ROCS, SEM of China, and the National Basic
Research Program of China (973 Program) under Contract No.
2007CB815004.


\end{document}